\documentclass[11pt,a4paper,titlepage,twoside,openany,hidelinks]{book}

\usepackage{amsmath}
\usepackage{amsbsy}
\usepackage{bm}
\usepackage{tensor}

\DeclareMathAlphabet{\mathsfit}{\encodingdefault}{\sfdefault}{m}{sl}
\SetMathAlphabet{\mathsfit}{bold}{\encodingdefault}{\sfdefault}{bx}{n}

\usepackage{booktabs}
\usepackage{forest}

\usepackage{url}

\usepackage{outlines}
\usepackage{multirow}
\usepackage{float} 
\usepackage[normalem]{ulem}
\usepackage{subcaption}
\usepackage{multirow}
\usepackage{booktabs}
\usepackage{tabularx}
\usepackage{multirow}
\usepackage{makecell}
\usepackage{framed,enumitem} 
\usepackage{amsmath,amssymb}
\usepackage{lscape} 

\usepackage{makecell} 

\usepackage{algorithm}%
\usepackage{algorithmicx}%
\usepackage{algpseudocode}%
\usepackage{graphicx}
\usepackage{amssymb} %
\usepackage{bbding}
\usepackage{wrapfig}
\usepackage{adjustbox}

\usepackage{array}
\usepackage{rotating}

\newcommand{\RNum}[1]{\uppercase\expandafter{\romannumeral #1\relax}}

\newcolumntype{C}[1]{>{\centering\arraybackslash}m{#1}}
\newcolumntype{R}[1]{>{\raggedleft\arraybackslash}m{#1}}
\newcolumntype{L}[1]{>{\raggedright\arraybackslash}m{#1}}

\usepackage{listings}

\usepackage{listings}

\usepackage{verbatim}

\usepackage[final,babel]{microtype}

\usepackage{emptypage}
\usepackage[pass]{geometry}
\usepackage{ebgaramond}
\usepackage[T1]{fontenc}
\usepackage[ngerman, greek,english]{babel}
\usepackage[utf8]{inputenc}
\usepackage{standalone}
\usepackage[colon,round,authoryear]{natbib}
\usepackage{bibentry}
\usepackage{graphicx}
\usepackage{latexsym}
\usepackage{amsmath}
\usepackage{amssymb}
\usepackage{multirow}
\usepackage{tabularx}
\usepackage{pbox}
\usepackage[commandnameprefix=always]{changes}
\usepackage{booktabs}
\usepackage{multicol}
\usepackage{bbm}
\usepackage{colortbl}
\usepackage{ulem}
\usepackage{enumitem}
\usepackage{amsthm}
\usepackage{microtype}
\usepackage{comment}
\usepackage{subcaption}
\usepackage{tipa}
\usepackage[hyperfootnotes=false]{hyperref}
\usepackage[nameinlink,capitalise]{cleveref}
\usepackage{listings}
\usepackage{makecell}
\usepackage{arydshln}
\usepackage{xspace}
\usepackage{teubner}
\usepackage{syllogism}
\usepackage{sectsty}
\usepackage{tcolorbox}
\usepackage{fancyhdr}
\usepackage{epigraph}
\usepackage[acronym, nomain, toc, automake,  sort=standard]{glossaries}
\usepackage[section]{placeins}

\usepackage{rotating}
\usepackage{tocloft}  

\usepackage{dcolumn}
\newcolumntype{d}[1]{D{.}{.}{#1}}
\makeatletter
\newcolumntype{B}[3]{>{\boldmath\DC@{#1}{#2}{#3}}c<{\DC@end}}
\makeatother

\usepackage{inconsolata}

\makeatletter
\newcommand{\srcsize}{\@setfontsize{\srcsize}{5pt}{5pt}}
\makeatother

\makeatletter
\newcommand{\mathsize}{\@setfontsize{\mathsize}{0.5pt}{0.5pt}}
\makeatother

\definecolor{Gray}{gray}{0.95}
\definecolor{GreenRegion}{RGB}{0,160,139}
\definecolor{EncoderBlue}{RGB}{67,114,196}
\definecolor{DecoderOrange}{RGB}{238,124,50}
\newcolumntype{g}{>{\columncolor{Gray}}c}

\newlength{\EqNumInset}
\newcommand{\coleq}[2]{
  \refstepcounter{equation}\label{#1}%
  \vspace{\abovedisplayskip}\noindent
  \makebox[\linewidth]{%
    \hfil$\displaystyle #2$\hfil
    \llap{\hspace{\EqNumInset}\normalfont(\theequation)}%
  }%
  \par\vspace{\belowdisplayskip}%
}
\makeglossaries

\newcounter{savefootnote}
\newcounter{symfootnote}
\newcommand{\symfootnote}[1]{%
   \setcounter{savefootnote}{\value{footnote}}%
   \setcounter{footnote}{\value{symfootnote}}%
   \ifnum\value{footnote}>8\setcounter{footnote}{0}\fi%
   \let\oldthefootnote=\thefootnote%
   \renewcommand{\thefootnote}{\fnsymbol{footnote}}%
   \footnote{#1}%
   \let\thefootnote=\oldthefootnote%
   \setcounter{symfootnote}{0}%
   \setcounter{footnote}{\value{savefootnote}}%
}

\newcommand{\symfootnotemark}{%
   \setcounter{savefootnote}{\value{footnote}}%
   \setcounter{footnote}{\value{symfootnote}}%
   \ifnum\value{footnote}>8\setcounter{footnote}{0}\fi%
   \let\oldthefootnote=\thefootnote%
   \renewcommand{\thefootnote}{\fnsymbol{footnote}}%
   \footnotemark[1]%
   \let\thefootnote=\oldthefootnote%
   \setcounter{symfootnote}{0}%
   \setcounter{footnote}{\value{savefootnote}}%
}

\addto\extrasenglish{%
}
\crefname{paragraph}{\S}{\S\S}
\crefname{subsection}{\S}{\S\S}

\newcommand{\myName}{Kristian Kolthoff}
\newcommand{\myTitle}{From Natural Language Requirements to Graphical User Interfaces: Automated Prototyping and Verification with Pretrained Language Models}

\hypersetup{%
    colorlinks=true, linktocpage=true, pdfborder={0 0 0}, 
    breaklinks=true, pdfpagemode=UseNone, pageanchor=true, pdfpagemode=UseOutlines,%
    plainpages=false, bookmarksnumbered,   bookmarksopenlevel=0, bookmarksopen=true,
    hypertexnames=true, pdfhighlight=/O,
    urlcolor=black, linkcolor=black, citecolor=black, 
    pdftitle={\myTitle},%
    pdfauthor={\myName}
    pdfproducer={LaTeX with hyperref and classicthesis}
} 

\chapterfont{\textsc}

\renewcommand{\chaptermark}[1]{%
\markboth{\thechapter.\ #1}{}}
\usepackage{caption, setspace}
\begin{document}
\normalem
\pagenumbering{roman}
\newgeometry{centering}
\hypersetup{pageanchor=false}

\newcommand{\titlespacing}{\vspace{3cm}}

\begin{titlepage}
  \vspace*{2cm}
  \begin{center}
    {\fontsize{17.2}{19}\linespread{1.4}\selectfont \textsc{\myTitle}\\}
    
    \titlespacing
    
    {D i s s e r t a t i o n\\
    zur Erlangung des Doktorgrades\\
    der Naturwissenschaften\\}
    
    \vspace{4cm}
    
    {vorgelegt von\\
    \vspace{0.0cm}
    \Large \textsc{\myName}\\
    \vspace{0.1cm}
    \large \textnormal{aus Nagold}\\}
    
    \vspace{1cm}
    
    {genehmigt von der Fakultät für\\
    Naturwissenschaften, Mathematik und Informatik\\
    der Technischen Universität Clausthal\\}
    
    \vspace{1cm}
    
    {Tag der mündlichen Prüfung\\
    29.04.2026}
    
  \end{center}
\end{titlepage}

\restoregeometry
\hypersetup{pageanchor=true}

\clearpage
\addtocounter{page}{1}  
\begin{table}[b]
\vspace{17.5cm}



\begin{tabular}{@{}ll@{}}
\textbf{Dekan} & Prof. Dr. René Wilhelm, TU Clausthal\\
\textbf{\begin{tabular}[b]{@{}l@{}}
Vorsitzender der\\
Promotionskommission
\end{tabular}}
& Prof. Dr. Rüdiger Ehlers, TU Clausthal\\
\textbf{Betreuer} & Prof. Dr. Christian Bartelt, TU Clausthal\\
\textbf{Gutachter} & Prof. Dr. Jörg P. Müller, TU Clausthal\\
\textbf{Gutachter} & Prof. Dr. Simone Paolo Ponzetto, Universität Mannheim\\
\end{tabular}

\end{table}
\chapter*{Acknowledgements}
\setstretch{1.02}

Throughout the course of my doctoral journey, I have received invaluable support, guidance and encouragement from many individuals, for which I am deeply grateful.

First and foremost, I wish to express my sincere gratitude to my two supervisors, Prof. Dr. Christian Bartelt and Prof. Dr. Simone Paolo Ponzetto. From the beginning of my doctoral journey, Christian contributed substantially to shaping my research direction through countless hours of engaging discussions, provided continuous guidance and support, and offered steady encouragement, particularly during challenging periods. Beyond his significant influence on my academic development, Christian's mentorship also played an important role in my personal growth, for which I am sincerely thankful. Simone likewise supported my research journey in many meaningful ways through insightful discussions, consistently constructive feedback and encouragement, which profoundly improved the quality of my research and this thesis.

I am also deeply grateful to Prof. Dr. Alexander Mädche from the Karlsruhe Institute of Technology for the inspiring collaboration and insightful feedback over the final two years of my doctoral journey. In particular, I wish to express my sincere gratitude to Dr. Felix Kretzer for the close collaboration during this period, which had a substantial impact on this thesis. I greatly appreciated both the professional and personal exchange, and I am especially thankful for his remarkable commitment and perseverance during demanding phases marked by tight deadlines, late-night work and countless discussions.

I am also grateful to my friends and former colleagues at the Institute for Enterprise Systems, University of Mannheim, and my friends and colleagues at the Clausthal University of Technology, Institute for Software and Systems Engineering, for their contributions to this thesis through insightful discussions and diverse perspectives.

I would also like to acknowledge the many researchers I encountered over the years at conferences for the insightful discussions, inspiring perspectives and valuable exchanges. These interactions helped broaden my perspective and refine many of my research ideas.

I am also deeply grateful to my parents, Cornelia Kolthoff and Dr. med. Dr. phil. Helmut Kolthoff, whose constant support and encouragement throughout my life provided the foundation for my academic journey and made this work possible.

Lastly, my deepest gratitude goes to my family -- my wonderful wife Yaru, and my son, Lio -- for your unconditional love, understanding and endless support, which carried me through every stage of this journey. Without your love, constant encouragement and many sacrifices, pursuing a PhD and completing this work would not have been possible. 
\cleardoublepage

\newcommand{\challone}{\textbf{C1}}
\newcommand{\challtwo}{\textbf{C2}}
\newcommand{\challoneone}{\textbf{C1-1}}
\newcommand{\challonetwo}{\textbf{C1-2}}
\newcommand{\challonethree}{\textbf{C1-3}}
\newcommand{\challonefour}{\textbf{C1-4}}
\newcommand{\challonefive}{\textbf{C1-5}}
\newcommand{\challtwoone}{\textbf{C2-1}}
\newcommand{\challtwotwo}{\textbf{C2-2}}

\selectlanguage{english}
\chapter*{Abstract}
Requirements elicitation is a crucial activity before and during the development of interactive software systems, ensuring that the resulting product meets the needs of the stakeholders. Typically, the process of requirements elicitation relies heavily on natural language (NL) communication between requirements analysts and stakeholders, being prone to misunderstandings due to the inherent ambiguity of NL. While formal software specifications serve as ambiguity mitigation, they necessitate technical knowledge for proper interpretation. Therefore, Graphical User Interface (GUI) prototyping has emerged as an effective requirements elicitation and validation technique, providing tangible, visual artifacts representing the implemented requirements and acting as a common language between analysts and stakeholders. In particular, employing GUI prototyping improves the elicitation and validation of requirements by enabling fruitful discussions and integrating stakeholders more closely into the development process. Although GUI prototyping offers these advantages in comparison to other elicitation and validation approaches, it simultaneously necessitates increased manual effort, which is time-consuming and costly, especially for high-fidelity GUI prototypes that enable stakeholders to provide feedback of higher quality. While requirements validation ensures that the software system aligns with the actual needs of the stakeholders, verification of requirements is equally important, ensuring that the implementation of requirements in the software system and prototypes aligns with the specification. However, similar to validation, verification is still often a manual and effort-demanding task, while existing solutions offer mostly merely rule-based, static techniques and approaches.

In this work, we propose several approaches to address the mentioned challenges, namely (\challone{}) reducing the effort required to transform NL requirements into GUI prototypes to facilitate requirements elicitation and validation, and (\challtwo{}) reducing the effort required for conducting requirements verification in GUI applications and prototypes. Specifically, we tackle the first challenge (\challone{}) by initially presenting novel NL-based GUI retrieval and reranking approaches, enabling to rapidly map NL requirements to tangible GUI prototypes, while creating new benchmarks and achieving state-of-the-art performance. Furthermore, we propose several methods for efficiently adapting LLMs to GUI generation and devise novel techniques to efficiently enable LLM-based generation for proprietary GUI representations, allowing to quickly translate NL requirements into customizable GUI prototypes. We evaluated these techniques on a novel benchmark encompassing an extensive collection of human annotations created via crowdsourcing, showing their high effectiveness for GUI generation. To tackle the second challenge (\challtwo{}), we initially present novel LLM-based methods enabling the verification of semantically complex NL requirements on static GUI prototypes. To assess their performance, we created a high-quality benchmark, which links NL requirements to GUI prototypes, indicating high effectiveness for requirements verification on GUI prototypes. Moreover, we propose a multi-modal (M)LLM-based agent approach that enables verification of complex functional and non-functional NL requirements of fully-fledged, highly dynamic GUI prototypes and applications by automatically creating and evaluating meaningful interaction trajectories, showing high effectiveness on a novel benchmark of human-annotated NL requirements and GUI applications. To summarize, by providing effective methods for rapidly mapping NL requirements to GUI prototypes, we facilitate automated GUI prototyping and substantially increase prototyping productivity, positively impacting elicitation and validation activities. Moreover, with our novel and effective MLLM-based agent approach, we substantially improve and facilitate the automated verification of complex NL requirements in GUIs.
 
\selectlanguage{ngerman}

\chapter*{Zusammenfassung}
Die Erhebung von Anforderungen ist eine entscheidende Tätigkeit vor und während der Entwicklung interaktiver Softwaresysteme und stellt sicher, dass das resultierende Produkt den Bedürfnissen der Stakeholder entspricht. Typischerweise stützt sich der Prozess der Anforderungserhebung stark auf die Kommunikation in natürlicher Sprache (NL) zwischen Anforderungsanalysten und Stakeholdern und ist aufgrund der inhärenten Mehrdeutigkeit der natürlichen Sprache anfällig für Missverständnisse. Während formale Softwarespezifikationen zur Reduzierung von Mehrdeutigkeiten dienen, erfordern sie technisches Fachwissen für eine korrekte Interpretation. Daher hat sich das Prototyping grafischer Benutzeroberflächen (Graphical User Interface (GUI)) als effektive Technik zur Erhebung und Validierung von Anforderungen etabliert, da es greifbare, visuelle Artefakte bereitstellt, die die implementierten Anforderungen repräsentieren und als gemeinsame Sprache zwischen Analysten und Stakeholdern fungieren. Insbesondere verbessert der Einsatz von GUI-Prototyping die Erhebung und Validierung von Anforderungen, indem er fruchtbare Diskussionen ermöglicht und Stakeholder enger in den Entwicklungsprozess einbindet. Obwohl GUI-Prototyping diese Vorteile gegenüber anderen Erhebungs- und Validierungsansätzen bietet, erfordert es gleichzeitig einen erhöhten manuellen Aufwand, der zeitaufwendig und kostspielig ist – insbesondere bei hochausgereiften GUI-Prototypen, die es Stakeholdern ermöglichen, qualitativ hochwertigeres Feedback zu geben. Während die Anforderungsvalidierung sicherstellt, dass das Softwaresystem den tatsächlichen Bedürfnissen der Stakeholder entspricht, ist die Verifikation von Anforderungen ebenso wichtig, da sie gewährleistet, dass die Umsetzung der Anforderungen im Softwaresystem und in Prototypen mit der Spezifikation übereinstimmt. Allerdings ist die Verifikation – ähnlich wie die Validierung – nach wie vor häufig eine manuelle und aufwändige Aufgabe, während bestehende Lösungen meist lediglich regelbasierte, statische Techniken und Ansätze bieten.

In dieser Arbeit schlagen wir mehrere Ansätze vor, um die genannten Herausforderungen zu adressieren, nämlich (\challone{}) den Aufwand zur Transformation von NL-An\-for\-de\-run\-gen in GUI-Prototypen zu reduzieren, um die Erhebung und Validierung von Anforderungen zu erleichtern, und (\challtwo{}) den Aufwand für die Durchführung der Anforderungsverifikation in GUI-Anwendungen und -Prototypen zu verringern. Konkret adressieren wir die erste Herausforderung (\challone{}), indem wir zunächst neuartige NL-basierte Ansätze zur GUI-Suche und -Neubewertung (Reranking) vorstellen, die eine schnelle Abbildung von NL-Anforderungen auf greifbare GUI-Prototypen ermöglichen, während wir neue Benchmarks erstellen und eine Leistungsfähigkeit auf dem Stand der Technik erreichen. Darüber hinaus schlagen wir mehrere Methoden zur effizienten Anpassung großer Sprachmodelle (LLMs) an die GUI-Generierung vor und entwickeln neuartige Techniken, um die LLM-basierte Generierung von proprietären GUI-Repräsentationen effizient zu ermöglichen, sodass NL-Anforderungen schnell in anpassbare GUI-Prototypen übersetzt werden können. Wir evaluierten diese Techniken anhand eines neuartigen Benchmarks, der eine umfangreiche Sammlung von durch Crowdsourcing erstellten menschlichen Annotationen umfasst, und zeigen deren hohe Effektivität für die GUI-Generierung. Zur Bewältigung der zweiten Herausforderung (\challtwo{}) präsentieren wir zunächst neuartige LLM-basierte Methoden, die die Verifikation semantisch komplexer NL-Anforderungen auf statischen GUI-Prototypen ermöglichen. Zur Bewertung ihrer Leistungsfähigkeit erstellten wir einen hochwertigen Benchmark, der NL-Anforderungen mit GUI-Prototypen verknüpft und eine hohe Effektivität für die Anforderungsverifikation auf GUI-Prototypen zeigt. Darüber hinaus schlagen wir einen multimodalen (M)LLM-basierten Agentenansatz vor, der die Verifikation komplexer funktionaler und nicht-funktionaler NL-Anforderungen vollständig ausgereifter, hochdynamischer GUI-Prototypen und -Anwendungen ermöglicht, indem automatisch sinnvolle Interaktionspfade erzeugt und evaluiert werden. Dieser Ansatz zeigt eine hohe Effektivität auf einem neuartigen Benchmark mit von Menschen annotierten NL-Anforderungen und GUI-Anwendungen. Zusammenfassend erleichtern wir durch die Bereitstellung effektiver Methoden zur schnellen Abbildung von NL-Anforderungen auf GUI-Prototypen das automatisierte GUI-Prototyping und steigern die Produktivität des Prototypings erheblich, was sich positiv auf Erhebungs- und Validierungsaktivitäten auswirkt. Darüber hinaus verbessern und erleichtern wir mit unserem neuartigen und effektiven MLLM-basierten Agentenansatz die automatisierte Verifikation komplexer NL-Anforderungen in GUIs erheblich.

\selectlanguage{english}

\tableofcontents
\cleardoublepage
\chapter*{List of Publications}
\markboth{List of Publications}{}
\addcontentsline{toc}{chapter}{List of Publications}
\begingroup
\hypersetup{urlcolor=black}
This thesis is based on work that has been previously published in proceedings of premier international software engineering conferences including the \emph{IEEE/ACM International Conference on Automated Software Engineering} and the corresponding \emph{Automated Software Engineering Journal}, \emph{IEEE/ACM International Conference on Software Engineering}, \emph{ACM CHI Conference on Human Factors in Computing Systems} and the \emph{IEEE International Requirements Engineering Conference}. These publications are annotated with their respective conference rankings (A*, a leading conference in the research area and A, a highly respected conference in the research area) based on the CORE Conference Ranking, which is a popular and widely adopted ranking system for conferences in computer science. The publications encompass text, tables and figures and are listed in chronological order regarding their publication time. Subsequent chapters will reference these publications accordingly. Additionally, the list includes work that has been submitted to top-tier conferences but is currently undergoing peer review, which we denote accordingly.

\vspace{1em}

\noindent\textbf{Kolthoff, Kristian}. Automatic Generation of Graphical User Interface Prototypes from Unrestricted Natural Language Requirements. In \emph{Proceedings of the 34th IEEE/ACM International Conference on Automated Software Engineering (ASE, A*)}, San Diego, CA, USA, January 2020, pages 1234--1237. IEEE.

\vspace{1em}

\noindent\textbf{Kolthoff, Kristian}, Bartelt, Christian, and Ponzetto, Simone Paolo. GUI2WiRe: Rapid Wireframing with a Mined and Large-Scale GUI Repository using Natural Language Requirements. In \emph{Proceedings of the 35th IEEE/ACM International Conference on Automated Software Engineering (ASE, A*)}, Melbourne, Australia (Virtual Event), January 2021, pages 1297--1301. ACM.

\vspace{1em}

\noindent\textbf{Kolthoff, Kristian}, Bartelt, Christian, and Ponzetto, Simone Paolo. Automated Retrieval of Graphical User Interface Prototypes from Natural Language Requirements. In \emph{Proceedings of the 26th International Conference on Applications of Natural Language to Information Systems (NLDB)}, Saarbrücken, Germany (Virtual Event), June 2021, pages 376--384, Cham: Springer International Publishing.

\vspace{1em}

\noindent\textbf{Kolthoff, Kristian}, Bartelt, Christian, and Ponzetto, Simone Paolo. Data-Driven Prototyping via Natural-Language-based GUI Retrieval. \emph{Automated Software Engineering}, March 2023, 30(1), 13, pages 1--34, Springer.

\vspace{1em}

\begingroup
\renewcommand{\thefootnote}{\fnsymbol{footnote}} 

\noindent\textbf{Kolthoff, Kristian\footnotemark[1]}, Kretzer, Felix\footnotemark[1], Bartelt, Christian, Maedche, Alexander, and Ponz-etto, Simone Paolo. Interlinking User Stories and GUI Prototyping: A Semi-Automatic LLM-based Approach. In \emph{Proceedings of the 32nd IEEE International Requirements Engineering Conference (RE, A)}, Reykjavik, Iceland, June 2024, pages 380--388. IEEE.

\footnotetext[1]{Authors contributed equally.}

\endgroup

\vspace{1em}

\noindent\textbf{Kolthoff, Kristian}, Bartelt, Christian, Ponzetto, Simone Paolo, and Schneider, Kurt. Self-Elicitation of Requirements with Automated GUI Prototyping. In \emph{Proceedings of the 39th IEEE/ACM International Conference on Automated Software Engineering (ASE, A*)}, Sacramento, CA, USA, October 2024, pages 2354--2357. ACM.

\vspace{1em}

\noindent Kretzer, Felix\textsuperscript{*}, \textbf{Kolthoff, Kristian\textsuperscript{*}}, Bartelt, Christian, Ponzetto, Simone Paolo, and Maedche, Alexander. Closing the Loop Between User Stories and GUI Prototypes: An LLM-based Assistant for Cross-Functional Integration in Software Development. In \emph{Proceedings of the 2025 CHI Conference on Human Factors in Computing Systems (CHI, A*)}, Yokohama, Japan, April 2025, pages 1--19. ACM.

\vspace{1em}

\noindent\textbf{Kolthoff, Kristian\textsuperscript{*}}, Kretzer, Felix\textsuperscript{*}, Bartelt, Christian, Maedche, Alexander, and Ponz-etto, Simone Paolo. GUIDE: LLM-Driven GUI Generation Decomposition for Automated Prototyping. In \emph{2025 IEEE/ACM 47th International Conference on Software Engineering: Companion Proceedings (ICSE-Companion) (ICSE, A*)}, Ottawa, ON, Canada, April 2025, pages 1--4. IEEE.

\vspace{1em}

\noindent\textbf{Kolthoff, Kristian}, Kretzer, Felix, Bartelt, Christian, Maedche, Alexander, and Ponzetto, Simone Paolo. GUI-ReRank: Enhancing GUI Retrieval with Multi-Modal LLM-based Reranking. In \emph{Proceedings of the 40th IEEE/ACM International Conference on Automated Software Engineering (ASE, A*)}, Seoul, Republic of Korea, January 2026, pages 1--4. ACM. (in press)

\vspace{1em}

\noindent\textbf{Kolthoff, Kristian\textsuperscript{*}}, Kretzer, Felix\textsuperscript{*}, Bartelt, Christian, Maedche, Alexander, and Ponz-etto, Simone Paolo. GUISpector: An MLLM Agent Framework for Automated Verification of Natural Language Requirements in GUI Prototypes. In \emph{Proceedings of the 48th International Conference on Software Engineering: Companion Proceedings (ICSE-Companion)}, April 2026, 1-4, Rio de Janeiro, Brazil. ACM. (in press)

\vspace{1em}
\noindent Furthermore, we list papers that are currently available as ArXiv preprints and are undergoing review at the time of thesis submission.
\vspace{0.4cm}

\noindent\textbf{Kolthoff, Kristian\textsuperscript{*}}, Kretzer, Felix\textsuperscript{*}, Fiebig, Lennart, Maedche, Alexander, Ponzetto, Simone Paolo, and Bartelt, Christian. Zero-Shot Prompting Approaches for LLM-based Graphical User Interface Generation. arXiv preprint arXiv:2412.11328, 2024. The paper is currently under review for the  \emph{2026 ACM Joint European Software Engineering Conference and Symposium on the Foundations of Software Engineering (ESEC/FSE)}.

\vspace{1em}

\endgroup
\cleardoublepage

\begingroup
\hypersetup{citecolor=black}
\addcontentsline{toc}{chapter}{List of Figures}
\listoffigures
\endgroup
\cleardoublepage
\addcontentsline{toc}{chapter}{List of Tables}
\listoftables
\cleardoublepage

\renewcommand*{\glstreenamefmt}[1]{\textbf{#1}\hspace{0.85em}}

\newacronym{nlp}{NLP}{Natural Language Processing}
\newacronym{nlu}{NLU}{Natural Language Understanding}
\newacronym{mlm}{MLM}{Masked Language Modeling}
\newacronym{rs}{RS}{Response Selection}
\newacronym{ds-tod}{DS-TOD}{\textbf{D}omain \textbf{S}pecialization for \textbf{T}ask \textbf{O}riented-\textbf{D}ialog}
\newacronym{plm}{PLM}{Pretrained Language Model}
\newacronym{tod}{TOD}{Task-oriented Dialog}
\newacronym{dst}{DST}{Dialog State Tracking}
\newacronym{rr}{RR}{Response Retrieval}
\newacronym{nn}{NN}{Neural Networks}
\newacronym{rnn}{RNN}{Recurrent Neural Networks}
\newacronym{sg}{SG}{Skip-Gram}
\newacronym{cbow}{CBOW}{Continuous Bag-of-Words}
\newacronym{oov}{OOV}{Out-of-Vocabulary}
\newacronym{bpe}{BPE}{Byte-Pair Encoding}
\newacronym{wpa}{WPA}{Word-Piece Algorithm}
\newacronym{ner}{NER}{Named Entity Recognition}
\newacronym{nli}{NLI}{Natural Language Inference}
\newacronym{mtl}{MTL}{Multi-Task Learning}
\newacronym{spa}{SPA}{Sentence Piece Algorithm}
\newacronym{nsp}{NSP}{Next Sentence Prediction}
\newacronym{xlmr}{XLM-R}{XLM-RoBERTa}
\newacronym{tlm}{TLM}{Translation Language Modeling}
\newacronym{mt}{MT}{machine translation}
\newacronym{gui}{GUI}{Graphical User Interface}
\newacronym{re}{RE}{Requirements Engineering}
\newacronym{rel}{REl}{Requirements Elicitation}
\newacronym{rver}{RVer}{Requirements Verification}
\newacronym{rval}{RVal}{Requirements Validation}
\newacronym{rd}{RD}{Requirements Documentation}
\newacronym{se}{SE}{Software Engineering}
\newacronym{sd}{SD}{Software Development}
\newacronym{st}{ST}{Software Testing}
\newacronym{sver}{SVer}{Software Verification}
\newacronym{sval}{SVal}{Software Validation}
\newacronym{nl}{NL}{Natural Language}
\newacronym{nlr}{NLR}{Natural Language Requirements}
\newacronym{fr}{FR}{Functional Requirements}
\newacronym{nfr}{NFR}{Non-Functional Requirements}
\newacronym{ml}{ML}{Machine Learning}
\newacronym{ir}{IR}{Information Retrieval}
\newacronym{llm}{LLM}{Large Language Model}
\newacronym{slm}{SLM}{Small Language Model}
\newacronym{mllm}{MLLM}{Multi-Modal Large Language Model}
\newacronym{html}{HTML}{Hypertext Markup Language}
\newacronym{css}{CSS}{Cascading Style Sheets}
\newacronym{js}{JS}{JavaScript}
\newacronym{zs}{ZS}{Zero-Shot}
\newacronym{fs}{FS}{Few-Shot}
\newacronym{os}{OS}{One-Shot}
\newacronym{amt}{AMT}{Amazon Mechanical Turk}
\newacronym[
  plural=USs,
  longplural=User Stories,
  firstplural=User Stories (USs)
]{us}{US}{User Story}
\newacronym{ac}{AC}{Acceptance Criteria}
\newacronym{prf}{PRF}{Pseudo-Relevance Feedback}
\newacronym{aqe}{AQE}{Automatic Query Expansion}
\newacronym{kld}{KLD}{Kullback-Leibler Divergence}
\newacronym{bert}{BERT}{Bidirectional Encoder Representations from Transformers}
\newacronym{sbert}{SBERT}{Sentence Bidirectional Encoder Representations from Transformers}
\newacronym{ltr}{LTR}{Learning-To-Rank}
\newacronym{ser}{SER}{Self-Elicitation of Requirements}
\newacronym{pd}{PD}{Prompt Decomposition}
\newacronym{rag}{RAG}{Retrieval-Augmented Generation}
\newacronym{ragg}{RAGG}{Retrieval-Augmented GUI Generation}
\newacronym{sc}{SC}{Self-Critique}
\newacronym{ood}{OoD}{Out-of-Domain}
\newacronym{dsl}{DSL}{Domain-Specific Language}
\newacronym{cot}{CoT}{Chain-of-Thought}
\newacronym{mcp}{MCP}{Model Context Protocol}
\newacronym{icl}{ICL}{In-Context Learning}
\newacronym{ce}{CE}{Context Engineering}
\newacronym{pe}{PE}{Prompt Engineering}
\newacronym{uml}{UML}{Unified Modeling Language}
\newacronym{sdlc}{SDLC}{Software Development Life Cycle}
\newacronym{sud}{SUD}{System Under Development}
\newacronym{UAT}{UAT}{User Acceptance Testing}
\newacronym{jad}{JAD}{Joint Application Development}
\newacronym{uims}{UIMS}{User Interface Management System}
\newacronym{ucd}{UCD}{User-Centered Design}
\newacronym{json}{JSON}{JavaScript Object Notation}
\newacronym{svg}{SVG}{Scalable Vector Graphics}
\newacronym{kld}{KLD}{Kullback-Leibler Divergence}
\newacronym{cnf}{CNF}{Conjunctive Normal Form}
\newacronym{bow}{BoW}{Bag-of-Words}
\newacronym{nbow}{nBoW}{Neural Bag-of-Words}
\newacronym{tf}{TF}{Term Frequency}
\newacronym{idf}{IDF}{Inverse Document Frequency}
\newacronym{tfidf}{TF-IDF}{Term Frequency-Inverse Document Frequency}
\newacronym{mhsa}{MHSA}{Multi-Head Self-Attention}
\newacronym{ffnn}{FFNN}{Feed-Forward Neural Network}
\newacronym{sts}{STS}{Semantic Textual Similarity}
\newacronym{sgd}{SGD}{Stochastic Gradient Descent}
\newacronym{sft}{SFT}{Supervised Finetuning}
\newacronym{rl}{RL}{Reinforcement Learning}
\newacronym{gpt}{GPT}{Generative Pretrained Transformer}
\newacronym{vit}{ViT}{Vision Transformer}
\newacronym{tot}{ToT}{Tree-of-Thoughts}
\newacronym{nlg}{NLG}{Natural Language Generation}
\newacronym{tot}{ToT}{Tree-of-Thoughts}
\newacronym{got}{GoT}{Graph-of-Thoughts}
\newacronym{rtm}{RTM}{Ready-to-Manufacture}
\newacronym{bm25}{BM25}{Best Matching 25}
\newacronym{rawi}{RaWi}{Rapid Wireframing}
\newacronym{ebm}{EBM}{Extended Boolean Model}
\newacronym{ascii}{ASCII}{American Standard Code for Information Interchange}
\newacronym{mle}{MLE}{Maximum Likelihood Estimates}
\newacronym{rest}{REST}{Representational State Transfer}
\newacronym{ap}{AP}{Average Precision}
\newacronym{map}{MAP}{Mean Average Precision}
\newacronym{ndcg}{NDCG}{Normalized Discounted Cumulative Gain}
\newacronym{dcg}{DCG}{Discounted Cumulative Gain}
\newacronym{mrr}{MRR}{Mean Reciprocal Rank}
\newacronym{sus}{SUS}{System Usability Scale}
\newacronym{hit}{HIT}{Human Intelligence Task}
\newacronym{iaa}{IAA}{Inter-Annotator Agreement}
\newacronym{dl}{DL}{Deep Learning}
\newacronym{sdim}{SDim}{Search Dimension}
\newacronym{api}{API}{Application Programming Interface}
\newacronym{prp}{PRP}{Pairwise Ranking Prompting}
\newacronym{ser}{SER}{Self-Elicitation of Requirements}
\newacronym{xml}{XML}{Extensible Markup Language}
\newacronym{cnn}{CNN}{Convolutional Neural Network}
\newacronym{s2w}{S2W}{Screen2Words}
\newacronym{pdf}{PDF}{Portable Document Format}
\newglossaryentry{uiux}{
  name={UI/UX},
  description={User Interface and User Experience}
}
\newacronym{rlhf}{RLHF}{Reinforcement Learning with Human Feedback}
\newacronym{url}{URL}{Uniform Resource Locator}
\newglossaryentry{ci}{
  name={CI},
  description={Confidence Interval}
}
\newacronym{gan}{GAN}{Generative Adversarial Network}
\newacronym{nasatlx}{NASA-TLX}{National Aeronautics and Space Administration – Task Load Index}
\newacronym{csi}{CSI}{Creativity Support Index}
\newacronym{irb}{IRB}{Institutional Review Board}
\newacronym{ux}{UX}{User Experience}
\newacronym{tn}{TN}{True Negative}
\newacronym{tp}{TP}{True Positive}
\newacronym{fp}{FP}{False Positive}
\newacronym{fn}{FN}{False Negative}
\newacronym{bdd}{BDD}{Behavior-Driven Development}
\newacronym{rsl}{RSL}{Requirements Specification Language}
\newacronym{lda}{LDA}{Latent Dirichlet Allocation}
\newacronym{cua}{CUA}{Computer Use Agent}
\newacronym{ide}{IDE}{Integrated Development Environment}
\newacronym{cdn}{CDN}{Content Distribution Network}
\newacronym{snn}{SNN}{Siamese Neural Network}
\setglossarystyle{alttree}
\glssetwidest{CRISP-EM}
\printglossary[type=\acronymtype,title={List of Acronyms}]
\renewcommand*{\glstextformat}[1]{\textcolor{black}{#1}}



\newpage
\thispagestyle{empty}
\null  

\newpage
\thispagestyle{empty}
\vspace*{\fill}
\begin{center}
{\fontsize{16}{20}\selectfont\emph{To Yaru \& Lio.}}
\end{center}
\vspace*{\fill}
\clearpage


\newpage
\thispagestyle{empty}
\null  

\pagenumbering{arabic}



\chapter{Introduction} 
\label{cha:introduction} 

\section{Motivation}

\gls{rel} represents a crucial activity within \gls{re} for \gls{sd} projects, which is characterized by gathering, interpreting and understanding requirements from respective requirements sources, encompassing stakeholders, documents and legacy systems in operation \citep{zowghi2005requirements, pohl2010requirements, pohl2016requirements}. This activity involves two important parties: the requirements engineer or requirements analyst\footnote{The terms \textit{requirements engineer}, \textit{requirements analyst} and the abbreviated form \textit{analyst} are used interchangeably in the remainder of this work, referring to the individual(s) responsible for conducting the \gls{re} activities within a \gls{sd} project.}, who closely speaks the language of the stakeholders, is responsible for becoming acquainted with the corresponding application domain and creates the requirements artifacts, such as requirements documents, among others \citep{pohl2016requirements}. Second, the stakeholders, who support the requirements analyst to become familiar with their application domain, provide the requirements to the analyst based on the requirements activities conducted together and help inspect collected requirements as well as rapidly communicate requirements changes \citep{pohl2016requirements}. 

To facilitate an effective \gls{rel} process, several elicitation\footnote{The terms \textit{requirements elicitation} and the abbreviated version \textit{elicitation} are used interchangeably in the remainder of this work, referring to the activities of gathering, interpreting and understanding requirements.} techniques and approaches have been developed and are applied in real-world projects, which can be specifically selected and conducted depending on multiple conditions \citep{goguen1993techniques, zowghi2005requirements, davis2006effectiveness, pohl2010requirements, pohl2016requirements, dar2018systematic, pacheco2018requirements}. For example, these techniques include \textit{surveying}, particularly in the form of unstructured, semi-structured and structured \textit{interviews} \citep{bano2019teaching} (e.g., in the form of \textit{laddering} \citep{corbridge1994laddering}) as well as \textit{questionnaires} \citep{foddy1993constructing} and \textit{introspection} \citep{zowghi2005requirements}, among others. Moreover, \textit{prototyping} represents another prominent elicitation technique, which is the construction of a preliminary, incomplete and simplified or abstracted representation of the envisioned software system, application or product, enabling stakeholders to experience tangible, often visual artifacts of the system to be developed, which facilitates effectively gathering detailed feedback and relevant information \citep{hickey1998prototyping, mannio2001requirements, zowghi2005requirements, pacheco2018requirements, abad2018loud}. Prototyping has also been shown to enable more effective capturing of \gls{fr} and \gls{nfr} compared to standalone elicitation meetings, with prototyping being absent \citep{abad2018loud}. Notably, most of these commonly employed elicitation techniques and approaches are heavily based on \gls{nl} communication between stakeholders and analysts. Furthermore, \gls{rd}, the activity of storing gathered information and requirements for the system obtained within the \gls{re} activities, is most commonly conducted by employing \gls{nlr} \citep{pohl2016requirements}. 

While \gls{nl} provides the advantage of representing a common language between the stakeholders and analysts as well as enabling the expression of arbitrary requirements due to the unrestricted nature of \gls{nl}, it inherently possesses ambiguity and provides the potential to create misinterpretations between the parties \citep{kamsties2000taming, kamsties2005understanding, kiyavitskaya2008requirements, shah2015resolving}. Apparently, this is most prevalent when conducting \gls{nl}-focused elicitation techniques, such as interviewing \citep{ferrari2016ambiguity}. Research has shown that various problems can arise from unclear, ambiguous and vague requirements, such as a failure to meet the expectations of customers, quality issues of the created software system as well as wasted effort \citep{bjarnason2011requirements}. Moreover, poor-quality and ambiguous requirements negatively affect all downstream activities in the \gls{sdlc}, including software architecture and design, the system implementation represented by source code and test artifacts. Defects detected only at later development stages are many times more expensive to resolve than defects identified early during the \gls{rel} phase \citep{boehm1988understanding, firesmith2007common}. This can lead to heavy rework, increased project costs as well as project delays and schedule overruns \citep{boehm1988understanding, bjarnason2011requirements}.

To mitigate issues arising from \gls{nlr} specified and documented during \gls{rel}, \gls{rval} plays a critical role, encompassing activities in \gls{re} intended to facilitate the detection of erroneous requirements, including ambiguity, vagueness, incompleteness as well as contradictions \citep{pohl2010requirements, pohl2016requirements}. In order to support \gls{rval}, various techniques have been proposed before and are employed in practice \citep{bilal2016requirements}. For example, review techniques include \textit{commenting}, where requirements are validated by expert co-workers \citep{pohl2016requirements}, \textit{inspections}, which represent a systematic process of error identification based on development artifacts \citep{gilb1993software, laitenberger2000encompassing}, and \textit{walk-throughs}, which are a lightweight variant of an inspection \citep{pohl2016requirements}. Moreover, \textit{prototyping} is one of the most popular approaches for \gls{rval} and a versatile technique, enabling not only effective elicitation by facilitating the discovery of novel requirements \citep{abad2018loud}, but also an effective validation\footnote{\glsreset{rval}The terms \textit{\gls{rval}} and the abbreviated version \textit{validation} are used interchangeably in the remainder of this work, referring to the conducted activities and processes of ensuring that the elicited and documented requirements correctly reflect the actual stakeholder needs of the software system.} procedure via tangible, interactive representations of the software system \citep{pohl2016requirements}. 

\noindent Particularly, employing \gls{gui} prototypes for \gls{rval} has been shown to be one of the most valuable validation techniques \citep{moore2000comparison, kamalrudin2011generating}, effectively bridging the gap between potentially ambiguous \gls{nlr} and the interpretation of the requirements by the analyst. Moreover, conducting \gls{rval} by employing \gls{gui} prototypes facilitates the incorporation of stakeholders closely into the system development activities, helps in sparking fruitful discussions and clarifying as well as refining existing requirements from previous elicitation iterations \citep{windsor1992prototyping, rudd1996low, ravid2000method, beaudouin2002prototyping, mukasa2008integration}. 

\gls{gui} prototypes can be broadly characterized by different dimensions, including the reusability or development life cycle intent such as \textit{throw-away} or \textit{evolutionary} prototypes \citep{zowghi2005requirements, sommerville2011software}, as well as the prototype fidelity \citep{rudd1996low, coyette2007multi}, which represents how closely a prototype resembles the envisioned final system, among others. While throw-away prototypes are solely employed for the specific purpose of \gls{rel} and \gls{rval} and subsequently discarded, evolutionary prototypes are reused in the process and continuously improved in later stages \citep{zowghi2005requirements, sommerville2011software}. Furthermore, the fidelity level of prototypes plays an important role, ranging from low-fidelity \gls{gui} prototypes, such as pen-and-paper prototypes \citep{snyder2003paper, davis2007sketchwizard}, which are highly abstracted representations with restricted functionality and typically limited interaction possibilities, to high-fidelity \gls{gui} prototypes, which offer more functionality as well as interactivity, providing a resemblance closer to the final system \citep{rudd1996low}. Especially high-fidelity \gls{gui} prototypes offer a more effective prototyping mechanism compared to their low-fidelity variants, since high-fidelity prototypes provide the foundation for higher-quality discussions between stakeholders and analysts, and more detailed feedback can be gathered while conducting validation sessions with stakeholders \citep{landay1994interactive, rudd1996low, coyette2007multi}. However, the benefits of high-fidelity \gls{gui} prototypes are accompanied by the shortcomings of requiring more time and effort to create, therefore entailing higher production costs, as well as requiring more technical experience to create them \citep{rudd1996low}. Moreover, since the creation of these prototypes necessitates time, multiple meetings between stakeholders and analysts are required, often spanning several weeks or months with long breaks in between, leading to outdated requirements and project delays \citep{schneider2007generating, debnath2021ideas}. 

While several approaches have been developed over the years to facilitate the creation of \gls{gui} prototypes, providing support directly based on automated processing of \gls{nlr} has been largely neglected in research before. For example, traditionally, proprietary and commercially available \gls{gui} prototyping editors are widely adopted in practical prototyping environments, providing visual representations and editing functionality to improve the \gls{gui} prototyping efficiency and simplicity, e.g., \textit{Balsamiq} \citep{faranello2012balsamiq}, \textit{Mockplus} \citep{mockplus}, \textit{Adobe XD} \citep{adobexd}, \textit{Figma} \citep{figma} and \textit{Sketch} \citep{sketch}. These prototyping tools enable users to combine fundamental \gls{gui} components and building blocks as well as provide a small amount of predefined and reusable \gls{gui} templates. However, the employment of these tools still remains largely a manual, time-consuming process, additionally requiring prototyping expertise. Moreover, multiple \gls{gui} retrieval approaches have been proposed before, which, however, necessitate \gls{gui} sketches or screenshots as input to retrieve similar \glspl{gui}, such as \textit{GUIFetch} \citep{behrang2018guifetch}, \textit{Swire} \citep{huang2019swire}, \textit{VINS} \citep{bunian2021vins} and \textit{Screen2Vec} \citep{li2021screen2vec}. While these approaches help in rapidly obtaining similar design ideas for inspiration, they are incapable of directly processing \gls{nlr}. Likewise, the \gls{gui} prototyping assistant approach \textit{GUIComp} \citep{lee2020guicomp} supports prototype developers via image similarity retrieval and usability metrics computations during the creation, but cannot support the creation of \gls{gui} prototypes from \gls{nlr}. Therefore, providing effective approaches for automating and facilitating the rapid mapping of \gls{nlr} to \gls{gui} prototypes represents the first major challenge (\challone) addressed in this work.

Besides \gls{rval}, \gls{sver} represents another vital activity conducted in the \gls{sdlc}, closely related to the validation activities \citep{wallace1989software, rakitin2001software, post2009linking}. However, while validation focuses on ensuring alignment of the explicitly specified requirements with the stakeholders' actual needs and expectations, verification\footnote{The terms \glsreset{sver}\textit{\gls{sver}} and the abbreviated version \textit{verification} are used interchangeably in the remainder of this work, referring to activities and processes ensuring that the implemented software system aligns with the requirements and system specification.} is concerned with ensuring that the final and intermediate development artifacts, such as software components, and the resulting software system, adhere to their specification, i.e. correctly implement the explicitly specified requirements \citep{wallace1989software}. Specifically, since \gls{sver} can be associated with different aspects of a software system, we focus on the verification of \gls{gui} applications, namely, the fully implemented version of the earlier developed \gls{gui} prototype used for \gls{rel} and \gls{rval}.


Testing represents the most common verification technique, while \gls{gui} testing is particularly time-consuming and effort-demanding \citep{memon2001comprehensive, memon2007event}. To achieve verification for \gls{gui} applications and thereby improve the \gls{gui} quality and the correctness of the implementation, a plethora of testing approaches has been proposed in research \citep{xie2007designing, amalfitano2012using, mao2016sapienz, gu2019practical}. For example, these approaches include manual testing via human-based interaction with the \gls{gui} application \citep{marick1998should}, traditionally the most popular approach, writing and employing test scripts, which rely on interacting with specifically named \gls{gui} components in a predefined manner \citep{xie2007designing}, automated random exploration of the \gls{gui} application \citep{amalfitano2012using, mao2016sapienz}, also referred to as \textit{fuzzing}, such as in \textit{GUIFuzz++} \citep{otto2025guifuzz}, replay techniques \citep{hicinbothom1993tool, memon2003advances} as well as model-based techniques \citep{gu2019practical}. However, these \gls{gui} testing approaches necessitate the creation of test artifacts, which in turn require technical expertise. Moreover, they are not capable of directly processing \gls{nlr} to automatically verify the \gls{gui} application and require experts to create respective test representations by translating them from \gls{nlr}. Hence, providing support to automate the verification of \gls{gui} applications based on \gls{nlr} represents the second major challenge (\challtwo).

Despite their practical relevance, both introduced challenges are inherently difficult to address. On the one hand, \gls{nlr} are inherently ambiguous, incomplete and context-dependent. Moreover, \gls{nlr} are often heterogeneous (a mix of functional and non-functional aspects) and unstructured, lacking a fixed syntax and semantics, which renders parsing and transformation into representations optimized for automatic machine-based computations difficult. On the other hand, \glspl{gui} are complex, multi-dimensional objects, encompassing structural information (such as layout and component hierarchy), functional aspects (expressed through \gls{gui} components and their interactions), visual appearance and often complex, dynamic behavior (in the case of \gls{gui} applications). Therefore, the complexity and multi-dimensionality of \glspl{gui} render an optimized representation for automatic computations similarly difficult. This shows that an alignment between \gls{nlr} and \glspl{gui} is inherently difficult, due to their individual complexity and the lack of a shared representation space. However, recent advances in \gls{nlp} -- through powerful foundational language processing models such as \glspl{plm} and \glspl{llm} -- provide large potential to tackle these challenges effectively. Therefore, this work explores the potential of such models for \gls{nlr}-based automated \gls{gui} prototyping and verification.

\section{Problem Statement}

\begin{figure}[!t]
\centering
  \includegraphics[width=1.0\textwidth]{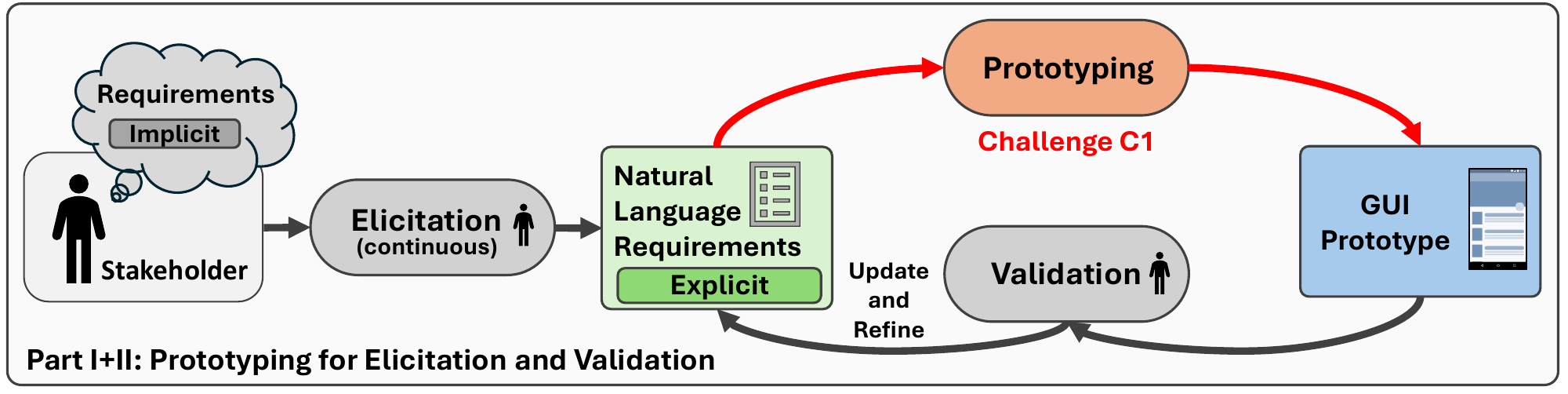}
  \caption[Schematic overview of requirements elicitation, validation and \gls{gui} prototyping]{Schematic overview of the parties, artifacts, activities and their interplay for \textit{elicitation} of requirements (\gls{rel}), \textit{validation} of requirements (\gls{rval}) and \textit{\gls{gui} prototyping}.}
	\label{fig:chapter_1_chall_1}
    \vspace{-0.4cm}
\end{figure}

Despite the existence of numerous approaches aimed at simplifying and supporting the \gls{gui} prototyping and testing activities, a substantial research gap remains for facilitating and automating these processes directly from elicited \gls{nlr}. While prior research has primarily focused on different techniques for creating \gls{gui} prototypes to support \gls{rel} and \gls{rval} as well as developing \gls{gui} testing approaches for \gls{sver}, these methods are incapable of supporting \gls{nlr} directly. Consequently, this work is particularly concerned with the following two guiding challenges:

\begin{enumerate}[label=\textbf{(C\arabic*)}]
    \item \textit{Mapping \gls{nlr} to \gls{gui} Prototypes:} the creation of high-fidelity \gls{gui} prototypes from \gls{nlr} is a labor-intensive, costly endeavor, requiring expert knowledge and experience. How can we automate the \gls{gui} prototype creation based on \gls{nlr}? 
    \item \textit{Verification of \gls{gui} Applications from \gls{nlr}}: the verification of \gls{gui} applications against \gls{nlr} remains predominantly a time-consuming, manual process, with existing solutions requiring expert knowledge and experience to be employed effectively. How can we automate the verification of \gls{nlr} in \gls{gui} applications?
\end{enumerate}


\begin{figure}[!t]
\centering
  \includegraphics[width=1.0\textwidth]{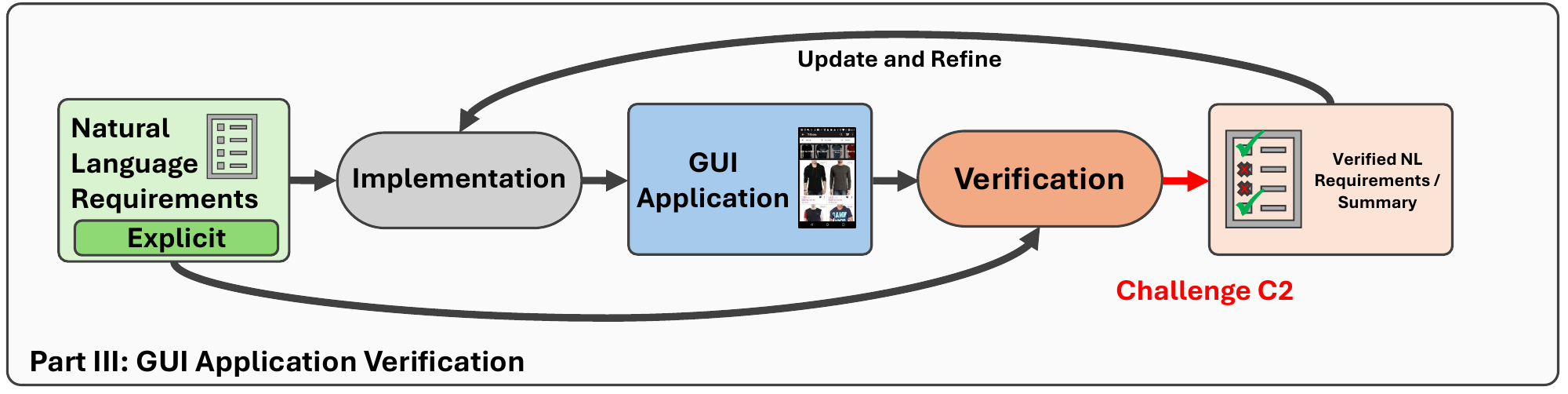}
  \caption[Schematic overview of \gls{gui} verification]{Schematic overview of the artifacts, activities and their interplay for verification of a \gls{gui} application based on \gls{nlr} (verification drives iterative implementation updates).}
	\label{fig:chapter_1_chall_2}
    \vspace{-0.4cm}
\end{figure}

\noindent A schematic and simplified overview of the \textit{elicitation}, \textit{validation} and \textit{\gls{gui} prototyping} activities, together with their respective inputs and outputs, is presented in Figure \ref{fig:chapter_1_chall_1}. From the initial elicitation (e.g., in the form of interviews), explicit \gls{nlr} are gathered and documented. Subsequently, \gls{gui} prototypes are developed and, together with the stakeholders, discussed in validation sessions, leading to updates and refinements (i.e. addition, modification and removal) of the explicit \gls{nlr}. Furthermore, verification is performed by employing the \gls{nlr} and an implemented \gls{gui} application to assess conformance with the specified requirements, as shown in Figure \ref{fig:chapter_1_chall_2}. In the presence of deviations between them, the \gls{gui} implementation is accordingly updated and refined.

To tackle the two major challenges introduced, we develop several approaches. The approach for automated mapping from \gls{nlr} to \gls{gui} prototypes (\challone) is divided into two parts: \textit{(I)} first, we approach the mapping as a \textit{retrieval} problem, i.e. given the \gls{nlr}\glsreset{sud}\footnote{When we refer to the terms \gls{nlr} or requirements from now on, we specifically refer to functional requirements (\gls{fr}) in natural language (\gls{nl}) of the \gls{sud} of an interactive, user-facing software system for the remainder of this work, if not explicitly stated otherwise.} as input, we exploit a large-scale \gls{gui} repository to automatically compute a ranking over \gls{gui} prototypes according to their level of correspondence with the \gls{nlr} (\textit{Part \ref{part:gui_retrieval}} of the thesis). Second, \textit{(II)} we leverage an \gls{llm} to frame the mapping as a \textit{translation} problem, with \gls{nlr} as source and \gls{gui} code as target language (\textit{Part \ref{part:gui_generation}} of the thesis). Moreover, \textit{(III)} we approach the automated verification of \gls{gui} applications from \gls{nlr} by leveraging an \gls{llm} as well as proposing a \gls{mllm}-based agent (\textit{Part \ref{part:verification}} of the thesis). Each of these approaches is accompanied by its own sub-challenges, which we describe next. An overview of the addressed challenges and their relatedness to the posed main challenges is depicted in Figure \ref{fig:chapter-1-challenges}. First, we state the sub-challenges regarding \gls{nlr}-based \gls{gui} retrieval approaches \textit{(Part \ref{part:gui_retrieval}}):

\begin{figure}
\centering
  \includegraphics[width=1.0\textwidth]{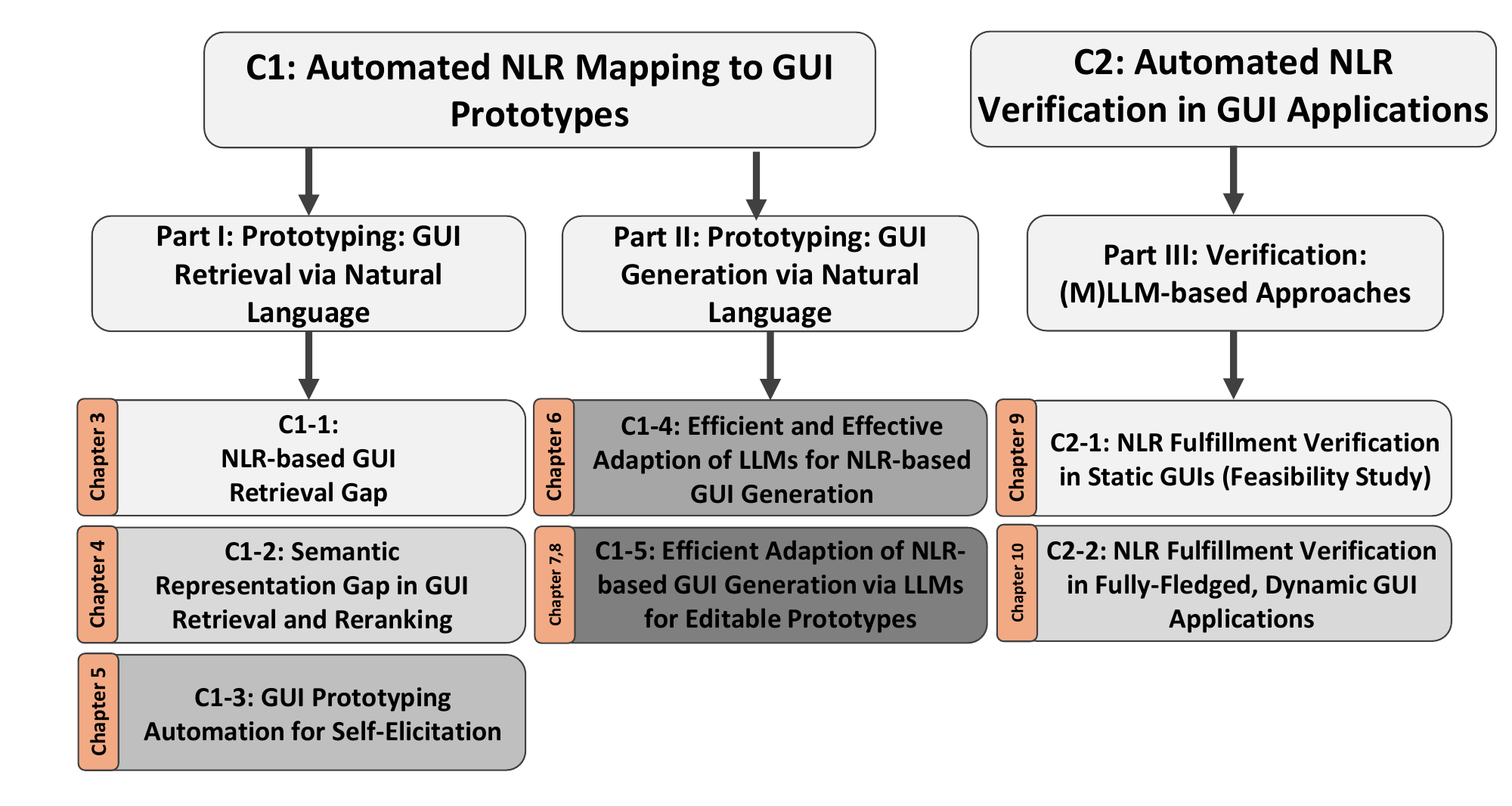}
  \caption[Overview and relatedness of the challenges addressed in the thesis]{Overview and relatedness of the seven individual sub-challenges tackled in this work with regard to the two major challenges posed, where challenge \challone{} is divided into \gls{nlr}-based \gls{gui} retrieval and generation, and \challtwo{} into \gls{nlr}-based \gls{gui} app. verification.}
	\label{fig:chapter-1-challenges}
\end{figure}

\begin{enumerate}[label={}]
    \item[\textbf{C1-1}] \textit{\gls{nlr}-based \gls{gui} Retrieval Gap:} traditional \gls{ir} techniques are mainly focused on retrieving text artifacts and cannot be readily employed due to the semantic gap between unstructured \gls{nlr} and complex \gls{gui} prototypes. How can we adapt and optimize text-based retrieval methods and techniques to enable more effective \gls{nlr}-based \gls{gui} retrieval?
     \item[\textbf{C1-2}] \textit{Semantic Representation Gap in \gls{gui} Retrieval and Reranking:} while optimized text-based \gls{ir} methods improve the ranking of \gls{gui} prototypes based on \gls{nlr} over standard \gls{ir} methods, the multi-dimensionality of \gls{gui} prototypes is mostly neglected. How can the semantic representation gap between \gls{nlr} (including \gls{nfr}) and multi-dimensional \gls{gui} prototypes be reduced to enable more effective \gls{gui} ranking?
     \item[\textbf{C1-3}] \textit{\gls{gui} Prototyping Automation for Self-Elicitation:} while retrieval-based approaches enable rapid mapping from \gls{nlr} to \gls{gui} prototypes, these techniques assume sufficient prototyping process knowledge to be employed effectively and provide no guidance for the prototyping and elicitation process. How can we provide automatic assistance for self-elicitation of requirements with \gls{gui} prototyping?
\end{enumerate}

\noindent Second, we list the challenges regarding \gls{nlr}-based \gls{gui} prototype generation \textit{(Part \ref{part:gui_generation})}:

\begin{enumerate}[label={}]
    \item[\textbf{C1-4}] \textit{Efficient and Effective Adaptation of \glspl{llm} for \gls{nlr}-based \gls{gui} Generation:} while pretrained \gls{llm}s offer general capabilities to generate \gls{gui} prototypes from \gls{nlr}, optimizing them for more effective \gls{gui} generation without inefficient and resource-intensive fine-tuning of \gls{llm}s remains open. How can we efficiently optimize \gls{llm}s for more effective \gls{gui} prototype generation?
     \item[\textbf{C1-5}] \textit{Efficient Adaptation of \gls{nlr}-based \gls{gui} Generation via \gls{llm}s for Editable Prototypes:} while pretrained \gls{llm}s have profound capabilities of generating \gls{gui}s in widely available code formats, for example, \gls{html}, these representations are not well suited for customization. How can \gls{llm}s be efficiently adapted to generate editable \gls{gui} prototype representations from \gls{nlr}?
\end{enumerate}

\noindent Third, we list the challenges regarding \gls{nlr}-based verification of \gls{gui} artifacts \textit{(Part \ref{part:verification})}:

\begin{enumerate}[label={}]
    \item[\textbf{C2-1}] \textit{\gls{nlr} Fulfillment Verification in Static \glspl{gui}:} existing verification techniques are predominantly rule-based and require technical expertise, limiting their ability to handle semantically complex \gls{nlr}. How can \gls{nlr}-based verification be automated for simplified, static \gls{gui} representations as a foundation for more complex \gls{gui} verification tasks?
     \item[\textbf{C1-2}] \textit{\gls{nlr} Fulfillment Verification in Fully-Fledged, Dynamic \gls{gui} Applications:} similarly to the previous challenge, existing approaches are not capable of supporting highly dynamic, fully-fledged \gls{gui} application verification based on complex \gls{nlr}. How can \gls{nlr}-based verification be automated for highly dynamic, fully-fledged \gls{gui} applications? 
\end{enumerate}

\noindent In summary, we address seven individual challenges: optimizing \gls{nlr}-based \gls{gui} retrieval approaches from text representations, embracing the multi-dimensionality of \gls{gui} prototypes and efficiently adapting \gls{llm}s for \gls{gui} generation regarding the first challenge \challone{}. Moreover, we address the \gls{nlr}-based verification of simplified \gls{gui} representations and highly dynamic, fully-fledged \gls{gui} applications regarding the major challenge \challtwo{}.

\section{Solution Overview}

\begin{figure}
\centering
  \includegraphics[width=1.0\textwidth]{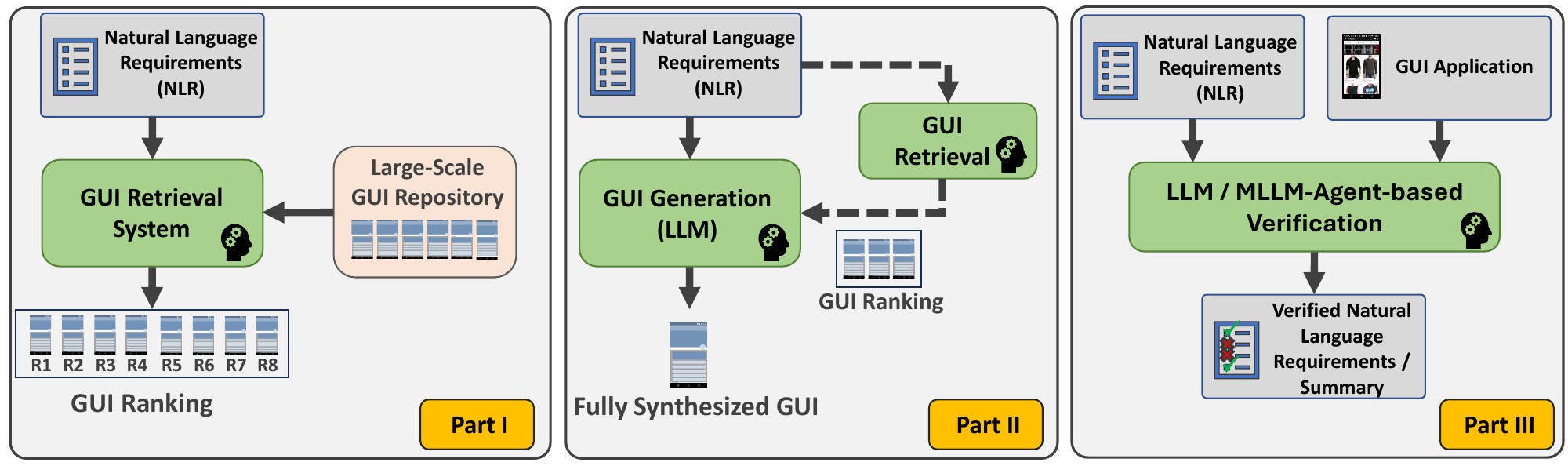}
  \caption[Overview of solution approaches]{General solution overview for the two challenges divided into three parts in the thesis, including \gls{nlr}-based \gls{gui} retrieval (\textit{Part \ref{part:gui_retrieval}}), \gls{nlr}-based \gls{gui} generation (\textit{Part \ref{part:gui_generation}}) and \gls{llm} as well as \gls{mllm}-agent-based verification of \gls{gui} applications (\textit{Part \ref{part:verification}}).}
	\label{fig:chapter-solution-overview}
\end{figure}

To address the challenges delineated in the previous section, we divided them into three different parts, as depicted in Figure \ref{fig:chapter-1-challenges}. For each of these parts, we briefly outline the general solution notion subsequently. In addition, we provide a visual overview of these ideas in Figure \ref{fig:chapter-solution-overview}, covering \gls{nlr}-based \gls{gui} retrieval (\textit{Part \ref{part:gui_retrieval}}), \gls{nlr}-based \gls{gui} generation (\textit{Part \ref{part:gui_generation}}) and \gls{llm} as well as \gls{mllm}-agent-based verification of \gls{gui} applications (\textit{Part \ref{part:verification}}). Particularly, for both challenges \challone{} and \challtwo{}, we focus on \glspl{gui} from either \textit{mobile} (specifically the \textit{Android} platform) or \textit{web} (i.e. implemented based on \textit{\gls{html}}, \textit{\gls{css}} and \textit{\gls{js}}) applications, since these software platforms are most prevalent and ubiquitous.

First, \textit{(I)} we approach \gls{nlr}-to-\gls{gui} prototype mapping as a \textit{retrieval} problem. To this end, a large-scale \gls{gui} repository encompassing over 72,000 individual \gls{gui}s is exploited, enabling matching to a vast number of different \gls{nlr}. Given the \gls{nlr}, we compute a ranking score for each \gls{gui} encompassed in the repository and provide the best matches to users. Therefore, while restricted to the existing \gls{gui} repository, we enable the fast mapping from \gls{nlr} to \gls{gui} prototypes for a broad spectrum of real-world \gls{gui}s. Specifically, we develop a text extraction approach from these \gls{gui}s, investigate the performance of different text-based \gls{ir} methods, adapt existing methods and finetune a \gls{plm} for improved \gls{gui} ranking performance. Moreover, we investigate \gls{mllm}-based \gls{gui} reranking, reducing the semantic representation gap between \gls{nlr} and multi-dimensional \gls{gui} prototypes. Finally, we provide a dialogue-based approach (with retrieval) for simplifying and guiding the \gls{gui} prototyping process for \gls{rel} and \gls{rval}.

Second, \textit{(II)} we tackle the \gls{nlr}-to-\gls{gui} prototype mapping as a \textit{translation} problem by employing a pretrained \gls{llm}, which encompasses vast amounts of general, programming and prototyping knowledge. In particular, we investigate several approaches to efficiently optimize a pretrained \gls{llm} to more effectively generate \gls{gui} prototypes. To this end, we follow a \gls{zs} prompting approach and leverage a large-scale \gls{gui} repository combined with a \gls{gui} retrieval approach (developed in \textit{Part \ref{part:gui_retrieval}}) to efficiently extend the prototyping knowledge of the \gls{llm}. Additionally, we investigate several other prompting approaches to more effectively access and employ the \gls{gui} prototyping knowledge already stored in the parameters of the \gls{llm}. Furthermore, in a similar manner, we efficiently adapt an \gls{llm} via retrieval of proprietary \gls{gui} component libraries, enabling the \gls{llm}-based generation of fully editable \gls{gui} prototypes from \gls{nlr}, integrated into a popular \gls{gui} prototyping editor with fully-fledged editing functionality.

Third, \textit{(III)} we approach the \gls{nlr}-based automatic verification of \gls{gui} applications in two stages. To begin with, we investigate an easier variant of the problem as a feasibility study by examining simplified, static \gls{gui} representations. Specifically, we focus on fixed \gls{gui} hierarchy data, encompassing the \gls{gui} components and their relationships, while excluding dynamic \gls{gui} behavior. To this end, we investigate different prompting techniques, optimize an \gls{llm} to compute binary predictions of \gls{nlr} fulfillment and match \gls{nlr} to corresponding \gls{gui} components given a static \gls{gui} representation. In the second step, we examine \gls{nlr}-based verification of fully-fledged, highly dynamic web-based \gls{gui} applications by adapting an \gls{mllm}-based \gls{cua}. This agent is capable of interacting with complex \glspl{gui}, predicting the most likely next interaction steps, executing them on the \gls{gui} application and obtaining the updated \gls{gui} states to repeat the same process again. Hence, complex \gls{nlr} (both \gls{fr} and \gls{nfr}) can be verified automatically in dynamic, web-based \gls{gui} applications with high effectiveness.



\section{Contributions}

After motivating our research, presenting both challenges with their sub-challenges regarding \gls{nlr}-based \gls{gui} prototyping and \gls{nlr}-based \gls{gui} application verification, and briefly delineating our solution approaches, including \gls{gui} retrieval, \gls{gui} generation and \gls{mllm}-based verification, we build on top of these insights and outline our contribution to the field of \gls{re} and \gls{se}. We evaluated our newly proposed \gls{gui} prototyping and verification approaches on several novel datasets and conducted multiple insightful analyses, showing the effectiveness of automating \gls{gui} prototyping and verification using \gls{nlr}. Subsequently, we present our contribution divided into created and published datasets (for fostering future research) and novel approaches proposed.

\vspace{1em}

\noindent\textbf{Datasets.} During our research, we created several datasets spanning both challenges of \gls{nlr}-based \gls{gui} prototyping (\challone{}) and verification (\challtwo{}) for enabling effective training as well as comprehensive evaluation of approaches focusing on \glspl{plm}, \glspl{llm} and \glspl{mllm}.

\begin{enumerate}

    \item \textit{\gls{nlr}-based \gls{gui} Retrieval and Reranking Gold Standard:} to evaluate the \gls{gui} retrieval and reranking methods (see Chapters \ref{cha:nl_gui_retrieval} and \ref{cha:gui_rerank}), we created a novel crowd-sourced benchmark and training dataset, consisting of \gls{nlr} combined with \gls{gui} prototypes from the \textit{Rico} \gls{gui} dataset \citep{deka2017rico} -- obtained from \gls{ir}-based pooling -- and corresponding three-level relevance annotations. This dataset consists of 450 \gls{nlr}, each combined with 20 \glspl{gui}, and we gathered over 40,500 relevance annotations using the crowd-sourcing platform \gls{amt}. This represents the first dataset for the problem and can be employed to both evaluate \gls{gui} ranking models on the benchmark and train novel models on the training dataset. Particularly, we employed these datasets to optimize and improve text-based \gls{gui} retrieval (\challoneone{}) as well as to reduce the semantic representation gap between \gls{nlr} and multi-dimensional \gls{gui} prototypes (\challonetwo{}).
    
     \item \textit{\textit{Rico} Annotation and Embeddings Dataset:} to support the multi-dimensionality of \gls{gui} prototypes during \gls{nlr}-based retrieval (\challonetwo{}), we annotated the \textit{Rico} \gls{gui} dataset \citep{deka2017rico} by employing an \gls{mllm} to obtain textual descriptions for each prototype regarding the aspects of \textit{functionality}, \textit{design}, \textit{domain}, \textit{\gls{gui} components} and \textit{displayed text}. Moreover, we provide 3,072-dimensional \textit{OpenAI} \citep{openai2023gpt4} embeddings for each annotated dimension and \gls{gui} prototype. Both datasets enable fostering future research by, for example, fueling other related downstream \gls{ml} tasks such as \gls{gui} classification, advanced \gls{gui} retrieval techniques as well as semantic understanding of \gls{gui} prototypes.
     
     \item \textit{\gls{gui} Description and \gls{gui} Prototype Dataset:} to assess \gls{llm}-based \gls{gui} generation approaches (\challonefour{}), we constructed a novel, high-quality \gls{gui} description dataset with more complex, information-dense and longer descriptions compared to our previously introduced benchmark and training dataset for \gls{nlr}-based \gls{gui} retrieval, carefully gathered in a controlled lab-based setting. Moreover, we created a \gls{gui} prototype dataset (web-based prototypes consisting of \textit{\gls{html}}, \textit{\gls{css}} and \textit{\gls{js}}) corresponding to the gathered \gls{nlr}, which has been generated by different \gls{zs} prompting methods investigated in Chapter \ref{cha:zs_gui_generation}. This dataset provides potential for various future \gls{ml} approaches (requiring high-quality \gls{gui} requirements) as well as analyses, including error analyses of \gls{llm}s in \gls{gui} generation, among others.
     
    \glsreset{us}
     \item \textit{Interlinked \gls{us} and \gls{gui} Components Dataset:} to conduct our feasibility study for verification of \gls{nlr} in simplified, static \gls{gui} representations, we created a novel dataset, encompassing \gls{nlr} in the popular form of \glsreset{us}\glspl{us}, widely adopted in industry and practice to document requirements, again gathered in a controlled, lab-based environment, ensuring high quality. Each \gls{us} was collected with respect to a specific \gls{gui} from the \textit{Rico} \gls{gui} dataset \citep{deka2017rico}, where a \gls{gui} consists of multiple \gls{us} annotations. In addition, this dataset provides a \textit{one-to-many} mapping from \gls{us} to individual \gls{gui} components in the \textit{Rico} \glspl{gui}, enabling requirements verification research (\challtwoone{}). Furthermore, this dataset fosters requirements traceability research efforts as well as other approaches, where detailed mappings between \glspl{us} and \gls{gui} components are of interest.

     \item \textit{\gls{nlr}-based Verification of \gls{gui} Applications Gold Standard:} aiming at assessing \gls{nlr}-based verification of highly dynamic, complex \gls{gui} applications (\challtwotwo{}), we finally constructed a dataset consisting of \glspl{us}, each combined with multiple, detailed \gls{ac}. Overall, this dataset encompasses five \gls{gui} applications (web-based applications consisting of \textit{\gls{html}}, \textit{\gls{css}} and \textit{\gls{js}}) from different domains (\textit{Park-and-Pay, Budget Tracker, Recipe Generator, Fitness Quests, Cleaning Booking}), where each application is mapped to a collection of \glspl{us} with their \gls{ac}. In addition, the \glspl{us} are annotated with a three-class label, indicating whether a \gls{us} and its \gls{ac} are \textit{met}, \textit{unmet} or \textit{partially met}. This is the first dataset available in research offering this combination of requirements and application mapping, potentially fueling further research for agent-based \gls{gui} verification.
\end{enumerate}


\noindent\textbf{Approaches.} In order to address the introduced challenges, we explore and propose multiple novel data-driven approaches and techniques, largely based on pretrained \gls{llm}s.
\begin{enumerate}
    \glsunset{rawi}
    \item First, we develop a new text extraction approach based on \gls{gui} hierarchy data, fueling text-based \gls{gui} retrieval from \gls{nlr}. In this context, we adapt a \gls{prf} method based on \gls{kld} scoring for optimized \gls{aqe} regarding \gls{gui}s. Moreover, we are the first to optimize a \gls{bert}-\gls{ltr} model to enhance the effectiveness of text-based \gls{gui} ranking (\challoneone). Finally, the proposed \gls{gui} retrieval techniques are integrated into a novel, fully data-driven \gls{gui} prototyping editor (\textit{\gls{rawi}}).

    \item To reduce the semantic representation gap between \gls{nlr} (including \gls{nfr}) and multi-dimensional \gls{gui} prototypes (\challonetwo), we propose a novel \gls{gui} retrieval and reranking approach, supporting multi-dimensional, negative constraints (exclusion criteria). Particularly, \gls{nlr} are deconstructed by a novel \gls{llm}-based query decomposition technique. Furthermore, we propose an \gls{mllm}-based \gls{gui} reranking approach, enabling efficient (text-based) or more effective (image-based) \gls{gui} reranking, achieving state-of-the-art performance for \gls{nlr}-based \gls{gui} ranking. These techniques are integrated into a novel \gls{gui} retrieval tool (\textit{\gls{gui}-ReRank}), which additionally enables the simple adaptation to arbitrary \gls{gui} (image) datasets.

    \item To support users in the \gls{gui} prototyping and elicitation process (\challonethree), we propose a novel approach based on the notion of \gls{ser} \citep{rietz2019ladderbot}, enabling effective \gls{gui} prototyping via a dialogue-based \gls{nl} interface. The approach integrates previously developed \gls{nlr}-based \gls{gui} retrieval, a novel \gls{gui} feature retrieval and an \gls{llm}-based \gls{gui} feature recommendation approach, proactively stimulating stakeholder-driven elicitation of requirements. Furthermore, we propose a novel retrieval method which incorporates a \gls{gui} feature scoring mechanism, ensuring that \gls{gui} prototypes with more relevant features are scored higher in the ranking. The entire approach is implemented and integrated into a novel \gls{gui} prototyping tool (\textit{\gls{ser}\gls{gui}}).

    \item For tackling the efficient adaptation of \gls{llm}s to effectively generate \gls{gui} prototypes (\challonefour), we propose multiple \gls{zs} prompting techniques. In particular, we adapt \gls{pd} \citep{khot2022decomposed}, which dissects the generation procedure into predefined subtasks with intermediate reasoning outputs. Moreover, we introduce \gls{ragg}, a novel technique which combines the advantages of retrieval techniques and language generation approaches, by injecting potentially relevant \gls{gui} prototypes into the \gls{llm} context, thereby extending the prototyping knowledge of the \gls{llm}, which is particularly valuable for Out-of-Domain (OoD) problems. Furthermore, we adapt \gls{sc} \citep{saunders2022self}, which iteratively generates new versions of the prototype and creates critiques, continuously improving the prototype.

    \item We are the first to efficiently adapt \gls{llm}s to generate fully editable \gls{gui} prototypes (\challonefive), integrating them into fully-fledged \gls{gui} prototyping editors. To this end, we adapt an \gls{llm} to generate a proprietary \gls{dsl} by employing a two-stage \gls{rag} approach on custom \gls{gui} component libraries. Moreover, by creating mappings between \gls{nlr} (in particular \glspl{us}) and their implementations, expensive regenerations of entire prototypes can be avoided. The proposed approach is integrated into a novel \textit{Figma} \citep{figma} plugin, enabling practitioners to easily apply these new techniques.

    \item To enable verification of \gls{nlr} (specifically \glspl{us}) in \gls{gui} applications (\challtwoone), we initially propose an \gls{llm}-based approach, where \glspl{us} are automatically matched against \gls{gui} components of the static \gls{gui}, thereby verifying the requirements fulfillment. We investigate different prompting paradigms, including \gls{zs} \citep{kojima2022large} and \gls{fs} prompting \citep{brown2020language} as well as the \gls{cot} technique \citep{kojima2022large, wei2022chain}.

    \item Last, we propose a novel \gls{mllm}-agent-based approach to enable \gls{nlr}-based verification of highly dynamic, fully-fledged \gls{gui} applications (\challtwotwo), by leveraging a \gls{cua} \citep{cua2025}, which is adapted to the \gls{gui} verification problem. To this end, the agent iteratively receives \gls{gui} states in the form of images and predicts next actions to best verify whether an \gls{ac} of a requirement is fulfilled. These actions are executed in the \gls{gui} environment and updated states are provided to the agent in turn. In addition, we provide a \gls{mcp} server implementation for the verification agent, thereby being the first to enable completely autonomous implementation-verification loops for \gls{gui} applications with \gls{llm}-based programming agents. Finally, these techniques are integrated into an intuitive tool (\textit{\gls{gui}Spector}), facilitating rapid adoption by practitioners.
\end{enumerate}






\section{Outline}

First, we provide a brief overview of the theoretical background (Chapter \ref{cha:background}), outlining the foundations from the domains of \gls{re}, \gls{se}, \gls{gui} prototyping and \gls{nlp}. Subsequently, Part \ref{part:gui_retrieval} introduces text-based \gls{gui} retrieval (Chapter \ref{cha:nl_gui_retrieval}), \gls{mllm}-based \gls{gui} reranking (Chapter \ref{cha:gui_rerank}) and the \gls{ser} approach (Chapter \ref{cha:self_elicitation}). Part \ref{part:gui_generation} is composed of efficiently adapting \gls{llm}s to \gls{gui} generation (Chapter \ref{cha:zs_gui_generation}) and generation of editable \gls{gui}s with a proprietary \gls{dsl} (Chapters \ref{cha:closing} and \ref{cha:guide}). Part \ref{part:verification} introduces \gls{llm} and \gls{mllm}-agent-based approaches for \gls{gui} verification (Chapters \ref{cha:interlinking} and \ref{cha:agent}). Each of these chapters provides a motivation, approach details, evaluation methodology, results, limitations and related work. Finally, Part \ref{part:discussion} provides an overall discussion and then concludes the thesis (Chapters \ref{cha:discussion} and \ref{cha:conclusion}).


\clearpage
\newpage
\thispagestyle{empty}
\null  

\chapter{Theoretical Background} 
\label{cha:background}

\glsreset{re}
\glsreset{se}
\glsreset{ml}
\glsreset{ir}
\glsreset{plm}
In this chapter, we present the foundational topics and concepts underlying this thesis: \textit{(1)} foundations of \gls{re} and \gls{se} (Section \ref{sec:re_and_se}), \textit{(2)} the history, process and important concepts of \gls{gui} prototyping (Section \ref{sec:cha-2-prototpying}) and \textit{(3)} relevant computational methods and techniques (Section \ref{sec:computationl_methods}), spanning \gls{ml}, \gls{ir}, \glspl{plm}, \gls{icl}, \gls{pe} and then \gls{ce}.

\section{Requirements and Software Engineering (RE/SE)}
\label{sec:re_and_se}
\glsreset{re}
\gls{re} is an active research area and simultaneously an important industry practice ensuring a systematic approach for eliciting, specifying and documenting, validating and verifying as well as managing requirements for the development of a software system \citep{pohl1997process, pohl2010requirements, pohl2016requirements}, and ``[...] the part of development in which people attempt to discover what is desired'' \citep{gause1989exploring}. The significance of \gls{re} is highlighted by earlier research indicating that approximately 60 \% of software system failures originate in the early stages of projects, due to issues associated with the requirements \citep{boehm1984software}, primarily as a result of unclear, ambiguous, incomplete or erroneous requirements, among others \citep{walia2009systematic, de2010ambiguity}. Specifically, a failure is a deviation of the software system behavior with respect to the expected requirements \citep{walia2009systematic}, or ``[t]he inability of a system or component to perform its required functions within specified performance requirements'' \citep{ieee-glossary} (IEEE 610.12-1990). The cost escalation of fixing errors associated with requirements in later development stages is substantial: the cost increase of detecting them during the design phase is estimated to be 3--8-fold higher, during the development phase 7--16 times higher, during the test and integration phase 21--78 times higher and during the operational phase 29--1500 times higher, in comparison to identifying them in the early \gls{rel} phase \citep{stecklein2004error}. 

\glsreset{nlr}
\begin{figure}
\centering
  \includegraphics[width=1.0\textwidth]{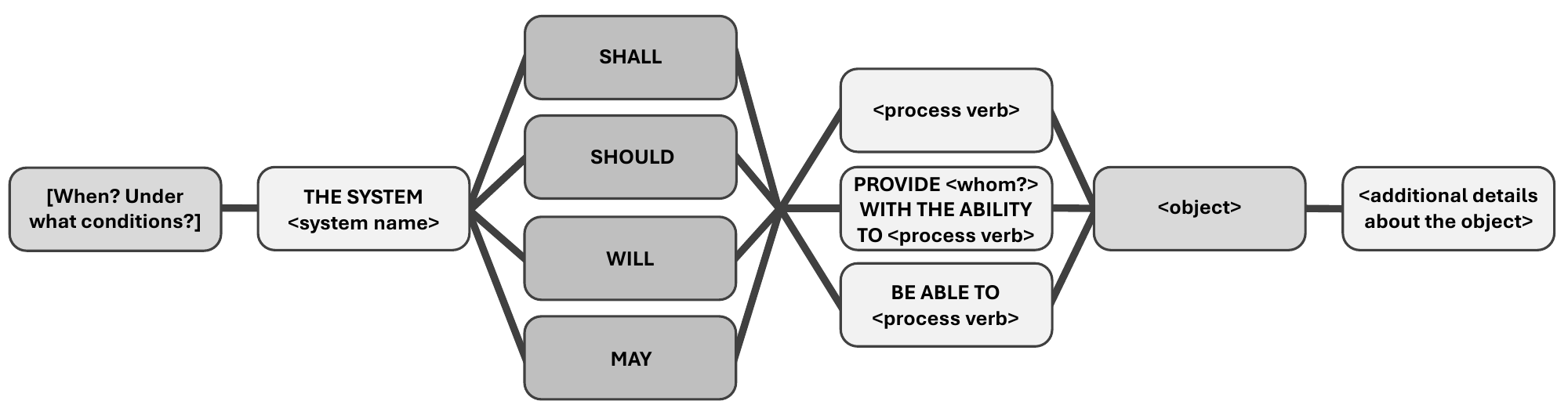}
  \caption[Requirements template for structured documentation of \glsentrylong{nlr} (\glsentryshort{nlr}) (adapted from \cite{pohl2016requirements})]
  {Requirements template for structured documentation of \gls{nlr} according to \cite{pohl2016requirements}, including temporal or logical constraints, the system name, a modal verb to distinguish between the different kinds of requirements, the process verb structure, the object of the requirement and additional details for the object, which can vary depending on the respective object (adapted from \cite{pohl2016requirements}).}
	\label{fig:chapter-2-template}
\end{figure}

\glsreset{sud}
Particularly, a requirement is defined as: ``\textit{(1)} A condition or capability needed by a user to solve a problem or achieve an objective. \textit{(2)} A condition or capability that must be met
or possessed by a system or system component to satisfy a contract, standard, specification, or other formally imposed documents.
\textit{(3)} A documented representation of a condition or capability as in \textit{(1)} or \textit{(2)}'' \citep{ieee-glossary} (IEEE 610.12-1990), as part of the standard glossary of \gls{se} terminology. Moreover, requirements are differentiated by their type, namely \textit{\gls{fr}}, which represent the actual functionality of the \gls{sud}, with regard to a result of a system behavior provided by a function of the software system \citep{pohl2010requirements, pohl2016requirements, glinz2011glossary}. In addition, \textit{\gls{nfr}}, also called \textit{quality requirements}, refer to quality-related aspects of the \gls{sud}, such as performance, scalability and availability, among others \citep{pohl2010requirements, pohl2016requirements, glinz2011glossary}. Further, \textit{constraints} are requirements which restrict the solution space for \gls{fr} and \gls{nfr} \citep{pohl2010requirements, pohl2016requirements, glinz2011glossary}. The overall \textit{functionality} of a system is defined by the collection of the respective \gls{fr} \citep{glinz2011glossary}. The entirety of requirements for a system is directly or indirectly influenced by the \textit{stakeholders}, such as the users and operators of the system, customers and testers, among others \citep{pohl2010requirements, pohl2016requirements}. Identifying relevant stakeholders is a crucial aspect of \gls{re}, since this directly affects the quality of the requirements \citep{glinz2007guest}.

\glsreset{ac}
\gls{re} encompasses a set of core activities, including \gls{rel}, \gls{rd} and specification, requirements negotiation, \gls{rval}, requirements verification and management \citep{kotonya1998requirements, pohl2010requirements, pohl2016requirements}. First, the main goal of \gls{rel} is to extract and gather all requirements from the stakeholders and related sources (e.g., legacy systems, existing documents), as well as refine requirements \citep{pohl2010requirements, pohl2016requirements}. To this end, stakeholders and \textit{system boundaries} need to be identified \citep{glinz2007guest}, separating the \gls{sud} from the \textit{system context}, which is the part of the environment that is relevant to the system \citep{nuseibeh2000requirements}. Second, elicited requirements are documented by employing different techniques, most commonly using \gls{nl}, but also via concept models \citep{pohl1996requirements, pohl2016requirements}. This requirements specification represents the input to subsequent \gls{sd} phases, which should be traceable, consistent and comprise the perspectives of all stakeholders \citep{pohl1996requirements}. In order to reduce limitations inherent to \gls{nl} while documenting requirements, such as incompletely specified conditions or process verbs, \textit{requirements templates} are often employed, which act as syntactical blueprints for the requirements, as depicted in Figure \ref{fig:chapter-2-template} \citep{pohl2016requirements}. Another widely applied and closely related technique of documenting \gls{nlr} is the employment of \glspl{us}, which are comprehensible, small functional descriptions that facilitate participatory design to tightly integrate users of the system into the elicitation and are primarily combined with agile methods \citep{cohn2004user, lucassen2016use}. Most often, the specification of \glspl{us} is conducted by employing the \textit{Connextra} template, namely, \textit{``As a <role>, I want to <action>, [so that <benefit>]''} \citep{lucassen2016use, dalpiaz2018agile}. When \glspl{us} are employed for specification, they are typically combined with \gls{ac}, which represent conditions for satisfaction, and are employed often in the form of the \textit{Given-When-Then template}, composed of three main building blocks, namely, \textit{``Given that <aCondition> when <anAction> then <aDesiredConsequence>''} \citep{ferreira2022towards}. In contrast to the \gls{nl}-based requirements specification, conceptual models are employed, for example, by using the \gls{uml} \citep{specification2007omg, pohl2016requirements}, providing technically more precise representations. Third, negotiation of requirements should ensure agreement among stakeholders regarding the gathered requirements \citep{pohl1996requirements}. Fourth, \gls{rval} is concerned with ensuring that the correct requirements have been elicited and specified, i.e. ensuring that a software system is developed which is aligned with the actual needs and intentions of the stakeholders, while \gls{rver} refers to checking the requirements regarding several aspects such as \textit{consistency}, ensuring the absence of internal contradictions in the requirements, and \textit{correctness}, referring to the absence of apparent mistakes in the requirements, among others \citep{pohl1996requirements, pohl2010requirements, pohl2016requirements}. Finally, requirements management is perceived as a continuous activity conducted orthogonally to other \gls{re} activities, ensuring the structuring of elicited and specified requirements, preparing requirements for later use and maintaining consistency as well as ensuring the implementation of the requirements \citep{pohl2016requirements}.

\begin{figure}
\centering
  \includegraphics[width=1.0\textwidth]{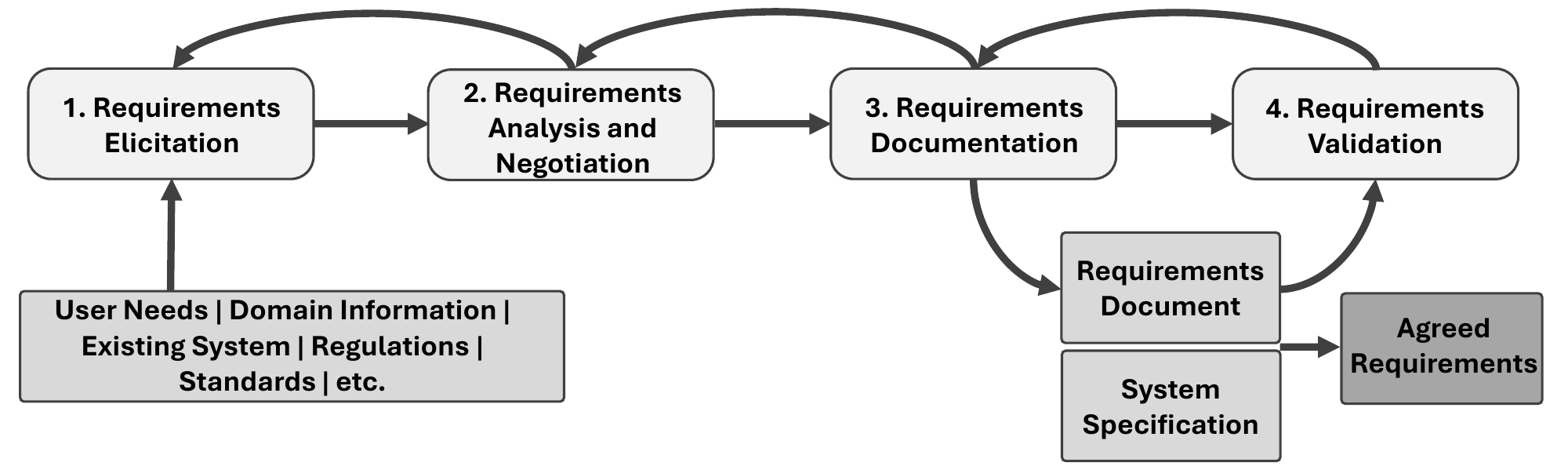}
  \caption[Linear and incremental \gls*{re} process model (adapted from \cite{kotonya1998requirements})]{\gls*{re} process model with different requirements artifacts, activities and their relationships: \textit{(1)} \gls*{rel}, \textit{(2)} requirements analysis and negotiation, \textit{(3)} \gls*{rd} and \textit{(4)} \gls*{rval}, depicted as a linear, incremental process (adapted from \cite{kotonya1998requirements}).}
	\label{fig:chapter-2-re-process-linear}
\end{figure}

Prior research described different \gls{re} process models, encompassing the various \gls{re} activities and depicting their relationships. For example, \cite{kotonya1998requirements} describe a linear, incremental process starting from elicitation, analysis and negotiation and progressing to documentation and validation, while activities allow interactions with preceding activities, as illustrated in Figure \ref{fig:chapter-2-re-process-linear}. Elicitation is fed by user needs, domain information, existing systems, regulations and standards, among others \citep{kotonya1998requirements}. Moreover, requirements documentation leads to a manifested requirements specification document, which is inspected during validation. In addition, the system specification is based on the created requirements specification. After successful requirements validation, the process model depicts an agreed requirements specification as the \gls{re} process model output \citep{kotonya1998requirements}. 

\begin{figure}
\centering
  \includegraphics[width=1.0\textwidth]{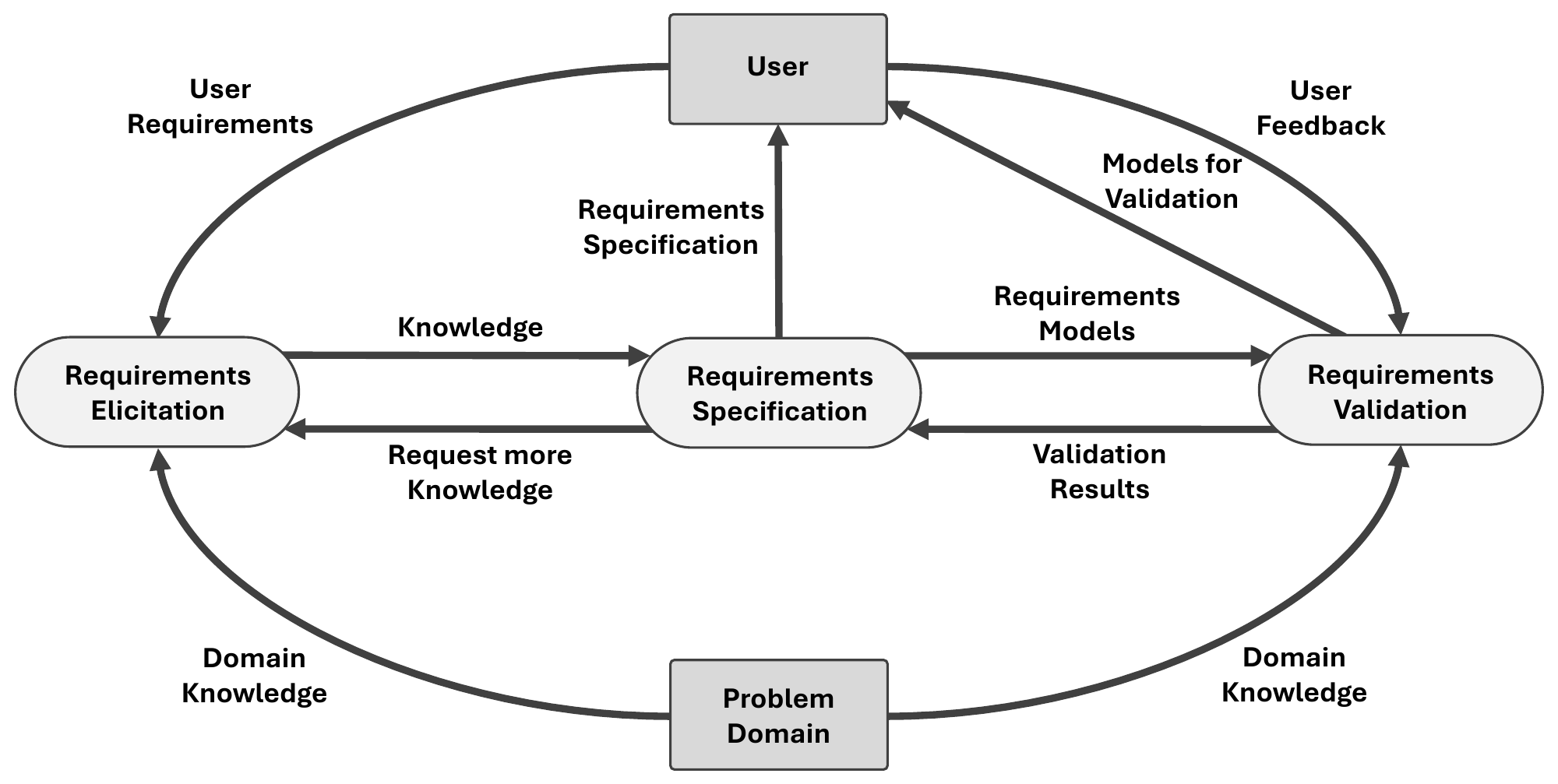}
  \caption[Iterative \gls*{re} process model (adapted from \cite{loucopoulos1995system})]{\gls*{re} process model illustrating the user and domain, information flow and \gls{re} activities of \gls*{rel}, requirements specification and \gls*{rval}, depicted as a non-linear, highly connected, iterative process model (adapted from \cite{loucopoulos1995system}).}
	\label{fig:chapter-2-process-iterative}
\end{figure}

In contrast, \cite{loucopoulos1995system} describe the \gls{re} relationships between activities as a non-linear, iterative process, showing the \gls{re} activities of \gls{rel}, requirements specification and \gls{rval}, as depicted in Figure \ref{fig:chapter-2-process-iterative}. Knowledge from the elicitation phase fuels the specification and, if necessary, additional elicitation is conducted to improve the specification \citep{loucopoulos1995system}. Requirements models are derived (e.g., conceptual models or prototypes), which fuel the validation in combination with the specification, by incorporating the domain knowledge and the user to obtain feedback, which is fed back to the specification and, if necessary, updates to the specification are conducted. Moreover, user requirements directly fuel the elicitation, additionally supported by incorporating domain knowledge \citep{loucopoulos1995system}. As can be observed from the possible interactions and relationships between the \gls{re} activities, their process model emphasizes a highly non-linear, iterative procedure.

While we presented two different \gls{re} process models, other similar processes are discussed in the research literature. For example, \cite{escalona2003requirements} describe the \gls{re} process using an \gls{uml} activity diagram, indicating iterative interrelations between elicitation, specification and validation activities. Furthermore, \cite{sawyer1997requirements} present the \gls{re} process as a spiral, emphasizing the iterative nature of \gls{re}, and integrate elicitation, analysis and validation, as well as requirements negotiation into the process. Both the cost and the quality of the requirements information increase with successive iterations \citep{sawyer1997requirements}. On the contrary, \cite{macaulay2012requirements} describe a simple, linear \gls{re} process including conceptualization, problem analysis, feasibility, analysis and modeling as well as requirements documentation. Other research also indicates that \gls{re} activities are highly intertwined, yet argues that the \gls{re} process should rather be considered as a collection of different process chunks, instead of a monolithic process \citep{houdek2000analyzing}. Finally, while different \gls{re} process models are described in research with varying characteristics, most share commonalities regarding the encompassed \gls{re} activities and stress the heavy relatedness of activities as well as the iterative nature of \gls{re}.

\begin{figure}
\centering
  \includegraphics[width=1.0\textwidth]{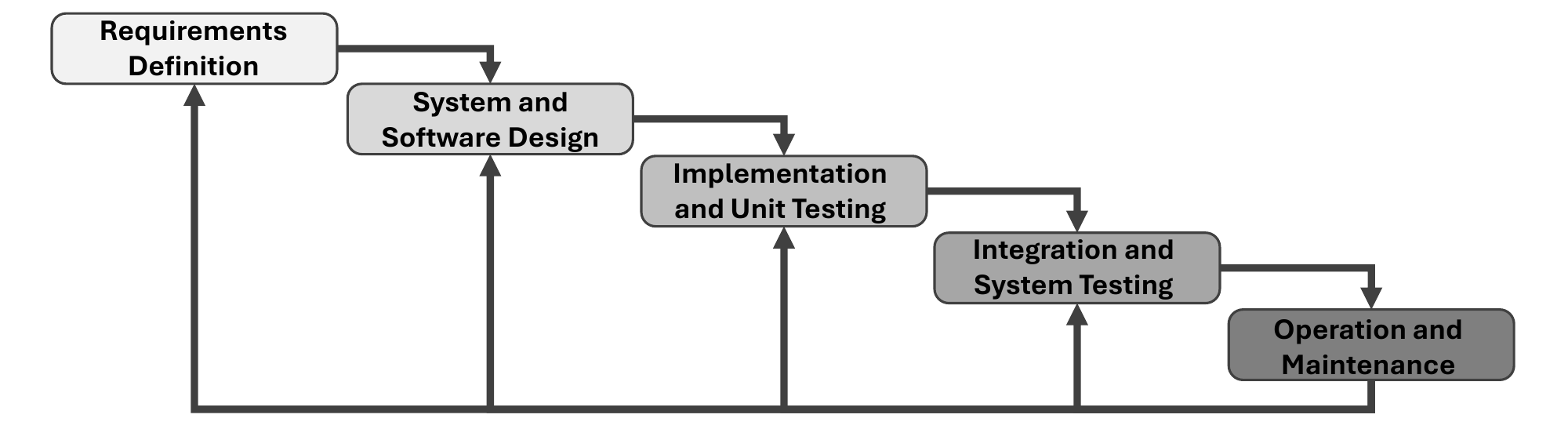}
  \caption[Waterfall model for \gls{sd} (adapted from \cite{sommerville2011software})]{\glsreset{sd}Waterfall Model for \gls{sd} including requirements definition, system and software design, implementation and unit testing, integration and system testing as well as operation and maintenance (adapted from \cite{sommerville2011software}).}
	\label{fig:chapter-2-waterfall-se}
\end{figure}

Before diving deeper into the \gls{re} activities most important to this thesis, namely \gls{rel} and \gls{rval}, we first provide a brief overview of the discipline of \gls{se} and its connection to \gls{re}. \gls{se} represents a systematic approach to the creation of software systems, encompassing a particular set of activities and techniques \citep{sommerville2011software}, or \gls{se} is described as ``[...] [t]he application of a systematic, disciplined, quantifiable approach to the development, operation, and maintenance of software [...]'', as defined by the standard \gls{se} glossary \citep{ieee-glossary} (IEEE 610.12-1990). Specifically, the \gls{se} process is characterized by activities of \textit{software specification}, ensuring that requirements are gathered and defined together with the stakeholders, \textit{\gls{sd}}, which is concerned with designing and programming the software system, \textit{software validation}, ensuring that the \gls{sud} is aligned with the actual stakeholder needs and \textit{software evolution}, ensuring a continuous adaptation and maintenance of the software to reflect changing requirements \citep{sommerville2011software}. Over the years, different \gls{se} processes have been described in research, commonly referred to as \gls{sdlc}, describing the usual phases from early planning to late maintenance of software systems, where connections between these activities depend on the concrete development approach \citep{ieee-glossary} (IEEE 610.12-1990). For example, the \textit{waterfall model} describes the \gls{se} process as a sequential, incremental one, where each phase depends on the completion of the preceding phase, as shown in Figure \ref{fig:chapter-2-waterfall-se} \citep{sommerville2011software}. This particular variant of the waterfall model described by \cite{sommerville2011software} allows for feedback loops to previous stages and consists of five phases, including requirements definition, system and software design, implementation and unit testing, integration and system testing as well as operation and maintenance, while other variants such as the original one from \cite{royce1987managing} comprise seven phases and are strictly linear. As apparent from the described \gls{se} process, requirements definition represents the first phase and illustrates the interrelation between \gls{re} and \gls{se}. The previously described discipline of \gls{re} is an important part of \gls{se}, ensuring an understanding of the needs of the stakeholder and providing appropriate requirements specifications, while \gls{se} is concerned with building the software, encompassing the actual design and architecture, implementation, testing and integration activities.


\glsreset{sver}

On the contrary, agile \gls{se} processes emerged, emphasizing a more flexible, iterative approach \citep{fowler2001agile, abad2018loud}. Within agile \gls{se} processes, the individual iterations are considered self-contained, spanning all previously described \gls{se} activities from requirements definition and analysis to implementation and testing, resulting in a release, which represents an evolving subset of the final product \citep{boehm2007survey}. Each iteration is usually conducted in a short time, enabling the quick incorporation of stakeholder feedback based on the previous and current product evolution and therefore rapid adaptation to changing requirements \citep{boehm2007survey}. While different agile approaches exist, for example, \textit{extreme programming} \citep{beck2000extreme} and \textit{adaptive software development} \citep{highsmith2013adaptive}, the most popular approach is \textit{Scrum} \citep{schwaber2001agile}.

\subsection{Requirements Elicitation (REl)}

\begin{figure}
\centering
  \includegraphics[width=1.0\textwidth]{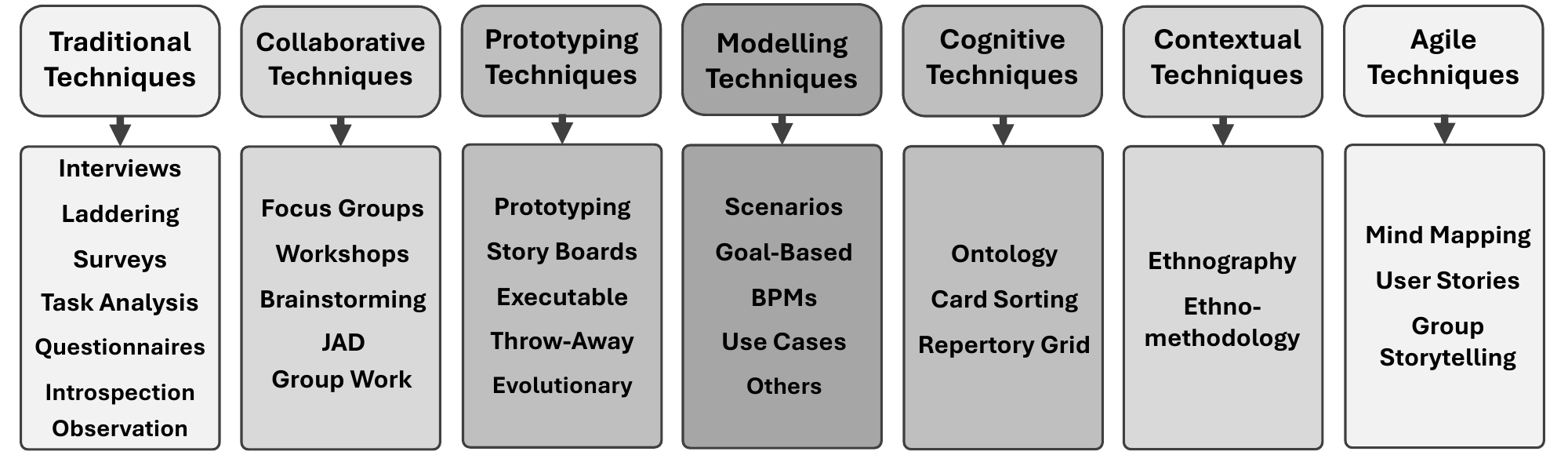}
  \caption[\gls{rel} techniques categorization (adapted from \cite{pacheco2018requirements})]{Categorization of \gls{rel} techniques covering traditional, collaborative, prototyping, modeling, cognitive, contextual and agile (adapted from \cite{pacheco2018requirements}).}
	\label{fig:chapter-2-process-elicitation}
\end{figure}

\gls{rel} represents an important activity comprised in the \gls{re} process, focusing on understanding the actual requirements of different stakeholders, including uncovering, extracting and refining requirements and providing respective requirements specifications as a means of communication to the developers \citep{zowghi2005requirements, pohl2010requirements, pohl2016requirements}. For conducting effective \gls{rel}, initially, the application domain needs to be explored, investigated and understood, providing a foundation for subsequent \gls{rel} activities \citep{zowghi2005requirements}. Next, identifying and analyzing all relevant requirements sources is instrumental, which has a large effect on the quality of the elicited requirements \citep{glinz2007guest}, and includes various stakeholders, existing documents, such as standards or domain- and organization-specific ones, as well as systems in operation, such as legacy systems \citep{alexander2002writing, zowghi2005requirements, pohl2010requirements, pohl2016requirements}. Afterwards, the appropriate \gls{rel} approaches, techniques and tools are selected, which may vary depending on the problem context, domain, constraints and project characteristics \citep{zowghi2005requirements, pohl2016requirements}, since no universal methodology or technique for effective elicitation exists \citep{hickey2003elicitation, pohl2010requirements, pohl2016requirements}. Most often, a combination of different elicitation techniques is employed, depending, for example, on the phase of the development, the types of requirements to be elicited, the experience of the analyst with a specific technique as well as budget and time constraints \citep{zowghi2005requirements, pohl2010requirements, pohl2016requirements}. The elicitation activities are typically conducted in an iterative manner, where each iteration provides an increasing level of detail and quality of the requirements specification \citep{zowghi2005requirements}, embracing the natural evolution and co-creation of requirements during elicitation \citep{ferrari2022requirements}. As discussed earlier, the \gls{re} and \gls{se} activities are highly intertwined and iterative.

To effectively elicit requirements, a plethora of different approaches has been proposed in research before and a comprehensive categorization of these techniques is shown in Figure \ref{fig:chapter-2-process-elicitation} \citep{pacheco2018requirements}. This includes traditional techniques, for example, \textit{interviews} \citep{agarwal1990knowledge} conducted in an unstructured, structured or semi-structured approach, with varying levels of guidance based on predefined questions or templates \citep{zowghi2005requirements}. A particular interview technique is represented by \textit{laddering}, asking a series of probing questions and arranging the gathered knowledge in a hierarchical manner \citep{hinkle1965change, corbridge1994laddering}. Moreover, \textit{task analysis} represents a top-down technique decomposing tasks into subtasks, to uncover all events and actions of processes the users and systems conduct \citep{carlshamre1996usability}. In addition, \textit{Joint Application Development (JAD)} represents a structured approach incorporating all stakeholders to jointly discuss problems and solutions, facilitating rapid decision making and resolution of issues \citep{wood1995joint}. While \textit{prototyping} is another effective elicitation technique leveraging abstracted versions of the \gls{sud} \citep{sommerville2011software}, on which we focus in this work, we provide an introduction in Section \ref{sec:cha-2-prototpying}. For a summary of other techniques, such as \textit{scenarios}, \textit{repertory grids} and \textit{ethnographies}, we refer the interested reader to \cite{zowghi2005requirements} and \cite{pacheco2018requirements}.


\subsection{Requirements Validation (RVal)}

\gls{rval} is the \gls{re} activity concerned with ensuring that the elicited and specified requirements correctly reflect the actual needs of the involved stakeholders \citep{boehm1984verifying, pohl1996requirements, pohl2010requirements, pohl2016requirements}. In contrast, requirements verification encompasses procedures and techniques to ensure that the requirements specification is correct by detecting internal errors, such as contradictions or inconsistencies \citep{sakthivel1991survey}. To increase validation quality, \gls{rval} incorporates multiple principles: ensuring that the adequate stakeholders are involved, decoupling the detection and correction of deviations, conducting validation from multiple perspectives, employing different requirements document types with varying advantages and weaknesses (e.g., text-based and graphical models), constructing development artifacts (e.g., test cases) and repeating validation, embracing the continuous evolution of the knowledge, understanding and the needs of stakeholders \citep{pohl2016requirements}.

\begin{figure}
\centering
  \includegraphics[width=1.0\textwidth]{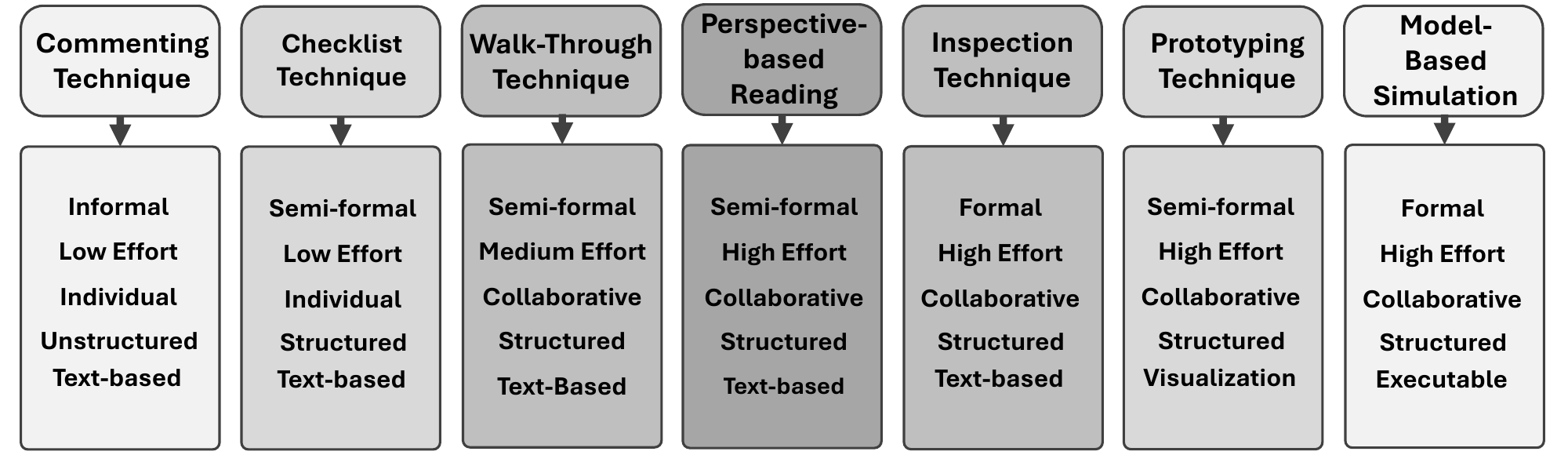}
  \caption[Overview of \gls{rval} techniques and their characteristics]{Overview of \gls{rval} techniques, including commenting, checklist, walk-through, perspective-based reading, inspection, prototyping and model-based simulation, each with characteristics of formality, effort, collaboration, structuredness and artifact type.}
	\label{fig:chapter-2-validation}
\end{figure}

\noindent To ensure effective \gls{rval}, numerous techniques have been proposed in research before \citep{pohl1996requirements}. An overview of well-known and widely studied validation techniques, each with their main characteristics, is shown in Figure \ref{fig:chapter-2-validation}. For example, \textit{commenting} is a technique where the author of a requirement gathers reviews from others (e.g., co-workers) and represents an informal, low-effort and text-based validation method \citep{pohl2010requirements, pohl2016requirements}. Moreover, \textit{inspections} represent a structured approach for collaboratively investigating a predetermined set of requirements for their correctness \citep{laitenberger2000encompassing, pohl2016requirements}, while \textit{walk-throughs} represent a lightweight version of an inspection, enforcing a less systematic and strictly organized error investigation approach, yet enabling collaborative improvement of the understanding of the requirements and detection of requirements errors \citep{pohl2016requirements}. In addition, \textit{prototyping} enables stakeholders to experience the software system as a tangible artifact, enabling more detailed feedback \citep{jones2007estimating, pohl2010requirements, pohl2016requirements}. Apparently, the prototyping technique that we previously discussed in the context of \gls{rel} can similarly be applied to the problem of \gls{rval}. This indicates the versatility of incorporating simplified, abstracted versions of the software system in the \gls{re} process, to facilitate both \gls{rel} and \gls{rval}. Next, we discuss the related concepts of both \gls{sval} and \gls{sver}.




\subsection{Software Validation (SVal) and Verification (SVer)}

\begin{figure}
\centering
  \includegraphics[width=1.0\textwidth]{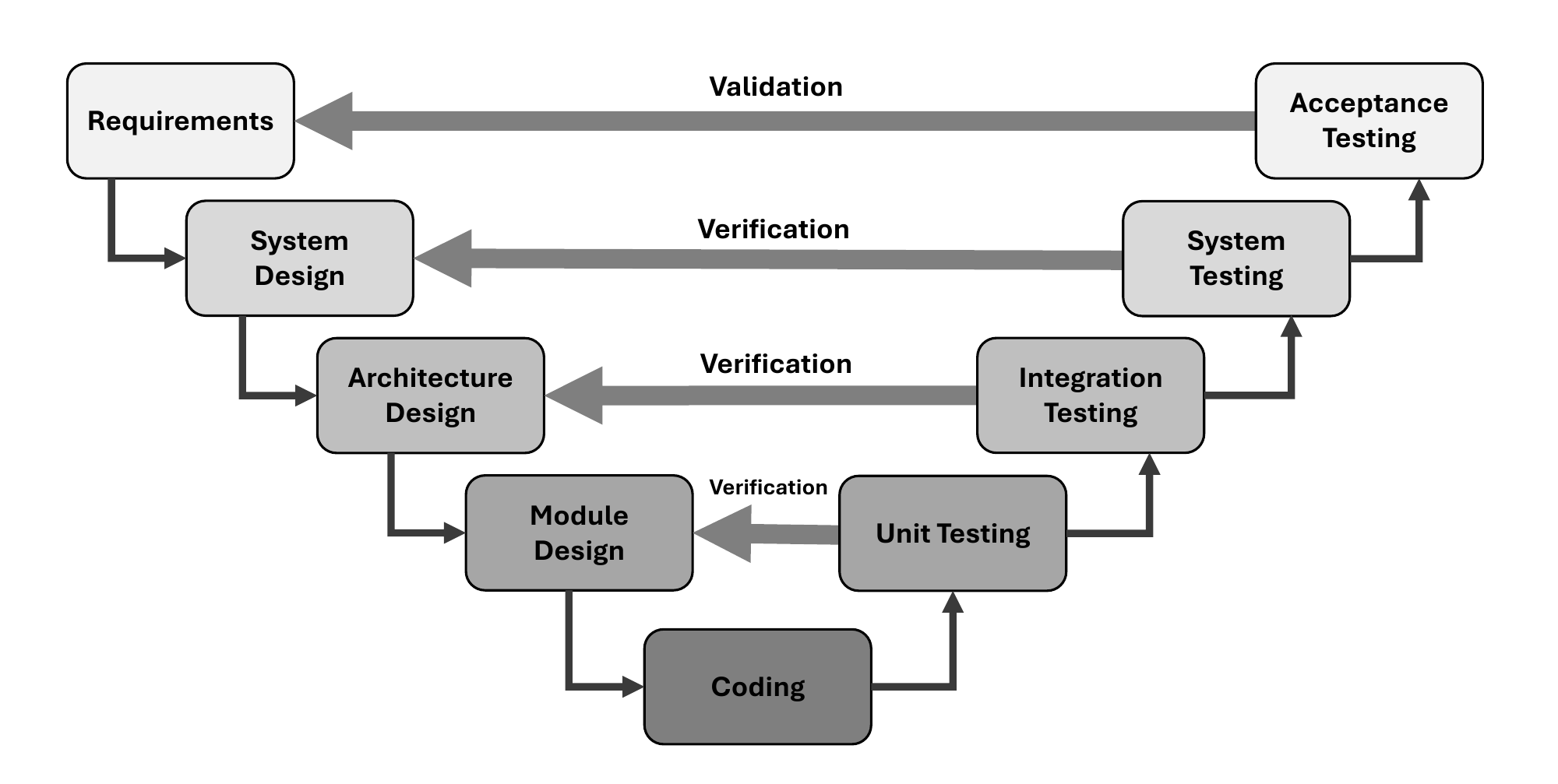}
  \caption[V-Model for \gls{sd} (adapted from \cite{sommerville2011software})]{V-model with multiple abstraction levels from requirements definition, system specification, system architecture to component design on the right-hand side, and respective testing activities, namely component, integration, system and acceptance testing, on the left-hand side, with verification and validation (adapted from \cite{sommerville2011software}).}
	\label{fig:chapter-2-v-model}
\end{figure}

\gls{sval} and \gls{sver} (often also abbreviated by the acronym \textit{V $\mathit{\&}$ V} \citep{ieee-glossary} (IEEE 610.12-1990)) represent two related activities in the \gls{se} process, ensuring that the developed software system performs according to specification as well as aligns with the actual stakeholder needs \citep{adrion1982validation, boehm1984verifying, sommerville2011software}. In particular, validation is defined as ``[...] [t]he process of evaluating a system or component during or at the end of the development process to determine whether it satisfies specified requirements [...]'', and verification is defined as ``[...] [t]he process of evaluating a system or component to determine whether the products of a given development phase satisfy the conditions imposed at the start of that phase [...]'', in the standard \gls{se} glossary \citep{ieee-glossary} (IEEE 610.12-1990). Moreover, according to \cite{william1996guide} (PMBOK), validation is defined as ``[t]he assurance that a product, service, or system meets the needs of the customer and other identified stakeholders [...]'' and ``[...] [i]t often involves acceptance and suitability with external customers [...]'', whereas verification is defined as ``[t]he evaluation of whether or not a product, service, or system complies with a regulation, requirement, specification, or imposed condition [...]'' and as ``[...] [i]t is often an internal process [...]'' \citep{william1996guide}. More informally, \cite{boehm1984verifying} defines validation as ``[Are we] building the right product?'' and verification as ``[Are we] building the product right?''. From these various definitions, it can be observed that validation refers more to an external process of ensuring alignment with the actual needs of the stakeholders, while verification is rather an internal process, ensuring that the software products or intermediate development artifacts adhere to their specification. In addition to the previously discussed \gls{se} process models, or the \gls{sdlc} models, another popular \gls{se} process representation is the \textit{V-model}, clearly representing the validation and verification processes in their connection to other \gls{sdlc} activities \citep{forsberg1991relationship, johansson1999v, sommerville2011software}, which is illustrated in Figure \ref{fig:chapter-2-v-model}. The \gls{sdlc} activities in the V-model start on the left-hand side with the requirements specification, followed by the derived system design, the architecture design and the module design and finish at the implementation level, representing the lowest, most detailed representation, and each step represents a further manifestation of the specification from the preceding level. The vertical axis from top to bottom represents an increasing level of detail, while the horizontal axis represents the increasing time \citep{sommerville2011software}. Each activity on the left-hand side is directly linked to verification and validation activities on the right-hand side. They start from low-level unit testing of the modules implemented according to the module specification and proceed to higher-level integration testing based on the architecture specification, to system testing given the system specification and finally to acceptance testing, which involves actual users. While the acceptance testing represents validation, involving the users to confirm that the system implements their actual needs, the remaining lower-level testing activities are considered verification, checking an implemented version of a specification against the actual specification at different granularity levels \citep{sommerville2011software}.

Apparently, similar to the previously discussed \gls{sdlc} models, the V-model closely integrates the \gls{re} activities into the process, where lower-level specifications of the system are all directly derived from the high-level requirements specification. However, the notion of validation and verification in the \gls{re} discipline and \gls{sdlc} models is slightly different. While \gls{rval} and \gls{sval} share the same idea of validating if the specified requirements align with actual stakeholder needs, they differ in the artifact that is being validated, namely, requirements or early prototypes for \gls{rval} and the actual software product in later stages of development for \gls{sval}. Within \gls{rver}, the requirements are inspected for internal consistency and correctness, while \gls{sver} focuses on testing the conformity of an implemented system or system component with its respective specification. Next, we discuss \gls{gui} prototyping, covering history, process models and prototype characteristics.








\section{\gls{gui} Prototyping}
\label{sec:cha-2-prototpying}

\gls{gui} prototyping represents one of the most effective techniques for the elicitation and validation of requirements, facilitating the communication and discussion of specified and understood requirements via tangible, visual artifacts of the \gls{sud} \citep{baumer1996user, pohl1996requirements, zowghi2005requirements, pohl2010requirements, teixeira2014requirements, pohl2016requirements}. \gls{gui} prototyping refers to a collection of techniques and processes for creating a simplified, abstracted, small-scale version of an actually complex software system, thereby reducing the costs and risks compared to developing the entire software system \citep{szekely1994user, baumer1996user}. Prototyping techniques range from creating highly abstract, simple pen-and-paper sketch-based prototypes to drawing more detailed mockups, defined as ``[...] model[s] or replica[s] of a machine or structure, used for instructional or experimental purposes'' \citep{dictionary1989oxford} and creating sophisticated high-fidelity prototypes that possess a look-and-feel close to the final product \citep{szekely1994user, baumer1996user}. For \gls{rel}, \gls{gui} prototyping is particularly relevant for interactive and user-facing software systems, often used in conjunction with other elicitation techniques \citep{zowghi2005requirements} and helps to elicit more requirements \citep{abad2018loud}. Moreover, \gls{rval} is enhanced due to the \gls{gui} prototype providing a less ambiguous representation as a visual manifestation of the specified requirements \citep{jones2007estimating, pohl2010requirements, pohl2016requirements}. Additionally, this facilitates the integration of stakeholders early in the development, reducing potential misunderstandings \citep{pohl2010requirements, pohl2016requirements}. Next, we provide an overview of the history of \gls{gui} prototyping tools and editors, discuss common \gls{gui} prototyping process models from the research literature, and introduce \gls{gui} prototype characteristics such as different fidelity levels and types of prototypes as well as typical \gls{gui} representations.

\subsection{History of \gls{gui} Prototyping Tools}

\begin{figure}
\centering
  \includegraphics[width=1.0\textwidth]{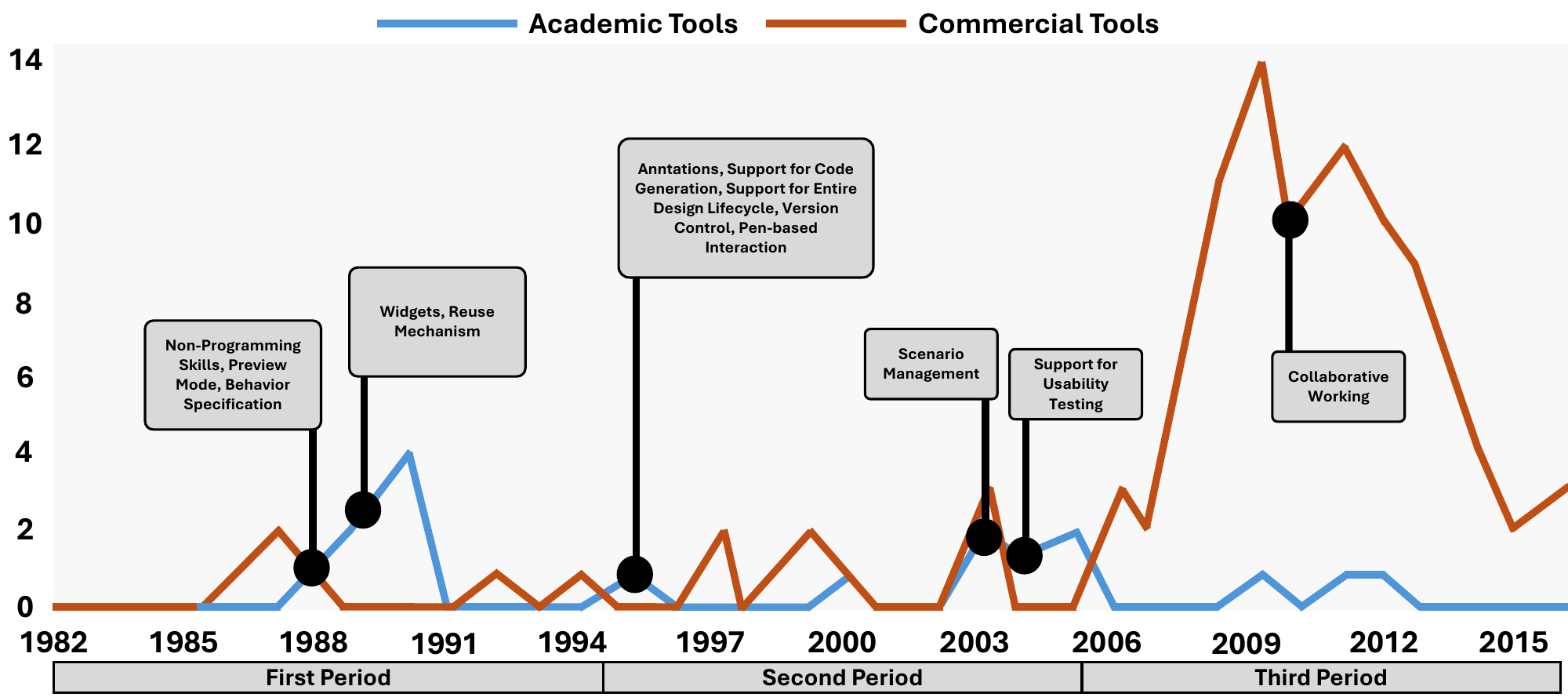}
  \caption[History overview of the development of \gls{gui} prototyping editors and tools (adapted from \cite{silva2017comparative})]{History overview (1982 to 2015) of the development of \gls{gui} prototyping editors and tools differentiated by academic and commercial tools, annotated with multiple milestones in the development of prototyping tools (adapted from \cite{silva2017comparative}).}
	\label{fig:chapter-2-history-of-prototyping-tools}
\end{figure}

The interest in providing support for \gls{gui} prototyping through tools and editors both in research and for commercial use dates back around forty years, with early approaches for rapid \gls{gui} prototyping such as \textit{Mirage} \citep{mcdonald1988mirage}. \cite{silva2017comparative} provide a detailed historic overview of the development of \gls{gui} prototyping approaches, which is depicted in Figure \ref{fig:chapter-2-history-of-prototyping-tools}, showing the yearly occurrence of novel tools both in research and for commercial use as well as milestones in the development. The rise of \gls{gui} prototyping approaches was largely based on the interest in User Interface Management Systems (UIMS) \citep{kasik1982user}, which tried to decouple the \gls{gui} representation from the business logic of the software system \citep{silva2017comparative}. 

In the beginning, the focus of proposed approaches lay heavily on supporting non-programming skills for the creation of \gls{gui} prototypes, enabling a shift of the focus to the content and functionality of the \gls{gui} away from implementation details, thereby simplifying the overall prototyping procedure, such as in \textit{DENIM} \citep{lin2000denim, silva2017comparative}. To this end, one central notion was to enable users to directly manipulate the visual representation of the \gls{gui} \citep{silva2017comparative}, informally often referred to as \textit{WYSIWYG (what you see is what you get)}. Another line of early \gls{gui} prototyping tools focused on supporting pen-based interaction, enabling the recording of hand-drawn sketches of the \gls{gui} \citep{silva2017comparative}, while other approaches such as \textit{SILK} \citep{landay1995interactive} and \textit{SketchiXML} \citep{coyette2006sketchixml} also support the translation of sketches into less ambiguous widget-based prototypes with automated recognition techniques. \gls{gui} widgets represent predefined, essential \gls{gui} components, basically the smallest building blocks of \gls{gui} prototypes. Early tools, such as \textit{Lapidary} \citep{zanden1991lapidary} and most modern prototyping tools, provide large widget libraries to choose from, accelerating and simplifying the prototyping process \citep{silva2017comparative}. Widget libraries can incorporate more abstract widgets (e.g., basic shapes) for building low-fidelity prototypes, such as in \textit{Balsamiq} \citep{faranello2012balsamiq}, or more detailed widgets (e.g., buttons, text input fields, drop-down selections, list items) possessing a visual representation close to the final system for high-fidelity prototypes, such as in \textit{Mockplus} \citep{mockplus}.

Another major aspect introduced early into \gls{gui} prototyping tools was the ability to specify dynamic behavior, typically characterized by a collection of \gls{gui} prototype states reachable via defined transitions, in addition to the otherwise static nature of \gls{gui} prototypes \citep{silva2017comparative}. For example, prototyping approaches based on sketches such as \textit{Marvel} \citep{silva2017comparative} enable simplistic dynamic behavior by employing \textit{hotspots}, which represent rectangular shapes drawn on top of the sketch that can be linked to various actions and transition behaviors. Similarly, tools like \textit{SILK} \citep{landay1995interactive} and \textit{DENIM} \citep{lin2000denim, silva2017comparative} support merely fundamental wireframe (a basic, simplified layout and structure of the \gls{gui} prototype) interactions, represented by direct links between prototype states. More advanced prototyping tools such as \textit{Figma} \citep{figma} that rely on building \gls{gui} prototypes with widget libraries typically offer widget-specific event handlers, often also allowing the definition of variables and conditions to support more complex dynamic behavior specifications, but simultaneously require more effort to create.

An additional important development of \gls{gui} prototyping editor functionality includes the preview mode, rapidly enabling users to explore an interactive representation of the prototype based on the predefined dynamic behavior, implemented already in early tools such as \textit{Lapidary} \citep{zanden1991lapidary}, \textit{Mirage} \citep{mcdonald1988mirage}, \textit{SILK} \citep{landay1995interactive} and most modern editors such as \textit{Figma} \citep{figma}. Another key functionality introduced in early prototyping editors was a reuse mechanism, for example, facilitated by the employment of widget libraries, predefined \gls{gui} templates and prespecified behaviors \citep{silva2017comparative}, fostering the acceleration of \gls{gui} prototyping. Much later introduced in prototyping tools was the ability to create \gls{gui} prototype annotations to capture additional important information such as user feedback, problem reports or design preferences, during the prototype creation and usability testing phases \citep{silva2017comparative}. At the same time, prototyping tools started to encompass code generation capabilities to directly transform the \gls{gui} prototype specification into executable program code, thereby reducing the development gap between prototyping and implementation stages \citep{silva2017comparative}. In later stages, prototyping editors incorporated scenario management, the ability to integrate written scenarios into built \gls{gui} prototypes, the support for usability testing, in which users are asked to conduct several tasks with the prototype and data is recorded for future analysis, and eventually collaborative working modes, which allow multiple developers to simultaneously work on an integrated and distributed \gls{gui} prototype \citep{silva2017comparative}.

To summarize, the first period of \gls{gui} prototyping editors was characterized by the discussed UIMS tools, decoupling the \gls{gui} code from the rest of the software system with a focus on non-programming skills, behavior specification and widget libraries \citep{silva2017comparative}. The second phase was mainly characterized by the introduction of additional useful features, including annotations, version control and code generation, while in the third stage there was a large increase in available commercial \gls{gui} prototyping tools \citep{silva2017comparative}. The three different phases are depicted in Figure \ref{fig:chapter-2-history-of-prototyping-tools}. More recently, many commercial editors supporting \gls{gui} prototyping exist, such as \textit{Sketch} \citep{sketch}, \textit{Adobe XD} \citep{adobexd}, \textit{Mockplus} \citep{mockplus}, \textit{Figma} \citep{figma} and \textit{Balsamiq} \citep{faranello2012balsamiq}.

\subsection{\gls{gui} Prototyping Processes and Approaches}
\label{subsec:gui-prototyping-processes}

\begin{figure}
\centering
  \includegraphics[width=1.0\textwidth]{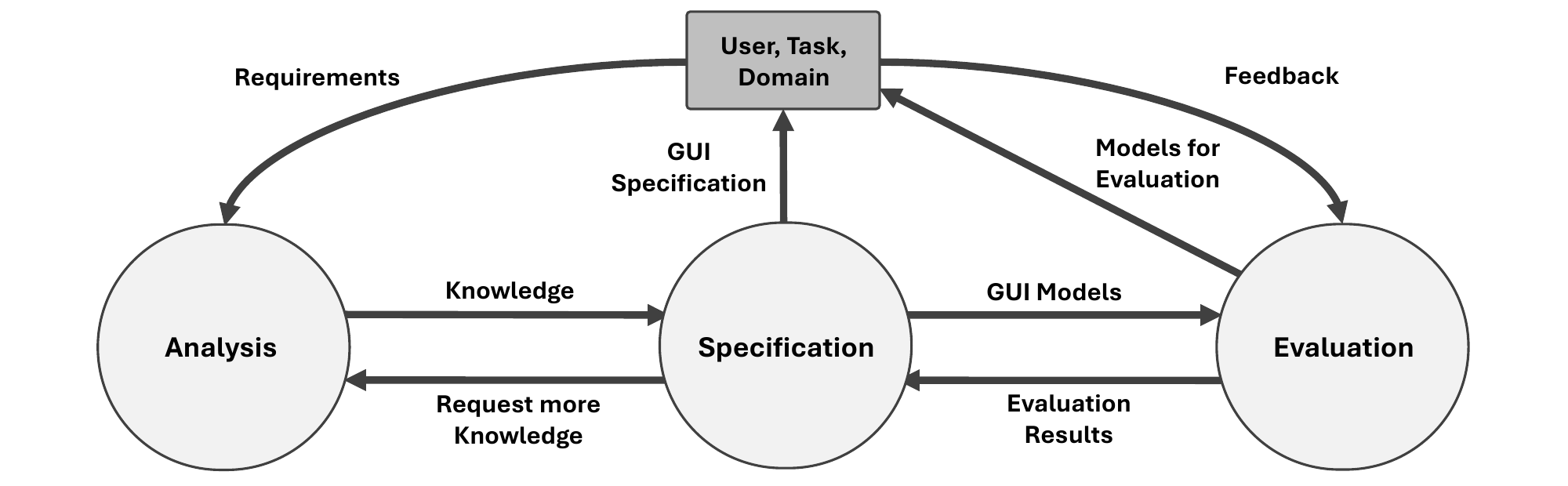}
  \caption[Iterative \gls{gui} prototyping process model (adapted from \cite{van2001human})]{Iterative \gls{gui} prototyping process model including activities of requirements analysis, \gls{gui} specification and \gls{gui} model evaluation based on obtaining detailed feedback from the user, task and domain (adapted from \cite{van2001human}).}
	\label{fig:chapter-2-prototyping-process-basic}
\end{figure}

Multiple \gls{gui} prototyping process models were discussed in research before, describing the key activities and artifacts involved in the development of interactive systems. For example, \cite{van2001human} present an iterative \gls{gui} prototyping process model that includes the requirements analysis from the user, task and domain, the \gls{gui} specification and evaluation of \gls{gui} models based on their feedback, as depicted in Figure \ref{fig:chapter-2-prototyping-process-basic}. As can be observed, this prototyping process model closely resembles the \gls{re} process model from \cite{loucopoulos1995system}, discussed as part of Section \ref{sec:re_and_se} in Figure \ref{fig:chapter-2-process-iterative}, while key aspects are adapted to \gls{gui} models. Requirements are gathered and analyzed in a task-centered manner, based on existing inadequacies as well as articulation by the users. Accordingly, a \gls{gui} specification is derived, \gls{gui} models are created and evaluated based on feedback from the user, task and domain \citep{van2001human}.

\begin{figure}
\centering
  \includegraphics[width=1.0\textwidth]{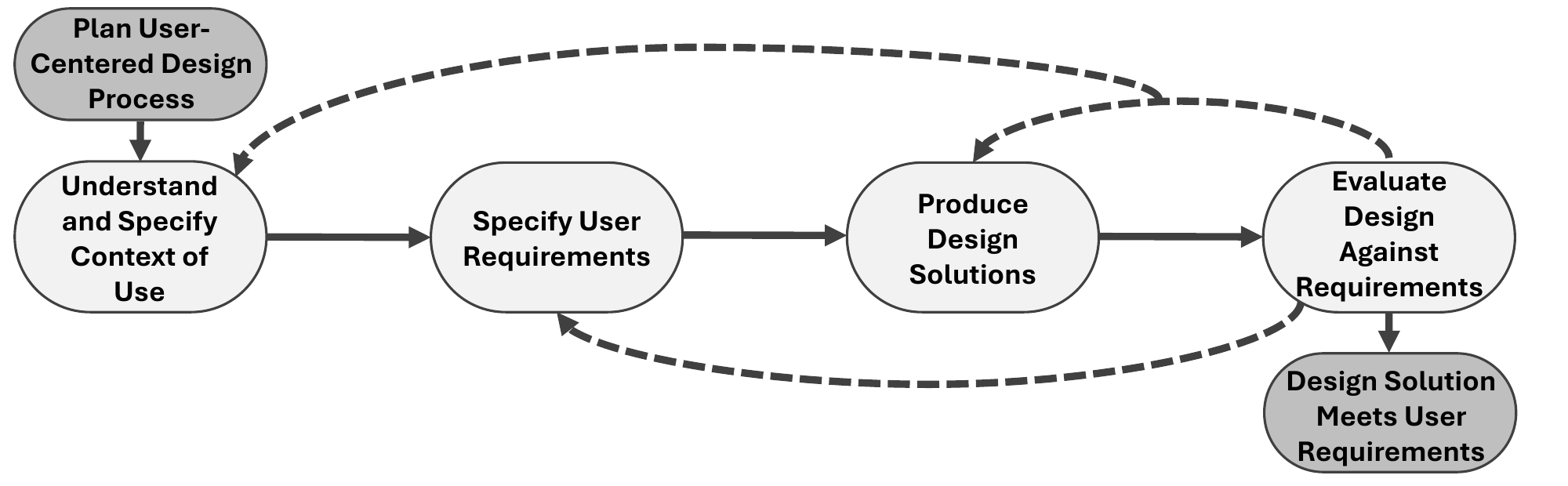}
  \caption[\glsentrylong{ucd} process model (adapted from \cite{zimmermann2007requirement})]{Iterative \gls{ucd} process model including the planning of the design process, understanding and specification of context of use, specification of user requirements, creation of design solutions and finally the evaluation of design solutions against the previously specified requirements (adapted from \cite{zimmermann2007requirement}).}
	\label{fig:chapter-2-prototyping-process-ucd}
\end{figure}

\begin{figure}
\centering
  \includegraphics[width=1.0\textwidth]{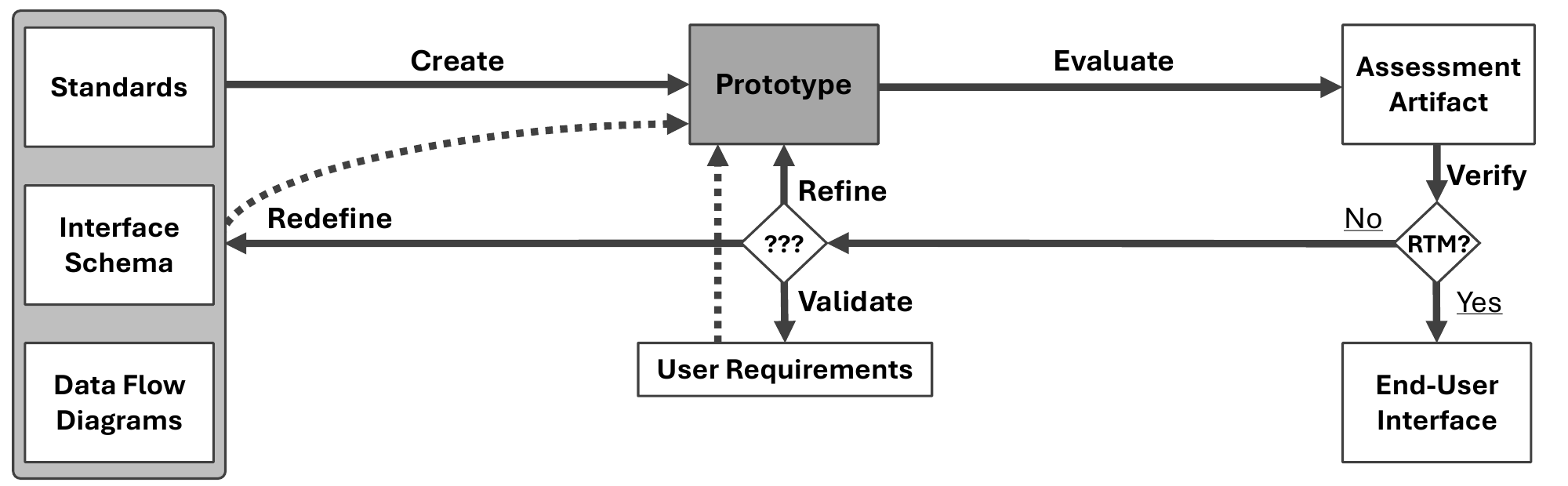}
  \caption[\gls{gui} prototyping process model (adapted from \cite{weichbroth2015user})]{Iterative \gls{gui} prototyping process model including the creation of the prototype based on standards, interface schema and data flow diagrams, evaluating the \gls{gui} prototype artifact, validating requirements and accordingly refining the \gls{gui} prototype as well as the respective \gls{gui} specifications (adapted from \cite{weichbroth2015user}).}
	\label{fig:chapter-2-prototyping-process-newer}
    \vspace{-0.5cm}
\end{figure}

A similar process model is defined by \gls{ucd} for interactive systems (DIN EN ISO 9241-220) \citep{din_ucd}, which encompasses the planning of the \gls{ucd} process, understanding and specifying the context of use followed by the specification of the user requirements, the creation of design solutions and finally the evaluation of the design solutions against the requirements \citep{zimmermann2007requirement, din_ucd}, as depicted in Figure \ref{fig:chapter-2-prototyping-process-ucd}. Apparently, this process model is similar to the \gls{re} process model from Section \ref{sec:re_and_se} in Figure \ref{fig:chapter-2-re-process-linear}, while the focus here lies on design creation and evaluation. It represents an iterative and complementary process model to the presented \gls{re} and \gls{se} process models, particularly with a human-centered perspective \citep{zimmermann2007requirement}. In addition, a similar but simplified variant of this model was presented by \cite{greenberg1996teaching}, representing a cyclic, iterative process with activities of designing, prototyping and user testing with evaluation. \cite{weichbroth2015user} present a \gls{gui} prototyping process model which displays the creation of prototypes based on \gls{gui} standards, such as navigation and structure patterns, the interface schema with primary components and data flow diagrams, as illustrated in Figure \ref{fig:chapter-2-prototyping-process-newer}. Subsequently, the \gls{gui} prototypes are evaluated and verified whether they are Ready-to-Manufacture (RTM). Otherwise, the prototypes are validated with users and refined, and the \gls{gui} specifications are redefined \citep{weichbroth2015user}. To summarize, most \gls{gui} prototyping process models presented in the research literature agree on an inherently iterative procedure, with many overlapping activities. Moreover, many similarities exist between the \gls{gui} prototyping and \gls{re} process models.

A simple yet effective \gls{gui} prototyping technique is paper-based prototyping, often in the form of hand-drawn sketches \citep{snyder2003paper}. In particular, this technique is cost-effective and prototypes can be created rapidly. However, the resulting prototypes are low-fidelity, i.e. highly abstracted with reduced information and a lack of interactivity. Further, low-fidelity prototypes might also be represented as wireframes, which encompass merely rough layouts and reduced information \citep{rudd1996low}. Moreover, high-fidelity \gls{gui} prototypes are typically constructed with the help of previously discussed prototyping editors, benefiting from advanced levels of detail and realism, which is advantageous for obtaining higher-quality feedback during user tests \citep{rudd1996low}. By employing the \textit{Wizard-of-Oz} technique, interactions can be mimicked by a human, creating the impression of a fully implemented system \citep{bernsen1994wizard} or the interactions can be fully implemented through predefined behaviors in the corresponding \gls{gui} prototype.

\subsection{\gls{gui} Prototype Characteristics}

\gls{gui} prototypes are simplified representations of the final systems' \gls{gui} and can be characterized by various properties, which we will discuss in the following. \cite{hartson1991rapid} identify three orthogonal dimensions for \gls{gui} prototypes. First, they categorize prototypes based on their relation to the final product: \textit{revolutionary}, which describes prototypes employed solely for their specific purpose (e.g., elicitation) and discarded afterwards (also referred to as \textit{throw-away} \citep{kordon2002introduction}), and \textit{evolutionary}, which represents prototypes that are iteratively modified and refined, evolving into an implementation of the final software product \citep{hartson1991rapid}. Second, they differentiate between building the interface only and creating the entire software system, including the actual computational components \citep{hartson1991rapid}. The third property according to \cite{hartson1991rapid} is related to execution: \textit{intermittent}, if the prototype can demonstrate the software behavior merely at cyclic times during development, when the system has been completely created, and \textit{continuous}, if the prototype is not dependent on development-specific times to be exercised. According to them, prototypes can possess any combination of properties along the discussed dimensions.

\begin{figure}
\centering
  \includegraphics[width=1.0\textwidth]{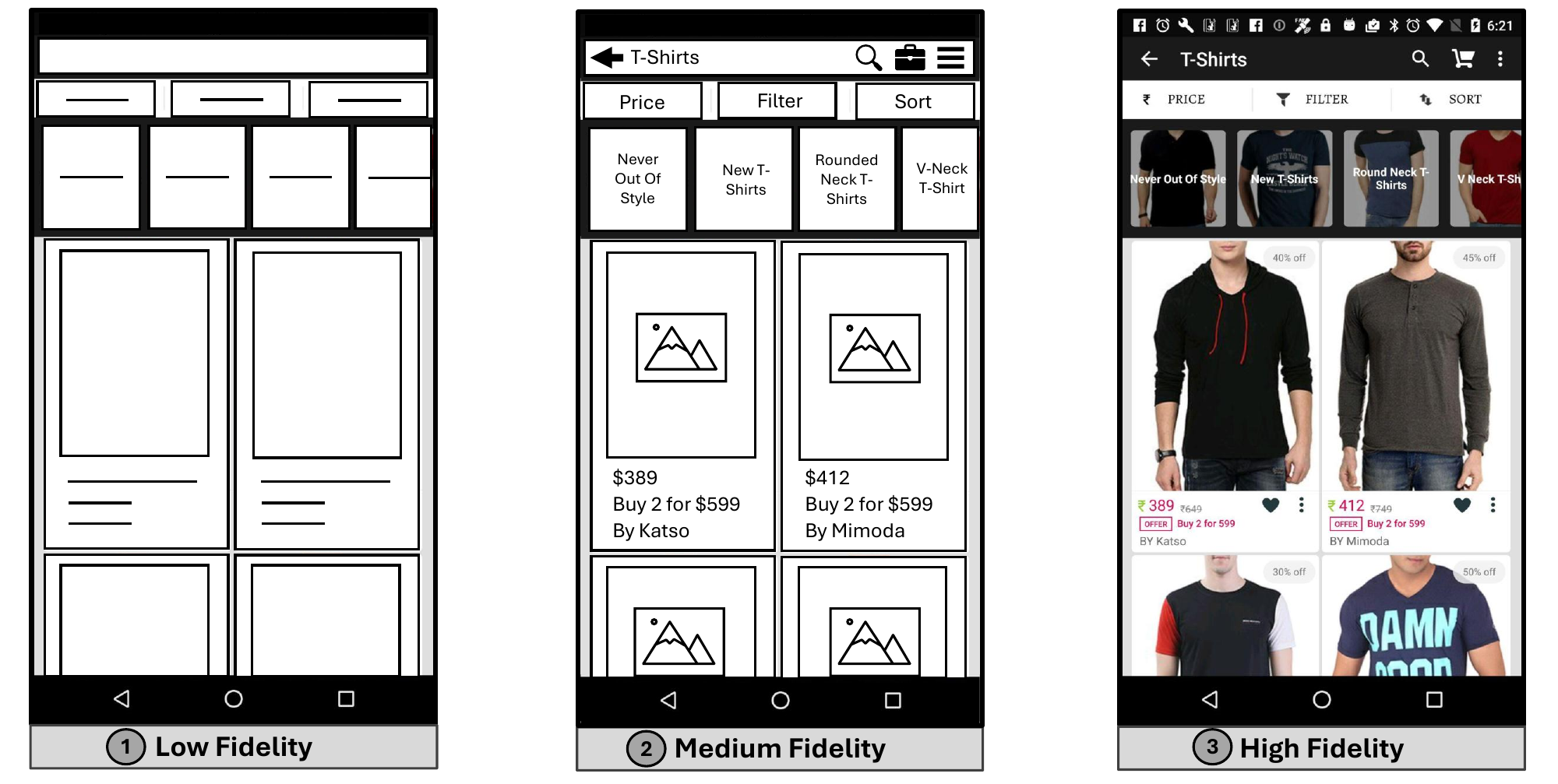}
  \caption[Three-level \gls{gui} prototype fidelity overview (adapted from \cite{feng2023designing})]{Three-level \gls{gui} prototype fidelity overview including low-, medium- and high-fidelity \gls{gui} prototype examples (adapted from \cite{feng2023designing}), while the high-fidelity \gls{gui} shows a screenshot extracted from the \textit{Rico} \gls{gui} dataset \citep{deka2017rico}.}
	\label{fig:chapter-2-fidelity}
    \vspace{-0.5cm}
\end{figure}

Moreover, the most defining characteristic of \gls{gui} prototypes is the level of fidelity, describing the similarity of the prototype in comparison to the final \gls{gui} \citep{rudd1996low, mccurdy2006breaking, feng2023designing}. Commonly, low-fidelity and high-fidelity levels are described \citep{rudd1996low}, while others encompass a third medium-fidelity level \citep{feng2023designing}, as illustrated in Figure \ref{fig:chapter-2-fidelity}. Low-fidelity prototypes provide a simple, abstract starting point for the \gls{gui}, neglecting details of scaling and pixel-accuracy as well as detailed functional information \citep{rudd1996low,feng2023designing}. Moreover, medium-fidelity prototypes provide more specifics compared to their low-fidelity counterparts, including pixel-accuracy for layouts, semantic icons and rough text descriptions, enabling increased possible interactivity \citep{feng2023designing}. Finally, high-fidelity \gls{gui} prototypes are already close to the final system in terms of their look-and-feel \citep{rudd1996low, feng2023designing}. However, we emphasize that \gls{gui} prototype fidelity should be regarded rather as a spectrum instead of fixed categories. Moreover, \cite{mccurdy2006breaking} argue that prototypes possess five independent dimensions, while each of these dimensions can possess different fidelity levels. Specifically, the first dimension is represented by the level of visual refinement, where hand-drawn sketches reside on the low-fidelity side, while pixel-accurate mockups are defined as high-fidelity. Second, the breadth of functionality describes whether a highly functional prototype includes approximations for a large portion of the most important \gls{fr}. Third, the depth of the functionality represents the implementation detail for particular features. Fourth, the richness of interactivity, where prototypes with transitions between \gls{gui} states and enabled user input represent high-fidelity variants. Finally, the fifth dimension is the richness of the data model, representing the closeness of the employed data (such as names, etc.) in the prototype compared to the final system \citep{mccurdy2006breaking}.

\gls{gui} prototypes can be represented in a large variety of different internal formats, ranging from commonly applied web standards (e.g., \textit{\gls{html}} and \textit{\gls{css}}, \textit{\gls{json}}, \textit{Scalable Vector Graphics (SVG)}) to proprietary hierarchical or binary structures. These internal representation models define the different aspects of \gls{gui} prototypes, such as the \gls{gui} components or widgets, their layout, design and interaction behavior. Often, \gls{gui} prototypes are represented as tree-like hierarchical structures internally, which reflects the inherently nested, visual \gls{gui} component structure, where frames represent the roots, while they contain grouping containers and individual components.

\vspace{-0.3cm}
\section{Computational Methods and Techniques}
\label{sec:computationl_methods}
\vspace{-0.2cm}

In this section, we introduce key computational methods fundamental to our work. In particular, we introduce the basics of \gls{ml} (Subsection \ref{chapter-2:sec-ml}), followed by classical document scoring and ranking methods as well as text representation methods for \gls{ir} (Subsection \ref{chapter-2:sec-ir}). Next, we introduce main concepts of \glspl{plm} and \glspl{llm} (Subsection \ref{chapter-2:sec-pml}), including the Transformer architecture and related models. Finally, we discuss multiple methods for efficiently adapting \glspl{llm}, such as \gls{pe} and \gls{ce} (Subsection \ref{chapter-2:sec-pe}).

\vspace{-0.2cm}
\subsection{Machine Learning (ML)}
\label{chapter-2:sec-ml}

\gls{ml} represents a subfield of artificial intelligence, which, in comparison to explicitly programmed rules or instructions, aims at learning and inferring computational models from experience (data) in order to make predictions or decisions for unseen data points and create representations of the underlying data \citep{mitchell1997does}. More formally, given a task $\mathcal{T}$, an \gls{ml} model improves the performance measure $\mathcal{P}$ with exposure to additional experience $\mathcal{E}$ (data). Typically, \gls{ml} techniques are differentiated into \textit{unsupervised}, \textit{supervised}, \textit{semi-supervised}, \textit{self-supervised} and \textit{\gls{rl}} paradigms, each referring to different forms of supervisory information employed for deriving the model \citep{bishop2006pattern, rani2023self}. 

In particular, for \textit{(1)} unsupervised learning, supervisory information is absent and the performance measure $\mathcal{P}$ must rely on intrinsic properties of the data and the task itself, such as the intra-cluster similarity within a document clustering task. Next, for \textit{(2)} supervised learning, a supervisory signal is available for optimizing the model, often based on human annotations. Specifically, given a dataset $D = \{(x_i, y_i)\}_{i=1}^{N}$, with $x \in X$ being the input data and $y \in Y$ being the respective labels, the task $\mathcal{T}$ is to learn a parametrized function $f_{\theta} : X \to Y$ mapping input examples to corresponding output labels. Binary classification represents a special instance of this problem, where, for example, $Y = \{0,1\}$ represents a binary label set, and a potential \gls{ml} model typically predicts a scalar score $s_{\theta}(x)$, converted into a probability with the sigmoid function:

\begin{equation}
P_{\theta}(y = 1 \mid x)
= \sigma\!\left( s_{\theta}(x) \right)
= \frac{1}{1 + e^{-s_{\theta}(x)}}.
\end{equation}

\noindent Afterwards, a threshold is applied to create the binary class prediction. To optimize the model for the prediction task, different loss functions or learning objectives can be applied, depending on the concrete task $\mathcal{T}$. For example, to optimize the binary classification model, one objective is to minimize the \textit{binary cross-entropy} \citep{ho2019real}. By employing this loss, the prediction model is encouraged to maximize the prediction probabilities for the correct class across the entire training dataset. Particularly, the binary cross-entropy loss function \citep{ho2019real} is defined as 

\begin{equation}
\mathrm{L_{BCE}}(\theta)
= -\frac{1}{N} \sum_{i=1}^{N} 
\big[\, y_i \log P_{\theta}(x_i) 
      + (1 - y_i)\log\!\big(1 - P_{\theta}(x_i)\big) \,\big]
\end{equation}

\noindent While for binary prediction tasks the shown binary cross-entropy loss is commonly employed for optimizing \gls{ml} models, for multi-class prediction tasks, i.e. with the label set $Y=\{1,2,…,K\}, K>2$, encompassing $K$-many labels, the \textit{multinomial cross-entropy} loss function \citep{golik2013cross} is typically employed and minimized as:

\begin{equation}
\mathrm{L_{CE}}(\theta)
= -\frac{1}{N} \sum_{i=1}^{N} \log P_{\theta}(y_i \mid x_i).
\end{equation}

\noindent Furthermore, \textit{(3)} self-supervised learning is a form of representation learning, exploiting inherent correlations of the data as a supervisory signal in comparison to human-created annotations \citep{liu2021self}. This paradigm plays a significant role in \glspl{plm}, which are typically pretrained on self-supervised tasks, such as contextual or next word predictions. In addition, and similar to cross-entropy, another important measure of divergence between two probability distributions is the \textit{\gls{kld}} \citep{kullback1951kullback}, defined as

\begin{equation}
D_{\mathrm{KL}}(P \,\|\, Q)
= \sum_{x \in \mathcal{X}} P(x) \log \frac{P(x)}{Q(x)} .
\end{equation}

\noindent where $P$ represents the true probability distribution and $Q$ refers to the approximating probability distribution, with both distributions defined over the respective support $\mathcal{X}$.

\subsection{Information Retrieval (IR)}
\label{chapter-2:sec-ir}

\vspace{-0.1cm}

\gls{ir} is concerned with retrieving relevant documents or resources given an information need (query) \citep{singhal2001modern, schutze2008introduction}. More formally, given a corpus of documents $D = \{ d_1, \ldots, d_N \}$ and a set of possible queries $Q$, a modern \gls{ir} system provides a scoring function $s : Q \times D \to \mathbb{R}$, which assigns a relevance score $s(q, d)$ to each document $d$ associated with query $q$, ultimately creating a ranking from most to least relevant according to $s(q, d)$ over the entire collection $D$ as $\pi(q) = \operatorname*{argsort}_{d \in D} s(q, d)$. We refer to the set of relevant documents with respect to a query $q$ as $R(q) \subseteq D$. However, classical \gls{ir} methods are based on the \textit{Boolean Model} \citep{lancaster1973information}, where documents are represented as collections of individual index terms $t_i \in V$, and queries are Boolean expressions in Conjunctive Normal Form (CNF) over $V$ such as $q = (t_1 \lor \lnot t_2) \land (t_3 \lor t_4)$, allowing solely exact matches \citep{lancaster1973information}.

\vspace{-0.1cm}

\paragraph{\gls{bow}.} To compute similarities between queries and documents, the textual information needs to be represented in a form which enables effective processing by computational methods. A simple representation of text documents is provided by the \gls{bow} model, which essentially is a text feature extraction technique. In this model, text is represented as an unordered collection of words, where each word of the vocabulary $V$ corresponds to a dimension in a vector representing document $d$ as $\mathbf{x}_d = (c_{d,1},\, c_{d,2},\, \ldots,\, c_{d,|V|})$, where $c_{d,i}$ represents the frequency of word $t_i$ within document $d$. However, this representation neglects word order and semantics entirely.
\vspace{-0.1cm}

\paragraph{\gls{tfidf}.} To improve the simple \gls{bow} model and adjust the importance of terms with respect to their occurrence in the corpus, the \gls{tf} is adjusted by the \gls{idf}: 

\begin{equation}
w_{t,d}
=
\frac{c_{d,t}}{\sum_{t'} c_{d,t'}} \,
\log\!\left(\frac{N}{1 + n_t}\right)
\end{equation}

\noindent with $w_{t,d}$ being the weight for word $t$ in document $d$, $c_{d,t}$ the frequency of word $t$ in document $d$, $N = |D|$ being the number of documents in the corpus and $n_t$ the number of documents encompassing word $t$. Similarity between queries and documents can then be computed based on cosine similarity, i.e. $\mathrm{sim}(d_1, d_2)=\frac{x_{d_1} \cdot x_{d_2}}{\lVert x_{d_1}\rVert \lVert x_{d_2}\rVert}$, with $x_{d_{1}}$ and $x_{d_{2}}$ representing the \gls{tfidf} vectors of documents $d_1$ and $d_2$, respectively. \gls{tfidf} scoring ensures that frequently occurring words (e.g., stop words, providing little extra information for distinction) accordingly receive lower weights \citep{salton1988term}. 
\vspace{-0.1cm}
\vspace{-0.1cm}

\paragraph{Okapi \gls{bm25}.} A more advanced scoring function of the \gls{tfidf} family is represented by Okapi \gls{bm25}, which also incorporates both \gls{tf} and \gls{idf} components with adjusted weightings \citep{robertson1995okapi}, and is defined as

\begin{equation}
\mathrm{BM25}(q,d)
=
\sum_{t \in q}
\mathrm{IDF}(t)\,
\frac{c_{d,t}(k_1 + 1)}
     {c_{d,t} + k_1 \left( 1 - b + b \frac{|d|}{\mathrm{avgdl}} \right)}
\end{equation}

\noindent with $t \in q$ referring to the words $t$ contained in the query $q$, $IDF(t)$ being a \gls{bm25}-specific variant of the \gls{idf} for term $t$, while it should be noted that multiple \gls{idf} variants exist, $c_{d,t}$ the frequency of word $t$ in document $d$, $|d|$ the number of terms $t$ in document $d$, also referred to as the length of the document, $avgdl$ the average document length across the corpus, while $k_1 \in [1.2,2.0]$ and $b=0.75$ are adjustable weighting terms.

\vspace{-0.1cm}

\begin{figure}[!t]
\centering
  \includegraphics[width=1.0\textwidth]{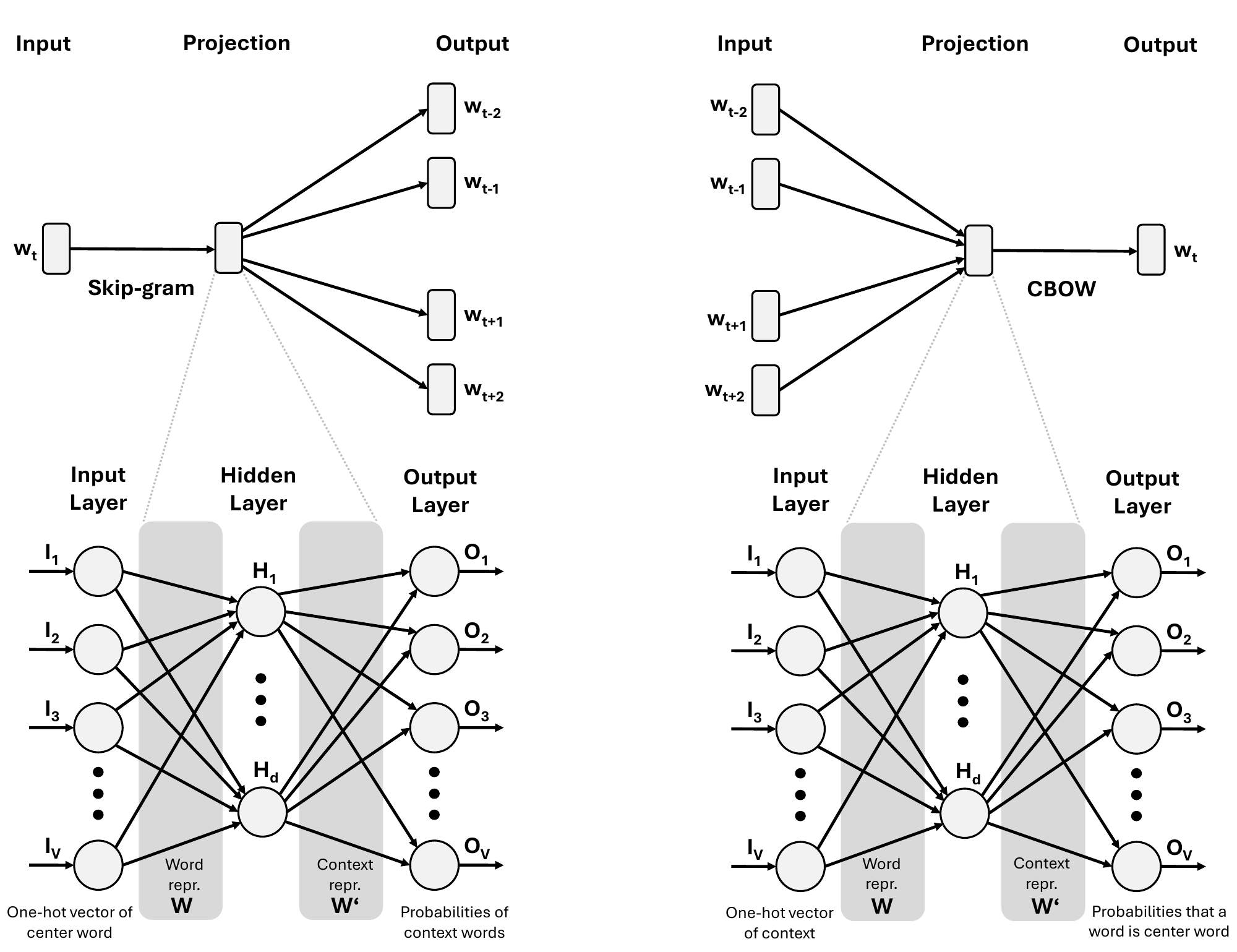}
  \caption[\textit{Skip-gram} and \textit{\gls{cbow}} architectures for learning semantic word representations (adapted from \cite{mikolov2013efficient} and \cite{lauscher2021language})]{Two approaches for learning semantic word representations: \textit{(1)} \textit{Skip-gram} architecture (on the left), predicting the context words given the one-hot encoded vector of the center word and \textit{(2)} \textit{\gls{cbow}} (on the right), predicting the center word given one-hot encoded context words, both with a \gls{ffnn} architecture (adapted from \cite{mikolov2013efficient} and \cite{lauscher2021language}).}
	\label{fig:chapter-2-word2vec}
\end{figure}

\paragraph{word2vec.} \textit{word2vec} is one of the early approaches to enhance word representations over the simple sparse models, enabling them to capture semantic relationships \citep{mikolov2013distributed, mikolov2013efficient}. Fundamental to the approach is the \textit{distributional hypothesis}, which notes that words appearing in similar contexts tend to have similar meanings \citep{harris1954distributional}. Particularly, in such a word vector space, each word $w_i$ is represented as a static, real-valued, $d$-dimensional vector $\mathbf{e}(w_i) \in \mathbb{R}^d$. To compute these embeddings, two common approaches exist. First, \textit{Skip-gram} maximizes the average log probability
\vspace{-0.1cm}

\begin{equation}
\frac{1}{T} \sum_{t=1}^{T} \sum_{j=-c}^{c} \log P(w_{t+j} \mid w_t)
\end{equation}

\noindent for predicting the surrounding context words $w_{t-c}, ...,w_{t-1},w_{t+1}, ..., w_{t+c}$ given the center word $w_t$, where $c$ denotes the context size, while $T$ represents the number of words in the sequence, which is depicted in Figure \ref{fig:chapter-2-word2vec} \citep{mikolov2013efficient, lauscher2021language}. The probability $P(w_{t+j} \mid w_t)$ is computed with the \textit{softmax} function as

\begin{equation}
P(w_{t+j} \mid w_t)
=
\frac{\exp\!\left( {x'_{t+j}}^{\top} x_t \right)}
     {\sum_{i=1}^{|V|} \exp\!\left( {x'_i}^{\top} x_t \right)}
\end{equation}

\noindent with $x_i$ and $x_i^{'}$ denoting the actual $d$-dimensional word and context vector representations, while $V$ denotes the vocabulary. In contrast, the \gls{cbow} approach predicts the center word $w_t$ given the sequence of surrounding context words $w_{t-c}, ..., w_{t-1}, w_{t+1}, \allowbreak ..., \allowbreak w_{t+c}$ by optimizing the following average log probability \citep{mikolov2013distributed}

\begin{equation}
\frac{1}{T}\sum_{t=1}^{T}
\log P\!\left( w_t \mid w_{t-c}, \dots, w_{t+c} \right)
\end{equation}

\paragraph{\gls{nbow}.} Since the discussed \textit{word2vec} embeddings are static and computed for single words only, a simple approach to compute embeddings for a sequence of words, such as in document $d$, is to employ the \textit{\gls{nbow}} technique \citep{sheikh2016learning}, by computing a weighted average over the individual word embeddings as

\begin{equation}
x_d = \sum_{i=1}^{L} \alpha_i\, \mathbf{e}(w_i)
\end{equation}

\noindent with $L$ representing the number of words in the sequence $w_1, ..., w_L$, $\mathbf{e}(w_i) \in \mathbb{R}^d$ being the $d$-dimensional embedding of word $w_i$ and $\alpha_i$ its respective weight. As meaningful weights, the normalized \gls{tfidf} scores can be applied, i.e. $\alpha_i = w_{i,d}$, emphasizing embeddings from more relevant words, while discounting less informative embeddings.

\vspace{-0.2cm}

\subsection{Pretrained Language Models (PLMs)}
\label{chapter-2:sec-pml}

In contrast to the previously introduced static word representations, \glspl{plm} aim to learn generalized and contextualized text representations by leveraging large corpora for \textit{pretraining}, which can be adapted to various downstream \gls{nlp} problems and tasks via subsequent \textit{finetuning} approaches \citep{wang2023pre}. In particular, the objective employed to optimize \glspl{plm} requires them to predict \textit{tokens} based on their context. Modern \glspl{plm} rely on deep neural network architectures, especially the \textit{Transformer} architecture \citep{vaswani2017attention}, which we present subsequently, followed by different \gls{plm} variants.

\vspace{-0.1cm}

\paragraph{Transformer Architecture.} The most popular approach underlying \glspl{plm} is based on the \textit{Transformer} architecture \citep{vaswani2017attention}, particularly on the decoder component, which comprises multiple identical layers. Each of these layers encompasses a \textit{self-attention} and a \gls{ffnn} component. Relying on \textit{self-attention}, the \textit{Transformer} overcomes prior efficiency limitations of recurrent neural architectures \citep{sutskever2011generating}. Specifically, the \textit{self-attention} usually employs the \textit{scaled dot-product (multiplicative) attention} \citep{vaswani2017attention}. The \textit{self-attention} mechanism operates on three matrices, namely, queries $\mathbf{Q}$, keys $\mathbf{K}$ and values $\mathbf{V}$, which are obtained from the input representations. Subsequently, the \textit{self-attention} is computed using a \textit{softmax} function as

\begin{equation}
\text{Attention}(\mathbf{Q}, \mathbf{K}, \mathbf{V})
=
\mathrm{softmax}\!\left(
\frac{\mathbf{Q}\mathbf{K}^{\top}}{\sqrt{d_k}}
\right)\mathbf{V}
\end{equation}

\noindent with $\frac{1}{\sqrt{d_k}}$ being a scaling factor. Hence, every token in the sequence is enabled to attend to every other token for long-range dependency coverage. To enable capturing diverse relationships across the different representation spaces, \textit{Multi-Head Self-Attention (MHSA)} is introduced, which employs multiple \textit{self-attention heads} in parallel. The computed outputs of each \textit{self-attention head} are concatenated and projected with matrix $\mathbf{W^O}$ as

\begin{equation}
\mathrm{MHSA}(\mathbf{Q}, \mathbf{K}, \mathbf{V})
=
\mathrm{Concat}(head_1, \dots, head_H)\, \mathbf{W^O}
\end{equation}

\noindent In particular, this mechanism enables the model to simultaneously attend to diverse information from multiple representation spaces from various positions, creating richer contextual representation modeling \citep{vaswani2017attention}. Each $head_i$ is computed as

\begin{equation}
\text{head}_i
=
\mathrm{Attention}(\mathbf{Q} \mathbf{W_i^Q},\; \mathbf{K} \mathbf{W_i^K},\; \mathbf{V} \mathbf{W_i^V})
\end{equation}

\noindent and projection matrices as
$\mathbf{W}_i^{Q} \in \mathbb{R}^{d_{\text{model}} \times d_k}$,
$\mathbf{W}_i^{K} \in \mathbb{R}^{d_{\text{model}} \times d_k}$,
$\mathbf{W}_i^{V} \in \mathbb{R}^{d_{\text{model}} \times d_v}$,
and $\mathbf{W}^{O} \in \mathbb{R}^{h d_v \times d_{\text{model}}}$, with the dimension $d_k$ for queries and keys, and dimension $d_v$ for values. While modern \glspl{plm} are based on the \textit{Transformer} architecture as the state-of-the-art approach, they typically rely on decoder-only or encoder-only architectures.

\begin{figure}
\centering
  \includegraphics[width=1.0\textwidth]{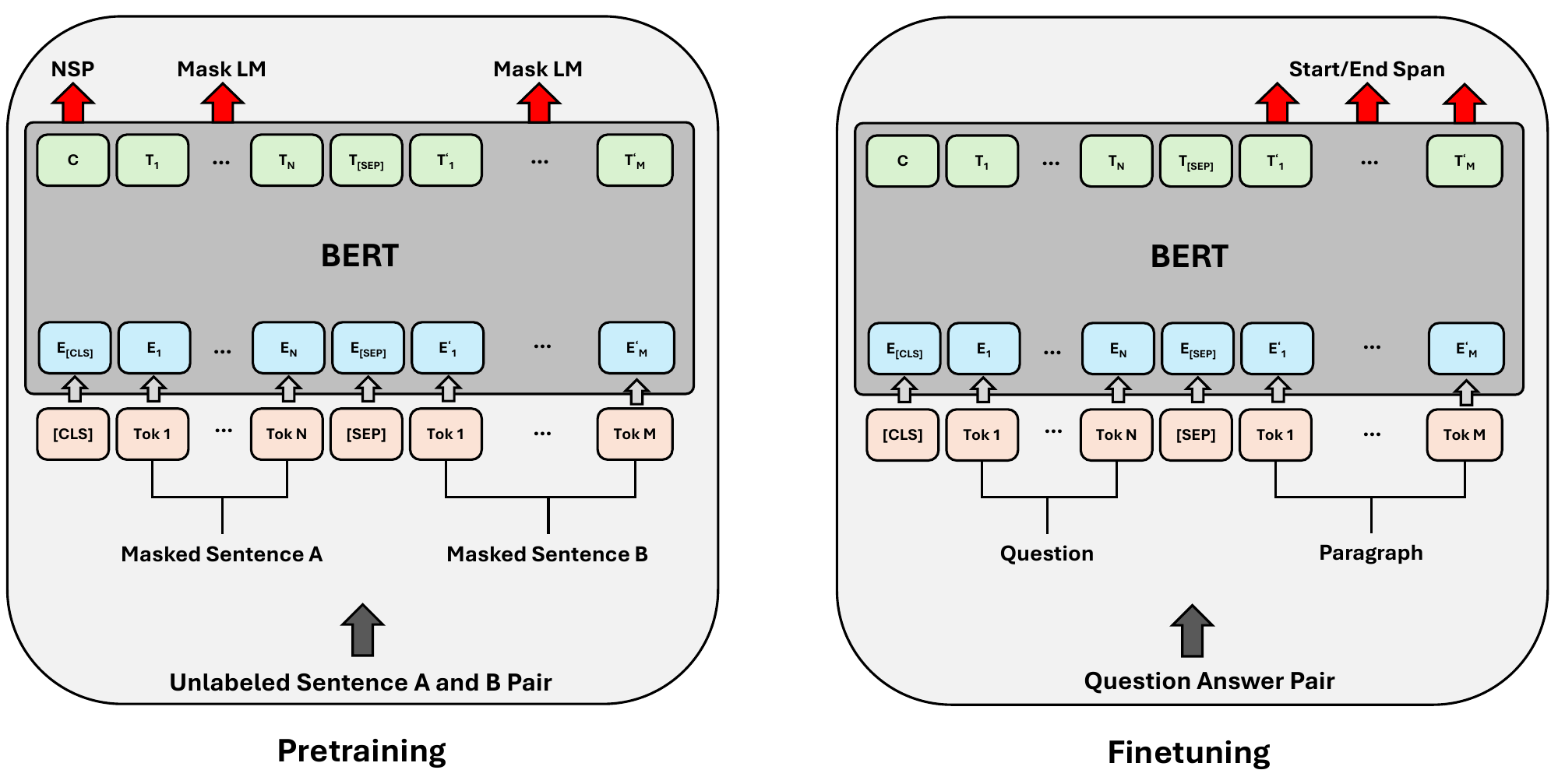}
  \caption[\gls{bert} approach with pretraining and finetuning of a \textit{Transformer} encoder (adapted from \cite{devlin2019bert})]{\textit{\gls{bert}} approach based on a \textit{Transformer} encoder with \textit{(1)} pretraining phase, which includes two training objectives: \gls{mlm} and NSP, and \textit{(2)} the finetuning stage with the same parameters from pretraining for \gls{nlp} tasks (e.g., question answering or query-document scoring) (adapted from \cite{devlin2019bert}).}
	\label{fig:chapter-2-bert}
\end{figure}

\glsreset{mlm}
\glsreset{nsp}
\glsreset{bert}
\paragraph{\gls{bert}.} An early adaptation of the \textit{Transformer} architecture as an encoder-only bidirectional model for creating language representations is \gls{bert}, which is trained on large-scale text corpora by incorporating the left and right context across all layers \citep{devlin2019bert}. In particular, the model is pretrained with two training objectives: \textit{(1)} a \gls{mlm} token-level task, which asks the model to predict randomly masked tokens in the sequence and \textit{(2)} a Next Sentence Prediction (NSP) sentence-level task, which requires the model to predict whether two sentences are adjacent in the text, as depicted in Figure \ref{fig:chapter-2-bert}. The input to the model is divided into token, segment and position embeddings \citep{devlin2019bert}. Specifically, \gls{bert} employs the two loss functions together in the pretraining stage as $\mathcal{L}_{\text{BERT}} = \mathcal{L}_{\text{MLM}} + \mathcal{L}_{\text{NSP}}$, with $\mathcal{L}_{\text{MLM}}$ and $\mathcal{L}_{\text{NSP}}$ defined as follows:

\begin{equation}
\mathcal{L}_{\text{MLM}}
= - \sum_{t \in \mathcal{M}} \log P_{\theta}(x_t \mid \tilde{X}).
\end{equation}

\begin{equation}
\mathcal{L}_{\text{NSP}}
= - \left[ 
y \log P_{\theta}(\text{1}) 
+ (1 - y) \log P_{\theta}(\text{0})
\right].
\end{equation}

\noindent with $\mathcal{M}$ being the set of randomly masked tokens, $x_t$ the masked token and $\tilde{X}$ the masked input token sequence. \gls{bert} utilizes two special input tokens: \textit{(1)} \texttt{[CLS]}, representing the starting token, whose hidden state $C \in \mathbb{R}^{H}$ represents the aggregated sequence embedding, which can be used for sentence-level classification tasks (e.g., NSP), and \textit{(2)} \texttt{[SEP]}, which represents the separation between the first (e.g., sentence A) and the second (e.g., sentence B) input. \gls{bert} is available in different sizes, but a popular variant is $\mathbf{BERT_{BASE}}$ with L=12 layers, H=768 hidden dimension size, A=12 self-attention heads and 110M parameters in total \citep{devlin2019bert}. To finetune \gls{bert} for different tasks, the input sequence requires preparation encompassing the special tokens. For example, considering a ranking task, query-document pairs $(q, d_i)$ can be inputted to \gls{bert}, while a prediction head (e.g., a simple \gls{ffnn}) processes the encoded \gls{bert} state to compute ranking score predictions. The entire \gls{bert} model can be finetuned end-to-end, which updates all parameters, or solely the prediction head can be finetuned for the problem.

\begin{figure}
\centering
  \includegraphics[width=1.0\textwidth]{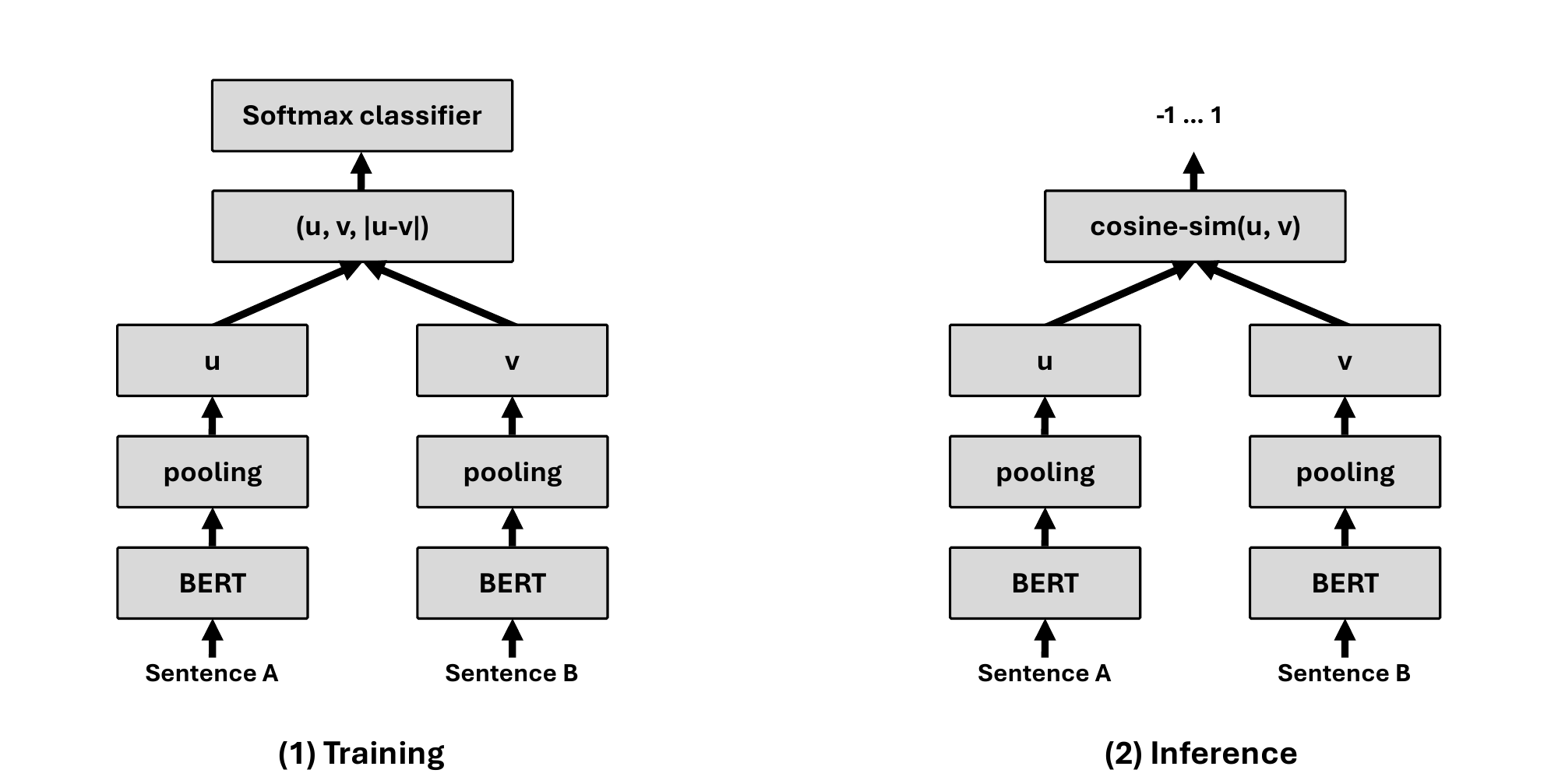}
  \caption[\gls{sbert} approach with training and inference architecture based on the \gls{bert} model (adapted from \cite{reimers2019sentence})]{\textit{\gls{sbert}} approach with \textit{(1)} training architecture (on the left), showing a siamese network structure with shared parameters based on the \gls{bert} model and \textit{(2)} inference architecture with \textit{cosine similarity} output (on the right) (adapted from \cite{reimers2019sentence}).}
	\label{fig:chapter-2-sbert}
\end{figure}

\glsunset{gpt}
\paragraph{\gls{sbert}.} While \gls{bert} provided state-of-the-art performance across various \gls{nlp} tasks (before \glspl{llm}), the computational overhead for tasks such as computing Semantic Textual Similarity (STS) is immense \citep{reimers2019sentence}. To mitigate this, \cite{reimers2019sentence} proposed \gls{sbert}, which enables the computation of semantic sentence embeddings for downstream tasks, such as similarity scoring and text retrieval problems. The model is based on a Siamese Neural Network (SNN) \citep{chicco2021siamese}, which replicates its structure and conducts parameter sharing for both input sentences, as shown in Figure \ref{fig:chapter-2-sbert}. Each sentence is fed through the \gls{bert} model and the outputs are pooled based on different strategies (e.g., employing the \textit{[CLS]} token, \textit{mean} over output vectors), concatenated with element-wise difference $|u-v|$, which are then multiplied by trainable weights $W_t$ and fed through $\text{softmax}(W_t (u,\, v,\, |u - v|))$ to optimize cross-entropy loss. Finally, the text representations obtained from \gls{sbert} can be meaningfully applied to compare text via \textit{cosine similarity}, i.e. $\text{sim}(u, v) = \frac{u^\top v}{\lVert u \rVert \, \lVert v \rVert}$, useful for ranking.

\definecolor{lightgray}{gray}{0.92}
\begin{table}[t!]
\small
\setlength{\tabcolsep}{5pt}
\centering
\begin{tabular}{lccccc}
\hline
\textbf{Model Name} & \textbf{Organization} & \textbf{Parameters} & 
\textbf{Context} & \textbf{MM} & \textbf{Reference} \\
\hline
GPT-3 & OpenAI & 175B & 2,048 & $\times$ & \citep{brown2020language} \\
\rowcolor{lightgray}
GPT-4o & OpenAI & UD & 128k & \checkmark & \citep{openai_gpt4o_docs} \\

GPT-4.1 & OpenAI & UD & 1M & \checkmark & \citep{openai_gpt41_docs} \\

\rowcolor{lightgray}
GPT-5 & OpenAI & UD & 400k & \checkmark & \citep{openai_gpt5_docs} \\

LLaMA 2 & Meta & 7,13,70B & 4k & $\times$ & \citep{meta_llama2_docs} \\
\rowcolor{lightgray}
LLaMA 3.2 & Meta & 3,11,90B & 128k & \checkmark & \citep{meta_llama3_docs} \\

Gemini 2.5 Pro & Google & UD & 1M & \checkmark & \citep{google_gemini_models_2_5_pro} \\

\rowcolor{lightgray}
Gemini 3 Pro & Google & UD & 1M & \checkmark & \citep{google_gemini_models} \\

Claude Sonnet 4.5 & Anthropic & UD & 200k,1M & \checkmark & \citep{anthropic_claude_models_overview} \\
\rowcolor{lightgray}
Claude Opus 4.5 & Anthropic & UD & 200k & \checkmark & \citep{anthropic_claude_models_overview} \\

Qwen 2 & Alibaba & 0.5,1.5,7,72B & 32k & $\times$ & \citep{team2024qwen2} \\

\rowcolor{lightgray}
Qwen 3  & Alibaba & 4,8,14,32B & 128k & $\times$ & \citep{yang2025qwen3} \\

DeepSeek & DeepSeek & 7,64B & 4k & $\times$ & \citep{bi2024deepseek} \\

\rowcolor{lightgray}
DeepSeek-VL & DeepSeek & 1.3,7B & 4k & \checkmark & \citep{lu2024deepseek} \\
\hline
\end{tabular}

\caption[Overview of popular \glspl{llm}/\glspl{mllm}]{\gls{llm} overview (OpenAI, Meta, Google, Anthropic, Alibaba, DeepSeek) with organization, parameters, context, multi-modal (MM) support (UD = Undisclosed).}
\label{tab-chapter-2:llms}
\end{table}

\vspace{-0.1cm}

\paragraph{\glspl{llm}.} Based on the large success of the \textit{Transformer} architecture, its adaptation for state-of-the-art \glspl{plm} and the insight that scaling these architectures up leads to improved capabilities \citep{kaplan2020scaling}, the term \glspl{llm} has been coined, referring to massive variants of \glspl{plm}, typically decoder-only, which encompass billions of parameters and are trained on vast amounts of text data \citep{zhao2023survey}. In particular, \glspl{llm} are initially optimized via self-supervised pretraining on large token sequences by optimizing a common (autoregressive) language model objective \citep{radford2018improving} as follows

\vspace{-0.1cm}

\begin{equation}
\sum_i \log P_\theta(u_i \mid u_{i-k}, \ldots, u_{i-1})
\end{equation}

\vspace{-0.1cm}

\noindent with $u_i$ referring to a token, $k$ being the context size and $\theta$ the parameters of a neural network trained with \gls{sgd}-based optimization \citep{amari1993backpropagation}. Similar to the discussed \gls{bert} model, \glspl{llm} typically undergo \gls{sft} for specific \gls{nlp} problems after generic pretraining using an objective such as

\begin{equation}
\sum_{(x, y)} \log P\big( y \mid x_1, \ldots, x_m \big)
\end{equation}

\noindent with $x_1, \ldots, x_m$ referring to an input token sequence, while $y$ denotes a corresponding label for the task \citep{radford2018improving}. While increasing the parameter size of the models constantly improves their performance under scaling laws \citep{kaplan2020scaling}, the ability to follow human instructions is restricted \citep{ouyang2022training}. Therefore, \cite{ouyang2022training} proposed an adaptation pipeline consisting of three phases: \textit{(1)} conduct \gls{sft} of the \gls{llm} given high-quality task completions with respect to a prompt, \textit{(2)} train a reward model based on sampling prompts and multiple model completions, which are subsequently ranked by annotators and \textit{(3)} optimize a policy with regard to the reward model by employing \gls{rl}. The resulting model shows increased ability to follow provided instructions over previous models. Table \ref{tab-chapter-2:llms} provides an overview of well-known proprietary and open-source \glspl{llm}, showing their parameter size (while often undisclosed for many proprietary \glspl{llm}), the context size and their multi-modal support. Most prominent is \textit{OpenAI}, providing one of the earliest \glspl{llm}, with the \gls{gpt} model series, achieving state-of-the-art results across diverse \gls{nlp} tasks. Next, we discuss \glspl{mllm}. 

\begin{figure}
\centering
  \includegraphics[width=1.0\textwidth]{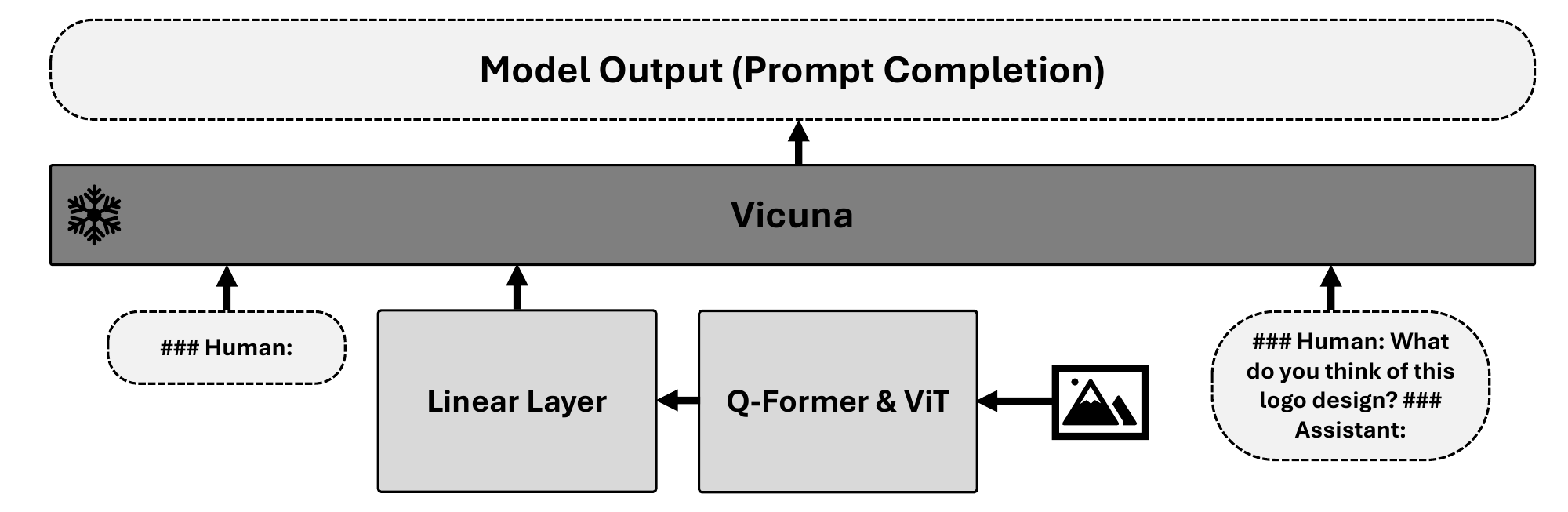}
  \caption[\gls{mllm} architecture of the \textit{MiniGPT-4} model (adapted from \cite{zhu2023minigpt})]{Example \textit{\gls{mllm}} architecture of \textit{MiniGPT-4} employing a vision encoder based on a pretrained \textit{Vision Transformer (ViT)} model and \textit{Q-Former} network, a linear projection layer, which is integrated into \textit{Vicuna} \gls{llm} (adapted from \cite{zhu2023minigpt}).}
	\label{fig:chapter-2-mllm}
\end{figure}

\paragraph{MLLMs.} While the increasing capabilities of \glspl{llm} for text-based tasks are promising and achieve state-of-the-art results, incorporating other modalities such as images represents an important problem. Specifically, \glspl{mllm} extend \glspl{llm} with multi-modal processing capabilities. \cite{zhu2023minigpt} integrated a visual encoder with frozen weights into an advanced \gls{llm} via a single projection layer, as depicted in Figure \ref{fig:chapter-2-mllm}. Particularly, the results indicate that, when visual features are properly aligned with an \gls{llm}, the resulting model possesses multi-modal abilities, such as image captioning or generating website code given a respective screenshot \citep{zhu2023minigpt}. Instead of creating and training a \gls{mllm} from scratch end-to-end, \cite{wu2023visual} propose \textit{Visual-ChatGPT} by integrating various ViT models into \glspl{llm} via a high-level prompt manager, which provides the \gls{llm} with access to the image information via conversion by utilizing ViT.

\subsection{Prompt and Context Engineering (PE/CE)}
\label{chapter-2:sec-pe}

With increasing parameter size, \glspl{llm} have been shown to possess emergent abilities such as \gls{icl}, which refers to their capability of task adaptation during inference time through examples in their context, requiring no parameter updates or training \citep{brown2020language, liu2023pre}. More formally, given the context $C = \{ I,\, (x_1, y_1),\, \ldots,\, (x_k, y_k) \}$ of $k$ examples as input to the \gls{llm}, with $I$ denoting optional instructions in natural language, while $(x_i, y_i)$ refers to an example pair of input text $x_i$ with the respective label $y_i$, the \gls{llm} predicts the label for the next unseen example $x_{k+1}$ \citep{dong2024survey} as

\begin{equation}
P_{\theta}(y_{k+1} \mid x_{k+1}, C) = \mathrm{LLM}_{\theta}\!\left(C,\, x_{k+1}\right).
\end{equation}

\glsreset{pe}
\glsreset{ce}
\noindent by computing the respective $\hat{y}$ that achieves the maximum probability among all possible solutions based on its parameters $\theta$, i.e. $\hat{y} = \arg\max_{y_j \in Y} P_\theta(y_j \mid x)$. Following this paradigm, \cite{liu2023pre} coined the overall approach as \textit{``pre-train, prompt, and predict''} or \textit{``prompt-based learning''}. This enables efficient adaptation of \glspl{llm} to a plethora of different \gls{nlp} tasks, including classification and task-specific \gls{nlg} \citep{dong2022survey}. Furthermore, research has shown that \glspl{llm} can be effectively adapted to various tasks in the complete absence of examples, solely by providing meaningful instructions in natural language \citep{wei2021finetuned, kojima2022large}. To optimize the effectiveness of \glspl{llm} in such settings, the discipline of \textit{\gls{pe}} has emerged \citep{marvin2023prompt, schulhoff2024prompt}, referring to the process of creating, refining and optimizing instructions given to the \gls{llm} in order to solve a specific task effectively, thereby leveraging the full potential of the \gls{llm}. More recently, the term \textit{\gls{ce}} \citep{haseeb2025context} has been coined, referring to the systematic construction of the input context of an \gls{llm}, by providing all task-specific information and external knowledge relevant for effectively solving a task, including necessary tools. Subsequently, we provide an overview of effective \gls{pe} techniques proposed in research.

\glsreset{zs}
\paragraph{\gls{zs}.} Research has shown that providing solely meaningful instructions in \gls{nl} can be sufficient for \glspl{llm} to effectively solve certain tasks \citep{wei2021finetuned, kojima2022large}. More formally, this approach is referred to as \gls{zs} and the input context $C = \{ I\}$ to the \gls{llm} consists merely of the instruction $I$. Typically, \textit{prompt templates} are employed to replace example-specific information, while retaining generic textual instructions, i.e. context $C = \{ I(x_1, ... , x_k)\}$, with $x_i$ referring to task-specific inputs.

\glsreset{fs}
\paragraph{\gls{fs}.} In contrast to \gls{zs} (only instructions and task-specific information), \gls{fs} prompting relies on providing multiple input-output pair examples in the context, i.e. $C = \{ I,\, (x_1, y_1),\, \ldots,\, (x_k, y_k) \}$ with $k \geq 2 $ \citep{brown2020language}. This approach can be conducted both with and without task-specific instructions $I$ given in the context $C$.

\glsreset{os}
\paragraph{One Shot (OS).} A particular instance of \gls{fs} prompting is OS prompting, which provides merely a single example in the input context of the \gls{llm}, i.e. $C = \{ I,\, (x, y) \}$ \citep{brown2020language}. However, \gls{zs} and \gls{fs} are more widely used in research and practice.

\begin{figure}
\centering
  \includegraphics[width=1.0\textwidth]{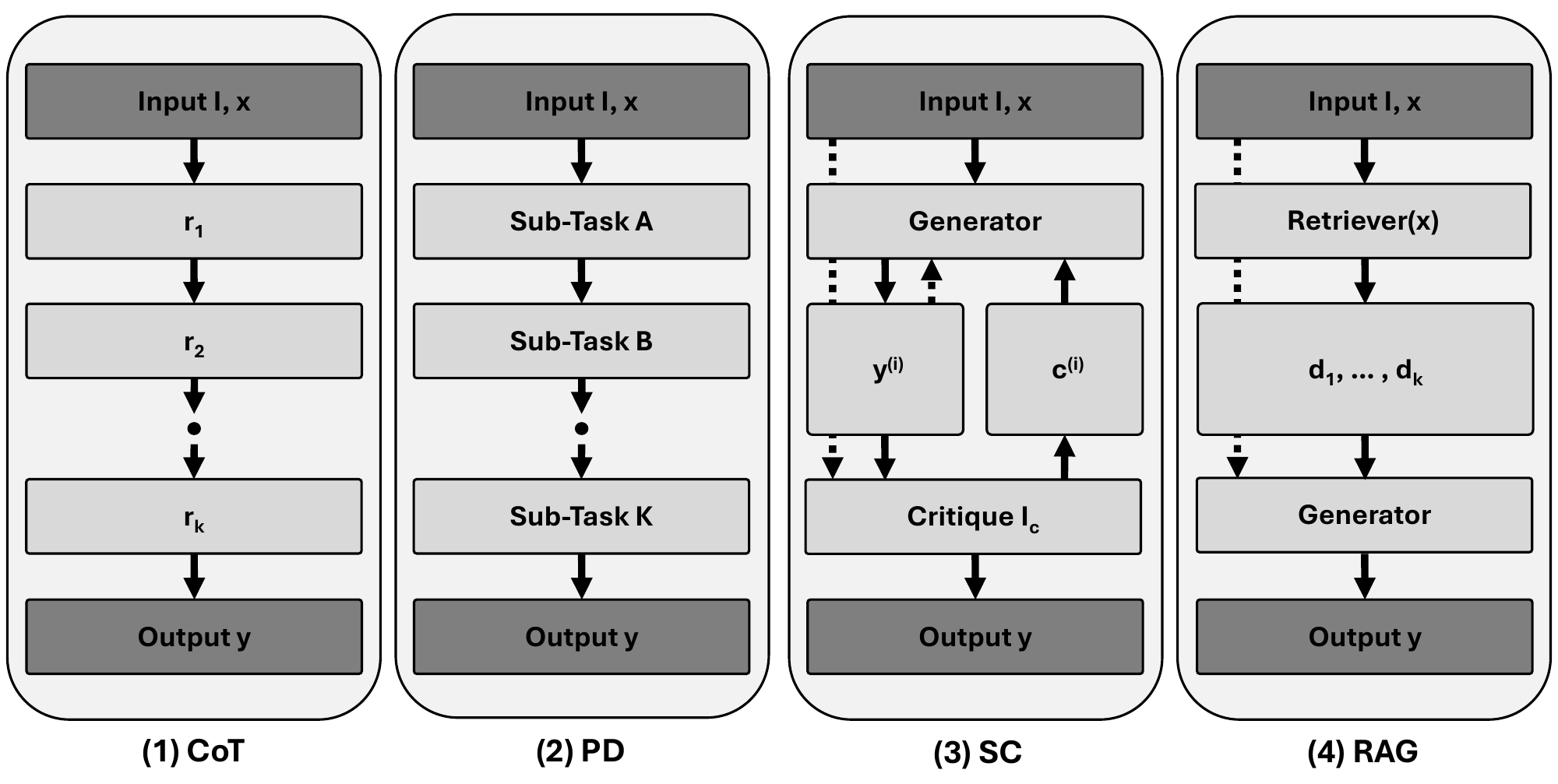}
  \caption[\textit{\glsentrylong{pe} (\glsentryshort{pe})} techniques overview]{\textit{\gls{pe}} techniques overview encompassing four well-known and popular approaches proposed in research before such as \textit{(1)} \textit{OS-\gls{cot}}, \textit{(2)} \gls{pd}, \textit{(3)} \gls{sc} and \textit{(4)} \gls{rag}.}
	\label{fig:chapter-2-pe}
    \vspace{-0.2cm}
\end{figure}

\glsreset{cot}
\paragraph{\gls{cot}.} An early approach to improve the reasoning capabilities of \glspl{llm} by \gls{pe} techniques alone is represented by \gls{cot} \citep{wei2022chain}, which asks the \gls{llm} to introduce a sequence of intermediate reasoning steps instead of immediately providing the output answer, as depicted in Figure \ref{fig:chapter-2-pe} (1). More formally, the input context of the \gls{llm} is constructed as $C = \{ I,\, (x_1, y_1, r_1),\, \ldots,\, (x_k, y_k, r_k) \}$, with $x_i$ being the input problem of the $i$-th example, $y_i$ denoting the correct output for the $i$-th example and $r_i = \{ r_i^1,\, \ldots,\, r_i^l \}$ providing $l$-many intermediate reasoning steps for the $i$-th example. Multiple variations of \gls{cot} have been proposed, such as the \gls{fs} variant (\gls{fs}-\gls{cot}), an OS variant with $k=1$ (OS-\gls{cot}) and a \gls{zs} variant (\gls{zs}-\gls{cot}), which directly instructs the model to think step-by-step without examples \citep{kojima2022large}.

\glsreset{pd}
\paragraph{\gls{pd}.} To enable \glspl{llm} to handle more complex problems, \cite{khot2022decomposed} proposed a novel \gls{pe} technique referred to as \gls{pd}. In comparison to \gls{cot}, \gls{pd} divides the overall problem into smaller subtasks, with each subtask being delegated to a collection of task-specific prompting strategies, as shown in Figure \ref{fig:chapter-2-pe} (2). 

\paragraph{\gls{sc}.}\glsreset{sc} Another important \gls{pe} technique to improve the task-solving effectiveness of \glspl{llm} is \textit{\gls{sc}} \citep{saunders2022self}, depicted in Figure \ref{fig:chapter-2-pe} (3). Instead of generating an answer $y$ directly by the \gls{llm} from the input context, i.e. $y = LLM_\theta(I, x)$, \gls{sc} first \textit{(1)} generates an initial solution response $y^{(0)} = LLM_\theta(I, x)$, which \textit{(2)} is followed by a \textit{self-critique} as $c^{(0)} = LLM_\theta(I, x, y^{(0)}, I_c)$ with critique instructions $I_c$ to \textit{(3)} produce an enhanced response $y^{(1)} = LLM_\theta(I, x, y^{(0)}, c^{(0)})$. In particular, as indicated by the indices, \gls{sc} can also be conducted in an iterative manner, consisting of multiple rounds of output and critique generation, depending on the problem context.

\vspace{-0.2cm}
\paragraph{\gls{rag}.} Since the previously discussed \gls{pe} techniques mostly rely on static, predefined input context, \glspl{llm} might miss contextually relevant external information for a given problem and example. Furthermore, without access to external knowledge, \glspl{llm} are forced to solely rely on information stored in their parameters during pretraining. To mitigate this problem, \textit{\gls{rag}} \citep{lewis2020retrieval, li2022survey} has been proposed in research, exploiting external knowledge bases to inject context-specific information into the input context, i.e. $C = \{ I,\, x,\, d_1, \ldots,\, d_k\}$, with $D = \{d_1, \ldots,\, d_k\}$ referring to the top-$k$ external documents or chunks retrieved by a respective retrieval model $R(x) = D$, depicted in Figure \ref{fig:chapter-2-pe} (4). This \gls{pe} technique proves particularly effective for problems that tackle domains and knowledge outside of the \glspl{llm}' pretraining data.

\vspace{-0.2cm}
\paragraph{Other \gls{pe} Techniques.} While we discussed several of the most important \gls{pe} techniques, particularly relevant in the context of this work, many other \gls{pe} approaches have been proposed in research, of which we briefly summarize a selection. For example, \textit{self-consistency} \citep{wang2022self} samples multiple, diverse reasoning paths and aggregates the results to provide an improved output response, while \textit{Tree-of-Thoughts (ToT)} \citep{long2023large, yao2024tree} generalizes the notion of \gls{cot} to tree-like structures, allowing \glspl{llm} to self-evaluate decision paths and conduct backtracking when necessary. Moreover, \textit{Graph-of-Thoughts (GoT)} represents a generalization of ToT, where intermediate thoughts or reasoning steps are interpreted as graph vertices, while edges between these nodes represent arbitrary dependencies, enabling the modeling of complex problems.




\definecolor{specialgray}{RGB}{240,240,240}

\newtcolorbox{myrqbox}{
  colback=specialgray,
  colframe=black, 
  boxrule=0pt, 
  toprule=1.5pt, 
  bottomrule=1.5pt, 
  leftrule=0pt, 
  rightrule=0pt, 
  sharp corners,
  left=2pt, right=2pt, top=2pt, bottom=2pt, 
  before skip=8pt, after skip=8pt
}
\clearpage
\newpage
\thispagestyle{empty}
\null  

\part[\sc{Prototyping: GUI Retrieval via Natural Language}]{\sc{Prototyping: GUI Retrieval via Natural Language}}
\label{part:gui_retrieval}

\clearpage
\newpage
\thispagestyle{empty}
\null  

\chapter{Natural-Language-Based GUI Retrieval}
\label{cha:nl_gui_retrieval}
\vspace{-0.2cm}

Given the context of the first main challenge (\challone{}), namely, mapping \gls{nlr} to \gls{gui} prototypes, in this chapter, we provide details about the first solution approach, the creation of a novel benchmark for evaluation and the results obtained. In particular, we investigate text-based retrieval methods with \gls{nlr} as input and rankings over \gls{gui} prototypes as the output target. The following sections are based on previous work and were already partially published before \citep{kolthoff2019automatic, kolthoff2020gui2wire, kolthoff2021automated, kolthoff2023data}\footnote{This section is adapted from: (1) \textbf{Kolthoff, Kristian}, Bartelt, Christian, and Ponzetto, Simone Paolo. Data-Driven Prototyping via Natural-Language-based GUI Retrieval. \emph{Automated Software Engineering}, March 2023, 30(1), 13, pages 1--34, Springer. Sections are directly reused with only minor adaptations (i.e. thesis Section \ref{cha:nl_gui_retrieval}.i corresponds to paper Section i) (2) \textbf{Kolthoff, Kristian}, Bartelt, Christian, and Ponzetto, Simone Paolo. GUI2WiRe: Rapid Wireframing with a Mined and Large-Scale GUI Repository using Natural Language Requirements. In \emph{Proceedings of the 35th IEEE/ACM International Conference on Automated Software Engineering (ASE, A*)}, Melbourne, Australia (Virtual Event), January 2021, pages 1297--1301. ACM. (3) \textbf{Kolthoff, Kristian}, Bartelt, Christian, and Ponzetto, Simone Paolo. Automated Retrieval of Graphical User Interface Prototypes from Natural Language Requirements. In \emph{Proceedings of the 26th International Conference on Applications of Natural Language to Information Systems (NLDB)}, Saarbrücken, Germany (Virtual Event), June 2021, pages 376--384, Cham: Springer International Publishing.}. Our interactive prototype, source code, datasets (including the novel gold standard) and demonstration video are all publicly available to foster future research for \gls{nlr}-based \gls{gui} retrieval\footnote{Materials for this chapter available at \url{https://github.com/kristiankolthoff/RaWi} and \url{https://github.com/kristiankolthoff/GUI2WiReTool}, video at \url{https://youtu.be/2nN-Xr2Hk7I}}.

\vspace{-0.1cm}

\paragraph{Personal Contribution.} I ideated and created the concept for the work, constructed the entire approach and architecture, implemented the \gls{bert}-\gls{ltr} model and data-driven tool prototype. Moreover, I designed and conducted the evaluation, including the novel gold standard, conducted data analysis and was the sole author of the manuscript.

\vspace{-0.2cm}

\section{Motivation}
As we already discussed in the introduction of this work, \gls{gui} prototyping represents an important technique to visualize the analysts' understanding of the \gls{nlr} from stakeholders, enable their validation by stakeholders as a tangible artifact, provide the foundation for incorporating stakeholders early into the application development and lead to fruitful discussions, clarifications and refinements of requirements \citep{windsor1992prototyping, rudd1996low, pohl1996requirements, ravid2000method, beaudouin2002prototyping, mukasa2008integration, pohl2010requirements, pohl2016requirements}. Especially high-fidelity \gls{gui} prototypes have been proven to be effective, since these prototypes provide the foundation for discussions of higher quality between stakeholders and analysts, and more detailed feedback can be obtained during user tests compared to low-fidelity approaches \citep{landay1994interactive, rudd1996low, coyette2007multi}. However, the benefits of high-fidelity \gls{gui} prototypes are accompanied by the disadvantages of increased time and experience required for their development \citep{rudd1996low}.

To facilitate and simplify \gls{gui} prototyping, a plethora of approaches has been proposed before. Many popular \gls{gui} prototyping tools are available and employed in practical prototyping environments such as \textit{Sketch} \citep{sketch}, \textit{Adobe XD} \citep{adobexd}, \textit{Mockplus} \citep{mockplus}, \textit{Figma} \citep{figma} and \textit{Balsamiq} \citep{faranello2012balsamiq}. These approaches typically allow users to combine basic \gls{gui} components and a small number of \gls{gui} templates. However, they are time-consuming and demand a wide prototyping experience for effective application. More recent \gls{gui} retrieval approaches such as \textit{GUIFetch} \citep{behrang2018guifetch}, \textit{Swire} \citep{huang2019swire}, \textit{Screen2Vec} \citep{li2021screen2vec} and \textit{VINS} \citep{bunian2021vins} use hand-drawn sketches or \gls{gui} screenshots as input for retrieving \glspl{gui} to support prototyping. However, they initially require a basic \gls{gui} representation and the retrieved \gls{gui} images cannot be directly reused or adapted. Overall, these approaches focus on supporting the prototyping of the final \gls{gui} design by showing GUI design alternatives, but cannot be applied for interactive \gls{gui} prototyping for \gls{rel} and \gls{rval}.

In addition, \textit{Guigle} \citep{bernal2019guigle} represents the first basic search engine for \glspl{gui} of mobile apps supporting a query language and stylistic searches. However, it proposes only a simple retrieval approach and the obtained \gls{gui} screenshots are not directly reusable. Specifically, \textit{Lucene} \citep{lucene} was employed, focusing on simple, traditional \gls{tfidf} and \gls{bm25} methods. However, these methods are optimized for the same modality and assume similar text representations between \textit{query} and \textit{documents} to work effectively. When considering the problem at hand, the input representations in the form of \gls{nlr} represent high-level, abstract concepts and descriptions, while \glspl{gui} are multi-dimensional objects, typically represented as layout hierarchies encompassing \gls{gui} components with a diverse set of properties. Therefore, instead of applying these methods without any adaptation to \gls{nlr}-based \gls{gui} retrieval emphasizing the representation mismatch, it is crucial to tailor them to this problem. Hence, this work is based on the leading research question of challenge \challoneone{}: \textit{How can we adapt and optimize text-based retrieval methods and techniques to enable more effective \gls{nlr}-based \gls{gui} retrieval?}

\noindent
\paragraph{\gls{nlr}-based \gls{gui} retrieval for \gls{gui} prototyping.} In this work, we introduce \textit{\gls{rawi}}, a data-driven rapid \gls{gui} prototyping approach (to create high-fidelity \gls{gui} prototypes) that \textit{(i)} exploits a large-scale semi-automatically created \gls{gui} repository of mobile applications by employing \gls{nlr}-based ad-hoc \gls{gui} retrieval and automatically derives editable \gls{gui} screens for reuse and \textit{(ii)} thus effectively leverages the \gls{gui} prototyping knowledge embodied in the repository to facilitate prototyping and improve prototyping productivity. Due to enabling rapid mapping of \gls{nl} fragments from users to \gls{gui} prototypes, our approach allows for significantly faster \gls{gui} prototyping compared to a traditional approach and thus can be applied more effectively by analysts to elicit requirements with stakeholders via interactive \gls{gui} prototyping. Moreover, we envision supporting novice analysts with little or no prior experience in \gls{gui} prototyping with our approach by providing easy access to plenty of reusable \glspl{gui} and domain knowledge embodied in the \gls{gui} repository. Due to the automatic derivation of partly editable \gls{gui} screens from a large-scale \gls{gui} screenshot repository, our approach is able to provide a vast number of reusable \gls{gui} screens, which eliminates the manual effort typically required for providing \gls{gui} templates in traditional prototyping approaches. \textit{\gls{rawi}} integrates both the improved \gls{gui} retrieval mechanism and editable \gls{gui} screen derivation techniques in a novel, data-driven and web-based graphical prototyping editor.

\noindent
\paragraph{Contributions.} With our work, we make the following four research contributions:

\begin{itemize}[leftmargin=6mm]
    \item \textit{\gls{bert}-\gls{ltr} model for \gls{gui} ranking:} we present a \gls{bert}-\gls{ltr} approach that is trained on large-scale \gls{gui} relevance data harvested using crowdsourcing techniques, demonstrating substantial improvements over traditional retrieval and ranking methods.
    \item \textit{Novel gold standard for \gls{nlr}-based \gls{gui} retrieval:} we create a novel and comprehensive gold standard for \gls{nlr}-based \gls{gui} retrieval with crowdsourcing techniques to enable a systematic evaluation of \gls{nlr}-based \gls{gui} ranking models and make it publicly available to foster further research.
    \item \textit{In-depth \gls{ir} method comparison:} we conduct the first in-depth analysis and evaluation of various traditional \gls{ir}, \gls{aqe} and trained \gls{bert}-based \gls{ltr} models for \gls{nlr}-based \gls{gui} ranking using the novel gold standard.
    \item \textit{Evaluation methodology:} we propose a comprehensive evaluation methodology including multiple metrics for measuring \gls{gui} prototyping productivity and conduct an extensive user study to assess the usefulness of the \gls{gui} prototyping approach in a practical rapid prototyping environment.
\end{itemize}

\begin{figure*}
 \includegraphics[width=\textwidth]{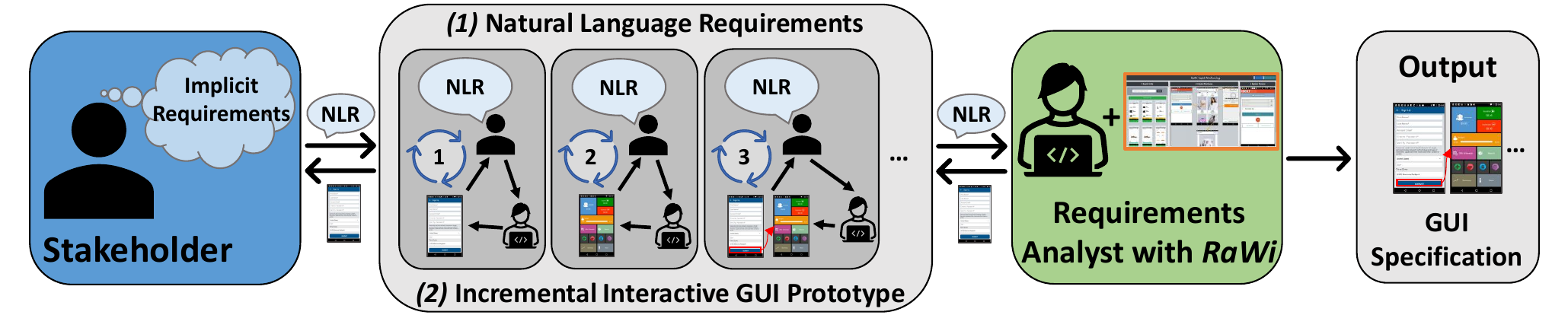}
  \caption[Iterative GUI prototyping process with \textit{RaWi}]{\gls{rel} and \gls{rval} via interactive \gls{gui} prototyping assisted by our approach \textit{RaWi} to rapidly map \gls{nlr} into \gls{gui} prototypes and iteratively create a \gls{gui} specification.}
	\label{fig:reqsoverview}
\end{figure*}

\noindent
\paragraph{Focus on the \gls{rel} process.} While \gls{gui} prototypes are employed with different characteristics for different scenarios, the focus of early-stage elicitation lies in uncovering the functionality of the system. Therefore, prototypes optimally should possess high-fidelity with regard to the functionality dimension, while the final \gls{gui} design plays a less significant role at that stage \citep{mccurdy2006breaking}. In this work, we particularly focus on supporting \gls{gui} prototyping with respect to the functionality of the system. As stated before, previous research showed that by employing high-fidelity \glspl{gui} for elicitation, the feedback quality and detail are improved \citep{rudd1996low, coyette2007multi, landay1994interactive}. In an interactive elicitation session with stakeholders, the available time for \gls{gui} prototyping is typically very limited and requires fast mapping of \gls{nl} fragments from stakeholders to \gls{gui} prototypes with the respective functionality, as shown in Figure \ref{fig:reqsoverview}. Here, our \textit{\gls{rawi}} approach provides automatic assistance especially for inexperienced analysts. Accordingly, in our conducted experiments we restricted the available time for \gls{gui} prototyping and quantified productivity in terms of selecting the necessary \gls{gui} components (see Section \ref{chapter-2:subsec:rq2-userstudy}). Similarly, our process of creating a gold standard for \gls{gui} retrieval using crowdsourcing techniques is meant to capture the scenario of stakeholders with no particular experience in \gls{gui} prototyping providing \gls{nl} queries, since participants on crowdsourcing platforms have a variety of backgrounds.

\noindent
\paragraph{Structure of the chapter.} The remainder of this chapter is structured as follows. In Section \ref{sec:rawi}, we provide a summary of our \gls{gui} retrieval and \gls{gui} prototyping approach \textit{RaWi}. Section \ref{sec:eval} presents the details on the experimental setting used for evaluation, whose results are presented in Section \ref{sec:results}. Threats to validity and limitations of this work are presented in Section \ref{sec:gui-retrieval-threats} and \ref{sec:limitations}, respectively. In Section \ref{sec:related_gui_retrieval}, we provide an overview of related work. Section \ref{sec:conclusions} covers the concluding remarks and presents future work.

\section{Approach: \gls{rawi}}
\label{sec:rawi}

\begin{figure*}
  \includegraphics[width=\textwidth]{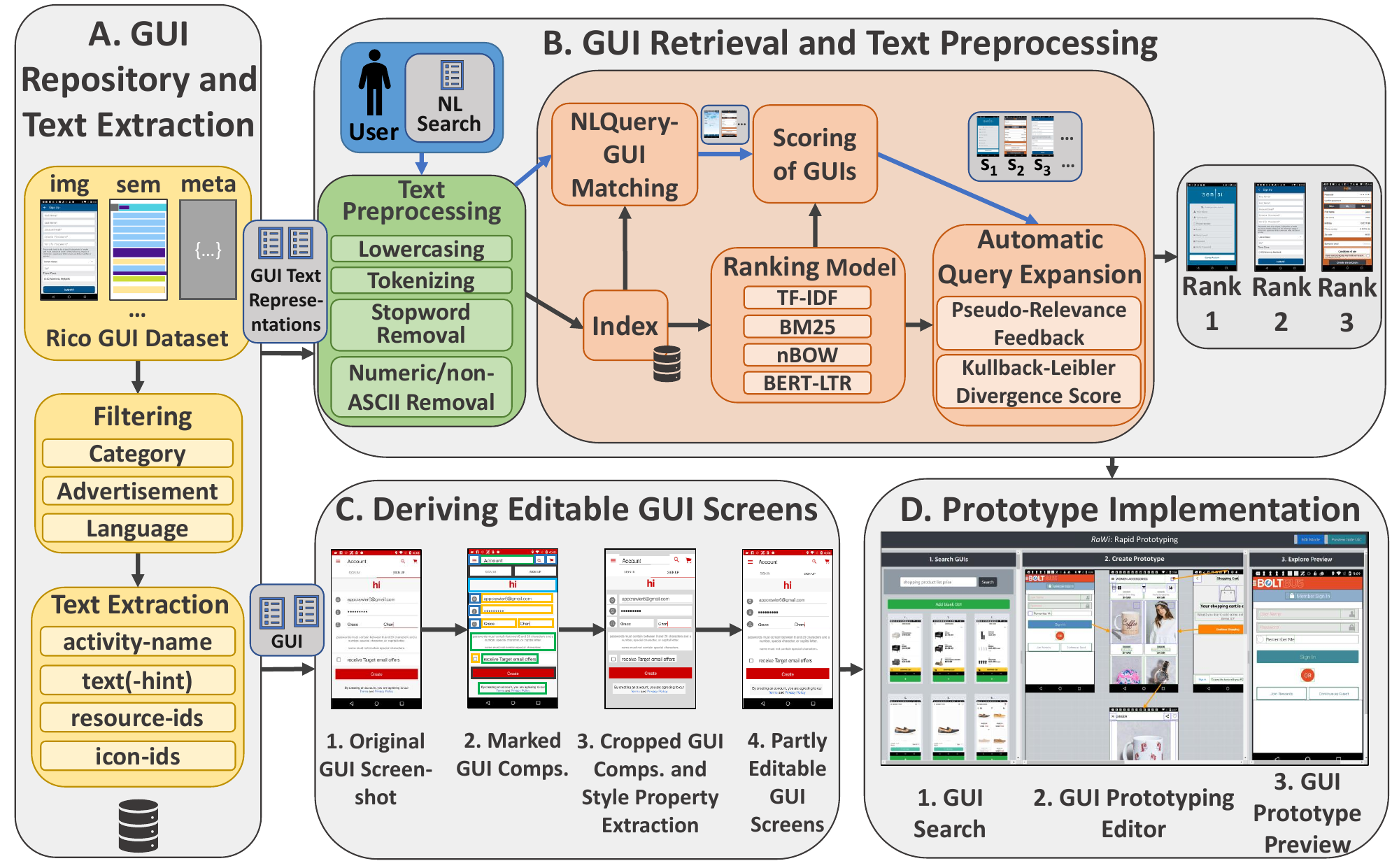}
  \caption[Architecture overview of \textit{RaWi}]{Architecture overview of \textit{RaWi} with four components: \textit{(A)} the large-scale \gls{gui} repository \textit{Rico} and \gls{gui} text extraction, \textit{(B)} \gls{gui} retrieval and text preprocessing pipeline, \textit{(C)} automatic \gls{gui} screen derivation and \textit{(D)} the rapid \gls{gui} prototyping editor.}
	\label{fig:overview}
    \vspace{-0.5cm}
\end{figure*}

The main goal of our approach is to \textit{(i)} enable fast \gls{gui} retrieval from a large-scale GUI repository via \gls{nlr}-based search queries to leverage the embodied \gls{gui} prototyping knowledge and \textit{(ii)} to allow users to rapidly build prototypes using automatically derived editable \gls{gui} screens. Fig. \ref{fig:overview} shows an overview of the architecture of \textit{RaWi} separated into four components. First, \textit{(A)} we employ the large-scale semi-automatically created \gls{gui} repository \textit{Rico} \citep{deka2017rico} for mobile applications as the basis for \gls{gui} retrieval. This dataset encompasses a large number of \textit{Android} applications that are crawled from the \textit{Google Play} store. To improve the quality of the \gls{gui} repository, we initially cleanse the \glspl{gui} based on multiple criteria. Afterwards, we create textual representations of the \glspl{gui} by extracting different text segments from the \gls{gui} hierarchy data. Second, \textit{(B)} we apply an identical pipeline of text preprocessing techniques to both the textual \gls{gui} representations and the \gls{nlr}-based search queries. In particular, in the context of this approach, \gls{nlr} refer to short, abstract descriptions representing the main functionality of a \gls{gui} (which could also be extracted from \glspl{us}). The retrieval architecture employed in \textit{RaWi} follows a multi-stage retrieval and reranking pipeline\footnote{While retrieval refers to selecting an initial candidate set of documents across the entire document collection with a fast, high-recall model, reranking refers to computing an improved ranking on a document subset (usually the top-$k$ documents) with a more expressive but computationally more expensive model.}. To this end, we compute an index over the \gls{gui} text representations and match the \gls{nlr}-based queries against it to obtain initial matches. Subsequently, we compute a ranking over the \gls{gui} matches using popular \gls{ir} and state-of-the-art \gls{bert}-based \gls{ltr} models. In addition, we experimented with multiple \gls{aqe} techniques. Third, \textit{(C)} we automatically create partly editable \gls{gui} screens by exploiting the \gls{gui} screenshots and \gls{gui} hierarchy data and extract basic style properties of some \gls{gui} components. Finally, \textit{(D)} we provide a web-based implementation of our approach that integrates all of the previously discussed components. Subsequently, we describe each component in detail.

\subsection{\gls{gui} Repository and Text Extraction}

To retrieve relevant \glspl{gui} for \gls{nlr}-based search queries, we first require a suitable \gls{gui} repository. Recent data-driven design research constructed and released several \gls{gui} datasets appropriate for our approach by automatically crawling a vast number of \textit{Android} applications from \textit{Google Play} such as \textit{ReDraw} \citep{moran2018machine}, \textit{ERICA} \citep{deka2016erica}, \textit{Rico} \citep{deka2017rico} and \textit{Enrico} \citep{leiva2020enrico}. \textit{ReDraw} collects \glspl{gui} by automatically exploring 5,416 \textit{Android} applications comprising a total of 14,382 unique \gls{gui} screenshots and hierarchy data using \gls{gui} testing automation techniques to simulate user input. Screenshots of the \glspl{gui} are captured either with third-party frameworks or platform-dependent utilities. In contrast, \textit{ERICA} solely relies on manual human-based \gls{gui} exploration and is primarily developed for capturing user interaction traces. \textit{ERICA} comprises around 18,600 unique \glspl{gui} harvested from 2,400 applications. \textit{Rico}, an extension of the \textit{ERICA} dataset, mines \gls{gui} screenshots, \gls{gui} hierarchy data, application metadata and interaction traces with both human-based and automatic exploration techniques and constitutes the largest design dataset of the discussed ones with 72,219 \glspl{gui} collected from 9,772 unique \textit{Android} applications (from 27 different application categories). \textit{Rico} employs a custom \textit{Android} service to access detailed information about all \gls{gui} components such as text, absolute position on the screen, visibility indicators, resource identifiers and the activity name, among others. For our \textit{\gls{rawi}} approach, we employ \textit{Rico} for multiple important reasons: \textit{(i)} the large number of \glspl{gui} encompassed, \textit{(ii)} the coverage of many diverse application domains and \textit{(iii)} the comprehensive textual information provided, particularly valuable for text-based \gls{gui} retrieval approaches.

\paragraph{\gls{gui} Filtering.} Due to the partly automated extraction of \glspl{gui} in \textit{Rico}, the \gls{gui} repository naturally contains erroneous \glspl{gui} inapplicable to our approach. We recognized multiple \gls{gui} types to be excluded through manual inspection of the dataset. First, \textit{(1)} \glspl{gui} of the \textit{entertainment} category (e.g. \textit{gaming}) are discarded since \textit{RaWi} should specifically support rapid prototyping of business and utility applications. Second, \textit{(2)} \glspl{gui} displaying \textit{advertisement overlays} and \textit{full-screen web views} are identified heuristically by particular patterns of component labels and discarded from the repository (based on the presence of a specific number of components and component types, e.g., only \textit{web view}, \textit{web view} and \textit{icons}, etc.). By employing this filtering heuristic, a common hierarchy-\gls{gui}-screenshot mismatch error type, in which the \gls{gui} screenshot appears to be a normal \gls{gui} with many low-level \gls{gui} components, but the \textit{Rico} view hierarchy data represents the entire \gls{gui} only as a single full-screen web view, is identified and removed. This mismatch occurs often since, e.g., developers directly embed their web apps into \textit{Android} apps instead of developing a native application. Third, \textit{(3)} we discard non-English \glspl{gui} identified through a language detection framework that computes language probabilities by accumulating character-level $n$-gram spelling feature probabilities \citep{shuyo2010language}. Other languages could be supported by adapting the retrieval models (e.g., by employing multilingual embeddings or simply translating the corpus).

After applying filtering, the \gls{gui} repository encompasses 57,764 unique \glspl{gui}. Furthermore, research identified several other error types encompassed in the \textit{Rico} \gls{gui} dataset \citep{leiva2020enrico}. Many of these errors are related to component-level mismatches (e.g., vertical offset, missing background image, wrong component type), which do not affect our \gls{gui} retrieval methods. However, the \textit{Rico} \gls{gui} dataset also rarely contains completely mismatched hierarchy-\gls{gui}-screenshot pairs, where the \gls{gui} screenshot shows completely different functionality compared to the available view hierarchy. These cases can potentially create \glspl{fp} during \gls{gui} retrieval. However, they cannot be identified easily. Available research provides no evaluation of the extent of their occurrences in \textit{Rico} and based on our manual inspection, the occurrence of these cases appears to be infrequent. Moreover, the potential negative impact of erroneous \glspl{gui} on the retrieval performance is mitigated by the vast number of available \glspl{gui} in the dataset.

\begin{figure}
  \includegraphics[width=\textwidth]{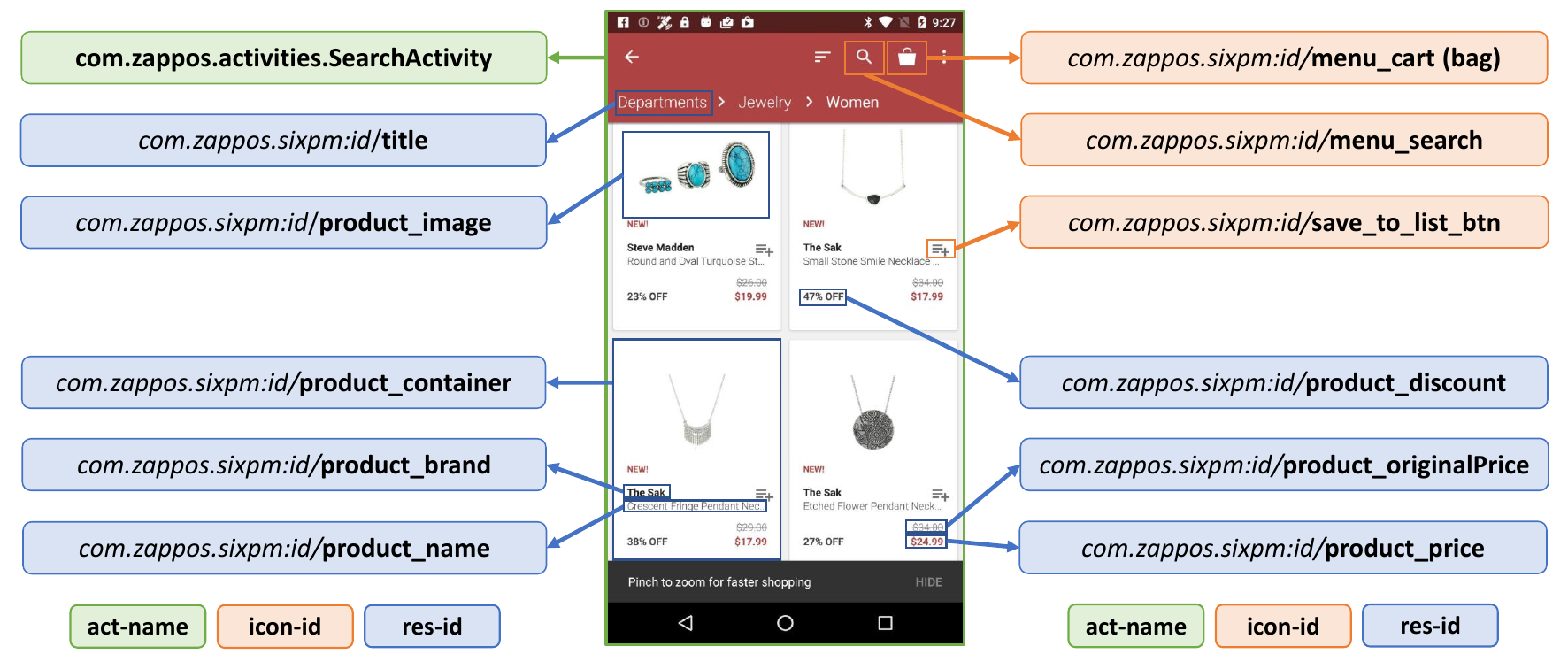}
  \caption[Extracting text segments from GUI prototypes]{Multiple \gls{gui} text segments (\textit{activity name (act-name)}, \textit{icon (icon-id)} and \textit{resource identifiers (res-id)}) that are extracted and preprocessed for \gls{nl}-based \gls{gui} retrieval.}
	\label{fig:textextraction}
    \vspace{-0.5cm}
\end{figure}

\vspace{-0.2cm}
\paragraph{\gls{gui} Text Extraction.} To support \gls{gui} retrieval from \gls{nlr}-based search queries over the \gls{gui} repository, we construct a textual representation for each \gls{gui} by extracting several text segments through \textit{XPath} \citep{olteanu2002xpath} expressions from the \gls{gui} hierarchy. First, we extract \textit{displayed text} and \textit{text hints} that are marked as being visible on the captured \gls{gui} screenshot. Based on initial experiments, we decided to neglect the extraction of text from components marked as non-visible on the captured \gls{gui} screenshot since these text sections often tend to mislead the retrieval models. Similarly, we exploit the \textit{full activity name} of the \gls{gui} and the \textit{resource identifier} of each \gls{gui} component since developers often provide semantically descriptive naming especially valuable for \gls{gui} retrieval. However, these strings (for example \textit{"com.sample.sens.register.\-CreateNewAccountActivity"}) require additional preprocessing through a tokenization pipeline to make them fully searchable. To this end, we apply \textit{punctuation}, \textit{camel} and \textit{snake case} tokenization and finally use a probabilistic tokenizer based on English Wikipedia unigram frequencies to extract tokens from the remaining parts. To clean the created tokens from non-descriptive and general tokens (for example \textit{"com"}, \textit{"main"} and \textit{"activity"}), we employ a custom domain-specific stopword list for filtering, containing the most frequent words in the corpus. Furthermore, we exploit the semantic labels of icons provided by \textit{Rico} due to their similarly meaningful semantic descriptions. Fig. \ref{fig:textextraction} shows an example \gls{gui} retrieved with \textit{RaWi} and multiple text segments that are extracted for retrieval, including the \textit{activity name}, the \textit{icon labels} and the \textit{resource identifiers}. By extracting these dimensions of text, our approach supports both the search for entire \glspl{gui} and the retrieval of \glspl{gui} that only contain parts relevant to the \gls{nlr} search query.

\subsection{\gls{gui} Retrieval and Text Preprocessing}

For text preprocessing of both the \gls{gui} text representations and the \gls{nlr} input query, we apply a standard preprocessing pipeline. First, we lowercase the text and apply tokenization. Tokens are then excluded by multiple filters. We discard \textit{stopwords} and words comprising \textit{numeric} or \textit{non-ASCII} \textit{(American Standard Code for Information
Interchange (ASCII))} characters. Subsequently, we describe the used baseline \gls{gui} retrieval techniques, the adapted \gls{aqe} techniques and the semantic \gls{bert}-based \gls{ltr} models.

\vspace{-0.2cm}
\paragraph{Baseline Retrieval Models.} For enabling retrieval of relevant \glspl{gui} with \gls{nlr}-based search queries from the introduced \textit{Rico} \gls{gui} repository, we adopt well-known baseline retrieval models that showed their effectiveness in various domains before and are established in general-purpose search engines \citep{manning2008introduction}. Therefore, we decided to adopt the \gls{tfidf} \citep{salton1988term}, \gls{bm25} \citep{robertson1995okapi} and a \gls{tfidf}-weighted \gls{nbow} method \citep{sheikh2016learning, galke2017word} using pretrained dense word embeddings for similarity scoring to establish a first baseline for \gls{gui} ranking.

\begin{figure}
  \includegraphics[width=\textwidth]{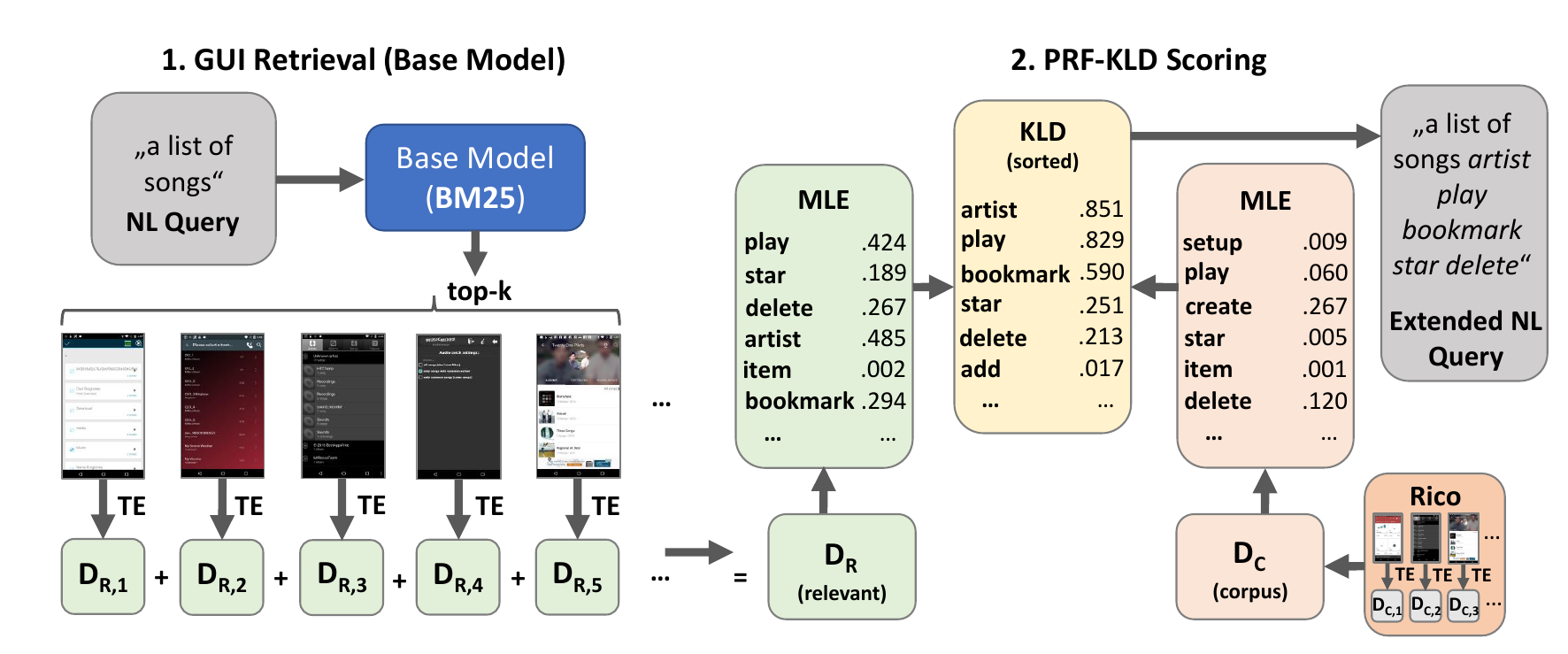}
  \caption[PRF-KLD scoring example for AQE]{Example of using the \gls{prf}-\gls{kld} scoring: \textit{(1)} Initial top-$k$ \gls{gui} retrieval with a base ranking model (e.g., \gls{bm25}) and \textit{(2)} \gls{prf}-\gls{kld} score calculation based on the \glspl{mle}.}
	\label{fig:aqe}
\end{figure}

\vspace{-0.2cm}
\glsreset{aqe}
\paragraph{\gls{aqe}.} To further enhance the retrieval performance and tackle the vocabulary mismatch problem (or synonymy), we additionally examined \gls{aqe} techniques \citep{manning2008introduction, azad2019query}. Synonyms such as \textit{"choose"} and \textit{"select"} cannot be matched by traditional \gls{ir} methods such as \gls{bm25} (although the later-described \gls{bert}-\gls{ltr} approach mitigates this problem). \gls{aqe} methods attempt to extend and refine the initial query automatically with relevant terms in order to improve the ranking performance. In the absence of user feedback, \gls{aqe} methods employ \textit{\gls{prf}}, where the initially retrieved top-$k$ documents (from a base model such as \gls{bm25}) are assumed to be relevant and thus used as input to compute the query expansion terms. For our \gls{gui} retrieval approach, we therefore focused on the incorporation and evaluation of \gls{prf} methods. Based on the notion of \gls{prf}, many expansion term scoring techniques have been devised in \gls{aqe} research before \citep{azad2019query}. These scoring techniques generally follow a similar procedure. By comparing the term distributions of term $t$ in the relevant documents $D_{R}$ and the entire corpus $D_{C}$, the importance of a term in the relevant documents $D_{R}$ can be estimated. The initial \gls{nlr} query can then be expanded with the top-$n$ terms according to their ranking score. For our \gls{gui} retrieval system, we compute the \gls{kld} score \citep{kullback1951kullback, carpineto2001information} for each term $t \in D_{R}$ as

	\begin{equation*}
  \label{eq:t}
      Score_{KLD}(t) = p(t \mid D_{R}) \cdot \log\frac{p(t \mid D_{R})}{p(t \mid D_{C})}
\end{equation*}

\noindent with $p(t \mid D_{R})$ and $p(t \mid D_{C})$ being the probability of term $t$ occurring in the relevant documents $D_{R}$ and in the entire document collection $D_{C}$, respectively. To compute these probabilities, we use the \textit{\gls{mle}}, i.e. $p(t \mid D_{x}) = \frac{f_{t,x}}{\sum_{d \in D_{x}}{\mid d \mid}}$, where $d$ is a document of $D_{x}$ with $\mid d \mid$ tokens and $f_{t,x}$ represents the frequency of term $t$ across all documents in $D_x$. Applying and evaluating the \gls{kld} score in our \gls{aqe} experiments is based on the notion that this term scoring showed its effectiveness compared to other expansion term scoring methods before \citep{carpineto2001information}. In addition to using this method for expansion term selection solely, we evaluated a second variant that includes the \gls{kld} score as a weight for the expanded terms in the retrieval model in order to control their effect on the \gls{gui} document ranking. Since the \gls{gui} prototypes are represented through multiple text segments, we adapt the \gls{prf}-\gls{kld} score to compute expansion candidates for each \gls{gui} text segment individually and aggregate the top-$n$ terms of the different text segments to obtain the query expansion terms. If duplicate top-$n$ terms from different text segments appear, we only incorporate them once in the expanded query. Similarly to the previously discussed method, we additionally evaluated a second variant of the \textit{text segment-wise} \gls{prf}-\gls{kld} by weighting the expansion terms with their respective scores to control their influence on the ranking. Figure \ref{fig:aqe} exemplifies our \gls{aqe} by \textit{(1)} retrieving the initial top-$k$ \glspl{gui} and extracting the text (TE) to obtain $D_{R}$, \textit{(2)} computing the \glspl{mle} for $D_{R}$ and $D_{C}$ to obtain the scores.

\paragraph{\gls{bert}-based \gls{ltr} Models.} The models presented previously can be applied to the \gls{gui} text documents, but as motivated earlier, \gls{gui} ranking differs from other text-based ranking tasks in multiple aspects. First, \glspl{gui} are not standard well-structured text documents, but typically provide solely \textit{brief text fragments} (e.g., button or short label text), \textit{named components} (often with proprietary abbreviations), \textit{many weakly relevant data items} (e.g., product descriptions or names) and \textit{no natural ordering} of the text. Accordingly, the semantic gap between the high-level, abstracted \gls{nlr} queries and the available \gls{gui} text representations is large. Therefore, more sophisticated semantic models are required to further improve the ranking performance. To this end, we finetune a state-of-the-art \gls{bert}-based \citep{devlin2019bert} \gls{ltr} model \citep{han2020learning}. \gls{bert} is a generally effective pretrained and large-scale language model that outperforms many traditional approaches across different \gls{nlp} tasks\footnote{Writing the original manuscript and publication occurred before the advent of more powerful \glspl{llm}.} (before the advent of \glspl{llm}).

In particular, \gls{ltr} approaches represent supervised models that learn a scoring function given a training set $T = \{(\mathbf{x},\mathbf{y}) \in \mathcal{X}^n \times \mathbb{R}_+^n\}$, with $\mathcal{X}$ representing the space of document-query pairs, $\mathbf{x}$ denoting an $n$-dimensional vector of individual items $x_i$ and $\mathbf{y}$ denoting an $n$-dimensional vector of non-negative real-valued relevance labels $y_i$ \citep{bruch2019analysis}. Here, an item $x_i$ refers to a representation of a query-document pair $(q,d)$, which is matched with a respective relevance label $y_i$. The scoring function of \gls{ltr} models $f_\theta: \mathcal{X}^n \to \mathbb{R}^n$ is learned by optimizing the parameters $\theta$ through employing the training data and should create scores that optimize the ranking of an ordered document list with respect to a query. To create query-document pair representations, we utilize \gls{bert}. Since the \gls{bert} model requires a specific input structure for the text, we prepare the \gls{bert} model by concatenating the \texttt{[CLS]} token (the first \gls{bert}-specific token), followed by the \gls{nlr} query, a separator \texttt{[SEP]} token (the second \gls{bert}-specific token) and the potentially truncated \gls{gui} document text, followed by a final \texttt{[SEP]} token as the input to the \gls{bert}-\gls{ltr} model. Next, this \gls{bert} model input structure is shown:

\begin{figure}
  \includegraphics[width=\textwidth]{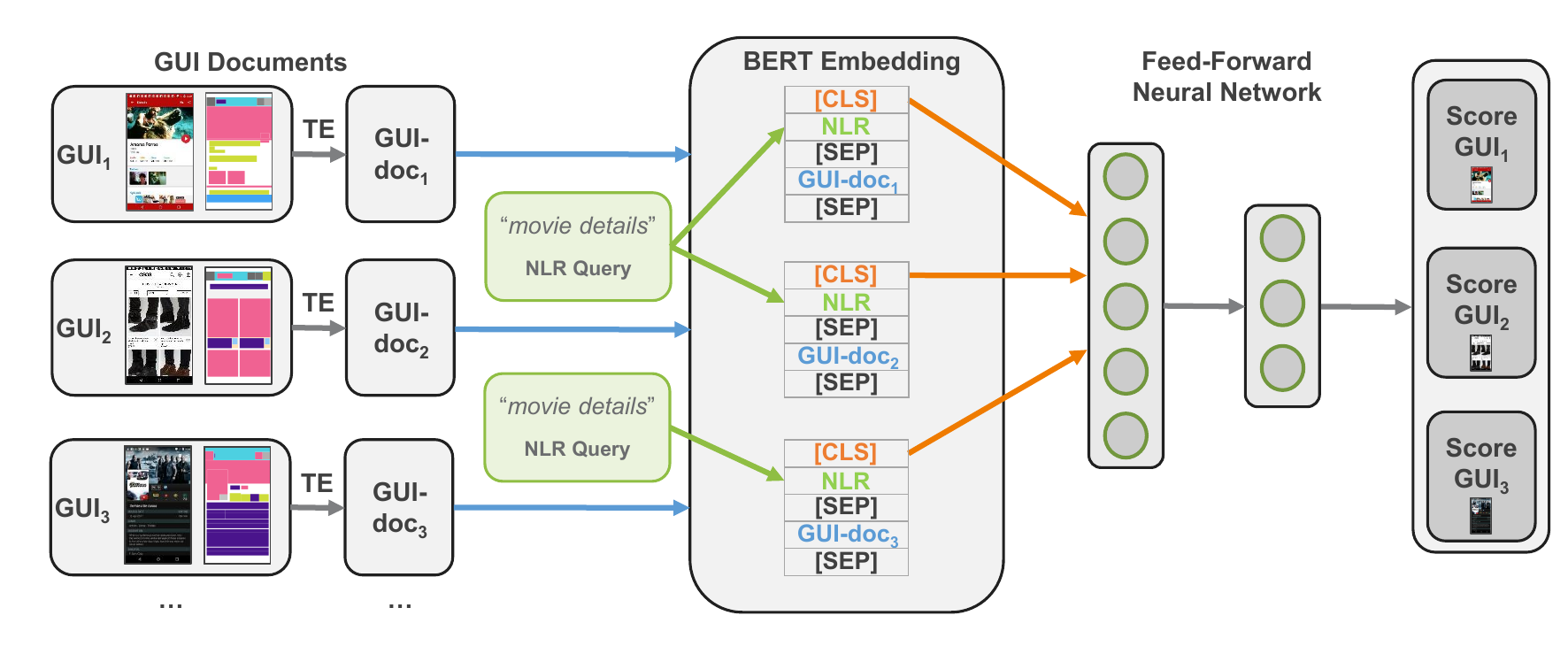}
  \caption[BERT-LTR model architecture]{\gls{bert}-\gls{ltr} model based on \cite{han2020learning} and adapted to the problem of \gls{nlr}-based \gls{gui} ranking using a pretrained \gls{bert} model with \gls{ffnn} output processing.}
	\label{fig:bertmodel}
\end{figure}

\begin{center}
\small
\begin{tabular}{ccccccccc}
    \texttt{[CLS]} & \texttt{NLR$_{t_{1}}$} & \texttt{NLR$_{t_{2}}$} & \texttt{...} & \texttt{[SEP]} &
    \texttt{GUI$_{t_{1}}$} & \texttt{GUI$_{t_{2}}$} & \texttt{...} & \texttt{[SEP]}
\end{tabular}
\end{center}

\vspace{-0.1cm}

\noindent where \texttt{NLR$_{t_{i}}$} refers to token $t_i$ of the \gls{nlr} input and \texttt{GUI$_{t_{j}}$} represents token $t_j$ of the \gls{gui} document input, provided by applying the \gls{bert} tokenizer to both inputs. Subsequently, the hidden representation $h_{\texttt{CLS}}$ of the special \gls{bert} token \texttt{[CLS]} is utilized as input to a ranking model based on a simple \gls{ffnn} to obtain a ranking score $s(q,d)$ for the pair of query $q$ (i.e. the \gls{nlr}) and document $d$ (i.e. the \gls{gui} text representation), i.e. $s(q,d) = FFNN(h_{\texttt{CLS}})$. To finetune \gls{bert} and train the ranking model, we experimented with three different approaches by varying the loss function. First, we utilized a \textit{pointwise} approach, which treats each query-document pair independently and forces the model to predict a single document score without comparison or contextualization to other documents in the list. To this end, we utilized the \textit{(1)} \textit{sigmoid cross-entropy loss} defined as

\vspace{-0.5cm}
\begin{equation}
\mathcal{L_{\textit{point}(\theta)}}
= -\frac{1}{|T|}\sum_T\sum_{i}\Big[y_i \log \sigma(s_i) + (1-y_i)\log\big(1-\sigma(s_i)\big)\Big]
\end{equation}

\vspace{-0.1cm}
\noindent with $|T|$ referring to the number of examples in the training set and $\sigma(s_i)$ denoting the \textit{sigmoid} function applied to the score $s_i$. For this particular loss function, the relevance labels should be binary, i.e. $y_i \in \{0,1\}$. Since this loss neglects the context of other documents, we experimented with a second variant, namely, a \textit{pairwise} \gls{ltr} approach. To this end, we applied a different loss function, which compares two documents and therefore incorporates a more relative ranking perspective in contrast to \textit{pointwise} \gls{ltr}. In particular, we employ a \textit{(2)} \textit{pairwise logistic loss} to learn the scoring function defined as


\begin{equation}
\mathcal{L}_{\textit{pair}}(\theta)
=
\frac{1}{|T|}\sum_T\sum_{i}\sum_{j}
\mathbb{I}\!\left[y_i > y_j\right]\,
\log\!\left(1 + \exp\!\left(-(s_i - s_j)\right)\right)
\end{equation}

\noindent with $|T|$ again referring to the number of examples in the training set, $\mathbb{I}(\cdot)$ denoting the \textit{indicator} function, which compares whether label $y_i$ is larger than label $y_j$, $s_i$ and $s_j$ referring to the scores of query-document pair $i$ and $j$ (having the same query), respectively. For query-document pairs with larger relevance labels compared to other pairs, the model should create scores that are higher in comparison. While comparing query-document pairs forces the model to better optimize the actual ranking between two documents with respect to a query, \textit{listwise} \gls{ltr} approaches incorporate the entire list of documents for comparison. To this end, we employ a \textit{softmax cross-entropy loss} \citep{bruch2019analysis}

\begin{equation}
\mathcal{L}_{\text{list}}(\theta)
= -\frac{1}{|T|}\sum_{T}\sum_{i}
y_i\log\!\left(
\frac{\exp(s_{i})}{\sum_{j}\exp(s_{j})}
\right)
\end{equation}

\noindent with $|T|$ again referring to the number of examples in the training set, while $s_i$ and $s_j$ denote the scores of query-document pair $i$ and $j$ (having the same query), respectively. For this loss function, labels $y_i$ have to be projected to a probability distribution, i.e. $\frac{y_i}{\sum_jy_j}$ \citep{bruch2019analysis}. The resulting values can be interpreted as probabilities that a particular document $x_i$ appears at the top position of the ordered list. This loss function was proposed as part of \textit{ListNet} \citep{cao2007learning}. We employ the \gls{bert}-\gls{ltr} model proposed by \citep{han2020learning} and illustrate its application for \gls{nlr}-based \gls{gui} ranking in Figure \ref{fig:bertmodel}. The respective \glspl{gui} are first transformed into \gls{gui} text documents, concatenated with the \gls{nlr} query and fed into the \gls{bert} model to obtain the \gls{bert} representations. To compute the ranking score for a \gls{gui}, the embedding is fed through a \gls{ffnn}. On this basis, the discussed \textit{pointwise}, \textit{pairwise} and \textit{listwise} losses can be computed to optimize the \gls{ffnn} and finetune the pretrained \gls{bert} model. Moreover, as an additional baseline for semantic models, we included a pretrained \gls{sbert} model \citep{reimers2019sentence} in our evaluation, in order to better understand the performance gained by the proposed \gls{bert}-\gls{ltr} models, which are trained specifically for the \gls{nlr}-based \gls{gui} ranking task. \gls{sbert} was also used as a \textit{text-only} baseline in \textit{Screen2Vec} \citep{li2021screen2vec}.

\subsection{Deriving Editable \gls{gui} Prototypes}

\begin{figure*} \includegraphics[width=\textwidth]{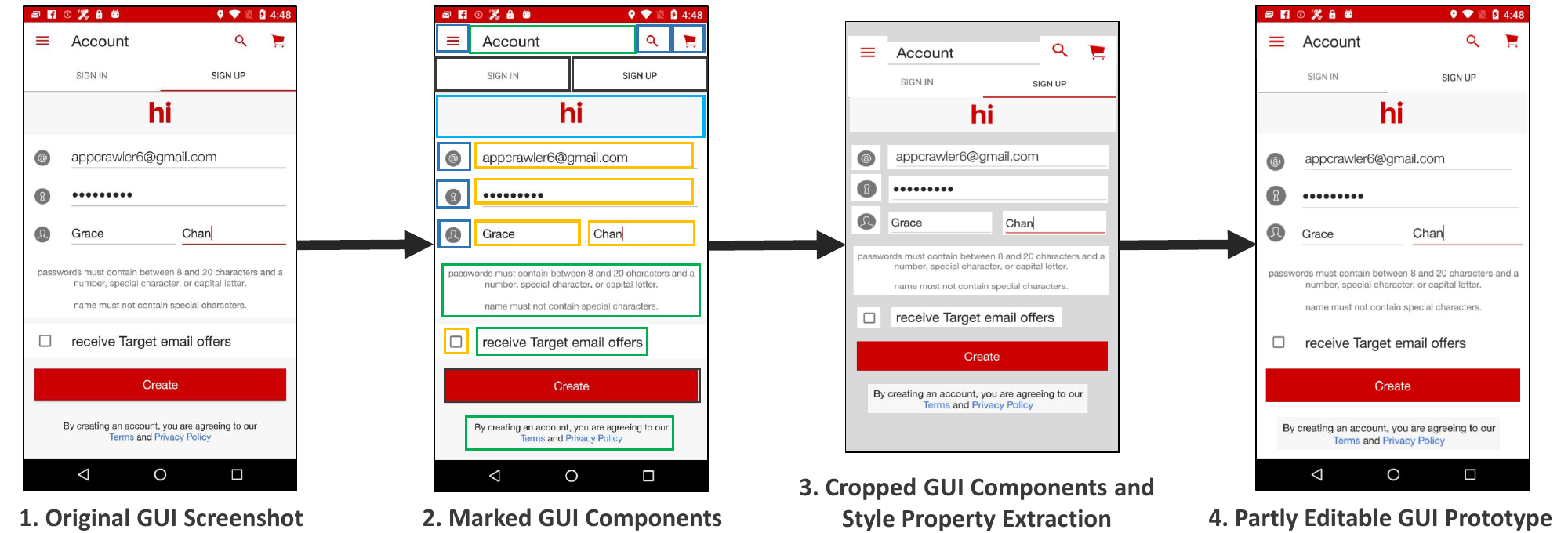}
  \caption[Example for deriving partly editable GUI prototypes]{Approach for automatically deriving partly editable \gls{gui} prototypes from \textit{Rico} \citep{deka2017rico} \gls{gui} screenshots and their corresponding \gls{gui} view hierarchies.}
	\label{fig:deriving_prototypes}
    \vspace{-0.2cm}
\end{figure*}

Since \textit{Rico} \citep{deka2017rico} provides solely \gls{gui} screenshots which are not directly reusable for rapid \gls{gui} prototyping and editing, we propose a simple yet effective algorithm to transform them into partly editable \gls{gui} screens. The automatic derivation of editable \glspl{gui} from the \gls{gui} repository is integrated into the graphical editor of \textit{\gls{rawi}}, which will be described in more detail in the subsequent section. Figure \ref{fig:deriving_prototypes} illustrates the entire procedure for creating partly editable \gls{gui} prototypes. In particular, we can exploit the \gls{gui} hierarchy data and semantic labels for \gls{gui} components encompassed in \textit{Rico} to extract them accordingly from the accompanying \gls{gui} screenshots. Therefore, \textit{(1)} we initially obtain the original \gls{gui} screenshot provided by \textit{Rico} and \textit{(2)} crop each contained \gls{gui} component from the screenshot based on its absolute position. We applied this procedure to both individual \gls{gui} components and additionally identified layout components. Third, \textit{(3)} for three \gls{gui} component types including \textit{labels}, \textit{buttons} and \textit{text-input}, we extract multiple style properties in our current prototype to provide more detailed editing capabilities. For the remaining components, we currently simply reuse their image crop in the editable \gls{gui} screens. For each layout group, we compute the background color as the top-ranked RGB color in its histogram. For \textit{labels}, we identify the font color by first obtaining the background color as before and then computing a normalized color distance to the background color for all colors in the histogram. The first top-ranked color in the histogram that exceeds a predefined distance threshold is selected as the font color, being a simple yet effective heuristic. In addition, we estimate the font size by comparing the bounding boxes from the hierarchy data and an instantiated bounding box containing the text. Since \textit{Rico} often provides only rough bounding boxes for the \textit{labels}, we compute refined bounding boxes using Tesseract-OCR \citep{smith2007overview}. For \textit{buttons} and \textit{text-inputs}, we similarly extract the background and font colors. We enrich \textit{Rico} with the identified style properties to enable proper \gls{gui} component instantiation in the editor later. Finally, \textit{(4)} the cropped \gls{gui} components are placed on their exact \gls{gui} screen positions. Subsequently, we describe our prototypical tool implementation.

\vspace{-0.1cm}

\subsection{Prototype Implementation}

\begin{figure*} \includegraphics[width=\textwidth]{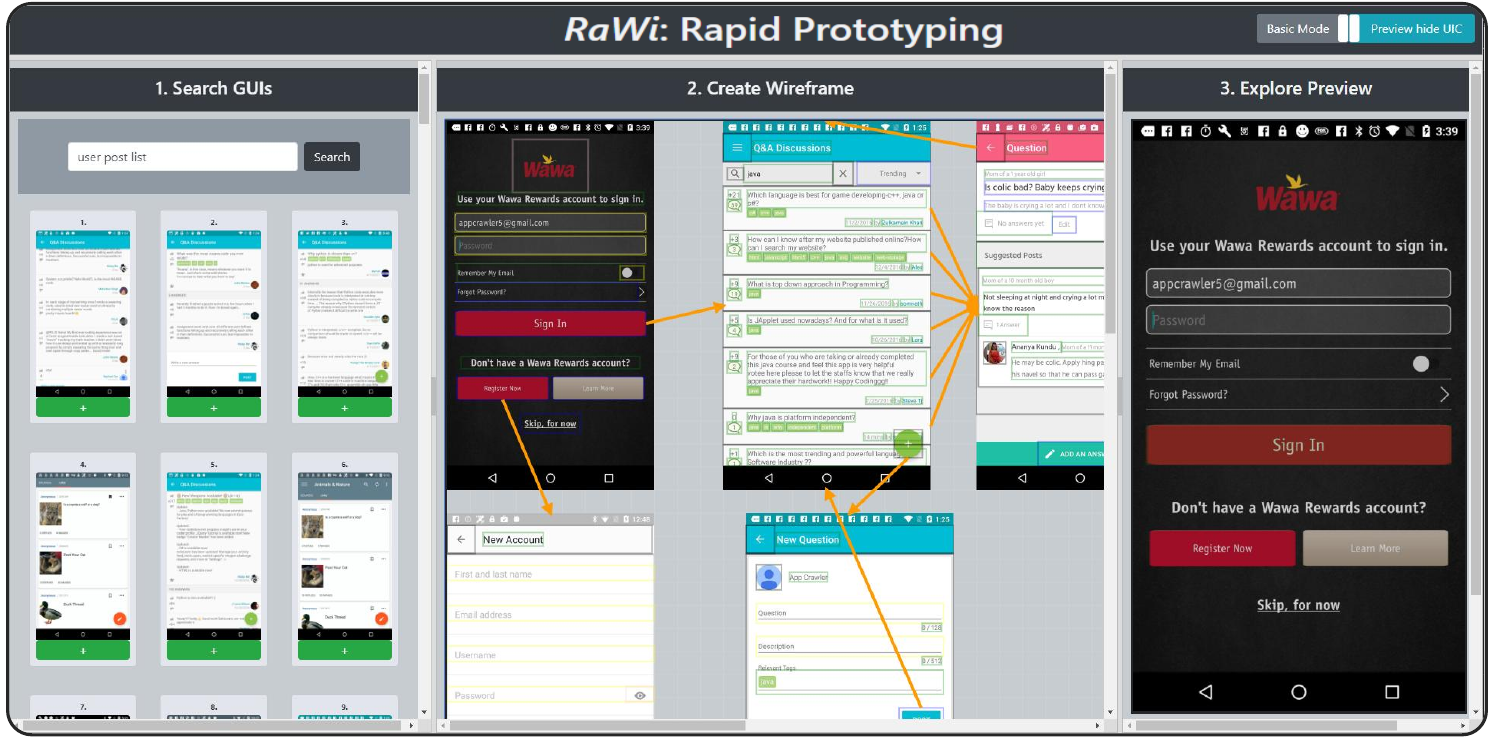}
  \caption[Web-based \gls{gui} prototyping editor of \textit{RaWi}]{Web-based \gls{gui} prototyping editor of our approach \textit{\gls{rawi}} with three components: \textit{(1)} the \textit{\gls{gui} search view}, \textit{(2)} the \textit{graphical editor} and \textit{(3)} the \textit{prototype preview}.}
	\label{fig:editor}
\end{figure*}

The presented \gls{gui} retrieval and \gls{gui} prototype derivation components are integrated in a tool prototype for a completely data-driven, web-based rapid prototyping editor using \gls{html}, \gls{css} and \gls{js}. Our current prototype provides \textit{(1)} a \textit{\gls{gui} search view}, \textit{(2)} a \textit{graphical prototyping editor} and \textit{(3)} an interactive and dynamically updated \textit{preview} of the created application, as illustrated in Figure \ref{fig:editor}. Subsequently, we provide more details about the features of our current prototypical tool implementation and each of the components.

\paragraph{\gls{nlr}-based \gls{gui} Search.} \gls{nlr}-based search queries can be submitted by users through a simple search interface with autocompletion capabilities. The \textit{Python}-based \gls{gui} retrieval component is accessible through a \textit{\gls{rest}} interface, which is implemented as a \textit{Django} application. For efficiency, the default is \gls{bm25}.

\paragraph{\gls{gui} Prototyping Editor.} Users are enabled to add retrieved \gls{gui} prototypes to the graphical prototyping editor. Here, the previously created enriched \gls{gui} hierarchy data is employed to create the editable \gls{gui} prototypes. Initially, we place all the image crops of layout and individual \gls{gui} components at their absolute positions on the \gls{gui} screen to preserve the complete original appearance of the \gls{gui}. By clicking on \textit{labels}, \textit{buttons} or \textit{text-input} components, an editable version of the \gls{gui} component is instantiated and basic properties such as \textit{text}, \textit{color} and \textit{size} can be modified. Furthermore, users can create custom \gls{gui} components directly or reuse different components from other retrieved \gls{gui} screens. In addition, we currently provide basic editing functionality: \gls{gui} components can be \textit{removed}, \textit{copied} and \textit{bound} to other screens. To create an interactive prototype, \textit{RaWi} provides the capabilities to create simple click-event transitions on \gls{gui} components. The editor is based on the \textit{2D-canvas} \gls{js} library \textit{KonvaJS} \citep{konvajs}.

\paragraph{\gls{gui} Prototype Preview.} We provide a dynamically updated preview of the created prototype showing all modifications similar to traditional prototyping approaches. The preview enables users to directly explore an interactive version of the created application prototype, which is based on the previously created \gls{gui} transitions as the foundation.

\section{Experimental Evaluation}
\label{sec:eval}
\vspace{-0.2cm}
In this section, we present the design and methodology of our experimental procedure to evaluate \textit{\gls{rawi}}. In particular, the main goals of our experimental evaluation are \textit{(i)} to measure the performance of the employed \gls{gui} retrieval techniques and conduct a comprehensive comparison between them on a newly created gold standard, \textit{(ii)} to assess the productivity improvements of \textit{\gls{rawi}} compared to a traditional rapid prototyping approach in a practical \gls{gui} prototyping environment and \textit{(iii)} to measure the perceived usefulness and gain user insights for our approach. To this end, we investigate the following three research questions in the subsequent experimental evaluation:
\vspace{-0.1cm}
\begin{itemize}
    \item \textbf{RQ$_{1}$}: \textit{Which method performs best for \gls{gui} retrieval on the basis of \gls{nlr} search queries?} To answer this question, we compare the presented baseline, adapted \gls{aqe} techniques (particularly \gls{prf}-\gls{kld} scoring methods), \gls{sbert} and the three \gls{bert}-\gls{ltr} variants. To evaluate the effectiveness, we utilize standard \gls{ir} metrics such as \textit{\gls{ap}} and \textit{\gls{mrr}}, among others.
    \item \textbf{RQ$_{2}$}: \textit{Does \textit{\gls{rawi}} increase the \gls{gui} prototyping productivity compared to a traditional prototyping tool?} To answer this question, we compare our proposed data-driven prototyping tool against \textit{MockPlus} \citep{mockplus}, which represents a popular, traditional prototyping approach. To measure productivity gains, we conduct a \textit{within-subjects} user study and compare the created \gls{gui} prototypes.
    \item \textbf{RQ$_{3}$}: \textit{Do users perceive \textit{\gls{rawi}} as useful for rapid high-fidelity \gls{gui} prototyping?} To answer this question, we asked participants several questions on a five-point Likert scale questionnaire regarding the perceived usefulness of the data-driven prototyping tool. Additionally, we computed the \textit{\gls{sus}}.
\end{itemize}
\vspace{-0.4cm}
\subsection{\texorpdfstring{RQ$_1$: \gls{gui} Retrieval Performance}{RQ-1: \gls{gui} Retrieval Performance}}
\label{rawi:rq1-setup}

\begin{figure*}   \includegraphics[width=\textwidth]{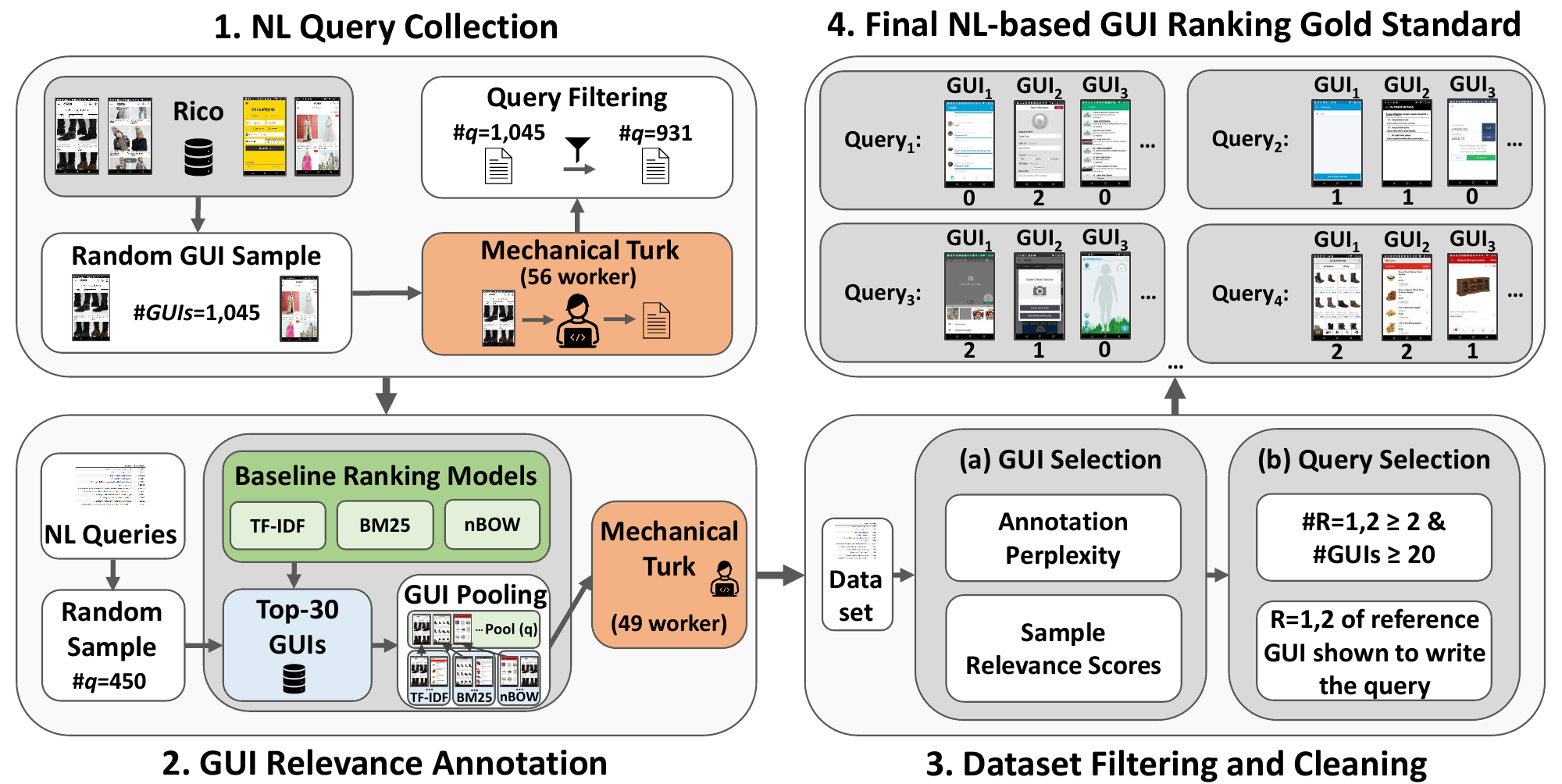}
  \caption[Overview of the gold standard creation pipeline for \gls{nlr}-based \gls{gui} retrieval]{Overview of the pipeline to create the gold standard for \gls{nlr}-based \gls{gui} retrieval using crowdsourcing with the three main steps: \textit{(1)} \gls{nlr} query collection through \gls{amt}, \textit{(2)} \gls{gui} relevance annotation through \gls{amt} and \textit{(3)} dataset filtering and cleaning.}
	\label{fig:goldstandard-construction}
    \vspace{-0.3cm}
\end{figure*}

\glsreset{amt}
\paragraph{Gold Standard.} Due to the absence of an evaluation dataset for \gls{gui} retrieval from \gls{nlr} queries, we decided to build a new and comprehensive gold standard through crowdsourcing techniques \citep{carvalho2011crowdsourcing} and publicly release it to foster future research. The construction of the retrieval gold standard is divided into three main phases: \textit{(i)} the collection of \gls{nlr} queries, \textit{(ii)} the collection of relevance annotations for potentially relevant \glspl{gui} with regard to \gls{nlr} queries and \textit{(iii)} the derivation of the final gold standard via data cleansing. To accomplish the large-scale data collection, we used the well-established crowdsourcing platform \gls{amt} \citep{paolacci2010running}. The \gls{gui} retrieval gold standard is available at our accompanying repository, including detailed descriptive statistics of the dataset (such as the \textit{distribution of contained app categories}, the \textit{average number of tokens per query}), which will also be discussed briefly at the end of this section. Subsequently, we explain the three phases of the gold standard construction in detail as shown in Figure \ref{fig:goldstandard-construction}.

\begin{figure*}
  \includegraphics[width=\textwidth]{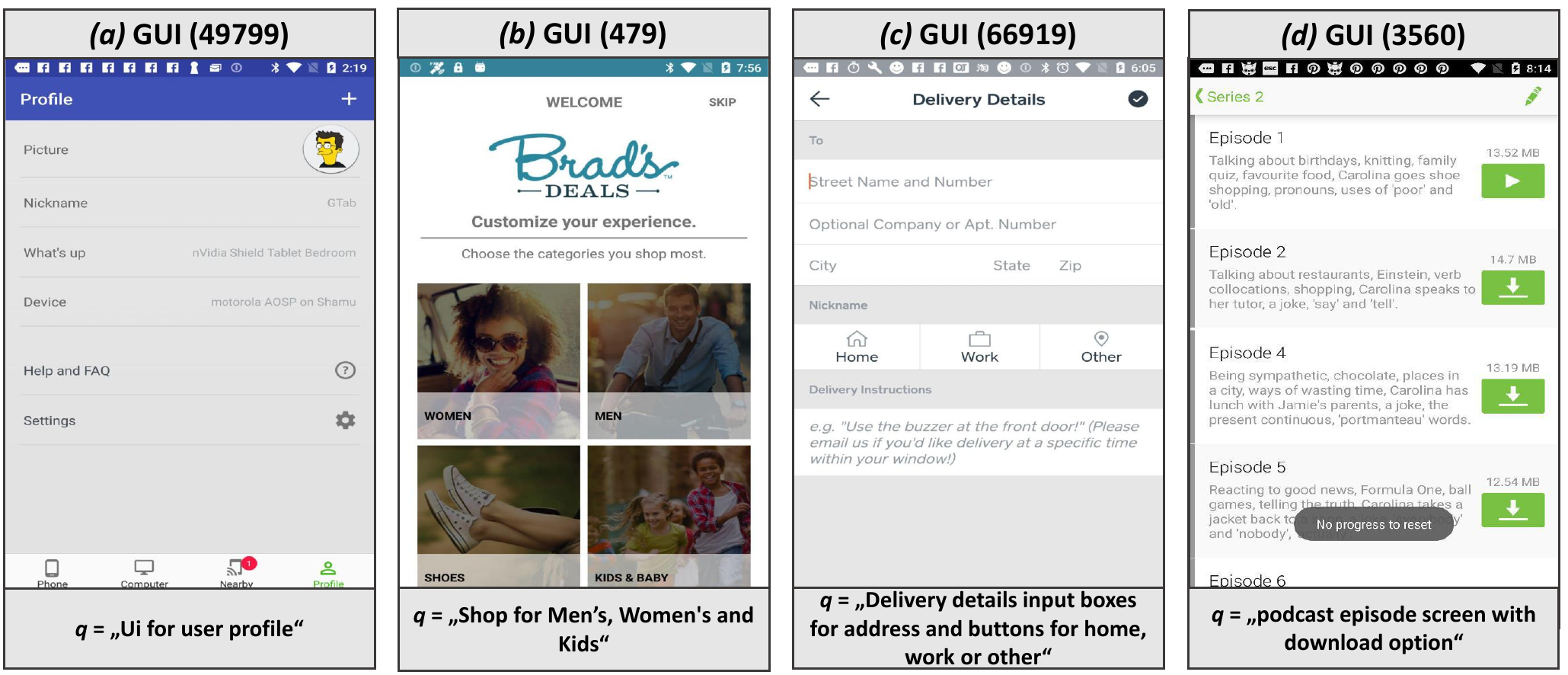}
  \caption[Four example \gls{gui}-\gls{nlr} pairs taken from the gold standard]{Four \textit{(a)}-\textit{(d)} example \gls{gui}-\gls{nlr}-query pairs taken from the gold standard with the \gls{nlr} queries gathered via crowdsourcing and the respective \textit{Rico} \gls{gui} indices.}
	\label{fig:goldstandard-gui-examples}
\end{figure*}

First, \textit{(i)} the collection of \gls{nlr} queries poses a challenge due to the difficulties of assuring high data quality for free-form \gls{nl}-based crowdsourcing tasks \citep{rashtchian2010collecting}. Therefore, we decided to employ \gls{amt} for data collection but restricted the tasks to workers that we instructed (hired) specifically for the completion of this task. These annotators represent a good proxy for the population in our scenario, since workers on \gls{amt} typically have diverse demographics \citep{ross2010crowdworkers, difallah2018demographics} (potential threats to validity on worker selection are discussed later). As the foundation for writing queries and to avoid bias, we randomly sampled \gls{gui} screenshots from the filtered \textit{Rico} dataset and asked the workers to write queries for them in a \textit{\gls{hit}}. Overall, for each of 1,045 unique \glspl{gui} from the sample, we gathered one unique \gls{nlr} query written by one of our overall 56 unique workers (resulting in 1,045 \gls{gui}-\gls{nlr}-query pairs). For additional quality assurance, we manually reviewed the queries and filtered six individual queries for \glspl{gui} that were erroneous or unusable screenshots from \textit{Rico} (e.g. \textit{white screen}, \textit{foreign language} not detected by the language detection framework, among others) and all queries from six workers that made systematic errors due to misunderstandings of the task. By filtering these queries, solely apparently erroneous \gls{nlr} queries were discarded. Thus, no bias is introduced, but the quality of the dataset is enhanced. After filtering, we retain 931 \gls{gui}-\gls{nlr}-query pairs with queries written by 50 unique workers and apply basic text preprocessing methods (\textit{dequoting}, \textit{stripping} and \textit{removal} of special characters). By employing \gls{amt} for query collection, we are enabled to gather a large number of different \gls{nlr} queries from a variety of annotators, thereby increasing the heterogeneity of the \gls{nlr} queries the retrieval and ranking models have to handle. Four \gls{gui}-\gls{nlr}-query pairs taken from the gold standard are shown in Figure \ref{fig:goldstandard-gui-examples}. The examples illustrate the diversity of the gathered \gls{nlr} queries ranging from short and general queries (e.g., Figure \ref{fig:goldstandard-gui-examples} \textit{(a)}) to long and detailed search queries (e.g., Figure \ref{fig:goldstandard-gui-examples} \textit{(c)}) across a broad selection of different application domains.

Second, \textit{(ii)} to accomplish the collection of relevance annotations, we initially require a collection of potentially relevant \gls{gui} documents to be annotated for each query. Due to the infeasibility of annotating the relevance of every \gls{gui} in \textit{Rico} for each query, we decided to heuristically employ our three baseline models (\textit{\gls{tfidf}}, \textit{\gls{bm25}} and \textit{\gls{nbow}}) to retrieve the potentially relevant top-30 \glspl{gui} for each query. To retain a diverse and heterogeneous \gls{gui} collection for each query, we apply the pooling technique \citep{jones1976information, teufel2007overview, tonon2015pooling, lipani2016fairness} to the \gls{gui} retrieval results of the three models by iterating over them in an alternating fashion and adding the top-ranked \gls{gui} not yet contained in the pool to the \gls{gui} collection. Moreover, we directly added the reference \gls{gui} that we previously showed to workers during the \gls{nlr} query harvesting to the pool of potentially relevant \gls{gui} documents. The pooling technique entails the disadvantage of potentially excluding some relevant documents in the collection. However, it represents a standard procedure for constructing large \gls{ir} test collections by overcoming the infeasibility of annotating the large-scale corpus entirely.

To keep the data collection feasible while ensuring a sufficient amount of data for a valid retrieval evaluation, we randomly sampled 450 queries from our previously created query collection, computed the pooled top-30 \glspl{gui} for each query with the approach described previously and published another \gls{hit} on \gls{amt} that asked workers to provide relevance annotations. Due to the subjectivity of the relevance \textbf{R}, each \gls{gui} was annotated by three different workers on a relevance scale of $\mathbf{R=0}$ (\textit{low relevance}), $\mathbf{R=1}$ (\textit{medium relevance}) and $\mathbf{R=2}$ (\textit{high relevance}). All 30 \glspl{gui} of an \gls{nlr} query were annotated exactly three times, resulting in 90 \gls{gui} relevance annotations per \gls{nlr} query by three distinct annotators. As guidance for the annotators, we provided multiple annotation examples for the relevance scale and included them in our accompanying repository. To assure high annotation quality, we restricted the task solely to workers that met multiple qualifications (\textit{\#Approved-\glspl{hit}} $\geq$ 1,000, \textit{Approval-Rate} $\geq$ 90\%, Residence in an English-speaking country (\textit{USA}\slash \textit{GB}\slash \textit{AU})), as previous research found these metrics particularly effective for enhancing the annotation quality \citep{akkaya2010amazon, peer2014reputation}. In addition to these generic \gls{amt} platform requirements calculated based on the workers' history, we created a custom qualification test to assess the comprehension of the relevance criteria and the overall task of the annotators on a small set of \gls{gui}-\gls{nlr}-query pairs, which simultaneously acted as a training phase. Workers were required to meet a certain performance threshold to be allowed to work on the annotation task. Furthermore, we informed the workers that submissions were rejected if a submission scored below $\frac{2}{3}$ agreement with the majority voting of a query. Overall, we gathered a total of 40,500 individual relevance annotations (for 450 \gls{nlr} queries with each of the 30 \glspl{gui} annotated three times) from 49 unique workers.

\begin{figure*} \includegraphics[width=\textwidth]{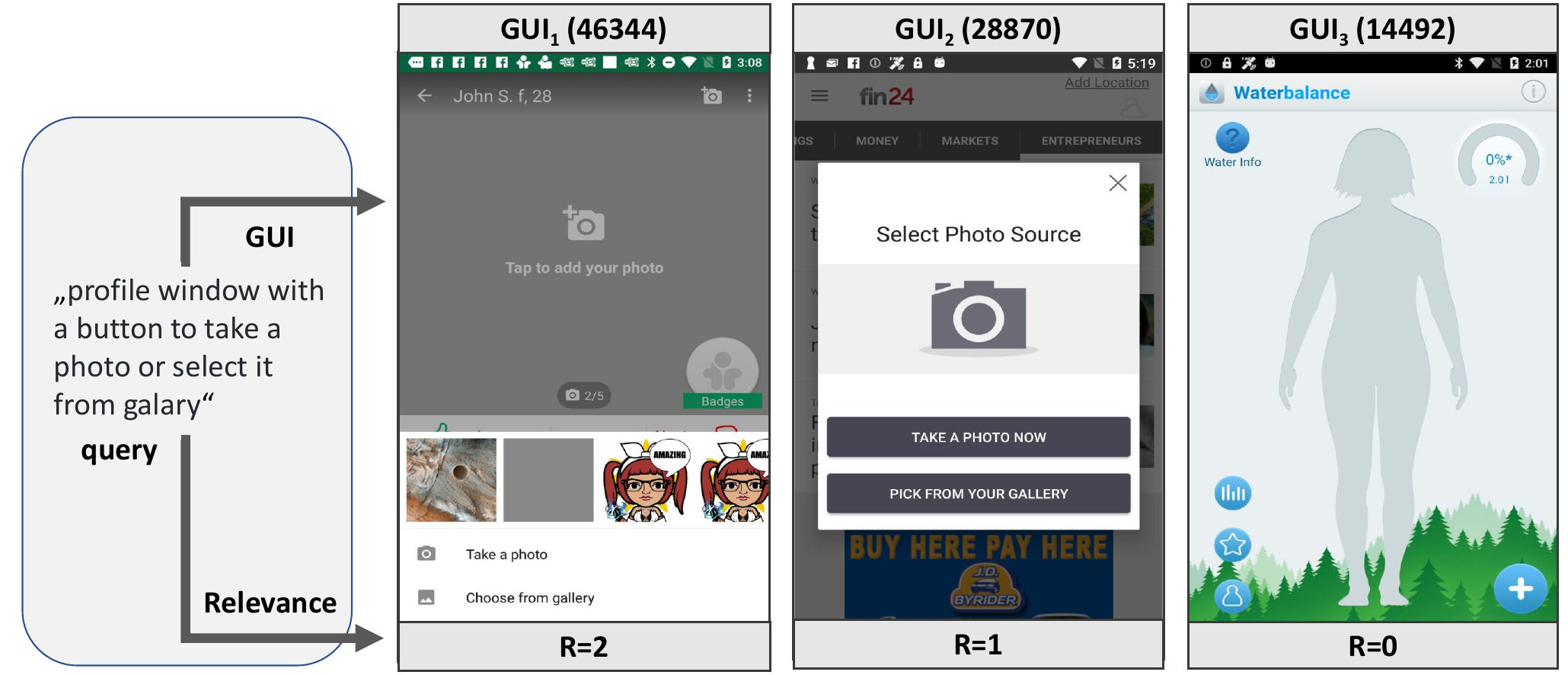}
  \caption[Example of an \gls{nlr} from the gold standard]{Example of an \gls{nlr} query from the gold standard with three \glspl{gui}, each annotated with their relevance, according to the majority voting from crowdsourcing.}
	\label{fig:goldstandard-full-example}
\end{figure*}

Third, \textit{(iii)} to obtain a high-quality gold standard from the previously harvested dataset, we run a pipeline of selection steps described subsequently. Due to the filtering, the final gold standard comprises a reduced set of 100 queries, each with their top-20 \glspl{gui}. To increase data quality, we removed \glspl{gui} with high annotation perplexity. In particular, we discarded \glspl{gui} with annotation ties or annotations that diverged to the extremes, for example, two votes for $\mathbf{R=0}$ and one vote for $\mathbf{R=2}$ or vice versa, since these annotations indicate high uncertainty in deciding for the ground truth. Annotators may have different opinions on relevance and are more or less strict in close annotation cases. Therefore, these annotation ties cannot be resolved and the respective \glspl{gui} are excluded. Afterwards, we set a suitable predefined ratio of different relevance scores per query: 14$\times$$\mathbf{R=0}$, 3$\times$$\mathbf{R=1}$ and 3$\times$$\mathbf{R=2}$. This provides a typical relevance distribution with mainly non-relevant \glspl{gui} while ensuring a small number of relevant \glspl{gui}. For each query, we randomly sampled \glspl{gui} with the respective relevance scores according to the previously defined quantities. If the quantity for a relevance score could not be reached, we randomly sampled from the other scores to obtain 20 \glspl{gui} per query. However, we only kept queries that have a minimum of two $\mathbf{R=1}$ \glspl{gui}, two $\mathbf{R=2}$ \glspl{gui} and 20 \glspl{gui} remaining. To obtain the same amount of \glspl{gui} per query, we decided to set the cutoff at the top-20. In addition, we discarded queries where the reference \gls{gui} that was used to write the query did not receive a relevance score of $\mathbf{R=1}$ or $\mathbf{R=2}$ from the human relevance annotators, as a query sanity check. The rationale for discarding these queries is to ensure high quality in the written \gls{nlr} queries. If the \gls{nlr} queries do not match the shown reference \glspl{gui} (based on the relevance annotators' judgments), the query is potentially of low quality (e.g., the annotator that wrote the query did not understand the \gls{gui}, provided an ambiguous query, etc.). From the remaining query collection, we randomly sampled 100 queries to obtain the final gold standard. To illustrate the gold standard, an example \gls{nlr} query with three \glspl{gui} and their respective ground truth relevance annotations is illustrated in Figure \ref{fig:goldstandard-full-example}. Each of the \glspl{gui} comes with a different relevance score with respect to the \gls{nlr} query. For example, \gls{gui}$_{1}$ is annotated with $\mathbf{R=2}$ since it fully corresponds to the functionality requested in the query. However, \gls{gui}$_{2}$ only receives a relevance of $\mathbf{R=1}$ since the \gls{gui} contains the requested photo selection options, but the functionality is not embedded in a user profile context. Finally, \gls{gui}$_{3}$ is annotated with a relevance of $\mathbf{R=0}$ since it is unrelated to the query showing a user profile for daily water management.

\begin{figure*}[!t]
  \centering
  \frame{
    \subfloat[\centering Relevance distribution for entire dataset]{
      \includegraphics[width=0.45\linewidth]{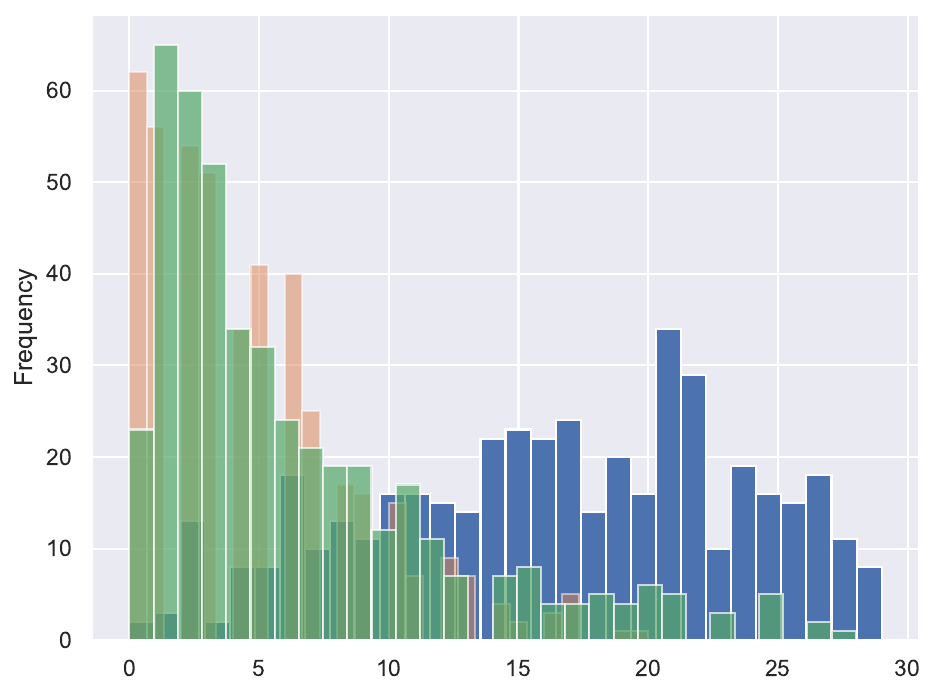}
      \label{fig:rel_dist_1}
    }
  }%
  \qquad
  \frame{
    \subfloat[\centering Relevance distribution for gold standard]{
      \includegraphics[width=0.45\linewidth]{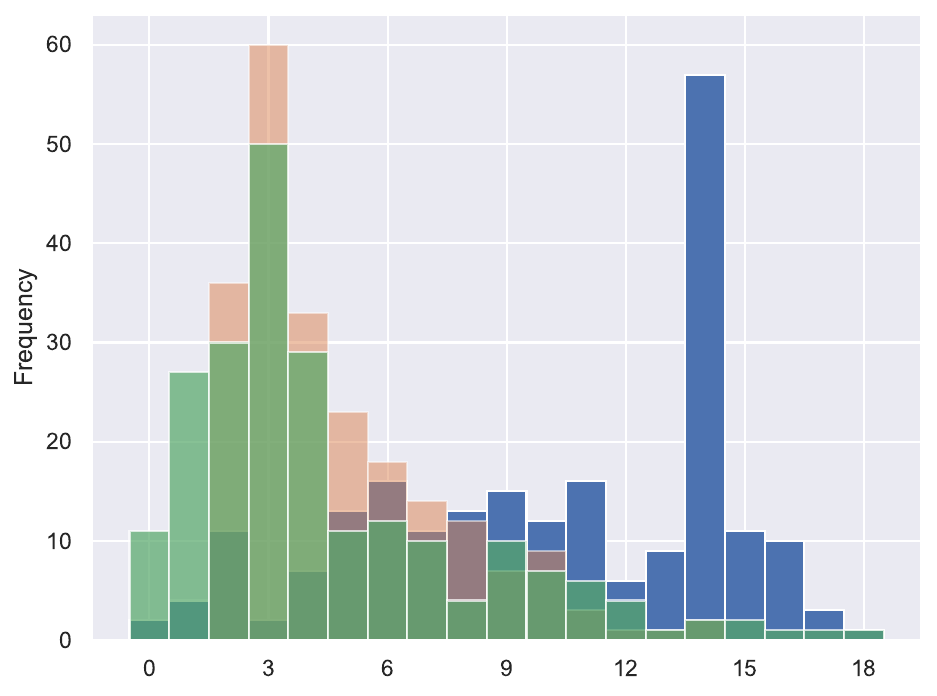}
      \label{fig:rel_dist_2}
    }
  }
  \caption[Relevance level distributions for the gold standard]{Distributions of the different relevance levels (\textit{blue} = \textit{low}, \textit{yellow} = \textit{medium}, \textit{green} = \textit{high}) for \textit{(a)} the entire dataset ($k=30$) and \textit{(b)} the final gold standard ($k=20$).}
  \label{fig:rel_dist}
\end{figure*}

\begin{table}[t]
\centering
\small
\setlength{\tabcolsep}{5pt}
\begin{tabular}{lcccccccc}
\toprule
\textbf{Variable} & \textbf{n} & \textbf{$\mathbf{\mu}$} & \textbf{$\mathbf{\sigma}$} & \textbf{min} & \textbf{Q1} & \textbf{med.} & \textbf{Q3} & \textbf{max} \\
\midrule
(1) \#Queries/Annotator       & 50  & 18.62  & 7.33   & 1 & 15.25 & 20.00  & 21.75  & 35 \\
(2) \#Annotations/Annotator   & 49  & 55.10  & 105.72 & 1 & 2.00  & 10.00  & 55.00  & 569 \\
(3) \#Words/Query            & 931 & 6.17   & 3.69   & 1 & 4.00  & 5.00   & 8.00   & 29 \\
(4) \#Components/GUI         & 931 & 17.52  & 13.75  & 1 & 7.00  & 13.00  & 24.00  & 104 \\
(5) \#Words/GUI               & 931 & 180.88 & 200.31 & 4 & 56.50 & 118.00 & 229.50 & 2464 \\
\bottomrule
\end{tabular}
\caption[Summary statistics for the \gls{nlr}-based \gls{gui} retrieval dataset]{Summary statistics (\textit{mean} $\mu$, \textit{standard deviation} $\sigma$, \textit{minimum}, \textit{first quartile} \textbf{Q1}, \textit{median}, \textit{third quartile} \textbf{Q3}, \textit{maximum}) for the collected dataset: \textit{(1)} number of queries written per annotator, \textit{(2)} number of relevance annotations per annotator, \textit{(3)} number of words per query, \textit{(4)} number of components per \gls{gui} and \textit{(5)} number of words per \gls{gui}.}
\vspace{-0.3cm}
\label{tab:dataset_stats}
\end{table}

To gain insights into the overall collected dataset and gold standard, we computed several summary statistics shown in Table \ref{tab:dataset_stats}. For example, annotators ($\mathbf{n}$=50) wrote 18.62 ($\sigma$=7.33) \gls{nlr} queries on average, while the other annotators ($\mathbf{n}$=49) provided 55.1 ($\sigma$=105.72) relevance annotations on average. In addition, the collected \gls{nlr} queries consist of 6.17 ($\sigma$=3.69) words on average in our dataset, which indicates higher information need complexity and specificity for \gls{nlr}-based \gls{gui} searches in comparison to typical search queries, where most of the queries consist of only up to three words \citep{phan2007understanding}. Moreover, sampled \glspl{gui} encompass 17.52 ($\sigma$=13.75) components on average, which provides a simplified proxy to \gls{gui} complexity. Furthermore, based on the applied \gls{gui} sampling approach, the distribution of included app categories is similar to the distribution of the original \textit{Rico} dataset, which corresponds to high diversity (26 app categories). High lexical diversity is also achieved among the queries, as indicated by a very small \textit{average pairwise Jaccard similarity} \citep{jaccard1901etude} (2.7\%) over all queries in the gold standard. Figure \ref{fig:rel_dist} shows the relevance level distributions for the entire dataset (Figure \ref{fig:rel_dist_1}) and the final gold standard (Figure \ref{fig:rel_dist_2}), both illustrating typical distributions with only a few queries having large numbers of \textit{high} (and \textit{medium}) relevance \glspl{gui}, but usually larger numbers of \textit{low} relevance \glspl{gui}. To assess the \textit{\gls{iaa}} of the gold standard, we computed \textit{Krippendorff's $\alpha$} \citep{krippendorff2011computing}, a chance-corrected \gls{iaa} metric supporting ordinal data (such as relevance levels). As a result, we obtain a score of $\alpha$=0.67, which indicates substantial agreement according to different score interpretation methods \citep{landis1977measurement, kraemer2012dsm}.

\vspace{0.2cm}
\paragraph{Ranking Model Parameters.} To evaluate the various retrieval and ranking models in our experiments, we utilized \gls{tfidf}, \gls{bm25} ($k_{1}$=1.5, $b$=0.75, $\epsilon$=0.25 as standard parameters \citep{manning2008introduction}) and \gls{nbow} (using 300-dimensional \textit{word2vec} embeddings \citep{mikolov2013distributed}) as baselines. The four considered \gls{aqe} methods based on \gls{prf}-\gls{kld} are evaluated using \gls{bm25} as the base ranking model. In particular, we restricted the number of documents employed for expansion term computation to $k$=10 and set the number of added expansion terms to $n$=10. We used a similar setup for the \textit{text-segment-wise} \gls{prf}-\gls{kld} variants, but restricted the number of expansion terms per text segment to $n_{ts}$=2. For brevity, we refer to the \textit{text-segment-wise} \gls{prf}-\gls{kld} method as \textit{(s)} and to the \textit{weighted} variants as \textit{(w)} in the evaluation results. For the \gls{bert}-\gls{ltr} models, we adopted the pretrained large \gls{bert}-base version \citep{devlin2019bert} with a vector size of 768 and applied finetuning using the relevance annotations of the remaining 350 queries (9,495 instances) not contained in the gold standard. We employed the \textit{TensorFlow}-Ranking \gls{bert}-Extension for training \citep{han2020learning} and trained each of the three model variants for 30,000 iterations (\textit{learning-rate} = $1\times10^{-5}$ and \textit{batch-size} = 1). For the additional pretrained, semantic \gls{sbert} baseline \citep{reimers2019sentence}, we employed the full pretrained 768-dimensional model with \textit{cosine similarity} to obtain a \gls{gui} ranking.
\vspace{0.2cm}
\glsreset{ap}
\paragraph{Evaluation Metrics.} To measure the performance of the \gls{gui} ranking models, we computed several well-known \gls{ir} metrics. First, \textit{(1)} we computed the \textit{Precision at rank} $k$ ($P@k$) as the fraction of relevant \gls{gui} documents $\vert R_{k} \vert$ over the number of all retrieved \gls{gui} documents $k$. Second, \textit{(2)} we computed the \textit{\gls{ap}}, defined as the average of $P@k$ values at all relevant \gls{gui} document positions with $rel_{k}$ being an indicator for the relevance at position $k$ and $\vert R \vert$ the number of relevant documents. In particular, the described $P@k$ and $AP$ metrics are computed as shown in the following:

\renewcommand{\arraystretch}{0.0}

\begin{tabular}{m{5cm}m{5.5cm}} 
        \centering
        \begin{equation}
            P@k = \frac{\vert R_{k} \vert}{k} 
        \end{equation}
         & \begin{equation}
            AP = \frac{\sum_{k=1}^{n}{P@k \times rel_{k}}}{\vert R \vert}
        \end{equation} 
\end{tabular}

\glsreset{mrr}

\noindent The $AP$ is averaged over the dataset (in research often denoted as \textit{\gls{map}}). Third, \textit{(3)} we computed the \textit{\gls{mrr}}, where $rank_{i}$ denotes the rank of the highest ranked relevant \gls{gui} document of the $i$-th query. In addition, \textit{(4)} we computed the $HITS@k$, indicating whether at least one relevant \gls{gui} document is ranked among the top-$k$ ranked documents. To compute these metrics \textit{(1)}-\textit{(4)} that require binary relevance annotations on the employed relevance scale, we binarize the relevance scores beforehand, thereby considering only \textit{high} relevance scores (i.e. $\mathbf{R=2}$) as relevant. In particular, the \textit{\gls{mrr}} and $HITS@k$ metrics are computed as follows:

\renewcommand{\arraystretch}{0.0}
\begin{tabular}{m{5cm}m{5.5cm}} 
   \centering 
 \begin{equation}
            MRR = \frac{1}{\vert Q \vert} \sum_{i=1}^{\vert Q \vert} \frac{1}{rank_{i}}
        \end{equation}
         & \begin{equation}
            HITS@k = \begin{cases} 
          1 & \vert R_{k} \vert > 0 \\
          0 & \vert R_{k} \vert = 0 
       \end{cases}
        \end{equation}  \\
\end{tabular}

\glsreset{ndcg}
\noindent with $R_k$ denoting the set of relevant \gls{gui} documents up to rank $k$. In order to further take into account the full annotation scale and the ranking position of \gls{gui} documents with \textit{high} and \textit{medium} relevance (i.e. $\mathbf{R=2}$ or $\mathbf{R=1}$), we additionally considered \textit{(5)} the \textit{\gls{dcg} at rank} $k$ (\textit{\gls{dcg}@k}). Relevant \gls{gui} documents ranked higher in the retrieval collection are more valuable, and the relevance score should accordingly be weighted more than the score of \gls{gui} documents at lower-ranked positions. Hence, in \textit{\gls{dcg}@k}, each relevance score is discounted by a weight (log-based smoothing), which is computed based on the ranking position and decreases with lower ranking positions. The \textit{\gls{dcg}@k} is normalized by the \textit{IDCG@k} (\textit{Ideal} \textit{\gls{dcg}@k}) to obtain \textit{(6)} the \textit{\gls{ndcg} at rank $k$ (\gls{ndcg}@k}), which is employed in our evaluation. The \textit{IDCG@k} is computed as the \textit{\gls{dcg}@k} of relevance scores sorted in descending order (perfect ranking). To compute the final results, the metrics are averaged over all queries of the gold standard. Subsequently, the \textit{\gls{dcg}@k} and \textit{\gls{ndcg}@k} metrics are shown, with $rel_{i}$ denoting the relevance score at the $i$-th rank:

\renewcommand{\arraystretch}{0.0}

\begin{tabular}{m{5.5cm}m{5.5cm}} 
\centering
         \begin{equation}
            DCG@k = \sum_{i=1}^{k}{\frac{rel_{i}}{\log_{2}(i + 1)}}
        \end{equation}
         & \begin{equation}
            NDCG@k = \frac{DCG@k}{IDCG@k}
        \end{equation}  \\
\end{tabular}

\noindent To enable detection of statistically significant differences, we apply the non-parametric \textit{Wilcoxon signed-rank test} \citep{woolson2007wilcoxon} with per-metric \textit{Holm} correction across model comparisons to account for multiple pairwise model comparisons. Since \gls{bm25} is considered one of the strongest baseline \gls{ir} models, we first compare \gls{bm25} to the other two baseline models \gls{tfidf} and \gls{nbow}. Second, to determine any improvement of the \gls{aqe} methods over \gls{bm25}, we compare the four \gls{prf}-\gls{kld} variants against \gls{bm25}. Third, to evaluate whether the \gls{bert}-based models improve over the strong baseline \gls{bm25}, we compare the four \gls{bert} models against \gls{bm25}. Since \gls{sbert} acts as a pretrained, semantic baseline, we compare for differences between \gls{sbert} and the \gls{bert}-\gls{ltr} variants.

\subsection{\texorpdfstring{RQ$_2$: Productivity of Rapid Prototyping}{RQ-2: Productivity of Rapid Prototyping}}
\label{chapter-2:subsec:rq2-userstudy}

\vspace{-0.1cm}
\paragraph{User Study Design.} To measure potential productivity improvements of our data-driven \gls{gui} prototyping approach, we conducted a controlled experiment with 19 participants as a \textit{within-subjects} study. In the experiment, we compared our approach to a traditional high-fidelity \gls{gui} prototyping approach. With regard to the discussed \gls{rel} scenario, participants in the role of analysts were asked to create two different application prototypes (consisting of two \glspl{gui} each) on the basis of typical \gls{nlr} requirements, using each approach separately to mitigate \textit{carry-over} effects. To avoid bias from both the tasks and the ordering of the approaches, we considered all four combinations of the approaches A\slash B and the tasks 1\slash 2 (A1B2\slash A2B1\slash B1A2\slash B2A1) and assigned the conditions to participants randomly while ensuring that the occurrence of conditions is evenly distributed. As the approach for comparison, we employed \textit{Mockplus} \citep{mockplus} since it can be regarded as a representative tool for traditional \gls{gui} prototyping approaches, providing a restricted number of manually crafted \gls{gui} screens to reuse and enabling users to create \glspl{gui} by combining a large number of searchable \gls{gui} components in a graphical editor. Overall, we recruited 19 participants with technical backgrounds (BSc:7\slash MSc:8\slash PhD:4) mainly having \textit{medium} to \textit{high} experience in software development (M:3.52\slash SD:1.02) and mainly having \textit{little} to \textit{medium} \gls{gui} prototyping experience (M:2.42\slash SD:0.96), as self-reported by the study participants on a five-point Likert scale questionnaire. Most of the participants (13) reported that they have used a prototyping tool before, with the most frequently mentioned ones being general-purpose tools such as \textit{PowerPoint} (9), \textit{Photoshop} (2) and \textit{Adobe XD} (2). Therefore, the participants are representative regarding the elicitation scenario, especially targeting support for inexperienced requirements analysts.

\vspace{-0.2cm}
\paragraph{User Study Tasks.} Both \gls{gui} prototyping tasks are based on real \textit{Android} applications within the top-50 apps on \textit{Google Play}, which are not included in the \textit{Rico} \gls{gui} repository to avoid bias. The first task represents a \textit{shopping browser} consisting of a start \gls{gui} asking the user to select from a list of countries (\textit{\#\gls{gui}-comps}: 25) and the actual main shopping browser \gls{gui} that allows users to search for websites and displays favorite websites (\textit{\#\gls{gui}-comps}: 33), among others. The second task represents a \textit{hotel booking} application consisting of a \gls{gui} that allows users to enter search details (\textit{\#\gls{gui}-comps}: 25) and a \gls{gui} that shows a list of search results (\textit{\#\gls{gui}-comps}: 29). We omit the detailed task descriptions and \gls{nlr} requirements for brevity. However, we provide all study instructions among the other publicly released materials in our accompanying repository. We decided to employ these \glspl{gui} in our experiments due to multiple reasons. First, \textit{(i)} the \glspl{gui} contain a large number of different \gls{gui} components, thus making them difficult to create, especially for novice users. Second, \textit{(ii)} both approaches \textit{\gls{rawi}} and \textit{Mockplus} cover the shopping and booking application domains with basic \gls{gui} prototypes allowing for a fair comparison.

\vspace{-0.2cm}
\paragraph{Experimental Procedure.} 
\label{sec:proc} In the study, we asked the participants to create a \gls{gui} prototype using each approach separately. Depending on the assigned condition, the participant started with the first approach and task. Before working on the actual task, we showed a short tutorial video of the approach with a \gls{gui} prototyping example and allowed the participants as much time as they needed to test the approach. Afterwards, the actual task was conducted. With regard to the interactive \gls{gui} prototyping scenario with stakeholders for elicitation, the available time for prototyping is strictly limited. Therefore, we accordingly restricted the available time for creating the \gls{gui} prototypes to several minutes, with a maximum of seven minutes. To obtain a fine-grained productivity measurement, we captured screenshots of the current \gls{gui} prototype state after each minute starting from three minutes and thus collected five screenshots per \gls{gui} prototype. These screenshots formed the basis for our post-experiment data analysis. After completing the first task, we continued with the identical procedure for the second approach and task combination. Overall, we gathered 20 \gls{gui} screenshots per participant.

\vspace{-0.3cm}
\paragraph{Evaluation Metrics.} From the gathered screenshots, we computed multiple metrics. However, no prior research proposed metrics for assessing \gls{gui} prototyping productivity before. In particular, a closely related problem on \gls{gui} implementation effort estimation \citep{lo1996sizing} identifies especially the number and type of \gls{gui} components as valuable factors for predicting the overall \gls{gui} implementation effort. In a similar fashion, \gls{gui} prototyping in the \gls{rel} phase asks analysts to quickly select and arrange \gls{gui} components that properly reflect the requirements specified by the stakeholder. Here, one or more \gls{gui} components represent a particular functional requirement and, therefore, we consider the number of \gls{gui} components as a proxy for \gls{gui} prototyping productivity.

Thus, we first \textit{(a)} count the number of \textit{correct} \gls{gui} components (based on the \gls{gui} component type, e.g., using a \textit{password field} instead of a plain \textit{text field} for the functionality to enter passwords) available in the considered \gls{gui} prototype at time $t$ that are appropriate for the given requirements as an approximation for productivity (denoted by \textit{\#\gls{gui}-comps}). \gls{gui} components that overlapped the \gls{gui} screen or other \gls{gui} components were not included in the count. Detailed design decisions (e.g., placing a button bar on top or bottom or similar) are apparently neglected in this metric, since in the elicitation phase, the focus lies on quickly representing the main functionality and not the final \gls{gui} design. Different working styles and other relevant individual factors of participants that influence the outcome are considered by our \textit{within-subject} design, comparing productivity only between the participants themselves. Second, \textit{(b)} we count the number of \gls{gui} components as before, but data elements (e.g., \textit{images} or \textit{text}) are only counted multiple times if they contain realistic and diverse data (denoted by \textit{\#\gls{gui}-comps-div}), accounting for data model fidelity (see \cite{mccurdy2006breaking}). We considered computing this corrected count since having high content diversity through realistic data is important for more realistic high-fidelity \gls{gui} prototypes. Third, \textit{(c)} we count the number of \textit{incorrect} \gls{gui} components available at time $t$ that were not included in the requirements (denoted by \textit{\#\gls{gui}-comps-neg}). Finally, \textit{(d)} we compute an adjusted count where including non-required components is punished by taking the count of required \gls{gui} components \textit{(a)} minus the count of \gls{gui} components not included in the requirements \textit{(c)}. We aim to compare the prototyping productivity between the two approaches at each time step $t$. To evaluate if the productivity is significantly different between approaches, we compute the \textit{Wilcoxon signed-rank test} \citep{woolson2007wilcoxon} over all pairs of study tasks (\textit{Holm}-corrected). We compare the productivity of each participant between both first \glspl{gui} of the tasks and both second \glspl{gui} of the tasks. Moreover, we compute \textit{Cliff's delta} $\delta$ \citep{macbeth2011cliff} to estimate the overall statistical effect size.

\vspace{-0.2cm}

\glsreset{sus}

\subsection{\texorpdfstring{RQ$_3$: Perceived Usefulness}{RQ-3: Perceived Usefulness}}
\vspace{-0.2cm}
In addition to measuring the \gls{gui} prototyping productivity, we assessed the perceived usefulness of our \gls{gui} prototyping approach by the study participants. Using a five-point Likert scale, we asked participants \textit{(a)} how much our approach supported them during the prototyping process, \textit{(b)} how relevant the retrieved \glspl{gui} were on average, \textit{(c)} how much the \gls{gui} retrieval results supported the users during the prototyping and \textit{(d)} how much the \gls{gui} search results helped better visualize how the \gls{gui} could be prototyped. Moreover, we computed the \textit{\gls{sus}} \citep{brooke1996sus}, which is a validated, reliable and well-known measurement for system usability. To gain further insights, we asked the participants for free-form feedback to potentially discover current issues and interesting ideas for improvements to our novel \gls{gui} prototyping approach.

\newcommand{\PkNdcgSepRule}{%
  \cmidrule(lr){1-1}\cmidrule(lr){2-5}\cmidrule(lr){6-9}
}

\begin{table*}[!t]
\footnotesize
\caption[Evaluation results of \gls{nlr}-based \gls{gui} retrieval approaches (1)]{Evaluation results overview of the different models on the \gls{nlr}-based \gls{gui} retrieval gold standard using $P@k$ and \textit{\gls{ndcg}@k} (\textbf{Bold} values indicate best metric score within result group and \underline{underlined} values indicate best metric score across all groups).}
\centering
\setlength\tabcolsep{6.5pt}
\renewcommand{\arraystretch}{1.1}

\begin{tabular}{l|cccc|cccc}
\toprule
\multicolumn{1}{l|}{} &
\multicolumn{4}{c|}{\textbf{P@k}} &
\multicolumn{4}{c}{\textbf{NDCG@k (N@k)}} \\
\cmidrule(lr){2-5}\cmidrule(lr){6-9}
\multicolumn{1}{l|}{} &
$\mathbf{P@3}$ & $\mathbf{P@5}$ & $\mathbf{P@7}$ & $\mathbf{P@10}$ &
$\mathbf{N@3}$ & $\mathbf{N@5}$ & $\mathbf{N@10}$ & $\mathbf{N@15}$ \\
\midrule

\textbf{TF-IDF} & 22.3 & 20.4 & 18.1 & 17.5 & 32.9 & 33.9 & 39.5 & 48.0 \\
\rowcolor{lightgray}
\textbf{BM25}   & \textbf{30.3} & \textbf{27.6} & \textbf{24.6} & \textbf{22.6} & \textbf{42.6} & \textbf{44.1} & \textbf{51.5} & \textbf{57.9} \\
\textbf{nBoW}   & 27.0 & 23.4 & 22.0 & 19.3 & 39.5 & 37.0 & 37.4 & 39.8 \\

\PkNdcgSepRule

\rowcolor{lightgray}
\textbf{BM25}      & 30.3 & 27.6 & 24.6 & 22.6 & 42.6 & 44.1 & 51.5 & 57.9 \\
\textbf{+PRF}      & 31.3 & 26.6 & 25.9 & 23.6 & 43.2 & 44.3 & 52.0 & 58.4 \\
\rowcolor{lightgray}
\textbf{+PRF (\textit{s})}  & 31.7 & \textbf{28.0} & \textbf{26.0} & 23.5 & \textbf{44.1} & \textbf{46.2} & \textbf{53.6} & \textbf{60.4} \\
\textbf{+PRF (\textit{w})}  & \textbf{32.0} & 27.6 & 24.4 & 23.1 & 43.9 & 41.7 & 42.1 & 44.6 \\
\rowcolor{lightgray}
\textbf{+PRF (\textit{sw})} & 31.7 & 28.0 & 25.6 & \textbf{23.7} & 42.9 & 41.6 & 42.6 & 45.6 \\

\PkNdcgSepRule

\textbf{SBERT} & 34.3 & 31.4 & 29.4 & 26.7 & 48.1 & 51.1 & 61.1 & 66.7 \\
\rowcolor{lightgray}
\textbf{BERT-LTR (\textit{1})}  & 37.7 & 35.0 & 30.7 & 26.9 & 53.0 & \textbf{\underline{56.0}} & 63.4 & 69.7 \\
\textbf{BERT-LTR (\textit{2})}  & \textbf{\underline{40.0}} & 34.0 & 30.4 & 28.1 & \textbf{\underline{54.3}} & 55.6 & 63.6 & \textbf{\underline{70.1}} \\
\rowcolor{lightgray}
\textbf{BERT-LTR (\textit{3})}  & 36.3 & \textbf{\underline{35.4}} & \textbf{\underline{31.7}} & \textbf{\underline{28.7}} & 51.7 & 55.4 & \textbf{\underline{64.6}} & 69.4 \\

\bottomrule
\end{tabular}
\vspace{-0.3cm}
\label{tab:results-1}
\end{table*}
\vspace{-0.4cm}

\section[Results \&\ Discussion]{Results \& Discussion}
\label{sec:results}

\vspace{-0.2cm}
\subsection{\texorpdfstring{RQ$_1$: \gls{gui} Retrieval Performance}{RQ-1: \gls{gui} Retrieval Performance}}

\vspace{-0.1cm}

Table \ref{tab:results-1} shows the evaluation results of the different baseline ranking models, \gls{aqe} techniques and \gls{bert}-based \gls{ltr} models for the $P@k$ and \textit{\gls{ndcg}@k} metrics, whereas Table \ref{tab:results-2} shows the evaluation results for the \textit{\gls{ap}}, \textit{\gls{mrr}} and $HITS@k$ metrics. Considering the three baseline ranking models, we can observe that \gls{bm25} outperforms the other two models substantially across most of the examined mean metric values, while statistically significant differences between \gls{bm25} and \gls{tfidf} are observed for \textit{\gls{ap}}, \textit{P@5,7,10} and \textit{\gls{ndcg}@5,10,15}, and between \gls{bm25} and \gls{nbow} for \textit{\gls{ap}} and \textit{\gls{ndcg}@10,15} (cf. Appendix \ref{chaptper:app-rawi} Tables \ref{tab:appendix-rawi-holm-1}, \ref{tab:appendix-rawi-holm-2} and \ref{tab:appendix-rawi-holm-3}). In particular, the \textit{\gls{mrr}} of the \gls{bm25} model indicates that the first relevant \gls{gui} appears at rank 1.92 on average over the gold standard. Moreover, the $HITS@5$ indicates that in 69\% of the queries, the \gls{bm25} model ranked at least one relevant \gls{gui} among the top-5. Thus, \gls{bm25} seems to be an effective baseline model for \gls{nlr}-based \gls{gui} retrieval. Comparing the plain \gls{tfidf} and \gls{nbow} models, we can observe that especially the mean $P@k$ and $HITS@k$ as well as the \textit{\gls{ndcg}@k} for small $k$ are higher for the \gls{nbow} model, indicating that pretrained word embeddings can provide additional benefit for \gls{gui} retrieval from \gls{nlr}-based searches.


\newcommand{\SepRule}{%
  \cmidrule(lr){1-1}\cmidrule(lr){2-2}\cmidrule(lr){3-3}\cmidrule(lr){4-9}
}

\begin{table*}[!t]
\footnotesize
\caption[Evaluation results of \gls{nlr}-based \gls{gui} retrieval approaches (2)]{Evaluation results overview of the different models on the \gls{nlr}-based \gls{gui} retrieval gold standard using \textit{\gls{ap}}, \textit{\gls{mrr}} and \textit{HITS@k} (\textbf{Bold} values indicate best metric score within result group, \underline{underlined} values indicate best metric score across all groups).}
\centering
\setlength\tabcolsep{7pt}
\renewcommand{\arraystretch}{1.1}

\begin{tabular}{l|c|c|cccccc}
\toprule
\multicolumn{1}{l|}{} &
\multicolumn{1}{c|}{\textbf{AP}} &
\multicolumn{1}{c|}{\textbf{MRR}} &
\multicolumn{6}{c}{\textbf{HITS@k (H@k)}} \\
\cmidrule(lr){2-2}\cmidrule(lr){3-3}\cmidrule(lr){4-9}
\multicolumn{1}{l|}{} &
\textbf{AP} & \textbf{MRR} &
$\mathbf{H@1}$ & $\mathbf{H@3}$ & $\mathbf{H@5}$ & $\mathbf{H@7}$ & $\mathbf{H@10}$ & $\mathbf{H@15}$ \\
\midrule

\textbf{TF-IDF} & 33.1 & 45.1 & 32.0 & 46.0 & 58.0 & 65.0 & 76.0 & 91.0 \\
\rowcolor{lightgray}
\textbf{BM25}   & \textbf{41.3} & \textbf{52.0} & \textbf{37.0} & \textbf{60.0} & \textbf{69.0} & \textbf{76.0} & \textbf{86.0} & 93.0 \\
\textbf{nBoW}   & 28.1 & 49.0 & 34.0 & 54.0 & 63.0 & 75.0 & 84.0 & \textbf{98.0} \\

\SepRule

\rowcolor{lightgray}
\textbf{BM25}               & 41.3 & 52.0 & 37.0 & 60.0 & 69.0 & 76.0 & 86.0 & 93.0 \\
\textbf{+PRF}               & 41.9 & 50.5 & 37.0 & 58.0 & 68.0 & 78.0 & 85.0 & 93.0 \\
\rowcolor{lightgray}
\textbf{+PRF (\textit{s})}  & \textbf{42.7} & 53.2 & 38.0 & \textbf{61.0} & \textbf{70.0} & \textbf{78.0} & \textbf{88.0} & \textbf{96.0} \\
\textbf{+PRF (\textit{w})}  & 32.5 & 52.3 & 38.0 & 58.0 & 70.0 & 72.0 & 86.0 & 93.0 \\
\rowcolor{lightgray}
\textbf{+PRF (\textit{sw})} & 33.3 & \textbf{53.3} & \textbf{39.0} & 59.0 & 70.0 & 77.0 & 87.0 & 93.0 \\

\SepRule

\textbf{SBERT} & 45.4 & 56.0 & 37.0 & 68.0 & 76.0 & 88.0 & 96.0 & 99.0 \\
\rowcolor{lightgray}
\textbf{BERT-LTR (\textit{1})} & 48.6 & 61.8 & \textbf{\underline{46.0}} & 71.0 & 86.0 & 92.0 & 98.0 & \textbf{\underline{100.0}} \\
\textbf{BERT-LTR (\textit{2})} & \textbf{\underline{50.1}} & \textbf{\underline{63.1}} & 44.0 & \textbf{\underline{75.0}} & \textbf{\underline{91.0}} & \textbf{\underline{96.0}} & 98.0 & \textbf{\underline{100.0}} \\
\rowcolor{lightgray}
\textbf{BERT-LTR (\textit{3})} & 49.9 & 62.6 & 45.0 & 73.0 & 86.0 & 94.0 & \textbf{\underline{100.0}} & \textbf{\underline{100.0}} \\

\bottomrule
\end{tabular}

\label{tab:results-2}
\end{table*}

\begin{figure*} \includegraphics[width=\textwidth]{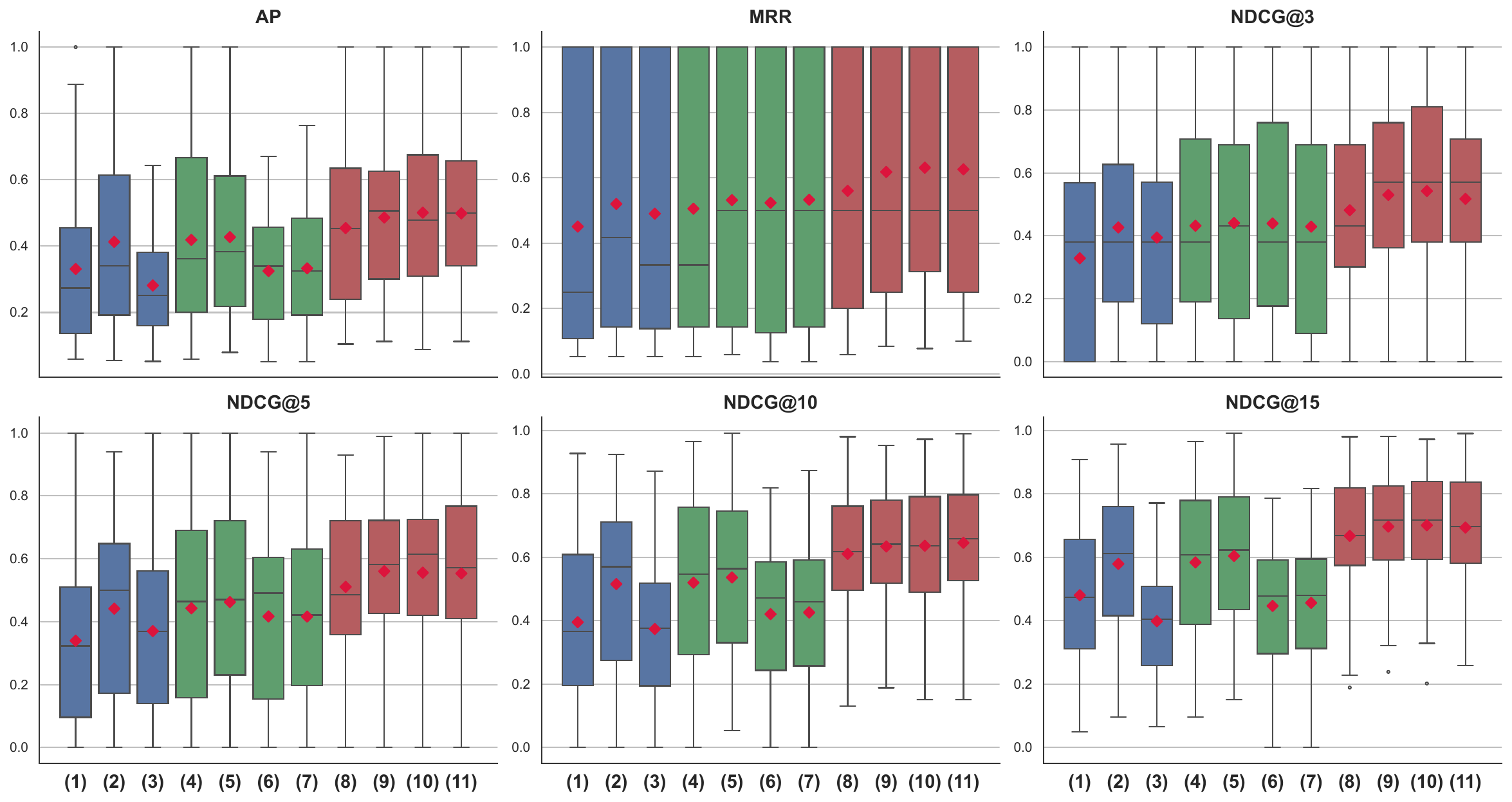}
  \caption[Boxplots for \gls{nlr}-based \gls{gui} retrieval effectiveness]{Multiple boxplots with \textit{\gls{ap}}, \textit{\gls{mrr}} and \textit{\gls{ndcg}@k} metrics across models: \textit{(1)} \gls{tfidf}, \textit{(2)} \gls{bm25}, \textit{(3)} \gls{nbow}, \textit{(4)} \gls{prf}, \textit{(5)} \gls{prf} \textit{(s)}, \textit{(6)} \gls{prf} \textit{(w)}, \textit{(7)} \gls{prf} \textit{(sw)}, \textit{(8)} \gls{sbert}, \textit{(9)} \gls{bert}-\gls{ltr} \textit{(1)}, \textit{(10)} \gls{bert}-\gls{ltr} \textit{(2)}, \textit{(11)} \gls{bert}-\gls{ltr} \textit{(3)}, red diamonds repr. means.}
	\label{fig:rawi-boxplots-models}
\end{figure*}

In addition, both tables show the results of the four different \gls{aqe} methods based on the \gls{prf}-\gls{kld} score using the strong baseline ranking model \gls{bm25} as the base ranker. The results show that the text-segment-wise \gls{prf} (s) method obtains the highest scores among most of the considered mean metric values and outperforms the \gls{bm25} base model across all mean metric values. Both weighted \gls{prf}-\gls{kld} variants can outperform the \gls{prf} (s) method for some individual metrics and mainly perform better for the binary metrics $P@k$ and $HITS@k$ compared to the \gls{bm25} base ranker. However, both are outperformed on the \textit{\gls{ndcg}@k} metric that takes into account the entire relevance scale and additionally requires ranking \glspl{gui} with \textit{medium} relevance at higher positions. Overall, the results indicate that \gls{aqe} techniques in the form of \gls{prf}-\gls{kld} can improve the performance for \gls{nlr}-based \gls{gui} retrieval compared to the baseline \gls{bm25} ranking model when considering mean metric values, especially the variant that takes the \gls{gui}-specific text segments into account. However, no statistically significant difference was observed (cf. Appendix \ref{chaptper:app-rawi} Tables \ref{tab:appendix-rawi-holm-1}, \ref{tab:appendix-rawi-holm-2} and \ref{tab:appendix-rawi-holm-3}). Additionally, we provide a more detailed per-query analysis of the \gls{aqe} evaluation results in our accompanying repository. For example, the \gls{mrr} and \textit{P@5} metrics were improved or at least as good in 69\% and 77\% of the queries with an average improvement of .19 and .07 compared to the base model \gls{bm25} by applying the \gls{prf} (s) method. In particular, for queries that have many relevant \glspl{gui} in the \gls{gui} repository and the initial retrieval results with the base model \gls{bm25} provide relevant and cohesive \glspl{gui}, the \gls{aqe} method can improve the results. For instance, for the query \textit{"a list of songs"} meaningful expansion words such as \textit{artist}, \textit{bookmark}, and \textit{play} can be extracted, which strengthen the query. Figure \ref{fig:rawi-boxplots-models} illustrates several boxplots for metrics of \textit{\gls{ap}}, \textit{\gls{mrr}} and \textit{\gls{ndcg}@k}.

\begin{figure*}[!t] \includegraphics[width=\textwidth,height=0.45\textheight,keepaspectratio]{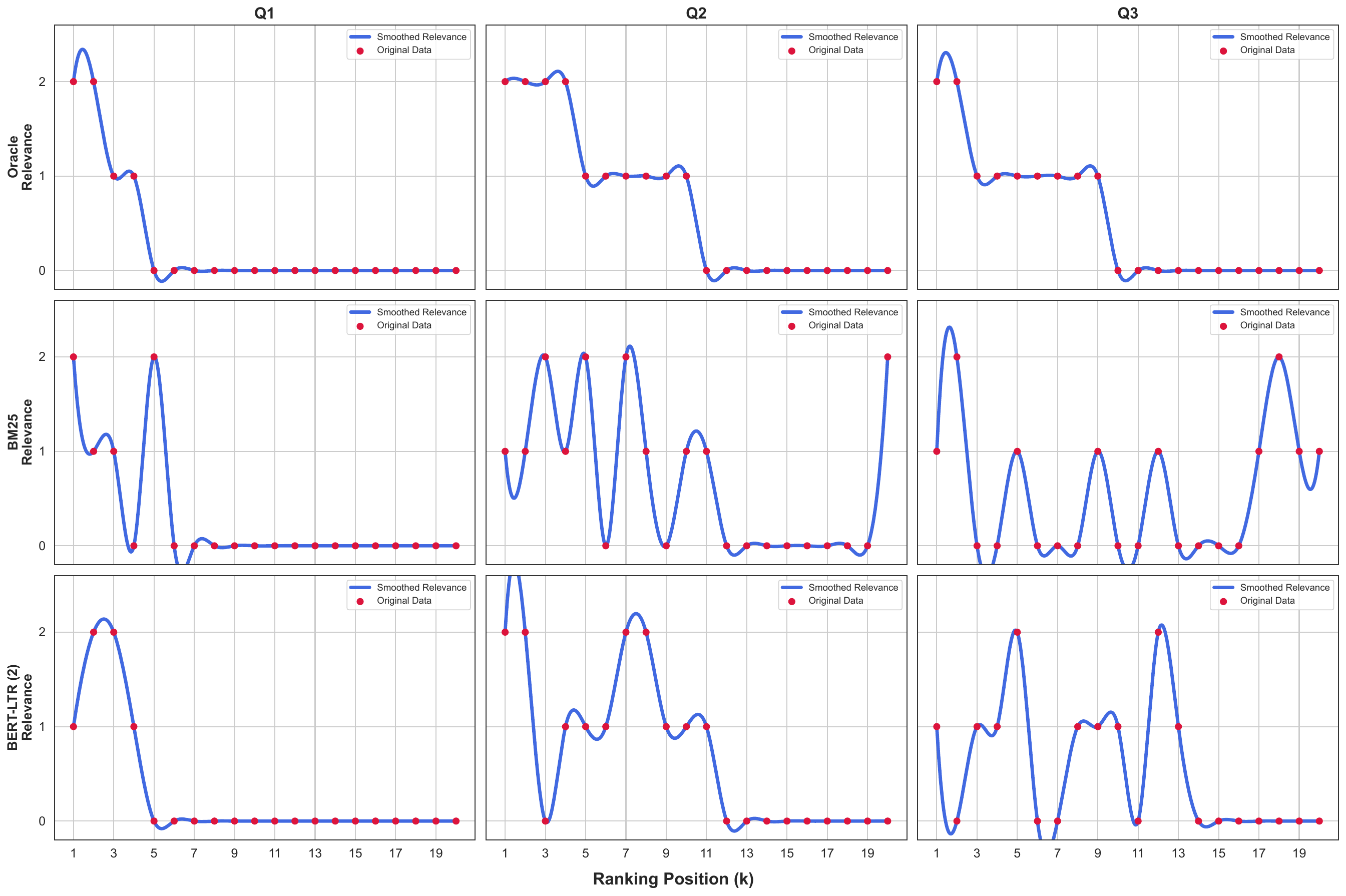}
  \caption[Ranking plots for Oracle, \gls{bm25} and \gls{bert}-\gls{ltr}]{Multiple ranking plots showing the relevance of a \gls{gui} at position $k$ for a particular model and query: Columns represent the same query, while rows represent the same model, enabling a comparison. The plot includes an \textit{Oracle} model (the perfect ranking), \gls{bm25} and \gls{bert}-\gls{ltr} \textit{(2)}. \textit{Q1}=\textit{``screen with sample movies and information on recording with SnapMovie''}, \textit{Q2}=\textit{``profile window with a button to take a photo or select it from galary [sic]''} and \textit{Q3}=\textit{``Weather details for city with option in Celsius and Fahrenheit''.}}
	\label{fig:raw-ranking-plots}
\end{figure*}

\begin{figure*} \includegraphics[width=\textwidth]{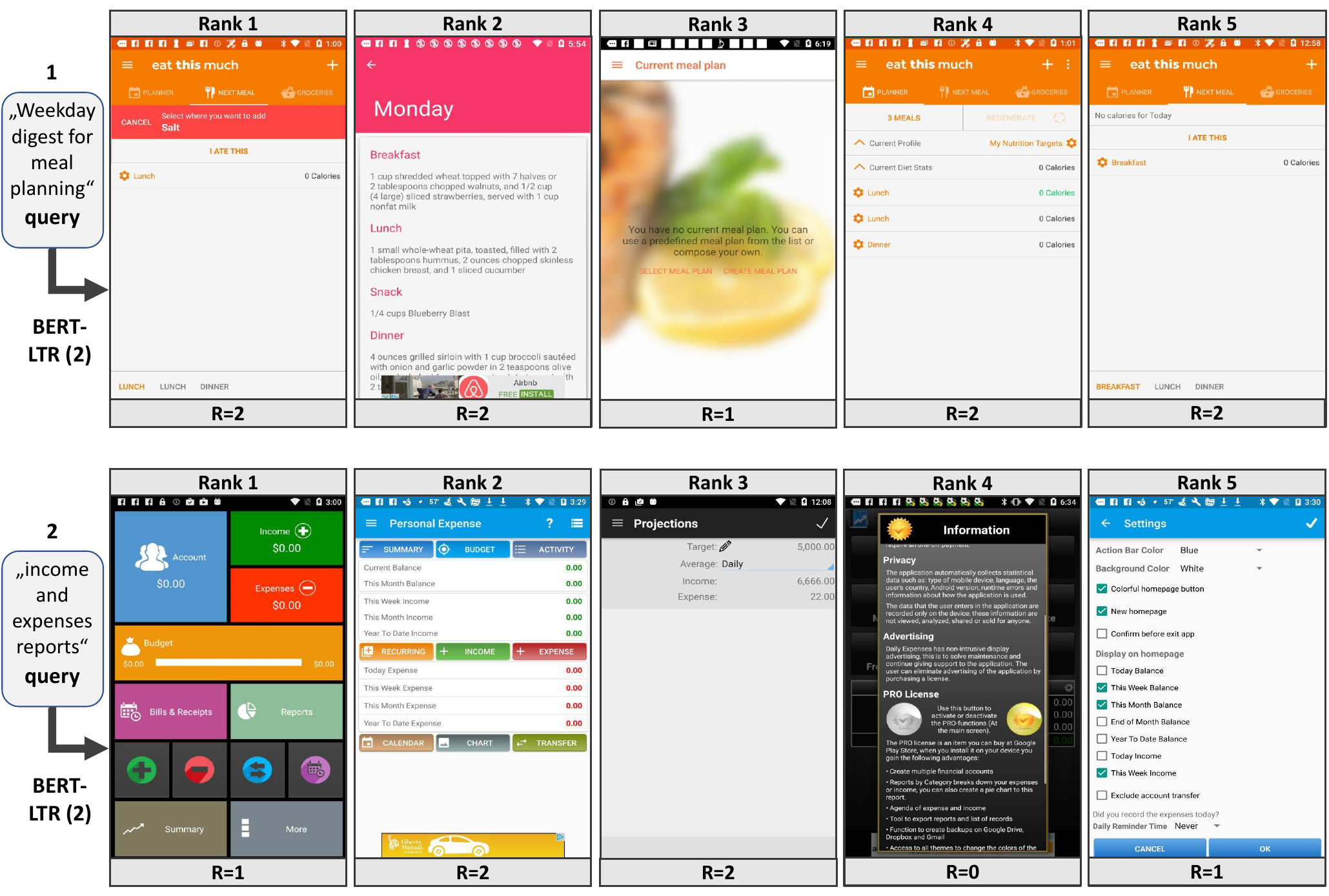}
  \caption[Example top-5 rankings with \gls{bert}-\gls{ltr}]{Example of two queries from the gold standard with the respective top-5 \gls{gui} rankings and their gold standard relevance scores, ranked by the \gls{bert}-\gls{ltr} (2) model.}
	\label{fig:goldstandard-top-5-example}
\end{figure*}

Considering the results of the trained \gls{bert}-\gls{ltr} models, we can observe that all three models consistently outperform the strong baseline \gls{bm25} across mean metric values, while statistically significant differences between \gls{bm25} and \gls{bert}-\gls{ltr} (2) (\textit{pairwise}) are observed for \textit{\gls{ap}}, \textit{H@5,10}, \textit{P@7,10} and \textit{\gls{ndcg}@3,5,10,15} (cf. Appendix \ref{chaptper:app-rawi} Tables \ref{tab:appendix-rawi-holm-1}, \ref{tab:appendix-rawi-holm-2} and \ref{tab:appendix-rawi-holm-3}). Between \gls{sbert} and \gls{bm25}, \textit{P@10} and \textit{HITS@10} are significantly improved. In our experiments, the \textit{pairwise} (\gls{bert}-\gls{ltr} (2)) and \textit{listwise} (\gls{bert}-\gls{ltr} (3)) models appear to mainly perform best among the examined models based on the mean metric values. While the \gls{bert}-\gls{ltr} (2) (\textit{pairwise}) model consistently outperforms the \gls{sbert} model across all mean metric values, statistically significant differences are observed for \textit{H@5}. Regarding the \textit{\gls{ndcg}@k} metrics, the \gls{bert}-\gls{ltr} models outperform the \gls{sbert} baseline substantially on the mean values, indicating that the models learned to rank \glspl{gui} with both \textit{medium} (\textbf{R=1}) and \textit{high} relevance (\textbf{R=2}) higher in the overall ranking. Overall, the results indicate that the pretrained and finetuned \gls{bert} language model helps to better represent the semantics of the query and \gls{gui} documents and that, in combination with a \gls{ltr} model trained specifically on the task of \gls{gui} ranking, it can be effectively applied to improve the \gls{nlr}-based \gls{gui} ranking performance compared to the strong \gls{bm25} baseline. In particular, these specifically trained semantic models are better at bridging the gap between the language used in \gls{nlr} queries and the \gls{gui} text representations. Figure \ref{fig:raw-ranking-plots} shows the ranking plots of an \textit{Oracle} model (i.e. perfect ranking), \gls{bm25} and \gls{bert}-\gls{ltr} (2) across three queries, where the \textit{x}-axis represents the position in the ranking ($k$=1--20), while the \textit{y}-axis shows the relevance level of the \gls{gui} at that rank position. In the illustrated examples, it can be observed that \gls{bert}-\gls{ltr} (2) is able to rank \glspl{gui} with \textit{medium} and \textit{high} relevance levels at positions higher in the ranking, while \textit{low} relevance \glspl{gui} are mostly placed at lower ranking positions. Moreover, Figure \ref{fig:goldstandard-top-5-example} shows two example queries taken from the gold standard with their respective top-5 \glspl{gui} as ranked by the trained \gls{bert}-\gls{ltr} (2) model. Each \gls{gui} is annotated with its ground truth relevance. For the queries, we can observe that only a single \gls{gui} is annotated with a \textit{low} relevance (due to an information overlay), whereas all other \glspl{gui} appear to have a \textit{medium} to \textit{high} relevance for the corresponding queries, which provides additional indication for the ranking strength of the finetuned \gls{bert}-\gls{ltr} models.

\begin{myrqbox}
\textbf{Answer to RQ$_1$:} The \gls{bert}-\gls{ltr} models consistently outperform the strong baseline model \gls{bm25} across all mean metric values, while statistically significant differences between \gls{bm25} and \gls{bert}-\gls{ltr} (2) (\textit{pairwise}) are observed for \textit{\gls{ap}}, \textit{H@5,10}, \textit{P@7,10} and \textit{\gls{ndcg}@3,5,10,15} metrics. Moreover, the \gls{bert}-\gls{ltr} models consistently outperform the \gls{sbert} model across mean metric values and statistically significant differences between \gls{bert}-\gls{ltr} (2) (\textit{pairwise}) and \gls{sbert} are observed for the \textit{HITS@5} metric.
\end{myrqbox}

\subsection{\texorpdfstring{RQ$_2$: Productivity of Rapid Prototyping}{RQ-2: Productivity of Rapid Prototyping}}

\begin{figure*}[!t]
 \centering
   \frame{ \subfloat[\centering \#\gls{gui}-comps correct based on \\ the requirements]{{\includegraphics[width=0.45\linewidth]{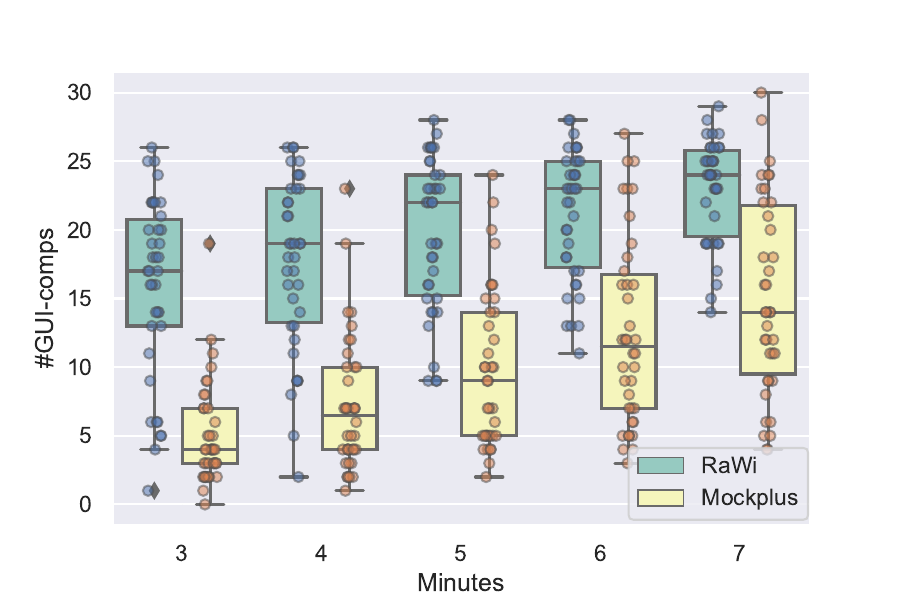} }
        \label{fig:text1}}
}%
    \qquad
   \frame{ \subfloat[\centering \#\gls{gui}-comps-div with \\ content diversity]{{\includegraphics[width=0.45\linewidth]{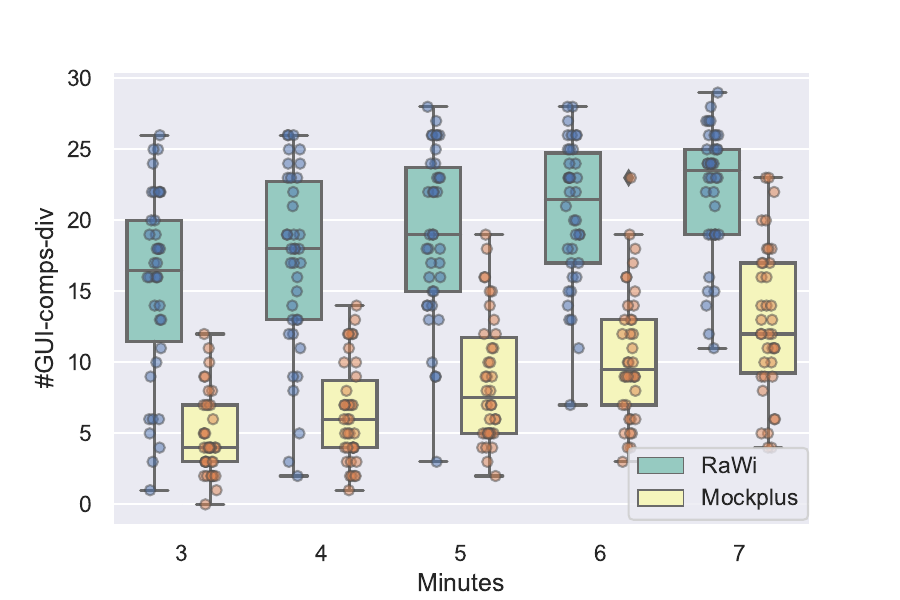} }
     \label{fig:retrieval_examples_1}}%
	\label{fig:annotationexample1}}
	
	\hspace{1cm}
	
   \frame{ \subfloat[\centering \#\gls{gui}-comps-neg that were \\ not requested]{{\includegraphics[width=0.45\linewidth]{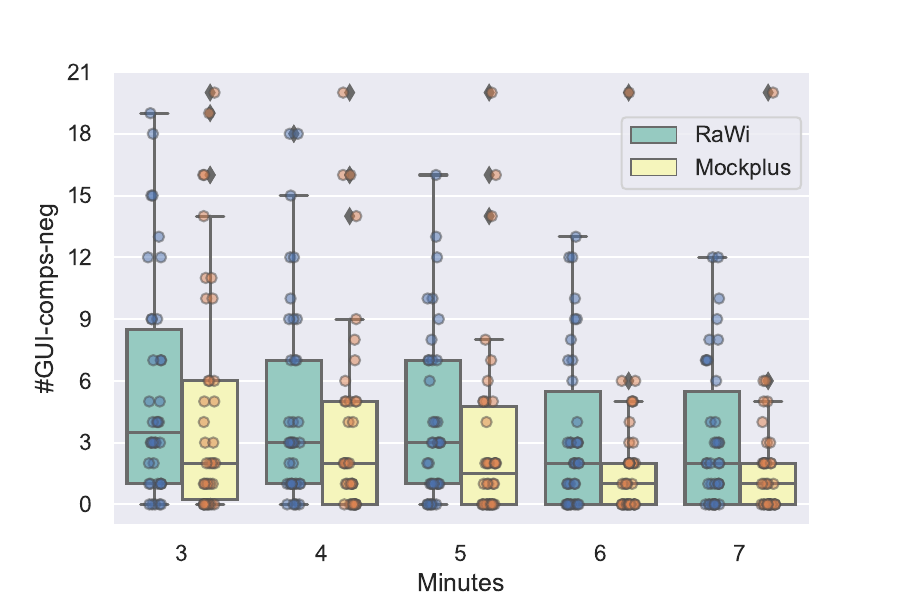} }
        \label{fig:text2}}
}%
    \qquad
  \frame{ \subfloat[\centering \#\gls{gui}-comps of (a) adjusted by \#\gls{gui}-comps-neg (c)]{{\includegraphics[width=0.45\linewidth]{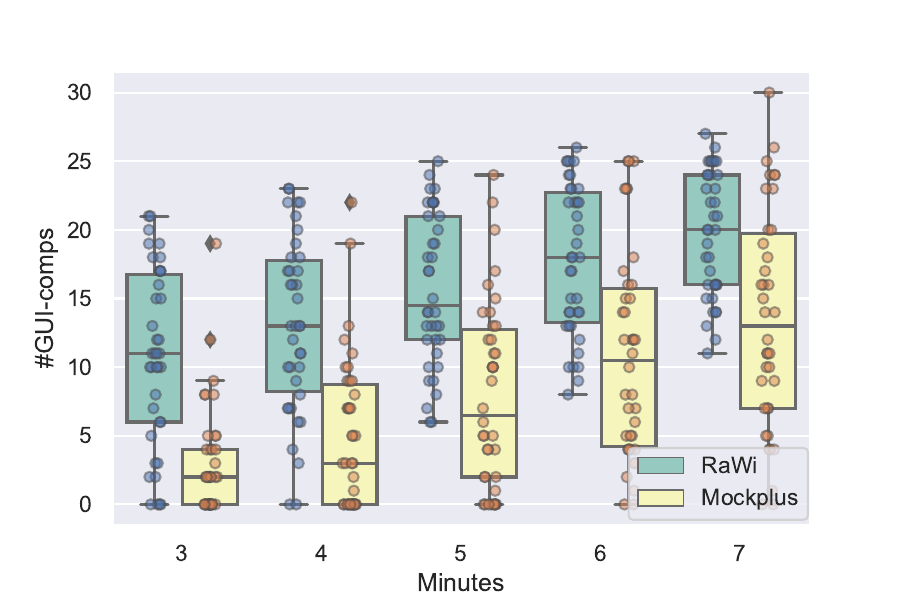} }
     \label{fig:retrieval_examples_2}}%
	\label{fig:annotationexample2}}
  \caption[Evaluation results of the \textit{RaWi} user study]{Evaluation results of our controlled experiment: each boxplot \textit{(a)}-\textit{(d)} shows one of the computed \#\gls{gui}-comp metrics that the participants achieved with each rapid \gls{gui} prototyping approach \textit{\gls{rawi}} and \textit{Mockplus} at each minute $t$ from minutes 3--7.}
	\label{fig:userstudy}
\end{figure*}

\noindent In Figure \ref{fig:userstudy}, we show the evaluation results of our controlled experiment. Each of the four boxplots \textit{(a)}-\textit{(d)} shows one of the four \textit{\#\gls{gui}-comp} metrics as a proxy for productivity as explained earlier. For the basic \textit{\#\gls{gui}-comp} count shown in Figure \ref{fig:userstudy}\textit{(a)}, we can observe that our approach outperforms the traditional approach substantially at each time step $t$ (cf. Appendix \ref{chaptper:app-rawi} Tables \ref{tab:productivity_1}--\ref{tab:mixed_effects_3}). The counts in both approaches increase with increasing $t$, but the differences between the two approaches become smaller with increasing $t$. Participants are often able to find suitable starting \gls{gui} screens via adequate searches in \textit{\gls{rawi}} and thus often begin with higher counts followed by smaller adjustments (\textit{adding} and \textit{removing} components). In contrast, in the traditional approach, it was often more difficult for participants to find an appropriate \gls{gui} at the beginning. However, repetitive component groups could be copied quickly after initial creation, which led to a strong increase in counts over time $t$. When working with \textit{\gls{rawi}}, participants were challenged to extract good queries on the basis of the given \gls{nlr} and select relevant starting \glspl{gui}. Here, participants were not always able to find optimal starting \glspl{gui}, but productivity improvements typically could still be observed in comparison to the traditional prototyping approach since \textit{\gls{rawi}} simplified the task.

For the \textit{\#\gls{gui}-comp-div} counts that take into account the content diversity, shown in Figure \ref{fig:userstudy}\textit{(b)}, we can observe a behavior similar to that in \textit{(a)}, but the differences between the two approaches are apparently larger. \glspl{gui} retrieved with \textit{\gls{rawi}} typically display already diverse and realistic data since they are derived from \gls{gui} screenshots of real applications. In contrast, participants often copied repetitive component groups and often were not able to provide diverse data in their \gls{gui} prototypes with the traditional approach. For the \textit{\#\gls{gui}-comp-neg} counts that measure the available but not requested components shown in Figure \ref{fig:userstudy}\textit{(c)}, we can observe that in both approaches the counts decrease over time $t$ due to the removal of components available in the initial \gls{gui} screens that were not required. However, at the same time, the \gls{gui} prototypes created with \textit{\gls{rawi}} tend to contain more unrequested components. In \textit{\gls{rawi}}, participants frequently found suitable screens in the beginning, yet often these \glspl{gui} included many additional components for other specific requirements that were not relevant for the requested prototype. However, we can observe a similar phenomenon in the traditional approach, where participants who started with unsuitable \gls{gui} screens obtained high counts indicated by the outliers. In particular, one participant used a \gls{gui} with many unrequested components and did not manage to remove them up to minute 7. Users of \textit{\gls{rawi}} tend to focus on adding more requested functionality not yet contained in the starting \glspl{gui} and neglect removing unrequested components initially. This could potentially be due to the fact that the unrequested components in \textit{\gls{rawi}} were possibly often perceived as less wrong (e.g., \textit{a star rating for each of multiple search results}) compared to unrequested features in the templates in \textit{Mockplus}, and therefore potentially neglected more initially.

For the adjusted \textit{\#\gls{gui}-comp} counts that represent the difference between the original count in \textit{(a)} and \textit{(c)} shown in Figure \ref{fig:userstudy}\textit{(d)}, we still can observe that our approach has higher counts at time $t$, but the differences become smaller when punishing the prototypes for containing unrequested components. In addition to the differences observed in the boxplots, the \textit{Wilcoxon signed-rank test} results indicate that the counts as a proxy for productivity using our approach are statistically significantly larger at all considered time steps $t$ ($p$-value $<$ 0.05) and all considered counts \textit{(a)}, \textit{(b)} and \textit{(d)}. Moreover, Cliff's $\delta$ indicates a \textit{large} difference between counts of the approaches also in the cases where the difference appears smaller in the boxplots. Overall, the results indicate that our approach can improve the prototyping productivity compared to a traditional \gls{gui} prototyping approach and therefore provides a better applicability for interactive \gls{gui} prototyping.

\begin{myrqbox}
\textbf{Answer to RQ$_2$:} Our prototyping approach, which integrates the effective \gls{nlr}-based \gls{gui} retrieval techniques, significantly outperforms the traditional \gls{gui} prototyping approach \textit{Mockplus} regarding the number of \textit{correct} \gls{gui} components (\textit{\#\gls{gui}-comp}), the data model diversity (\textit{\#\gls{gui}-comp-div}) and the adjusted number of components.
\end{myrqbox}

\subsection{\texorpdfstring{RQ$_3$: Perceived Usefulness}{RQ-3: Perceived Usefulness}}

\begin{figure*}[!t]
 \centering
   \frame{ \subfloat[\centering Support during \gls{gui} prototyping]{{\includegraphics[width=0.45\linewidth]{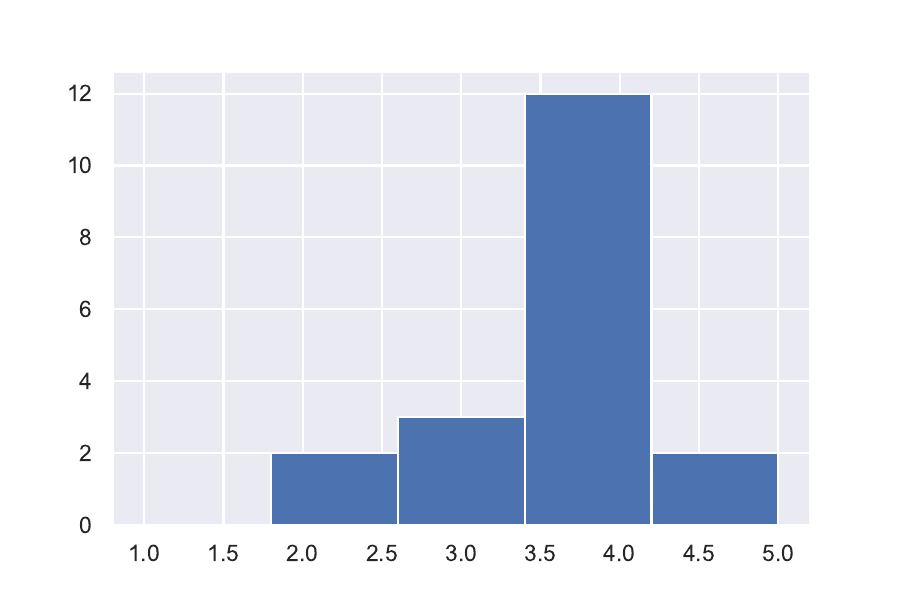} }
        \label{fig:text1}}
}%
    \qquad
   \frame{ \subfloat[\centering Average relevance of \glspl{gui} for query]{{\includegraphics[width=0.45\linewidth]{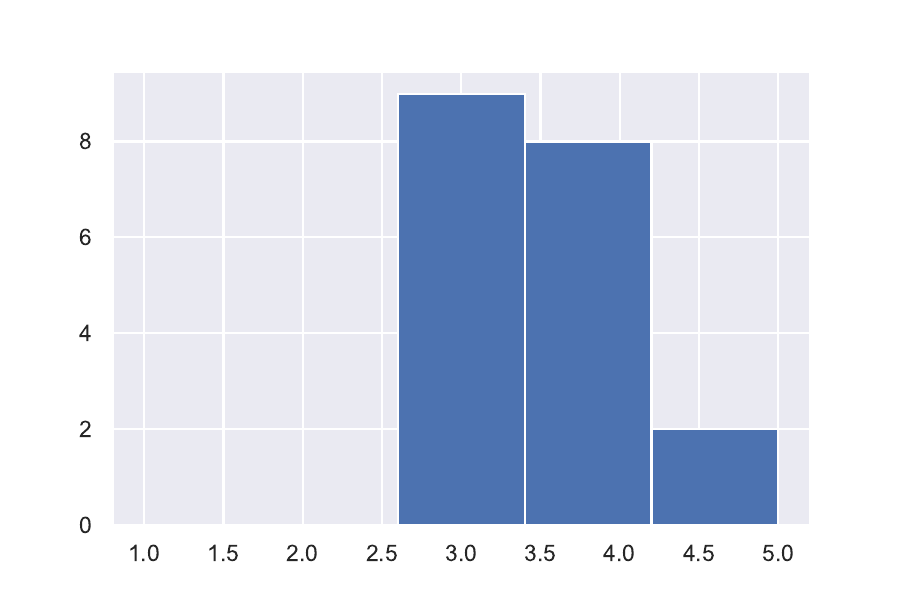} }
     \label{fig:retrieval_examples_1}}%
	\label{fig:annotationexample1}}
	
	\hspace{1cm}
	
   \frame{ \subfloat[\centering Support of \gls{gui} search results during \gls{gui} prototyping]{{\includegraphics[width=0.45\linewidth]{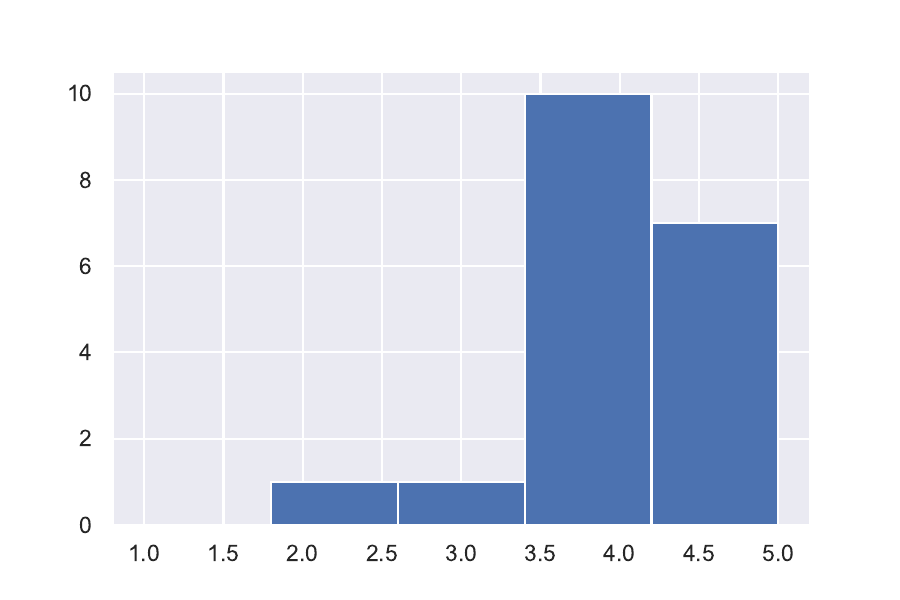} }
        \label{fig:text2}}
}%
    \qquad
  \frame{ \subfloat[\centering Support of \gls{gui} search results for visualizing how the screen could look]{{\includegraphics[width=0.45\linewidth]{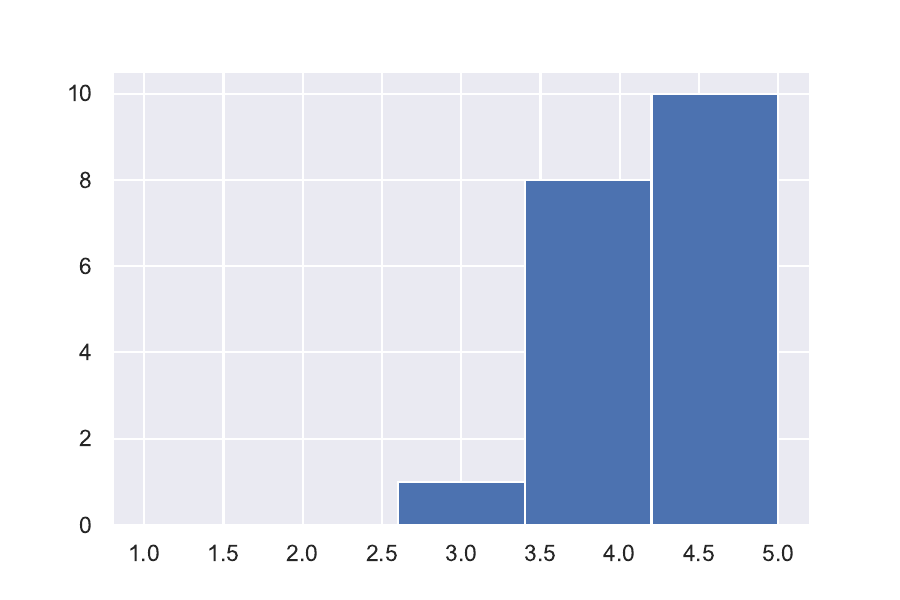} }
     \label{fig:retrieval_examples_2}}%
	\label{fig:annotationexample2}}
  \caption[Perceived usability histograms]{Four histograms showing the results of the four usability questions \textit{(a)}-\textit{(d)} regarding \textit{\gls{rawi}} for rapid \gls{gui} prototyping as perceived by the user study participants.}
	\label{fig:usability}
\end{figure*}

Fig. \ref{fig:usability} shows the evaluation results of the perceived usefulness of our \gls{gui} prototyping approach. The first question \textit{(a)} regarding the support during the prototyping process received a high score (\textit{M}:3.73\slash \textit{SD}:0.80). The second question \textit{(b)} regarding the perceived relevance of the search results received a similarly high score (\textit{M}:3.63\slash \textit{SD}:0.68). The third question \textit{(c)} regarding the support of the results received an even higher score (\textit{M}:4.21\slash \textit{SD}:0.78). Finally, the fourth question \textit{(d)} regarding the support of the search results for better visualization received the highest score (\textit{M}:4.47\slash \textit{SD}:0.61). Overall, we achieved a \gls{sus} of 81.57, which indicates that our approach has high usability compared to the average \gls{sus} of 68. Participants found our approach very easy to understand and use, even for inexperienced and novice users in \gls{gui} prototyping. In general, the participants reported further that they liked the large choice of available \gls{gui} screens to create a first prototype quickly. Concerning improvements to the approach, participants mainly reported that the editor functionalities should be extended to enable more sophisticated \gls{gui} editing.

\begin{myrqbox}
\textbf{Answer to RQ$_3$:} Our approach is perceived as useful, achieving a \gls{sus} of 81.57, which represents high usability. Moreover, the approach receives good scores regarding the support \textit{\gls{rawi}} provides during prototyping, the relevance of retrieved \glspl{gui}, the support by the retrieval results and their help with visualization of the \gls{gui} prototype.
\end{myrqbox}

\section{Threats to Validity}
\label{sec:gui-retrieval-threats}

\paragraph{Internal Validity.} First, we discuss threats to internal validity, which refers to potential bias present in our experimental evaluation. To reduce bias and subjectivity in the creation of the gold standard, we conducted several steps. Due to the absence of any prior accessible system that could be used to extract search queries for \glspl{gui}, we employed 50 workers from \gls{amt} to write queries on a random sample of the \textit{Rico} \glspl{gui}, to reduce bias during the query creation. \gls{amt} provides a large number of workers with diverse demographics, suiting the need for including a large bandwidth of different ways of formulating queries, use of language and vocabulary. However, there could be potential selection bias. For example, workers that are willing to work on the task may have a specific interest in apps and thus may have more knowledge on the topic compared to the average worker on \gls{amt}. Similarly, we asked 49 workers to provide relevance judgments and had each \gls{gui} annotated by three workers to reduce subjectivity. These workers were required to pass our task-specific qualification test, which served the purpose of increasing the relevance annotation quality. Final relevance annotations are based on majority voting and \glspl{gui} with high annotation perplexity were discarded due to the high uncertainty of the ground truth relevance for these instances. To obtain a high-quality gold standard, we applied several filters and randomly sampled from both the remaining queries and the \glspl{gui} to avoid selection bias. To reduce bias in our controlled experiment, we randomized the tasks and the ordering of the approaches. In addition, we ensured that the applications employed in the user study tasks are not included in \textit{Rico} and both approaches basically support the domains of the selected prototyping tasks. Moreover, for each approach, we ensured that we conducted the identical experimental procedure. 
\vspace{0.3cm}

\paragraph{External Validity.} Second, we discuss threats to external validity, which refers to the generalizability of the results obtained in our experimental evaluation. Similarly, due to the absence of any prior system for \gls{gui} search that could be used as a resource for obtaining real-world queries, we employed \glspl{gui} taken from \textit{Rico} for which to write queries. We already discussed the many reasons for employing \textit{Rico} and emphasize that \textit{Rico} is large-scale and covers many diverse domains, making it applicable in many scenarios. In addition to many domain-specific \glspl{gui}, \textit{Rico} provides a plethora of domain-independent \glspl{gui}. Using \glspl{gui} from \textit{Rico} as the basis for writing queries also assures that models could at least in principle retrieve a single relevant \gls{gui} for the query. We restricted our controlled experiment to two domains, but we hypothesize that we could obtain similar results for other domains covered by \textit{Rico}. However, more user evaluations are required that include more diverse application domains to confirm the obtained results. Moreover, since \textit{Rico} is a semi-automatic approach, it could be scaled up to harvest millions of \glspl{gui}, making our prototyping approach even more valuable for more domains.

\section{Limitations}
\label{sec:limitations}

Our retrieval is currently restricted to the \textit{Rico} dataset, which, however, is large-scale and covers many diverse domains already. To extend the \gls{gui} repository, other smaller \gls{gui} datasets (such as \textit{ReDraw}) could be integrated. Since prototypes are created on the basis of \glspl{gui} from different real-world applications in \textit{\gls{rawi}}, the design of the resulting application prototype may not be cohesive. However, the focus of our work lies on rapidly creating \gls{gui} prototypes that properly reflect \glspl{fr} of stakeholders, neglecting the final \gls{gui} design. Since the implementation of our \gls{gui} prototyping approach is an early prototype, the editing functionality is currently restricted. However, we plan to extend our prototype to a fully-fledged graphical \gls{gui} editor with more editing functionality.

\section{Related Work}
\label{sec:related_gui_retrieval}

\paragraph{\gls{gui} Retrieval} \textit{Guigle} \citep{bernal2019guigle} similarly exploits automatically crawled \textit{Android} apps from \textit{Google Play} \citep{moran2018machine} to index multiple parts from the crawled \gls{gui} hierarchy data such as the app name, text and type of \gls{gui} components, the screen color and employs a basic Boolean query language for relevant \gls{gui} retrieval. In contrast, our approach \textit{\gls{rawi}} enables users to specify \gls{nlr}-based searches to retrieve relevant \glspl{gui} and employs a more sophisticated ranking mechanism including \gls{bert}-based \gls{ltr} models. In addition, \textit{\gls{rawi}} enables users to directly employ retrieved \glspl{gui} to rapidly create \gls{gui} prototypes including the derivation of partly editable screens, whereas \textit{Guigle} stops after \gls{gui} image retrieval. However, a direct comparison of the retrieval performance is not possible due to the unavailability of the implementation and the employment of a different, much smaller \gls{gui} dataset in \textit{Guigle}. As stated in their paper, \textit{Guigle} is built on top of the open-source retrieval framework \textit{Lucene} \citep{lucene}. The \textit{Lucene} framework employs \gls{tfidf} and \gls{bm25} retrieval methods, which are included in our evaluation experiments as well-known standard \gls{ir} baselines. Therefore, although implementation details such as preprocessing and retrieval function parameters are not specified in \textit{Guigle}, the presented \gls{ir} baselines can be considered a good proxy for the retrieval methods employed in their approach. Moreover, to support further research and enable a direct comparison between approaches in the first place, we publicly provide the first high-quality gold standard for \gls{nlr}-based \gls{gui} retrieval based on a state-of-the-art \gls{gui} repository in this work.

\textit{Gallery D.C.} \citep{chen2019gallery} crawls a large number of real-world applications and automatically extracts \gls{gui} components such as various types of buttons from \gls{gui} screenshots and provides a multi-modal search supporting multiple dimensions such as width, height, main color and text for them. In addition, there is a plethora of \gls{gui} retrieval approaches that employ visual inputs such as screenshots, hand-drawn sketches or basic wireframes to retrieve similar \glspl{gui}, in contrast to our approach that focuses on \gls{nlr} input. For example, \cite{chen2020wireframe} train an image autoencoder to find relevant \glspl{gui} employing a basic wireframe of the \gls{gui} as their input. Similarly, \textit{VINS} \citep{bunian2021vins} expects visual input either as a basic wireframe prototype or as a fully designed \gls{gui} image to retrieve similar \glspl{gui} using a multi-modal embedding network. \textit{Swire} \citep{huang2019swire} also employs a visual embedding approach to find relevant \gls{gui} images on the basis of hand-drawn sketches provided by the user. \textit{GUIFetch} \citep{behrang2018guifetch} requires an entire \textit{Android} application sketch as input to retrieve similar apps from \textit{GitHub}. \textit{d.tour} \citep{ritchie2011d} uses stylistic keywords and stylistic similarity to find similar website designs. By using a web design as input and conducting structure matching, \textit{FaceOff} retrieves web \gls{gui} components fitting the design. 

Another approach proposes a neural translator for transforming \gls{gui} images to \gls{gui} skeletons \citep{chen2018ui}. \textit{Screen2Vec} \citep{li2021screen2vec} proposes a technique for learning multi-modal (textual content, visual design and layout patterns) \gls{gui} embeddings, which are useful for many \gls{gui}-related downstream tasks (e.g., retrieving similar \glspl{gui} using a \gls{gui} as the query). Due to the difference of the problem, these \gls{gui} embeddings cannot be directly reused in our approach for comparison. The \gls{nlr} query embeddings (e.g., based on \gls{bert}) and \textit{Screen2Vec} embeddings represent two separate embedding spaces that require alignment in order to enable \gls{nlr}-based \gls{gui} retrieval with \textit{Screen2Vec} \gls{gui} embeddings. However, we included the same text-only \gls{sbert} baseline in our experiments, also used in their approach.

The \textit{Screen2Words} \citep{wang2021screen2words} approach proposes an Encoder-Decoder architecture for translating \glspl{gui} into short text descriptions. The \gls{gui} encoder could potentially be employed to obtain \gls{gui} embeddings, but these embeddings similarly cannot be directly employed for retrieval based on \gls{nlr} queries as described earlier. Considering general-purpose search engines (e.g., \textit{Google}), a direct comparison with our approach is not meaningful since these engines cannot be evaluated on our gold standard. Especially due to the huge differences in the employed datasets (specialized \textit{Rico} \gls{gui} dataset vs. general \textit{Google} image dataset), it would be unclear whether differences between the retrieval performances are due to differences in the algorithms or due to the differences in the datasets. For example, since \textit{Google} uses a general image dataset, many image results would be obtained that are not related to \glspl{gui}, compared to the \textit{Rico} dataset specializing in \textit{Android} \glspl{gui}. To make the comparison between these approaches meaningful, the algorithm would need to be evaluated on the new gold standard directly.

\paragraph{Program Code Retrieval} Apart from \gls{gui} search approaches, retrieving code through \gls{nl} queries has been studied extensively in research before \citep{mcmillan2011exemplar, lv2015codehow, zhang2016bing, gu2018deep, cambronero2019deep}. These approaches range from the application of traditional \gls{ir} algorithms to specifically developed semantic Deep Learning (DL) architectures to find relevant program code for a given \gls{nl} query. More recently, the \textit{CodeSearchNet} challenge \citep{husain2019codesearchnet} released a dataset of \gls{nl} queries and expert relevance annotations of respective functions from various programming languages to provide a source for systematic evaluation of code search approaches. In contrast, we created the first \gls{gui} retrieval gold standard in this work to foster further research on retrieval methods specifically for \gls{nlr}-based \gls{gui} search.

\paragraph{\gls{gui} Prototyping} In addition, many traditional prototyping approaches exist and are employed by many designers, developers and analysts in their everyday work such as \textit{Balsamiq} \citep{faranello2012balsamiq}, \textit{Sketch} \citep{sketch}, \textit{Figma} \citep{figma} and \textit{Mockplus} \citep{mockplus}, among others. These approaches typically enable users to create prototypes in a graphical editor on the basis of a small number of hand-crafted templates and basic \gls{gui} components supporting low-fidelity or high-fidelity prototyping. Other approaches such as \textit{UISKEI} \citep{segura2012uiskei} and \textit{SketchiXML} \citep{coyette2006sketchixml} attempt to automatically identify \gls{gui} components from hand-drawn sketches and generate reusable \gls{gui} representations. Recent work on \gls{gui} prototyping assistance such as \textit{GUIComp} \citep{lee2020guicomp} supports novice users during the prototyping process via the recommendation of similar \glspl{gui}, provides various complexity metrics and visual attention maps based on the design currently created by the user. In contrast to these approaches, \textit{\gls{rawi}} enables users to quickly retrieve matching \glspl{gui} based on \gls{nlr} queries from a large-scale \gls{gui} repository and automatically provides partly editable \gls{gui} screens to reuse for rapid, interactive \gls{gui} prototyping, thereby facilitating requirements elicitation and validation.
\section{Conclusion}
\label{sec:conclusions}

This work was driven by challenge \challoneone{}: \textit{How can we adapt and optimize text-based retrieval methods and techniques to enable more effective \gls{nlr}-based \gls{gui} retrieval?} To answer this question, we presented \textit{\gls{rawi}}, a data-driven \gls{gui} prototyping approach based on exploiting a large-scale \gls{gui} repository via effective \gls{nlr}-based \gls{gui} retrieval methods and automatically deriving partly editable \gls{gui} screens for interactive \gls{gui} prototyping in the requirements elicitation phase. Particularly, we presented \gls{bert}-\gls{ltr} models, which improved the \gls{nlr}-based \gls{gui} ranking performance substantially over the strong text-based baseline model \gls{bm25}. In addition, our approach is able to improve the prototyping productivity in comparison to a traditional \gls{gui} prototyping approach.

\chapter{GUI-ReRank: Enhancing GUI Retrieval with Multi-Modal LLM-based Reranking}
\chaptermark{Enhancing GUI Retrieval with MLLM-based Reranking}
\label{cha:gui_rerank}

After introducing our first solution approach and evaluation for \gls{nlr}-based \gls{gui} retrieval, we continue with challenge \challone{}, namely, mapping \gls{nlr} to \gls{gui} prototypes. In this chapter, we introduce a novel \gls{mllm}-based \gls{gui} reranking approach and improve on the prior results achieved with the \gls{bert}-\gls{ltr} models. In particular, the work presented in this chapter is based on previously published research \citep{kolthoff2025guirerankenhancingguiretrieval}\footnote{This section is adapted from: \textbf{Kolthoff, Kristian}, Kretzer, Felix, Bartelt, Christian, Maedche, Alexander, and Ponzetto, Simone Paolo. GUI-ReRank: Enhancing GUI Retrieval with Multi-Modal LLM-based Reranking. In \emph{Proceedings of the 40th IEEE/ACM International Conference on Automated Software Engineering (ASE, A*)}, Seoul, Republic of Korea, January 2026, pages 1--4. ACM. (in press)}. Again, our interactive prototype, source code, evaluation results and video are all publicly available\footnote{Materials for this chapter are available at \url{https://github.com/kristiankolthoff/GUI-ReRank}, \url{https://zenodo.org/records/16451923} and demo video at \url{https://youtu.be/_7x9UCh82ug}}.

\paragraph{Personal Contribution.} I developed the idea and concept for the proposed approach, implemented the entire framework and tool prototype. The evaluation is based on the gold standard introduced in the previous chapter, which I employed to design the evaluation concept and conduct the data analysis. Finally, I wrote the entire manuscript. 

\section{Motivation}

While we investigated \gls{nlr}-based \gls{gui} retrieval methods that solely relied on textual representations of the \gls{gui} prototypes in the previous chapter and provided substantial effectiveness improvements based on the proposed \gls{bert}-\gls{ltr} models over the strong \gls{bm25} baseline, this approach carries one major shortcoming. Representing \gls{gui} prototypes merely through text fragments extracted from various \gls{gui} hierarchy components (such as \textit{displayed text of \gls{gui} components}, \textit{Android} specifics including \textit{activity names} and component-wise \textit{resource identifiers}, \textit{semantic labels for icons}) provides only a heavily simplified, abstracted representation, neglecting crucial semantic information encompassed in the \gls{gui} prototypes in \textit{Rico} \citep{deka2017rico}. Particularly, these \gls{gui} prototypes are multi-dimensional objects, which contain most of their information in the screenshot images themselves. These prototype screenshots convey functional and non-functional information through displayed text (e.g., component-specific text), component types (e.g., \textit{buttons}, \textit{text fields}, \textit{check boxes}), visual icons, component and group layouts, and the overall \gls{gui} design. Therefore, incorporating this important information while deciding on relevance given \gls{nlr} has the potential to enhance the \gls{gui} retrieval and ranking effectiveness. While the \gls{gui} design plays a less significant role with respect to prototypes employed for \gls{rel} and \gls{rval}, the remaining dimensions are crucial for determining relevance according to functional \gls{nlr}. Moreover, the previous approach heavily relies on specific \gls{gui} hierarchy information, which might not be available for other \gls{gui} datasets. This restricts the generalizability of the approach and impairs the practical adoption. Although \gls{gui} design is of less importance for early requirements elicitation and validation purposes, it still represents an important aspect of \gls{gui} prototypes (\gls{nfr}). To summarize, the following work is based on the leading research question of challenge \challonetwo{}: \textit{How can the semantic representation gap between \gls{nlr} (including \gls{nfr}) and multi-dimensional \gls{gui} prototypes be reduced to enable more effective \gls{gui} ranking?}

To tackle this challenge and close the gap, in this work, we introduce \textit{\gls{gui}-ReRank}, a novel framework that significantly advances the previous \gls{nlr}-based \gls{gui} retrieval and reranking models by integrating the information encompassed in \gls{gui} screenshot images with powerful, state-of-the-art \glspl{mllm}. In particular, \textit{\gls{gui}-ReRank} employs a two-stage retrieval and reranking approach, which introduces an embedding-based, constrained \gls{gui} retrieval architecture over multiple dimensions followed by an \gls{mllm}-based \gls{gui} reranking technique. \textit{\gls{gui}-ReRank} enables more expressive and complex \gls{nlr} queries including negation (i.e. \textit{exclusion criteria/requirements}) across multiple customizable dimensions such as \textit{domain}, \textit{functionality}, \textit{design} and \textit{GUI components} via the advanced \gls{nlu} and image processing capabilities of modern \glspl{mllm}. To facilitate the adoption for various information needs and \gls{gui} datasets (e.g., \textit{domain-}, \textit{industry-} or \textit{company-specific}), our approach offers a fully customizable \gls{gui} dataset annotation pipeline, which employs \glspl{mllm} to initially derive rich textual representations for each dimension of the \gls{gui} prototypes, enabling fast and computationally cheap multi-dimensional retrieval. To evaluate the effectiveness of the \gls{mllm}-based \gls{gui} reranking, we utilize the gold standard created in the previous chapter and additionally conduct a comprehensive cost analysis, showing the cost-effectiveness trade-offs.
\vspace{-0.1cm}
\noindent
\paragraph{Contributions.} With this work, we make the following four research contributions:
\vspace{-0.1cm}
\begin{itemize}[leftmargin=6mm]
\setlength{\itemsep}{1pt}
    \item \textit{\gls{mllm}-based \gls{gui} reranking approach:} we present a novel, \gls{mllm}-based \gls{gui} reranking approach that incorporates the information contained in the \gls{gui} images and significantly outperforms prior state-of-the-art models such as \gls{bert}-\gls{ltr}.
    \item \textit{Multi-dimensional embedding-based constrained \gls{gui} retrieval:} we propose a new approach to facilitate multi-dimensional constrained \gls{nlr}-based \gls{gui} retrieval via pretrained embeddings and also enable fully customizable \gls{gui} search dimensions.
    \item \textit{Detailed effectiveness and cost analysis:} we conduct a comprehensive evaluation of the proposed \gls{mllm}-based \gls{gui} reranking approaches by utilizing the previously created gold standard and investigate trade-offs between effectiveness and costs. 
    \item \textit{Rico \gls{gui} annotations and embeddings dataset:} we provide multi-dimensional text annotations for the \textit{Rico} \gls{gui} dataset created by an \gls{mllm} and additionally their respective 3,072-dimensional embeddings to foster future \gls{gui} retrieval research.
\end{itemize}

\vspace{-0.5cm}
\paragraph{Structure of the chapter} The remainder of this chapter is structured as follows. In Section \ref{sec:chapter-rerank:approach}, we present the novel multi-dimensional embedding-based constrained \gls{gui} retrieval and \gls{mllm}-based \gls{gui} reranking approach. Section \ref{sec:chapter-rerank:evaluation} introduces the research questions and evaluation setup, while Section \ref{sec:chapter-rerank:results} discusses the results. In Section \ref{sec:chapter-rerank:threats}, we present threats to validity, followed by limitations of our approach in Section \ref{sec:chapter-rerank:limitations}. Finally, we provide an overview of related work in Section \ref{sec:chapter-rerank:relwork} and a conclusion in Section \ref{sec:chapter-rerank:conclusion}.
\vspace{-0.2cm}
\section{Approach: \gls{gui}-ReRank}
\label{sec:chapter-rerank:approach}

\begin{figure}[!t]
\includegraphics[width=\textwidth]{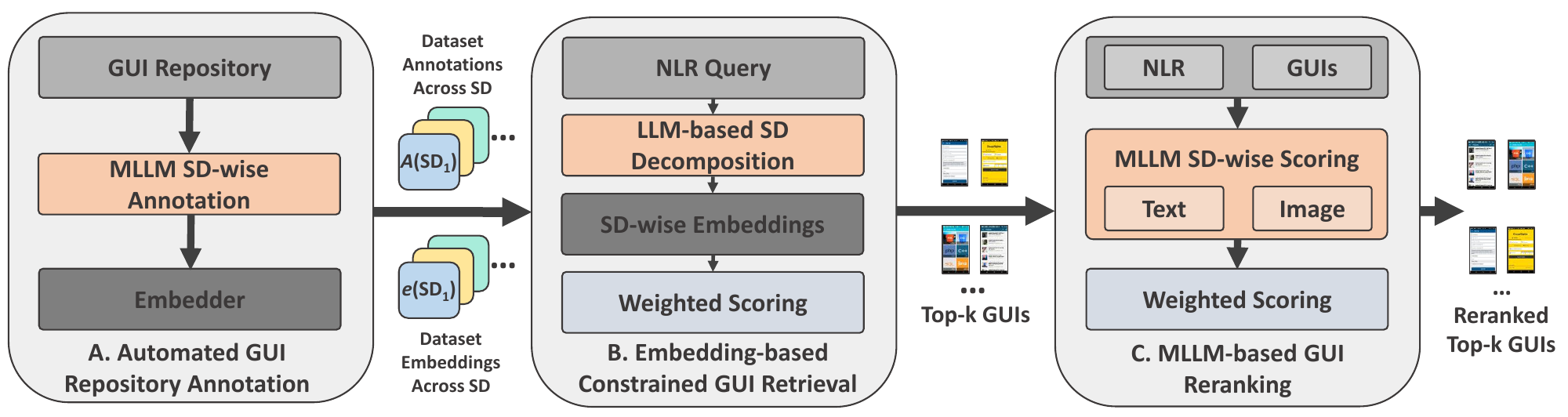}
  \caption[Overview of the \textit{\gls{gui}-ReRank} approach]{Overview of the \textit{\gls{gui}-ReRank} approach with \textit{(A)} the automated \gls{gui} repository annotation and embedding pipeline, \textit{(B)} the embedding-based constrained \gls{gui} retrieval approach and \textit{(C)} the \gls{mllm}-based \gls{gui} reranking via images or annotations.}
	\label{fig:gui-rerank-overview}
\end{figure}

\textit{\gls{gui}-ReRank} encompasses three main components: \textit{(A)} an \gls{mllm}-based, automated and customizable \gls{gui} repository annotation pipeline that creates textual \gls{gui} annotations and embeddings across predefined \glspl{sdim}, \textit{(B)} an embedding-based constrained \gls{gui} retrieval approach, which leverages the previously created embeddings and enables multi-dimensional, complex \gls{nlr} search queries (including negation) based on \gls{llm}-driven query decomposition and \textit{(C)} an \gls{mllm}-based \gls{gui} reranking technique, which utilizes both \textit{image} and \textit{text} representations of the \gls{gui} prototypes to compute a refined ranking score according to the predefined \glspl{sdim}. Figure \ref{fig:gui-rerank-overview} shows an overview of \textit{\gls{gui}-ReRank}, with the main components and their respective intermediate outputs.

\begin{figure*}
\includegraphics[width=\textwidth]{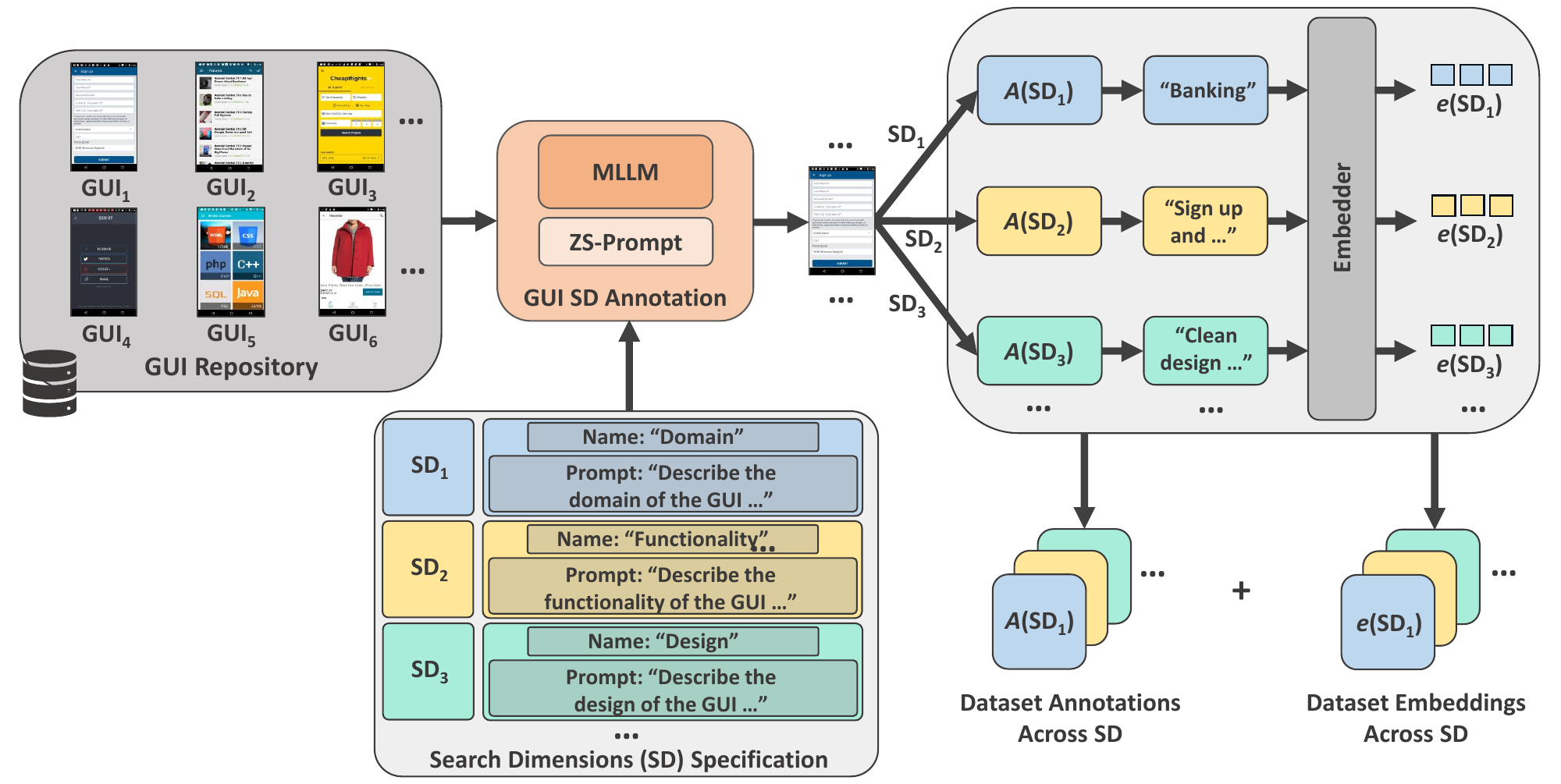}
  \caption[Overview of \gls{mllm}-based \gls{gui} annotation pipeline]{Overview of the automated \gls{gui} repository annotation pipeline, showing the input \gls{gui} repository (e.g., \textit{Rico}), a collection of customizable \glspl{sdim}, the \gls{zs}-prompted \gls{mllm} which creates \gls{sdim}-specific text annotations that are subsequently embedded.}
	\label{fig:overview_dataset_gui_rerank}
\end{figure*}

\vspace{-0.2cm}
\subsection{Automated \gls{gui} Repository Annotation}

Given a \gls{gui} repository as images and a collection of customizable \glspl{sdim}, \textit{\gls{gui}-ReRank} employs a \gls{zs}-prompted \gls{mllm} to generate textual annotations for each \gls{gui} according to the \glspl{sdim} (Figure \ref{fig:overview_dataset_gui_rerank}). Our framework provides a default collection of \glspl{sdim}, including \textit{domain}, \textit{functionality}, \textit{design}, \textit{\gls{gui} components}, and \textit{displayed text}, which are widely applicable to various \gls{fr} and \gls{nfr} search needs. Particularly, to support \gls{rel} and \gls{rval}, the default dimensions of \textit{domain}, \textit{functionality} and \textit{\gls{gui} components} are valuable, since these directly relate to functional aspects of the \gls{gui} prototypes. In addition, users are enabled to modify or extend the \glspl{sdim} by specifying a name and a brief \gls{nl} description of the \gls{sdim} to address user-specific requirements for the search (e.g., \textit{platform} or \textit{accessibility}). Subsequently, each of the created annotations is processed by an embedding model to derive dense, \gls{sdim}-specific embeddings. The generated text and embedding datasets provide the foundation for the following \gls{gui} retrieval and \gls{mllm}-based reranking approaches. To facilitate adoption, \textit{\gls{gui}-ReRank} includes the well-known, large-scale \gls{gui} dataset \textit{Rico} \citep{deka2017rico} ready to search, fully pre-annotated and embedded.

\subsection{Embedding-based Constrained GUI Retrieval}

\begin{figure*}
\includegraphics[width=\textwidth]{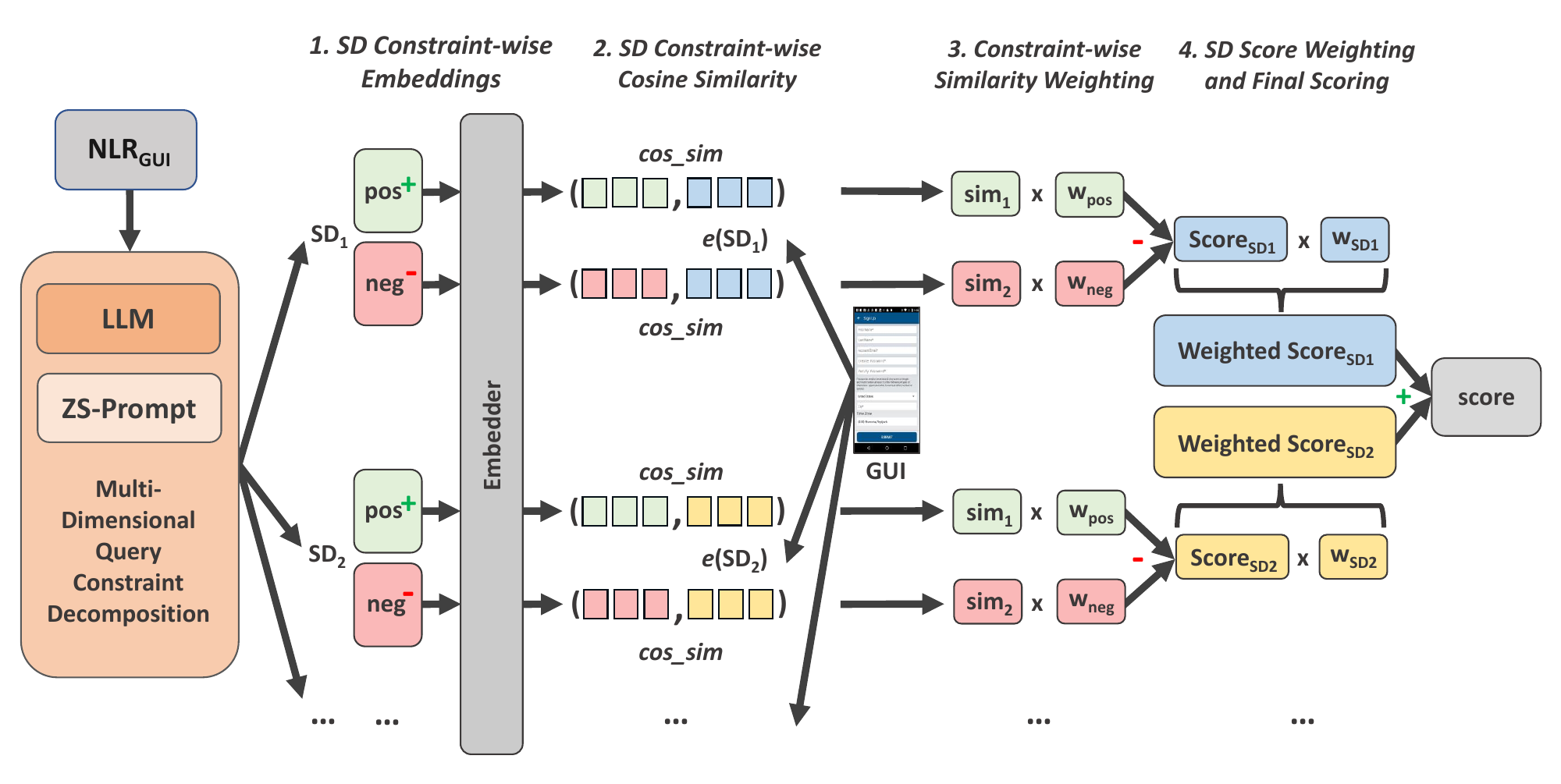}
  \caption[Overview of embedding-based constrained multi-dimensional \gls{gui} retrieval approach]{Overview of the embedding-based constrained \gls{gui} retrieval approach, showing \textit{(1)} the decomposition of \gls{nlr} into \gls{sdim}-wise positive and negative constraints which are then embedded, \textit{(2)} the computation of cosine similarity across \glspl{sdim} with the \gls{gui} embeddings, \textit{(3)} similarity weighting and combination into single \gls{sdim}-wise scores, and finally \textit{(4)} the computation of the final score as a weighted sum of \gls{sdim}-wise scores.}
	\label{fig:overview-gui-rerank-embedding}
    \vspace{-0.3cm}
\end{figure*}

\newcommand{\cosim}{\operatorname{cos\_sim}}
\newcommand{\emb}{\mathbf{e}}

To support complex semantic \gls{gui} searches with both positive and negative \gls{nlr} constraints across various \glspl{sdim}, \textit{\gls{gui}-ReRank} initially decomposes the input \gls{nlr} by employing a \gls{zs}-prompted \gls{llm}, which extracts both positive and negative constraints or requirements across all predefined \glspl{sdim} (e.g., $SD_{Design}$: $pos^{+}$(\textit{``modern''}), $neg^{-}$(\textit{``dark''}) or $SD_{Functionality}$: $pos^{+}$(\textit{``Social login''}), $neg^{-}$(\textit{``Google''})). Figure \ref{fig:overview-gui-rerank-embedding} illustrates an overview of the approach. More formally, given \glspl{sdim} $\mathcal{SD} = \{SD_1,\dots,SD_K\}$ and \gls{nlr}, the \gls{llm} derives positive $P_k = \{p_{k,1},\dots,p_{k,|P_k|}\}$ and negative $N_k = \{n_{k,1},\dots,n_{k,|N_k|}\}$ \gls{sdim}-wise constraints. Afterwards, the extracted constraints are embedded by the same embedding model which was utilized to create the \gls{gui} dataset embeddings, i.e. $\emb(p_{k,j}) = f_{\text{emb}}(p_{k,j})$ and $\emb(n_{k,j}) = f_{\text{emb}}(n_{k,j})$. We denote the \gls{sdim}-wise \gls{gui} embedding by $\emb(G_{i,k}) = f_{\text{emb}}(G_{i,k})$. Subsequently, we compute the matching score as the \textit{cosine similarity} between the positive and negative constraints for a particular \gls{sdim} $SD_k$ and the respective embeddings of \gls{gui} $G_i$ as

\vspace{-0.4cm}
\begin{align}
s^{+}_{i,k,j} &= \cosim\!\left(\emb(p_{k,j}), \emb(G_{i,k})\right)\\
s^{-}_{i,k,j} &= \cosim\!\left(\emb(n_{k,j}), \emb(G_{i,k})\right) \label{eq:cos}
\end{align}

\noindent Particularly, $s^{+}_{i,k,j}$ represents a score of how closely the $j$-th positive constraint for $SD_k$ matches the corresponding \gls{gui} embedding for the same $SD_k$. Next, we compute the average score for the positive and negative constraints, respectively, as follows

\begin{equation}
\mathrm{sim}^{+}_{i,k} = \frac{1}{|P_k|}\sum_{j=1}^{|P_k|} s^{+}_{i,k,j}
\qquad
\mathrm{sim}^{-}_{i,k} = \frac{1}{|N_k|}\sum_{j=1}^{|N_k|} s^{-}_{i,k,j}
\label{eq:aggregate}
\end{equation}

\noindent Afterwards, we compute the weighted sum of the overall positive and negative scores for each $SD_k$ and \gls{gui} $G_i$ as $\mathrm{Score}_{i,k}
= w_{\text{pos}} \, \mathrm{sim}^{+}_{i,k}- w_{\text{neg}} \, \mathrm{sim}^{-}_{i,k}$, followed by weighting the score for the search dimension $SD_k$ as $\mathrm{WScore}_{i,k} = w_{SD_k}\,\mathrm{Score}_{i,k}$. Finally, we compute the normalized sum across all weighted dimension scores to obtain the final score. This approach enables a fine-grained, weighted search across multiple dimensions.

\vspace{-0.2cm}
\subsection{Multi-Modal (M)LLM-based GUI Reranking}

\begin{figure*}
\includegraphics[width=\textwidth]{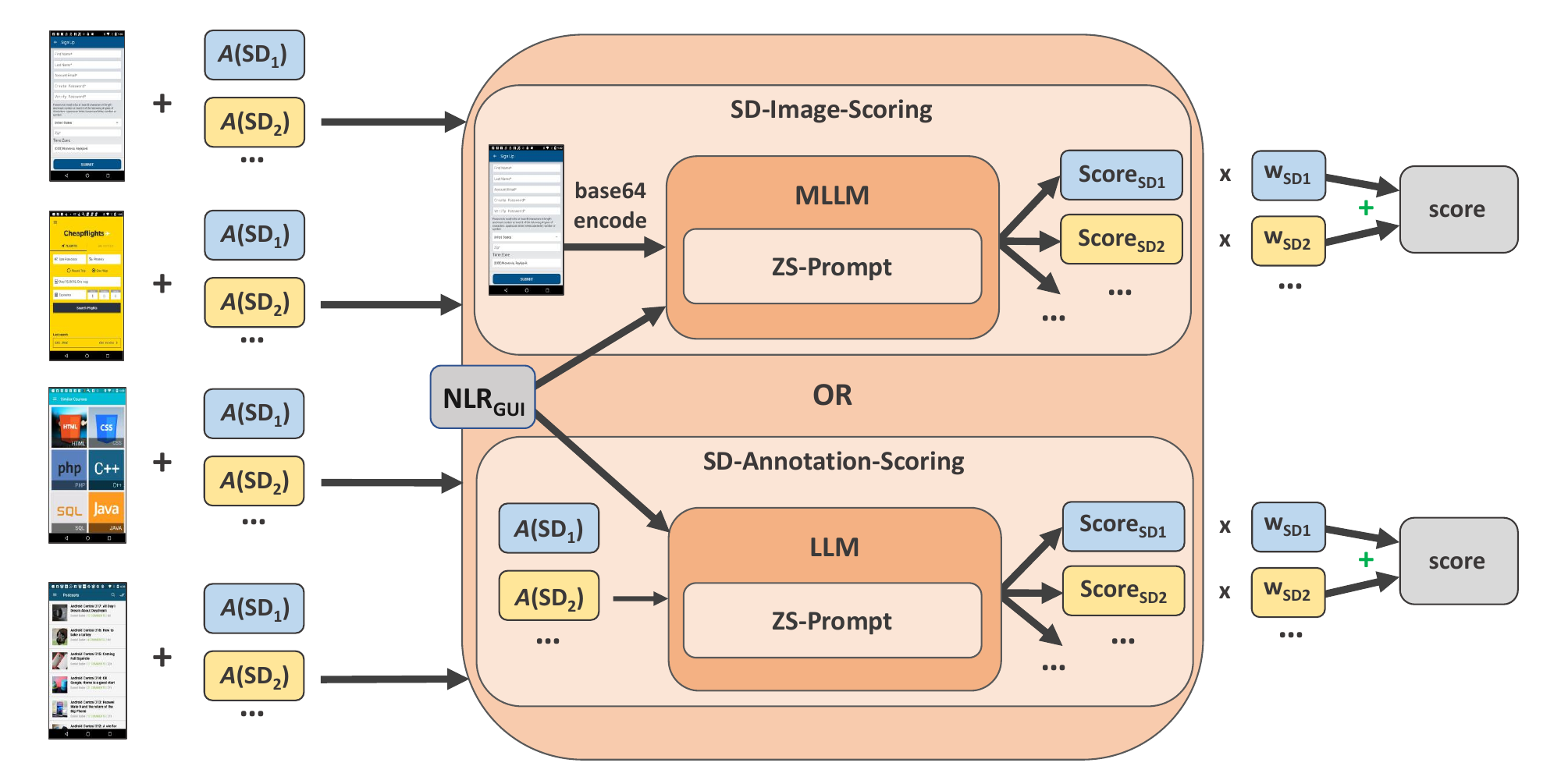}
  \caption[\gls{mllm}-based \gls{gui} reranking approach]{Overview of the \gls{mllm}-based \gls{gui} reranking showing both available variants: \textit{image-based scoring}, which uses the \textit{base64}-encoded image with the \gls{nlr} for scoring, and \textit{text-based scoring}, which uses the \gls{sdim}-wise text annotations with the \gls{nlr} for scoring.}
	\label{fig:gui_rerank_reranking}
    \vspace{-0.3cm}
\end{figure*}

Given the top-\textit{k} \glspl{gui} retrieved by the previously described model, we subsequently propose \gls{mllm}-based \gls{gui} reranking, which refines the top-\textit{k} \gls{gui} ranking by evaluating either the \gls{gui} screenshot or the \gls{sdim}-wise text annotations by using a \gls{zs}-prompted \gls{mllm} (Figure \ref{fig:gui_rerank_reranking}). For each \gls{sdim}, the \gls{mllm} computes a score between 0 (\textit{no match}) and 100 (\textit{perfect match}), i.e. $RScore_{d,i} = MLLM(SD_d, NLR, G_i)$, which reflects the relevance of this dimension for the given \gls{nlr} and the \gls{gui} $G_i$ (represented either as an image or a text annotation). By allowing both image and detailed text annotations as input to the reranking model, \textit{\gls{gui}-ReRank} offers the option for an effectiveness-cost trade-off. While image-based \gls{gui} reranking provides the most detailed and information-dense representation to the model for computing relevance, it also involves higher cost (i.e., more \textit{computational resources}, \textit{time} and \textit{monetary cost}). On the contrary, text-based \gls{gui} reranking provides higher cost efficiency, but the input representation is still an abstraction, being less information-dense in comparison to the actual image. Similar to the previous embedding-based retrieval, the dimension-specific scores are weighted and a normalized sum is computed to obtain the final ranking score. We also provide a non-weighted variant, where \gls{nlr} are matched against a single aggregated \gls{gui} summary.

\vspace{-0.2cm}
\subsection{Prototype Implementation}

The \textit{GUI-ReRank} prototype is implemented as a \textit{Django} web application with an \textit{HTML/CSS} and \textit{JavaScript} frontend, which is combined with \textit{MySQL} \citep{mysql_server_github} for data storage. To enable efficient and parallel processing of background tasks, for example, the \gls{gui} text annotation, embedding creation and batch-wise reranking of the top-\textit{k} \glspl{gui}, we utilize the \textit{Celery} framework \citep{celery_docs_introduction} distributed across multiple worker nodes, combined with \textit{Redis} \citep{redis_github} as the message broker. Our current implementation provides support for many popular \glspl{mllm}, including \textit{GPT} \citep{openai2023gpt4} from \textit{OpenAI} \citep{openai_gpt4o_docs, openai_gpt41_docs}, \textit{Google Gemini} \citep{google_gemini_models} and \textit{Anthropic Claude Sonnet} \citep{anthropic_claude_models_overview}, and is easily extensible and customizable for additional \glspl{mllm}.
\section{Experimental Evaluation}
\label{sec:chapter-rerank:evaluation}
In the following, we present our comprehensive evaluation of the proposed approach, which is based primarily on the \gls{nlr}-based \gls{gui} retrieval and ranking gold standard. In particular, we provide the experimental setup for the following two research questions:

\definecolor{lightgray}{gray}{0.92}
\newcolumntype{G}{c}

\newcommand{\PkNdcgSepRuleRQOne}{%
  \cmidrule(lr){1-2}\cmidrule(lr){3-6}\cmidrule(lr){7-10}
}

\begin{table*}[!t]
\footnotesize
\caption[Evaluation results of \gls{mllm}-based \gls{gui} reranking approaches (1)]{Evaluation results overview of the different models on the \gls{nlr}-based \gls{gui} ranking gold standard using $P@k$ and \textit{\gls{ndcg}@k} (\textbf{Bold} values indicate best metric score within result group, \underline{underlined} values indicate best metric score across all shown groups).}
\centering
\setlength\tabcolsep{4.5pt}
\renewcommand{\arraystretch}{1.1}

\begin{tabular}{G|l|cccc|cccc}
\toprule
\multicolumn{1}{c|}{} & \multicolumn{1}{l|}{} &
\multicolumn{4}{c|}{\textbf{P@k}} &
\multicolumn{4}{c}{\textbf{NDCG@k (N@k)}} \\
\cmidrule(lr){3-6}\cmidrule(lr){7-10}
\multicolumn{1}{c|}{} & \multicolumn{1}{l|}{} &
$\mathbf{P@3}$ & $\mathbf{P@5}$ & $\mathbf{P@7}$ & $\mathbf{P@10}$ &
$\mathbf{N@3}$ & $\mathbf{N@5}$ & $\mathbf{N@10}$ & $\mathbf{N@15}$ \\
\midrule

\multirow[c]{3}{*}{\cellcolor{white}\textbf{BL}}
 & \textbf{BERT-LTR-1}      & 37.7 & 35.0 & 30.7 & 26.9  & 53.0 & \textbf{56.0} & 63.4 & 69.7 \\
\rowcolor{lightgray}
\textbf{BL}  & \textbf{BERT-LTR-2}      & \textbf{40.0} & 34.0 & 30.4 & 28.1  & \textbf{54.3} & 55.6 & 63.6 & \textbf{70.1} \\
 & \textbf{BERT-LTR-3}      & 36.3 & \textbf{35.4} & \textbf{31.7} & \textbf{28.7}  & 51.7 & 55.4 & \textbf{64.6} & 69.4 \\

\PkNdcgSepRuleRQOne

\multirow[c]{7}{*}{\cellcolor{white}\textbf{Text}}
 & \textbf{GPT-4.1}             & \textbf{72.0} & \textbf{55.2} & \textbf{45.1} & \textbf{34.9}  & \textbf{84.1} & \textbf{82.3} & \textbf{86.6} & \textbf{89.1} \\
\rowcolor{lightgray}
 & \textbf{GPT-4.1 Mini}        & 67.3 & 54.6 & 44.6 & 34.4  & 81.7 & 81.0 & 84.4 & 88.0 \\
 & \textbf{GPT-4.1 Nano}        & 55.0 & 46.0 & 39.0 & 30.4  & 71.4 & 71.2 & 76.0 & 81.0 \\
\rowcolor{lightgray}
\textbf{Text}  & \textbf{Gemini-2.5 Flash}    & 66.0 & 51.8 & 43.1 & 34.1  & 78.4 & 77.9 & 82.6 & 86.3 \\
 & \textbf{Gemini-2.5 Pro}      & 69.0 & 54.0 & 44.1 & 34.1  & 82.6 & 80.5 & 84.0 & 87.9 \\
\rowcolor{lightgray}
 & \textbf{Claude Sonnet 3.7}   & 65.0 & 52.6 & 43.7 & 34.0  & 78.2 & 77.9 & 83.0 & 86.6 \\
 & \textbf{Claude Sonnet 4.0}   & 66.3 & 51.4 & 43.9 & 34.6  & 81.8 & 79.1 & 84.8 & 88.1 \\

\PkNdcgSepRuleRQOne

\multirow[c]{7}{*}{\cellcolor{white}\textbf{Image}}
 & \textbf{GPT-4.1}             & \textbf{\underline{75.0}} & \textbf{\underline{56.6}} & \textbf{\underline{46.3}} & \textbf{\underline{35.4}}
                               & \textbf{\underline{87.7}} & \textbf{\underline{85.2}} & \textbf{\underline{88.4}} & \textbf{\underline{90.8}} \\
\rowcolor{lightgray}
 & \textbf{GPT-4.1 Mini}        & 71.7 & 54.0 & 44.1 & 35.1  & 85.0 & 82.0 & 86.8 & 89.5 \\
 & \textbf{GPT-4.1 Nano}        & 49.0 & 40.8 & 33.4 & 27.7  & 62.1 & 62.2 & 66.7 & 72.4 \\
\rowcolor{lightgray}
\textbf{Image}  & \textbf{Gemini-2.5 Flash}    & 66.7 & 53.6 & 43.4 & 34.5  & 78.9 & 78.3 & 83.6 & 86.6 \\
 & \textbf{Gemini-2.5 Pro}      & 68.0 & 55.8 & \textbf{\underline{46.3}} & 34.6  & 81.4 & 80.9 & 84.5 & 87.8 \\
\rowcolor{lightgray}
 & \textbf{Claude Sonnet 3.7}   & 70.3 & 55.8 & 45.1 & 34.8  & 83.3 & 82.1 & 86.0 & 89.2 \\
 & \textbf{Claude Sonnet 4.0}   & 70.7 & 54.6 & 45.1 & 34.7  & 83.9 & 82.3 & 86.1 & 89.0 \\
\bottomrule
\end{tabular}
\vspace{-0.2cm}
\label{tab:gui_rerank_tab_1}
\end{table*}

\begin{itemize}
    \item \textbf{RQ$_{1}$}: \textit{How effective are \gls{mllm}-based approaches for \gls{nlr}-based \gls{gui} reranking?} To answer this question, we utilize the \gls{nlr}-based \gls{gui} retrieval and ranking gold standard created in the previous chapter, compute the same \gls{ir} metrics and compare multiple variants of \gls{mllm}-based \gls{gui} reranking to previous baselines.
    \item \textbf{RQ$_{2}$}: \textit{What are the effectiveness and cost trade-offs for \gls{mllm}-based \gls{gui} reranking?} To answer this question, we track different cost metrics while running the experiments across models, including \textit{token consumption}, \textit{time} and \textit{monetary cost}.
\end{itemize}
\vspace{-0.4cm}

\subsection{RQ$_{1}$: Effectiveness of \gls{mllm}-based \gls{gui} Reranking}


We evaluated the effectiveness of the \gls{mllm}-based \gls{gui} reranking based on the gold standard developed in the previous chapter. To summarize, this benchmark comprises 100 \gls{nlr} queries, each paired with 20 \glspl{gui} from \textit{Rico} \citep{deka2017rico} and enables direct comparison with state-of-the-art \gls{bert}-\gls{ltr} models (cf. Chapter \ref{cha:nl_gui_retrieval}). In addition, we employed the same \gls{ir} metrics utilized in the previous chapter (\textit{\gls{ap}}, \textit{\gls{mrr}}, \textit{P@k}, \textit{HITS@k} and \textit{\gls{ndcg}@k}) to measure the ranking effectiveness. We used the non-weighted reranking and experimented with different models including three capacity variants of \textit{\gls{gpt}-4.1} (the \textit{Original}, \textit{Mini} and \textit{Nano}) \citep{openai_gpt41_docs}, two variants of \textit{Gemini-2.5} (\textit{Flash} and \textit{Pro}) \citep{google_gemini_models, google_gemini_models_2_5_pro} and two variants of \textit{Claude Sonnet} (\textit{3.7} and \textit{4}) \citep{anthropic_claude_models_overview, claude4}. For determining statistically significant differences between models, we applied the \textit{Wilcoxon signed-rank test} \citep{woolson2007wilcoxon} with per-metric \textit{Holm} correction across model comparisons. In particular, we compare the three \gls{bert}-\gls{ltr} models as baselines against the most powerful non-reasoning \textit{\gls{gpt}-4.1} model (according to different benchmarks at the time of release \citep{openai_gpt41_release}) in the \textit{text} and \textit{image} reranking variants. To generate the \gls{gui} text annotations, we used \textit{\gls{gpt}-4.1}. To investigate significant differences between \textit{text} and \textit{image} input, we compare the \textit{\gls{gpt}-4.1} model in both variants. Finally, to determine significant differences between different model capacities, we compare the full \textit{\gls{gpt}-4.1} model (\textit{Original}) against its smaller variants (\textit{Mini}, \textit{Nano}).

\definecolor{lightgray}{gray}{0.92}
\newcolumntype{G}{c}

\newcommand{\ApMrrHitsSepRuleRQOne}{%
  \cmidrule(lr){1-2}\cmidrule(lr){3-4}\cmidrule(lr){5-9}
}

\begin{table*}[!t]
\footnotesize
\caption[Evaluation results of \gls{mllm}-based \gls{gui} reranking approaches (2)]{Evaluation results overview of the different models on the \gls{nlr}-based \gls{gui} ranking gold standard using \textit{\gls{ap}}, \textit{\gls{mrr}} and \textit{HITS@k} (\textbf{Bold} values indicate best metric score within result group, \underline{underlined} values indicate best metric score across all groups).}
\centering
\setlength\tabcolsep{6pt}
\renewcommand{\arraystretch}{1.1}

\begin{tabular}{G|l|c|c|ccccc}
\toprule
\multicolumn{1}{c|}{} & \multicolumn{1}{l|}{} &
\multicolumn{1}{c|}{\textbf{AP}} &
\multicolumn{1}{c|}{\textbf{MRR}} &
\multicolumn{5}{c}{\textbf{HITS@k (H@k)}} \\
\cmidrule(lr){3-3}\cmidrule(lr){4-4}\cmidrule(lr){5-9}
\multicolumn{1}{c|}{} & \multicolumn{1}{l|}{} &
$\mathbf{AP}$ & $\mathbf{MRR}$ &
$\mathbf{H@1}$ & $\mathbf{H@3}$ & $\mathbf{H@5}$ & $\mathbf{H@10}$ & $\mathbf{H@15}$ \\
\midrule

\multirow[c]{3}{*}{\cellcolor{white}\textbf{BL}}
 & \textbf{BERT-LTR-1} & 48.6 & 61.8 & \textbf{46.0} & 71.0 & 86.0 & 98.0 & \textbf{\underline{100.0}} \\
\rowcolor{lightgray}
\textbf{BL}  & \textbf{BERT-LTR-2} & \textbf{50.1} & \textbf{63.1} & 44.0 & \textbf{75.0} & \textbf{91.0} & 98.0 & \textbf{\underline{100.0}} \\
 & \textbf{BERT-LTR-3} & 49.9 & 62.6 & 45.0 & 73.0 & 86.0 & \textbf{\underline{100.0}} & \textbf{\underline{100.0}} \\

\ApMrrHitsSepRuleRQOne

\multirow[c]{7}{*}{\cellcolor{white}\textbf{Text}}
 & \textbf{GPT-4.1}           & \textbf{81.3} & 92.7 & 87.0 & \textbf{\underline{100.0}} & \textbf{\underline{100.0}} & \textbf{\underline{100.0}} & \textbf{\underline{100.0}} \\
\rowcolor{lightgray}
 & \textbf{GPT-4.1 Mini}      & 79.7 & 92.3 & 87.0 & 98.0 & \textbf{\underline{100.0}} & \textbf{\underline{100.0}} & \textbf{\underline{100.0}} \\
 & \textbf{GPT-4.1 Nano}      & 66.2 & 84.2 & 75.0 & 91.0 & 96.0 & \textbf{\underline{100.0}} & \textbf{\underline{100.0}} \\
\rowcolor{lightgray}
\textbf{Text}  & \textbf{Gemini-2.5 Flash}  & 77.0 & 90.1 & 83.0 & 97.0 & \textbf{\underline{100.0}} & \textbf{\underline{100.0}} & \textbf{\underline{100.0}} \\
 & \textbf{Gemini-2.5 Pro}    & 80.5 & \textbf{\underline{94.8}} & \textbf{\underline{91.0}} & \textbf{\underline{100.0}} & \textbf{\underline{100.0}} & \textbf{\underline{100.0}} & \textbf{\underline{100.0}} \\
\rowcolor{lightgray}
 & \textbf{Claude Sonnet 3.7} & 76.9 & 92.3 & 87.0 & 97.0 & 99.0 & \textbf{\underline{100.0}} & \textbf{\underline{100.0}} \\
 & \textbf{Claude Sonnet 4.0} & 78.3 & 92.3 & 88.0 & 95.0 & \textbf{\underline{100.0}} & \textbf{\underline{100.0}} & \textbf{\underline{100.0}} \\

\ApMrrHitsSepRuleRQOne

\multirow[c]{7}{*}{\cellcolor{white}\textbf{Image}}
 & \textbf{GPT-4.1}           & \textbf{\underline{84.0}} & 92.8 & 87.0 & \textbf{99.0} & \textbf{\underline{100.0}} & \textbf{\underline{100.0}} & \textbf{\underline{100.0}} \\
\rowcolor{lightgray}
 & \textbf{GPT-4.1 Mini}      & 81.1 & \textbf{93.0} & 88.0 & 98.0 & \textbf{\underline{100.0}} & \textbf{\underline{100.0}} & \textbf{\underline{100.0}} \\
 & \textbf{GPT-4.1 Nano}      & 58.2 & 79.0 & 66.0 & 91.0 & 99.0 & \textbf{\underline{100.0}} & \textbf{\underline{100.0}} \\
\rowcolor{lightgray}
\textbf{Image}  & \textbf{Gemini-2.5 Flash}  & 77.5 & 89.7 & 83.0 & 96.0 & 99.0 & \textbf{\underline{100.0}} & \textbf{\underline{100.0}} \\
 & \textbf{Gemini-2.5 Pro}    & 79.8 & 90.8 & 85.0 & 96.0 & 99.0 & \textbf{\underline{100.0}} & \textbf{\underline{100.0}} \\
\rowcolor{lightgray}
 & \textbf{Claude Sonnet 3.7} & 81.2 & 91.9 & 86.0 & 97.0 & \textbf{\underline{100.0}} & \textbf{\underline{100.0}} & \textbf{\underline{100.0}} \\
 & \textbf{Claude Sonnet 4.0} & 80.7 & 91.8 & 85.0 & \textbf{99.0} & \textbf{\underline{100.0}} & \textbf{\underline{100.0}} & \textbf{\underline{100.0}} \\
\bottomrule
\end{tabular}

\label{tab:gui_rerank_tab_2}
\end{table*}

\vspace{-0.1cm}
\subsection{RQ$_{2}$:  \gls{mllm}-based Reranking Effectiveness-Cost Trade-offs}

To evaluate the effectiveness-cost trade-offs associated with \gls{mllm}-based reranking, we tracked multiple metrics including \textit{token consumption}, \textit{time} and \textit{monetary cost} during the execution of the experiments. We report the different values as averaged values obtained from the gold standard experiment runs and provide effectiveness(\textit{\gls{ap}})-cost tradeoff plots.
\begin{figure*}[!t] \includegraphics[width=\textwidth]{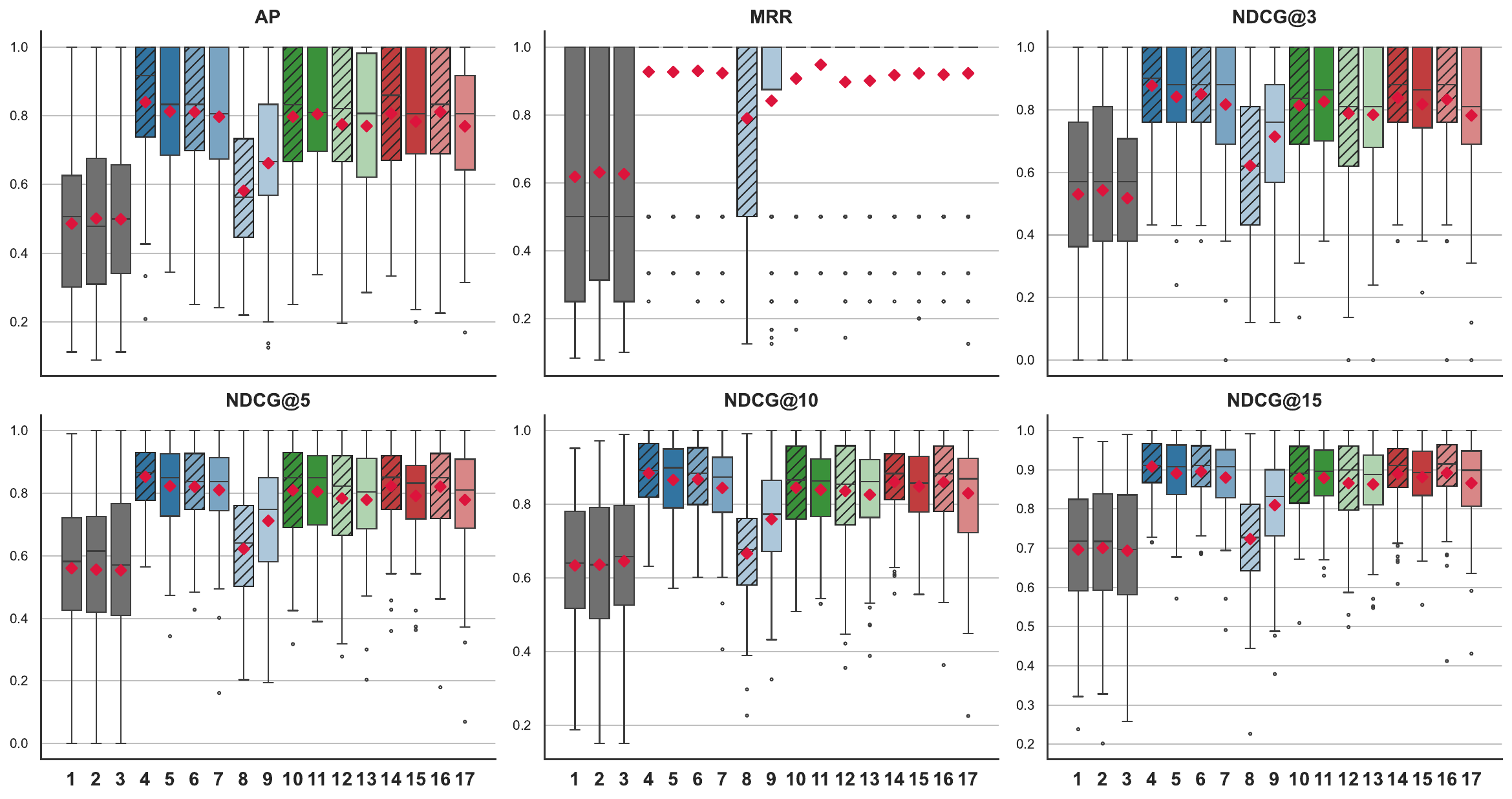}
  \caption[Boxplots for \gls{nlr}-based \gls{gui} reranking effectiveness]{\textit{\gls{ap}}, \textit{\gls{mrr}} and \textit{\gls{ndcg}@k} across models: \textit{(1--3)} BERT-LTR (1--3), \textit{(4) \gls{gpt}-4.1 (I)}, \textit{(5) \gls{gpt}-4.1 (T)}, \textit{(6) \gls{gpt}-4.1 Mini (I)}, \textit{(7) \gls{gpt}-4.1 Mini (T)}, \textit{(8) \gls{gpt}-4.1 Nano (I)}, \textit{(9) \gls{gpt}-4.1 Nano (T)}, \textit{(10) Gemini-2.5 Pro (I)}, \textit{(11) Gemini-2.5 Pro (T)}, \textit{(12) Gemini-2.5 Flash (I)}, \textit{(13) Gemini-2.5 Flash (T)}, \textit{(14) Claude Sonnet 4 (I)}, \textit{(15) Claude Sonnet 4 (T)},
  \textit{(16) Claude Sonnet 3.7 (I)}, \textit{(17) Claude Sonnet 3.7 (T)}, I=Image, T=Text, red diamonds=means.}
	\label{fig:gui-rerank-boxplots-models}
    \vspace{-0.3cm}
\end{figure*}

\vspace{-0.6cm}
\section[Results \&\ Discussion]{Results \& Discussion}
\label{sec:chapter-rerank:results}
\vspace{-0.2cm}
\subsection{RQ$_{1}$: Effectiveness of \gls{mllm}-based \gls{gui} Reranking}

Table \ref{tab:gui_rerank_tab_1} and Table \ref{tab:gui_rerank_tab_2} show the ranking performance across the different \gls{ir} metrics for \gls{bert}-\gls{ltr} baselines (BL), text- and image-based \gls{mllm} approaches. As can be observed, all \gls{mllm}-based approaches (text and image variants) substantially outperform the \gls{bert}-\gls{ltr} models in terms of mean metric values. For example, the \textit{\gls{ap}} achieved by the \textit{pairwise} \gls{bert}-\gls{ltr} (2) model is improved by 67.66\% with the \textit{\gls{gpt}-4.1 (Image)} reranking. Moreover, the high \textit{HITS@1} for \textit{\gls{gpt}-4.1 (Image)} indicates that in 87\% of cases, a relevant \gls{gui} is ranked at the top position, which represents an 89.13\% improvement over the best \gls{bert}-\gls{ltr} model. Statistically significant differences between all \gls{bert}-\gls{ltr} models and \textit{\gls{gpt}-4.1} (both \textit{Image} and \textit{Text}) were observed across all metrics except \textit{HITS@10,15} (cf. Appendix \ref{chapter:app-gui-rerank} Tables \ref{tab:gui-rerank-appendix-llm-holm-1}, \ref{tab:gui-rerank-appendix-llm-holm-2} and \ref{tab:gui-rerank-appendix-llm-holm-3}). These results and the substantial improvements in mean metric values clearly illustrate the high effectiveness of the \gls{mllm}-based \gls{gui} reranking technique. Figure \ref{fig:gui-rerank-boxplots-models} illustrates multiple boxplots across the considered models as well as \textit{\gls{ap}}, \textit{\gls{mrr}} and \textit{\gls{ndcg}@k} metrics.

\begin{figure*}[!t] \includegraphics[width=\textwidth,height=0.45\textheight,keepaspectratio]{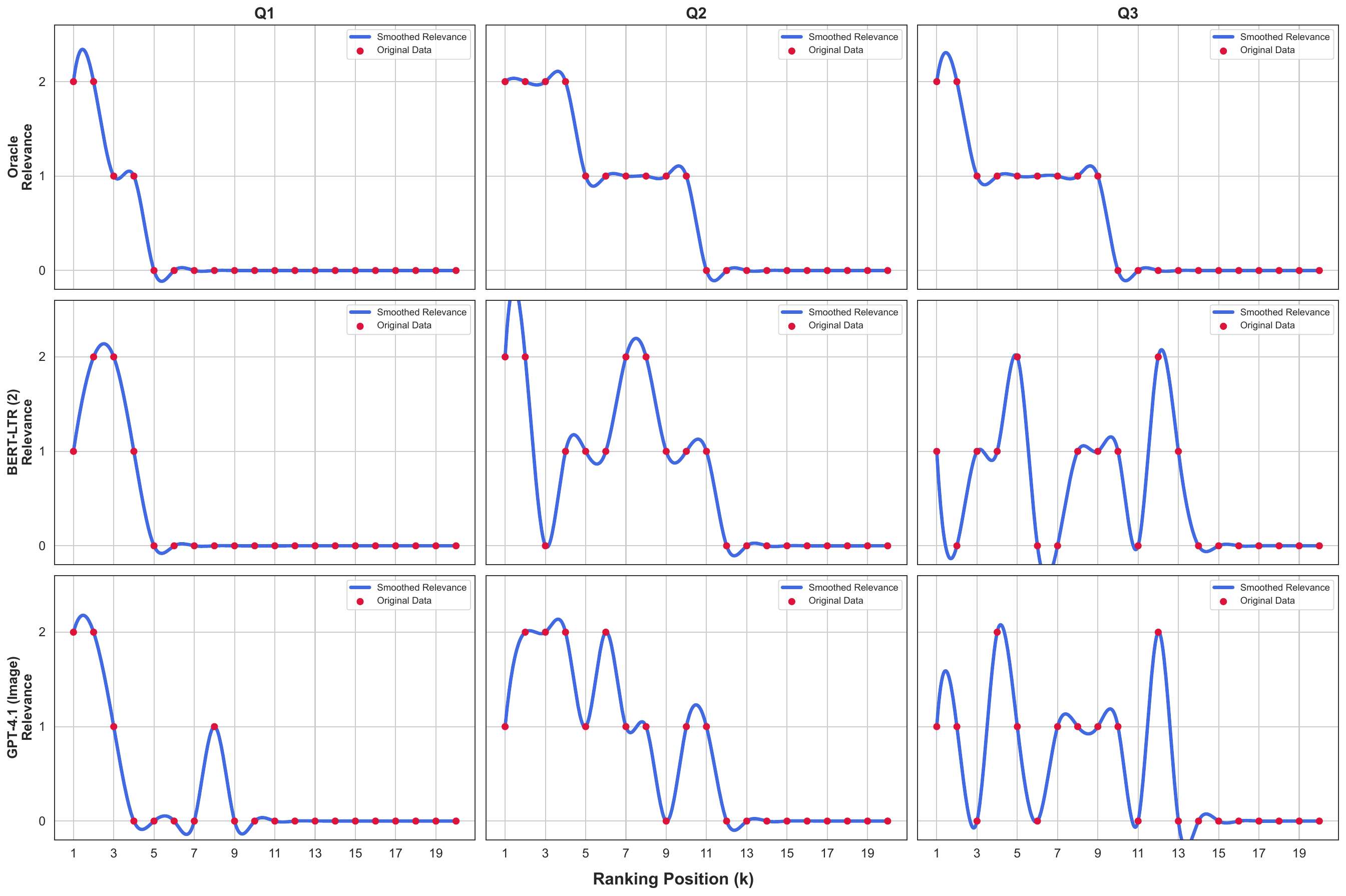}
  \caption[Ranking plots for Oracle, \gls{bert}-\gls{ltr} and \gls{mllm}-based reranking]{Ranking plots showing the relevance of a \gls{gui} at position $k$ for a particular model and query: Columns represent the same query, while rows represent the same model, enabling a comparison. The plots include an \textit{Oracle} model (the perfect ranking), \gls{bert}-\gls{ltr} \textit{(2)} and \gls{gpt}-4.1 (Image). \textit{Q1}=\textit{``screen with sample movies and information on recording with SnapMovie''}, \textit{Q2}=\textit{``profile window with a button to take a photo or select it from galary [sic]''} and \textit{Q3}=\textit{``Weather details for city with option in Celsius and Fahrenheit''.}}
	\label{fig:gui-rerank-raw-ranking-plots}
\end{figure*}

Considering differences between input modalities by comparing text-based and image-based \textit{\gls{gpt}-4.1} models for \gls{gui} reranking, we can observe that the image variant consistently outperforms the text variant across mean \textit{P@k}, \textit{\gls{ndcg}@k}, \textit{\gls{ap}} and \textit{\gls{mrr}}, while differences are statistically significant solely for the \textit{\gls{ndcg}@k} metric (cf. Appendix \ref{chapter:app-gui-rerank} Tables \ref{tab:gui-rerank-appendix-llm-holm-1}, \ref{tab:gui-rerank-appendix-llm-holm-2} and \ref{tab:gui-rerank-appendix-llm-holm-3}). This illustrates that the higher information density of image-based \gls{gui} representations improves the effectiveness over text-based representations, even for mainly functional \gls{nlr} (as in the gold standard). While most of the image-based models outperform their text-based counterparts across many mean metric values, we can observe some exceptions. Most notably, the image variant of \textit{\gls{gpt}-4.1 (Nano)} is substantially outperformed across all mean metric values (except \textit{HITS@5}) by the corresponding text variant. This indicates that, while \gls{gui} screenshots carry higher information density, the image understanding capabilities of the \gls{mllm} are required to achieve a certain level of effectiveness. Otherwise the model is not capable of extracting relevant features from the \gls{gui} prototype image. In such a case, the model potentially can more easily extract relevant information from the detailed text annotations. Moreover, comparing differences between model capacities, we observed that the larger variant of \textit{\gls{gpt}-4.1} consistently and statistically significantly outperformed its respective counterparts (e.g., \textit{Original} vs. \textit{Mini}, \textit{Original} vs. \textit{Nano}, \textit{Mini} vs. \textit{Nano}) for both text and image variants across \textit{\gls{ndcg}@3,5,10,15} (except for \textit{\gls{ndcg}@5} between \textit{Original} and \textit{Mini}). Furthermore, statistically significant differences were also observed between both image and text variants (\textit{Original} vs. \textit{Nano} and \textit{Mini} vs. \textit{Nano}) for \textit{P@3,5,7,10}, \textit{\gls{ap}}, \textit{\gls{mrr}} and \textit{HITS@1}. Figure \ref{fig:gui-rerank-raw-ranking-plots} shows multiple ranking plots for three queries and the \textit{Oracle} (perfect ranking), \gls{bert}-\gls{ltr} (2) and \textit{\gls{gpt}-4.1 (Image)}, which also indicates the improved performance of the \gls{mllm}-based \gls{gui} reranking, with more relevant \glspl{gui} at higher rank positions.

\begin{myrqbox}
\textbf{Answer to RQ$_1$:} \gls{mllm}-based \gls{gui} reranking substantially improves the ranking performance compared to prior \gls{bert}-\gls{ltr} baselines, statistically significant across all metrics (except \textit{HITS@10,15}). The image variant consistently outperforms the text variant across mean values (except for \textit{Nano}) and significantly for \textit{\gls{ndcg}@k}. The higher capacity models significantly outperformed the lower capacity models for \textit{\gls{ndcg}@k}.
\end{myrqbox}

\definecolor{lightgray}{gray}{0.92}

\newcolumntype{G}{>{\columncolor{white}}c}

\newcommand{\TokCostTimeSepRule}{%
  \cmidrule(lr){1-2}\cmidrule(lr){3-4}\cmidrule(lr){5-6}\cmidrule(lr){7-8}
}

\begin{table}[!t]
\footnotesize
\setlength\tabcolsep{8.4pt}
\renewcommand{\arraystretch}{1.1}
\centering
\begin{tabular}{G|l|c|c|c|c|c|c}
\toprule
 &  & \multicolumn{2}{c|}{\textbf{\#Tokens (k=1)}} & \multicolumn{2}{c|}{\textbf{Cost (k)}} & \multicolumn{2}{c}{\textbf{Time (k)}} \\
\cmidrule(lr){3-4} \cmidrule(lr){5-6} \cmidrule(lr){7-8}
 &  & Input & Output & 100 & 500 & 100 & 500 \\

\TokCostTimeSepRule

\multirow[c]{7}{*}{\cellcolor{white}\textbf{Text}}
 & \textbf{GPT-4.1}             & 179.77  & 6.00  & \$.041 & \$.204  & 2.4s  & 12s   \\
\rowcolor{lightgray}
 & \textbf{GPT-4.1 Mini}        & 179.77  & 6.00  & \$.008 & \$.041  & 2.6s  & 13s   \\
 & \textbf{GPT-4.1 Nano}        & 179.77  & 7.11  & \$.002 & \$.010  & 2s    & 10s   \\
\rowcolor{lightgray}
\textbf{Text} & \textbf{Gemini-2.5 Flash}    & 183.15  & 6.86  & \$.007 & \$.036  & 11s   & 55s   \\
 & \textbf{Gemini-2.5 Pro}      & 183.15  & 10.20 & \$.033 & \$.165  & 10.1s & 50.5s \\
\rowcolor{lightgray}
 & \textbf{Claude Sonnet 3.7}   & 195.49  & 6.60  & \$.069 & \$.343  & 5.5s  & 27.5s \\
 & \textbf{Claude Sonnet 4.0}   & 201.29  & 81.63 & \$.183 & \$.914  & 20.2s & 101s  \\

\TokCostTimeSepRule

\multirow[c]{7}{*}{\cellcolor{white}\textbf{Image}}
 & \textbf{GPT-4.1}             & 1089.02 & 6.00  & \$.223 & \$1.113 & 12.4s & 62s   \\
\rowcolor{lightgray}
 & \textbf{GPT-4.1 Mini}        & 2154.96 & 6.00  & \$.087 & \$.436  & 9s    & 45s   \\
 & \textbf{GPT-4.1 Nano}        & 3242.41 & 5.95  & \$.032 & \$.163  & 8s    & 40s   \\
\rowcolor{lightgray}
 \textbf{Image} & \textbf{Gemini-2.5 Flash}    & 1088.37 & 6.34  & \$.034 & \$.171  & 12.2s & 61s   \\
 & \textbf{Gemini-2.5 Pro}      & 1088.37 & 9.54  & \$.146 & \$.728  & 16.2s & 81s   \\
\rowcolor{lightgray}
 & \textbf{Claude Sonnet 3.7}   & 1439.31 & 6.60  & \$.442 & \$2.208 & 42.8s & 214s  \\
 & \textbf{Claude Sonnet 4.0}   & 1445.31 & 96.60 & \$.578 & \$2.892 & 31s   & 155s  \\
\bottomrule
\end{tabular}
\caption[Cost-related performance metrics for \gls{mllm}-based \gls{gui} reranking]{\textit{Token usage}, \textit{cost} and \textit{runtime} for the MLLM rerankers (means over gold standard (\textit{\#Queries=100}), $k$ = number of retrieved GUIs to rerank, \textit{\#Celery workers} = 10).}
\label{tab:gui-rerank-token_cost_runtime}
\end{table}

\subsection{RQ$_{2}$: \gls{mllm}-based Reranking Effectiveness-Cost Trade-offs}

\begin{figure*}[t!]
\includegraphics[width=\textwidth]{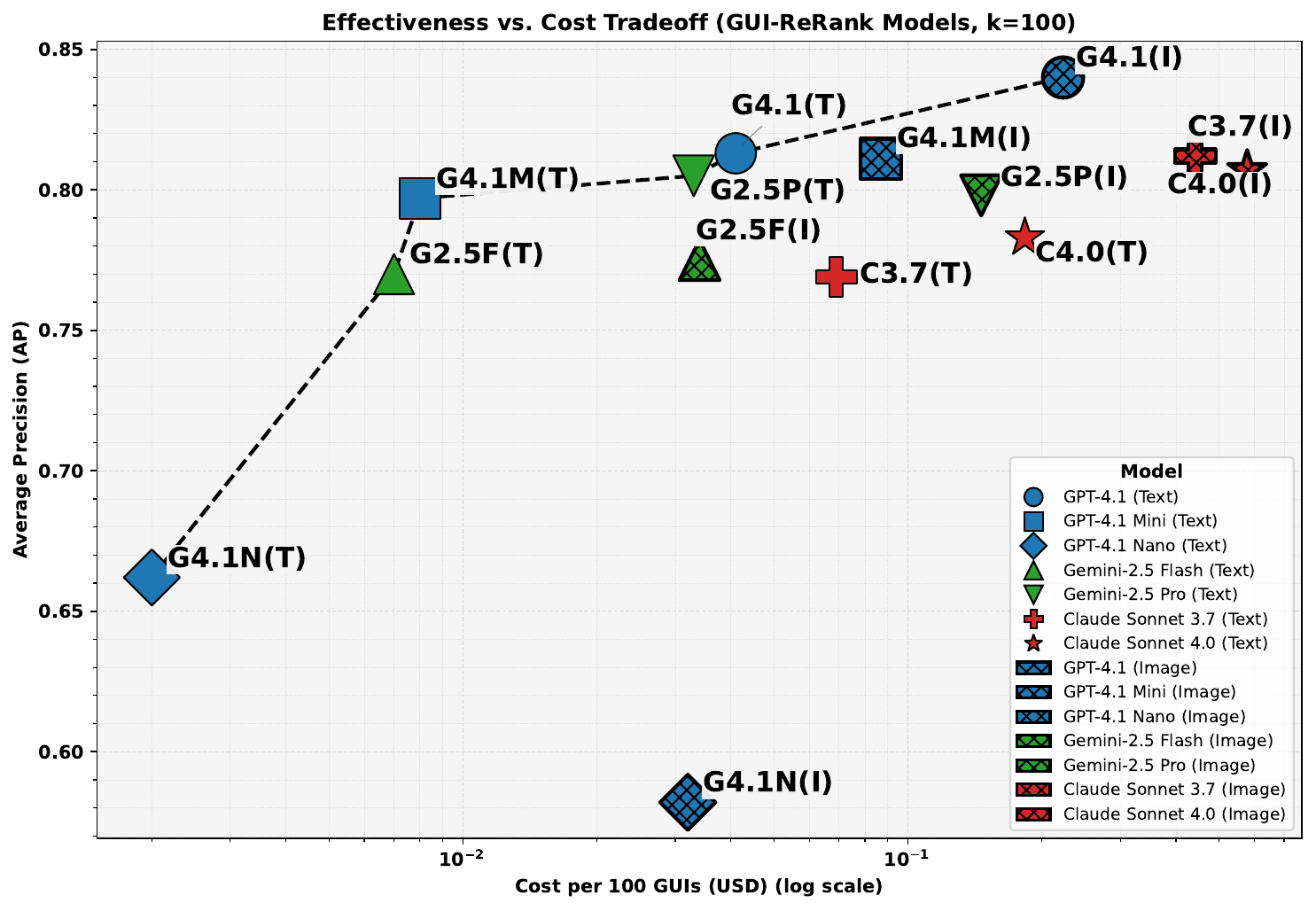}
  \caption[Effectiveness-cost (\gls{api} costs) trade-off for \gls{mllm}-based reranking]{Effectiveness (as \textit{\gls{ap}}) and \textit{monetary \gls{api} cost} ($k$=100) (log scale) trade-off plot with \textit{Pareto frontier}: all text-based \textit{\gls{gpt}-4.1} and \textit{Gemini-2.5} models, as well as image-based \textit{\gls{gpt}-4.1} represent \textit{Pareto}-efficient solutions.}
	\label{fig:gui-rerank-tradeoff-1}
    \vspace{-0.0cm}
\end{figure*}
\begin{figure*}
\includegraphics[width=\textwidth]{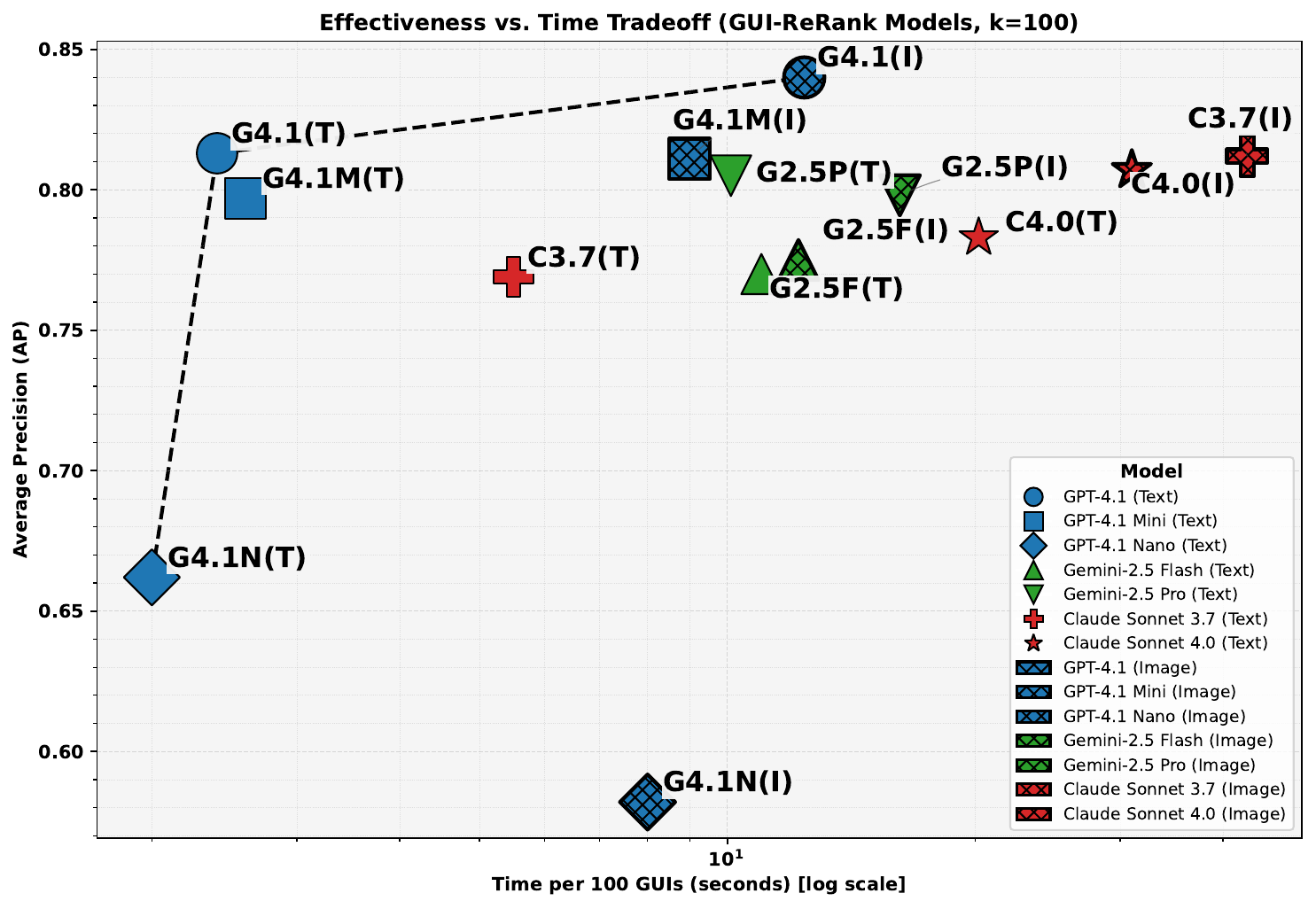}
  \caption[Effectiveness-cost (time) trade-off for \gls{mllm}-based \gls{gui} reranking]{Effectiveness (as \textit{\gls{ap}}) and \textit{time} required for reranking (in \textit{seconds} ($k$=100 and \textit{\#Celery workers}=10)) (log scale) trade-off plot with \textit{Pareto frontier}: text-based \textit{\gls{gpt}-4.1} (\textit{Original} and \textit{Nano}) and image-based \textit{\gls{gpt}-4.1} models provide \textit{Pareto}-efficient solutions.}
	\label{fig:gui-rerank-tradeoff-2}
\end{figure*}

 \glsreset{api}
Table \ref{tab:gui-rerank-token_cost_runtime} shows multiple cost metrics associated with different \gls{mllm}-based \gls{gui} reranking approaches, including the \textit{input} and \textit{output tokens} required for reranking a single \gls{gui} ($k$=1), the \textit{monetary \gls{api} costs} according to the providers\footnote{The monetary \gls{api} costs are reported based on the \textit{input} and \textit{output token} price per 1M tokens as of December 2025 for \textit{OpenAI} \citep{openai_gpt41_docs}, \textit{Google} \citep{google_gemini_models} and \textit{Anthropic} \citep{claude4}.} ($k$=100 and $k$=500) and the required time in seconds ($k$=100 and $k$=500, with 10 parallel \textit{Celery} workers). As can be observed, when comparing the text-based \glspl{mllm} to their respective image-based variants model-wise, there is a substantial increase across all considered performance metrics. Therefore, while the image-based models outperform the text-based models in reranking effectiveness, their application entails much higher cost. For example, the image-based \textit{\gls{gpt}-4.1} improves the mean \textit{\gls{ap}} by 3.32\%, while the associated \textit{\gls{api}} costs ($k$=100) increase by 543.9\%. This shows that substantial costs can be saved by applying text-based variants, while losing only little effectiveness according to the \textit{\gls{ap}}. Figures \ref{fig:gui-rerank-tradeoff-1} and \ref{fig:gui-rerank-tradeoff-2} show trade-off plots with the \textit{Pareto frontier} (\textit{Pareto}-efficient solutions), illustrating \textit{\gls{ap}} in relation to \gls{api} cost ($k$=100) and time ($k$=100), respectively. When considering \textit{\gls{api} cost}, all text-based \textit{\gls{gpt}-4.1} and \textit{Gemini-2.5} models lie on the \textit{Pareto frontier}, and, additionally, \textit{\gls{gpt}-4.1} is the only image variant. This indicates that text inputs for \gls{mllm}-based \gls{gui} reranking offer a valuable effectiveness-cost trade-off. However, it should be noted that the employed gold standard focuses on functional \gls{nlr} (the primary objective for \gls{rel} and \gls{rval}). The gains could be larger in scenarios where non-functional \gls{nlr} (\gls{nfr}) (such as the \gls{gui} design) play a significant role, since the expressiveness of detailed \gls{gui} text annotations might be limited for these dimensions.

\vspace{-0.0cm}
\begin{myrqbox}
\textbf{Answer to RQ$_2$:} While image variants of \gls{mllm}-based \gls{gui} reranking outperform text variants in effectiveness, text variants are considerably more cost-efficient (e.g., \textit{\gls{gpt}-4.1 (Image)} improves mean \textit{\gls{ap}} over \textit{\gls{gpt}-4.1 (Text)} by 3.32\% while entailing 543.9\% higher \textit{\gls{api} costs} ($k$=100)). Moreover, all text-based \textit{\gls{gpt}-4.1} and \textit{Gemini-2.5} models provide \textit{Pareto}-efficient solutions for \gls{mllm}-based \gls{gui} reranking from \gls{nlr}.
\end{myrqbox}

\section{Threats to Validity}
\label{sec:chapter-rerank:threats}

\paragraph{Internal Validity.} Threats related to internal validity mainly lie in the employed \gls{nlr}-based \gls{gui} retrieval gold standard. Particularly, bias regarding the creation of queries, the pooling of potentially relevant \glspl{gui}, the subjectivity and inconsistency of relevance judgments are critical. However, we discussed mitigation strategies already (Section \ref{sec:gui-retrieval-threats}).

\paragraph{External Validity.} Threats related to external validity again mainly are represented by the employed \gls{nlr}-based \gls{gui} retrieval gold standard, including the generalizability of encompassed \gls{nlr} queries, the utilized \textit{Rico} \gls{gui} repository diversity and the population of crowdsourcing workers. Again, we discussed mitigation strategies before (Section \ref{sec:gui-retrieval-threats}).
\section{Limitations}
\label{sec:chapter-rerank:limitations}

While the presented \textit{\gls{gui}-ReRank} approach achieves high effectiveness across many evaluation metrics by utilizing \gls{mllm}-based \gls{gui} reranking, it also exhibits some limitations. First, even in low-temperature settings for non-reasoning \glspl{mllm}, the output might still encompass some variance and, therefore, reranking results might slightly vary between different runs. However, we manually evaluated multiple examples across runs and found few deviations in the \gls{gui} rankings. Second, employing \glspl{mllm} for \gls{gui} reranking entails high cost in terms of monetary \gls{api} cost, token consumption and time (latency). However, as we have shown in our experiments, text-based reranking offers high effectiveness with substantially reduced costs. Finally, while the multi-dimensional retrieval and reranking with \textit{\gls{gui}-ReRank} is customizable, manually tuning for appropriate weights requires some expertise and experience to fully leverage the available search capabilities.

\section{Related Work}
\label{sec:chapter-rerank:relwork}

As described earlier, many \gls{gui} retrieval approaches have been proposed in research before. For example, \textit{GUIFetch} \citep{behrang2018guifetch} and \textit{Swire} \citep{huang2019swire} both propose sketch-based \gls{gui} retrieval from large-scale \gls{gui} repositories by utilizing \gls{ml} approaches. Similarly, \textit{VINS} \citep{bunian2021vins} conducts wireframe-based \gls{gui} retrieval via neural image and \gls{gui} component embedding models. \textit{Gallery DC} \citep{chen2019gallery} devises a \gls{gui} component retrieval approach, with components automatically extracted from a large-scale \gls{gui} collection. However, prior research neglected to investigate multi-dimensional and multi-modal \gls{nlr}-based \gls{gui} retrieval and reranking with \glspl{mllm}. For a more detailed discussion of the related work on \gls{gui} retrieval, please refer to the previous chapter (see Section \ref{sec:related_gui_retrieval}).

Moreover, utilizing \glspl{llm} for reranking text documents and passages has been investigated in research before. For example, \textit{Pairwise Ranking Prompting (PRP)} \citep{qin2024large} represents a simplified \gls{llm}-based text reranking approach by comparing document pairs. \cite{mozafari2025good} conduct a comprehensive evaluation of text reranking approaches, showing that \gls{llm} reranking achieves higher effectiveness for familiar queries. Furthermore, \textit{RagVL} \citep{chen2024mllm} represents an instruction-tuned \gls{mllm}-based reranking approach for images, which enhances the image filtering and \gls{rag} capabilities. While prior research mainly focused on \gls{llm}-based reranking for text documents, \gls{mllm}-based reranking is increasingly explored in recent work. However, we were the first to propose and explore multi-dimensional \gls{mllm}-based \gls{gui} reranking.
\section{Conclusion}
\label{sec:chapter-rerank:conclusion}

The proposed approach in this chapter was driven by the challenge \challonetwo{}: \textit{How can the semantic representation gap between \gls{nlr} (including \gls{nfr}) and multi-dimensional \gls{gui} prototypes be reduced to enable more effective \gls{gui} ranking?} To tackle this challenge, we proposed \textit{\gls{gui}-ReRank}, a novel multi-dimensional embedding-based constrained \gls{gui} retrieval and \gls{mllm}-based \gls{gui} reranking approach from \gls{nlr}. In particular, the \gls{mllm}-based \gls{gui} repository annotation pipeline provides multi-dimensional textual representations of \glspl{gui} for enhanced retrieval. Moreover, by employing \glspl{mllm} with advanced \gls{nlu} and image understanding capabilities, the multi-dimensionality of \glspl{gui} can be more effectively reflected and evaluated for rank scoring. This stands in contrast to previous approaches, which relied solely on basic text fragments extracted from the \gls{gui} hierarchy data and thus neglected relevant information contained in the \gls{gui} images.


\chapter{Self-Elicitation of Requirements with Automated GUI Prototyping}
\chaptermark{Self-Elicitation with Automated GUI Prototyping}
\label{cha:self_elicitation}
The first chapter of the thesis introduced novel approaches for \gls{nlr}-based \gls{gui} ranking with a focus on \gls{bert}-\gls{ltr} models, followed by the second chapter, which introduced \gls{mllm}-based \gls{gui} reranking, achieving state-of-the-art performance. In this chapter, we continue with challenge \challone{}, namely, mapping \gls{nlr} to \gls{gui} prototypes. While the previous two approaches focused on providing \gls{gui} prototyping support for the requirements analyst or prototype developer via retrieval and reranking techniques, this approach emphasizes enabling stakeholders themselves to conduct \gls{rel} and \gls{rval} via automated \gls{gui} prototyping assistance, which integrates \gls{gui} and feature retrieval techniques. In particular, this work has been previously published \citep{kolthoff2024self}\footnote{This section is adapted from: \textbf{Kolthoff, Kristian}, Bartelt, Christian, Ponzetto, Simone Paolo, and Schneider, Kurt. Self-Elicitation of Requirements with Automated GUI Prototyping. In \emph{Proceedings of the 39th IEEE/ACM International Conference on Automated Software Engineering (ASE, A*)}, Sacramento, CA, USA, October 2024, pages 2354--2357. ACM. Sections are directly reused with only minor adaptations (i.e. thesis Section \ref{cha:self_elicitation}.i matches paper Section i)}. Our interactive prototype, code, evaluation results, and demonstration video are publicly available\footnote{Materials for this chapter are available at \url{https://github.com/kristiankolthoff/SERGUI-Prototyping} and the prototype demonstration video at \url{https://youtu.be/pzAAB9Uht80}}.
\vspace{-0.7cm}
\paragraph{Personal Contribution.} The core idea and concept for this work were proposed by me. Moreover, I implemented all the techniques and the prototype. I designed the evaluation concept, the user study, and conducted the data analysis. Finally, I wrote the manuscript. 
\vspace{-0.2cm}
\section{Motivation}

\vspace{-0.2cm}

The \gls{nlr}-based \gls{gui} retrieval and reranking techniques proposed in the previous two chapters enable rapid mapping of \gls{nlr} to relevant \gls{gui} prototypes. Moreover, these techniques were integrated into a rapid prototyping approach \textit{\gls{rawi}}, which provides capabilities for adapting the \gls{gui} prototype to particular requirements through a data-driven editor. This method substantially improves \gls{gui} prototyping productivity, as demonstrated through experimental evaluation, and focuses on facilitating the creation of \gls{gui} prototypes, while also assuming that users possess sufficient knowledge to conduct the \gls{gui} prototyping, elicitation and validation process. In particular, \textit{\gls{rawi}} provides neither guidance nor a structured approach for the overall \gls{gui} prototyping process, which restricts its direct applicability to stakeholder-driven \gls{rel} and \gls{rval}.

Prior research proposed various approaches aiming to simplify the \gls{gui} prototyping procedure. For instance, the \textit{Fast Feedback} technique \citep{schneider2007generating} reduces the number of necessary elicitation sessions with stakeholders by providing tool-based support to rapidly and interactively create pen-and-paper prototypes combined with use cases, directly allowing stakeholder feedback to be incorporated. However, an experienced analyst is required to conduct the initial elicitation. In the \gls{rel} approach proposed by \cite{teixeira2014requirements}, analysts create initial \gls{gui} prototypes based on an initial meeting with stakeholders, subsequently enabling the stakeholders to access and modify the \gls{gui} prototype through a web-based \gls{gui} prototyping tool, but the approach provides no automatic guidance for prototyping and elicitation. Similarly, the \textit{Graphical Requirements Collector} \citep{moore2000comparison} requires users to directly create their own \gls{gui} prototypes by constructing them bottom-up and augmenting them with textual requirements. This integrates stakeholders closely into the \gls{gui} prototyping process, but, due to the lack of guidance, requires them to have experience with the \gls{gui} prototyping process. Moreover, \textit{LadderBot} \citep{rietz2019ladderbot} enables automatic dialogue-based elicitation with stakeholders by applying \textit{laddering}, a structured interview technique. However, this automatic dialogue-based elicitation approach focuses solely on the collection of simple textual requirements and therefore cannot be applied to the automated initial elicitation with \gls{gui} prototyping. Hence, to summarize, the work presented in this chapter is based on the leading research question of challenge \challonethree{}: \textit{How can we provide automatic assistance for self-elicitation of requirements with \gls{gui} prototyping?}

\begin{figure*}
 \includegraphics[width=\textwidth]{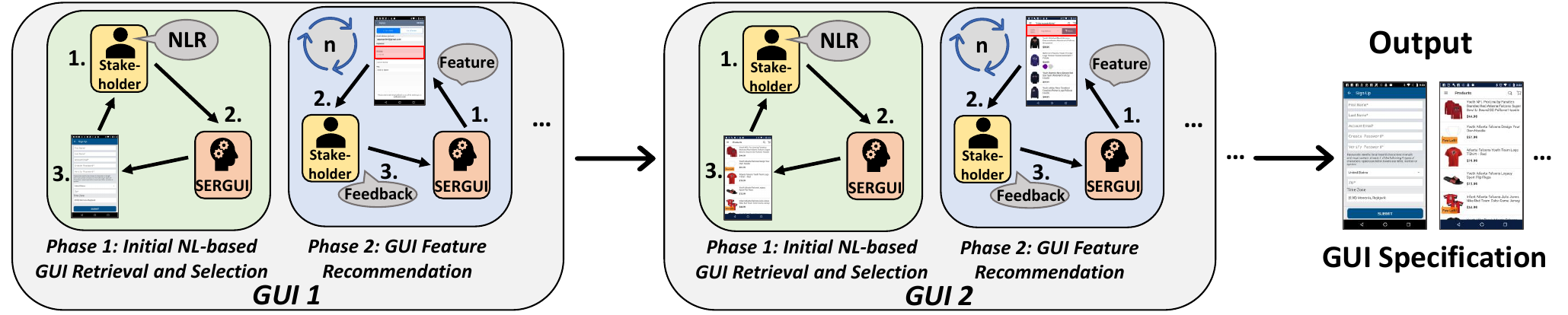}
  \caption[Overview of the interactive \gls{gui} prototyping procedure with \textit{\gls{ser}\gls{gui}}]{\gls{rel} via interactive \gls{gui} prototyping assisted by our \textit{\gls{ser}\gls{gui}} approach enabling to rapidly map \gls{nlr} into relevant \gls{gui} prototypes (\textit{Phase 1}) followed by proactively stimulating \gls{rel} and \gls{rval} via prompting-based \gls{gui} feature recommendations (\textit{Phase 2}).}
	\label{fig:re_process_overview}
\end{figure*}

To tackle this challenge, we propose \textit{SER\gls{gui}}, a novel \gls{gui} prototyping approach that enables stakeholders to perform \textit{\gls{ser}} for interactive software systems through automated \gls{nlr}-based \gls{gui} prototyping assistance. Furthermore, we provide a \gls{llm}-based contextualized \gls{gui} feature recommendation mechanism to promote the elicitation of requirements. The general notion of \textit{\gls{ser}} refers to guiding stakeholders to uncover their own requirements and has originally been introduced by \textit{LadderBot} \citep{rietz2019ladderbot}, thus reducing the effort of analysts in the initial \gls{rel} phase, thereby closely integrating stakeholders into the \gls{rel} process. By extending the notion of \textit{\gls{ser}} with automated \gls{gui} prototyping, we exploit the benefits of \gls{gui} prototypes as a requirements specification artifact, facilitating the obtainment of fast stakeholder feedback and achieving early clarification of requirements. \textit{\gls{ser}\gls{gui}} enables users to rapidly recombine and easily adapt existing \glspl{gui} from the large-scale \gls{gui} repository \textit{Rico} \citep{deka2017rico} to incrementally build their own \gls{gui} prototypes, as shown in Figure \ref{fig:re_process_overview}. In \textit{\gls{ser}\gls{gui}}, stakeholders are guided through the \gls{gui} prototyping process by a rule-based dialogue assistant, which provides relevant \gls{gui} screens and proactively recommends contextually relevant \gls{gui} features. Therefore, \textit{\gls{ser}\gls{gui}} enables stakeholders to autonomously build their \gls{gui} prototypes requiring little prior \gls{gui} prototyping process experience. Overall, \textit{\gls{ser}\gls{gui}} produces a \gls{gui} prototype output in the form of selected \glspl{gui} and additional \textit{aspect}-\glspl{gui}, which focus on and highlight particular \gls{gui} features (i.e. \textit{aspects}). The produced \gls{gui} prototypes are intended to serve as an initial requirements specification artifact for analysts.

To evaluate \textit{\gls{ser}\gls{gui}}, we conducted a controlled user study. In particular, we asked 12 participants to create several \gls{gui} prototypes with our approach and evaluated the effectiveness of the \gls{gui} feature recommendation mechanism, the matching of features in \glspl{gui} for visualization, the \gls{gui} reranking performance and overall usability of the \textit{\gls{ser}\gls{gui}} approach. The results of the evaluation indicate that our approach can effectively support inexperienced users in rapidly creating \gls{gui} prototype specifications and is capable of recommending relevant features in combination with high usability (\gls{sus}).

\vspace{-0.2cm}
\paragraph{Contributions.} With this approach, we make the following research contributions:
\begin{itemize}[leftmargin=6mm]
\setlength{\itemsep}{1pt}
    \item \textit{Novel \gls{gui} prototyping assistance approach:} we present \textit{\gls{ser}\gls{gui}}, the first approach for \glsreset{ser}\textit{\gls{ser}} with \glspl{gui} enabled by an automatic \gls{gui} prototyping assistant to support fast feedback, driven by a \gls{nlr}-based \gls{gui} retrieval and feature-based \gls{gui} reranking mechanism and the large-scale \gls{gui} repository \textit{Rico}.
    \item \textit{\gls{llm}-based \gls{gui} feature recommendation:} \gls{llm}-based contextualized \gls{gui} feature recommendation approach to proactively stimulate \gls{rel}, integrated with the \gls{gui} repository \textit{Rico} to facilitate the visualization of recommended \gls{gui} features for \gls{rval}.
    \vspace{-0.2cm}
\end{itemize}
\section{Approach: \gls{ser}\gls{gui}}
\glsreset{us}
This section delineates the \textit{\gls{ser}\gls{gui}} approach for enabling \gls{ser} based on automated \gls{gui} prototyping. The traditional \gls{rel} process with \gls{gui} prototyping typically consists of the following main phases: To begin with, an initial elicitation interview is conducted in a joint session between the stakeholder and the analyst, followed by initial requirements modeling (e.g., in the form of elementary \textit{use cases} or \textit{\glspl{us}}). Afterwards, an iterative process is conducted in the steps of \textit{(a)} requirements visualization in the form of \gls{gui} prototypes, \textit{(b)} gathering stakeholder feedback related to the tangible \gls{gui} prototype and \textit{(c)} the subsequent adaptation of the requirements and update of the respective \gls{gui} prototype. After several iterations, an adapted and validated \gls{gui} prototype is produced as output (see Chapter~\ref{cha:background}, Section~\ref{subsec:gui-prototyping-processes}, for \gls{gui} prototyping process model details). However, this procedure necessitates numerous synchronous meetings between stakeholders and requirements analysts, often with multiple weeks in between, potentially delaying the overall application development process \citep{schneider2007generating}.

\begin{figure*}
\includegraphics[width=\textwidth]{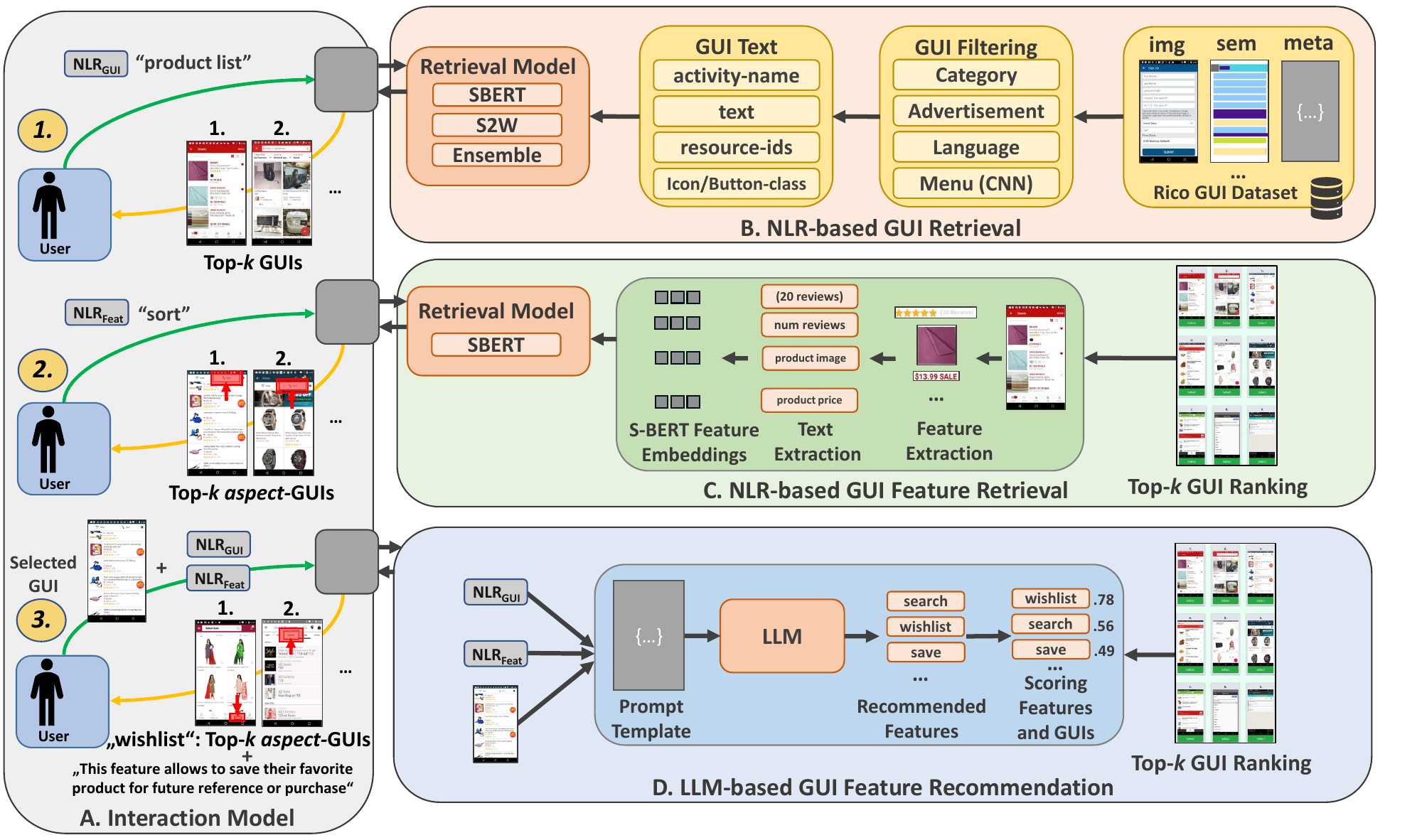}
  \caption[Overview of the \textit{\gls{ser}\gls{gui}} architecture]{Overview of the \textit{\gls{ser}\gls{gui}} approach with \textit{(A)} the interaction model showing the main interaction mechanisms, \textit{(B)} \gls{nlr}-based \gls{gui} retrieval approach (\textit{Rico}), \textit{(C)} the \gls{nlr}-based \gls{gui} feature retrieval and \textit{(D)} the \gls{llm}-based \gls{gui} feature recommendation.}
	\label{fig:overview-sergui}
\end{figure*}

\textit{\gls{ser}\gls{gui}} is an approach to automating the initial \gls{rel} phase by leveraging a comprehensive \gls{gui} repository in combination with a novel feature-based \gls{gui} reranking and \gls{gui} feature recommendation technique. Subsequently, an overview of our approach is given and depicted in Figure \ref{fig:overview-sergui}. \textit{\gls{ser}\gls{gui}} is divided into multiple components: First, \textit{(A)} an interaction model encompassing the essential interaction mechanisms between the user and the automatic \gls{gui} prototyping assistance. Second, \textit{(B)} a \gls{nlr}-based \gls{gui} retrieval technique building on previously presented work (see Chapter \ref{cha:nl_gui_retrieval}). Third, \textit{(C)} a \gls{nlr}-based \gls{gui} feature retrieval mechanism. Fourth, \textit{(D)} an \gls{llm}-based \gls{gui} feature recommendation mechanism to proactively suggest potentially relevant \gls{gui} features and simultaneously illustrate them by matching them to the \glspl{gui} from the top-\textit{k} ranking.

\subsection{Interaction Model}

\textit{\gls{ser}\gls{gui}} is built as an interactive dialogue-based approach providing guidance through the \gls{gui} prototyping process, facilitating the close integration of stakeholders. The interaction model encompasses three essential patterns that are repeated for each \gls{gui} in the application prototype. Initially, users can specify their \gls{nlr} for a particular \gls{gui} (denoted by \gls{nlr}$_{GUI}$), to which the approach responds with a top-\textit{k} \gls{gui} ranking, as illustrated in Figure \ref{fig:overview-sergui} \textit{A1}. This ranking represents the best matches as computed by the \gls{nlr}-based \gls{gui} retrieval model. The \gls{gui} ranking showcases many potentially relevant \glspl{gui} encompassing numerous variations, which already stimulates \gls{rel}. Users can reformulate their \gls{nlr} in case of an inadequate relevance of the ranked \gls{gui}.

Second, users are enabled to specify additional \gls{nlr} for individual features (denoted by \gls{nlr}$_{Feat}$), to which the approach responds with a top-$k$ ranking of \textit{aspect}-\glspl{gui}, as depicted in Figure \ref{fig:overview-sergui} \textit{A2}. An \textit{aspect}-\gls{gui} is a \gls{gui} containing one particular highlighted \gls{gui} feature (i.e. \textit{aspect}) that is considered relevant to the user. Therefore, the approach presents a ranking of matching \glspl{gui} potentially containing the formulated \gls{gui} feature. If the user finds a relevant \textit{aspect}-\gls{gui} matching their needs, then it can be selected. Subsequently, the \gls{gui} ranking is recomputed based on the new aspects (potentially improving the ranking) and the \textit{aspect}-\gls{gui} will be saved as part of the \gls{gui} prototype.

Third, when the user initially selects a \gls{gui} already satisfying many features plus the specified \textit{aspect}-\glspl{gui}, then the approach utilizes the \gls{nlr}$_{GUI}$, collection of \gls{nlr}$_{Feat}$ and the selected \gls{gui} to proactively recommend potentially relevant \gls{gui} features, as illustrated in Figure \ref{fig:overview-sergui} \textit{A3}. In an iterative fashion, the \textit{\gls{ser}\gls{gui}} approach will then present each \gls{gui} feature with a short textual explanation and the top-\textit{k} ranking of potentially matching \textit{aspect}-\glspl{gui} encompassing and visualizing the respective \gls{gui} feature. Similarly to before, a relevant \textit{aspect}-\gls{gui} can directly be selected, which will then be added to the overall \gls{gui} prototype specification and exploited for \gls{gui} reranking. Moreover, if the \gls{gui} feature is relevant but no matching \textit{aspect}-\gls{gui} could be retrieved, then the \gls{gui} feature will be added to the prototype specification in textual form. \textit{\gls{ser}\gls{gui}} allows users to walk through these three patterns and eventually, the overall \gls{gui} specification encompasses the finally selected \gls{gui}, potentially a collection of \textit{aspect}-\glspl{gui} and a collection of textual \gls{gui} requirements. This \gls{gui} specification is then added to a simple linear application prototype, allowing users to briefly skip through the entire \gls{gui} prototype.

\subsection{\gls{nlr}-Based \gls{gui} Retrieval}
\label{subsec:sergui_dataset}

\begin{figure*}
 \includegraphics[width=\textwidth]{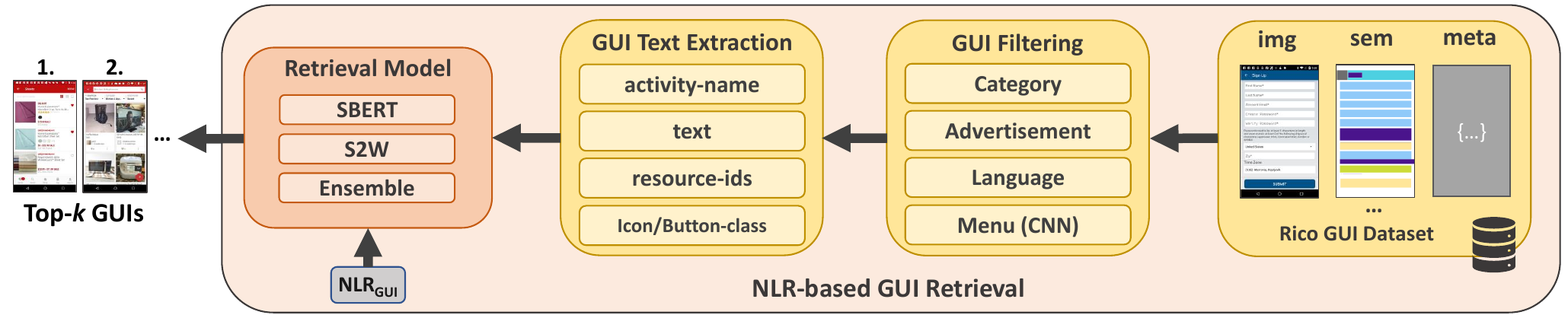}
  \caption[\gls{nlr}-based \gls{gui} retrieval approach]{\gls{nlr}-based \gls{gui} retrieval approach: \textit{Rico} \gls{gui} dataset is utilized as the foundation, multiple \gls{gui} filtering steps are conducted, followed by the text extraction from the \gls{gui} hierarchy data and the retrieval model computes a score to produce top-$k$ \glspl{gui}.}
	\label{fig:sergui_gui_retrieval_model}
\vspace{-0.3cm}
\end{figure*}

In order to achieve \gls{nlr}-based \gls{gui} retrieval, we mainly adopt the work presented in Chapter \ref{cha:nl_gui_retrieval} and provide an extension for filtering and retrieval models. As before, the \gls{gui} retrieval mechanism exploits the \gls{gui} repository \textit{Rico} \citep{deka2017rico}. For \textit{Rico}, the applied filtering steps and the extended \gls{gui} retrieval model, we provide a brief summary. Figure \ref{fig:sergui_gui_retrieval_model} depicts the overall procedure for the \gls{nlr}-based \gls{gui} retrieval component.

\paragraph{Rico \gls{gui} Dataset.} \textit{Rico} \citep{deka2017rico} represents the most sophisticated \gls{gui} dataset for mobile applications currently available, compared to other research-based \gls{gui} datasets such as \textit{ReDraw} \citep{moran2018machine} and \textit{ERICA} \citep{deka2016erica}. It is \textit{(i)} the largest semi-automatically harvested \gls{gui} dataset (9,772 unique \textit{Android} applications from 27 categories encompassing 72,219 \glspl{gui}), \textit{(ii)} incorporates a wide range of heterogeneous application domains and \textit{(iii)} provides multiple artifacts including \gls{gui} screenshots, \gls{xml}-based \gls{gui} hierarchy and metadata. For \gls{nlr}-based \gls{gui} ranking, we exploit the \gls{gui} hierarchy, extract and preprocess multiple text segments including the \textit{activity name}, for each \gls{gui} component the \textit{displayed text}, the \textit{resource-id} as an internal name given by the developers and special semantic attributes included in the \textit{Rico} \gls{gui} dataset such as \textit{semantic button} and \textit{icon categories}.
\vspace{-0.2cm}
\paragraph{Dataset Filtering.} To improve the overall quality of the \gls{gui} repository, we applied a filtering pipeline adopted from Chapter \ref{cha:nl_gui_retrieval}. This pipeline includes removal of \glspl{gui} from the \textit{entertainment (game)} category, \glspl{gui} with \textit{advertisement overlays} based on a heuristic and \textit{non-English} \glspl{gui} utilizing a language detection framework \citep{shuyo2010language}. In addition, manual inspection of \gls{gui} rankings indicated that \textit{Rico} contains a large portion of \glspl{gui} showing an opened side menu, providing no benefit for elicitation and additionally decreasing \gls{gui} retrieval effectiveness. To automatically remove these \glspl{gui}, we manually constructed a dataset consisting of \glspl{gui} with and without opened menus. Afterwards, we trained a \textit{\gls{cnn}} classifier \citep{gu2018recent} (three \textit{convolution} and \textit{pooling} layers) and provided as input a combination of both the original \gls{gui} screenshot image and the semantic annotation image as grayscale variants. We trained the \textit{\gls{cnn}} model for 6 epochs (\textit{adagrad} optimizer and \textit{binary cross-entropy} as loss) and achieved a satisfying performance on a separate test set (\textit{Precision}=.9818 / \textit{Recall}=.7012). We focused on achieving high precision to avoid removal of adequate \glspl{gui}. We filtered 23,817 \glspl{gui} (9,363 by the \textit{\gls{cnn}}) resulting in 48,402 \glspl{gui} remaining.

\vspace{-0.2cm}
\paragraph{Adapted \gls{gui} Retrieval Model.} As a \gls{gui} retrieval approach, we adopt the strong pretrained embedding-based \gls{sbert} model presented in Chapter \ref{cha:nl_gui_retrieval}\footnote{Note that the earlier presented \gls{bert}-\gls{ltr} models are primarily reranking approaches, which cannot be utilized for initial \gls{gui} retrieval due to computational overhead and latency, whereas \gls{sbert} is suitable. In addition, \textit{\gls{gui}-ReRank} was published after this work, but was presented earlier for coherence with \textit{\gls{rawi}}.}. This model computes the ranking using \textit{cosine similarity} between the embedded query $\emb(NLR_{GUI}) = f_{\text{emb}}(NLR_{GUI})$ and embedded text representation of the \gls{gui} $\emb(G_{i}) = f_{\text{emb}}(G_{i})$ as

\begin{equation}
\textbf{S$_{1}$}(NLR_{GUI}, G_i) = \cosim(\emb(NLR_{GUI}), \emb(G_i))
\end{equation}

\noindent with $\emb$ referring to the respective \gls{sbert} embedding. Moreover, we extend this \gls{gui} ranking score by incorporating the \textit{\gls{s2w}} dataset \citep{wang2021screen2words}, adding another representation to the \gls{gui} prototypes to score against. \textit{\gls{s2w}} represents a large collection of manually crafted high-level \gls{nl} descriptions of \textit{Rico} \glspl{gui}, providing five descriptions per \gls{gui} from different annotators. To effectively utilize the \textit{\gls{s2w}} data for \gls{gui} $G_i$ as $S2W(G_i)=\{d_{i,1},..., d_{i,5}\}$, we similarly compute the average over the \textit{cosine similarity} between the query embedding $\emb(NLR_{GUI})$ and the embeddings of all five descriptions $d_{i,j}$, namely, $\emb(d_{i,j}) = f_{\text{emb}}(d_{i,j})$ as
\vspace{-0.2cm}
\begin{equation}
\textbf{S$_{2}$}(NLR_{GUI}, G_i) = \frac{1}{5}\sum_{j=1}^{5}\cosim(\emb(NLR_{GUI}), \emb(d_{i,j}))
\end{equation}

\noindent By incorporating the average scores across five high-level descriptions from different annotators, we obtain a more robust retrieval score. Subsequently, we compute an ensemble between the two \gls{gui} scores to combine them as follows
\vspace{-0.1cm}
   \begin{equation}
        \textbf{S}(NLR_{GUI}, G_i) = \alpha \textbf{S$_{1}$}(NLR_{GUI}, G_i) + (1 - \alpha) \textbf{S$_{2}$}(NLR_{GUI}, G_i)
    \end{equation}

\noindent with $\alpha \in [0,1]$. By creating an ensemble \gls{gui} ranking model using both the extracted \gls{gui} text representations and the high-level descriptions of \textit{\gls{s2w}}, we facilitate obtaining a more flexible and robust \gls{gui} ranking in contrast to utilizing a single representation only. 

\vspace{-0.1cm}
\subsection{\gls{nlr}-based \gls{gui} Feature Retrieval}

\begin{figure}[!t]
 \includegraphics[width=\textwidth]{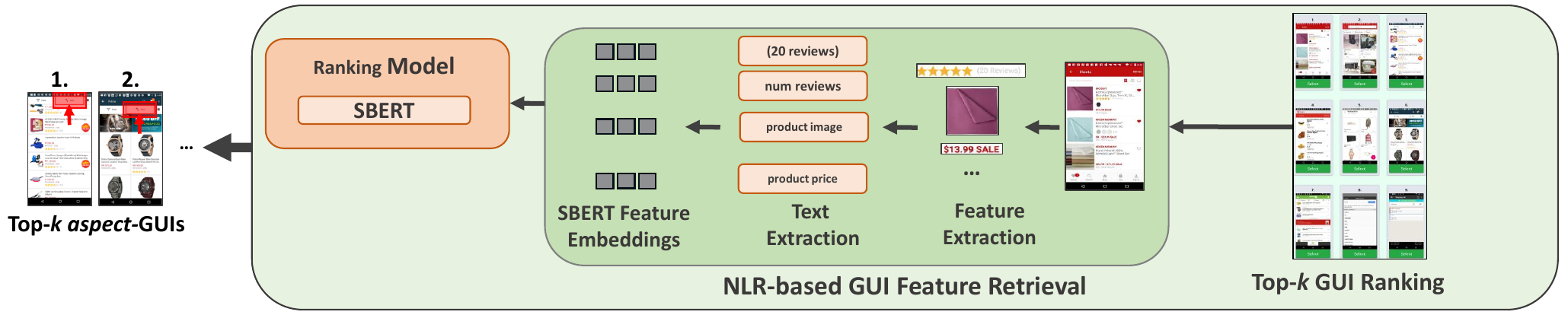}
  \caption[\gls{nlr}-based \gls{gui} feature retrieval approach]{\gls{nlr}-based \gls{gui} feature retrieval approach: starting from the top-$k$ \glspl{gui}, individual \gls{gui} features are extracted (\gls{gui} components), followed by extracting text representations and then computing \gls{sbert} embeddings to produce feature rankings. Each of the features is represented as part of an \textit{aspect}-\gls{gui}, highlighting its appearance.}
	\label{fig:sergui-feature-retrieval}
\end{figure}

To achieve \gls{nlr}-based \gls{gui} feature retrieval, we employ the top-\textit{k} \glspl{gui} representing potentially relevant context for matching the features, as depicted in Figure \ref{fig:sergui-feature-retrieval}. Here, we restrict \gls{gui} features to individual \gls{gui} components. From the top-\textit{k} \glspl{gui}, textual representations of the \gls{gui} components are extracted, including multiple texts for each component (such as \textit{displayed text}, \textit{resource-id} and \textit{semantic classes}). The text representation of the $j$-th feature of \gls{gui} $G_i$ is subsequently embedded with \gls{sbert} as $\emb(G_{i,j}) = f_{\text{emb}}(G_{i,j})$ to compute the score against the corresponding $NLR_{Feat}$ \gls{sbert} embedding $\emb(NLR_{Feat}) = f_{\text{emb}}(NLR_{Feat})$ based on the \textit{cosine similarity} as

\begin{equation}
\textbf{S$_{F}$}(NLR_{Feat}, G_{i,j}) = \cosim(\emb(NLR_{Feat}), \emb(G_{i,j}))
\end{equation}

\subsection{\gls{llm}-based \gls{gui} Feature Recommendation}
\label{sec:sergui-feature-rec}

\begin{figure*}[!t]
 \includegraphics[width=\textwidth]{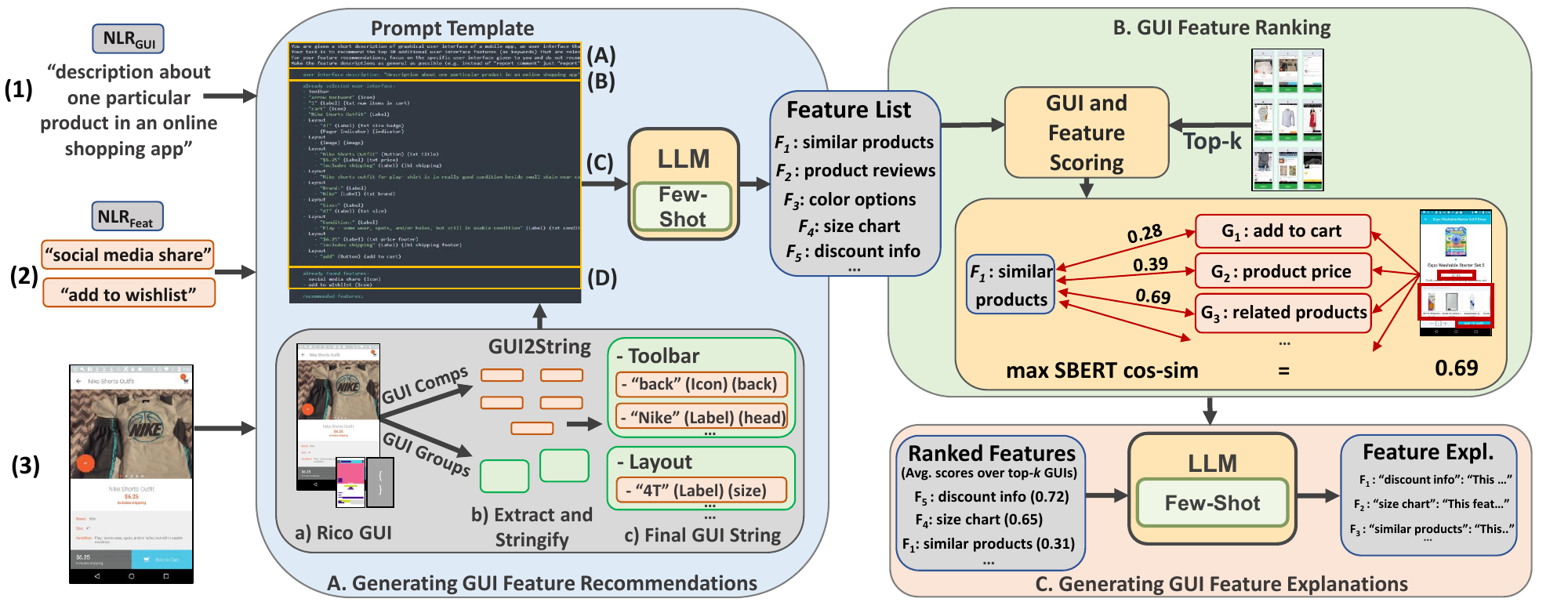}
  \caption[\gls{llm}-based \gls{gui} feature recommendation approach]{\gls{gui} feature recommendation approach overview of \textit{\gls{ser}\gls{gui}} with \textit{(A)} generating \gls{llm}-based \gls{gui} feature recommendations via \gls{fs} prompting, \textit{(B)} scoring recommended features and top-$k$ \glspl{gui}, and \textit{(C)} generating \gls{llm}-based feature explanations.}
	\label{fig:llm-featrec}
\end{figure*}

\glspl{llm} recently gained popularity due to their ability to rapidly learn and adapt to new tasks solely based on \gls{fs} examples \citep{brown2020language, kojima2022large}. Particularly, these models can essentially be adapted to a plethora of specific tasks through a technique called \textit{prompting} \citep{liu2023pre}. In prompting, solely a textual instruction for the task and several examples of input and expected output are provided to the \gls{llm} (see Chapter \ref{cha:background} Section \ref{chapter-2:sec-pe}). Due to the vast amount of knowledge (including technical and domain knowledge) embodied in the \glspl{llm} and its accessibility via prompting, exploiting these models within \textit{\gls{ser}\gls{gui}} holds large potential to facilitate automatic elicitation. Subsequently, we introduce our novel approach for \gls{llm}-based \gls{gui} feature recommendation. An overview of the approach architecture is illustrated in Figure \ref{fig:llm-featrec}.

\paragraph{Generating \gls{gui} Feature Recommendations.} Within the \textit{\gls{ser}\gls{gui}} approach, the \gls{gui} feature recommendations are based on the context of \textit{(1)} the initial textual requirements for the \gls{gui} denoted by $NLR_{GUI}$, \textit{(2)} a collection of already textually specified features by the user as $NLR_{Feat}$ and \textit{(3)} an initially selected \gls{gui} from the top-\textit{k} \gls{gui} ranking. With these contextual inputs, we fill in a \textit{prompt template} encompassing the following sections. First, \textit{(A)} we provide the \textit{task instructions} asking the model to recommend the top-\textit{30} \gls{gui} features given the described context. We emphasize that the model should thoroughly examine the provided \gls{gui} and features to avoid recommending already included features. Moreover, we instruct the model to focus on recommending features for the particular \gls{gui} at hand instead of more general application-level features (e.g., \textit{help}, \textit{user support}). Next, \textit{(B)} the template displays the initial requirements as $NLR_{GUI}$. Third, \textit{(C)} the initially selected \gls{gui} is provided to the model\footnote{Note that at the time of implementing the approach, more powerful \glspl{mllm}, which have highly effective image understanding capabilities, were not readily available yet (such as \textit{\gls{gpt}-4o} \citep{openai_gpt4o_docs}).}. Since the original \gls{xml}-based \gls{gui} hierarchy contains a large amount of information, often spanning several thousand lines, transforming the \gls{gui} hierarchy into a more abstract and focused representation is necessary. This both helps to reduce consumed context length in the \gls{llm} and enables the model to focus on the most important aspects. To this end, we first produce a string representation for each \gls{gui} component in the following abstract format, accompanied by three concrete examples (\textit{Label}, \textit{Button} and \textit{Text Input}):

\begin{center}
\textit{"uicomp-text"} \textbf{(uicomp-type)} \texttt{(uicomp-name)} \\
\textit{"+7.10"} \textbf{(Label)} \texttt{(price Change TV)}\\
\textit{"Install App"} \textbf{(Button)} \texttt{(native Ad Call To)} \\
\textit{"Example: 'New York'"} \textbf{(Text Input)} \texttt{(location)}
\end{center}
\begin{sloppypar}
\noindent Additionally, we extract grouping components from the semantic annotations of \textit{Rico} \citep{deka2017rico} including types such as \textit{List Item}, \textit{Card} and \textit{Toolbar}, among others. To also find groups for remaining components not grouped by the previous groups, we further extracted layout groups from the original \gls{gui} hierarchy by matching them to the \gls{gui} components. Finally, we construct the \gls{gui} representation as two-level bullet points, the outer level being the layout groups and the inner level being their respective \gls{gui} components. Before producing the string representation, the layout groups are sorted based on their bounds from top-left to bottom-right and similarly the \gls{gui} components within each group. Fourth, \textit{(D)} the already specified $NLR_{Feat}$ is added to the template as additional bullet points. Following the \gls{fs} technique for \gls{icl}, we provided several recommendation examples as additional context for each \gls{gui} feature recommendation.
\end{sloppypar}

\paragraph{\gls{gui} Feature Ranking.} To compute a \gls{gui} feature ranking, each of the recommended features is matched against the features in each \gls{gui} in the top-$k$ \gls{gui} ranking. To obtain a confidence score that a \gls{gui} $G_i$ contains a predicted feature ($NLR_{Feat}$) and to determine which feature within the \gls{gui} matches best (as \textit{aspect}-\gls{gui}), we compute

\begin{equation}
\mathbf{S}_{g}(NLR_{Feat}, G_i) = \max\limits_{j} \mathbf{S}_{F}(NLR_{Feat}, G_{i,j})
\end{equation}

\noindent where $G_{i,j}$ refers to feature $j$ of \gls{gui} $G_i$ and $\mathbf{S}_{F}$ uses \textit{cosine similarity}, as illustrated in Figure \ref{fig:llm-featrec}. Additionally, for each recommended feature, we compute a feature score as

\begin{equation}
\mathbf{S}_{pf}(NLR_{Feat}) = \frac{1}{k} \sum_{i=1}^{k} \mathbf{S}_{g}(NLR_{Feat}, G_{i})
\end{equation}

\noindent over the top-$k$ \gls{gui} ranking to estimate the feature coverage among top-$k$ \glspl{gui}. Predicted features are then recommended, sorted based on $\textbf{S$_{pf}$}$ and shown as the top-$k$ \textit{aspect}-\glspl{gui}.

\paragraph{Generating \gls{gui} Feature Explanations.} To convey the meaning of a predicted \gls{gui} feature, we already present users with the top-$k$ \textit{aspect}-\gls{gui} ranking as a visualization (as a form of validation), sorted by the score $\textbf{S$_{g}$}$. Additionally, we employ a second \gls{fs} prompt and instruct an \gls{llm} to generate a short textual explanation for each \gls{gui} feature, as illustrated in Figure \ref{fig:llm-featrec}. This explanation is then utilized in the dialogue with the user.

\subsection{Feature-based \gls{gui} Reranking}

\begin{figure}
  \includegraphics[width=\textwidth]{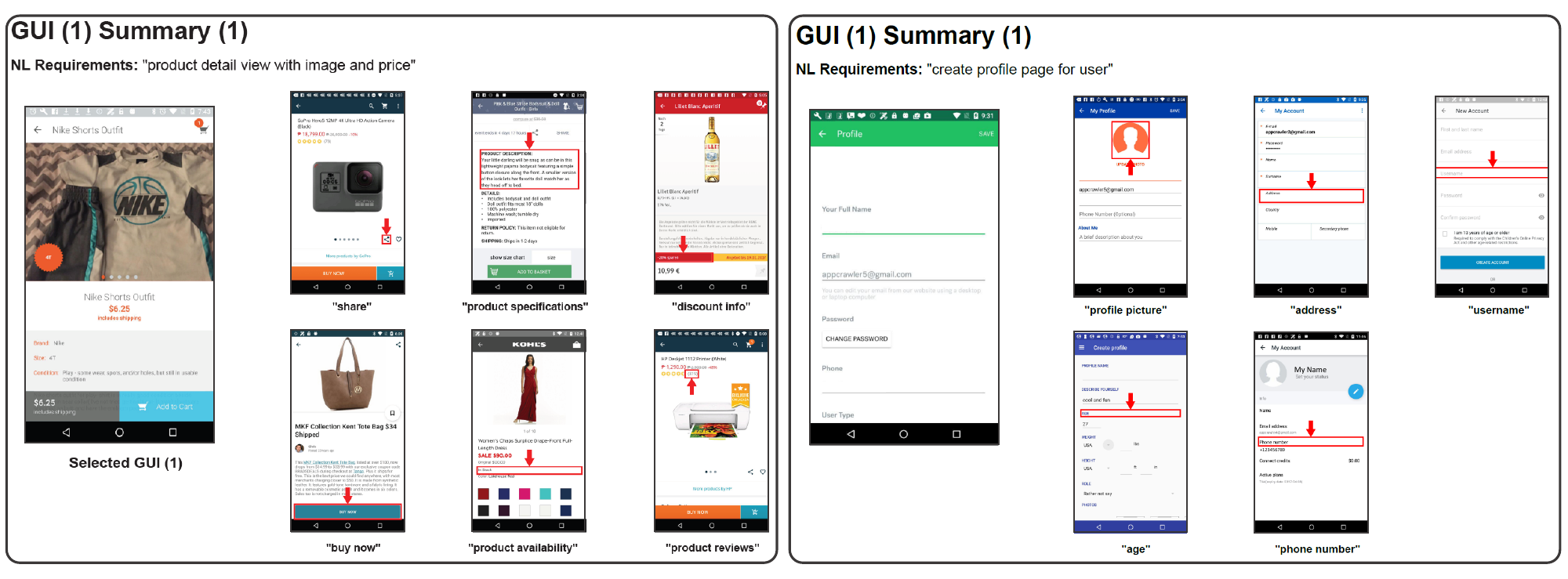}
  \caption[Example \gls{gui} summary outputs from \textit{\gls{ser}\gls{gui}}]{Example output of \textit{\gls{ser}\gls{gui}} showing two \gls{gui} summaries each with the selected \gls{gui} and several \textit{aspect}-\glspl{gui} for a \textit{product detail view} and a \textit{create profile page}.}
	\label{fig:app_summary}
\end{figure}

Based on positive feedback from the user by selecting an \textit{aspect}-\gls{gui} from the top-\textit{k} \textit{aspect}-\gls{gui} ranking, we can compute a reranking of the \glspl{gui}. Users often explore no more than the top-\textit{20} or top-\textit{30} \glspl{gui} in the ranking, missing potentially relevant \gls{gui} features. By reranking the top-$k$ \glspl{gui} based on additionally selected \gls{gui} features, users potentially find more relevant \glspl{gui} at the top of the \gls{gui} ranking or they can keep their original choice. We compute the reranking $\textbf{S$_{RR}$}$ of \gls{gui} $G_i$ with $\beta \in [0,1]$ as follows

   \begin{equation}
        \textbf{S$_{RR}$}(G_i) = \beta \textbf{S}(NLR_{GUI}, G_i) + (1 - \beta) \frac{1}{|F|}\sum_{NLR_{Feat} \in F}{\textbf{S$_{g}$}(NLR_{Feat}, G_i)}
    \end{equation}

\noindent The reranking score $\textbf{S$_{RR}$}(G_i)$ is an ensemble between the initial \gls{gui} ranking based on $NLR_{GUI}$ and the normalized sum over all feature scores $\textbf{S$_{g}$}(NLR_{Feat}, G_i)$. Only feature requirements $NLR_{Feat}$ are added to the set $F$ if the user selected one \gls{gui} from the top-$k$ \textit{aspect}-\gls{gui} ranking. Next, the original $\textbf{S$_{g}$}(NLR_{Feat}, G_i)$ scores are adapted as follows: The score of the selected \gls{gui} will be set to 1, scores of \glspl{gui} ranked higher than the selected \gls{gui} will be set to 0 and the scores of \glspl{gui} ranked lower are unchanged.

\subsection{\gls{gui} Prototype Output}

Since \textit{\gls{ser}\gls{gui}} is based on fixed \textit{Rico} \glspl{gui}, users potentially are not able to find a \gls{gui} perfectly fitting their requirements. Hence, we introduced the idea of \textit{aspect}-\glspl{gui}, enabling users to choose a single already well-matching \gls{gui} (supported by feature recommendations and reranking) plus a collection of \textit{aspect}-\glspl{gui}, which contain only a single aspect relevant for the user. At the end of the \gls{gui} prototyping process, \textit{\gls{ser}\gls{gui}} produces an application summary including \textit{(i)} a visualization of the overall linear application wireframe and \textit{(ii)} for each \gls{gui} the main selected \gls{gui}, the collection of \textit{aspect}-\glspl{gui} as illustrated in Figure \ref{fig:app_summary} and an additional collection of textual requirements, representing recommended features that were relevant but no relevant \textit{aspect}-\gls{gui} was found. This \gls{gui} prototype can then be employed by analysts as a starting point for further elicitation.

\subsection{Prototype Implementation}

\begin{figure}
  \includegraphics[width=\textwidth]{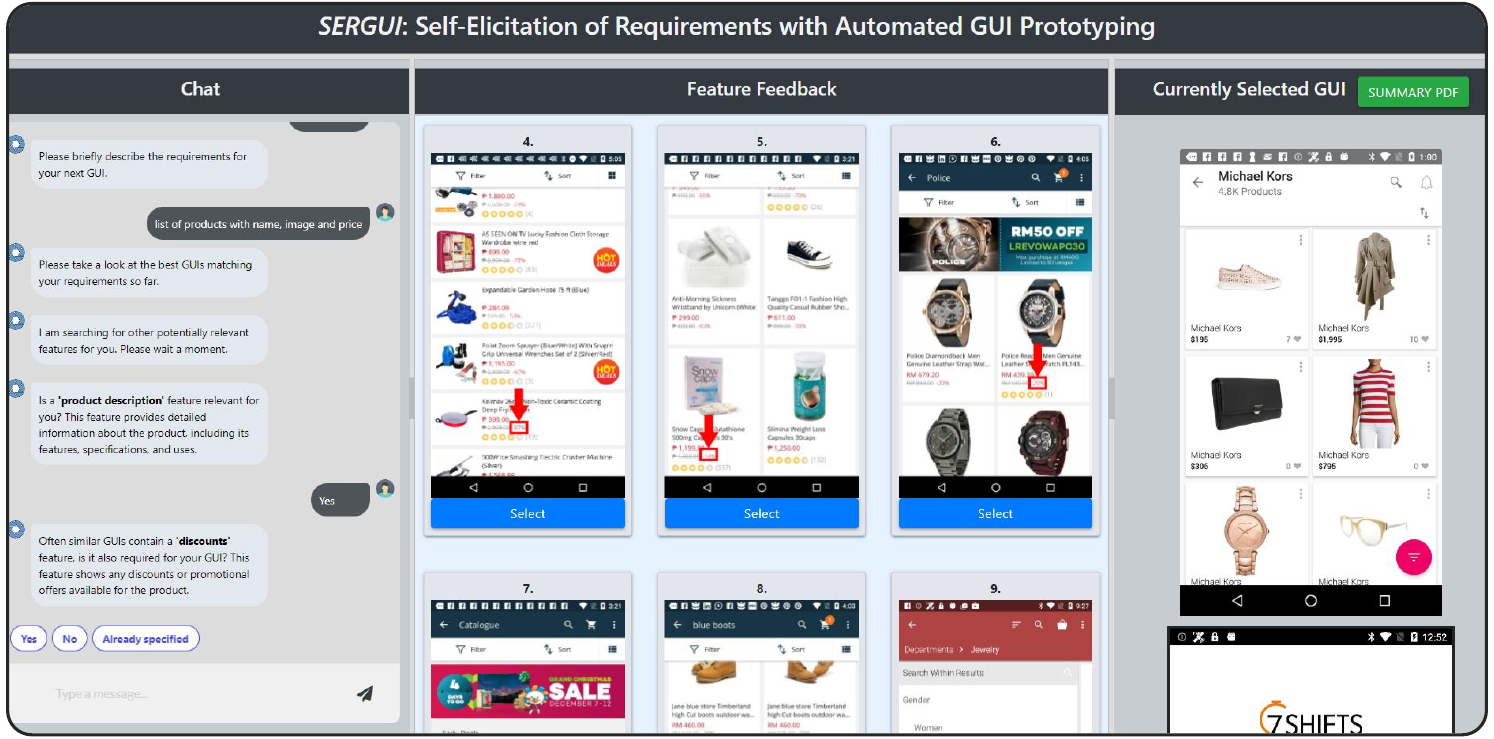}
  \caption[Web-based \textit{\gls{ser}\gls{gui}} tool prototype]{Web-based prototype implementation of the approach with a \textit{ chat-section}, \textit{workbench} and \gls{gui} \textit{prototype preview}, currently showing the top-$k$ \textit{aspect}-\gls{gui} ranking.}
	\label{fig:tool-sergui}
\end{figure}

We implemented the prototype of the presented \textit{\gls{ser}\gls{gui}} approach as a web application using \gls{html}, \gls{css} and \gls{js}. The application consists of a \textit{chat-section}, a \textit{workbench} for showing the top-\textit{k} \glspl{gui} or \textit{aspect}-\glspl{gui}, respectively. The implementation provides a \gls{gui} \textit{prototype preview}. Figure \ref{fig:tool-sergui} illustrates \textit{\gls{ser}\gls{gui}} with a top-$k$ \textit{aspect}-\gls{gui} ranking.

\section{Experimental Evaluation}

In this section, we delineate the design of our experiments to evaluate \textit{\gls{ser}\gls{gui}}. The main goal of our experimental evaluation is to assess the overall capabilities of the approach to stimulate \gls{rel} and facilitate \gls{ser}. This is divided into \textit{(i)} assessing the relevance of the \gls{gui} feature recommendations as a means to stimulate the automatic elicitation of interesting features with users, \textit{(ii)} measuring the relevance of the ranked \textit{aspect}-\glspl{gui} on the basis of \gls{nlr}, \textit{(iii)} assessing the performance of the feature-based \gls{gui} reranking and \textit{(iv)} measuring the overall perceived usefulness and gaining additional user insights. The first two points evaluate our second contribution, while the last two points evaluate our first contribution. To this end, we formulate the following four research questions:

\begin{itemize}
    \item \textbf{RQ$_{1}$}: \textit{How relevant are the contextualized \gls{llm}-based \gls{gui} feature recommendations?} To answer this question, we conducted a user study where participants employed the approach to create several prototypes and provided relevance annotations for the recommended \gls{gui} features. Given these relevance annotations, we computed several standard binary \gls{ir} metrics including \textit{\gls{ap}}, \textit{\gls{mrr}}, and \textit{P@k}.
    \item \textbf{RQ$_{2}$}: \textit{How relevant are the visualizations of the matched \gls{gui} features (i.e. \textit{aspect}-\glspl{gui})?} To answer this question, we collected binary relevance annotations during the user study and computed standard \gls{ir} metrics, including \textit{\gls{mrr}} and \textit{HITS@k}.
    \item \textbf{RQ$_{3}$}: \textit{Can users find more relevant \glspl{gui} at the top ranks through the feature-based \gls{gui} reranking?} To answer this question, we compared the ranks of the selected \gls{gui} prototype before and after the reranking based on the relevance annotations of recommended features and report how often users found better matches.
    \item \textbf{RQ$_{4}$}: \textit{Do users perceive \gls{ser}\gls{gui} as useful?} To answer this question, we asked participants of the user study several questions on a five-point Likert scale questionnaire regarding the perceived usefulness of the approach and report the \textit{\gls{sus}}.
\end{itemize}

\subsection{User Study}

\paragraph{User Study Design.} To evaluate our approach, we conducted a user study and asked participants to employ \textit{\gls{ser}\gls{gui}} to create several \gls{gui} prototypes. In particular, we recruited 12 participants possessing technical backgrounds (BSc.:7/MSc.:5). In addition, the recruited participants had medium to high experience in software development (Mean:3.50/SD:0.90) and mostly low experience in \gls{gui} prototyping (Mean:2.16/SD:0.83). These demographics were collected as self-reported by the participants on a five-point Likert scale. Moreover, a large fraction of participants (75\%) reported that they had never utilized a \gls{gui} prototyping tool. The remainder of participants reported that they previously had used two different \gls{gui} prototyping tools, either \textit{MockPlus} or \textit{Wireframe}. In particular, we recruited participants with low experience in \gls{gui} prototyping, matching the characteristics of the envisioned target audience of our \gls{gui} prototyping approach \textit{\gls{ser}\gls{gui}}. Furthermore, the user study encompassed three \gls{gui} prototyping tasks and each participant worked on two of them. Each combination of tasks was employed and the occurrence was distributed evenly and randomly among the participants. In summary, the 12 participants created 72 \glspl{gui} distributed among 24 applications, which provided the foundation for evaluating the proposed \gls{ser}\gls{gui} approach and answering the RQs.

\paragraph{User Study Tasks.} The user study consisted of three applications from diverse domains (\textit{shopping}, \textit{news} and \textit{social}), each encompassing three \glspl{gui}, resulting in nine different \glspl{gui} overall. Each application was presented to the participants as a low-fidelity \gls{gui} prototype, containing solely a minimal collection of features to enable participants to recognize the notion of the \gls{gui} and overall application. On this basis, participants were asked to create the respective \gls{gui} prototypes utilizing \textit{\gls{ser}\gls{gui}} and extending them with typical and meaningful additional features not contained in the rudimentary mockup provided to them. The first task \textit{Task-1} is a \textit{shopping} app encompassing the following \glspl{gui}: \textbf{(1-1)} an overview of products, each containing an image, name and price as features, \textbf{(1-2)} a subsequent product detail view with image, name, price and description and \textbf{(1-3)} a checkout page with one item (image, name and price), total cost and a confirmation button. Second, \textit{Task-2} is a \textit{news} app consisting of \textbf{(2-1)} a list of news, each with image and short heading, \textbf{(2-2)} a single news article with image, heading and description and \textbf{(2-3)} a comment section for the news article, each comment consisting of a user image, name and the text content of the comment. Third, \textit{Task-3} is a \textit{social} app containing more general \glspl{gui}, in particular, \textbf{(3-1)} a user registration form with name, gender and birthday inputs accompanied by a submit button, \textbf{(3-2)} a user profile overview with image, name and age and \textbf{(3-3)} a collection of contacts, each displaying a contact name and number.

\paragraph{Experimental Procedure.} Participants were first asked to read a user study instructions document, explaining their role and providing an example prototyping task and solution. Afterwards, a brief demonstration of \textit{\gls{ser}\gls{gui}} was given (always same for each participant), which was followed by the participants completing their first and second task. After each \gls{gui} prototype was completed, a summary (in the form of a Portable
Document Format (PDF)) produced by the tool was saved. Simultaneously, the interaction and feedback data provided by participants while completing their prototyping tasks were saved to a database. The user study instructions, tasks, created \gls{gui} prototype summaries and datasets are all publicly available at our accompanying \textit{GitHub} repository.

\subsection{RQ$_{1}$: Feature Recommendation Relevance}

To answer RQ$_{1}$, we evaluated the ability of \textit{\gls{ser}\gls{gui}} to recommend meaningful features. On the basis of the initial $NLR_{GUI}$, already specified features $NLR_{Feat}$ and the initially selected \gls{gui}, we recommended the top-\textit{10} features as predicted by our approach to the participants and the respective top-\textit{15} ranked \textit{aspect}-\glspl{gui}. In \textit{\gls{ser}\gls{gui}}, participants can provide positive feedback on a recommended \gls{gui} feature either by selecting an \textit{aspect}-\gls{gui} or by answering with \textit{"yes"}. Negative feedback could be provided either by selecting \textit{"no"} or \textit{"already specified"}, meaning that this \gls{gui} feature was already included either in the selected \gls{gui}, $NLR_{Feat}$ or in earlier recommendations. We transformed the iterative feedback into a relevance array and computed \textit{\gls{ap}}, \textit{\gls{mrr}} and \textit{P@k} (see Chapter \ref{cha:nl_gui_retrieval} Section \ref{rawi:rq1-setup}). To conduct the experiments, we used the most recent \textit{\gls{gpt}-4} model\footnote{Note that at the time of implementing the approach, \textit{\gls{gpt}-4} was recently published and achieved state-of-the-art results across many tasks \citep{openai2023gpt4}. Therefore, we decided to employ this model.} \citep{openai2023gpt4} (8,192 token context length, \textit{temperature}=0, \textit{top\_p}=1 (enabling nucleus sampling), accessed in July, 2023) as the \gls{llm} and employed the top-\textit{300} \glspl{gui} from the initial \gls{gui} ranking to compute the \gls{gui} feature ranking. Finally, we used the most recent and largest available \gls{sbert} model (768-dimensional embeddings) \citep{reimers2019sentence}.

\subsection{RQ$_{2}$: \gls{gui} Feature Retrieval Relevance}

To answer RQ$_{2}$, we evaluated the ability of \textit{\gls{ser}\gls{gui}} to retrieve \textit{aspect}-\glspl{gui} based on textual requirements of the features. Basically, we employed the identical setup as described for RQ$_{1}$ (top-\textit{15} \textit{aspect}-\glspl{gui}) and for each relevant feature (either answered with \textit{yes} or by picking an \textit{aspect}-\gls{gui}), we employed the rank of the selected \textit{aspect}-\gls{gui} to compute respective ranking metrics. If none was picked, it counted zero towards the metric. Since we only have at most one relevant \textit{aspect}-\gls{gui}, we computed the \textit{\gls{mrr}} and $HITS@k = 1 \text{ if } \vert R_{k} \vert > 0, \text{ and } 0 \text{ otherwise}$, with $R_{k}$ representing the set of relevant \textit{aspect}-\glspl{gui} positioned up to rank $k$. We computed these metrics for both the \gls{gui} features actively searched by participants and the \gls{llm}-based recommended features.

\subsection{RQ$_{3}$: \gls{gui} Reranking Performance}

To answer RQ$_{3}$, we evaluated the reranking performance of the proposed feature-based \gls{gui} reranking mechanism on the basis of the feature feedback provided by the participants. After completing the feedback on \gls{gui} feature recommendations, participants were shown an updated \gls{gui} ranking. Participants could either keep their original choice or, in case of a better matching \gls{gui} including more of the features marked as relevant, they could also select a new main \gls{gui}. In cases where participants picked a better matching \gls{gui}, we computed the reranking metric for the picked \gls{gui} as follows: 

\begin{equation}
    RR(G_i) = R_{initial}(G_i) - R_{updated}(G_i)
\end{equation}

\noindent with $R(G_i)$ being the rank of the \gls{gui} $G_i$. Since participants can only choose a single \gls{gui} in \textit{\gls{ser}\gls{gui}}, we cannot evaluate the entire top-\textit{k} reranking performance. Thus, we focus on the single selected \gls{gui}. For the initial \gls{gui} ranking score $\textbf{S}(G_i)$ based on the \gls{sbert} and \textit{\gls{s2w}} ensemble, we employed the average ($\alpha=0.5$). Similarly, for the \gls{gui} reranking score $\textbf{S$_{RR}$}(G_i)$, we also computed the corresponding score average ($\beta=0.5$).

\subsection{RQ$_{4}$: Perceived Usefulness}
\glsreset{sus}
To answer RQ$_{4}$, we evaluated the overall perceived usefulness of our approach. On the basis of a five-point Likert scale, we asked participants \textit{(a)} how much \textit{\gls{ser}\gls{gui}} supported them during the \gls{gui} prototyping process, \textit{(b)} how relevant the shown \glspl{gui} were on average for their initial requirements, \textit{(c)} how relevant the recommended \gls{gui} features were on average, \textit{(d)} how relevant the visualizations of \gls{gui} features (i.e. \textit{aspect}-\glspl{gui}) were on average, \textit{(e)} how much the feature recommendations helped them add relevant features that they did not think about initially and \textit{(f)} how much the visualizations of feature recommendations helped them better understand the \gls{gui} feature. Furthermore, we employed the \gls{sus} \citep{brooke1996sus} as a reliable, validated measurement of system usability. In order to gain further insights, we gathered additional free-form feedback from participants regarding potential improvements for \gls{ser}\gls{gui}.
\section[Results \&\ Discussion]{Results \& Discussion}


\subsection{RQ$_{1}$: Feature Recommendation Relevance}

\definecolor{lightgray}{gray}{0.92}

\newcommand{\MapMrrPAtKSepRule}{%
  \cmidrule(lr){1-1}\cmidrule(lr){2-3}\cmidrule(lr){4-9}
}

\definecolor{lightgray}{gray}{0.92}

\newlength{\ResultsTabColSep}
\setlength{\ResultsTabColSep}{9pt} 

{\renewcommand{\arraystretch}{1.0}
\begin{table}[!t]
\footnotesize
\caption[Evaluation results of \gls{llm}-based \gls{gui} feature recommendation]{Evaluation results overview of \gls{llm}-based \gls{gui} feature recommendation per \gls{gui} and on average using \textit{\gls{ap}}, \textit{\gls{mrr}} and \textit{P@k} (\textbf{Bold} values indicate best metric score within result group and \underline{underlined} values indicate best metric score across all groups).}
\centering
\setlength\tabcolsep{\ResultsTabColSep}

\begin{tabular}{c|c|c|cccccc}
\toprule
\multicolumn{1}{c|}{} &
\multicolumn{1}{c|}{\textbf{AP}} &
\multicolumn{1}{c|}{\textbf{MRR}} &
\multicolumn{6}{c}{\textbf{P@k}} \\
\cmidrule(lr){2-2}\cmidrule(lr){3-3}\cmidrule(lr){4-9}
 &
$\mathbf{AP}$ & $\mathbf{MRR}$ &
$\mathbf{P@1}$ & $\mathbf{P@2}$ & $\mathbf{P@3}$ & $\mathbf{P@5}$ & $\mathbf{P@7}$ & $\mathbf{P@10}$ \\
\midrule

\textbf{\mbox{GUI 1-1}} & 76.4 & 91.7 & 87.5 & 68.8 & 62.5 & 67.5 & 67.9 & 56.2 \\
\rowcolor{lightgray}
\textbf{\mbox{GUI 1-2}} & \textbf{\underline{82.4}} & \textbf{\underline{100.0}} &
\textbf{\underline{100.0}} & \textbf{\underline{87.5}} & \textbf{\underline{83.3}} & \textbf{72.5} & \textbf{71.4} & \textbf{72.5} \\
\textbf{\mbox{GUI 1-3}} & 80.4 & 93.8 & 87.5 & 81.2 & \textbf{\underline{83.3}} & \textbf{72.5} & 64.3 & 60.0 \\

\MapMrrPAtKSepRule

\rowcolor{lightgray}
\textbf{\mbox{GUI 2-1}} & \textbf{81.1} & \textbf{87.5} & \textbf{75.0} & \textbf{75.0} & \textbf{70.8} & 75.0 & \textbf{\underline{76.8}} & \textbf{\underline{77.5}} \\
\textbf{\mbox{GUI 2-2}} & 76.3 & 79.2 & 62.5 & 62.5 & 66.7 & \textbf{\underline{77.5}} & 73.2 & 70.0 \\
\rowcolor{lightgray}
\textbf{\mbox{GUI 2-3}} & 69.9 & 62.9 & 37.5 & 56.2 & 66.7 & 70.0 & 62.5 & 63.8 \\

\MapMrrPAtKSepRule

\textbf{\mbox{GUI 3-1}} & \textbf{71.2} & \textbf{79.2} & \textbf{62.5} & \textbf{62.5} & \textbf{70.8} & \textbf{65.0} & \textbf{60.7} & \textbf{62.5} \\
\rowcolor{lightgray}
\textbf{\mbox{GUI 3-2}} & 60.0 & 65.8 & 50.0 & 43.8 & 58.3 & 55.0 & 51.8 & 53.8 \\
\textbf{\mbox{GUI 3-3}} & 69.2 & 74.4 & \textbf{62.5} & \textbf{62.5} & 58.3 & 60.0 & 57.1 & 57.5 \\

\midrule
\textbf{Avg.} & 74.1 & 81.6 & 69.4 & 66.7 & 69.0 & 68.3 & 65.1 & 63.8 \\
\bottomrule
\end{tabular}

\label{tab:sergui-results_rq1}
\end{table}
}

\begin{figure*}
 \includegraphics[width=\textwidth]{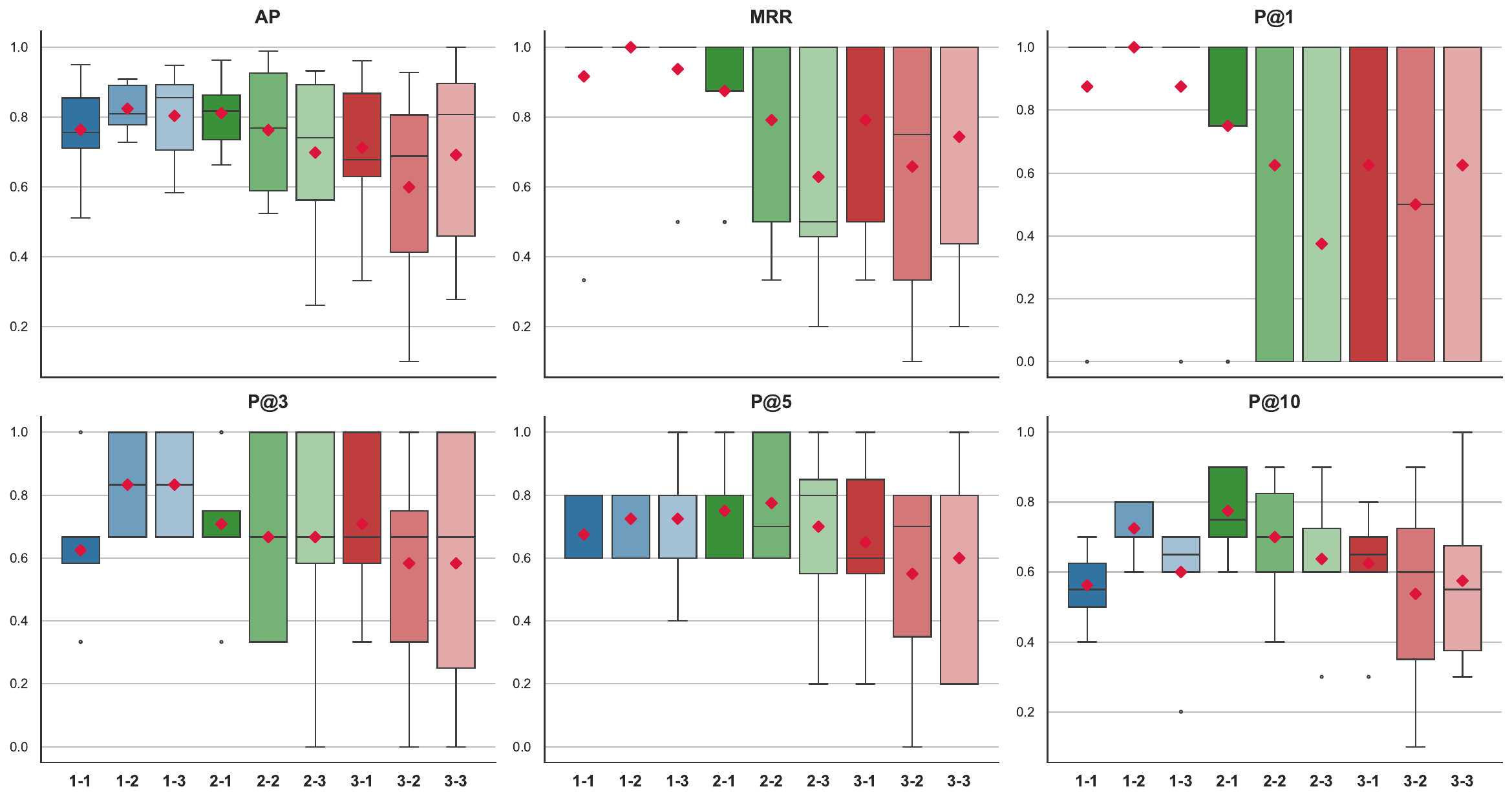}
  \caption[Boxplots for \gls{llm}-based \gls{gui} feature recommendation]{Boxplots with \textit{\gls{ap}}, \textit{\gls{mrr}} and \textit{P@k} across \textit{shopping app} (1-1, 1-2, 1-3) (\textit{blue}), \textit{news app} (2-1, 2-2, 2-3) (\textit{green}) and finally the \textit{social app} (3-1, 3-2, 3-3) (\textit{red}), diamonds=means.}
	\label{fig:sergui-boxplots}
    \vspace{-0.3cm}
\end{figure*}

Table \ref{tab:sergui-results_rq1} illustrates the evaluation results for RQ$_{1}$, showing the \textit{\gls{ap}}, \textit{\gls{mrr}} and \textit{P@k} values for each \gls{gui} individually and on average across all \glspl{gui}. Moreover, Figure \ref{fig:sergui-boxplots} shows multiple boxplots. An average \textit{\gls{ap}} of 74.1 indicates a strong feature recommendation performance as approximately three out of four features are marked as relevant on average. Likewise, the \textit{\gls{mrr}} of 81.6 shows that the first relevant feature occurs at rank 1.22 on average, indicating that the top-features are relevant. In addition, the \textit{P@k} values also indicate a substantial recommendation performance ranging from 69.1 (\textit{P@1}) to 63.8 (\textit{P@10}). This also shows that the feature scoring mechanism accordingly ranks more relevant \gls{gui} features higher as intended. From the features marked as non-relevant, 26.05\% were marked as \textit{already specified}. In these cases, the subsequent root causes could be identified and are discussed subsequently.


First, the model itself sometimes predicted similar features, for example, for \gls{gui} \textbf{(1-2)} the features \textit{product details} and \textit{product description} were recommended often. Through the explanations, the model might differentiate moderately between these two features (e.g. more general details in contrast to specifications such as weight, dimensions, etc.), but participants often identified them as identical. Another similar example from \gls{gui} \textbf{(1-3)} is represented by the features \textit{coupon code} and \textit{gift card}. Second, some recommended features were already encompassed in the initial \gls{gui}. Often, these features were not sufficiently represented in the \textit{Rico} \gls{gui} hierarchy data and subsequently in the generated \gls{gui} abstraction. For example, \textit{shipping information} on \gls{gui} \textbf{(1-2)} was often encapsulated in an image without a \textit{resource-id}. Hence, the feature was only encoded as an image component without further semantic information. 


In other cases, the feature was hard to identify for the model as being encoded as more complex semantics. For example, for task \gls{gui} \textbf{(1-2)} some selected \glspl{gui} contained a \textit{market price} and a \textit{shop price} (which was cheaper) and therefore participants marked the recommendation \textit{discount} as already specified. Similarly, for \gls{gui} \textbf{(2-3)} often the feature \textit{comment threading} was predicted, which typically is already included by most selected \glspl{gui}. However, this feature is also more difficult to recognize. Finally, other features marked as non-relevant often included general features such as \textit{help} or \textit{customer support}, which might be meaningful for the overall app but are not specific enough for the particular \gls{gui}. Figure \ref{fig:sergui-examples} depicts two examples of $NLR_{GUI}$ with selected \glspl{gui} and their feature predictions and top-three \textit{aspect}-\gls{gui} matches. The two illustrated examples represent two \glspl{gui} from the user study tasks (\textit{Rico} dataset \citep{deka2017rico}), namely, \gls{gui} \textbf{(2-3)}, a \textit{comment section} of a news article page and \gls{gui} \textbf{(1-2)}, a \textit{product details view}.

\begin{myrqbox}
\textbf{Answer to RQ$_{1}$:} \textit{\gls{ser}\gls{gui}} benefits from a strong \gls{llm}-based \gls{gui} feature recommendation performance, indicated by both a mean \textit{\gls{ap}} of 74.1 and an \textit{\gls{mrr}} of 81.6.
\end{myrqbox}

\vspace{-0.3cm}
\subsection{RQ$_{2}$: \gls{gui} Feature Retrieval Relevance}

\begin{figure*}
 \includegraphics[width=\textwidth]{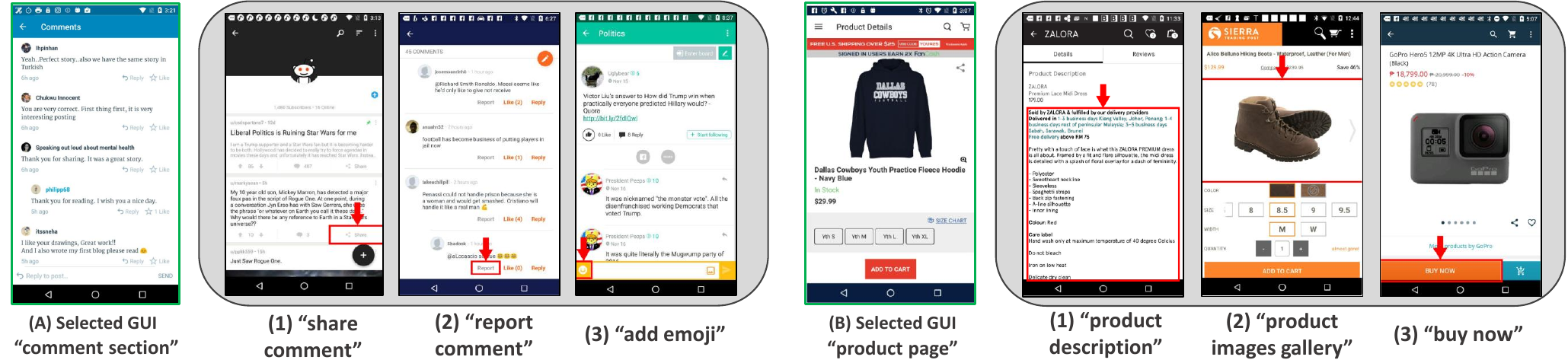}
  \caption[Example \gls{gui} feature recommendations]{Two \gls{gui} feature recommendations and their top-three \textit{aspect}-\gls{gui} matches: \textit{(A)} \textit{"comment section"} with recommended \gls{gui} features \textit{share}, \textit{report} and \textit{add emoji} and \textit{(B)} \textit{"product page"} with features \textit{product description}, \textit{product image gallery} and \textit{buy now}.}
	\label{fig:sergui-examples}
\end{figure*}

Table \ref{tab:sergui-results_rq2} depicts the evaluation results for RQ$_{2}$, showing the \textit{\gls{mrr}} and \textit{HITS@k} values for each \gls{gui} separately and averaged across all \glspl{gui}. An \textit{\gls{mrr}} of 39.0 shows that the first relevant \textit{aspect}-\gls{gui} appears at rank 2.56 on average, indicating a substantial performance in finding relevant \textit{aspect}-\gls{gui} matches. Moreover, an average \textit{HITS@1} of 27.0 represents that on average in 27\% a relevant \textit{aspect}-\gls{gui} could be found at the top-1 position. A \textit{HITS@5} of 52.7 indicates that in 52.7\% of the cases a relevant \textit{aspect}-\gls{gui} could be found within the top-5 ranking positions. Finally, a \textit{HITS@15} of 69.1 represents that on average our matching approach could find a relevant \textit{aspect}-\gls{gui} in 69.1\% within the top-15 \textit{aspect}-\gls{gui} ranking. Therefore, for a large number of recommended features, our feature matching approach is capable of finding a relevant \textit{aspect}-\gls{gui} visualizing the \gls{gui} feature. Additionally, for active feature searches our approach performs similarly with an overall better coverage (\textit{\gls{mrr}} = 26.2 and \textit{HITS@15} = 78.6). Figure \ref{fig:sergui-examples} illustrates the recommended \gls{gui} features and their respective top selected \textit{aspect}-\glspl{gui}.


For example, for \gls{gui} \textbf{(2-3)} \textit{"comment section"} the matching of recommended features performs well and includes features such as \textit{share comment}, enabling sharing on social media, \textit{report comment} to directly report a single comment and an \textit{add emoji} feature enabling users to add an emoji to their comment. Likewise, example \gls{gui} \textbf{(1-2)} \textit{"product page"} contains interesting top selected \textit{aspect}-\gls{gui} matches encompassing features such as \textit{product description} providing detailed descriptions of the product, \textit{product image gallery} enabling the users to quickly browse multiple images of the product (both matched by their \textit{resource-ids}) and a \textit{buy now} component, enabling fast purchasing without checkout. If \textit{aspect}-\glspl{gui} could not be matched, the root causes were identified.


First, sometimes the feature recommendation model generated over-specified text descriptions of the features, e.g., including non-valuable domain words. For example, instead of generating \textit{rating} only, sometimes \textit{product rating} was generated or instead of solely generating \textit{source}, sometimes \textit{news source} was generated by the \gls{llm}. These additional unspecific domain words often led to less precise matches. Additionally, the matching performance decreased with increasing feature recommendation rank, since the feature ranking is based on the matching scores. The initial $NLR_{GUI}$ also has a high impact on the matching performance. For example, when the initial top-\textit{k} \gls{gui} ranking is heterogeneous (e.g., containing both \textit{product overviews} and \textit{product detail views}), then it also becomes more difficult to properly match the features with \gls{sbert}. Overall, we obtained a good matching performance of the \textit{aspect}-\glspl{gui} as indicated by the metrics.

\begin{myrqbox}
\textbf{Answer to RQ$_{2}$:} \textit{\gls{ser}\gls{gui}} is able to retrieve relevant \textit{aspect}-\glspl{gui} for visualization with a good performance, indicated by an \textit{MRR} of 39.0 and an \textit{HITS@15} of 69.1. 
\end{myrqbox}

\subsection{RQ$_{3}$: \gls{gui} Reranking Performance}

Based on the positive feature feedback (selecting an \textit{aspect}-\gls{gui}), we computed a reranking of the \glspl{gui}. As per the design of \textit{\gls{ser}\gls{gui}}, participants could initially select a \gls{gui} and then were presented with the feature recommendations. Afterwards, participants could either choose a new \gls{gui} from the updated ranking or keep their previous choice. In 68.05\% of the cases, the participants selected a better matching \gls{gui} from the \gls{gui} reranking. For these cases, the reranking score is +61.89 ranks on average (SD:79.82/Min:-26/Max:335), indicating that the feature-based reranking method can substantially improve the ranks of relevant \glspl{gui}. We only computed the reranking score for cases where participants reselected the \gls{gui}, due to the fact that participants simply could keep their previous choices if no better \gls{gui} could be found. In addition, by comparing the \textit{\gls{mrr}} of the individual \textit{\gls{sbert}} (\textit{\gls{mrr}} = 17.2) and \textit{S2W} (\textit{MRR} = 21.0) retrieval models with the ensemble model (\textit{\gls{mrr}} = 42.9), we can observe that the ensemble considerably improves the \gls{gui} ranking performance of each separate model. Therefore, participants often picked initial \glspl{gui} only from the very top ranks (rank 2.38 on average) and rarely examined more than the top-\textit{20} \glspl{gui} in the overall ranking. Achieving an average reranking score of +61.89 thus could help participants find better matching \glspl{gui} at the very top of the \gls{gui} reranking.

\begin{myrqbox}
\textbf{Answer to RQ$_{3}$:} \gls{gui} feature feedback could be employed to compute a reranking score resulting in 68.05\% re-selections and +61.89 ranks improvement on average.
\end{myrqbox}

\definecolor{lightgray}{gray}{0.92}

\setlength{\ResultsTabColSep}{8pt} 

\newcommand{\MrrHitsSepRule}{%
  \cmidrule(lr){1-1}\cmidrule(lr){2-2}\cmidrule(lr){3-9}
}

{\renewcommand{\arraystretch}{1.0}
\begin{table}[!t]
\footnotesize
\caption[Evaluation results of \gls{gui} feature (\textit{aspect}-\gls{gui}) retrieval]{Evaluation results overview of our \gls{gui} feature (\textit{aspect}-\gls{gui}) retrieval approach for each \gls{gui} and on average using \textit{\gls{mrr}} and \textit{HITS@k} (\textbf{Bold} values indicate best metric score within result group, \underline{underlined} values indicate best metric score across all groups).}
\centering
\setlength\tabcolsep{\ResultsTabColSep}

\begin{tabular}{c|c|ccccccc}
\toprule
\multicolumn{1}{c|}{} &
\multicolumn{1}{c|}{\textbf{MRR}} &
\multicolumn{7}{c}{\textbf{HITS@k}} \\
\cmidrule(lr){2-2}\cmidrule(lr){3-9}
 &
$\mathbf{MRR}$ &
$\mathbf{H@1}$ & $\mathbf{H@2}$ & $\mathbf{H@3}$ & $\mathbf{H@5}$ & $\mathbf{H@7}$ & $\mathbf{H@10}$ & $\mathbf{H@15}$ \\
\midrule

\textbf{\mbox{GUI 1-1}} & 35.0 & 20.0 & 35.6 & 42.2 & 55.6 & 60.0 & 66.7 & 75.6 \\
\rowcolor{lightgray}
\textbf{\mbox{GUI 1-2}} & \textbf{41.5} & \textbf{27.6} & \textbf{43.1} & \textbf{51.7} & \textbf{56.9} & \textbf{65.5} & \textbf{67.2} & 72.4 \\
\textbf{\mbox{GUI 1-3}} & 28.9 & 16.7 & 25.0 & 31.2 & 43.8 & 54.2 & 60.4 & 66.7 \\

\MrrHitsSepRule

\rowcolor{lightgray}
\textbf{\mbox{GUI 2-1}} & \textbf{37.7} & 24.2 & \textbf{41.9} & \textbf{43.5} & \textbf{46.8} & \textbf{54.8} & \textbf{62.9} & \textbf{75.8} \\
\textbf{\mbox{GUI 2-2}} & 36.7 & \textbf{26.8} & 35.7 & 46.4 & \textbf{48.2} & \textbf{55.4} & \textbf{57.1} & \textbf{58.9} \\
\rowcolor{lightgray}
\textbf{\mbox{GUI 2-3}} & 30.8 & \textbf{25.5} & 31.4 & 31.4 & 35.3 & 39.2 & 45.1 & 47.1 \\

\MrrHitsSepRule

\textbf{\mbox{GUI 3-1}} & \textbf{\underline{63.7}} & \textbf{\underline{52.0}} & \textbf{\underline{66.0}} & \textbf{\underline{74.0}} & \textbf{\underline{78.0}} & \textbf{\underline{84.0}} & \textbf{\underline{84.0}} & \textbf{\underline{86.0}} \\
\rowcolor{lightgray}
\textbf{\mbox{GUI 3-2}} & 46.1 & 27.9 & 39.5 & 65.1 & 76.7 & 83.7 & 83.7 & \textbf{\underline{86.0}} \\
\textbf{\mbox{GUI 3-3}} & 30.1 & 21.7 & 30.4 & 37.0 & 37.0 & 41.3 & 45.7 & 54.3 \\

\midrule
\textbf{Avg.} & 39.0 & 27.0 & 39.0 & 46.8 & 52.7 & 59.5 & 63.4 & 69.1 \\
\bottomrule
\end{tabular}

\label{tab:sergui-results_rq2}
\end{table}
}

\subsection{RQ$_{4}$: Perceived Usefulness}

\begin{figure}[!t]
  \includegraphics[width=\textwidth]{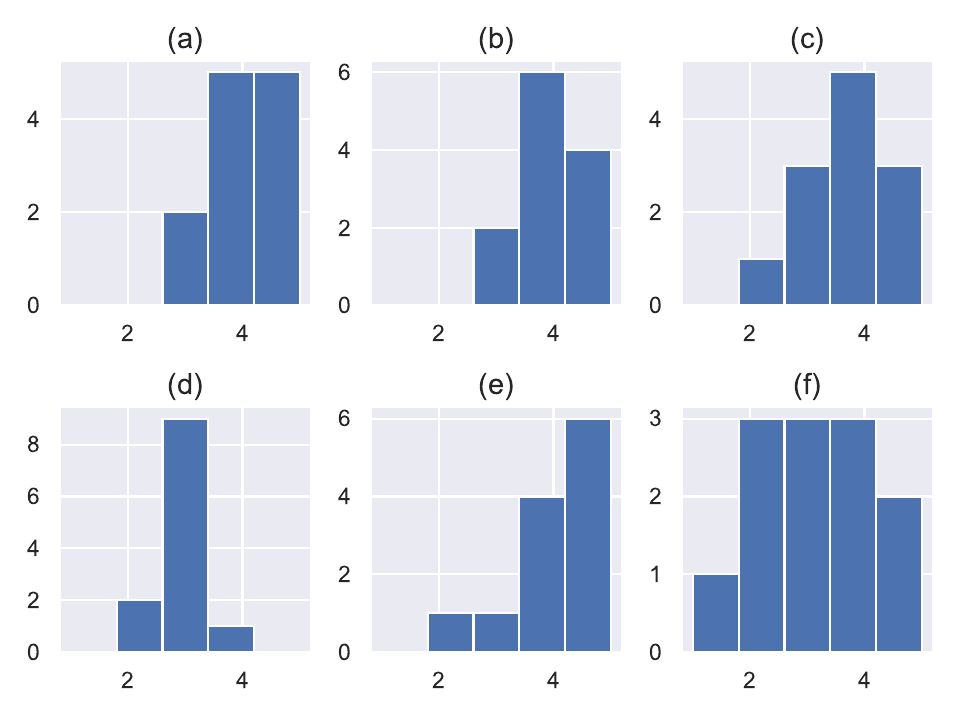}
  \caption[Perceived usability histograms]{Six histograms showing the evaluation results for the posed usability questions with reference to employing \textit{\gls{ser}\gls{gui}} for \gls{gui} prototyping as perceived by participants.}
	\label{fig:sergui-usability}
 \vspace{-0.3cm}
\end{figure}

Fig. \ref{fig:sergui-usability} depicts the evaluation results of the perceived usefulness of \textit{\gls{ser}\gls{gui}} as reported by the participants. First, \textit{(a)} the support provided by \textit{\gls{ser}\gls{gui}} during the prototyping process received a high score (M:4.25/SD:0.75). Second, \textit{(b)} the average relevance of the shown \glspl{gui} for the initial requirements received a similar high score (M:4.16/SD:0.71). Third, \textit{(c)} the average relevance of the recommended \gls{gui} features also received a comparably high score (M:3.83/SD:0.93). Fourth, \textit{(d)} the average relevance of the visualizations of \gls{gui} features received a medium score (M:2.9/SD:0.51). Fifth, \textit{(e)} how much the feature recommendations helped them add relevant features they did not think about initially also received a high score (M:4.25/SD:0.96). Lastly, \textit{(f)} how much the visualizations of the recommended features helped them better understand the feature received a medium score (M:3.16/SD:1.26). In summary, \textit{\gls{ser}\gls{gui}} accomplished a \gls{sus} of 84.16, indicating that our approach possesses high perceived usability particularly in comparison with the average \gls{sus} of 68. In general, participants reported that they found it very easy to get started quickly with \textit{\gls{ser}\gls{gui}} and they required little to no prior knowledge to be able to utilize our approach. Moreover, participants mostly found the feature recommendations helpful to integrate new relevant and interesting features into their prototype, which they were not aware of initially. For example, one participant reported specifically that a \textit{notification bell} and a \textit{subscription} feature in the context of the \textit{social} app was stimulated to be included by our recommendation approach. Regarding improvement directions for the proposed approach, participants mostly reported that the respective \gls{gui} feature visualizations sometimes did not match the corresponding recommended \gls{gui} features.

\begin{myrqbox}
\textbf{Answer to RQ$_{4}$:} \textit{\gls{ser}\gls{gui}} showcases a high perceived usability as indicated by a high \gls{sus} of 84.16, substantially higher than the average \gls{sus} value of 68 \citep{brooke1996sus}.
\end{myrqbox}

\section{Threats to Validity}

\textbf{Internal Validity.} To ensure internal validity, we conducted the experimental procedure including instructions, tasks and a tutorial in the same manner for each participant. To avoid bias from the tasks, we solely provided low-fidelity mockups of the \glspl{gui}. This ensures that participants have to formulate their own textual requirements for the \glspl{gui} and features. The annotation of relevance for selected \glspl{gui}, feature recommendations and \textit{aspect}-\glspl{gui} is inherently subjective. To mitigate potential bias from the annotation procedure, we therefore conducted the experiments with 12 different participants.

\vspace{9pt}

\noindent \textbf{External Validity.} To improve the generalizability of our evaluation, we included nine different \glspl{gui} from three diverse domains. Since \textit{Rico} encompasses many diverse domains and \glspl{gui}, we hypothesize that similar results could be obtained for other well-covered domains in \textit{Rico}. However, to increase the generalizability further, more evaluation should be conducted to confirm the findings. To improve external validity with regard to the user study participants, we included users with low \gls{gui} prototyping experience.

\section{Limitations}

Our approach possesses notable limitations which provide the foundation for future work. Due to the data-driven nature of our approach, the potentially supported domains and applications are constrained. \textit{\gls{ser}\gls{gui}} cannot support the \gls{ser} process for very special and custom \glspl{gui}. However, the \gls{gui} repository \textit{Rico} already encompasses many diverse domains with a large number of \glspl{gui}, enabling versatile application scenarios. Similarly, \textit{\gls{gpt}-4} has been trained on highly diverse text data \citep{brown2020language, openai2023gpt4}, encoding knowledge from a plethora of domains. 

Another limitation of our approach lies in the current inflexibility and the determined process flow of the interaction model. Therefore, enabling more flexibility with reference to, for example, interweaving more closely the active search for own features and getting features recommended could enhance the overall usability. Furthermore, the currently possible feedback on proposed \gls{gui} features is restricted to either including or excluding the feature in a binary fashion. This is sufficient in many cases, but there might be situations for which a fine-grained \gls{nl} response is more appropriate. For example, in cases where including a \textit{payment methods} feature is relevant, customers would potentially provide more detailed feedback directly such as the specific payment methods that should be supported. By integrating \glspl{llm}, with their impressive \gls{nlu} capabilities, more closely into the process, such \gls{nl} feedback could be processed additionally. 

\glsreset{dl}
Moreover, the \gls{llm} sometimes recommends similar features with different granularity, e.g., \textit{payment methods} and \textit{PayPal} or semantically related features such as \textit{add contact} and \textit{delete contact} at different ranks. Grouping these hierarchical and semantically related features could also improve the usability. Furthermore, the sometimes heterogeneous matching of \textit{aspect}-\glspl{gui} might mislead users. Improving the feature matching performance could solve this limitation, e.g., by optimizing a Deep Learning (DL) model specifically for this task. Finally, the current implementation of the tool misses standard features (e.g., reverting actions in case of user errors) and solely shows the initially selected \gls{gui} and the current app prototype. Extending the view of the current \gls{gui} to include all chosen \textit{aspect}-\glspl{gui} and text features could further facilitate utilizing the novel approach.
\section{Related Work}

In Chapter \ref{cha:nl_gui_retrieval} of this thesis, we introduced \textit{\gls{rawi}}, proposing \gls{nlr}-based \gls{gui} retrieval approaches. In that work, the focus lies on investigating and evaluating various methods for \gls{nlr}-based \gls{gui} retrieval and providing support for the analyst in a \gls{rel} scenario. With our proposed \textit{\gls{ser}\gls{gui}} approach, we extend this work to enable \gls{ser} with \glspl{gui} for users. In particular, we additionally propose an automatic \gls{gui} feature recommendation approach, the visualization of recommended features, feature search and a feature-based \gls{gui} reranking technique. Our approach is integrated into a user-friendly tool utilizing chat-based guidance, requiring no prior prototyping knowledge or experience. Furthermore, our approach integrates the selected \glspl{gui} and chosen \gls{gui} features into a novel \gls{gui} prototyping artifact that can be utilized by analysts.

Another similar approach for \gls{nl}-based \gls{gui} retrieval is \textit{Guigle} \citep{bernal2019guigle}, exploiting \textit{Android} apps automatically crawled from \textit{Google Play} and providing search functionality based on simple retrieval models from \textit{Lucene} \citep{lucene}. In contrast, our \textit{\gls{ser}\gls{gui}} approach utilizes more sophisticated embedding-based \gls{gui} ranking and reranking models plus \gls{gui} feature recommendation and matching techniques. Moreover, \textit{Guigle} can merely be employed by designers or analysts, but cannot be utilized for \gls{ser} with stakeholders. Many other \gls{gui} retrieval approaches utilizing visual input have been proposed before. For example, \textit{Swire} \citep{huang2019swire} employs visual embeddings for retrieving \glspl{gui} based on hand-drawn sketches, \textit{GUIFetch} \citep{behrang2018guifetch} enables retrieval of entire applications on the foundation of an entire \textit{Android} application sketch, and \textit{VINS} \citep{bunian2021vins} provides \gls{gui} retrieval by utilizing either a simple wireframe prototype or a fully designed \gls{gui} as input. Finally, \textit{Gallery D.C.} \citep{chen2019gallery} represents an approach for providing multi-faceted (e.g., \textit{width}, \textit{height} and \textit{color}) search functionality for designers to enable the retrieval of individual \gls{gui} components such as buttons. In contrast, \textit{\gls{ser}\gls{gui}} utilizes an embedding-based \gls{gui} component ranking to enable semantic feature search and is capable of matching a variety of \gls{gui} component types. 

Prior feature recommendation approaches as presented in \citep{chen2019recommending, wang2022missing} mainly provide the ability to predict potentially relevant features on the basis of a high-fidelity \gls{gui} as input. These bottom-up feature recommendation approaches merely extract simple features from the \glspl{gui} restricted to clickable components (mainly \textit{buttons} and \textit{icons}). In contrast, the feature recommendation problem in the context of \textit{\gls{ser}\gls{gui}} is substantially different, involving multiple inputs such as fine-grained unstructured \gls{nlr} for the \glspl{gui}, a collection of actively specified individual features as \gls{nlr} and an initially selected \gls{gui}. Moreover, our top-down \gls{llm}-based feature recommendation approach exploits the vast knowledge embedded in \glspl{llm} and enables the recommendation of \gls{gui} features that the mentioned works are not capable of, e.g., simple features such as a \textit{comment count} or \textit{promotion badge} and semantically more complex features such as \textit{comment threading} or \textit{related articles}. Further, we match recommended features with the \glspl{gui} from the top-\textit{k} \gls{gui} ranking to create \textit{aspect}-\glspl{gui}, enabling visualizations of features and integrate the recommendations into the whole, automated \gls{ser} approach.
\section{Conclusion}

This work was driven by challenge \challonethree{}: \textit{How can we provide automatic assistance for self-elicitation of requirements with \gls{gui} prototyping?} To answer this question, in this work, we proposed \textit{\gls{ser}\gls{gui}} as an initial approach towards enabling \gls{ser} with \gls{gui} prototyping. \textit{\gls{ser}\gls{gui}} allows users to rapidly and independently create their own \gls{gui} specifications, requiring little prototyping experience through multiple interaction mechanisms. The evaluation indicates that \textit{\gls{ser}\gls{gui}} is able to effectively support users during the \gls{gui} prototyping process in terms of providing relevant \glspl{gui} from \gls{nlr}, recommending relevant \gls{gui} features with visualizations and achieving high perceived usability (\gls{sus}).

\part[\sc{Prototyping: GUI Generation via Natural Language}]{\sc{Prototyping: GUI Generation via Natural Language}}
\label{part:gui_generation}

\clearpage
\newpage
\thispagestyle{empty}
\null  

\chapter{Tailoring LLMs to Prototyping: Zero-Shot GUI Generation with Retrieval-Augmentation and Self-Critique Prompting}
\chaptermark{Zero-Shot GUI Generation with Prompting Approaches}
\label{cha:zs_gui_generation}
\vspace{-0.5cm}
After presenting multiple \gls{nlr}-based \gls{gui} retrieval and reranking solution approaches for the first challenge (\challone{}) to rapidly map \gls{nlr} to \gls{gui} prototypes, we provide additional novel solution approaches through \gls{llm}-based \gls{gui} generation. In this chapter, we introduce several \gls{zs} prompting approaches to efficiently adapt pretrained \glspl{llm} for the problem of \gls{gui} generation from \gls{nlr}, requiring no training or finetuning. The following sections are based on previous work \citep{kolthoff2024zero}\footnote{This section is adapted from: \textbf{Kolthoff, Kristian\textsuperscript{*}}, Kretzer, Felix\textsuperscript{*}, Fiebig, Lennart, Maedche, Alexander, Ponzetto, Simone Paolo, and Bartelt, Christian. Zero-Shot Prompting Approaches for LLM-based Graphical User Interface Generation. arXiv preprint arXiv:2412.11328, 2024. The paper is currently under review for the \emph{2026 ACM Joint European Software Engineering Conference and Symposium on the Foundations of Software Engineering (ESEC/FSE)}. Sections are directly reused with adaptations (i.e. thesis Section \ref{cha:zs_gui_generation}.i matches paper Section i+1 starting from the \textit{Approach} section \ref{sec:zs-approach}, since the \textit{Background} section from the original paper has been removed. See Chapter \ref{cha:background} regarding background details.}. The source code, datasets and experimental resources are all publicly available to foster future research\footnote{Materials for this chapter, including source code, datasets and experimental resources, are available at our corresponding \textit{GitHub} repository \url{https://github.com/kristiankolthoff/ZS-Prompting}}.

\paragraph{Personal Contribution.} I ideated and created the concept for the work. While I created the final implementation for the prompting techniques and framework, \textit{Lennart Fiebig} created the initial implementation draft for the self-critique technique. \textit{Felix Kretzer} and I contributed equally to the evaluation design, while the data analysis was conducted by \textit{Felix Kretzer}. The manuscript was mainly written by me, while \textit{Felix Kretzer} wrote the \textit{\gls{gui} description dataset} section and created the result tables and figures.

\section{Motivation}

We previously presented \gls{nlr}-based \gls{gui} retrieval techniques, including \gls{bert}-\gls{ltr} models, improved the \gls{gui} reranking effectiveness with \gls{mllm}-based reranking and proposed an approach for \gls{ser} with automated \gls{gui} prototyping that integrates \gls{gui} retrieval techniques. While \gls{gui} retrieval enables rapid mapping of \gls{nlr} to corresponding \gls{gui} prototypes, these approaches have one major shortcoming. Practically, retrieval methods are bound to the available \gls{gui} datasets with static \gls{gui} prototypes, which restricts support for arbitrary \gls{nlr} input and \gls{gui} prototype customizability. While the proposed data-driven \gls{gui} prototyping editor \textit{\gls{rawi}} simplifies the process, it still requires effort to create novel \gls{gui} prototypes adapted to custom \gls{nlr} that are not directly supported by an individual \gls{gui} prototype. Furthermore, pretrained \glspl{llm} have gained popularity over recent years, showing impressive performance across many \gls{nlp} and \gls{nlu} tasks. In particular, \glspl{llm} possess high effectiveness for code generation \citep{jiang2024survey}. Hence, adapting \glspl{llm} for the problem of \gls{nlr}-based \gls{gui} generation offers significant potential. Furthermore, recent trends such as \textit{prompting-based coding} \citep{sapkota2025vibe, ray2025review} highlight the increasing need for automated \gls{gui} generation. In prompting-based coding, users experiment with design and functionality through iterative and conversational interactions. Since this approach often attracts participants who lack advanced prototyping skills \citep{sapkota2025vibe, ray2025review}, enhancing \gls{llm}-based \gls{gui} generation to convert \gls{nl} descriptions directly into high-fidelity \gls{gui} prototypes is particularly valuable.

\glsreset{dsl}
Recent research proposed several approaches for automatically generating \gls{gui} prototypes. For example, \textit{Instigator} \citep{brie2023evaluating} is based on \textit{minGPT} \citep{minGPT} trained on a vast number of crawled web pages to produce low-fidelity \gls{gui} layouts based on text descriptions and user-selected \gls{gui} component types. An alternative to training a task-specific \gls{gpt} model from scratch involves fine-tuning a pretrained LLM \citep{feng2023designing}, leveraging the \textit{Rico} \gls{gui} repository \citep{deka2017rico}. However, both approaches not only require resource-intensive training but are also limited to producing low-fidelity layouts in a \gls{dsl}, which is hard to integrate into practical \gls{gui} prototyping workflows or link to individual prototyping tools. Furthermore, \textit{MAxPrototyper} \citep{yuan2024maxprototyper} enables the creation of \gls{gui} prototypes by prompting \glspl{llm}. However, instead of relying solely on \gls{nlr} as input, their approach requires a fully developed \gls{gui} layout, generates a proprietary \gls{dsl} and neglects creating actual \gls{gui} prototype functionality. Therefore, the work presented in this chapter is driven by the leading research question of challenge \challonefour{}: \textit{How can we efficiently optimize \gls{llm}s for more effective \gls{gui} prototype generation?}

\begin{figure}
  \centering
  \includegraphics[width=0.85\textwidth]{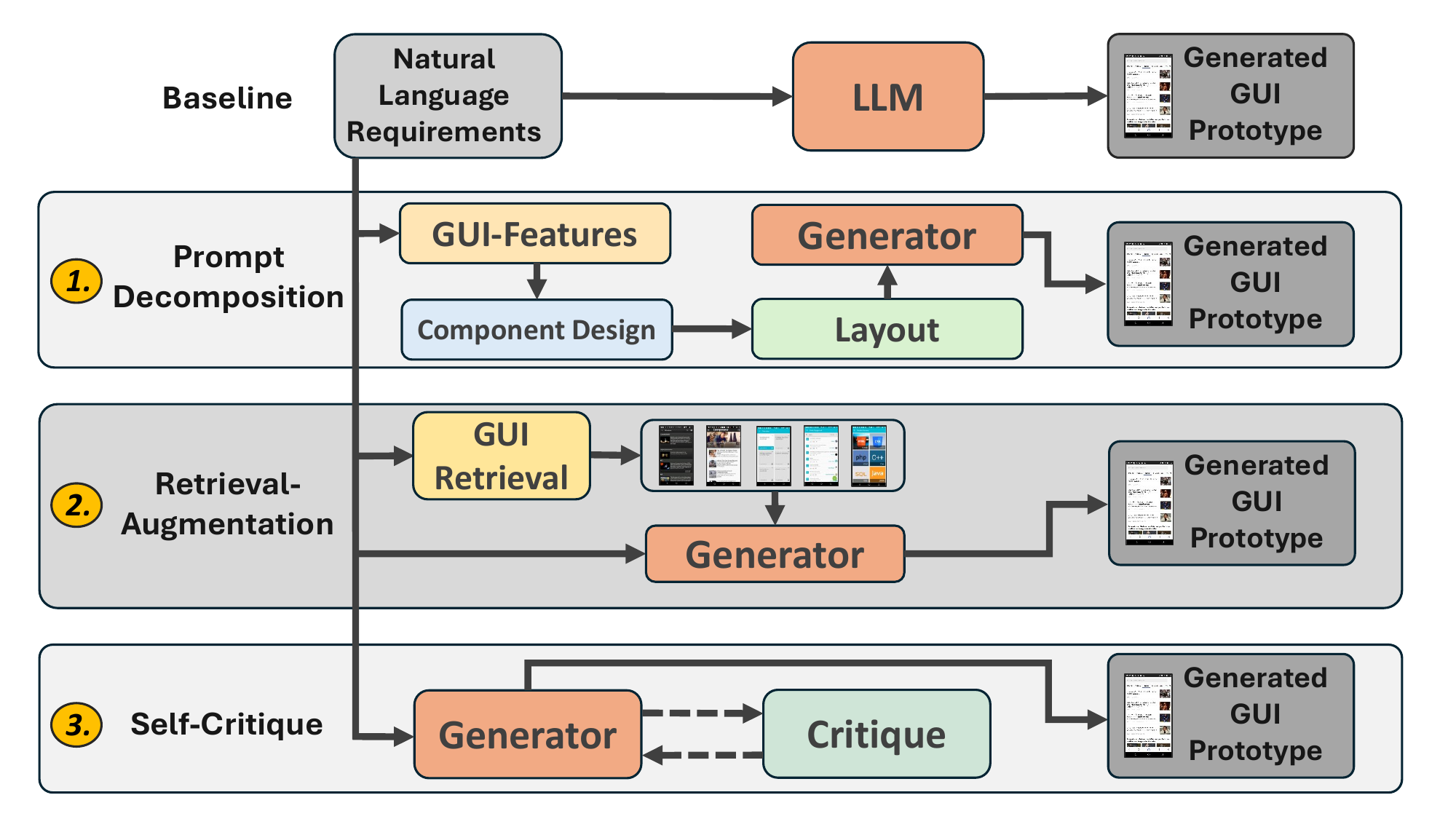}
  \caption[Overview of \gls{zs} prompting approaches]{\gls{zs} prompting approaches for \gls{gui} generation from \gls{nlr}: \textit{(1)} \gls{pd}, \gls{rag} and \gls{sc}.}
  \label{fig:overview_high_level}
\end{figure}

\glsreset{ragg}
\glsreset{sc}
\glsreset{pd}
While existing research has focused mainly on resource-intensive training and fine-tuning to generate low-fidelity \gls{gui} prototypes, a comprehensive investigation into less resource-intensive approaches based on \gls{zs} prompting techniques for generating high-fidelity \gls{gui} prototypes is currently lacking. In order to address this research gap, we explore the potential and effectiveness of different \gls{zs} prompting approaches for generating high-fidelity \gls{gui} prototypes from \gls{nlr} in \gls{html}/\gls{css}, as shown in Figure \ref{fig:overview_high_level}. Our focus lies particularly on \gls{zs} prompting because \textit{(i)} training or fine-tuning \glspl{llm} is resource-intensive and \textit{(ii)} \gls{fs} prompting struggles with very large context windows required for \gls{icl} \citep{dong2022survey, li2024long}, which would be necessary for \gls{gui} generation, since high-quality \glspl{gui} typically consist of thousands of tokens. In particular, we propose \gls{ragg}, which combines the advantages of \gls{gui} retrieval for rapid access to vast prototyping knowledge embodied in large \gls{gui} repositories with the reasoning and adaptation capabilities of \glspl{llm}. To this end, we utilize the presented \gls{gui} retrieval and reranking approaches. Moreover, we investigate \gls{pd} \citep{khot2022decomposed} for \gls{gui} generation, enabling the \gls{llm} to generate meaningful intermediate reasoning outputs, instead of directly generating low-level \gls{html}/\gls{css} from high-level \gls{nlr}. This approach more closely follows a human expert process and ensures that computational capabilities of the \gls{llm} are used more effectively. Finally, we investigate \gls{sc} \citep{saunders2022self} for \gls{gui} generation, employing the \gls{llm} itself in a \gls{gui} prototyping and feedback loop.

\glsreset{pd}
\glsreset{rag}
\glsreset{sc}

To assess the proposed approaches, we conducted an extensive evaluation consisting of over 20,400 \gls{gui} annotations from 101 crowd-workers with \gls{uiux} experience regarding the accuracy and self-reported assessments of generated \glspl{gui}. The results suggest that especially the \gls{ragg} and \gls{sc} prompting approaches can substantially enhance \gls{gui} prototypes across multiple evaluation metrics in comparison with multiple \gls{zs} baselines.

\paragraph{Contributions.} With this work, we make the following two research contributions:
\glsreset{ragg}
\glsreset{sc}
\glsreset{pd}
\begin{itemize}[leftmargin=0.15in]

\item \textit{\gls{ragg} and \gls{sc}:} we present the first pipeline that retrieves example \glspl{gui} from a large-scale \gls{gui} repository, applies advanced \gls{mllm}-based reranking techniques to them, extracts their feature and layout knowledge, and conditions an \gls{llm} to generate executable, high-fidelity \gls{html}/\gls{css} prototypes. Moreover, we provide the first systematic adaptation of \gls{pd} and \gls{sc} prompting techniques for end-to-end \gls{nlr}-based \gls{gui} generation.

\item \textit{Large-scale, reproducible evaluation and insights:} we conduct a large-scale human study with 20,400 annotations from 101 \gls{uiux} practitioners over 50 \gls{gui} generation tasks and release code, prompts and data and analyze common \gls{llm}-induced \gls{gui} defects.
\end{itemize}

\section{Approach}
\label{sec:zs-approach}

\begin{figure*}
  \centering
 \includegraphics[width=1\textwidth]{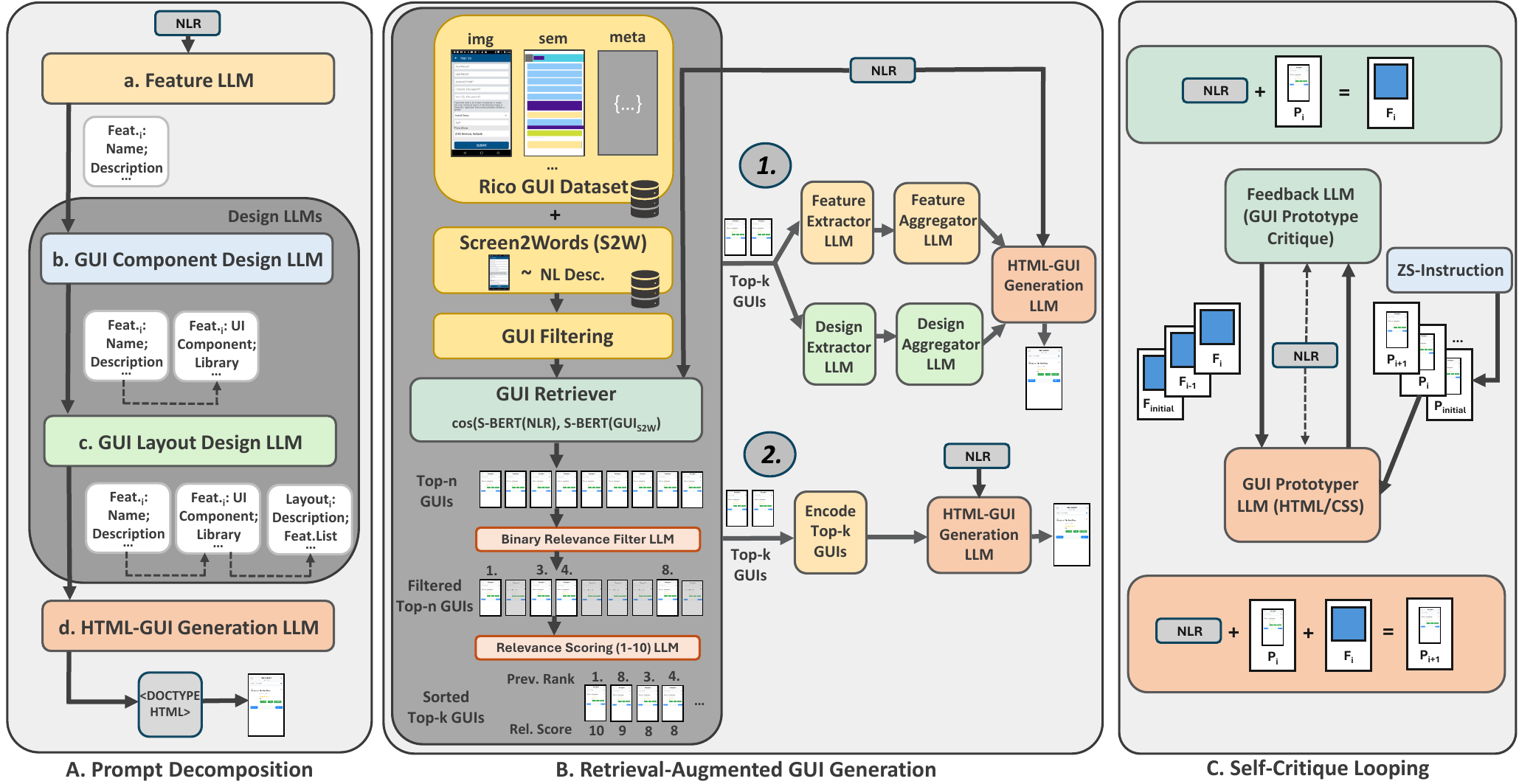}
  \caption[Overview of \gls{zs} prompting approaches architecture]{Overview of \gls{zs} prompting approaches for generating \gls{gui} prototypes from \gls{nlr}: \textit{(A)} \gls{pd}, which divides the generation task into smaller sub-tasks, \textit{(B)} \gls{ragg}, which exploits the \textit{Rico} \gls{gui} dataset and previously developed \gls{gui} retrieval and reranking techniques to augment the prompt context with \glspl{gui} relevant to the provided \gls{nlr}, and \textit{(C)} \gls{sc}, which employs generation-and-critique loops to iteratively improve the \gls{gui} prototypes.}
    \label{fig:overview-zero-shot}
\end{figure*}

We present the different \gls{zs} prompting methods for generating \gls{gui} prototypes. First, we introduce the baselines to benchmark the more comprehensive approaches. Next, we present \gls{pd}, \gls{ragg} and \gls{sc} techniques for \gls{gui} generation, as illustrated in Figure \ref{fig:overview-zero-shot}.

\glsreset{pd}
\glsreset{ragg}
\glsreset{sc}

\subsection{Baselines: \gls{zs}-Instruction and \gls{zs}-\gls{cot}}

As our prompting baselines, we first employ a \gls{zs} instruction prompt, including a clear description of the base task to create a mobile page in \gls{html}/\gls{css} according to the provided \gls{nlr}. We selected \gls{html}/\gls{css} as the target language to generate due to its widespread use as a \gls{gui} description language, support for interaction and extensive pretraining of \glspl{llm} on this language. To improve the alignment of \gls{llm} responses for user interaction, \glspl{llm} are often optimized using \textit{Reinforcement Learning with Human Feedback (RLHF)} \citep{ouyang2022training}, leading to \glspl{llm} providing explanations and a structured representation of the response. In particular, the \glspl{llm} for \gls{gui} generation have a tendency to provide explanations and separate the \gls{html}/\gls{css} into multiple markdown blocks. To avoid this, we additionally instruct the \gls{llm} to directly output the non-separated code without explanation. In addition, we employ a \gls{zs}-\gls{cot} prompt \citep{wei2022chain} as our second baseline, instructing the model to provide a self-created step-by-step reasoning sequence with intermediate computation steps, instead of directly mapping high-level \gls{nlr} to low-level \gls{html}/\gls{css}, which is effective for several tasks.

\glsreset{pd}
\subsection{\gls{pd} for \gls{gui} Generation}

Regarding the notion of enabling the model to perform step-by-step reasoning for generating \glspl{gui} instead of directly outputting code, we extend this idea by decomposing the instruction into multiple separate tasks and prompts \citep{khot2022decomposed} for \gls{gui} generation, which also follows a human expert approach more closely, as shown in Figure \ref{fig:overview-zero-shot} \textit{(A)}. Usually, \gls{rel} takes place initially, where human experts utilize knowledge and previous experience to gain deeper insights into the requirements. 

Therefore, in the first step, we employ a feature extraction prompt, asking the \gls{llm} to provide a more detailed collection of \gls{gui} features with textual descriptions based on the high-level input. Second, we instruct the \gls{llm} to derive implementation and design ideas for each of the features, i.e. which \gls{gui} components and libraries to employ for implementing the feature. Third, we instruct the \gls{llm} to provide an overall layout structure and page design given the previously generated feature collection. With each step of the decomposed prompt pipeline, the initial more implicit \gls{nlr} is extended by more and more information to form a more explicit representation of the \gls{gui} generation problem. Subsequently, this representation serves as the input to the final \gls{llm} which is tasked with the translation of the entire text specification of the \gls{gui} into \gls{html}/\gls{css}. In addition to decomposing the task into subtasks to provide more computational capabilities for each step, we also created a second variant of the described process by combining all instructions into a single prompt. This represents a custom \gls{zs}-\gls{cot} prompt with predefined domain-specific reasoning steps. The stepwise process mirrors human expert reasoning, encouraging logical and contextually grounded responses. As the code is generated from an explicit, incrementally refined \gls{gui} prototype specification, the likelihood of propagating inaccurate information into the final prototype is substantially reduced.

\glsreset{ragg}
\subsection{\gls{ragg}}
\label{sec:ragg}

The notion behind \gls{ragg} lies in integrating \glspl{llm}, possessing impressive text generation and reasoning capabilities, with the advantage of retrieval approaches --- namely, providing rapid access to a vast number of potentially relevant documents for effectively completing a generation task, which achieves state-of-the-art results for many \gls{nlp} problems, especially knowledge-intensive tasks \citep{weston2018retrieve, lewis2020retrieval, li2022survey}. Usually, the retrieved documents are provided as part of the prompting context in a \gls{zs} setting, enabling the \gls{llm} to complete tasks requiring knowledge which has not implicitly been stored in the \gls{llm} during the pretraining phase. We adopt this idea and propose \gls{ragg} by leveraging the vast \gls{gui} prototyping knowledge embodied in a large-scale \gls{gui} repository for \gls{llm}-based \gls{gui} generation. Subsequently, we briefly delineate the employed \gls{gui} repository and its preprocessing, a novel \gls{llm}-based \gls{gui} reranking and filtering approach to ensure high relevance of retrieved \glspl{gui}, and our prompting approaches to integrate knowledge embodied in the \gls{gui} screens for \gls{gui} generation. An overview of the proposed \gls{ragg} approach is provided in Figure \ref{fig:overview-zero-shot} \textit{(B)}.

\paragraph{\gls{gui} Retrieval Approach} To enable \gls{ragg}, we employ the previously developed \gls{nlr}-based \gls{gui} retrieval approach. Particularly, the approach relies on the filtered \textit{Rico} \gls{gui} repository \citep{deka2017rico} and computes ranking scores over the dataset with \gls{sbert} by matching \gls{nlr} against the \textit{\gls{s2w}} data (see Chapter \ref{cha:self_elicitation} Section \ref{subsec:sergui_dataset} for details).

\paragraph{Adapted \gls{gui} Reranking.} As shown in Figure \ref{fig:overview-zero-shot} \textit{(B)}, we first retrieve the top-\textit{n} relevant \glspl{gui} with \gls{sbert}, then apply a binary relevance filter using \gls{zs} prompting. In contrast to \textit{\gls{gui}-ReRank} (see Chapter \ref{cha:gui_rerank}), we initially conduct binary relevance filtering to ensure that only highly relevant \gls{gui} prototypes are provided to the generation context. We \textit{base64}-encode each top-\textit{n} \gls{gui} screenshot and provide it to an \gls{mllm} as part of the prompt, then instruct the \gls{mllm} to decide whether the \gls{gui} is relevant given the requirements. To ensure high precision (thus avoiding inputting \glspl{fp}), we instructed the model to be critical and rather vote for non-relevant in cases of ambiguity. The binary decisions obtained from the \gls{llm} were observed to be consistent across multiple runs. Since the \gls{sbert} ranks of \glspl{gui} do not necessarily correspond with fine-grained relevance, we run a second-level \gls{zs} prompt instructing the \gls{llm} to provide a finer-grained relevance score (from 1 (\textit{not relevant}) to 10 (\textit{relevant})) and sort the binary-filtered \glspl{gui}. To avoid inconsistencies in the relevance scores (which is a finer-grained task compared to binary relevance and potentially more challenging), we sampled the \gls{llm} multiple times and computed the average and standard deviation over the obtained relevance scores, which indicated consistent ratings. The filtered and sorted \gls{gui} screenshots are used as input to the generation, ensuring that solely relevant \glspl{gui} are provided in the \gls{mllm} context.

\paragraph{Prompting Approaches (\gls{gui}-enriched Context).} We propose two distinct approaches for utilizing the retrieved \glspl{gui} for \gls{llm}-based \gls{gui} generation in combination with the requirements, as shown in Figure \ref{fig:overview-zero-shot} \textit{(B)}. First, \textit{(1)} we employ a \gls{zs} prompt for each \gls{gui} by \textit{base64}-encoding the screenshot in the prompt context of an \gls{mllm} and instructing the model to extract a feature collection from the \gls{gui}. Afterwards, we employ a second \gls{zs} prompt instructing the model to summarize and aggregate the \textit{k} feature collections of the top-\textit{k} \glspl{gui}. Semantically similar features should be aggregated and ranked higher in the collection based on their frequency. In addition, we employ a similar pipeline to extract design and layout for each of the top-\textit{k} \glspl{gui} and subsequently aggregate them. Aggregated feature and design collections are then provided in a \gls{zs} prompt instructing the model to generate \gls{html}/\gls{css} using the \gls{nlr} and the provided collections. Second, \textit{(2)} we directly \textit{base64}-encode the top-\textit{k} \glspl{gui} in the context of a \gls{zs} prompt instructing the \gls{mllm} to implicitly use the provided \glspl{gui} as inspiration for features, design, and layout while generating \gls{html}/\gls{css} for the prototype from \gls{nlr}.

\glsreset{sc}
\subsection{\gls{sc} Looping for GUI Generation}

\gls{sc} aims at enhancing the effectiveness of \gls{llm} task solving by employing the \gls{llm} itself to provide feedback on an \gls{llm}-generated solution (i.e. criticize one's own prior output) \citep{saunders2022self}. Particularly, the effectiveness improves when providing critiques compared to directly \gls{zs} tasking the \gls{llm} to refine the response. For enhancing \gls{zs}-based \gls{gui} generation, we adapt the concept and propose \gls{sc} looping for \gls{gui} generation. An overview of the approach is illustrated in Figure \ref{fig:overview-zero-shot} \textit{(C)}. Our self-critique approach consists of two main components. First, \textit{(1)} a feedback \gls{mllm} (i.e. \gls{gui} prototype critique) that utilizes the \gls{nlr} and \gls{gui} prototype \textbf{P$_{i}$} to generate feedback (i.e. critique) \textbf{F$_{i}$}. This feedback \gls{mllm} is based on a \gls{zs} prompt, instructing the model to provide feedback on additional relevant \gls{gui} features that the current prototype neglects, improvements to the implementation of the current features and improvements to the design and layout of the overall prototype. In addition, we instruct the \gls{mllm} to solely focus on textual descriptions and avoid providing code. Second, \textit{(2)} a \gls{gui} prototyping \gls{llm} that utilizes the requirements, prior \gls{gui} prototype \textbf{P$_{i}$} and the respective critique \textbf{F$_{i}$} to generate the subsequent \gls{gui} prototype \textbf{P$_{i+1}$}. In the prompt, we instruct the model to improve the current \gls{gui} prototype by integrating the provided feedback and respond with revised \gls{html}/\gls{css} code. To summarize, the self-critique approach employs two components:
\begin{align}
  \textit{Feedback-MLLM:} \quad \mathbf{F}_i &= \textbf{MLLM}(\mathbf{P}_i, \text{NLR}) \\
  \textit{Prototyper-LLM:} \quad \mathbf{P}_{i+1} &= \textbf{LLM}(\mathbf{P}_i, \text{NLR}, \mathbf{F}_i)
\end{align}

\noindent This approach requires an initial prototype \textbf{P$_{0}$} in \gls{html}/\gls{css} to be able to provide the initial critique. Therefore, we provide the \gls{zs} instruction prototype as a starting point (i.e. the \gls{zs} baseline prompt) to further refine the generated prototypes over \textit{k} iterations.

\subsection{\gls{llm}-based \gls{gui} Content Generation}

\begin{figure}
  \centering
  \includegraphics[width=0.85\textwidth]{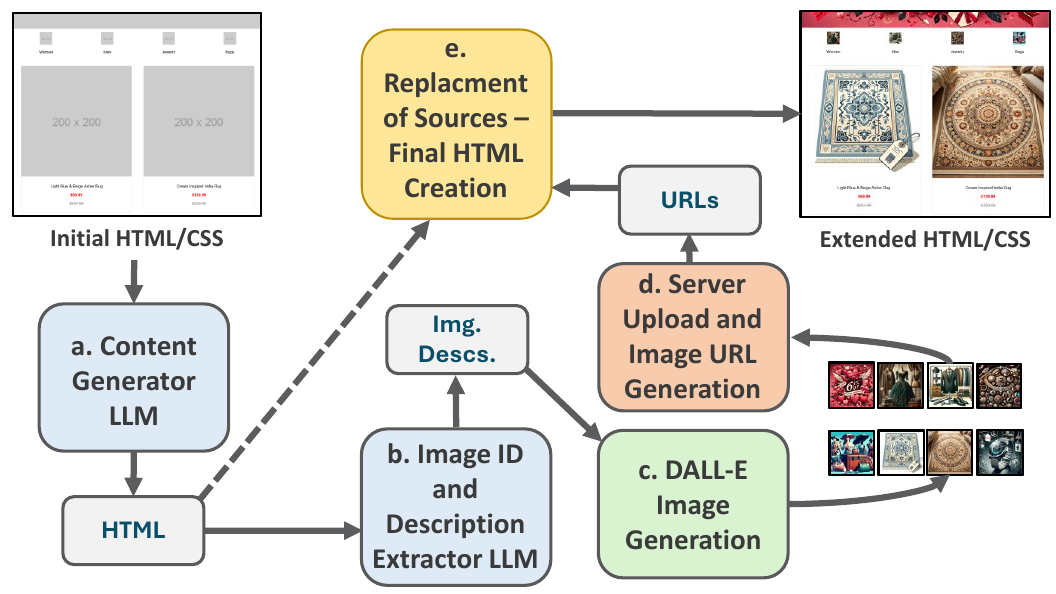}
  \caption[Overview of the \gls{llm}-based content generation pipeline]{Overview of the \gls{llm}-based content generation pipeline, which augments the initial prototype with content, generates images with \textit{DALL-E-3} and integrates them.}
  \label{fig:content_generation}
\end{figure}

To further improve the high-fidelity \gls{gui} prototypes, we propose a novel \gls{llm}-based \gls{gui} content generation pipeline illustrated in Figure \ref{fig:content_generation}. Often, generated \glspl{gui} from the proposed \gls{zs} approaches do not include realistic data (or sufficient example data at all) and contain generic content\footnote{At the time of implementing the proposed approach, we utilized the most advanced \gls{mllm} available from \textit{OpenAI}, namely \textit{\gls{gpt}-4o} \citep{hurst2024gpt}, which is an extension over prior \textit{\gls{gpt}-4} \citep{openai2023gpt4}.}. To enhance the realism and the appearance of the \gls{gui} prototypes using content, we first provide the base \gls{html}/\gls{css} to an \gls{llm} and instruct it to extend the prototype with realistic data and add \textit{id} attributes to each \textit{$<$img$>$} tag (e.g., \textit{add list items}, \textit{related products}, etc.). Subsequently, we instruct another \gls{llm} to extract a collection of all images contained in the \gls{html} and provide more detailed descriptions. We utilize an \gls{llm} for this step instead of directly extracting the \textit{$<$img$>$} tags with a library to enable the \gls{llm} to exploit the context to create better descriptions. These descriptions are then employed to generate corresponding images with \textit{DALL-E-3} \citep{betker2023improving}. The generated images are uploaded to a server, and the \glspl{url} are incorporated into the \gls{html}/\gls{css} by matching the extracted \textit{ids}.

\section{Experimental Evaluation}

This section presents the design and methodology of our experimental evaluation. The main goal of our evaluation is to assess the effectiveness of the different prompting approaches for \gls{gui} generation. Hence, we state the following four research questions:

\begin{itemize}[leftmargin=0.2in, rightmargin=0in, itemsep=0.1in]

    \item[-] \textbf{RQ$_{1}$}: \textit{Which \gls{llm} prompting approach is most effective for generating \glspl{gui} from \gls{nlr}?} We compare different prompting strategies to determine which yields the most accurate and user-aligned \gls{gui} prototypes. Effectiveness is assessed through human ratings on multiple qualitative criteria for the generated prototypes. In addition, we investigate the token consumption of the prompting techniques.
    
    \item[-] \textbf{RQ$_{2}$}: \textit{How does the number of examples in \gls{ragg} impact its effectiveness?} This question examines whether providing more \gls{gui} examples enhances the quality of \gls{llm}-generated \glspl{gui} for \gls{ragg}. We evaluate this by varying the number of input examples and analyzing changes in human ratings for the quality of \gls{gui} prototypes.
    
    \item[-] \textbf{RQ$_{3}$}: \textit{How does the number of loops in the \gls{sc} for \gls{gui} generation impact its effectiveness?} We assess if iterative \gls{sc} cycles lead to incremental improvements in generated \glspl{gui}. The impact is measured by comparing the quality of \glspl{gui} after multiple \gls{sc} iterations.
    
    \item[-] \textbf{RQ$_{4}$}: \textit{How does \gls{llm}-based content generation for \glspl{gui} influence the overall \gls{gui} quality?} We explore whether automatically generated content increases the realism and quality of \gls{gui} prototypes, as perceived by human annotators. In particular, we compare user ratings of \glspl{gui} with and without \gls{llm}-generated content and images.

\end{itemize}

\subsection{\gls{gui} Requirements Dataset}

\begin{figure*}
  \centering
 \includegraphics[width=1\textwidth]{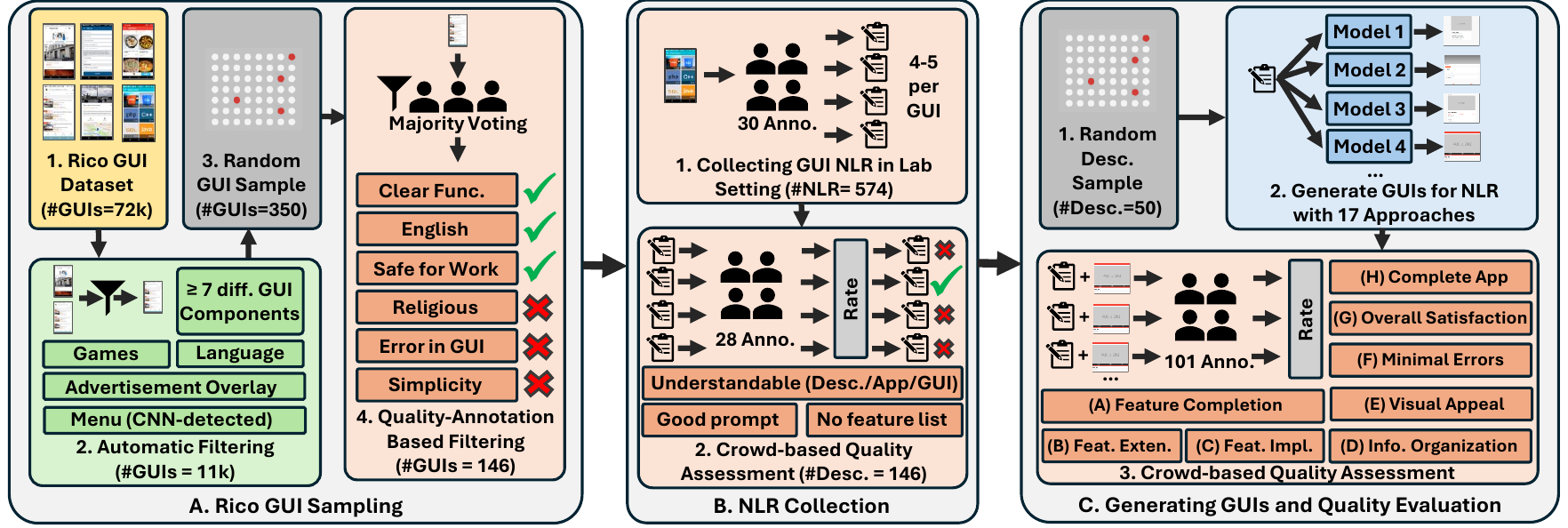}
  \caption[Overview of the evaluation procedure]{Overview of the evaluation: \textit{(A)} Sampling and filtering \glspl{gui} from the \textit{Rico} \gls{gui} dataset, \textit{(B)} collecting and assessing \gls{nlr} for corresponding \glspl{gui}, and \textit{(C)} generating and evaluating \glspl{gui} using multiple models and their crowd-based quality assessment.}
  \label{fig:OverviewEvaluation-zs}
  \vspace{-0.3cm}
\end{figure*}

In this section, we present the conducted procedure to collect the \gls{gui} \gls{nlr} dataset to evaluate the research questions, as illustrated in Figure \ref{fig:OverviewEvaluation-zs}. First \textit{(1)}, \textit{Rico} \citep{deka2017rico} \glspl{gui} were sampled, for which requirements were collected in a lab-based setting in the following step \textit{(2)} and then evaluated by crowd-workers to increase data quality \textit{(3)}.

\vspace{-0.2cm}
\paragraph{Rico \gls{gui} Sampling.} A \gls{gui} prototype selection was required before participants could create \gls{nlr} for them. Again, we used the established \textit{Rico} \gls{gui} dataset \citep{deka2017rico}. To build the collection of \textit{Rico} \glspl{gui} serving as a foundation for the following data collection, we performed multiple steps to increase the quality and diversity of \glspl{gui}. First, we applied basic \gls{gui} filtering (see Chapter \ref{cha:self_elicitation} Section \ref{subsec:sergui_dataset}). To ensure the inclusion of feature-wise more comprehensive \glspl{gui} (to increase the difficulty for the \gls{gui} generation), we removed \glspl{gui} with fewer than seven different \gls{gui} component types. Then, three research assistants rated each \gls{gui} for six predefined categories 
(\textit{Clear Funct.}, 
\textit{English}, 
\textit{Safe for Work},
\textit{Errors}, 
\textit{Religious Content}, 
\textit{Simplicity}), generating 1,050 ratings in total. Research assistants were provided with clear instructions, as well as positive and negative examples for each of the dimensions. On average, each \gls{gui} received three sets of ratings. These categories were selected to ensure ethical and practical suitability: \textit{Safe for Work} and \textit{Religious Content} excluded problematic or sensitive screens, while \textit{Errors} addressed technical issues common in \textit{Rico} \glspl{gui} \citep{leiva2020enrico}. \textit{English} and \textit{Clear Functionality} ensured participants could understand the \gls{gui} without extra context, and \textit{Simplicity} helped us focus on more challenging, feature-rich \glspl{gui} for robust evaluation. Participants rated each criterion on a binary \textit{yes/no} basis. We decided to use majority voting for all aspects (i.e., 2/3 participants had to approve the respective category), except for the \textit{Safe for Work} category. Here, a \gls{gui} was already screened out if this category was rated accordingly by one of the three annotators.
Consequently, 146 \glspl{gui} were admitted.

\paragraph{Collecting \gls{nlr} for \gls{gui} Prototypes.} We recruited 30 participants (21 male, nine female), with an average age of 23.67 years ($\sigma$ = 3.25), from a university panel to create \gls{nlr} of the previously randomly sampled and filtered \glspl{gui} that later served as the basis for our automated \gls{gui} prototype generation.
We chose a lab-based setting instead of sourcing crowd-workers for this step to ensure that participants were not using \glspl{llm} to generate the \gls{gui} prototype \gls{nlr}, since recent work \citep{veselovsky2023artificialintelligence} has shown that crowd-workers broadly use \glspl{llm} for text-generation tasks. Each participant created 20 \gls{gui} prototype \gls{nlr}. 600 \gls{gui} prototype \gls{nlr} were created in total, with an average of 4.11 \gls{nlr} for each of the 146 individual \glspl{gui}.
We then had a research assistant examine the created \gls{gui} prototype \gls{nlr} to identify \gls{nlr} we would have to sort out (e.g., because of sensitive content or personal data). Consequently, 26 \gls{nlr} were finally sorted out. Our final collection amounted to 574 \gls{gui} \gls{nlr}. Next, data quality assurance followed.

\paragraph{Evaluating \gls{nlr} for \gls{gui} Prototypes.} Subsequently, we asked crowd-workers to rate the created \gls{gui} \gls{nlr}, filtering for quality and ensuring understandability. Additionally, we expected the \gls{gui} \gls{nlr} to not solely represent a listing of single features since such a granular feature level is usually unavailable when generating first prototypes. We also required \gls{nlr} that allow understanding of not only the individual \gls{gui} page (essential functionality), but also the overall app.
To rate the previous \gls{gui} \gls{nlr}, we invited 30 crowd-workers from \textit{Prolific} \citep{palan2018prolific}, requiring \textit{English skills (B2)} and \textit{experience in \gls{uiux} design}. We excluded two participants for failing our attention checks, leaving us with 28 crowd-workers (20 male, seven female, and one non-binary). Participants had an average age of 30.50 ($\sigma$ = 7.13), self-reported 3.79 ($\sigma$ = 3.55) years of experience creating and 2.71 years ($\sigma$ = 3.08) of experience evaluating visual design.

Crowd-workers rated 574 \gls{gui} \gls{nlr} for 146 \glspl{gui} (on average, 3.93 \gls{nlr} for one \gls{gui}). Participants, on average, created 1.95 ratings per \gls{gui} \gls{nlr} on 9-point Likert scales (including \textit{Description Understandability}, \textit{App Purpose Understandability},
\textit{\gls{gui} description mainly a list of features},
\textit{\gls{gui} description a prompt for Gen AI to create a first \gls{gui} draft},
\textit{Understandable overall app's purpose},
\textit{Understandable individual \gls{gui} screen}).
To select a single suitable \gls{nlr} per \gls{gui}, we created weighted averages of ratings for each created \gls{gui} \gls{nlr}. For each \gls{gui}, we then selected the \gls{nlr} which was rated the highest. From the 146 high-quality \gls{gui} \gls{nlr} with an average of 18.77 ($\sigma$ = 8.18) words per \gls{nlr}, we randomly sampled ten examples as our validation dataset, which we used to conduct internal experiments. Next, we randomly sampled 50 \gls{nlr} from the remaining examples to obtain our final test dataset. For the feasibility of conducting the expensive and labor-intensive annotation of generated \glspl{gui} by crowd-workers with \gls{uiux} experience, we restricted the test set to 50 examples, while still encompassing diverse \gls{nlr} for \glspl{gui}.

\subsection{RQ$_{1}$: Effectiveness of Prompting for \gls{gui} Generation}
\label{sec:rq2}

\paragraph{Annotation of Generated \gls{gui} Prototypes.} The evaluation of the approaches focuses on assessing the created \glspl{gui}. We utilized working prototypes in \gls{html}/\gls{css} for this assessment, and humans evaluated the \glspl{gui} based on multiple metrics given the \gls{gui} \gls{nlr}, as illustrated in Figure \ref{fig:OverviewEvaluation-zs} \textit{(C)}. For this purpose, we hired crowd-workers with \gls{uiux} experience on the crowd-working platform \textit{Prolific} \citep{prolific_academic_ltd_prolific_nodate}. Participants on \textit{Prolific} offer better annotation quality in comparison to other crowd-working platforms \citep{douglas2023data} and were required to have a high \textit{approval rate} ($>99\%$), \textit{minimum number of previously completed tasks} ($>30$), \textit{experience in \gls{uiux} design} (self-reported), and \textit{English skills}.

Out of 126 survey submissions, we meticulously screened out 25 participants, ensuring a high-quality dataset (eight participants did not complete the survey, and 17 participants failed one or more attention checks). The following data is reported for the 101 participants (73 male, 27 female, and one non-binary) accepted to our study. On average, participants reported an age of 30.86 years ($\sigma$ = 9.21), 4.60 years of experience creating visual design ($\sigma$ = 4.61), and 3.58 years of experience evaluating visual design ($\sigma$ = 3.72). 
49\% of our participants reported \textit{somewhat high} to \textit{very high} (85\% \textit{medium} or higher) knowledge of creating \gls{gui} prototypes and 39\% \textit{somewhat high} to \textit{very high} (78\% \textit{medium} or higher) skills in creating \gls{gui} prototypes.

\definecolor{lightgray}{rgb}{0.93, 0.93, 0.93}


\definecolor{lightgray}{gray}{0.92}

\newlength{\ItemRowHeight}
\setlength{\ItemRowHeight}{2.0\baselineskip} 

\newcommand{\QCell}[1]{%
  \parbox[c][\ItemRowHeight][c]{\hsize}{#1}%
}
\newcommand{\ACell}[1]{%
  \parbox[c][\ItemRowHeight][c]{0.28\linewidth}{%
    \itshape\raggedright%
    \hyphenpenalty=10000\exhyphenpenalty=10000 
    #1%
  }%
}

\begin{table}[t]
\footnotesize
\caption[Likert-scale items and self-reported assessments for evaluating \gls{llm}-generated \gls{gui} prototypes]{Eight Likert-scale items and corresponding self-reported assessments \textit{(A)}---\textit{(H)} employed to evaluate \gls{gui} prototypes generated by different \gls{zs} prompting approaches.}
\label{tab:questions_liker}

\setlength{\tabcolsep}{5pt}
\renewcommand{\arraystretch}{1.0}
\rowcolors{2}{lightgray}{white}

\begin{tabularx}{\linewidth}{l X p{0.15\linewidth}}
\toprule
\rowcolor{white}
\textbf{ID} & \textbf{Likert-scale question} & \textbf{Assessment} \\
\midrule

\rowcolor{white}
\textbf{A} & \QCell{This \gls{gui} prototype fulfills all functions defined in the \gls{gui} task description.}
          & \ACell{Feature completion} \\
\textbf{B} & \QCell{The \gls{gui} prototype's feature set is extensive, clearly exceeding the \gls{gui} task description by implementing additional useful functions.}
          & \ACell{Feature extensiveness} \\
\textbf{C} & \QCell{The features were perfectly implemented for the given \gls{gui} prototyping task.}
          & \ACell{Feature implementation} \\
\textbf{D} & \QCell{The organization of information on the \gls{gui} prototype page is clear.}
          & \ACell{Information organization} \\
\textbf{E} & \QCell{The visual design of the \gls{gui} prototype is appealing.}
          & \ACell{Visual appeal} \\
\textbf{F} & \QCell{This \gls{gui} prototype has only minimal errors.}
          & \ACell{Minimal errors} \\
\textbf{G} & \QCell{Overall, I am satisfied with this \gls{gui} prototype.}
          & \ACell{Overall satisfaction} \\
\textbf{H} & \QCell{The presented \gls{gui} prototype looks like a screen from a complete app.}
          & \ACell{Complete app} \\
\bottomrule
\end{tabularx}

\rowcolors{1}{}{} 
\end{table}

We conducted a comprehensive evaluation of three baseline models: \gls{zs} Instruction, \gls{zs}-\gls{cot} and \gls{zs}-\gls{cot} with content generation, alongside 14 prompting-based approaches. The latter include \gls{pd} variants (\gls{zs}, \gls{zs}-\gls{cot} and \gls{zs}-\gls{cot} with content generation), \gls{ragg} with direct encoding for different numbers of retrieved examples ($k = 1, 3, 5, 7$), \gls{ragg} with content generation ($k = 3$), and \gls{ragg} with explicit feature and layout extraction ($k=3$). Additionally, we evaluated \gls{sc} models with varying numbers of critique loops ($k = 1, 2, 3, 4$) and SC with content generation ($k = 3$). For each of these 17 models, we generated interactive \glspl{gui} for all 50 sampled \gls{gui} \gls{nlr}. Each \gls{gui} was then annotated three times using eight metrics, resulting in 20,400 ($17 \times 50 \times 3 \times 8$) annotations.
For conducting the annotation, participants first received a detailed explanation followed by comprehension checks. Participants were presented with the respective \gls{gui} \gls{nlr}, eight Likert scale items and a link to the interactive \gls{gui} prototype. We randomized the \gls{gui} selection and order in which \glspl{gui} were shown to each participant while ensuring that no participant received multiple \glspl{gui} generated from the same \gls{gui} \gls{nlr} (but different model) so that similar \glspl{gui} would not bias participants. Since each crowd-worker received 25.5 \glspl{gui} on average generated from 17 models (based on 50 \gls{gui} \gls{nlr}), each crowd-worker was randomly assigned a few \glspl{gui} generated from the same model, but not based on the same \gls{nlr}.
This rigorous evaluation allows us to systematically compare the strengths and limitations of each prompting-based approach and its parametrization for prototype generation via \gls{nlr}.

\vspace{-0.1cm}
\paragraph{Evaluation Metrics.} To assess the quality of prototypes, we evaluated \glspl{gui} based on eight measures. Table \ref{tab:questions_liker} presents the measured Likert items (9-point scale from \textit{Strongly Disagree} to \textit{Strongly Agree}) including items regarding the features and overall satisfaction.

\vspace{-0.1cm}
\paragraph{Model Setup.} We used the most recent \textit{GPT-4o}\footnote{At the time of implementing the proposed approach, we utilized the most advanced \gls{mllm} available from \textit{OpenAI}, namely \textit{\gls{gpt}-4o} \citep{hurst2024gpt}, which is an extension over prior \textit{\gls{gpt}-4} \citep{openai2023gpt4}.} model  (128k token context length, accessed in July, 2024), which is a multi-modal extension of the previous \textit{GPT-4} model \citep{openai2023gpt4}. We employed a consistent temperature of $t = .50$ for all models to balance creativity and consistency in the generated outputs (same for RQ2-4), which is an important factor for creative tasks such as \gls{gui} generation. To compare the models for significant differences, we conducted two analyses. The first analysis included the two \gls{zs} baselines, two \gls{pd} models and \gls{sc} ($k=4$) on the full dataset of 50 \gls{nlr}. The second analysis additionally included \gls{ragg} ($k=7$) and was conducted on a reduced (\textit{cleaned}) dataset of 15 \gls{nlr}, since at least 7 \glspl{gui} could be retrieved for only 15 \gls{nlr}. For both analyses, we then conducted the following procedure: to obtain paired data across models, we first aggregated (median) over multiple annotators per same \gls{gui} and prompting method (ensures paired observations across models). Subsequently, we initially computed \textit{Friedman tests} \citep{sheldon1996use} per metric to identify whether there are any significant differences between the set of models. For metrics with significant \textit{Friedman tests}, we computed post-hoc paired \textit{Wilcoxon signed-rank tests} (two-sided) \citep{woolson2007wilcoxon} to identify significant differences between individual models per metric and finally applied \textit{Holm} correction across model comparisons per evaluation metric.

\vspace{-0.3cm}
\subsection{RQ$_{2}$: Effectiveness of Example Number on \gls{ragg}}

To evaluate the influence of the number \textit{k} of retrieved \glspl{gui} for \gls{ragg}, we employed the approach with direct encoding of top-\textit{k} \glspl{gui} with a varying number of \glspl{gui} ($k = 1,3,5,7$). We restricted the maximum number tested in this setup, since only a smaller fraction of the entire dataset would be influenced by it (only 30\% of the top-20 retrieval sets of the test set contain seven or more relevant \glspl{gui} after applying \gls{mllm}-based filtering and reranking). As in RQ$_{1}$, we computed the results on the \textit{cleaned} dataset (15 \gls{nlr}) and computed \textit{Wilcoxon signed-rank tests} with \textit{Holm} correction within metrics.
\glsreset{sc}

\vspace{-0.3cm}
\subsection{RQ$_{3}$: Effectiveness of Number of Loops in \gls{sc}}

To evaluate the impact of the number of loops in the \gls{sc} approach on the effectiveness of \gls{gui} generation, we employed the \gls{zs} instruction output as initial input and the identical \textit{\gls{gpt}-4o} configuration as described before (see RQ$_1$ for details). Then, we evaluated this \gls{sc} model with five different settings ($k = 0,1,2,3,4$) of loop numbers. We computed \textit{Wilcoxon signed-rank tests} with \textit{Holm} correction within metrics on the full \gls{nlr} dataset.

\vspace{-0.3cm}
\subsection{RQ$_{4}$: Effectiveness of \gls{gui} Content Generation}

To evaluate the influence of the \gls{gui} content generation approach on the effectiveness of generated \gls{gui} prototypes, we employed two identical prompting methods \textit{with} and \textit{without} content generation, allowing us to measure statistically significant differences with the \textit{Wilcoxon signed-rank test}. We employed \textit{(i)} the \gls{zs}-\gls{cot} model for the baseline prompting approach, \textit{(ii)} the decomposed prompting approach (\gls{pd}-\gls{zs}), \textit{(iii)} \gls{ragg}$_{k=3}$ with direct encoding and \textit{(iv)} \gls{sc}$_{k=3}$ with \gls{zs} instruction output as the initial prototype.

\vspace{-0.6cm}
\section[Results \&\ Discussion]{Results \& Discussion}

\subsection{RQ$_{1}$: Effectiveness of Prompting for \gls{gui} Generation}
\glsreset{sc}
\glsreset{pd}
\glsreset{ragg}
\begin{figure*}
 \includegraphics[width=\textwidth]{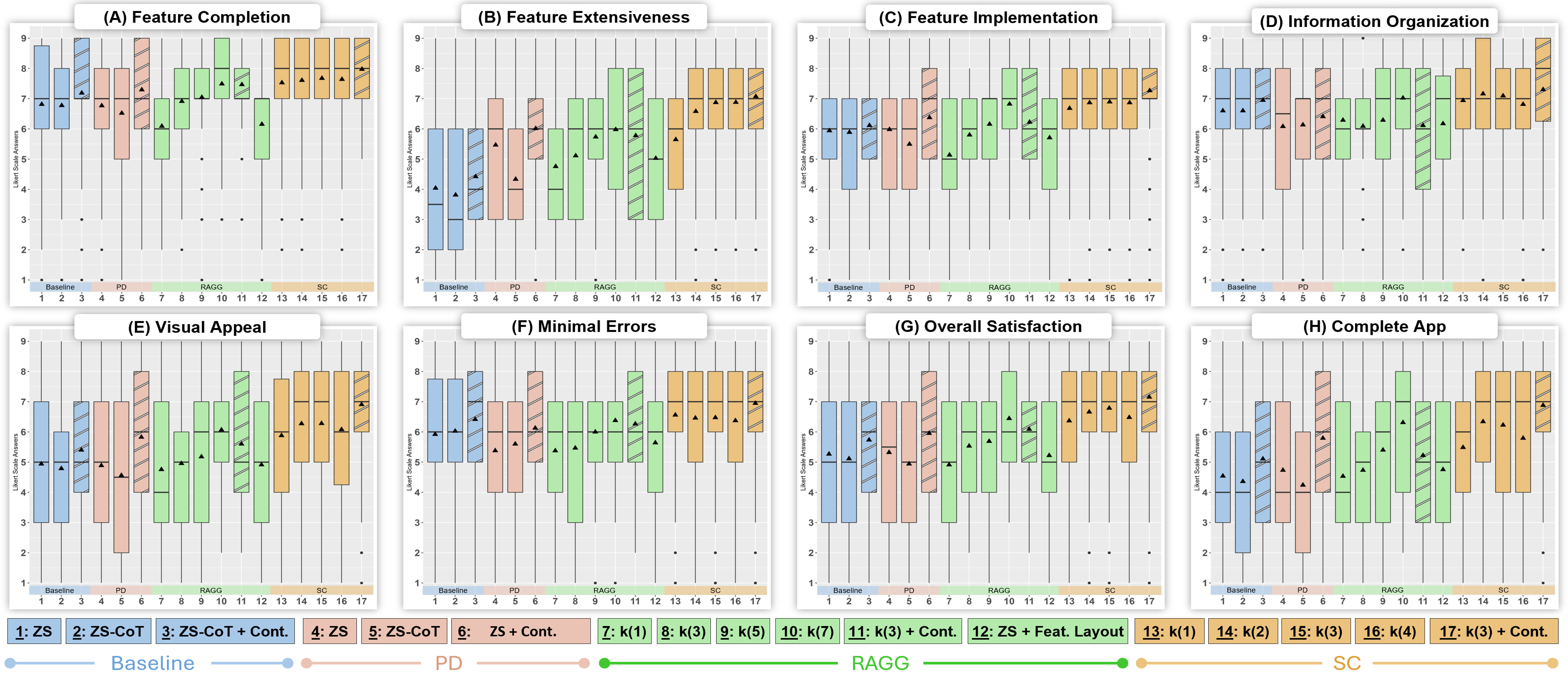}
  \caption[Boxplots for \gls{llm}-based \gls{gui} generation effectiveness]{Crowd-workers' \gls{gui} ratings across eight different \gls{gui} quality metrics and prompting approaches: baseline (\textit{1-3, blue}), \gls{pd} (\textit{4-6, red}), \gls{ragg}  (\textit{7-12, green,  cleaned set}), \gls{sc} (\textit{13-17, orange}). Hatched bars repr. models using content generation, triangles means.}
  \label{fig:results_main-zs}
\end{figure*}

To assess the effectiveness of our approaches for \gls{gui} generation, we evaluated the \gls{gui} annotations collected from crowd-workers. Figure \ref{fig:results_main-zs} presents boxplots for the ratings across the eight different metrics considered. Subsequently, we briefly summarize the results for the baselines (\gls{zs}, \gls{zs}-\gls{cot}), \gls{pd}, \gls{ragg} ($k = 7$) and \gls{sc} (loop number $k=4$) approaches (the two analyses), not considering \gls{gui} content generation.

\gls{pd} models, particularly when using the sequential prompts variant (\gls{pd}-\gls{zs}), significantly outperformed the baseline in \textit{(B)} \textit{Feature Extensiveness}. However, both \gls{pd} approaches did not show statistically significant improvements over the baselines across other metrics such as \textit{(D)} \textit{Information Organization} (cf. Appendix \ref{app:zs-generation} Table \ref{tab:analysisA_posthoc_holm}).

\gls{ragg} significantly surpassed at least one baseline on multiple metrics, with higher ratings overall (see Figure \ref{fig:results_main-zs}), though differences in \textit{(C)} \textit{Feature Implementation}, \textit{(D)} \textit{Information Organization}, and \textit{(F)} \textit{Minimal Errors} were not significant. Given that \gls{ragg}-generated \glspl{gui} are more comprehensive, showing statistically significant improvements across \textit{(A)} \textit{Feature Completion}, \textit{(B)} \textit{Feature Extensiveness}, \textit{(E)} \textit{Visual Appeal}, \textit{(G)} \textit{Overall Satisfaction}, and \textit{(H)} \textit{Complete App}, maintaining performance on \textit{(C)} \textit{Feature Implementation}, \textit{(D)} \textit{Information Organization}, and \textit{(F)} \textit{Minimal Errors} at least on par with the baselines is notable given the increased complexity (cf. Appendix \ref{app:zs-generation} Table \ref{tab:analysisB_posthoc_holm}).

\gls{sc} models outperformed baselines and \gls{pd} across nearly all criteria, except for \textit{(D)} \textit{Information Organization} and \textit{(F)} \textit{Minimal Errors} (and \textit{(H)} \textit{Complete App} for \gls{pd}-\gls{zs}). In particular, \gls{sc} achieved significantly higher ratings for \textit{(A)} \textit{Feature Completion}, \textit{(B)} \textit{Feature Extensiveness}, \textit{(C)} \textit{Feature Implementation}, \textit{(E)} \textit{Visual Appeal}, \textit{(G)} \textit{Overall Satisfaction}, and \textit{(H)} \textit{Complete App} (except for \gls{pd}-\gls{zs}) (cf. Appendix \ref{app:zs-generation} Table \ref{tab:analysisA_posthoc_holm}). Between \gls{sc} and \gls{ragg} models, we could not observe significant differences, except for \textit{(H)} \textit{Complete App}. Notably, for this metric, \gls{ragg} ($k=7$) achieved higher ratings overall compared to the \gls{sc} ($k=4$) approach (cf. Appendix \ref{app:zs-generation} Table \ref{tab:analysisB_posthoc_holm}).

\gls{ragg} might not improve over \gls{sc} in our experiments largely because \glspl{llm} might already encode substantial latent \gls{gui} prototyping knowledge from pretraining, particularly in general domains. The \gls{sc} process effectively unlocks and applies this expertise, enabling the model to refine its outputs without relying on external examples. Therefore, \gls{ragg} may offer limited gains when the intrinsic knowledge of the \gls{llm} is already sufficient. However, the improvements of \gls{ragg} over the baseline indicate that \glspl{llm} can leverage contextual \gls{gui} information when required, suggesting that retrieval-based augmentation may be especially valuable in niche or less familiar domains, where the pre-existing knowledge of the \gls{llm} is less comprehensive.

\begin{figure*}[!t]
  \centering
 \includegraphics[width=1\textwidth]{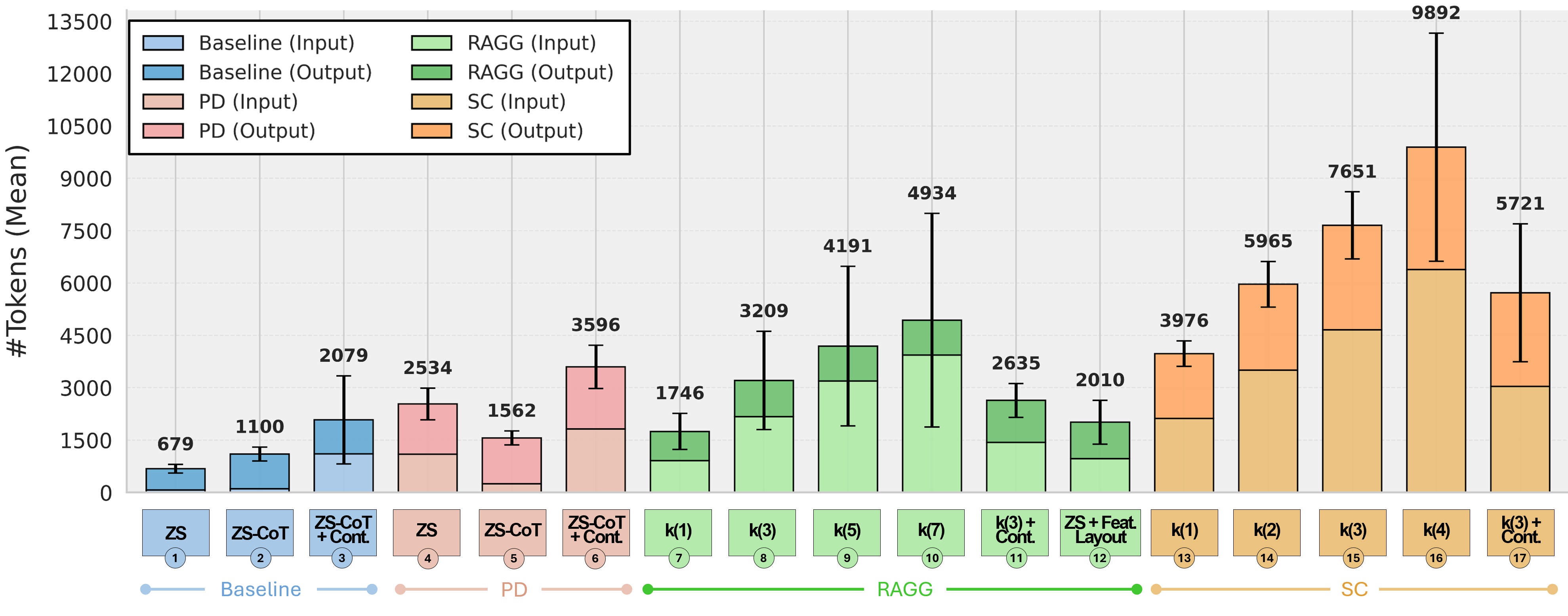}
  \caption[Mean input and output token usage for \gls{llm}-based \gls{gui} generation]{Mean input and output token usage and standard deviation per model configuration. Bars are stacked by input (\textit{lighter shade}) and output (\textit{darker shade}) tokens, grouped by method (\textit{Baseline, \gls{pd}, \gls{ragg}, \gls{sc}}). Error bars indicate one standard deviation for the total token count. Exact total mean tokens per prompting approach above bars.}
  \label{fig:EvaluationTokenConsumption}
\end{figure*}

Figure~\ref{fig:EvaluationTokenConsumption} illustrates the mean input and output token usage and the respective standard deviation for each prompting approach (for \gls{sc} the token usage per individual round). For \gls{ragg}, input token consumption increases with the number of retrieved examples, while output tokens remain relatively stable, reflecting that more context is provided to the \gls{llm}, but the generated amount of tokens for the \gls{gui} prototypes does not grow proportionally. In contrast, both input and output tokens increase with additional \gls{sc} iterations, as each round involves generating and critiquing new prototypes. In the baseline models, most tokens are output tokens because little prompting context is provided and the response is dominated by the \gls{html}/\gls{css} code. Although \gls{sc} and \gls{ragg} approaches consume more tokens than the baseline and \gls{pd}, the absolute token usage remains moderate and the associated costs are negligible in practice, especially since most performance gains for \gls{sc} are achieved after the first iteration. This suggests that even advanced prompting strategies can be employed effectively and cost-efficiently in real-world \gls{gui} prototyping workflows, providing higher quality of generated \glspl{gui}.

\begin{figure*}
  \centering
 \includegraphics[width=0.97\textwidth]{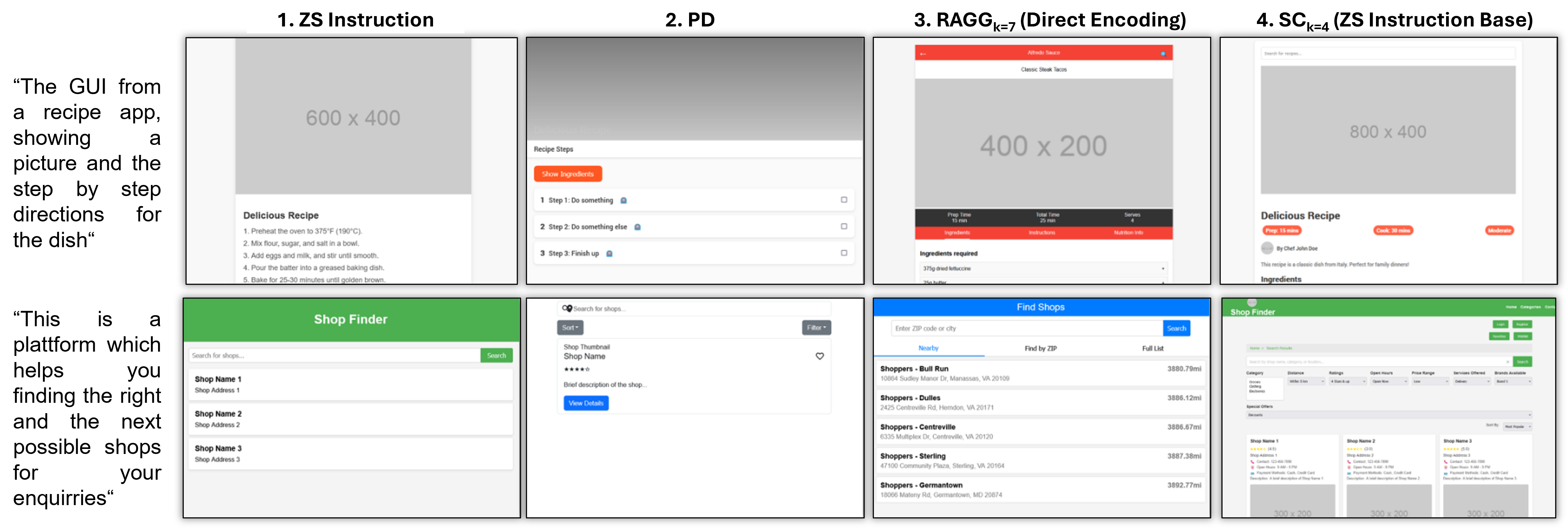}
  \caption[\gls{gui} examples from \gls{llm}-based \gls{gui} generation]{Two example requirements from the evaluation dataset with generated \gls{gui} prototypes for \textit{(1)} \gls{zs} instruction, \textit{(2)} \gls{pd}, \textit{(3)} \gls{ragg}$_{k=7}$ and \textit{(4)} \gls{sc}$_{k=4}$ illustrating an overall quality improvement towards the more comprehensive prompting approaches.}
  \label{fig:examples-zs}
  \vspace{-0.3cm}
\end{figure*}

To better understand the defects of \gls{llm}-generated \gls{gui} prototypes, we manually investigated a sample of \glspl{gui} with the lowest \textit{(F)} \textit{Minimal Errors} scores. The identified defects can be structured into three potentially non-exhaustive categories. First, \textit{(i)} defects related to the \textit{layout} occurred frequently. This includes \textit{misalignment of \gls{gui} components} (e.g., between labels and their respective check boxes), \textit{component overlapping} (e.g., labels and radio buttons or icons with a search bar), \textit{overflow of content} (e.g., images and text that exceed their grouping components) and \textit{underflow of content} (e.g., list items that only span half of the container width). Second, \textit{(ii)} defects related to \textit{function} included \textit{misinterpretation of requirements} (e.g., a navigation app provides buttons to navigate through the app), \textit{missing features} (e.g., upcoming shows not represented in a calendar although requirement explicitly stated in the description, no header or footer), \textit{unusual implementation of features} (e.g., implementing a save or share functionality with text buttons instead of more commonly used icons) and \textit{broken features} (e.g., broken image or icon source links). Third, \textit{(iii)} defects related to inconsistent styling of the \gls{gui} (e.g., inconsistent color schema or layouts). Figure \ref{fig:examples-zs} illustrates two requirements taken from our test set and their respective generated \gls{gui} prototypes for models from each group.

\begin{myrqbox}
    \textbf{Answer to RQ$_{1}$:} \gls{sc} significantly outperforms the baselines and \gls{pd} across most considered evaluation metrics.
Furthermore, \gls{ragg} models with seven examples ($k=7$) showed statistically significant improvements over at least one baseline in most metrics, but no statistically significant differences were observed between both \gls{ragg} ($k=7$) and \gls{sc} ($k=4$) techniques across all metrics, except for \textit{(H)} \textit{Complete App}.
\end{myrqbox}

\subsection{RQ$_{2}$: Effectiveness of Example Number on \gls{ragg}}

Considering the number of \gls{gui} examples in \gls{ragg}, we can observe a significant improvement between $k=1$ and $k=5$ for \textit{(A) Feature Completion}, but no significant improvement over the remaining metrics (cf. Appendix \ref{app:zs-generation} Table \ref{tab:ragg_k_posthoc_holm}). This suggests that the \gls{llm} can effectively utilize the examples to extract meaningful functionality relevant to the given requirements. Moreover, considering $k=1$ and $k=7$, we can observe statistically significant improvements for \textit{(A) Feature Completion}, \textit{(C) Feature Implementation}, \textit{(E) Visual Appeal}, and \textit{(H) Complete App}. This indicates that even more examples help the \gls{llm} not only to improve the feature-related aspects more (\textit{(A)}, \textit{(C)}, \textit{(H)}), but also non-functional aspects of the prototype, such as the \gls{gui} design \textit{(E)}. In addition, considering the increase of $k$, the mean ratings increased for each subsequent step (except for \textit{(D) Information Organization}), which suggests that -- while mostly not significant -- more examples potentially contribute to enhancing the overall \gls{gui} quality.

\begin{myrqbox}
\textbf{Answer to RQ$_{2}$:} A larger number of \gls{gui} examples in the prompt context for \gls{ragg} leads to increased mean ratings across metrics (except \textit{(D) Information Organization}), while significance is mainly observed between $k=1$ and $k=7$ for \textit{(A) Feature Completion}, \textit{(C) Feature Implementation}, \textit{(E) Visual Appeal}, and \textit{(H) Complete App}.
\end{myrqbox}

\vspace{-0.2cm}
\subsection{RQ$_{3}$: Effectiveness of Number of Loops in \gls{sc}}

In the first iteration ($k=1$, with $k=0$ representing our baseline), the improvements are significant for all metrics except \textit{(D)} \textit{Information Organization} and \textit{(F)} \textit{Minimal Errors} (cf. Appendix \ref{app:zs-generation} Table \ref{tab:sc_k_posthoc_holm}). These results show that both feature-related metrics and visual aspects are improved, while achieving better \textit{(D)} \textit{Information Organization} and \textit{(F) Minimal Errors} might be more difficult with an increasingly comprehensive feature collection. In addition, by comparing $k=2,3$ with $k=1$, we could observe a significant improvement for \textit{(B)} \textit{Feature Extensiveness}, \textit{(E)} \textit{Visual Appeal}, and \textit{(H)} \textit{Complete App}. This indicates that the \gls{sc} model performs plenty of improvements from the initial prototype to the first iteration and afterwards mainly new features are added, the visual design is improved and the \gls{gui} is extended for more \textit{(H) Complete App} (e.g., adding navigation). This could potentially be improved by dynamically focusing the feedback \gls{mllm} more on currently underdeveloped parts of the \gls{gui}. Overall, we recommend a single \gls{sc} iteration ($k=1$) as the default, considering the increasing token consumption.
\vspace{-0.2cm}

\newcommand{\ci}{\text{CI}}

\vspace{0cm}
\begin{myrqbox}
\textbf{Answer to RQ$_{3}$:} The first round ($k = 1$) of \gls{sc} leads to significant improvements in most aspects of \gls{gui} quality, while further iterations ($k>1$) mainly enhance \textit{(B) Feature Extensiveness}, \textit{(E)} \textit{Visual Appeal}, and \textit{(H) Complete App}. Most gains from \gls{sc} are achieved early, with diminishing returns later (token consum. can be kept small).
\end{myrqbox}

\subsection{RQ$_{4}$: Effectiveness of \gls{gui} Content Generation}

Content generation showed a statistically significant improvement for \gls{zs}-\gls{cot} across multiple metrics, except for \textit{(C) Feature Implementation} and \textit{(D) Information Organization} (cf. Appendix \ref{app:zs-generation} Table \ref{tab:content_diversification_wilcoxon}). Similarly, significant improvements are observed for the \gls{pd}-\gls{zs} method, with exceptions of \textit{(B) Feature Extensiveness}, \textit{(C) Feature Implementation}, and \textit{(D) Information Organization}. These results indicate that both methods benefit from the content generation approach. While no statistically significant differences were observed between the \gls{ragg} and \gls{sc} (except for \textit{(H) Complete App}) methods, the mean metric ratings are consistently higher for the content variant across all considered metrics, as shown in Figure \ref{fig:results_main-zs}. The results suggest that content generation mainly benefits the simpler methods, while \gls{ragg} already benefits from multiple relevant \gls{gui} examples to improve its content and \gls{sc} potentially enhances its content by its internal critique.

\begin{myrqbox}
\textbf{Answer to RQ$_{4}$:} \gls{llm}-based \gls{gui} content generation significantly enhances \gls{zs}-\gls{cot} and \gls{pd}-\gls{zs} across metrics (except for \textit{(B) Feature Extensiveness}, \textit{(C) Feature Implementation}, and \textit{(D) Information Organization}). \gls{ragg} and \gls{sc} achieve consistently higher mean ratings, but differences are not significant. These findings highlight the value of adding realistic, context-aware content and images for better \gls{gui} prototype quality.
\end{myrqbox}
\section{Threats to Validity}

\paragraph{Internal Validity.} One potential threat to internal validity is the relationship between measured items. In particular, a larger number of features in the generated prototypes might increase the difficulty of properly organizing the information, increase the number of feature implementation errors and increase the difficulty of creating appealing visual designs. Since the simpler baseline methods usually only implement a small number of features and the other metrics are not normalized on feature count or complexity, this could potentially favor the simpler methods in the mentioned items. Another threat is the selection of participants to evaluate the generated \gls{gui} prototypes. To reduce bias, we solely included participants with self-reported experience in \gls{gui} design, included only participants that were fluent in English, had at least 30 prior submissions, and had an approval rate of over 99\%. In addition, recent research showed that crowdworkers on \textit{Prolific} create better data quality in comparison to other popular crowdworking platforms \citep{douglas2023data}. Furthermore, we excluded annotations from crowdworkers that failed one or more of overall five attention checks, to improve data annotation quality.

\paragraph{External Validity.} One threat to external validity is the description dataset that we employed in our experiments. Since the descriptions are based on the \textit{Rico} \gls{gui} dataset, the generalizability of our findings might be restricted. For example, \glspl{gui} from special domains which are rare might not be included in the \textit{Rico} dataset. For these rare \glspl{gui}, the descriptions we could obtain from users might also contain more ambiguity or obscurity. To ensure a representative sample of functionally more complex \glspl{gui} employed for creating the descriptions, we filtered the \glspl{gui} by containing at least seven unique \gls{gui} component types and then applied random sampling. Furthermore, we asked annotators to detect functionally simple \glspl{gui} (e.g., \textit{login}), which we then excluded by majority voting. This increased the inclusion of more complex \glspl{gui}, increasing the difficulty of the \gls{gui} generation task. In addition, for more than 20\% of the descriptions, the retrieval approach was not able to find relevant \glspl{gui}, showing the complexity of descriptions and the inclusion of rare \glspl{gui}. Another threat to external validity is that we employed only a single \gls{llm} in our experiments. However, at the time of evaluation, this model achieved state-of-the-art performance on many benchmarks, meaning that the baseline capabilities of the vanilla \gls{llm} were already very strong. As a result, demonstrating further improvements with shown \gls{zs} techniques may be more challenging, and observed gains should be interpreted in the context of an already high-performing baseline \gls{llm}.
\section{Limitations}

While we presented several advantages and discussed the effectiveness of our proposed \gls{zs} prompting approaches for \gls{gui} generation, they also possess several limitations. First, the \gls{llm}-based \gls{gui} reranking approaches significantly outperform state-of-the-art methods (see Chapter \ref{cha:gui_rerank}), but still produce \glspl{fp} in some cases. Although the \glspl{fp} are usually not entirely irrelevant to the posed problem, these false instances might still negatively influence the \gls{gui} generation effectiveness of the \gls{ragg} approach. Therefore, \gls{ragg} is highly dependent on the retrieval method and underlying \gls{gui} repository. For requirements for which neither the retrieval model nor the \gls{mllm}-based filtering and reranking approach is able to find relevant \glspl{gui} or the \gls{gui} repository lacks relevant \glspl{gui} overall, the \gls{gui} generation falls back to the baseline \gls{zs} instruction approach. In addition, \gls{ragg} would potentially work most effectively if the knowledge embodied in the external \gls{gui} repository and internally stored in the \gls{llm} is complementary to each other. Due to the fact that \gls{gui} retrieval on \textit{Rico} as well as the \gls{llm} are both data-driven models, the distribution of seen \gls{gui} knowledge could be similar. However, in a more particular domain with custom characteristics of the \gls{gui} prototypes, which the \gls{llm} probably will less likely have stored, the \gls{ragg} approach might be able to outperform \gls{llm}-only methods. While \gls{sc} shows high effectiveness for generating \glspl{gui} as well, it consumes a larger amount of tokens while providing the critique and re-generating the prototype compared to baselines. This could potentially be enhanced in the future by a more sophisticated re-generation approach, e.g., updating only affected \gls{gui} segments.

\section{Related Work}
\label{sec:zs-related-work}

Earlier research proposed several \gls{ml}-based \gls{gui} generation approaches. \textit{Variational transformer networks} \citep{arroyo2021variational} and \textit{LayoutTransformer} utilizing \textit{self-attention} \citep{gupta2021layouttransformer} have been proposed to generate high-level layouts, including basic low-fidelity \gls{gui} layouts with abstract \gls{gui} component rectangles. In line with previous approaches, mixing \textit{transformer encoder-decoder} with \textit{Gaussian mixture models} \citep{bishop1994mixture} for generating abstract low-fidelity \glspl{gui} from text has been proposed before \citep{huang2021creating}. While these approaches can provide coarse design guidance, the generated layouts lack functional and design details compared to the interactive prototypes generated by our approach. Moreover, \textit{\gls{gui}GAN} \citep{zhao2021guigan} utilizes \textit{Generative Adversarial Networks (GANs)} \citep{goodfellow2020generative} to generate \gls{gui} prototypes based on \gls{gui} subtrees, i.e. image-based excerpts extracted from \textit{Rico} \glspl{gui}. While the \gls{gui} prototypes generated by their approach possess a realistic appearance, they are image-based and therefore not interactive, which restricts the usefulness compared to the \gls{gui} prototypes generated by our approach.

More recent approaches focused on \glspl{llm} for generating \gls{gui} prototypes of different fidelity. To generate low-fidelity \gls{gui} layouts based on high-level textual layout descriptions and a collection of selected \gls{gui} component types, \textit{Instigator} \citep{brie2023evaluating} trains a full \textit{minGPT} \citep{minGPT} model from scratch exploiting a large-scale repository of automatically scraped web pages, which are then transformed to low-fidelity layouts for training. Instead of training a task-specific \gls{gpt} model from scratch, another approach has been proposed which finetunes a pretrained \gls{llm} for generating \glspl{gui} exploiting the \textit{Rico} \gls{gui} repository \citep{feng2023designing}. While both approaches generate \gls{gui} layouts or prototypes from text, they require resource-intensive training, create low-fidelity prototypes and generate merely \glspl{dsl}.

Similarly, \textit{MAxPrototyper} \citep{yuan2024maxprototyper} generates a custom \gls{dsl} to represent \gls{gui} prototypes from short text descriptions and a prespecified \gls{gui} layout using an \gls{llm}. However, their approach focuses mainly on generating \gls{gui} content in the form of text and images matching the provided text description. \textit{UIDiffuser} \citep{wei2023boosting} utilizes \textit{stable diffusion} \citep{rombach2022high} to directly generate \gls{gui} prototype images from short text descriptions. While this approach can provide coarse design inspirations, the generated images lack clarity and do not represent functioning \gls{gui} components.
\section{Conclusion}

The proposed approaches in this chapter were driven by the challenge \challonefour{}: \textit{How can we efficiently optimize \glspl{llm} for more effective \gls{gui} prototype generation?} To tackle this challenge, we introduced several \gls{zs} prompting methods for high-fidelity \gls{gui} generation. Based on 20,400 expert annotations, \gls{sc} and \gls{ragg} mostly outperformed \gls{zs} baselines. \gls{ragg} benefited from more retrieved examples, and \gls{sc} delivered most of its gains after a single iteration, with both methods incurring moderate token costs. However, \gls{ragg} would perform most effectively when the \gls{llm} knowledge and \gls{gui} dataset are more complementary to each other. \gls{llm}-based content generation further improved perceived \textit{visual quality}, \textit{completeness}, and \textit{overall satisfaction}, underscoring that \gls{zs} prompting is a practical alternative to training or finetuning for generating interactive \gls{gui} prototypes.

\chapter{Closing the Loop Between User Stories and GUI Prototypes: An LLM-based Assistant for Cross-Functional Integration in Software Development}
\chaptermark{Closing the Loop Between User Stories and GUI Prototypes}
\label{cha:closing}

After presenting \gls{nlr}-based \gls{gui} retrieval and reranking approaches in the first part of the thesis, followed by \gls{zs}-based prompting approaches for efficiently adapting \glspl{llm} for \gls{nlr}-based \gls{gui} generation at the beginning of the second part, we continue with challenge \challone{}. In particular, we provide a novel approach for efficiently adapting \glspl{llm} to generate fully editable \gls{gui} prototypes in proprietary representations, which the \glspl{llm} have not been pretrained on. The following sections are based on previously published work \citep{kretzer2025closing}\footnote{This section is adapted from: Kretzer, Felix\textsuperscript{*}, \textbf{Kolthoff, Kristian\textsuperscript{*}}, Bartelt, Christian, Ponzetto, Simone Paolo, and Maedche, Alexander. Closing the Loop Between User Stories and GUI Prototypes: An LLM-based Assistant for Cross-Functional Integration in Software Development. In \emph{Proceedings of the 2025 CHI Conference on Human Factors in Computing Systems (CHI, A*)}, Yokohama, Japan, April 2025, pages 1--19. ACM. *Authors contributed equally. Sections are rearranged and directly reused with only minor adaptations.}. Our prompts and demonstration video are publicly available\footnote{Materials for this chapter are available at \url{https://github.com/kristiankolthoff/Closing-the-Loop-US-GUI} and the respective demo video at \url{https://youtu.be/QQd007gJLHQs}}.

\paragraph{Personal Contribution.} \textit{Felix Kretzer} and I contributed equally to the ideation and creation of the concept for the work. While I implemented the \gls{llm}-based approaches and backend functionality, \textit{Felix Kretzer} implemented the \gls{gui} and frontend of the approach. \textit{Felix Kretzer} and I contributed equally to the evaluation design, while \textit{Felix Kretzer} conducted the evaluation and analysis of results. I wrote the section on \gls{llm}-based approaches (and created the architecture overview figure) and related work, while \textit{Felix Kretzer} wrote the rest of the original paper and created respective figures and plots.


\vspace{-0.3cm}
\section{Motivation}

\begingroup
\renewcommand{\thefootnote}{\fnsymbol{footnote}} 

\endgroup

While \gls{nlr}-based \gls{gui} generation methods with \glspl{llm} from the previous chapter are effective in creating \gls{gui} prototypes of high quality in common representations, on which the \glspl{llm} are pretrained, such as the investigated \gls{html}/\gls{css}, these representations restrict direct editing and customization via commonly employed visual \gls{gui} prototyping editors, which enable direct manipulation of visual prototype components. Therefore, recent \gls{llm}-based approaches for \gls{gui} generation may be effective, but are currently not integrated into typical development workflows\footnote{At the time of writing the paper, to the best of our knowledge, we were the first to investigate the integration of \glspl{llm} into typical \gls{gui} prototyping workflows via a novel \gls{rag}-based integration approach.}.

\glsreset{us}
In practice, \gls{uiux} designers create prototypes based on requirements provided in different forms. For example, requirements are sometimes only discussed verbally and then quickly turned into paper prototypes. However, a prominent approach is to explicitly articulate requirements, for example, as \glspl{us} \citep{cohn2004user}. Requirements are sometimes formalized after building initial \gls{gui} prototypes, leading to ambiguity during early \gls{gui} prototyping stages. Software developers typically work with \gls{gui} prototypes and formalized requirements, where changes during the technical implementation can create a synchronization effort for both the formalized requirements and \gls{gui} prototypes. We found that updating requirements in prototypes can be overlooked, making the implemented version the de facto latest version.

Moreover, the creation of \gls{gui} prototypes typically necessitates the collaboration of \gls{uiux} designers, requirements analysts (e.g., product owners), and software developers. In practice, \gls{sd} teams often encounter different tools that are only partially integrated \citep{palani2022}. Although efforts are being made to integrate workflows (e.g., \textit{Figma's} developer mode \citep{figma_tool}), core processes are still entirely separate. Furthermore, requirements are subject to continuous adaptation and extension during the development \citep{debnath_re}. Changing requirements leading to extensive communication effort has been described in the literature, and was one of the most frequently mentioned topics during interviews conducted as part of this study. Due to the sustained modification of requirements, not only are synchronization efforts between the roles in the development teams additionally increased, but also the effort to continuously update the respective \gls{gui} prototypes. To summarize, \gls{llm}-based approaches are effective in generating \gls{gui} representations that are not suited for visual prototype editing and are not well integrated into typical \gls{gui} prototyping workflows. Thus, the work in this chapter is driven by the leading research question of challenge \challonefive{}: \textit{How can \gls{llm}s be efficiently adapted to generate editable \gls{gui} prototype representations from \gls{nlr}?}

\noindent To tackle this challenge, we devise a novel two-stage \gls{rag} approach, which enables the efficient integration of proprietary \gls{gui} component libraries (such as \textit{Material Design} \citep{material_design_kit}) into \gls{llm}-based \gls{gui} generation, thereby allowing the generation of proprietary \gls{gui} representations. In particular, we provide an implementation of the approach for the popular \gls{gui} component library \textit{Material Design} \citep{material_design_kit} within the well-known prototyping editor \textit{Figma} \citep{figma}. Moreover, we propose a novel \textit{Figma} plugin that integrates the novel \gls{rag}-based approach for generating proprietary \gls{gui} prototypes. In addition, it closely integrates requirements specifications (e.g., \gls{us} collections) and the corresponding \gls{gui} prototype implementations through several \gls{llm}-based matching techniques. Finally, we evaluated the proprietary \gls{gui} representation generation approach by crowdsourcing-based annotation of generated artifacts and assessed the effectiveness of our overall approach through a user study, indicating that our approach is able to effectively support users during the \gls{gui} prototyping process.  

\paragraph{Contributions.} With this work, we make the following three research contributions: 

\begin{itemize}[left=0.1cm]
    \item \textit{Two-stage \gls{rag} approach for adapting \glspl{llm} to proprietary \gls{gui} representations:} we present a two-stage \gls{rag} approach, which enables the efficient integration of proprietary \gls{gui} component libraries for generating proprietary \gls{gui} prototype representations. In addition, we provide an implementation of this approach for the popular \textit{Material Design} \citep{material_design_kit} \gls{gui} component library within \textit{Figma} \citep{figma}.
    \item \textit{Novel Figma plugin closely integrating \glspl{llm} into \gls{gui} prototyping:} we present a novel plugin for \textit{Figma} with role-specific functionalities. The interface for \gls{uiux} designers integrates an approach within a prototyping environment (e.g., \textit{Figma}) that utilizes a state-of-the-art \gls{llm} to \textit{(i)} assess whether a requirement is completed, \textit{(ii)} identify \gls{gui} components completing the requirement (i.e. \gls{gui} component and \gls{nlr} matching) and \textit{(iii)} generate recommendations for \gls{uiux} components fulfilling a given requirement. 
    \item \textit{Comprehensive evaluation and insights:} we conduct an empirical evaluation of the functionalities of our proposed \gls{gui} prototyping assistant for \gls{uiux} designers and the resulting prototype quality and \gls{us} completion.
  
\end{itemize}

\section{Approach}
\label{sec:initalSystemDesign}

We first present the \gls{llm}-based approaches utilized in the novel \textit{Figma} plugin, with a focus on the two-stage \gls{rag} approach for proprietary \gls{gui} representation generation. Subsequently, we provide an overview of the functionality of the developed \textit{Figma} plugin.

\subsection{\gls{llm}-Based Approaches}

\begin{figure*}[t!]
  \includegraphics[width=\textwidth]{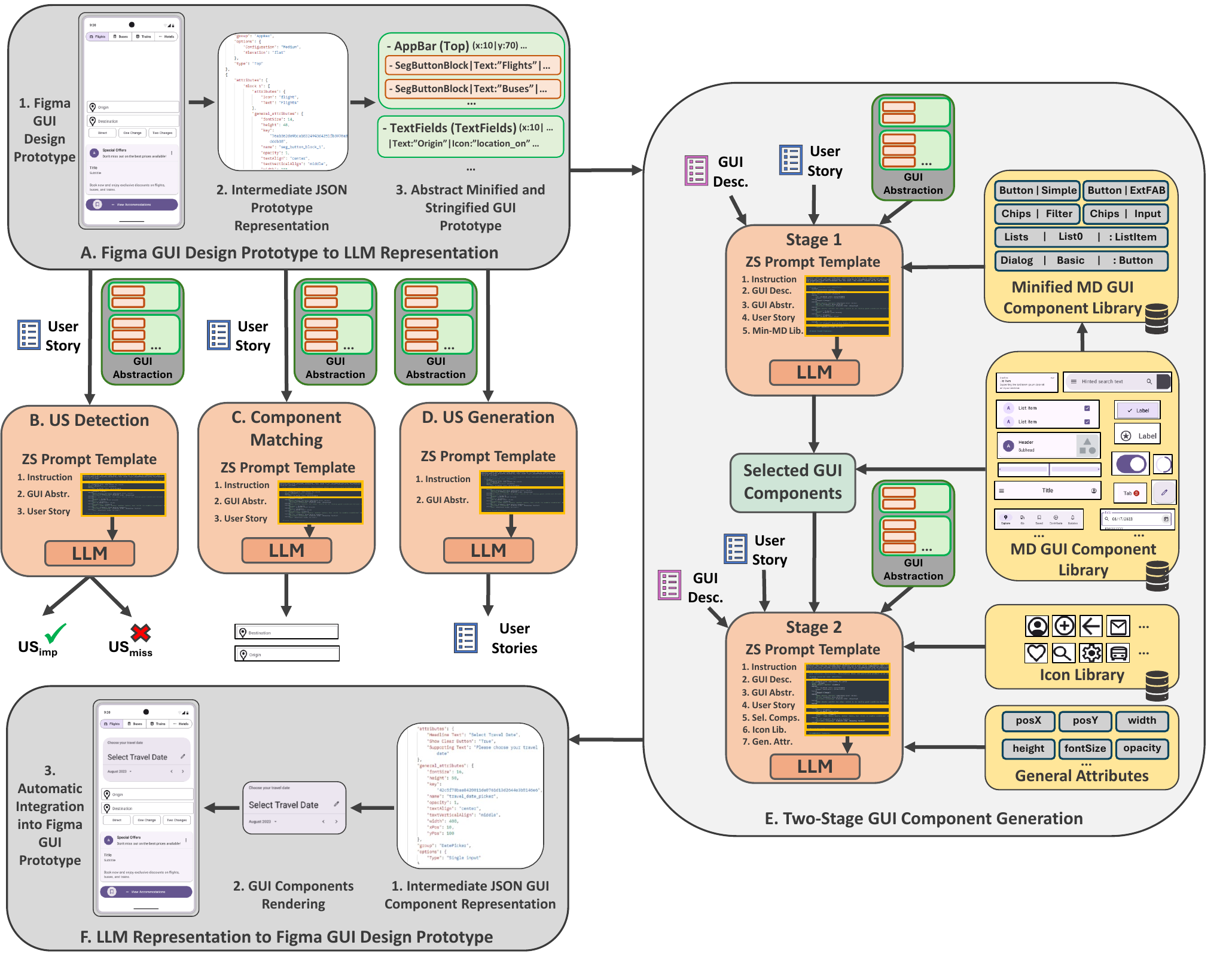}
  \caption[Overview of the \gls{llm} framework integrated into \textit{Figma}]{Overview of the \gls{llm} framework for automatically detecting \gls{us} implementation, matching \gls{gui} components to \gls{us}, generating \gls{us} for \gls{gui} prototypes and performing two-stage \gls{gui} component generation for \gls{us} including the transformation of \gls{gui} prototypes in \textit{Figma} to a compressed representation for inputting into the \gls{llm} and rendering of generated \gls{gui} components within the prototype. The figure contains \textit{Material 3 Design Kit} \citep{material_design_kit} components from \textit{Google}, under \textit{CC BY 4.0}.}

  \label{fig:llm_approaches}
\end{figure*}

Our underlying \gls{llm} framework is composed of several components. An overview of the approach is illustrated in Figure \ref{fig:llm_approaches}. In alignment with the figure, we describe the steps of the approach sequentially in a top-down manner. Since \gls{us} represent an important way of specifying requirements, our approach focuses on \gls{nlr} in the form of \gls{us}. First, \textit{(A)} we propose a component transforming the \gls{gui} prototype represented by \textit{Material Design} components in \textit{Figma} to an abstract and compressed string representation as the input to the \gls{llm}. Second, \textit{(B)} the \gls{us} detection component consists of a \gls{zs}-prompted \gls{llm} for deciding whether a provided \gls{us} is implemented in the \gls{gui} prototype, which is especially helpful for the prototype developers under rapidly, continuously changing and updating requirements, ensuring that all requirements are represented in the prototype. Third, \textit{(C)} the \gls{us} matching component enables the coupling of \gls{gui} components corresponding to a given \gls{us}, which provides traceability from the \gls{nlr} to the particular implementation within the \gls{gui} prototype. In addition, \textit{(D)} the \gls{us} generation component for automatically creating \glspl{us} for the \gls{gui} prototype, ensuring the availability of \gls{nlr} for each implemented feature in the \gls{gui} prototype. Fifth, \textit{(E)} the two-stage \gls{gui} component generation approach enables the rapid creation of implementations for a provided \gls{us}. Finally, \textit{(F)} the component transforms the generated intermediate prototype representation back to a rendered \gls{gui} component and places it appropriately in the \gls{gui} prototype within \textit{Figma}. All detailed prompts used in this work can be found in our supplementary materials. Subsequently, we present each of the shown components of our \gls{llm} framework for supporting the \gls{gui} prototyping process in more detail.

\paragraph{\gls{gui} Prototype Representation (A).} Our approach focuses on \gls{gui} prototypes created within the popular prototyping tool \textit{Figma} \citep{figma} and supports prototypes implemented with the \textit{Material Design} \citep{material_design_kit} component library. This extensive component library encompasses over 85 distinct \gls{gui} components ranging from elementary individual components such as \textit{Checkbox}, \textit{Slider}, and \textit{Button} to more intricate \gls{gui} elements assembled from multiple individual \gls{gui} components such as \textit{Dialogues}, \textit{List-Item}, \textit{Search-Bar}, and \textit{Card}. Each library \gls{gui} component possesses numerous configuration options including, for example, \textit{Icon} and \textit{Main-Text}, whereas the assembled components encompass several sub-components in a multi-level fashion. To further enhance the versatility of supported \gls{gui} components, we incorporated additional more generic component types (e.g., \textit{Label}, \textit{Image-Placeholder}, \textit{Rectangle}, and \textit{Icon}). To enable the previously discussed \gls{llm}-based assistance, the prototype initially needs to be transformed to an abstract and minified string representation to provide an efficient and effective input to the \gls{llm}. An intermediate \gls{json} representation of the prototype is reduced to the abstract variant, ensuring not only substantial increases in token efficiency, but also the removal of details unnecessary for the proposed tasks to potentially increase effectiveness. The abstract string representation is constructed as a multi-level bullet point list, with each \gls{gui} component being represented as an individual item using an abstract pattern, providing basic information of the component such as the \textit{group}, \textit{type}, \textit{position}, and \textit{size} as well as a list of component-specific attributes represented as

\begin{center}
\small
\begin{minipage}{0.85\linewidth}
\centering
\textit{component-group} \textbf{(component-type)} \texttt{(position)} \texttt{(size)} \texttt{|attribute name:"attribute value"|} \texttt{(id=\#)}\\[6pt]
\end{minipage}
\end{center}

\noindent This representation includes all necessary information for the problems considered in our approach. For example, a concrete instantiation of this pattern for a \textit{Button} could be

\begin{center}
\small
\begin{minipage}{0.85\linewidth}
\centering
\textit{Button} \textbf{(SimpleButton)} \texttt{(x:25|y:790)} \texttt{(width:359|height:54)}\\
\texttt{|Icon:"add"|Label Text:"Search"|Style:"Filled"|State:"Enabled"|Show Icon:"True"|} \texttt{(id=27)}
\end{minipage}
\end{center}

\noindent Multiple levels are introduced in the bullet point list when assembled elements contain sub-components and form nested structures to ensure and retain the appropriate grouping of the components, closely reflecting the grouping of the prototype in \textit{Figma}.

\paragraph{Implementation Detection (B).} The second component within our approach is represented by the automatic implementation detection of \gls{nlr} within a \gls{gui} prototype in \textit{Figma}. However, the core ideas behind this approach and evaluation results are presented in Chapter \ref{cha:interlinking} (Part \ref{part:verification} of the thesis), as part of the \gls{mllm}-based verification\footnote{Note that the \gls{llm}-based implementation detection approach has been published as part of a previous work \citep{kolthoff2024interlinking}. However, in the chapter, the focus lies on \gls{llm}-based generation of proprietary \gls{gui} representations and integration of these \gls{llm}-based components into a \gls{gui} prototyping assistance plugin for \textit{Figma}. For coherence, the detection approach with evaluation will be presented in Chapter \ref{cha:interlinking}.}.

\paragraph{\gls{gui} Component Matching (C).} Similarly as above, the third component in our approach is represented by the automatic component matching of \gls{nlr} within a prototype in \textit{Figma}. However, the core ideas behind this approach and evaluation results are presented in Chapter \ref{cha:interlinking} (Part \ref{part:verification} of the thesis), as part of the \gls{mllm}-based verification\footnote{Note that the \gls{llm}-based component matching approach has been published as part of a previous work \citep{kolthoff2024interlinking}. However, in the chapter, the focus lies on \gls{llm}-based generation of proprietary \gls{gui} representations and integration of these \gls{llm}-based components into a \gls{gui} prototyping assistance plugin for \textit{Figma}. For coherence, the detection approach with evaluation will be presented in Chapter \ref{cha:interlinking}.}.

\paragraph{User Story Generation (D).} To enable the creation of \glspl{us} from a fraction of or an entire \gls{gui} prototype currently not covered by any \gls{us} within the requirements collection, we also employ \gls{zs} prompting with an \gls{llm}. In the corresponding prompt template, we \textit{(i)} clearly instructed the model to extract all present \glspl{us} with their respective \gls{gui} components and \textit{(ii)} provided the abstract \gls{gui} representation with \gls{gui} component identifiers. In particular, the \gls{llm} is tasked with creating a \gls{json}, providing multiple objects with the \gls{us} text and a list of corresponding \gls{gui} components as attributes.

\paragraph{\gls{gui} Component Generation (E).} In addition to the detection, matching, and \gls{us} generation methods, we propose to facilitate the prototyping process by enabling the contextualized \gls{gui} component generation for a given \gls{us} based on the current \gls{gui} prototype in proprietary representations, thereby creating editable prototypes in visual prototyping environments. To tackle the challenge of generating \gls{gui} component recommendations for a given \gls{us}, we employed another \gls{zs} prompting approach utilizing an \gls{llm}. While the \gls{llm} is pretrained on large numbers of text corpora from the web and potentially includes various information about \textit{Material Design} utilized in our approach, the \gls{llm} lacks the specific \textit{Material Design} component library and configuration options employed in our approach. Therefore, we manually constructed the comprehensive component library as input to the model. However, to increase token efficiency for the \gls{llm}, we propose a two-stage \gls{rag}-based \gls{gui} component generation method.

As depicted in Figure \ref{fig:llm_approaches}\textit{E}, we first derive a minified \textit{Material Design} component library containing only the information about available component types and sub-component references. Afterwards, we construct our first stage \gls{zs} prompt template by instructing the \gls{llm} to select all required components from the minified library and encompassing \textit{(i)} the \gls{us} to generate the implementation for, \textit{(ii)} the abstract \gls{gui} representation, \textit{(iii)} a brief textual description of the main functionality of the \gls{gui}, and \textit{(iv)} the minified \textit{Material Design} component library. Afterwards, we construct the second stage ZS prompt template instructing the \gls{llm} to generate the intermediate \gls{gui} component representation by utilizing the same information as before, but instead of the minified \textit{Material Design} library, we additionally provide \textit{(i)} the full specifications from the \textit{Material Design} \gls{gui} component library retrieved for the components selected by the \gls{llm}, \textit{(ii)} an icon library, and \textit{(iii)} specifications for general attributes that are shared among all \gls{gui} components (such as \textit{positionX}, \textit{positionY}, \textit{width}, and \textit{height}).

By conducting this two-stage \gls{rag} approach, we avoid inputting the entire large \textit{Material Design} component library for each generation and instead reduce it to a minimal representation. To ensure correctness of the generated component specifications of the first and second stages, we set the temperature of the \gls{llm} to zero for a more probable and deterministic output. Moreover, we implemented an automatic verification of the generated specifications by matching them against the \textit{Material Design} specifications. The component position and size are generated by the \gls{llm} via the general attributes and influenced by the components and their positions in the current prototype. Finally, the intermediate representation is rendered within our \textit{Figma} plugin and the generated \gls{gui} component can directly be integrated into the existing \gls{gui} prototype design, including the automatic positioning of the component within the \gls{gui} prototype. Although the focus of our approach lies on generating functional \gls{gui} prototypes, the \gls{llm} is also able to generate different component styles (e.g., different buttons such as \textit{IconButton} or \textit{FloatingActionButton}), since different styles are represented as different components in the \textit{Material Design} specification. In the future, we plan to integrate more fine-grained styling options (e.g., \textit{background color}, \textit{font color}, \textit{font style}) by extending the general attributes. With this functionality, custom \textit{corporate identity} such as a color palette or special fonts could be incorporated into the generation process by enabling users to provide textual style requirements in addition to the textual functional requirements.

\paragraph{\gls{llm} Configuration.} As the \gls{llm} in our approach, we use the most recent \textit{GPT-4o} model\footnote{At the time of implementing the proposed approach, we utilized the most advanced \gls{mllm} available from \textit{OpenAI}, namely \textit{\gls{gpt}-4o} \citep{hurst2024gpt}, which is an extension over \textit{\gls{gpt}-4} \citep{openai2023gpt4}.} \citep{hurst2024gpt}  with 128k token context (\textit{accessed in August 2024}), which represents an advancement over \textit{\gls{gpt}-4} \citep{openai2023gpt4}. We decided on \textit{\gls{gpt}-4o} since it provides state-of-the-art performance across many \gls{nlu} and \gls{nlp} tasks \citep{gpt-4o}.

\vspace{-0.2cm}

\subsection{\gls{gui} Prototyping Plugin (Control)}
\vspace{-0.1cm}

\begin{figure}[!t]
  \centering
  \includegraphics[width=1\textwidth]{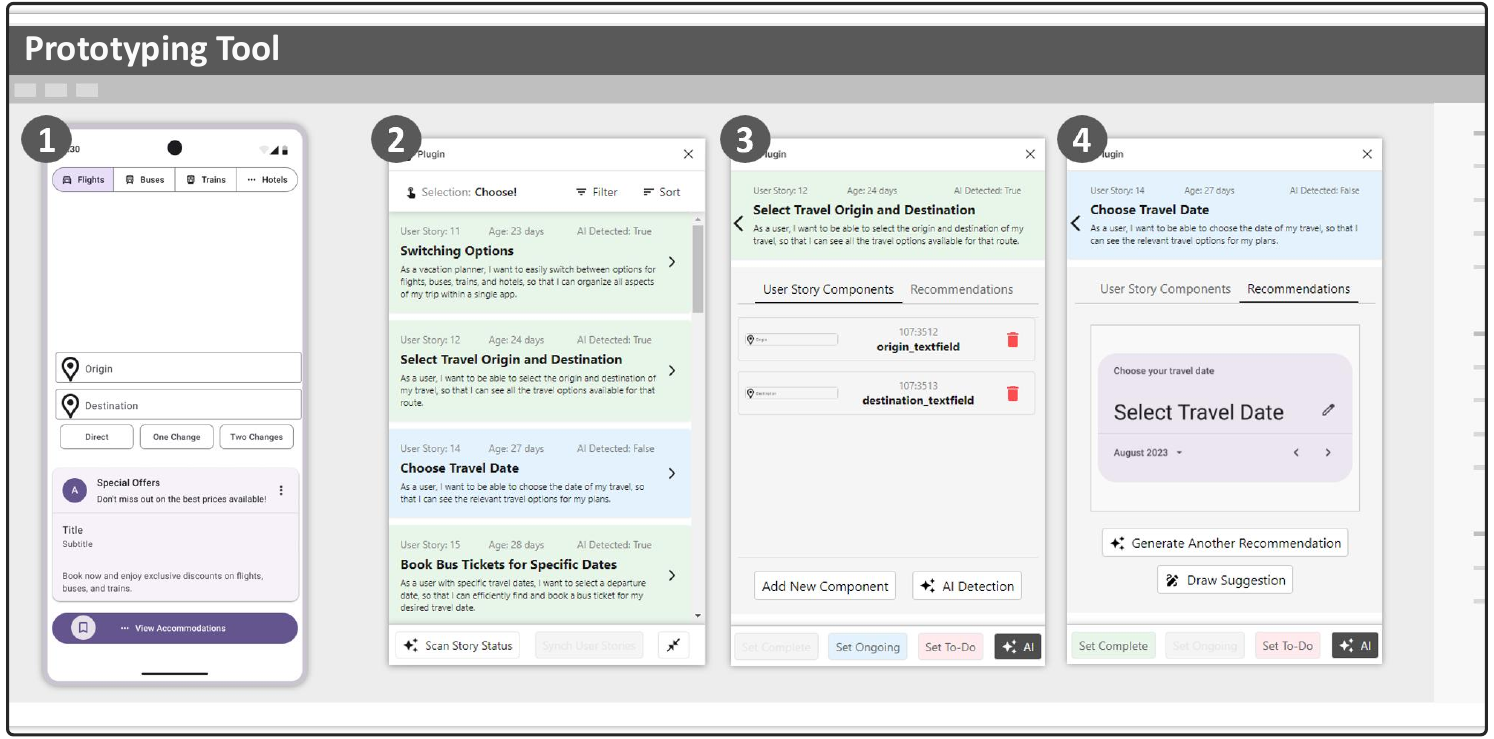}
  \caption[Proposed \gls{llm}-based \textit{Figma} \gls{gui} prototyping plugin]{\gls{gui} prototype \textit{(1)} and three views \textit{(2---4)} of our assistant for \gls{gui} prototype designers integrated as a plugin into a prototyping tool. Our assistant displays \glspl{us} \textit{(2)} imported from collaboration tools (e.g., \textit{JIRA}) for prototype designers to reference while working. The assistant can automatically detect (bottom bar in \textit{3} and \textit{4}) whether a \gls{us} is implemented in the \gls{gui} prototype, can detect \gls{gui} components completing a \gls{us} \textit{(3)} and can generate and draw \gls{gui} components fulfilling a given \gls{us} \textit{(4)}. Figure contains \textit{Material 3 Design Kit} \citep{material_design_kit} components from \textit{Google}, under \textit{CC BY 4.0}.}
  \label{fig:teaser-figma}
\end{figure}

Figure \ref{fig:teaser-figma} shows the \textit{Figma} plugin of our proposed approach. In particular, we developed two variants of our \gls{gui} prototyping assistant to evaluate its effectiveness in our lab study: a \textit{control} and a \textit{treatment} configuration. The \textit{treatment} configuration builds on the baseline functionality of the \textit{control} version by introducing advanced features enabling \gls{llm}-based automation. In the following, we describe the shared functionality of the \textit{control} configuration, followed by the unique capabilities of the \textit{treatment} configuration. Table \ref{tab:feature-comparison} provides a comparative overview of these differences. The following section outlines the baseline functionality of the \gls{gui} prototyping assistant, as implemented in the \textit{control} configuration. These core features are also present in the \textit{treatment} configuration.

\begin{figure*}[h]
  \includegraphics[width=\textwidth]{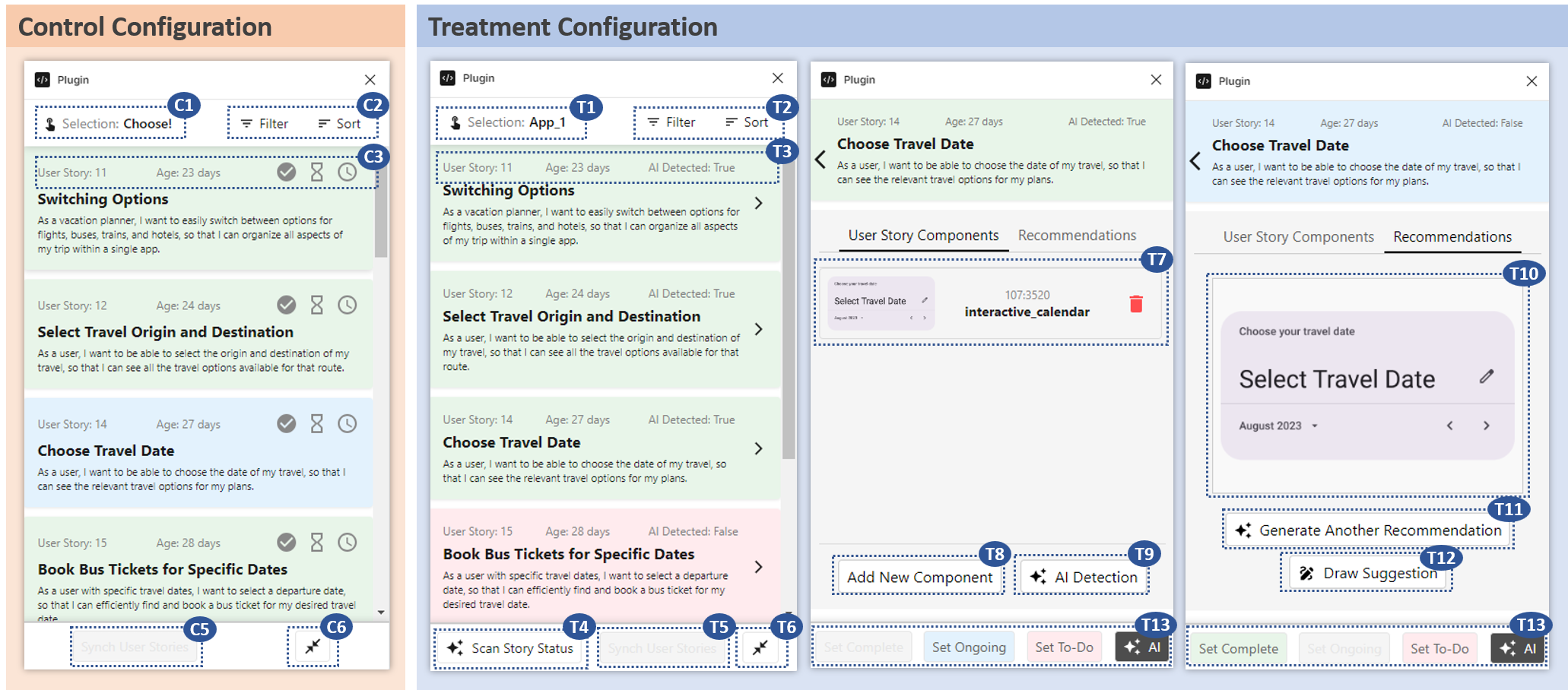}
  \caption[\textit{Control} and \textit{treatment} \gls{gui} prototyping assistant configuration]{Assistant for \gls{gui} prototype designers in the \textit{control} (\textit{left side}) configuration and \textit{treatment} (\textit{right side}) configuration. In both configurations, the assistant shows \glspl{us} based on a selected frame in the prototyping tool \textit{(C1/T1}), allows to filter and sort \glspl{us} (\textit{C2/T2}), can synchronize the \glspl{us} with a database (\textit{T4}) and be minimized (\textit{C6/T6}). \glspl{us} can be marked \textit{completed}, \textit{ongoing} or as a \textit{to-do} (\textit{C3 and T13}). In the \textit{treatment} configuration, the assistant can identify components completing a \gls{us} (\textit{T7}), manually added (\textit{T8}) or identified using an \gls{llm} (\textit{T9}). Additionally, an \gls{llm} can detect the completion status of a \gls{us} (\textit{T13}), can recommend \gls{gui} components completing a \gls{us} (\textit{T11 and T10}) and draw them into the \gls{gui} prototype (\textit{T12}).  Figure contains \textit{Material 3 Design Kit} \citep{material_design_kit} components from \textit{Google}, utilized under the \textit{CC BY 4.0 license} as previously.}
  \label{fig:assistantdetails}
\end{figure*}

\paragraph{Tool Integration.} Our assistant directly integrates into the popular prototyping tool \textit{Figma}. We based the size of the plugin on popular plugins for \textit{Figma} and integrated a minimization function (\textit{C6} and \textit{T6} in Figure \ref{fig:assistantdetails}), allowing the plugin to shrink and display only numerical values for \textit{open}, \textit{ongoing}, and \textit{completed} \glspl{us} to optimize screen space. This feature is particularly useful during tasks such as graphical finetuning of design components, where workspace is a priority (see \textit{C6} and \textit{T6} in Figure \ref{fig:assistantdetails}). With the assistant's integration as a plugin, we aim at supporting \gls{uiux} designers, product owners, and software developers within tools of practice. By facilitating tasks in tools like \textit{Figma} such as synthesizing textual descriptions (i.e. \gls{gui} prototype requirements) into \gls{gui} components, and picking and contextualizing fitting \gls{gui} components, we aim at allowing \gls{uiux} designers to focus more on the creative and conceptual aspects of their work, such as sketching rough designs and crafting the overall prototype user experience.

\paragraph{Listing User Stories in Figma.} The assistant lists \glspl{us} directly in the plugin, with their status visually highlighted as \textit{completed} \textit{(green)}, \textit{ongoing} \textit{(blue)}, or still a \textit{to-do} \textit{(red)}. By default, all \glspl{us} for all app screens currently open in \textit{Figma} are displayed. If a single frame (i.e., the screen of an app) is selected (selection indicator: \textit{C1} and \textit{T1} in Figure \ref{fig:assistantdetails}), only the \glspl{us} associated with that frame are shown in the plugin. 
Additional information, such as the age and identifier of \glspl{us}, is displayed alongside status indicators. Users can update the status of a \gls{us} (\textit{completed}, \textit{ongoing}, or \textit{to-do}) directly in the plugin (\textit{C3} and \textit{T13} in Figure \ref{fig:assistantdetails}).
Based on the interviews, we implemented filtering (based on \textit{State} and \textit{AI Detected} in the \textit{treatment} configuration) and sorting (based on \textit{age} and \textit{state} of the \gls{us}).

\begin{figure}[t!]
\centering
  \includegraphics[width=0.8\textwidth]{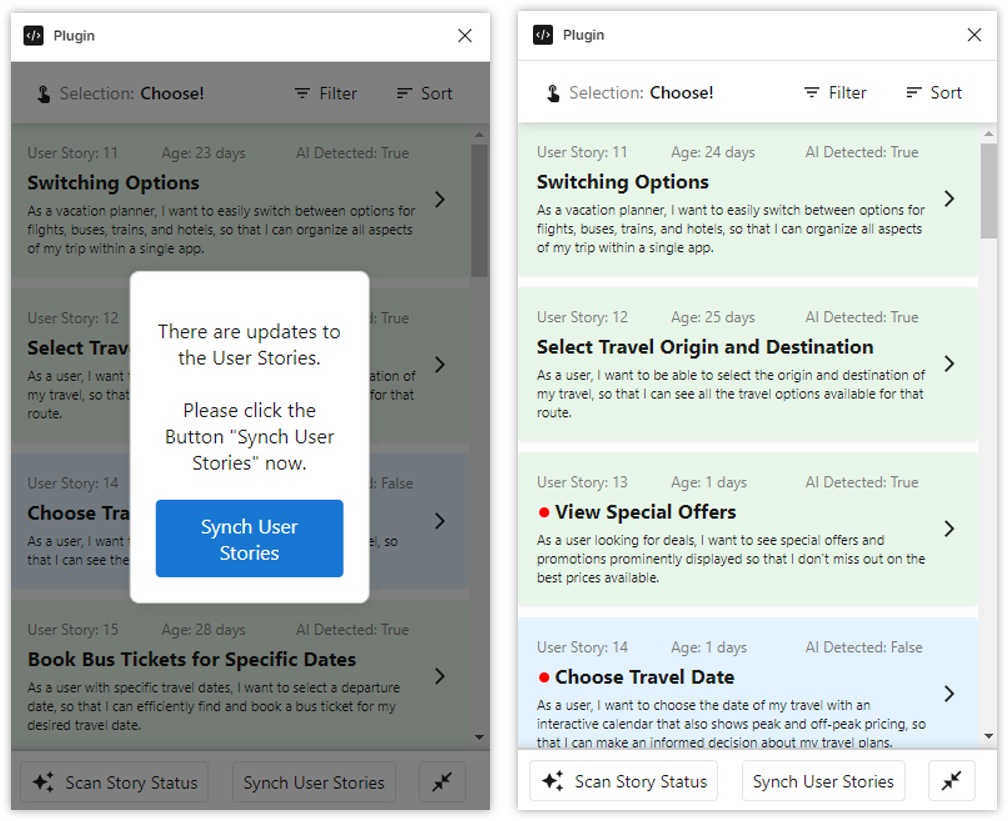}
  \caption[New and changed \gls{us} intervention]{New and changed \glspl{us} intervention shown to participants after 30 minutes in both the \textit{treatment} and \textit{control} configuration in lab experiments (shown: \textit{treatment}). Participants were shown a pop-up message and later red dots next to updated \glspl{us}, mimicking requirements updates and synchronization with collaboration tools (e.g., \textit{JIRA}).}
  \label{fig:storyupdates}
\end{figure}

\vspace{-0.2cm}
\paragraph{Change Intervention.} One identified design rationale guiding our implementation is the assistant's ability to deal with new and constantly changing requirements. To fulfill this rationale, we developed a feature synchronizing \glspl{us} (e.g., with an external requirements database). For the evaluation conducted in this paper, we developed an intervention, as shown in Figure \ref{fig:storyupdates}. As part of this intervention, users were notified of updates after a specified interval, when the “\textit{Sync User Stories}” button became active, simulating a synchronization with \gls{us} databases (such as \textit{JIRA}). Clicking the button led to the new or updated \glspl{us} being displayed in the plugin and highlighted with red dots.

\vspace{-0.3cm}
\subsection{\gls{gui} Prototyping Plugin (Treatment)}

The \textit{treatment} configuration builds upon the baseline features of the \textit{control} configuration by incorporating advanced functionalities guided by specific design rationales (from formative interviews). In this section, we describe the unique features implemented exclusively in the \textit{treatment} configuration, as illustrated on the right side of Figure \ref{fig:assistantdetails}.

\paragraph{User Story Detection.} The assistant can automatically detect whether a \gls{us} has already been implemented in a \gls{gui} prototype. Leveraging our \gls{llm}-based detection approach, the method automatically analyzes the implementation status of the \gls{us} in a selected prototype. 
Users can apply this feature at two levels: individually for specific \glspl{us} (\textit{T13} in Figure \ref{fig:assistantdetails}) or collectively for all stories in a frame (\textit{T4} in Figure \ref{fig:assistantdetails}). Detected implementation states are visually indicated by color changes.
After clicking “\textit{Scan Story Status}” (\textit{T14} in Figure \ref{fig:assistantdetails}), discrepancies between user-assigned and \gls{llm}-detected states are flagged for review in a dialog box, where users can confirm or override the \gls{llm}'s suggestions.
To ensure transparency, each \gls{us} in the list displays a label indicating whether the completion state was assigned by the user or detected by the \gls{llm} (e.g., “\textit{AI Detected: True/False},” see \textit{T3} in Figure \ref{fig:assistantdetails}). By automating detection, this feature aims at minimizing manual effort and ensuring a reliable overview of the \gls{gui} prototype implementation progress.

\begin{table}[!t]
\footnotesize
\caption[Feature comparison of \textit{control} and \textit{treatment} configuration of the \gls{gui} prototyping assistant]{Comparison of assistant's features in \textit{control} and \textit{treatment} configurations.}
\label{tab:feature-comparison}
\setlength{\tabcolsep}{5pt}
\renewcommand{\arraystretch}{1.0}

\newlength{\FeatRowHeight}
\setlength{\FeatRowHeight}{3.2\baselineskip} 

\newcommand{\HCell}[1]{%
  \parbox[c][\FeatRowHeight][c]{\hsize}{\centering\bfseries #1}%
}
\newcommand{\FCell}[1]{%
  \parbox[c][\FeatRowHeight][c]{\hsize}{%
    \raggedright
    \hyphenpenalty=10000\exhyphenpenalty=10000
    #1%
  }%
}
\newcommand{\BCell}[1]{%
  \parbox[c][\FeatRowHeight][c]{\hsize}{%
    \raggedright
    \hyphenpenalty=10000\exhyphenpenalty=10000
    #1%
  }%
}

\rowcolors{2}{lightgray}{white}

\begin{tabularx}{\linewidth}{p{0.30\linewidth} X X}
\toprule
\rowcolor{white}
\textbf{Feature} &
\cellcolor[HTML]{f8cbad}\HCell{Control Configuration} &
\cellcolor[HTML]{b4c7e7}\HCell{Treatment Configuration} \\
\midrule

\FCell{Tool Integration} &
\BCell{Plugin integrated in Figma} &
\BCell{Plugin integrated in Figma} \\

\FCell{User Story Listing} &
\BCell{Lists all user stories} &
\BCell{Lists all user stories} \\

\FCell{Change Intervention} &
\BCell{Synchronizes user story updates} &
\BCell{Synchronizes user story updates} \\

\FCell{User Story State Mang.\ / Detection} &
\BCell{Manual} &
\BCell{Manual and \gls{llm}-based} \\

\FCell{User Story Matching} &
\BCell{Not available} &
\BCell{\gls{llm}-based matching of \gls{gui} components to user stories} \\

\FCell{\gls{gui} Component Generation} &
\BCell{Manual} &
\BCell{\gls{llm}-based \gls{gui} components generation allowing user control over suggestions} \\
\bottomrule
\end{tabularx}

\rowcolors{1}{}{} 
\end{table}

\paragraph{User Story Matching.} For our assistant, we also implemented the function to automatically recognize which components of a \gls{gui} prototype in \textit{Figma} fulfill a \gls{us}, enabling the traceability of requirements within the implemented \gls{gui} prototype. This feature is shown in the third screen in Figure \ref{fig:assistantdetails}. 
Matched components are displayed in a dedicated field (\textit{T7} in Figure \ref{fig:assistantdetails}), showing their names as assigned in \textit{Figma} alongside an isolated image of each component. Users can manage these associations by adding components manually (“\textit{Add New Component},” \textit{T8}), using the \gls{llm} to detect components (“\textit{AI Detection},” \textit{T9}) or by removing incorrectly associated components.
The objective of this feature is to provide users with a better understanding of the components influencing the decision of the \gls{llm} for automatically assessing the \gls{us} implementation status and to enable the direct interlinking between \gls{us} and their counterparts in \gls{gui} prototypes. Thereby, it provides transparency regarding the features that still necessitate attention and enables the verification of whether the appropriate \gls{gui} components have been utilized.

\paragraph{\gls{gui} (Component) Generation.} Our assistant also allows users to generate components based on \glspl{us} in proprietary \gls{gui} representations, which can be directly integrated into \textit{Figma}. This feature is shown in Figure \ref{fig:assistantdetails} in the \textit{treatment} screens (fourth screen from the left). To ensure that users retain control over the decisions of the assistant and decide for themselves whether \gls{llm}-based component proposals are to be incorporated, recommendations are first previewed in the assistant (\textit{T10} in Figure \ref{fig:assistantdetails}). Users can then decide whether they want to use the button \textit{"Generate Another Recommendation"} (\textit{T11} in Figure \ref{fig:assistantdetails}) to generate another recommendation or use the button \textit{"Draw Suggestion"} (\textit{T12} in Figure \ref{fig:assistantdetails}) to insert these components as already correctly placed \gls{gui} components in \textit{Figma}. In this case, correctly contextualized and parameterized material design assets, material icons, labels, or rectangles are drawn into the existing \gls{gui} prototype. Our assistant utilizes the existing design as a context for correct dimensioning and positioning.

\section{Experimental Evaluation}
In this chapter, we present the underlying methodology of our evaluation studies. In particular, we first focused on measuring the ability of our approach to create relevant \gls{gui} components matching the \gls{nlr}. Subsequently, we focused on the question if our \gls{llm}-based assistant improves the effectiveness of \gls{gui} prototyping. In addition to the resulting quality of the \gls{gui} prototypes, we were also interested in the subjective satisfaction of the participants and the usability of our assistant. Finally, we focused on the question of how effective our approach is in generating \glspl{us} from components of \gls{gui} prototypes. Overall, we posed the following research questions for our evaluation:

\glsunset{nasatlx}
\glsunset{csi}
\begin{itemize}
    \item \textbf{RQ$_{1}$}: \textit{How effective are \gls{llm}-based approaches for generating \gls{gui} prototype components in proprietary representations completing a \gls{us}?} To answer this question, we used a previously created and published dataset encompassing \glspl{us} and corresponding \gls{gui} screens. With our two-stage \gls{rag} approach, we generated implementations for each \gls{us} and annotated their quality with \textit{Prolific} crowdworkers.
    \item \textbf{RQ$_{2}$}: \textit{How does our \gls{llm}-based assistant influence \gls{gui} prototype quality and \gls{us} completion?} To answer this question, we conducted a \textit{between-subjects} user study, asking participants to create several \gls{gui} prototypes in the \textit{control} or \textit{treatment} configuration of our approach and evaluated them with \textit{Prolific} crowdworkers.
    \item \textbf{RQ$_{3}$}: \textit{How does our \gls{llm}-based assistant influence perceived user experience?} To answer this question, participants were asked to answer multiple questionnaires after finishing the user study tasks. In particular, we measured common usability metrics including \textit{\gls{sus}}, \textit{\gls{nasatlx}}, and \textit{\gls{csi}} from participants.
    \item \textbf{RQ$_{4}$}: \textit{How effective are \gls{llm}-based approaches for generating user stories from components of \gls{gui} prototypes?} To answer this question, we employed a random sample of \gls{gui} screens (dataset from $RQ_1$) and generated \glspl{us} with the \gls{llm}-based approach. Afterwards, we evaluated the created \glspl{us} with \textit{Prolific} crowdworkers.
\end{itemize}

\begin{figure*}[]
  \includegraphics[width=\textwidth]{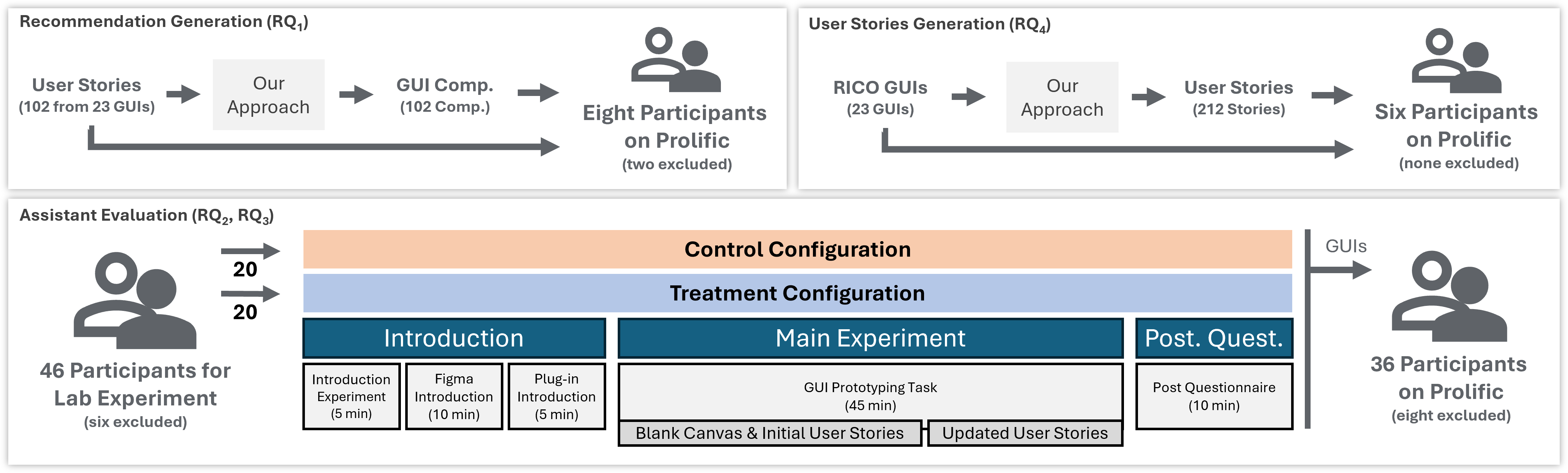}
  \caption[Overview of the evaluation procedure]{Overview of experiment procedures for evaluating the \gls{rag}-based \gls{gui} component generation in proprietary representations ($RQ_{1}$), the assistant ($RQ_{2}$ and $RQ_{3}$), and the \gls{us} generation ($RQ_{4}$). 
  For $RQ_{1}$, \gls{gui} components were generated with our approach from \glspl{us} and evaluated by crowdworkers on \textit{Prolific}.
  For $RQ_{2}$ and $RQ_{3}$, participants were recruited from a student panel, randomly assigned to either the \textit{control} or \textit{treatment} group to create \gls{gui} prototypes. They were then evaluated by a second set of participants sourced through \textit{Prolific} who provided prototype annotations. For $RQ_{4}$, \glspl{us} were generated with our approach and evaluated by crowdworkers on \textit{Prolific}.}
  \label{fig:evaluationsteps}
\end{figure*}

\noindent In the following, we will describe the underlying methods, procedures, and evaluation datasets employed to provide answers to the research questions articulated above by a series of evaluation studies. In Figure \ref{fig:evaluationsteps}, we present an overview of our three evaluations, investigating $RQ_{1}$, $RQ_{2}$ together with $RQ_{3}$ through a user study and finally $RQ_{4}$.

\subsection{RQ$_{1}$: \gls{gui} Component Generation}
\label{subsec:rq1-closing}
First, we present our evaluation which examines the extent to which the recommendations generated by our \gls{llm}-based approach are suitable for fulfilling the corresponding \glspl{us}.

\paragraph{Procedure.} In order to measure the extent to which the \gls{gui} component generations in proprietary representations, created by our two-stage \gls{rag} approach, match their corresponding \glspl{us}, we employed a previously created and published dataset, which provides a \gls{us} collection combined with \gls{gui} screens from \textit{Rico}\footnote{This dataset was created and previously published in prior work \citep{kolthoff2024interlinking} for evaluating the component detection and matching components and will be discussed in more detail in Chapter \ref{cha:interlinking}.}. Recommendations were then generated using our approach with the assistant in the prototyping tool, rendered as images and presented to crowdworkers on \textit{Prolific} \citep{prolific_academic_ltd_prolific_nodate, douglas2023data} for evaluation. As there was no existing \textit{Figma} context in this setup (no previous \gls{gui} prototype for which more recommendations are generated), a short, high-level text description as context created by students was utilized in addition to the \glspl{us} (e.g., \textit{“a screen from a travel app showing flight search results”}). Created \gls{gui} components were presented to the crowdworkers as they were created by the novel \gls{llm}-based assistant. 

\paragraph{Participants.} In particular, we invited eight crowdworkers from \textit{Prolific} with self-reported \gls{uiux} experience, more than 30 \textit{previous submissions} and a \textit{high approval rate} (>99\%). No crowdworker participated in any of our other studies. We had to exclude two participants based on failing our attention checks. The remaining participants (4 male, 2 female with an average age of 35.7 years) had, on average, 6.7 years ($\sigma$ = 5.59) of experience in creating and 3.7 ($\sigma$ = 3.40) years of experience in visual design evaluation (\gls{gui} prototyping). We did not collect information about participants’ \gls{llm} or generative AI experience, as such expertise was not required for evaluating the \gls{gui} components and revealing that the \gls{gui} components were \gls{llm}-generated could have introduced bias.

\paragraph{Data Collection.} The crowdworkers were shown the \gls{gui} components in this survey as a picture in combination with the respective \gls{us}. The crowdworkers were then asked whether the \gls{gui} components fully meet the functional requirements described in the \gls{us}, the \gls{gui} excerpt contains all the necessary components (e.g., \textit{buttons}, \textit{text input fields}) to fulfill the \gls{us} and whether textual descriptions within the \gls{gui} excerpt (e.g., \textit{labels}, \textit{instructions}, \textit{messages}) are clear and appropriate for fulfilling the \gls{us} (9-point Likert scale).

\subsection{RQ$_{2}$,  RQ$_{3}$: \gls{gui} Prototyping Assistant Evaluation}

To evaluate our assistant, we conducted a lab experiment in which participants were asked to create \gls{gui} prototypes based on predefined \glspl{us} in a controlled environment. Next, we evaluated the resulting \gls{gui} prototypes with crowdworkers. We explicitly decided against a remote structure (e.g., with crowdworkers) to better control boundary conditions (such as the context and environment as well as processing time for the \gls{gui} prototyping tasks).

\paragraph{Procedure.} We created two versions of the assistant (\textit{control} and \textit{treatment}) and evaluated both in a \textit{between-subjects} design experiment. For this purpose, \gls{gui} prototypes were created by participants in a lab from predefined \glspl{us}. \glspl{us} were derived from the publicly available \gls{us} dataset (same dataset as used in $RQ_1$) to employ high-quality \glspl{us} that have already been evaluated in previous studies. While the \glspl{us} are focused on single \glspl{gui} and do not span across multiple \glspl{gui} of an entire app, the dataset contains multiple coherent \glspl{us} describing different functionalities of a \gls{gui} screen. The participants in the lab had help from one or the other version of the assistant (\textit{control} or \textit{treatment}). The \gls{gui} prototypes created were then evaluated by crowdworkers with \gls{uiux} experience sourced on \textit{Prolific} \citep{prolific_academic_ltd_prolific_nodate, douglas2023data}. In the following, we present insights into the lab  experiment and crowdworking related procedures.

\vspace{0.15cm}
\textit{Lab Experiment.} In the 75-minute laboratory study, after agreeing to the data collection, participants read a briefing on the study task (5 min), watched a video explaining the \gls{gui} prototyping tool used (10 min) and a video demonstrating the assistant implemented as a plugin for the \gls{gui} prototyping tool (5 min). Since our assistant in the \textit{treatment} configuration generated \textit{Material Design} components, we therefore decided to give both groups an introduction to \textit{Figma}, including the use of \textit{Material Design} components. Then, participants started the \gls{gui} prototyping task (45 min) and finished the experiment with a post-hoc questionnaire (10 min). Participants worked with mobile screen templates and had pre-loaded icons and \textit{Material Design} assets available. For the \gls{gui} prototyping task, the laboratory study participants were challenged with creating up to three \gls{gui} prototypes based on \glspl{us}. For each \gls{gui} prototype, there were eight \glspl{us} to complete. After 30 minutes, there were updates for two existing \glspl{us} and two new \glspl{us} were added to the assistant. The update was intended to simulate changed requirements, as is usual in practice. Participants were instructed to start with the first \gls{gui} prototype and its first eight \glspl{us} for this \gls{gui} prototype and continue with the second and third \gls{gui} prototypes only when the previous prototype was completed. For our \textit{control} group, which used the assistant without generative component creation, producing three prototypes as part of the study task is clearly extensive. We deliberately opted for an extensive task so that no participant would finish early, even in the \textit{treatment} with generative component creation, and would continue to prototype in the 45-minute \gls{gui} prototyping phase. We measured the load in pretests to find the right amount of \gls{gui} prototyping tasks. The study included attention and comprehension checks. The study design was carefully evaluated with the university's Institutional Review Board (IRB).

\vspace{0.15cm}
\textit{Evaluation of \gls{gui} Prototypes.} We conducted an evaluation study on \textit{Prolific} with crowdworkers to evaluate the created \gls{gui} prototypes. We chose \textit{Prolific} since comparative studies indicated that crowdworkers on \textit{Prolific} produce higher data quality in comparison to other crowdworking platforms \citep{douglas2023data}. Participants from \textit{Prolific} received a questionnaire to evaluate the \gls{gui} prototypes created in the lab sessions. After consenting to data processing, participants received a study description and evaluated 20 \gls{gui} prototypes each, which were randomly drawn. On average, each participant received 13 \gls{gui} prototypes for the first \gls{gui} prototyping task (five for the second and two for the last \gls{gui} prototyping task) created with the assistant in the \textit{control} or \textit{treatment} configuration, respectively. The study lasted 65 minutes on average and included six attention checks. Furthermore, the study design was again carefully evaluated with the university's Institutional Review Board (IRB).

\paragraph{Participants} After describing the user study procedure, in the following, we describe the participants of both the lab study and the evaluation of the created \gls{gui} prototypes.

\vspace{0.15cm}
\textit{Lab Experiment.} We recruited 46 participants from a university panel for the lab study who were randomly assigned to the \textit{control} group (23 participants) or \textit{treatment} group (23 participants). In both groups, three participants had to be eliminated (one technical problem, three times failure to complete attention checks, two failures to meet the 45-minute \gls{gui} prototyping task time) ex-post. Eventually, 20 participants were admitted to the study in both groups. The participants in the \textit{control} group (14 male, 6 female) were, on average, 25.40 years ($\sigma$ = 4.65) old. The participants in the \textit{treatment} group (12 male, 8 female) were, on average, 23.85 years ($\sigma$ = 2.67) old. Participants in both groups studied, on average, a little over four years and had the same average experience with creating and evaluating visual design (such as \gls{gui} prototypes).

\vspace{0.15cm}
\textit{Evaluation of \gls{gui} Prototypes.} On \textit{Prolific}, we originally invited 36 participants with self-reported \gls{uiux} experience, more than 30 \textit{previous submissions} on \textit{Prolific}, and a \textit{high approval rate} (>99\%). Submissions from eight participants were excluded for failing one or multiple attention checks. Only data from the remaining 28 (22 male, 6 female) participants were considered. \textit{Prolific} participants were 28.25 years old ($\sigma$ = 6.78) and had 4.25 years ($\sigma$ = 5.87) of experience creating visual design and 3.86 ($\sigma$ = 4.04) years experience evaluating visual design on average. We did not collect information about participants’ \gls{llm} experience, as such expertise was not required for evaluating the \gls{gui} prototypes and revealing that the \gls{gui} prototypes' creation was partially assisted by our \gls{llm}-based assistant could have introduced bias.

\paragraph{Data Collection} We collected a range of data, both in the lab setting and on \textit{Prolific}.

\vspace{0.15cm}
\glsreset{csi}
\glsreset{sus}
\glsreset{nasatlx}
\textit{Lab Experiment.} During the lab experiment, we logged the usage data of the assistant (navigation in the assistant, clicks and web-calls in the \textit{treatment} group) and the \gls{gui} prototypes generated during the sessions. In addition, the participants filled out a post-hoc questionnaire in which task load with the \textit{\gls{nasatlx} Raw} \citep{hart_nasa-task_2006}, \textit{\gls{sus}} \citep{jordan_sus_1996}, and \textit{\gls{csi}} \citep{carroll_creativity_2009} were recorded. We asked for further Likert items (e.g., ease of use and effectiveness of the assistant) and surveyed subjective use of the features (detection of \gls{us} completion, recognition of the \gls{gui} components of a fulfilled \gls{us}, and recommendation of \gls{gui} components) in the \textit{treatment}, and in both groups positive and negative aspects of the assistant via open text.

\vspace{0.15cm}
\textit{Evaluation of resulting \gls{gui} Prototypes.} For each \gls{gui} and \gls{us} shown, the crowdworkers assessed the extent to which the \gls{gui} fully meets the functional requirements described in the corresponding \gls{us}, whether the \gls{gui} contains all the necessary \gls{gui} components (e.g., \textit{buttons}, \textit{text input fields}) to fulfill the \gls{us}, and whether text within the \gls{gui} (e.g., \textit{labels}, \textit{instructions}, \textit{messages}) is clear and appropriate for fulfilling the \gls{us}. Additionally, they rated each \gls{gui} for overall consistency with \glspl{us}, visually appealing design, clear information organization, intuitive interaction, whether the \gls{gui} looks like an app page, minimal prototype errors, and overall satisfaction with the \gls{gui} prototype.

\subsection{RQ$_{4}$: User Stories Generation}
We evaluated the extent to which our approach can create \glspl{us} from existing prototypes. 
\vspace{-0.2cm}
\paragraph{Procedure.} To evaluate the ability to create suitable \glspl{us}, research assistants recreated \textit{Rico} \glspl{gui} \citep{deka2017rico} for the prototyping tool. 23 \glspl{gui} were randomly chosen based on the \glspl{gui} employed in the previously described dataset (same dataset as used in $RQ_1$). We then utilized these \gls{gui} prototypes to create \gls{llm}-based \glspl{us} for the \gls{gui} components using our proposed assistant. Created \glspl{us} were presented to the crowdworkers as they were generated, i.e., without editing. These were then evaluated by \textit{Prolific} crowdworkers with \gls{uiux} experience. Each crowdworker rated ten prototypes and the corresponding \glspl{us}. On average, 9.2 \glspl{us} were created by the \gls{llm} for each \gls{gui}.
\vspace{-0.2cm}
\paragraph{Participants.} We recruited six crowdworkers through \textit{Prolific} \citep{prolific_academic_ltd_prolific_nodate, douglas2023data}, applying the same selection criteria as in the lab study. None of the crowdworkers had participated in any of our previous studies. No participant failed any of our attention checks. The final sample (6 male, average age of 33.83 years) had an average of 6.00 ($\sigma$ = 4.12) years of experience in creating and 4.33 ($\sigma$ = 2.98) years of experience in visual design evaluation (\gls{gui} prototyping). Once more, we did not inquire about participants’ \gls{llm} experience, since such was not required for assessing the \glspl{us}.
\vspace{-0.2cm}
\paragraph{Data Collection.} The \glspl{us} created and \gls{gui} prototypes were presented to crowdworkers for comparison. Next, they assessed the \glspl{us} in combination with the prototype for the following four items: \textit{(i) "The user story accurately describes functionality found in the presented GUI."}, \textit{(ii) "The user story is written with sufficient clarity and precision."}, \textit{(iii) "The user story is specific enough to describe a particular feature of the presented \gls{gui}."}, finally \textit{(iv) "The level of detail in this user story meets the standards in a professional project."}

\section{Results}
In this section, we provide the results on the capabilities of our \gls{llm}-based approach to create \gls{gui} components based on \glspl{us} through measuring how well the generated components fulfill the \glspl{us} ($RQ_{1}$). Afterwards, we present the results of the lab experiment using the assistant, including the evaluation of the lab results on \textit{Prolific} ($RQ_{2}$, $RQ_{3}$). In addition, we provide the results of the evaluation for automatically creating \glspl{us} ($RQ_{4}$).

\subsection{RQ$_{1}$: \gls{gui} Component Generation}

To maximize the diversity of our requirements, we took 102 \glspl{us} from 27 different \glspl{gui} in the publicly available dataset (for details, see Section \ref{subsec:rq1-closing}) and generated components for each \gls{us} using our \gls{llm}-based approach. These were presented individually (i.e. as single components with the \glspl{us}) to crowdworkers on a basic schematic mobile app outline. We collected 40 pairs of \gls{us} component ratings from each of the six crowdworkers, resulting in an average of 2.35 ratings per \gls{us} and \gls{gui} component.

\begin{figure*}[!t]
    \centering
    \includegraphics[width=0.90\textwidth]{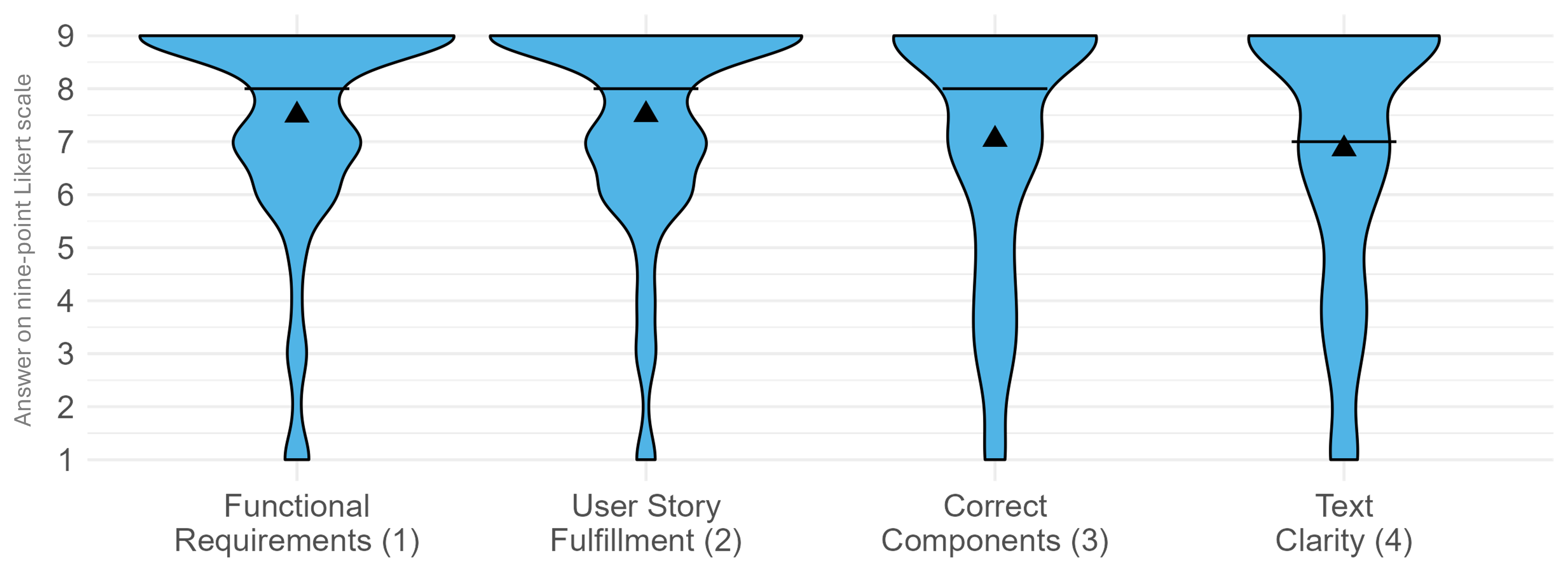}
    \caption[Violin plots for \gls{llm}-based \gls{gui} component generation effectiveness]{Crowdworker ratings (9-point Likert scale) for components generated by our \gls{llm}-based approach based on over 100 \glspl{us}. Crowdworkers rated \textit{(1)} functional requirements fulfillment, \textit{(2)} \gls{us} fulfillment, \textit{(3)} component correctness and \textit{(4)} text clarity for \glspl{us} and derived \gls{gui} components. Triangles represent means, black lines medians.}
	\label{fig:Comp_Re}
\end{figure*}

Crowdworkers rated \glspl{us} and generated \gls{gui} components to assess whether the components \textit{(1)} allow the user to perform the task as specified by the \gls{us}, \textit{(2)} all necessary components for fulfilling the \gls{us} are presented, \textit{(3)} appropriate components were chosen, and \textit{(4)} text of components was clear and appropriate. All ratings were obtained on nine-point Likert scales (1: \textit{Strongly Disagree}, 3: \textit{Disagree}, 5: \textit{Neutral}, 7: \textit{Agree}, 9: \textit{Strongly Agree}). Figure \ref{fig:Comp_Re} shows violin plots for each of the four assessed aspects. All four aspects were rated high, with mean (median) values of 7.5 (8) for functional requirement fulfillment, 7.51 (8) for \gls{us} fulfillment, 7.04 (8) for component correctness, and 6.86 (7) for textual clarity. In addition, to evaluate the efficiency benefits, we compared the proposed two-stage \gls{rag}-based \gls{gui} component generation approach against a similar one-stage \gls{gui} component generation \gls{zs} prompt, which always incorporates the full \textit{Material Design} component library. Over the 102 \glspl{us} dataset, the two-stage \gls{gui} generation approach accomplished a considerable 64.35\% reduction of the consumed input tokens (Mean \textit{\#One-Stage-Tokens}=8559.49|Mean \textit{\#Two-Stage-Tokens}=3050.78) and a significant 60.82\% reduction of the total consumed or generated tokens (Mean \textit{\#One-Stage-Tokens}=9359.44|Mean \textit{\#Two-Stage-Tokens}=3666.54)\footnote{At the time of writing the original manuscript, this reduction in token consumption led to a cost reduction from \$0.052 to \$0.024 per generated \gls{gui} component using the \textit{GPT-4o} model \citep{gpt-4o}.}. These evaluation results clearly highlight the strong token consumption efficiency improvements by using \gls{rag}.

\begin{myrqbox}
\textbf{Answer to RQ$_{1}$:} Our \gls{rag}-based approach for generating \gls{gui} components in proprietary representations achieves high effectiveness, as indicated by the high annotation ratings received for the generated \gls{gui} components. The two-stage \gls{rag} approach significantly reduces the mean total (input) token consumption by 60.82\% (64.35\%).
\end{myrqbox}

\begin{figure*}[!t]
    \centering
    \includegraphics[width=0.95\textwidth]{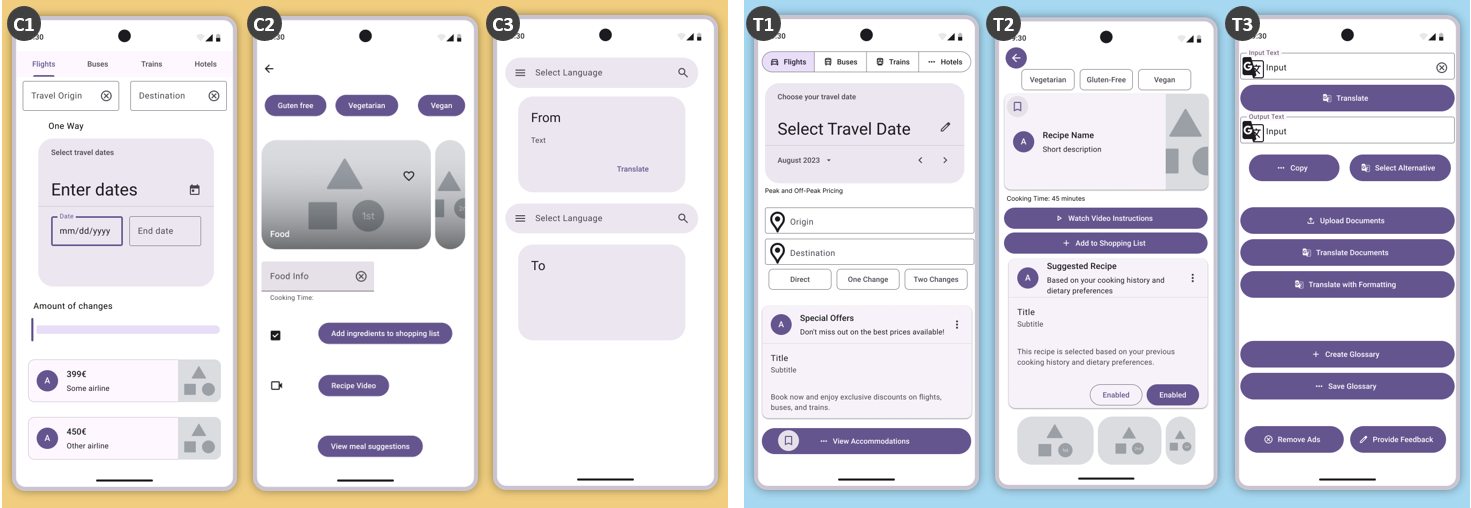}
    \caption[Example \gls{gui} prototypes created during the evaluation with the \textit{control} and \textit{treatment} \gls{gui} prototyping assistant configuration]{Examples of \gls{gui} prototypes created in the lab sessions with the \textit{control} (\textit{orange, left}) and \textit{treatment} (\textit{blue, right}) configuration of the assistant for all three tasks: \textit{(1)} travel \textit{booking app}, \textit{(2)} \textit{recipe app} and \textit{(3)} \textit{translation app} (\textit{Control}: \textit{C1} --- \textit{C3}, \textit{Treatment}: \textit{T1} --- \textit{T3}). \textit{Material 3 Design Kit} \citep{material_design_kit} components, \textit{Google}, \textit{CC BY 4.0}.}
	\label{fig:Examples_Star}
\end{figure*}

\subsection{RQ$_{2}$, RQ$_{3}$: \gls{gui} Prototyping Assistant Evaluation}
Our analysis of the data collected in the lab experiment is twofold. Firstly, we examine the completion of \glspl{us} and the quality of the \gls{gui} prototypes created by study participants, as assessed by crowdworkers on \textit{Prolific}. Afterwards, we report the self-assessments of the lab participants, including both qualitative and quantitative aspects. In addition, we evaluated the usage data of the assistant, highlighting which features and functions were used by the participants. Figure \ref{fig:Examples_Star} shows three \gls{gui} prototypes created by participants from the \textit{control} and \textit{treatment} groups for the first, second, and third prototyping tasks. 

\begin{figure*}[!t]
    \centering
    \includegraphics[width=0.98\textwidth]{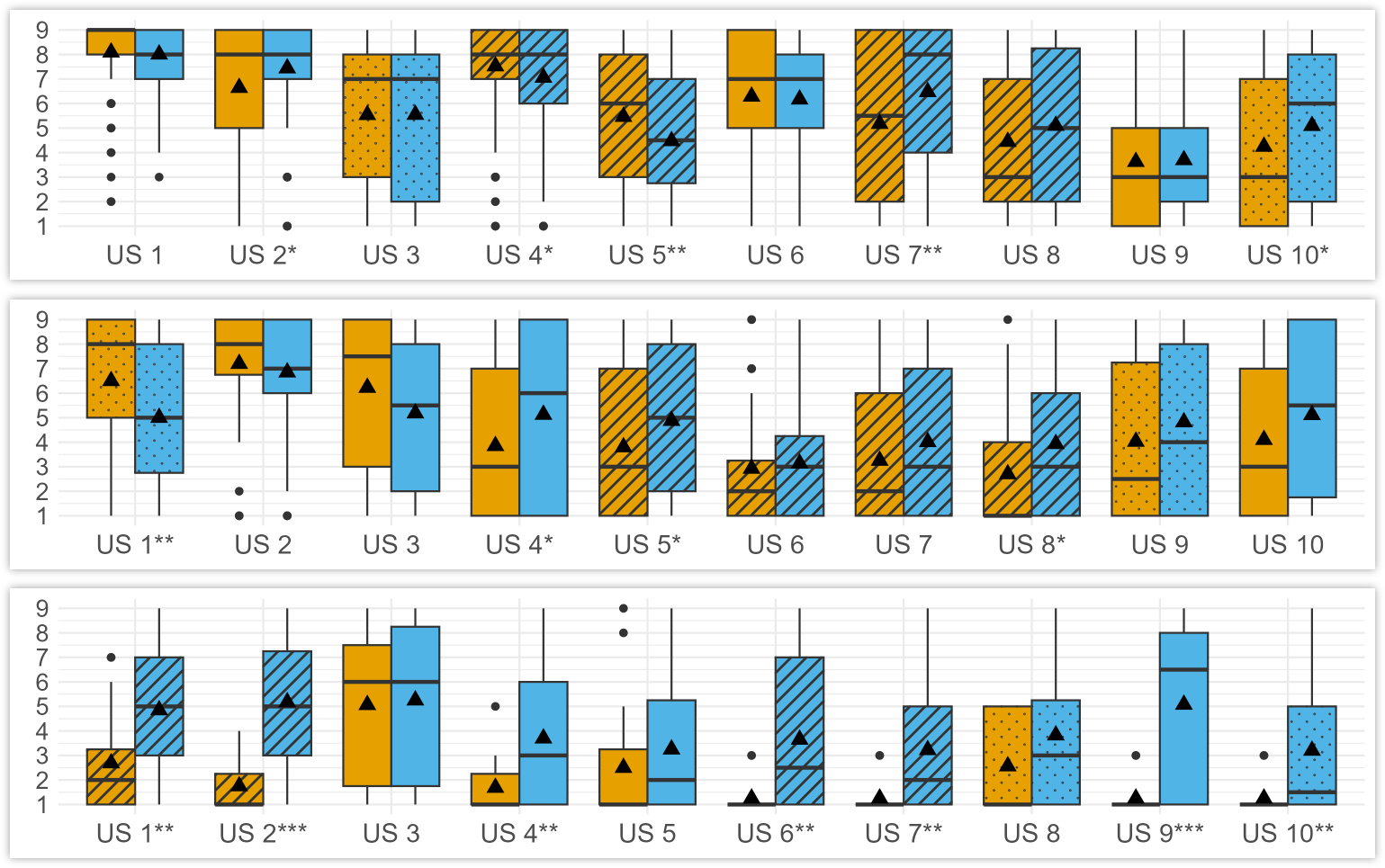}
    \caption[Boxplots for \gls{us} completion with \textit{control} and \textit{treatment} \gls{gui} prototyping assistant configuration]{Boxplots showing \gls{us} completion as rated by crowdworkers for the three experiment apps (\textit{top}: first app, \textit{middle}: second app, \textit{bottom}: third app) using the \textit{control} (\textit{orange}) or \textit{treatment} (\textit{blue}) configurations. For each app, ten \glspl{us} were to be created (\gls{us} 1 to \gls{us} 10). \textit{Dotted} (new \gls{us}) and \textit{dashed} (extended \gls{us}) bars represent \glspl{us} with an update intervention after 30 minutes. Triangles represent means. One, two, and three stars represent significance (<0.05, <0.01, <0.001) for \textit{two-sided Wilcoxon rank-sum tests}.}
    \label{fig:UserStories_Completion}
    \vspace{-0.4cm}
\end{figure*}

\paragraph{User Story Completion and Prototype Quality.} The results for \gls{us} completion and overall \gls{gui} prototype quality are based on data collected from crowdworkers on \textit{Prolific}. The results for evaluating the completion of the \glspl{us} are shown in Figure \ref{fig:UserStories_Completion}. The participants in the lab study were tasked with creating three app prototypes based on ten \glspl{us} each. The participants were told to start with the first app and move on to the next one once the \glspl{us} had been completed. As a result, a different number of \gls{gui} prototypes were created for each of the apps in the \textit{control} group and the \textit{treatment} group. In the \textit{control} group, all 20 participants created a first app, five created a second app, and two created a third one. In the \textit{treatment} group, 19 participants created the first app, 11 created the second one, and five participants created the third one. On average, 1.35 apps were created in the \textit{control} group and 1.75 apps in the \textit{treatment} group. To analyze the degree of fulfillment of the \glspl{us}, we solely investigated apps that participants had worked on. Apps that were not started were excluded from the evaluation in \textit{Prolific} (i.e. we did not count \glspl{us} for apps that were not started as unfulfilled). An app was classified as started when changes were made to the schematic app background in \textit{Figma} and \gls{gui} components were added (simply moving a schematic app background in \textit{Figma} would not count as starting the \gls{us}).

For the first app, for six (three significant)\footnote{Participants in the \textit{treatment} group started with just the schematic app background without any components, but could quickly use the recommendation feature to generate components. To give participants in the \textit{control} group an idea of how to pick and adjust \textit{Material Design} components, the introduction video demoed the use of an app bar demonstrating the creation of the first \gls{us}. This component was also already drawn to the schematic app background in \textit{Figma} when participants started. Since this component completes the first \gls{us}, technically participants in the \textit{control} group did not complete the first \gls{us} by themselves.} \glspl{us}, the \textit{treatment} group produced \gls{gui} prototypes with a higher rating for \gls{us} completion in contrast to four (two significant) \glspl{us} in the \textit{control} configuration that led to higher \gls{us} completion. 
For the second app, seven (three significant) \glspl{us} using the \textit{treatment} and three (one significant) \glspl{us} with the \textit{control} configuration were rated higher. 
Finally, for the third and last app, all ten (seven significant) \glspl{us} with the \textit{treatment} configuration were rated as more complete in contrast to the \textit{control} group.

For the two other aspects that were evaluated by crowdworkers for each \gls{us}, the following picture emerged: For the selection of the right components to fulfill the respective \gls{us}, the \textit{control} group was rated higher on average for nine \glspl{us}, and the \glspl{gui} created with the \textit{treatment} configuration for 21 cases. Regarding the selection of correct descriptions (e.g., \textit{displayed texts} or \textit{messages}), the results of the \textit{control} group were rated higher for eight \glspl{us} and the results of the \textit{treatment} group for 22 \glspl{us}.

Not all \glspl{us} were available to participants from the start. In our intervention, the \glspl{us} were updated after 30 minutes of processing. Two \glspl{us} were made available for the first time as new \glspl{us} (\textit{dotted} in Figure \ref{fig:UserStories_Completion}). For two \glspl{us}, an update of the previous \glspl{us} was displayed that expanded the scope of the previous \gls{us} (\textit{dashed} in Figure \ref{fig:UserStories_Completion}). In Figure \ref{fig:UserStories_Completion}, stories 4/5 and stories 7/8 represent pairs in the first app, in which the second \gls{us} represents the update. For app two (app three), these are 5/6 (1/2) and 7/8 (6/7). Congruent with our expectations, the updated \glspl{us} were, on average, rated as less complete in contrast to their initial versions, since the update of \glspl{us} merely extended the first version of the \gls{us}. In our discussion (Section \ref{subsec:completeness}), we provide insight into examples and discuss cases where the \textit{control} results were assessed as more complete.

\begin{table}[!t]
\centering
\caption[Mean ratings of Likert-scale items from crowdworkers for generated \gls{gui} prototypes]{Means of crowdworkers' ratings (Likert scale 1---9) of \gls{gui} prototypes for \textit{control} and \textit{treatment} (\textit{Mann-Whitney-U-Tests}). Crowdworkers were asked if \textit{(1)} \textit{the \gls{gui} design is consistent with the overall \glspl{us}}, \textit{(2)} \textit{whether the visual \gls{gui} design is appealing}, \textit{(3)} \textit{organization of information is clear}, \textit{(4)} \textit{the \gls{gui} allows for intuitive interaction}, \textit{(5)} \textit{the \gls{gui} prototype looks like a screen from a complete app}, \textit{(6)} \textit{the \gls{gui} prototype only has minimal errors}, and finally \textit{(7)} \textit{whether crowdworkers were satisfied with the \gls{gui} prototype}.}
\label{tab:likert-means}
\footnotesize
\setlength{\tabcolsep}{2pt}
\renewcommand{\arraystretch}{1.1}

\newcommand{\ColHead}[2]{\parbox[c]{\hsize}{\centering\bfseries\scriptsize #1\\\normalfont\scriptsize (#2)}}

\begin{tabularx}{\columnwidth}{l *{7}{>{\centering\arraybackslash}X}}
\toprule
& \ColHead{Consistent\\Design}{1}
& \ColHead{Appealing\\Design}{2}
& \ColHead{Information\\Organisation}{3}
& \ColHead{Intuitive\\Interaction}{4}
& \ColHead{Screen from\\Complete App}{5}
& \ColHead{Minimal\\Errors}{6}
& \ColHead{Overall\\Satisfied}{7} \\
\midrule
\textbf{Control}   & 4.583 & 4.454 & 4.820 & 4.528 & 3.465 & 4.190 & 3.819 \\
\textbf{Treatment} & 5.046 & 4.586 & 5.025 & 4.761 & 3.614 & 4.225 & 4.254 \\
\midrule
\textit{p}-value   & \textbf{0.0151*} & 0.5105 & 0.2600 & 0.1849 & 0.5540 & 0.7485 & \textbf{0.0423*} \\
\bottomrule
\end{tabularx}
\vspace{0.5cm}
\end{table}

In addition to completing the \glspl{us}, crowdworkers also independently evaluated whether \textit{(1) the \gls{gui} design is consistent with all \glspl{us} for this \gls{gui} prototype}, \textit{(2) the visual \gls{gui} design is appealing}, \textit{(3) organization of information is clear}, \textit{(4) the \gls{gui} prototype allows for intuitive interaction}, \textit{(5) the \gls{gui} prototype looks like a screen from a complete app}, \textit{(6) the \gls{gui} prototype has only minimal errors}, and \textit{(7) crowdworkers were satisfied with the \gls{gui} prototype} for each of the created \glspl{gui}. Results are shown in Table \ref{tab:likert-means}. For each aspect, means in the \textit{treatment} configuration were considerably higher, but solely the two metrics \textit{(1) Consistent Design} and \textit{(3) Overall Satisfaction} are statistically significant.

\begin{myrqbox}
\textbf{Answer to RQ$_{2}$:} Our approach mainly outperformed the \textit{control} configuration on \gls{us} completion and \gls{gui} prototype quality. Participants completed more prototypes with our approach (1.75 vs. 1.35 on average), especially for the third prototype where 7 (out of 10) \glspl{us} are rated as significantly higher for completion. Moreover, for perceived \gls{gui} quality, our approach outperformed the \textit{control} configuration across all mean metric values and is significant for \textit{(1) Consistent Design} and \textit{(3) Overall Satisfaction}.
\end{myrqbox}

\begin{table}[t]
\centering
\caption[Perceived usability results between \textit{control} and \textit{treatment}]{Rating means and respective \textit{p}-values (\textit{Mann-Whitney-U-Tests, one-sided}) for seven questions answered by the participants in lab experiment using Likert scales. Questions for \textit{(1)} \textit{satisfaction with final designs}, \textit{(2)} \textit{ease of use and navigation of plugin}, \textit{(3)} \textit{confidence in creating professional \gls{gui} designs}, \textit{(4)} \textit{efficiency in design process}, \textit{(5)} \textit{future use consideration}, \textit{(6)} \textit{impact on design quality}, and \textit{(7)} \textit{difficulty in creating \gls{gui} prototypes}.}
\label{tab:user_perceptions}
\footnotesize
\setlength{\tabcolsep}{2pt}
\renewcommand{\arraystretch}{1.05}

\newcommand{\ColHead}[2]{\parbox[c]{\hsize}{\centering\bfseries\scriptsize #1\\\normalfont\scriptsize (#2)}}

\begin{tabularx}{\columnwidth}{l *{7}{>{\centering\arraybackslash}X}}
\toprule
& \ColHead{Satisf.\\Designs}{1}
& \ColHead{Ease\\of Use}{2}
& \ColHead{Confidence\\Creating\\Prof.\ \glspl{gui}}{3}
& \ColHead{Efficiency\\in Design\\Proc.}{4}
& \ColHead{Future\\Use}{5}
& \ColHead{Impact\\Design\\Quality}{6}
& \ColHead{Difficulty\\Creating\\\gls{gui}\\Prototypes}{7} \\
\midrule

\textbf{Likert scale} &
\multicolumn{4}{c}{\textit{Strongly Disagree} (1) -- \textit{Strongly Agree} (9)} &
{\scriptsize\parbox[c]{\hsize}{\centering \textit{Definitely not} (1)\\-- \textit{Definitely yes} (5)}} &
{\scriptsize\parbox[c]{\hsize}{\centering \textit{Not at all} (1)\\-- \textit{Extremely} (5)}} &
{\scriptsize\parbox[c]{\hsize}{\centering \textit{Very easy} (1)\\-- \textit{Very difficult} (5)}} \\
\midrule

\textbf{Control}   & 3.00 & 6.20 & 4.35 & 5.35 & 3.25 & 2.90 & 3.75 \\
\textbf{Treatment} & 3.75 & 6.30 & 5.65 & 6.85 & 3.80 & 3.50 & 3.45 \\
\midrule
\textit{p}-values  & 0.2543 & 0.4400 & \textbf{0.0277*} & 0.0573 & \textbf{0.0392*} & \textbf{0.0462*} & 0.8408 \\
\bottomrule
\end{tabularx}
\end{table}

\paragraph{Participants' Perceptions.} We asked our participants qualitatively and quantitatively how they rated using the plugin in the \textit{control} and \textit{treatment} groups. Table \ref{tab:user_perceptions} shows the results for seven Likert scale items. The \textit{control} group achieved a lower mean value for each aspect, except for \textit{(7) Difficulty Creating \gls{gui} Prototypes}, where a higher value is associated with greater perceived difficulties. Particularly noteworthy are the significantly better values of the \textit{treatment} for \textit{(3) Confidence Creating Professional \gls{gui} Prototypes}, \textit{(5) Consideration for Future Use}, and \textit{(6) Impact on Design Quality}.

\begin{figure*}[!t]
    \centering
    \includegraphics[width=0.98\textwidth]{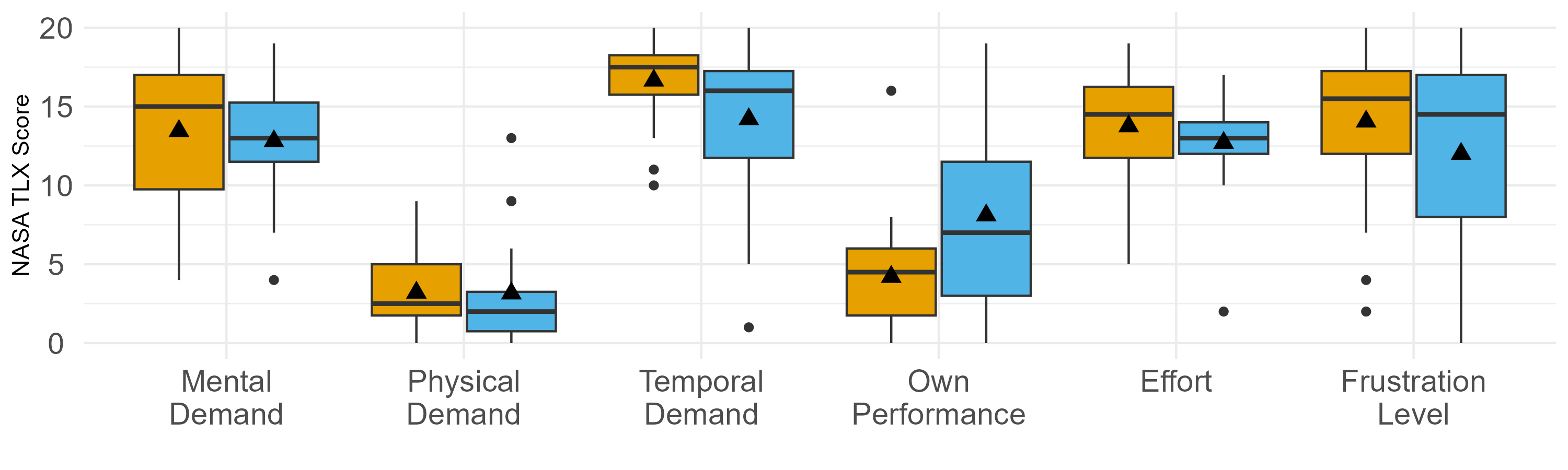}
    \caption[\gls{nasatlx} (raw) results for \textit{control} and \textit{treatment} \gls{gui} prototyping assistant configuration]{Results of \gls{nasatlx} (raw) \citep{hart_nasa-task_2006} for participants in \textit{control} (\textit{orange}) and \textit{treatment} (\textit{blue}). \textit{Own Performance} significant (5\% level). Triangles represent means.}
	\label{fig:NASATLX}

\end{figure*}

We also measured the task load via the \gls{nasatlx} (raw) questionnaire, as perceived by the user study participants. Figure \ref{fig:NASATLX} shows the values for the task load. On average, the \textit{treatment} configuration was rated lower for \textit{Mental Demand}, \textit{Physical Demand}, \textit{Temporal Demand}, \textit{Necessary Effort}, and \textit{Frustration Level}. The value for \textit{Own Performance} proved to be significantly higher for the \textit{treatment} configuration.
The measured \gls{sus} \citep{jordan_sus_1996} and \gls{csi} \citep{carroll_creativity_2009} showed no significant difference. However, for both scales the \textit{treatment} configuration rated higher than the \textit{control} configuration. For \gls{sus}, means were 52.13 for the \textit{control} and 55.50 for the \textit{treatment} configuration, reflecting a steep learning curve of participants using \textit{Figma} and \gls{gui} component libraries in \textit{Figma} (\textit{Material Design}). Considering the \gls{csi}, means were 47.63 and 56.13, respectively. 

\begin{myrqbox}
\textbf{Answer to RQ$_{3}$:} Participants rated all perception items higher for our approach on average, except for \textit{(7) Difficulty Creating \gls{gui} Prototypes}, for which lower values are considered to be better. Items \textit{(3) Confidence Creating Professional \gls{gui} Prototypes}, \textit{(5) Consideration for Future Use}, and the \textit{(6) Impact on Design Quality} are significantly better for our approach. Moreover, our approach received better mean ratings for the \gls{nasatlx} scores, with significant improvement for \textit{Own Performance}. Finally, our approach received a \gls{sus} score of 55.50 (vs. 52.13) and an \gls{csi} score of 56.13 (vs. 47.63).
\end{myrqbox}

\subsection{RQ$_{4}$: User Stories Generation}
To investigate the effectiveness of our \gls{llm}-based approach to create \glspl{us} from \glspl{gui}, we created 212 \glspl{us} from 23 \textit{Rico} \citep{deka2017rico} \gls{gui} prototypes directly in \textit{Figma} with our prototyping assistant (therefore, an average of 9.22 \glspl{us} per \gls{gui} prototype). Both were rated by our \textit{Prolific} experts on Likert scales (1: \textit{Strongly Disagree}, 3: \textit{Disagree}, 5: \textit{Neutral}, 7: \textit{Agree}, 9: \textit{Strongly Agree}). Participants rated the \gls{us} while being shown the \gls{gui} prototype for \textit{(1)} \textit{the user story being found in the \gls{gui} prototype}, as well as \textit{(2)} \textit{clarity and precision}, \textit{(3)} \textit{specificity}, and \textit{(4)} \textit{professionalism}. Figure \ref{fig:plots_gen_us} shows violin plots for each assessed aspect for all \glspl{us}. Means (medians) for each of the four questions were 6.70 (7), 6.63 (7), 6.53 (7), and 6.02 (6). However, while the total average and median values were high (medians of \textit{"Agree"} for the first three questions in the Likert scales), there was considerable variance of the ratings for the \glspl{us} derived from individual \glspl{gui}. Therefore, we included \gls{us} ratings for four particularly selected \glspl{gui} in the subsequent Figure \ref{fig:plots_gen_us}. 

\begin{myrqbox}
\textbf{Answer to RQ$_{4}$:} The \glspl{us} generated by our \gls{llm}-based approach from \gls{gui} prototypes achieved positive ratings across the four different considered metrics (mean scores from 6.02---6.70), but with substantial variance across individual \gls{gui} prototypes.
\end{myrqbox}

\section{Discussion}
Using our assistant in the \textit{treatment} configuration, participants created \gls{gui} components rated by crowdworkers as, on average, completing the \glspl{us} to a higher degree. Nevertheless, some \glspl{us} were rated as better completed in the \textit{control} configuration. Subsequently, we discuss some of the results and put them into context. In addition, we discuss recommendations for the design of automated \gls{llm}-based support systems for \gls{gui} prototyping.

\subsection{RQ$_{1}$: Effectiveness of Generating Prototype Components}
In our first study, we evaluated the ability of our \gls{llm}-based approach to generate \gls{gui} prototype components, which received high ratings from crowdworkers on functional requirements fulfillment, \gls{us} fulfillment, and component correctness (three times median of eight and one median of seven on a nine-point Likert scale). These results indicate that our approach successfully produced components that effectively fulfill functional requirements and align with \glspl{us}. This supports the thesis that \gls{llm}-based methods can assist early-stage prototyping by generating functional and relevant elements with minimal manual intervention.
However, slightly lower ratings for textual clarity suggest room for improvement in how text is generated for \gls{gui} components. We propose two potential avenues for refinement: \textit{(i)} technical improvements, such as domain-specific \gls{fs} prompting or finetuning, and \textit{(ii)} procedural improvements, such as integrating human feedback loops into the creation process.
The efficiency gains from the two-stage \gls{rag} approach hint at feasibility in large-scale design workflows where cost is a bottleneck.

\begin{figure*}[!t]
  \centering
 \includegraphics[width=\textwidth]{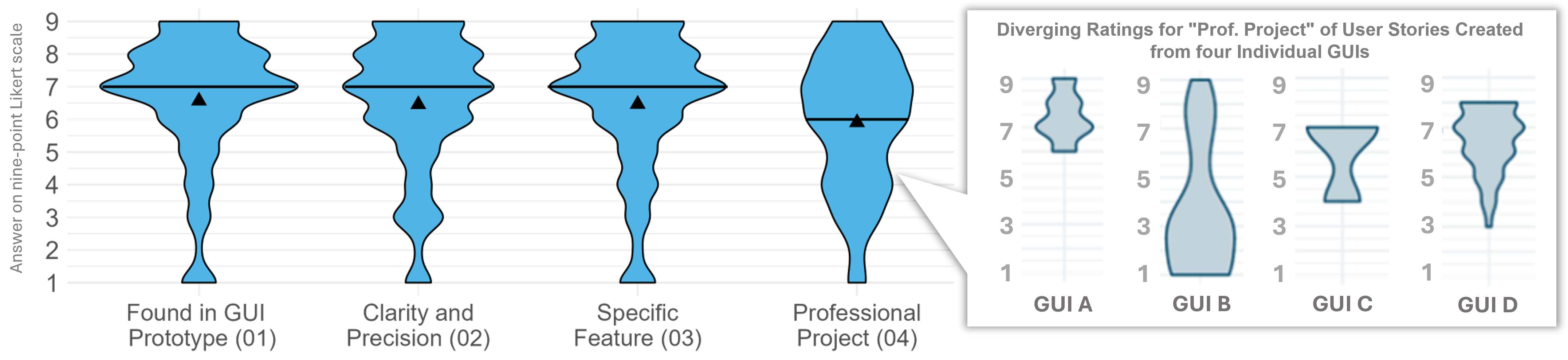}
  \caption[Violin plots for \gls{llm}-based \gls{us} generation]{\textbf{Left:} Violin plots for crowdworkers' rating of generated \glspl{us} from \gls{gui} prototypes for \textit{(1)} \textit{\glspl{us} accurately describing functions of the \gls{gui}}, \textit{(2)} \textit{\glspl{us} being precise}, \textit{(3)} \textit{for an identifiable feature}, and \textit{(4)} \glspl{us} \textit{formulated as in professional projects}. Triangles represent means, lines medians.
  \textbf{Right:} Plots for crowdworkers' rating of generated \glspl{us} from \gls{gui} prototypes for the fourth question, \gls{gui} prototypes (\textit{A---D}) and \gls{us} pairings.}
  \label{fig:plots_gen_us}
\end{figure*}
\glsunset{ux}
\subsection{RQ$_{2}$, RQ$_{3}$: \gls{llm}-Based Assistant Impact on \gls{gui} Quality}
\label{subsec:completeness}
\glsreset{ux}
Our investigation into how our \gls{llm}-based assistant affects the quality of \gls{gui} prototypes, \gls{us} completion, and perceived \gls{ux} showed that experts rated \gls{gui} prototypes created with the \gls{llm}-based assistant as having a higher quality, especially in selecting the right components and utilizing correct descriptions. \glspl{us} were also rated as completed to a higher degree for prototypes created using the \gls{llm}-based assistant (23 cases for the \textit{treatment} versus seven cases for the \textit{control} group). These results suggest that the assistant effectively supports task completion and improves the overall quality of \gls{gui} prototypes.
As expected, participants who used the \gls{llm}-based assistant started or completed more app screens as part of the study tasks. The automatic generation of \gls{gui} components likely contributed to this efficiency benefit and enabled faster task progression.
Participants also reported a higher perceived \gls{ux} when using the \gls{llm}-based assistant, such as significantly higher ratings for \textit{future use}, \textit{perceived impact on design quality}, and \textit{confidence in creating professional \glspl{gui}}. While \gls{nasatlx} scores for mental and temporal demands, effort, and frustration were lower for the \textit{treatment} group, these differences were not significant. However, both groups experienced high cognitive and temporal demands potentially due to the extent of the tasks (e.g., creating three \glspl{gui}, each with ten \glspl{us}). This intentional aspect of the study design was finetuned in pre-tests to ensure that the \textit{treatment} group supported by \gls{llm}-based \gls{gui} component generation would not run out of tasks.

\begin{figure*}[!t]
  \centering
 \includegraphics[width=0.98\textwidth]{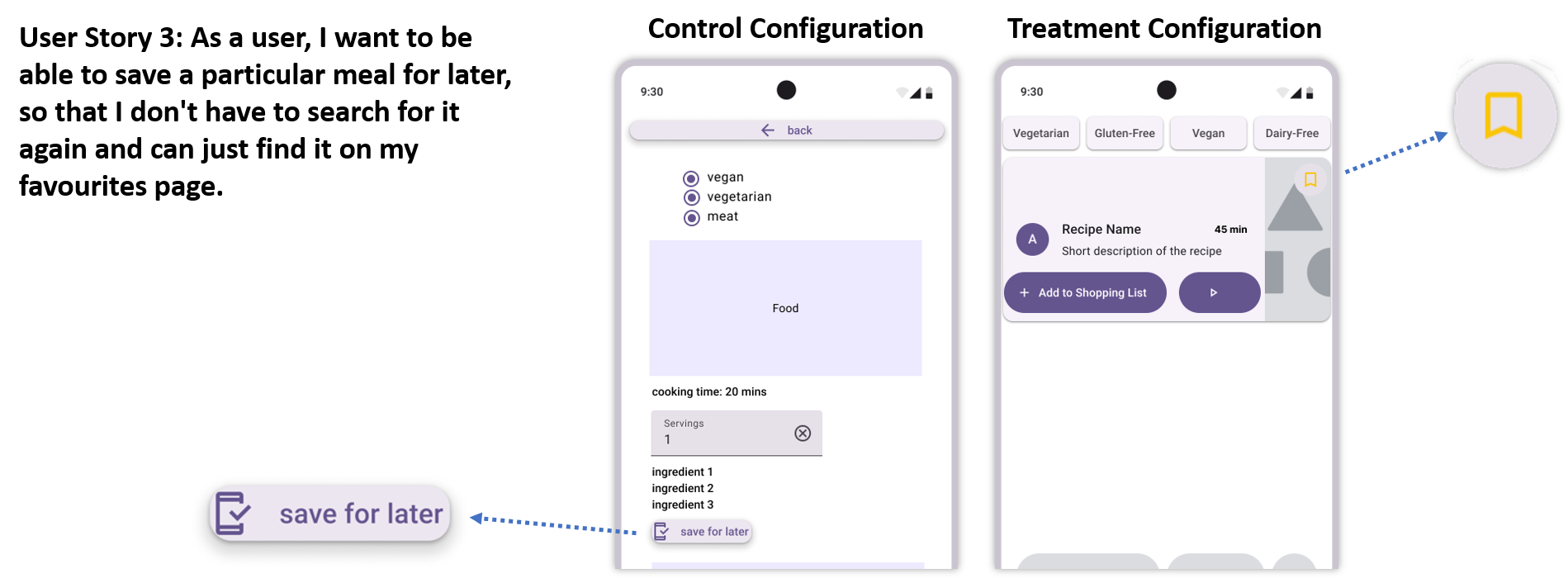}
  \caption[Discussion on example \gls{gui} prototypes created during the evaluation]{Example of an instance where a \gls{us} (taken from the \textit{second} app and \textit{third} \gls{us} of the user study) was rated more complete in the \textit{control} group. The \textit{control} group tended to use buttons with a clear call to action (e.g. \textit{“save for later”} here), whereas in the \textit{treatment} group our approach almost always created \textit{bookmark} icon buttons, representing a concise but more implicit implementation of this particular \gls{us}.  Figure contains \textit{Material 3 Design Kit} \citep{material_design_kit} components from \textit{Google}, used under \textit{CC BY 4.0 license}.}
    \label{fig:UserStories_Bookmarks}
\end{figure*}

When analyzing the cases in which crowdworkers rated \gls{gui} prototypes created in the \textit{control} group as completing the \gls{us} to a higher degree in comparison to the \textit{treatment} group, it is noticeable that the presence of a text label often played a crucial role. Figure \ref{fig:UserStories_Bookmarks} illustrates such an example where a button with a text label (\textit{"save for later"}) was created for the \gls{us} by a participant in the \textit{control} group (\textit{left}). At the same time, an icon button with a \textit{bookmark} icon was used in the \textit{treatment} group (\textit{right}). Almost every \gls{gui} prototype in the \textit{treatment} group shows a \textit{bookmark} icon for this \gls{us} and our tests have shown that our \gls{llm}-based approach generates \textit{bookmark} icon buttons almost consistently for this \gls{us}.

This motivates two discussion points. Firstly, the extent to which crowdworkers were influenced by explicit textual mentions in the components, especially when comparing them with aesthetically styled components such as the illustrated icon button. Secondly, how the temperature setting in our \gls{llm}-based approach led to the generation of similar recommendations when the solution space was somewhat limited and participants clicked the \textit{"Generate Another Recommendation"} button. We recommend balancing the generation of more diverse additional recommendations for further studies.
Here, multiple possible solutions come to mind: evaluating the extent to which the temperature can be changed in further iterations of generating the same component, enabling the approach to generate multiple implementations of the same \gls{us} while instructing it to produce distinct variants or allowing users to provide further \gls{llm} instructions in addition to the \gls{us}. 
Future users could restrict the next version of the same component generation by specifying instructions such as \textit{"include a label in the next generation"}. 
While participants expressed great appreciation for the automated generation of components, e.g. P12: \textit{"Straightforward to use and it gave good advice most of the time, excellent starting point for most stories"}, P19: \textit{"Really good recommendations."}, P10: \textit{"So much work was reduced"}, there were improvements mentioned in regard to the discussed variations of recommendations. Participants, such as P7: \textit{"[I] would have liked to have had an option to adjust the task that is sent to the AI"}, frequently mentioned the desire for more variation in the recommendations. In fact, this suggestion was among the most mentioned improvements by participants (only second to the time it took the \gls{llm} to generate recommendations).

\subsection{RQ$_{4}$: Effectiveness of Generating \gls{us} from Components}
In the third study, we investigated the effectiveness of our approach for automatically deriving \glspl{us} from \gls{gui} prototypes. Overall, crowdworkers rated the \glspl{us} created by the approach highly on relevance, clarity, precision, and specificity (all with a median of seven on a nine-point Likert scale), indicating that the generated stories generally align well with the functionality and purpose of the \glspl{gui}. However, ratings for professionalism (six-point median) were slightly lower, raising questions about, e.g., the tone and depth of the generated \glspl{us}.
While the median across all ratings and underlying prototypes is high, it is worth looking at the ratings of \glspl{us} derived from individual \glspl{gui}. In the results in Figure \ref{fig:plots_gen_us}, we have intentionally shown ratings of \glspl{us} that emerged from four individual \gls{gui} prototypes. Here, the presented \glspl{gui} resulted in variations in the ratings, showing that the quality of the \glspl{us} created varies depending on the underlying \gls{gui}. This variability may stem from differences in \gls{gui} complexity, ambiguity, or the \gls{llm}s' ability to generate \glspl{us} for specific tasks, which were more likely to be included in pretraining. Despite these limitations, the findings underscore the potential of \gls{llm}-based tools to support early-stage design workflows. Integrating such tools into iterative design processes could reduce manual effort, enabling focus on higher-level challenges.

\subsection{Generating \gls{gui} Prototypes from \glspl{us} or Vice-Versa?}
Our results show that our \gls{llm}-based approach enables the effective creation of \gls{gui} prototypes (components), mainly desired by \gls{uiux} designers and partly by product owners, and effectively derives \glspl{us} from \gls{gui} prototypes, which product owners mainly desired in our interviews. Readers may wonder that it is certainly not possible to do both simultaneously and might sense a certain chicken-and-egg problem. Based on our interviews, we identified two options for integrating both features. We see our assistant as a mediator for both roles through the two described features. While product owners define \glspl{us}, the \gls{gui} prototyping assistant could generate \gls{gui} components in parallel. Vice versa, the \gls{uiux} designer could create design drafts from which the assistant derives \glspl{us} for the product manager. In addition, \gls{gui} prototypes are often created in an iterative process, so it is conceivable that product owners formulate initial \glspl{us} and the assistant creates a design proposal, which \gls{uiux} designers subsequently refine. The created refinement is then used to generate more precise \glspl{us}. Therefore, we regard the features of automatic \gls{gui} creation and \gls{us} creation as one of the leading collaboration features.

\section{Threats to Validity}
In the following, we provide a collection of threats to internal validity, such as selection biases, and external validity, such as threats to generalizability. Furthermore, we briefly explain how they may influence the results of our paper and applied mitigation strategies.

\paragraph{Internal Validity.}

\vspace{0.15cm}
\textit{Measuring Treatment Completeness.} We evaluated several aspects such as the detection, matching, and component generation mechanisms as well as the overall design of the proposed plugin within a user study. Therefore, the contribution of each of these aspects to the quality of the created \gls{gui} prototypes is not entirely clear. To address this issue, we initially conducted the evaluation of \gls{llm}-based generation of \gls{gui} components based on \glspl{us} ex ante and showed its effectiveness, indicating that a large contribution to the improvement of the quality of the \gls{gui} prototypes and the improvements in terms of efficiency are mainly due to the automatic \gls{gui} generation.

\paragraph{External Validity.}
\vspace{0.15cm}
\textit{Artificial Lab Setting.} The user study was conducted in an artificial lab setting including a less representative population of participants as a proxy for \gls{uiux} designers regarding the group (i.e. undergraduate and graduate students) as well as the young age distribution (25.4 years on average). However, we primarily utilized this conducted user study to evaluate various usability aspects associated with the proposed approach. To address the cross-functional nature of \gls{gui} prototyping, we conducted interviews with stakeholders from different roles (\gls{ux}, Product Owner, \gls{sd}) to ensure that the assistant's design rationales reflect the perspectives of diverse stakeholders. To strengthen the evaluation of the quality of generated \glspl{us} and generated \gls{gui} components, we conducted an additional annotation of the generated artifacts with self-reported \gls{uiux} professionals on \textit{Prolific} \citep{prolific_academic_ltd_prolific_nodate}. The obtained results show the generation effectiveness. Nonetheless, we did not evaluate the assistant in a co-design situation involving multiple stakeholders at the same time, which represents an important direction for future work to explore its potential in collaborative prototyping contexts.

\vspace{0.15cm}
\textit{Artificial Functional User Stories.}
To evaluate our proposed approach, we heavily relied on the publicly available \gls{us} dataset (see Chapter \ref{cha:interlinking} for details), which is the only dataset available combining \glspl{us} with \gls{gui} prototypes. These \glspl{us} solely focus on functional aspects of a single \gls{gui} prototype. In more natural settings, \glspl{us} often span several \gls{gui} prototypes as well as encompass non-functional aspects. However, in our work, we decided to initially focus on functional \glspl{us} to reduce complexity, yet still provide valuable support and facilitate the creation of \gls{gui} prototypes. In addition, the employed US dataset is validated and filtered for quality (see Chapter \ref{cha:interlinking} for details).

\vspace{0.15cm} 
\textit{High-Level \gls{gui} Descriptions for Component Generation.}
To provide additional context for the generation of \gls{gui} components as part of its evaluation, we employed additional high-level \gls{gui} descriptions as an input to the \gls{zs} prompt for the \gls{llm}. Moreover, in the user study, the context of the current \gls{gui} prototype status could meaningfully be utilized by the \gls{llm}, for example, to adapt the generated \gls{gui} components to the \gls{gui} prototype and to predict accurate positioning of the \gls{gui} components. While these brief \gls{gui} descriptions can easily be obtained in practical \gls{gui} prototyping settings, in the conducted evaluation they were created by a research assistant and evaluated for accuracy and correctness by multiple paper authors.

\vspace{0.15cm} 
\textit{Participant Interviews.}
There is a potential sample bias through the recruitment of \gls{uiux} professionals, product owners, and software developers for the initial interviews, which may involve limited diversity in terms of company size, geographic location, and experience. However, we attempted to include multiple participants for each of the roles.

\section{Limitations}

\vspace{0.15cm} 
\textit{Mobile \glspl{gui} and Limited Component Library.}
Since our approach focuses solely on mobile \glspl{gui}, other \gls{gui} types with varying characteristics (e.g., screen size, domains, etc.) are disregarded. In addition, we heavily relied on mobile \glspl{gui} taken from the \textit{Rico} dataset \citep{deka2017rico}, which has a restricted scope and usefulness. However, the \textit{Rico} dataset is the largest publicly available \gls{gui} dataset and encompasses \glspl{gui} from over 27 different domains. In the future, we plan to extend our approach to more diverse \gls{gui} types. In addition, the considered \gls{gui} component library is restricted in size and configuration complexity. To enable meaningful support, we already included over 89 different \textit{Material Design} \gls{gui} components and 100 icons. However, we plan to further expand both libraries to additionally enhance the support capabilities.

\vspace{0.15cm} 
\textit{Static \gls{gui} Prototypes.}
In this work, we proposed a support approach for creating single static \gls{gui} prototypes, and the resulting format does not allow for interaction. Usually, the interactions considered in practice span multiple \glspl{gui} and maintaining visual design and interaction consistency across screens within an entire application is of high importance. In order to improve the usefulness and application scenarios of the proposed approach, we plan to further extend our \gls{llm}-based approaches and plugin to additionally enable support for \glspl{us} spanning across several \gls{gui} prototype screens.

\vspace{0.15cm} 
\textit{Functional User Stories.}
Since the current approach solely provides support for functional \glspl{us}, the application scenarios are limited. However, particularly for creating initial \gls{gui} prototypes for rapidly obtaining stakeholder feedback regarding the functionality, our approach can be applied. In such an initial elicitation scenario, where the \gls{gui} prototypes act as a functional requirements communication artifact, the detailed design aspects are of less significance. For future work, we plan to extend the support to also include non-functional \glspl{us}, for example, including styling aspects such as coloring and corporate design, among others. However, styling aspects of \gls{gui} components could also be integrated by replacing the component library within our approach. Therefore, the assistant could be extended to enable providing textual prompts for stylistic or design requirements. To increase flexibility, another extension could be that the assistant provides multiple variants of the generated component.

\textit{Handover, Versioning and Commenting.} 
One additional limitation is that we could only thoroughly implement selected collaboration requirements for the version of the assistant in this paper. In the interviews, both \gls{uiux} designers and product owners described the versioning of \glspl{us}, commenting on \glspl{us} combined with their components in the \gls{gui} prototype, and an improved handover (e.g., of parameters to software developers) compared to the existing functionalities in \textit{Figma} as further requirements. Versioning was addressed in our intervention, but the implementation of our approach in its current form is limited to creating new components and cannot adapt existing components. We plan to address this in the future. Commenting functionality was not implemented, but as \textit{Figma} already supports commenting components, it is feasible to link \glspl{us} to components and leverage this feature. Finally, the extraction and handover of data to developers were not included, but tools like \textit{Figma} already provide solutions for this.

\section{Related Work}
\label{sec:closing-rel-work}

To support the \gls{gui} prototyping process, different approaches have been proposed in research. For example, \textit{DesignScope} \citep{o2014learning, o2015designscape} facilitates the design process by actively recommending layout refinements and design suggestions. Moreover, \textit{SketchPlorer} \citep{todi2016sketchplore} represents a sketching approach that integrates an ad-hoc layout optimizer to rapidly provide layout improvement suggestions to users. In addition, \textit{GUIComp} \citep{lee2020guicomp} provides assistance for novices during \gls{gui} prototyping through a multi-faceted support system including the retrieval of similar \glspl{gui} from the \textit{Rico} dataset \citep{deka2017rico}. \textit{GUIComp} offers a visualization of several \gls{gui} complexity metrics such as \gls{gui} component alignment, balance, and density, and an attention map for the created \gls{gui} prototype. In contrast, our approach targets rapid generation of \gls{gui} prototypes (components) based on fine-grained \glspl{us} that can instantly be integrated into the current working \gls{gui} prototype in proprietary representations, enabling visual editing within the well-known visual editor \textit{Figma}. We close this loop and provide a tight integration of user requirements (in the form of \glspl{us}) and resulting \gls{gui} prototypes. To ensure consistency between user requirements and \gls{gui} prototypes, an ontology-based approach was proposed before \citep{silva_10.1007/978-3-030-24289-3_46}. However, this approach necessitates the availability of a respective ontology for the domain and enables mere verification of the implementation of requirements and thus cannot actively provide support during the creation of the \gls{gui} prototype. In contrast, with our research, we proposed an \gls{llm}-based approach for rapidly generating the implementations for \glspl{us} in an editable form as well as an integration of the \gls{llm}-based assistant in the form of a fully-fledged plugin in \textit{Figma}.

Moreover, multiple approaches for generating \gls{gui} prototypes from text descriptions have been proposed in research before. For example, \textit{Instigator} \citep{brie2023evaluating} trains a full \textit{minGPT} \citep{minGPT} model from scratch exploiting a large-scale repository of web pages to enable training of a task-specific \gls{gpt} model. Furthermore, approaches such as \textit{MAxPrototyper} \citep{yuan2024maxprototyper}, \textit{UIDiffuser} \citep{wei2023boosting}, and finetuning an \gls{llm} \citep{feng2023designing} have been proposed before (see Chapter \ref{cha:zs_gui_generation} Section \ref{sec:zs-related-work} for details). In contrast, our approach enables the generation of proprietary \gls{gui} representations through a novel two-stage \gls{rag} approach, which efficiently incorporates external \gls{gui} component libraries and closely integrates \glspl{llm} into prototyping tools.

\section{Conclusion}

The proposed approaches in this chapter were driven by the challenge \challonefive{}: \textit{How can \gls{llm}s be efficiently adapted to generate editable \gls{gui} prototype representations from \gls{nlr}?} To tackle this challenge, in this chapter, we proposed a novel two-stage \gls{rag} approach which enables the efficient generation of proprietary \gls{gui} representations and integration into well-known prototyping tools to generate visually editable \gls{gui} prototypes. In addition, we integrated other \gls{llm}-based approaches such as automatic detection of \gls{us} implementation, matching of requirements to their implementation in the \gls{gui} prototype, and generation of \glspl{us} from prototypes. Our evaluation indicates that the \gls{llm}-based approach can successfully generate \gls{gui} component recommendations from \glspl{us}.
Moreover, we present a novel \gls{llm}-based assistant as a plugin for \textit{Figma} with role-specific functionalities. In particular, we found that participants using our assistant achieved higher \gls{us} completion. Our assistant also led to significantly higher satisfaction of experts with the created \gls{gui} prototype designs. Using the \gls{us} detection, matching, and automated \gls{gui} (component) generation, our participants reported lower task load values, significantly higher confidence in creating professionally appearing \gls{gui} prototypes, high interest in future use of the assistant, and a higher impact on design quality.
Inspired by real-world requirements from product owners collected in our interviews, we successfully expanded the approach to derive \glspl{us} directly from \gls{gui} prototypes that can then be directly refined by product owners. In our evaluation, large parts of the \glspl{us} were rated as specific, explicit, and derived from identifiable \gls{gui} components. Our results show that \gls{llm}-based \gls{us} detection, matching, and the automated \gls{gui} (component) generation can be integrated directly into \textit{Figma} providing a resource-efficient prototyping assistant to increase \gls{us} completion and facilitate the integration of \glspl{us} and \gls{gui} prototypes.
\clearpage
\newpage
\thispagestyle{empty}
\null  

\chapter{GUIDE: LLM-Driven GUI Generation Decomposition for Automated Prototyping}
\chaptermark{LLM-Driven GUI Generation Decomposition for Prototyping}
\label{cha:guide}
In the previous chapter, we presented a novel two-stage \gls{rag} approach tackling challenge \challonefive{}, which enables the efficient integration of external \gls{gui} component libraries (such as \textit{Material Design}) into \glspl{llm} to generate proprietary \gls{gui} prototype representations and thereby create visually editable prototypes (inside the \textit{Figma} prototyping environment). While this approach requires a fine-grained requirements collection, such detailed individual requirements are often absent especially in the early stages of elicitation. Therefore, we build on the core \gls{rag} approach presented in the previous chapter, but extend it to enable the specification of high-level \gls{gui} prototype requirements. The work presented in this chapter has been previously published \citep{kolthoff2025guide}\footnote{This section is adapted from: \textbf{Kolthoff, Kristian\textsuperscript{*}}, Kretzer, Felix\textsuperscript{*}, Bartelt, Christian, Maedche, Alexander, and Ponzetto, Simone Paolo. GUIDE: LLM-Driven GUI Generation Decomposition for Automated Prototyping. In \emph{2025 IEEE/ACM 47th International Conference on Software Engineering: Companion Proceedings (ICSE-Companion) (ICSE, A*)}, Ottawa, ON, Canada, April 2025, pages 1--4. IEEE. *Authors contributed equally. Sections are rearranged and directly reused with only minor adaptations.}. Our source code, prototype (\textit{Figma} plugin), datasets, and demonstration video are all publicly available\footnote{Materials for this chapter are available at \url{https://github.com/kristiankolthoff/GUIDE-Prototyping} and the respective demonstration video at \url{https://youtu.be/C9RbhMxqpTU}}.

\paragraph{Personal Contribution.} \textit{Felix Kretzer} and I contributed equally to the ideation and creation of the concept for the work. While I implemented the \gls{llm}-based approaches and backend functionality, \textit{Felix Kretzer} implemented the \gls{gui} and frontend of the approach. \textit{Felix Kretzer} and I contributed equally to the evaluation design, while \textit{Felix Kretzer} conducted the evaluation and analysis of results. \textit{Felix Kretzer} wrote the section on the experimental evaluation setup and results (and created the evaluation results table), while I wrote the rest of the original manuscript and created respective figures and plots.

\section{Motivation}

With the approaches presented in the previous chapters, we introduced techniques for challenge \challone{}, namely, enabling rapid mapping of \gls{nlr} to \gls{gui} prototypes. The \gls{zs} prompting-based methods for efficiently adapting \glspl{llm} to more effectively generate \gls{gui} prototypes (see Chapter \ref{cha:zs_gui_generation}) provide a substantial performance. However, as discussed in the previous chapter, there is currently a gap between \glspl{llm} generating \gls{gui} code as text and established \gls{gui} prototyping workflows, namely visual editors such as \textit{Figma} \citep{figma_tool} widely employed by practitioners. Specifically, because \glspl{llm} are primarily designed for text generation, their outputs are not easily visualizable, posing a challenge for their deep integration into \gls{gui} prototyping. Nevertheless, these \glspl{llm} are pretrained on large amounts of \gls{html}/\gls{css} data, enabling them to generate prototype code that can be rendered in web browsers. However, modifications to \gls{gui} prototypes are generally performed by the prototype developers visually rather than through direct code manipulation. Moreover, when minor changes to generated \gls{gui} prototypes are requested by the user, \glspl{llm} typically inefficiently regenerate the entire \gls{gui} prototype. While we provided initial solution approaches in the previous chapter, these approaches assume the availability of a low-level, fine-grained requirements collection, which is often absent in early elicitation stages. More often, solely high-level, abstract \gls{nlr} are available. Therefore, the work in this chapter is still driven by the leading research question of challenge \challonefive{}, with the focus being on high-level, abstract \gls{nlr} in comparison to the fine-grained requirements as shown in the previous chapter: \textit{How can \gls{llm}s be efficiently adapted to generate editable \gls{gui} prototype representations from \gls{nlr}?}

To close this gap, in this work, we introduce \textit{GUIDE}, a novel \gls{llm}-driven \gls{gui} generation decomposition approach, which is tightly integrated into the popular prototyping tool \textit{Figma}. Our approach derives low-level \gls{gui} features from high-level \gls{nlr} provided by users and directly generates a corresponding editable \gls{gui} prototype using the popular \textit{Material Design} component library \citep{material_design_kit} by adopting the \gls{rag}-based approach developed in the previous chapter. \textit{GUIDE} facilitates the efficient modification of existing features or the addition of new ones within the \gls{gui} prototype, reducing the need to regenerate the entire prototype. Thus, this integration combines the \gls{llm} generation capabilities with the advantages of traditional visual \gls{gui} prototyping, enabling both more effective (\gls{gui} quality) and efficient (token consump.) \gls{gui} prototyping. Our approach can help prototype developers to rapidly and effectively create prototypes based on high-level \gls{nlr}. To evaluate our approach, we conducted a user study and asked participants to create \gls{gui} prototypes \textit{with} and \textit{without} \textit{GUIDE} and assessed the quality of created prototypes with crowdworkers from \textit{Prolific} \citep{prolific_academic_ltd_prolific_nodate, douglas2023data}. Our evaluation shows that the prototypes generated with our approach obtain significantly higher scores across all considered evaluation (\gls{gui} quality) metrics.

\paragraph{Contributions.} With this work, we make the following main research contribution: 

\begin{itemize}[left=0.1cm]

    \item \textit{High-level \gls{nlr} to editable \gls{gui} prototypes:} we extend our previous \gls{rag}-based approach --- which enables the efficient integration of external \gls{gui} component libraries to generate proprietary \gls{gui} prototype representations --- by an \gls{llm}-based approach that initially decomposes high-level \gls{nlr} into low-level \gls{gui} features. In addition, we implement a tool prototype of our proposed approach as a novel \textit{Figma} plugin.
     \item \textit{Comprehensive user study:} we conduct an experimental evaluation of our approach by comparing the quality of created \gls{gui} prototypes \textit{with} and \textit{without} our approach through a user study, showing significant prototype quality improvements for \textit{GUIDE}. 
\end{itemize}

\section{Approach: GUIDE}


\begin{figure*}[!t]
  \centering
 \includegraphics[width=1.0\textwidth]{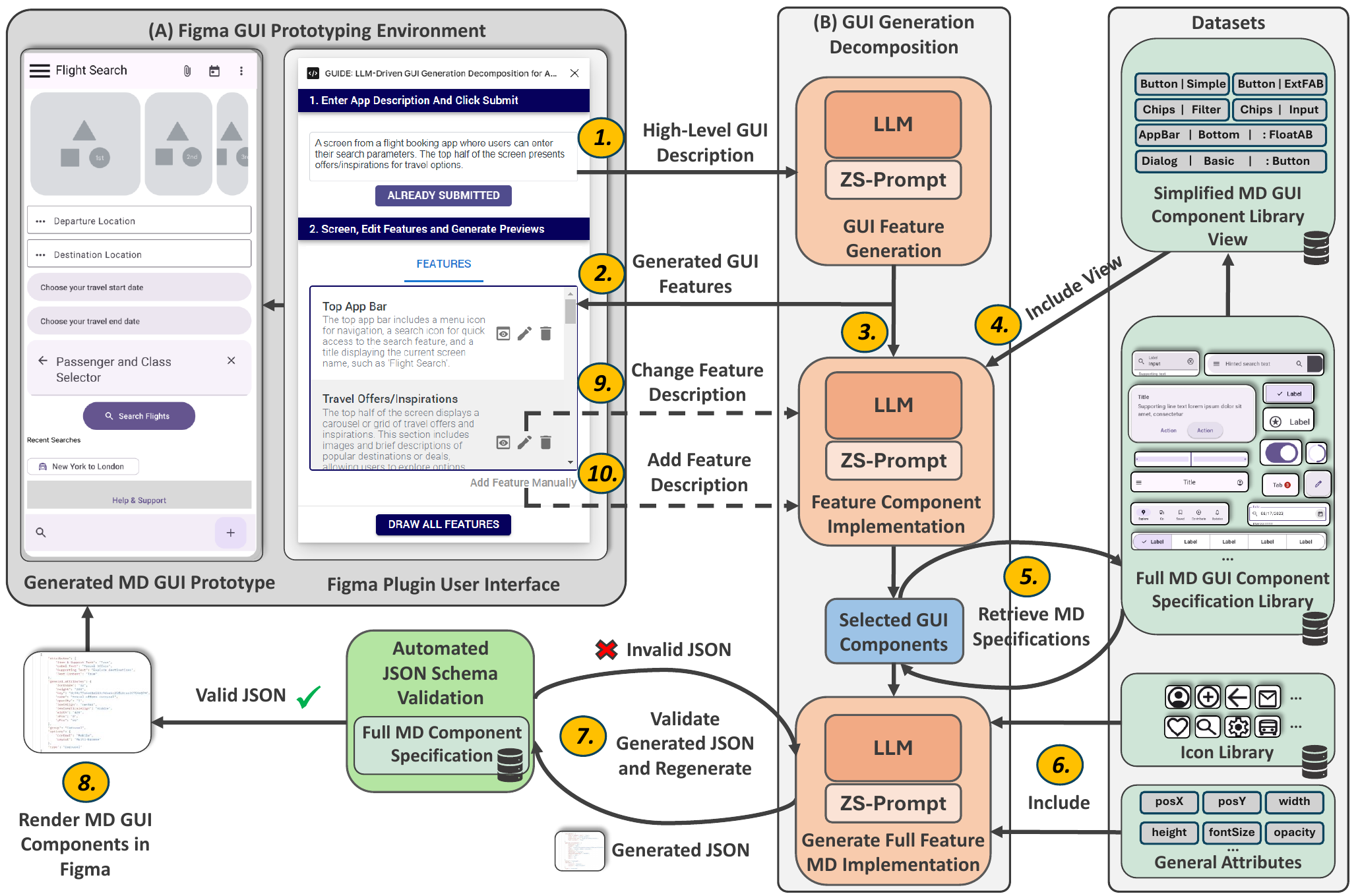}
  \caption[Overview of the \textit{GUIDE} architecture]{Overview of the \textit{GUIDE} architecture with \textit{(A)} \textit{Figma} plugin and \textit{(B)} \gls{gui} generation decomposition with a two-stage \gls{rag} approach and a \gls{gui} component library.}
	\label{fig:overview-guide}
 
\end{figure*}

\textit{GUIDE} is composed of two main components including \textit{(A)} the \textit{Figma} plugin enabling users to interact with our approach via an easy-to-use interface and \textit{(B)} the \gls{gui} generation decomposition, which is an \gls{llm}-based approach for \textit{(i)} decomposing the high-level \gls{nlr} provided by the user into a fine-grained \gls{gui} feature collection, \textit{(ii)} selecting relevant combinations of \gls{gui} component types for each \gls{gui} feature from a prespecified \gls{gui} component library, and \textit{(iii)} generating fully specified \gls{gui} component implementations (\gls{rag} approach). An overview of the \textit{GUIDE} architecture is illustrated in Figure \ref{fig:overview-guide}.

\subsection{Figma Plugin User Interface}

Initially, prototype developers can provide their short and high-level \gls{nlr} for the \gls{gui} prototypes they are tasked with creating directly into the user interface of \textit{GUIDE}, which is implemented as a plugin within \textit{Figma} \citep{figma}. Subsequently, \textit{GUIDE} derives a fine-grained \gls{gui} feature collection displayed in the plugin with a name and a short textual description of the feature. Our approach helps users to rapidly screen the generated feature list, directly edit the feature description, and delete or add additional features manually, controlling \gls{gui} generation. \textit{GUIDE} enables users to directly render the implementation of the suggested features as editable \gls{gui} components within \textit{Figma}.

\subsection{\gls{gui} Generation Decomposition}
\glsreset{pd}
To improve the generated \gls{gui} prototypes and enhance control over the generation process, we decompose the \gls{gui} generation task into multiple smaller steps, similar to \gls{pd} \citep{khot2022decomposed}. As research has shown before, this can not only improve the effectiveness of the \gls{llm} in completing complex tasks, but also follow a more human-like approach to creating \gls{gui} prototypes \citep{khot2022decomposed}. 

\paragraph{\gls{gui} Feature Generation.} Based on the high-level \gls{nlr} provided by the user, we employ an \gls{llm} in a \gls{zs} prompting approach \citep{radford2019language} to generate the corresponding \gls{gui} feature collection. In particular, we instruct the \gls{llm} to focus solely on recommending functional features that can be represented visually in a \gls{gui} prototype, therefore neglecting \gls{nfr} (such as \textit{responsive design}, \textit{accessibility}). In addition, we provide a \gls{json} schema to the \gls{llm} including a name and a short description of the requirements and instruct the model to directly generate the respective \gls{json} format. This representation can then easily be consumed by the frontend and subsequent steps.

\paragraph{Retrieval-Augmented \gls{gui} Feature Generation.} After generating a feature collection, we decompose the subsequent generation of the respective feature implementations as \gls{gui} components within \textit{Figma} into two separate steps, mainly adopted from the two-stage \gls{rag} approach from the previous chapter (see Chapter \ref{cha:closing}) and briefly summarized. Within \textit{GUIDE}, we integrated the popular \textit{Material Design (MD)} component library \citep{material_design_kit} to enable the generation of respective editable \gls{gui} prototypes. In particular, the library consists of over 59 different components, ranging from simple individual components (e.g., \textit{Button}, \textit{Checkbox}, and \textit{Label}) to more complex component groups (e.g., \textit{Search Bar}, \textit{Dialog}, and \textit{Top App Bar}). Although \glspl{llm} are trained on massive amounts of textual data, the specifically employed component library with the particular configuration options and syntax cannot directly be generated by the \gls{llm}. Since the full \gls{json} specification of each \gls{gui} component encompasses many different properties and typically only a small number of components are relevant for implementing a feature, incorporating the entire specification into the \gls{gui} generation as part of a \gls{zs} prompt for an \gls{llm} would be inefficient.

Therefore, we utilize the \gls{rag} \citep{lewis2020retrieval} approach for the implementation of \gls{gui} features presented in the previous chapter. In particular, we first automatically derive a simplified component library view from the full specification library (only including the component group (e.g., \textit{Button}) and component type (e.g., \textit{FloatingActionButton}) information), significantly reducing the number of required tokens (reduction of approximately 60\% of consumed tokens, see Chapter \ref{cha:closing} for details). Afterwards, we construct a \gls{zs} prompt that incorporates the simplified library in the context and instructs the \gls{llm} to select the relevant component types to implement the respective \gls{gui} feature. Subsequently, we employ the selected \gls{gui} components to retrieve the full \gls{json} specifications of the relevant \gls{gui} components and utilize them in a second \gls{zs} prompt that finally generates the actual \gls{gui} feature implementation with \textit{Material Design} components. In addition, we provide an icon collection and a set of general attributes (e.g., \textit{posX}, \textit{posY}, \textit{width}, and \textit{height}) that each component possesses in the generation process. The \gls{llm} is instructed to generate the full implementation for each \gls{gui} feature using a prespecified \gls{json} format, which can be used to render the components in \textit{Figma}. To improve the reliability of the \gls{json} generation, we additionally conduct an automated validation of the generated \gls{json} format by comparing it to the respective \gls{json} schema for correctness.

\subsection{Prototype Implementation}
\glsunset{rest}
The \textit{GUIDE} prototype is implemented as a \textit{Figma} plugin using \textit{TypeScript} for the frontend. To enable the integration of the \gls{llm} into generating prototypes in \textit{Figma}, we created a proprietary \gls{json} format that can be interpreted both by the \gls{llm} and the plugin for rendering. The backend of \textit{GUIDE} is implemented as a \textit{Python}-based \textit{Quart} app running within a highly scalable \textit{gunicorn}/\textit{nginx} server, providing a \gls{rest} \gls{api} for the described functionality. For the \gls{llm}, we employed the most recent \textit{\gls{gpt}-4o} model\footnote{At the time of implementing the proposed approach, we utilized the most advanced \gls{mllm} available from \textit{OpenAI}, namely \textit{\gls{gpt}-4o} \citep{hurst2024gpt}, which is an extension over \textit{\gls{gpt}-4} \citep{openai2023gpt4}.} from \textit{OpenAI} (128k token context, \textit{accessed in October 2024}), representing the multimodel extension of their prior \textit{\gls{gpt}-4} model with state-of-the-art performance \citep{openai2023gpt4}.
\vspace{-0.8cm}
\section{Experimental Evaluation}
In the following, we describe the setup of our evaluation. Particularly, the evaluation consists of a small \textit{between-subjects} lab-based study in which participants created \gls{gui} prototypes with (\textit{treatment}) and without (\textit{control}) our assistant based on \gls{nlr}. These \gls{gui} prototypes were later evaluated by crowdworkers with \gls{uiux} experience on \textit{Prolific}. In particular, we pose the following research question in our experimental evaluation:

\begin{itemize}
    \item \textbf{RQ$_{1}$}: \textit{How effective are \gls{llm}-based approaches for generating editable \gls{gui} prototypes from high-level \gls{nlr}?} To answer this question, we conducted an \textit{between-subjects} user study in which participants created \gls{gui} prototypes \textit{with} or \textit{without} our \textit{GUIDE} \gls{gui} prototyping assistance approach support. Subsequently, we evaluated the quality of created \gls{gui} prototypes with crowdworkers from \textit{Prolific}.
\end{itemize}

\subsection{RQ$_1$: Effectiveness of \gls{gui} Prototype Generation}

\paragraph{User Study Procedure.} Participants started the first sub-study with a questionnaire and watching an initial video introduction on the prototyping tool, including the component library \textit{Material Design 3} \citep{material_design_kit}, which was the component library used to create \gls{gui} prototypes by both groups. The \textit{treatment} group watched an introduction to the use of our assistant, including the core functions: \textit{(i)} allowing the assistant to create \gls{gui} features based on the \gls{nlr}, \textit{(ii)} editing existing features and adding custom features, \textit{(iii)} previewing the generation, and \textit{(iv)} drawing the generation in the prototyping tool. The participants then conducted a 45-minute \gls{gui} prototyping phase and were instructed to create four \gls{gui} prototypes for four \gls{nlr}. After the prototyping phase, the \textit{treatment} group answered multiple questions about using the assistant in the survey.

\paragraph{User Study Participants.} We recruited 11 participants (3 female, 8 male) from a university pool to create \gls{gui} prototypes in a lab environment. On average, participants were 26.64 years old ($\sigma=5.68$), studied for 5.05 years ($\sigma=1.44$), and were randomly assigned to either the \textit{control} group (six participants) or the \textit{treatment} group (five participants).

\newcommand{\ColHead}[2]{%
  \parbox[c]{\hsize}{%
    \centering\bfseries\normalsize #1\\%
    \normalfont\small (#2)%
  }%
}

\begin{table}[!t]
\centering
\caption[Evaluation results of \gls{llm}-driven \gls{gui} generation (1)]{Median, mean, and results of \textit{Wilcoxon rank-sum tests (two-sided)} for \gls{gui} ratings on Likert scales (\textit{Strongly Disagree (1)} to \textit{Strongly Agree (9)}) for created \glspl{gui} from the \textit{control} and \textit{treatment} prototyping groups. Crowdworkers on \textit{Prolific} rated \glspl{gui} on: \textit{(1) meets requirements from description}, \textit{(2) utilizes correct components}, \textit{(3) texts fit to GUI description}, \textit{(4) GUI design consistent with GUI description}, and \textit{(5) appealing GUI design}.}
\label{tab:results_prolific_1_colhead}

\normalsize
\setlength{\tabcolsep}{1.5pt}
\renewcommand{\arraystretch}{1.1}

\begin{tabularx}{\columnwidth}{>{\centering\arraybackslash}m{1.8cm} >{\raggedright\arraybackslash}l *{5}{>{\centering\arraybackslash}X}}
\toprule
& \textbf{Measure}
& \ColHead{Requi.}{1}
& \ColHead{Comp.}{2}
& \ColHead{Text}{3}
& \ColHead{Consist.}{4}
& \ColHead{Appeal.}{5} \\
\midrule

\multirow{2}{*}{\textbf{Control}}
  & \textbf{Median} & 5     & 5     & 6     & 5     & 4 \\
  & \textbf{Mean}   & 4.746 & 4.769 & 5.015 & 5.000 & 4.354 \\
\midrule

\multirow{2}{*}{\textbf{Treatment}}
  & \textbf{Median} & 8     & 7     & 7     & 7     & 7 \\
  & \textbf{Mean}   & 7.393 & 7.107 & 6.713 & 7.260 & 6.127 \\
\midrule

\multirow{3}{*}{\textbf{Wilcoxon}}
  & \textit{p}\textbf{-value} & \textbf{<0.001*} & \textbf{<0.001*} & \textbf{<0.001*} & \textbf{<0.001*} & \textbf{<0.001*} \\
  & \textbf{$r$-value}        & 0.545            & 0.470            & 0.371            & 0.480            & 0.376            \\
  & \textbf{\gls{ci}}         & \textbf{$[-3,-2]$} & \textbf{$[-3,-2]$} & \textbf{$[-2,-1]$} & \textbf{$[-3,-2]$} & \textbf{$[-2,-1]$} \\
\bottomrule
\end{tabularx}
\end{table}

\paragraph{\gls{gui} Prototype Evaluation Procedure.} To evaluate the \gls{gui} prototypes created in our experiment across quality metrics (9-point Likert scale), we asked crowdworkers with \gls{uiux} experience. In an annotation task, crowdworkers each evaluated ten randomly drawn \gls{gui} prototypes. Crowdworkers on \textit{Prolific} rated \glspl{gui} on \textit{(1) meets requirements from description}, \textit{(2) utilizes correct components}, \textit{(3) texts fit to GUI description}, \textit{(4) GUI design consistent with GUI description}, \textit{(5) appealing GUI design}, \textit{(6) clear information organization}, \textit{(7) intuitive interactions}, \textit{(8) minimal errors}, and \textit{(9) overall satisfaction}.

\paragraph{\gls{gui} Prototype Evaluation Participants.} We invited 30 participants through \textit{Prolific} with \textit{\gls{uiux} design experience}, a \textit{high approval rate} (>99\%) on \textit{Prolific}, and certain \textit{language skills} (\textit{English} and \textit{German}, since some \gls{gui} prototypes contained German texts). We excluded two participants for failing at least one attention check in our study, which we included to increase data annotation quality. Hence, we eventually considered data from 28 participants (12 female, 1 diverse, 15 male). Participants had an average age of 33.21 years ($\sigma=8.23$), 7.36 ($\sigma=7.71$) years of experience creating designs and 6.46 ($\sigma=6.97$) years of experience in judging design (spent 16 minutes (mean) on our survey).

\begin{table}[!t]
\centering
\caption[Evaluation results of \gls{llm}-driven \gls{gui} generation (2)]{Median, mean, and results of \textit{Wilcoxon rank-sum tests (two-sided)} for \gls{gui} ratings on Likert scales (\textit{Strongly Disagree (1)} to \textit{Strongly Agree (9)}) for created \glspl{gui} from the \textit{control} and \textit{treatment} groups. Crowdworkers rated \glspl{gui} on: \textit{(6) clear information organization}, \textit{(7) intuitive interactions}, \textit{(8) minimal errors}, and \textit{(9) overall satisfaction}.}
\label{tab:results_prolific_2_colhead}

\normalsize
\setlength{\tabcolsep}{1.5pt}
\renewcommand{\arraystretch}{1.1}

\begin{tabularx}{\columnwidth}{>{\centering\arraybackslash}m{1.8cm} >{\raggedright\arraybackslash}l *{4}{>{\centering\arraybackslash}X}}
\toprule
& \textbf{Measure}
& \ColHead{Inf. Org.}{6}
& \ColHead{Intui.}{7}
& \ColHead{Errors}{8}
& \ColHead{Satisf.}{9} \\
\midrule

\multirow{2}{*}{\textbf{Control}}
  & \textbf{Median} & 6     & 5     & 4     & 3 \\
  & \textbf{Mean}   & 5.154 & 4.762 & 4.392 & 4.015 \\
\midrule

\multirow{2}{*}{\textbf{Treatment}}
  & \textbf{Median} & 7     & 7     & 7     & 6.5 \\
  & \textbf{Mean}   & 6.973 & 6.967 & 6.227 & 6.187 \\
\midrule

\multirow{3}{*}{\textbf{Wilcoxon}}
  & \textit{p}\textbf{-value} & \textbf{<0.001*} & \textbf{<0.001*} & \textbf{<0.001*} & \textbf{<0.001*} \\
  & \textbf{$r$-value}        & 0.397            & 0.464            & 0.368            & 0.431            \\
  & \textbf{\gls{ci}}         & \textbf{$[-2,-1]$} & \textbf{$[-3,-2]$} & \textbf{$[-3,-1]$} & \textbf{$[-3,-2]$} \\
\bottomrule
\end{tabularx}
\end{table}

\vspace{-0.3cm}
\section[Results \&\ Discussion]{Results \& Discussion}

\subsection{RQ$_1$: Effectiveness of \gls{gui} Prototype Generation}

\begin{figure*}[!t]
  \centering
 \includegraphics[width=1.0\textwidth]{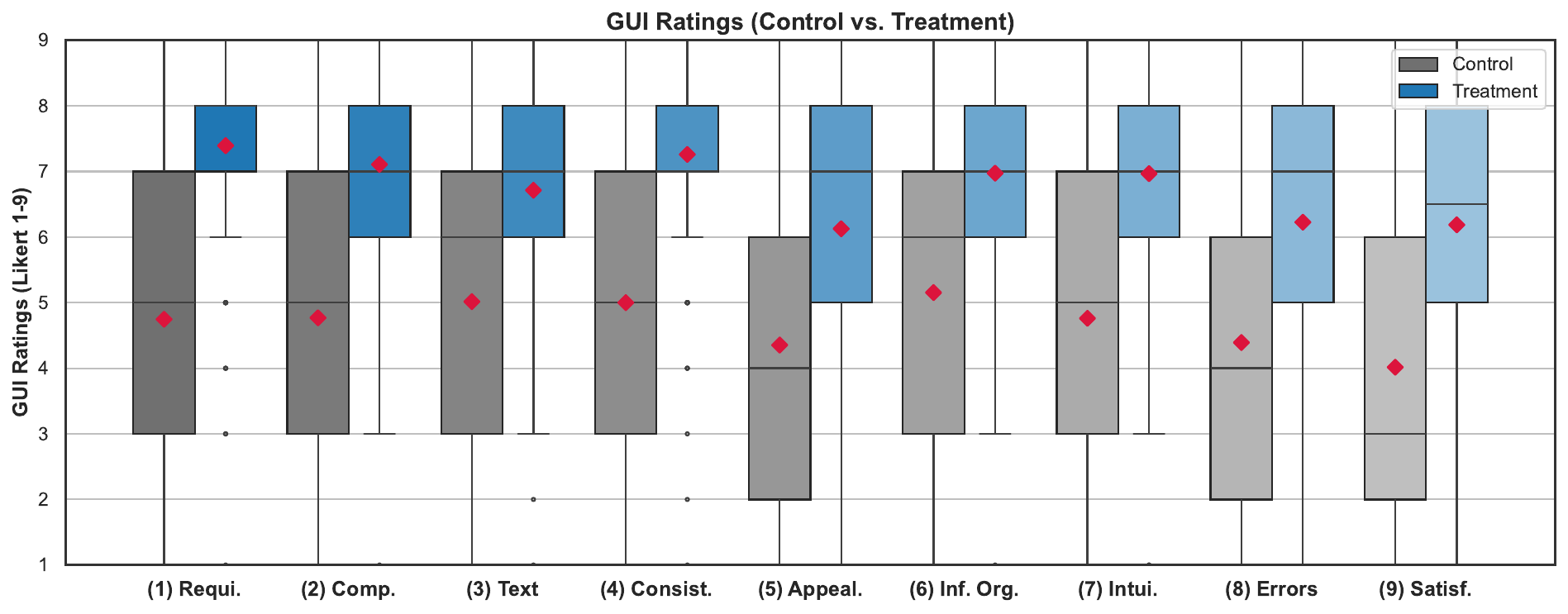}
  \caption[Boxplots for \textit{control} and \textit{treatment} for \gls{gui} prototype quality]{Multiple boxplots for \textit{control} (gray) and \textit{treatment} (blue) across nine \gls{gui} metrics (Likert-scale items (1--9)): \textit{(1) meets requirements from description}, \textit{(2) utilizes correct components}, \textit{(3) texts fit to GUI description}, \textit{(4) GUI design consistent with GUI description}, and \textit{(5) appealing GUI design}, \textit{(6) clear information organization}, \textit{(7) intuitive interactions}, \textit{(8) minimal errors}, and finally \textit{(9) overall satisfaction}, diamonds repr. means.}
	\label{fig:boxplots-guide}
\end{figure*}

Subsequently, we briefly present the results of both sub-studies. 
Although both groups had 45 minutes to create the \gls{gui} prototypes and received the identical short \gls{nlr}, participants started to work on significantly more \gls{gui} requirements while using \textit{GUIDE}, with 3.2 \gls{gui} prototypes per participant commenced in the \textit{treatment} group (16 \glspl{gui} with six participants in total) compared to 2.33 per participant (14 \glspl{gui} with five participants in total) in the \textit{control} group.
Table \ref{tab:results_prolific_1_colhead} and Table \ref{tab:results_prolific_2_colhead} show the medians, mean values, and the results of \textit{Wilcoxon rank-sum tests} for the crowdworker ratings, while Figure \ref{fig:boxplots-guide} illustrates boxplots across all considered evaluation metrics. For each of the asked questions, the mean and median ratings were higher for the \textit{treatment} group. Moreover, the \textit{Wilcoxon rank-sum tests (two-sided)} were significant for each of the evaluated questions.

For example, considering whether the created \gls{gui} prototypes meet the specified requirements \textit{(1)}, we observe a substantial improvement for \textit{GUIDE} with a median (mean) of 8 (7.393) compared to the \textit{control} with 5 (4.746). In addition, considering whether the correct \gls{gui} components are utilized to implement the requested functionality \textit{(2)}, we observe again a substantial improvement in the \textit{treatment} group, with a median (mean) of 7 (7.107) in contrast to 5 (4.769) in the \textit{control}. These two evaluation metrics focusing on the functional aspects of the prototypes are particularly important for early elicitation phases. This indicates a clear advantage and suitability of our approach in \gls{rel} scenarios. Furthermore, \gls{gui} prototypes created by participants using \textit{GUIDE} encompassed substantially fewer errors \textit{(8)} and design-related metrics also improved significantly (such as \textit{(4)}, \textit{(5)}). Overall, the results indicate that \textit{GUIDE} effectively generates editable prototypes.

\begin{myrqbox}
\textbf{Answer to RQ$_{1}$:} Our \textit{GUIDE} approach significantly outperforms the \textit{control} group without the \gls{gui} prototyping assistance across all nine considered \gls{gui} quality metrics.
\end{myrqbox}

\section{Threats to Validity}

\paragraph{Internal Validity.} One potential threat to internal validity is the subjectivity of \gls{gui} prototype quality annotations, which might be perceived differently by various human annotators. To mitigate this subjectivity and bias that could potentially be introduced, we invited 28 participants for annotation from \textit{Prolific} with multiple exclusion criteria to increase quality. On average, each \gls{gui} was annotated 9.3 times by distinct participants.

\vspace{0.1cm}

\paragraph{External Validity.} One potential threat to external validity is the restriction of the experiments to \textit{Figma}, \textit{Material Design}, and the consideration of only four different \gls{gui} prototypes, which might limit the generalizability. However, \textit{Figma} represents a widely employed and state-of-the-art prototyping tool, and other prototyping environments and component libraries typically possess similar editing functionality and complexity. To increase diversity in our dataset, we included \gls{gui} prototypes from different domains.
\section{Limitations}

The current implementation is limited to the popular \textit{Figma} prototyping platform and optimized for the \textit{Material Design} component library. To increase adoption of the approach, more prototyping tools and component libraries could be integrated. Moreover, our approach currently provides support with a focus on individual \gls{gui} prototypes. In the future, this could be extended to enable support for entire applications, including interactions and complex requirements dependencies between several \glspl{gui}.  
\section{Related Work}

To support the \gls{gui} prototyping process, several approaches have been proposed before. For example, \textit{DesignScope} \citep{o2014learning, o2015designscape}, \textit{SketchPlorer} \citep{todi2016sketchplore}, and \textit{GUIComp} \citep{lee2020guicomp} have been proposed. However, we discussed these approaches already in the previous chapter (see Chapter \ref{cha:closing} Section \ref{sec:closing-rel-work} for details). In addition, multiple approaches have been proposed before to automatically generate \glspl{gui} from text descriptions. Similarly, we also already discussed these \gls{gui} generation approaches in an earlier chapter of the thesis (see Section \ref{sec:zs-related-work} for more details).
\section{Conclusion}

The approach proposed in this chapter was driven by the challenge \challonefive{}: \textit{How can \gls{llm}s be efficiently adapted to generate editable \gls{gui} prototype representations from \gls{nlr}?} Specifically, we investigated the generation of editable \gls{gui} prototypes from high-level, abstract \gls{nlr}, whereas we already presented a solution approach for low-level, fine-grained requirements in the previous chapter. To tackle this specific challenge, we proposed \textit{GUIDE}, an approach combining the \gls{zs} \gls{gui} generation capabilities of recent \glspl{llm} with the advantages of traditional \gls{gui} prototyping tools such as \textit{Figma} by extending the \gls{rag} approach in the previous chapter by deriving low-level features via an \gls{llm} initially. This integration facilitates the creation of high-quality \gls{gui} prototypes and provides effective support for prototype developers. Our evaluation shows the effectiveness of our approach, which enables users to create significantly better prototypes across all metrics.
\clearpage
\newpage
\thispagestyle{empty}
\null  

\part[\sc{Verification: (M)LLM-based Approaches}]{\sc{Verification: (M)LLM-based Approaches}}
\label{part:verification}
\clearpage
\newpage
\thispagestyle{empty}
\null  

\chapter{Interlinking GUIs and User Stories: A Semi-Automatic LLM-Based Approach}
\chaptermark{Interlinking GUIs and User Stories: LLM-Based Approach}
\label{cha:interlinking}
\vspace{-0.5cm}
In the first two parts of the thesis, we provided solution approaches for challenge \challone{}, namely, to rapidly map \gls{nlr} to \gls{gui} prototypes. In particular, we investigated retrieval-based methods and \gls{llm}-based generation approaches. In the third part of the thesis, we focus on complementary challenge \challtwo{}, namely, automated verification of \gls{nlr} in \gls{gui} applications. Specifically, in this chapter, we conduct a feasibility study of automated \gls{llm}-based verification approaches by focusing on simplified, static \glspl{gui} without dynamic behavior. The work presented in this chapter has been previously published \citep{kolthoff2024interlinking}\footnote{This section is adapted from: \textbf{Kolthoff, Kristian\textsuperscript{*}}, Kretzer, Felix\textsuperscript{*}, Bartelt, Christian, Maedche, Alexander, and Ponzetto, Simone Paolo. Interlinking User Stories and GUI Prototyping: A Semi-Automatic LLM-based Approach. In \emph{Proceedings of the 32nd IEEE International Requirements Engineering Conference (RE, A)}, Reykjavik, Iceland, June 2024, pages 380--388. IEEE. *Authors contributed equally. Sections are rearranged and directly reused with adaptations.}. The source code, datasets, and results are all publicly available\footnote{Materials are available at \url{https://github.com/kristiankolthoff/IUS-GUI-Prototyping}}.

\paragraph{Personal Contribution.} \textit{Felix Kretzer} and I contributed equally to the ideation and creation of the concept for the work. I implemented the \gls{llm}-based automation approaches. \textit{Felix Kretzer} and I contributed equally to the evaluation design, while \textit{Felix Kretzer} conducted the lab-based data collection and I conducted the evaluation and analysis of results. \textit{Felix Kretzer} wrote the section on data collection, threats to validity, and half of the introduction. I wrote the remainder and created figures, tables, and plots.
\vspace{-0.2cm}

\section{Motivation}
\vspace{-0.1cm}

While previous chapters focused on rapidly mapping \gls{nlr} to \gls{gui} prototypes, which represents an important activity for \gls{rel} and \gls{rval}, the verification of implemented \gls{gui} applications also plays a significant role. Quickly and effectively communicating requirements and mitigating \gls{nl} ambiguity are the main goals of employing \gls{gui} prototypes. When implementing the actual software system, verification ensures the conformance of the system with its specification. One of the most commonly used verification techniques is testing, i.e. executing the software system with specific inputs and under certain conditions, then comparing its behavior and outputs against expected results according to the requirements specification. However, testing represents a time-consuming and effort-demanding activity \citep{myers2004art}. Moreover, \gls{gui} testing represents a particularly resource-intensive procedure, due to its special characteristics, such as graphical input and output, and the dynamic visual behavior of \gls{gui} applications \citep{memon2001comprehensive, memon2007event}.

To mitigate the time-consuming characteristic of testing, many approaches have been proposed in research before. Specifically, automated \gls{gui} testing approaches have evolved from automated test scripts \citep{xie2007designing}, random exploration \citep{mao2016sapienz}, and model-based techniques \citep{zeng2016automated, gu2019practical} to more scripted bug replay methods \citep{gomez2013reran, feng2022gifdroid}. Although these methods enhance the testing efficiency, they suffer from limited test coverage and brittleness, arising from the dynamic and semantic complexity of \gls{gui} applications. Moreover, these testing techniques primarily focus on identifying bugs (i.e., crashing the application), rather than verifying the functional correctness against \gls{nlr}. One major reason for the absence of automation methods for this particular scenario is represented by the high complexity and required \gls{nlu} capabilities of processing \gls{nlr} in combination with complex and dynamic \gls{gui} applications. While older \gls{nlp} approaches did not have sufficient capabilities, the advent of \glspl{llm} provides the potential to address this gap. Therefore, the work presented in this chapter is driven by the leading research question of challenge \challtwoone{}: \textit{How can \gls{nlr}-based verification be automated for simplified, static \gls{gui} representations as a foundation for more complex \gls{gui} verification tasks?}
 
To tackle this challenge, in this chapter, we conduct an initial feasibility study of employing \glspl{llm} for automated verification of \glspl{gui}. In particular, we evaluate the abilities of different \gls{llm}-based \gls{zs}, \gls{fs}, and \gls{cot} prompting approaches to \textit{(i)} identify whether \gls{nlr} (in the form of \glspl{us}) are implemented in a \gls{gui} and \textit{(ii)} which \gls{gui} components implement the requirements. While verification is usually conducted for implemented, complex, and highly dynamic \gls{gui} applications at later development stages, we initially focus on a simplified scenario, in which \glspl{gui} are static (no dynamic behavior) to evaluate the general feasibility of employing \glspl{llm}. To this end, we utilize the \textit{Rico} \gls{gui} dataset \citep{deka2017rico}, which encompasses individual \glspl{gui} screens from fully implemented applications, represented by comprehensive structural \gls{gui} component hierarchy data. Furthermore, the methods developed in this chapter have already been employed in our \textit{Figma} plugin (see Chapter \ref{cha:closing}) in the context of prototyping (identifying implementation, matching \gls{gui} components). As we have seen, employing these methods to enable conformance checking already at design time provides benefits, especially due to the large number of rapidly changing requirements in practice \citep{debnath_re}.

\vspace{-0.2cm}
\paragraph{Contributions.} With this work, we make the following three research contributions: 
\vspace{-0.1cm}
\glsreset{us}
\begin{itemize}[left=0.1cm]
    \item \textit{Novel \gls{llm}-based approaches for static \gls{gui} verification from \gls{nlr}:} we propose \gls{zs}, \gls{fs}, and \gls{cot} prompting approaches for enabling verification of static \glspl{gui} from \gls{nlr}. Specifically, we conduct this work as a first feasibility study and employ abstracted, simplified \gls{gui} representation from the \textit{Rico} dataset with detailed \gls{gui} hierarchy data.
    \item \textit{Interlinked \gls{us} and \gls{gui} components dataset:} we create a novel dataset through a lab-based study with human annotators, which represents the first dataset that combines \glspl{us} and \glspl{gui} from \textit{Rico} (matched to individual \gls{gui} components).
    \item \textit{Comprehensive evaluation and insights:} with the created \gls{gui}-\glspl{us} dataset, we conduct a comprehensive experimental evaluation of the proposed \gls{llm}-based approaches and conduct an error analysis to provide further insights into the \gls{llm}-based predictions.
\end{itemize}

\begin{figure}[t]
\centering
  \includegraphics[width=0.95\textwidth]{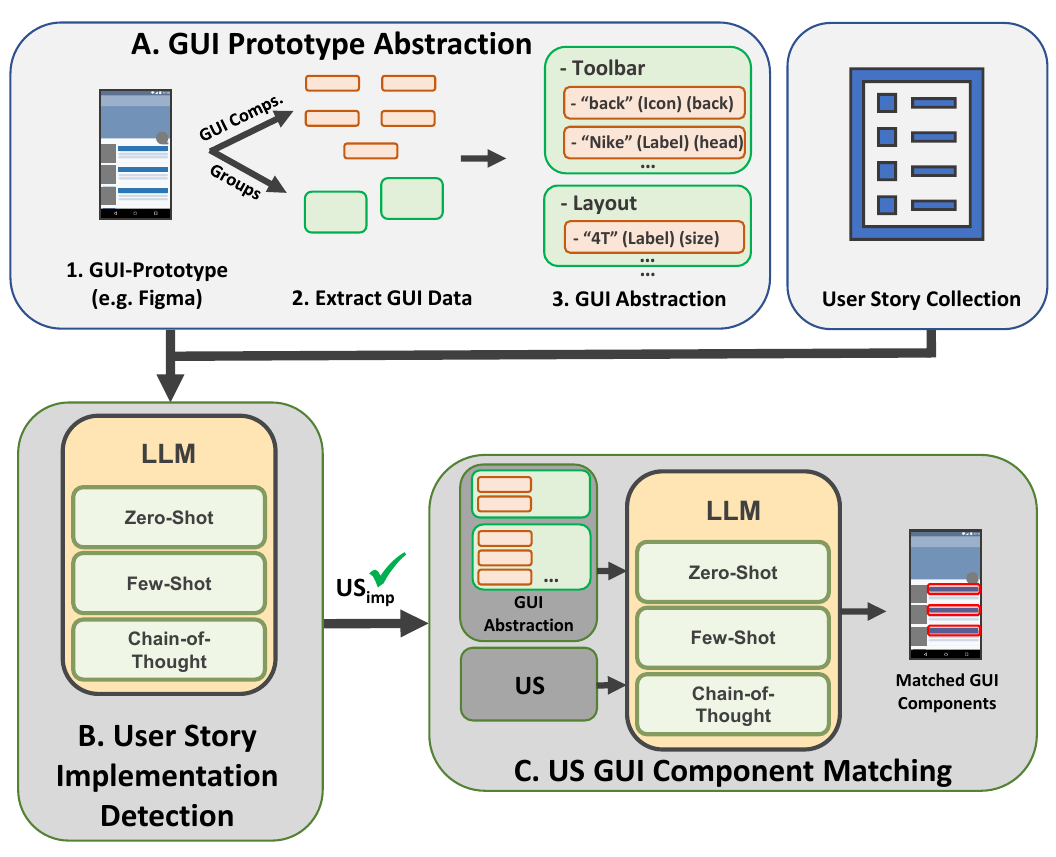}
  \caption[Overview of \gls{llm}-based approach for \gls{us} implementation detection and \gls{gui} component matching]{Overview of our approach consisting of \textit{(A)} creating a textual representation for \textit{Rico} \glspl{gui} to feed into an \gls{llm}, \textit{(B)} detecting the implementation of a \gls{us} in the \gls{gui} via \gls{llm}-based prompting and \textit{(C)} matching corresponding \gls{gui} components to the \gls{us}.}
	\label{fig:overview-interlinking}
    \vspace{-0.4cm}
\end{figure}

\vspace{-0.6cm}
\section{Approach}

Our approach is divided into several main components, as shown in Figure \ref{fig:overview-interlinking}. First, \textit{(A)} a \gls{gui} prototype abstraction component to transform the structural \textit{Rico} \gls{gui} hierarchy data to a textual representation, which can be fed into an \gls{llm}, \textit{(B)} an \gls{nlr} (here \gls{us}) verification component that utilizes the \gls{gui} abstraction and an \gls{us} collection in an \gls{llm}-based approach to classify whether \glspl{us} are already implemented in the given \gls{gui}, \textit{(C)} an additional component matching approach that identifies the \gls{gui} components that implement the \gls{us}. Next, we provide detailed descriptions for each of the components.

\subsection{\gls{gui} Prototype Abstraction}

As an input to the previously mentioned \gls{llm}-based methods, the \textit{Rico} \citep{deka2017rico} \gls{gui} needs to be transformed to an abstract textual representation to be fed into an \gls{llm}\footnote{Note that at the time of implementing the approach, more powerful \glspl{mllm}, which have highly effective image understanding capabilities, were not readily available yet (such as \textit{\gls{gpt}-4o} \citep{openai_gpt4o_docs}).}. In particular, we adopt the method already presented earlier (see Chapter \ref{cha:self_elicitation} Section \ref{sec:sergui-feature-rec}). For each extracted \gls{gui} component, we then create a textual representation using the following abstract pattern, followed by three examples created from respective \textit{Rico} \glspl{gui}:

\begin{center}
\textit{"uicomp-text"} \textbf{(uicomp-type)} \texttt{(uicomp-name)} \\
\textit{"+7.10"} \textbf{(Label)} \texttt{(price Change TV)}\\
\textit{"Install App"} \textbf{(Button)} \texttt{(native Ad Call To)} \\
\textit{"Example: 'New York'"} \textbf{(Text Input)} \texttt{(location)}
\end{center}

\noindent Specifically, the \textit{uicomp-text} refers to the displayed text of the component, the \textbf{uicomp-type} refers to the basic \gls{gui} component type (e.g., \textit{Label}, \textit{Button}, \textit{Checkbox}), and the \texttt{uicomp-name} refers to the internal name of the component (e.g., provided by developers). In the absence of any of the properties, the respective field is left empty. This textual representation of the \gls{gui} components encompasses relevant information from a functional perspective. Moreover, we derive clustering elements from the semantic annotations of \textit{Rico}, which incorporate categories such as \textit{List Item}, \textit{Card}, and \textit{Toolbar}, among others. To identify clusters for the remaining components not encompassed by the preceding groups, we further extracted layout clusters from the original \gls{gui} hierarchy by aligning them with the \gls{gui} components. Subsequently, we create the \gls{gui} representation as a two-tier bullet-point list, with the outer tier representing the layout groups and the inner tier denoting their corresponding \gls{gui} components. Prior to generating the string representation, the layout groups are arranged based on their boundaries from the top-left to the bottom-right and in a similar fashion, the \gls{gui} components within each group are organized to resemble the original \textit{Rico} \gls{gui} component structure closely.

\subsection{User Story Implementation Detection}

To tackle the problem of identifying whether a \gls{us} is implemented in a \gls{gui} prototype, we propose several \gls{llm}-based methods and approach it as a binary classification problem. Given the extensive knowledge encapsulated within \glspl{llm} and their \gls{icl} capabilities, the application for the detection and matching problems (verification) at hand is promising.

\glsreset{zs}
\paragraph{\gls{zs} Prompting.} In particular, we adopt the \gls{zs} prompting method \citep{kojima2022large} by creating a prompting template divided into \textit{(i)} a task instruction providing clear guidelines for the model, \textit{(ii)} the \gls{us} to validate, followed by \textit{(iii)} the generated \textit{Rico} \gls{gui} text representation. We instruct the model to predict a single token for the classification, and we extract the log probabilities for both labels, which provides a \gls{us} ranking mechanism. In particular, the extracted probability can be employed to estimate the certainty of the classification. This probability can be further exploited to rank the user stories from high to low probabilities (e.g., for later visualization to users).

\glsreset{fs}
\paragraph{\gls{fs} Prompting.} In addition, we adopt the \gls{fs} prompting method \citep{brown2020language} often showing enhanced performance for various tasks by providing examples. In \gls{fs} prompting, we basically follow the \gls{zs} pattern, but we additionally provide several \textit{input-output} pairs to guide the model for the specific task. For our experimental evaluation, we created multiple \gls{fs} prompting templates (each encompassing varying examples).

\glsreset{cot}
\paragraph{\gls{cot} Prompting.} Moreover, we adopt the \gls{cot} prompting method \citep{wei2022chain}, in which the \gls{llm} is instructed to create multiple intermediate reasoning steps before generating a prediction. Thus, we instruct the model to first generate an explanation providing reasoning whether the \gls{us} is implemented. In addition, this provides an interpretable explanation that can further be used for later error analysis.

\subsection{User Story \gls{gui} Component Matching}

In addition to solely predicting the coverage of a \gls{us} in a \textit{Rico} \gls{gui}, we further investigate the task of extracting all relevant components from the \gls{gui} that are required to fulfill the \gls{us}. This represents a natural extension of the previous task, enabling direct interlinking of the \gls{us} with its respective \gls{gui} components and gaining deeper insights into the \gls{llm} predictions. Similarly to the previous task, we adopt \gls{zs}, \gls{fs}, and \gls{cot} prompting models for this matching task. However, we extend the \gls{gui} abstraction by adding identifiers to each \gls{gui} component and correspondingly adapt the instructions and prompting templates to enable the \gls{llm} to output a parsable collection of the component identifiers. We evaluate \gls{zs}$_{A}$ (plain task desc.) and \gls{zs}$_{B}$ (focus on extracting all relevant components).
\vspace{-0.6cm}
\section{Experimental Evaluation}
This section delineates the design of our evaluation of the proposed approach. In particular, we focus the evaluation on two main aspects of the approach including the \gls{us} implementation detection and \gls{gui} component matching methods. For enabling this evaluation, we constructed a gold standard of \glspl{us} with corresponding \gls{gui} component annotations. To this end, we formulate the subsequent two research questions:

\vspace{0.1cm}
\begin{itemize}
       \item \textbf{RQ$_{1}$}: \textit{How effective are \gls{llm}-based approaches for detecting \gls{us} implementation in static \glspl{gui}?} To answer this question, we employ the created gold standard and create a test dataset for the binary classification problem. To construct negative instances, we remove the \gls{gui} components for the respective \gls{us} and for positive instances, we simply keep them. To assess the effectiveness of the approaches, we compute standard classification metrics including \textit{precision}, \textit{recall}, and \textit{F1-measure}.
    \item \textbf{RQ$_{2}$}: \textit{How effective are \gls{llm}-based approaches for extracting \gls{gui} components fulfilling a \gls{us} from static \glspl{gui}?} To answer this question, we use the gold standard and compare the extracted collection of \gls{gui} components by the \gls{llm} approaches with the reference collection. Again, we compute \textit{precision}, \textit{recall}, and \textit{F1-measure}.
\end{itemize}

\subsection{Data Collection}
To evaluate our approach, a dataset of \glspl{gui} and associated \glspl{us} was required. While there are established datasets of \glspl{gui} available (e.g., \textit{Rico} \citep{deka2017rico}), there exists no dataset combining \glspl{gui} with \glspl{us}. Therefore, we decided to collect \glspl{us} for existing \glspl{gui} from the \textit{Rico} dataset. Subsequently, we introduce the collection and preprocessing of the dataset by presenting existing \glspl{gui} to annotators writing \glspl{us} and matching them.

\paragraph{\gls{gui} Sample.} In order to get a broad selection of different \textit{Rico} \glspl{gui}, our initial \gls{gui} sample was randomly drawn from ten different domains. Following our exclusion criteria, we then selected valid \glspl{gui} from our random sample. We decided ex-ante to exclude interfaces with \textit{non-English text}, \textit{personal data displayed}, \textit{overlays} (such as \textit{pop-ups}) shown, \textit{components without annotations in the \textit{Rico} dataset}, \textit{trivial} \glspl{gui} (e.g., simple log-in screens resulting in the same repetitive \glspl{us}), and to exclude interfaces with \textit{unclear functionality}. Our final sample included 59 \glspl{gui} from the domains: \textit{Shopping} (8), \textit{Health \& Fitness} (11), \textit{Education} (5), \textit{News} (4), \textit{Sports} (5), \textit{Travel} (6), \textit{Books} (5), \textit{Music} (6), \textit{Finance} (4), and \textit{Food} \& \textit{Drink} (5). We additionally created \gls{gui} versions where each component was annotated with a number so that participants could associate their \glspl{us} with $\geq1$ \gls{gui} components.

\paragraph{Procedure and Survey.} \glspl{us} were collected using a questionnaire. The participants were presented with information about the study, data protection, and conditions of participation. They then learned how to write functional \glspl{us}. Learning content was supported with examples of \glspl{us} and concluded with comprehension checks. Participants had two chances for each comprehension check before a screenout took place. Participants then created \glspl{us} for nine consecutively shown \glspl{gui}. We instructed participants to create three to five functional \glspl{us} per \gls{gui} describing features already implemented and provided a \glspl{us} template for guidance. After creating \glspl{us} for nine \glspl{gui}, the final task was to specify for each \gls{us} which components of the \gls{gui} belong to the respective \gls{us}. The participants were shown the identical nine \glspl{gui} one after the other with the already created \glspl{us}, but the \glspl{gui} now contained numbers that identified each component.

\paragraph{Participants.} We selected 8 participants (2 female, 6 male) from a student pool. Participants were on average 24.3 years old, had 1.1 years of experience creating and 1.0 years of experience in evaluating visual design. Each participant generated 4.5 \glspl{us}/\gls{gui} (mean).

\begin{figure*}[!t]
  \centering
  \frame{
    \subfloat[\centering \#Word/\gls{us} distribution for dataset]{
      \includegraphics[width=0.44\linewidth]{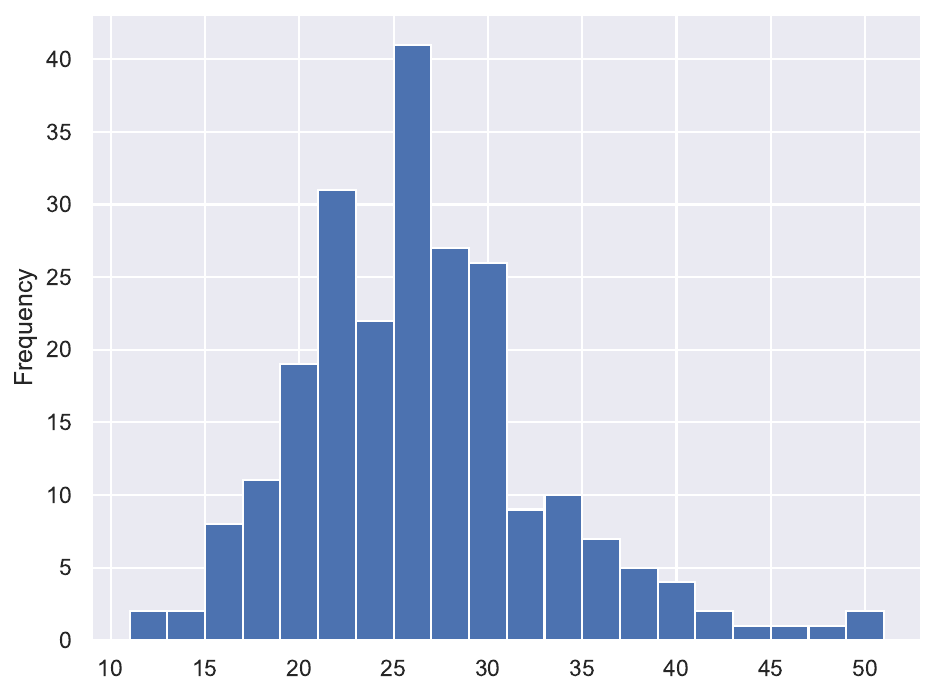}
      \label{fig:word_dist_inter}
    }
  }
  \qquad
  \frame{
    \subfloat[\centering \#Comps./\gls{us} distribution for dataset]{
      \includegraphics[width=0.44\linewidth]{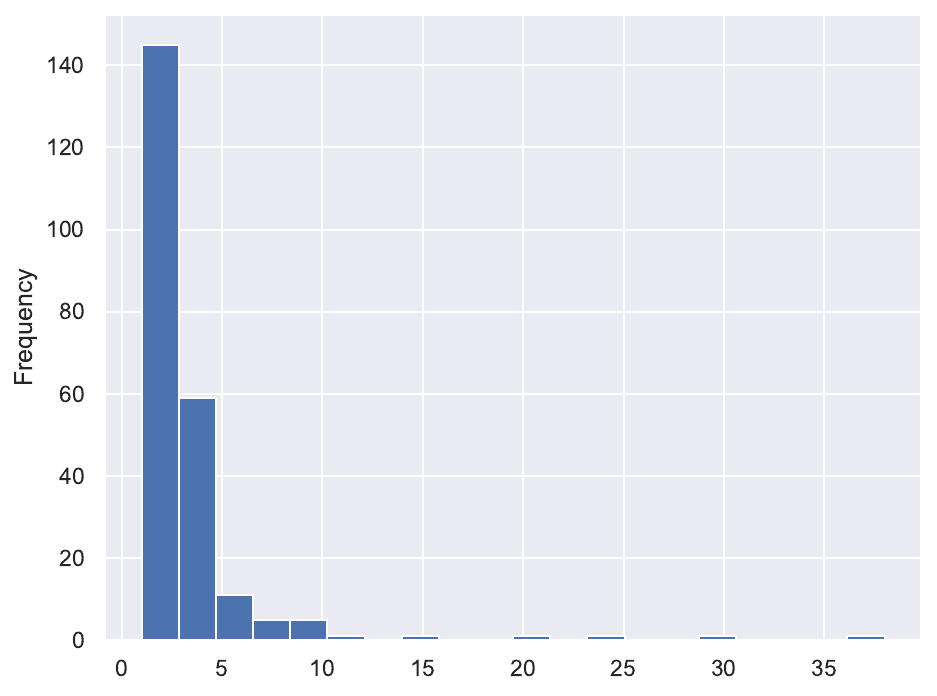}
      \label{fig:comp_dist_inter}
    }
  }
  \caption[\#Word/\gls{us} and \#Comps/\gls{us} distributions across the dataset]{Distributions for the created dataset combining \glspl{us} with \glspl{gui}. In particular, each \gls{us} is represented by a one-to-many mapping ($\geq1$) to respective \gls{gui} components.}
  \label{fig:interlinking_data}
\end{figure*}

\glsunset{iaa}
\paragraph{Data Processing.} Overall, the participants created 327 \glspl{us}, with duplicate \glspl{us} and varying quality (e.g., despite explicit instructions, some \glspl{us} were written for \textit{not implemented} features). In order to obtain a usable dataset, the \glspl{us} were cleaned up. Therefore, two paper authors labeled each \gls{us} independently. For this purpose, the first step was to determine whether the \gls{us} \textit{a)} fully meets the requirements, \textit{b)} contains one or more errors, or \textit{c)} is a duplicate of a previous \gls{us}. After the separate labeling, the \gls{iaa} was calculated (Cohen's Kappa $\kappa = .548$). After resolving disputes, 231 \glspl{us} (with their respective \glspl{gui}) were included in the final data set, whereas 96 \glspl{us} (16 duplicates, 80 different exclusion criteria) were not considered further. The applied exclusion criteria are provided in our accompanying materials. Figure \ref{fig:interlinking_data}\textit{(a)} shows the distribution of the number of words per \gls{us}, indicating complex requirements due to many words included per \gls{us} (26.08 on average). Moreover, Figure \ref{fig:interlinking_data}\textit{(b)} shows the distribution of the number of components matched to respective \glspl{us}. In addition, Table \ref{tab:us_dataset_stats_interlinking} provides summary statistics across these two aspects and the number of \glspl{us} per \gls{gui} in the cleaned dataset. Furthermore, Figure \ref{fig:matching_examples-interlinker} illustrates multiple example annotations taken from the gold standard, with \textit{Rico} \glspl{gui},  the \glspl{us} written by participants, and corresponding \gls{gui} component annotations.

\begin{table}[t]
\centering
\small
\setlength{\tabcolsep}{5pt}
\begin{tabular}{lcccccccc}
\toprule
\textbf{Variable} & \textbf{n} & \textbf{$\mathbf{\mu}$} & \textbf{$\mathbf{\sigma}$} & \textbf{min} & \textbf{Q1} & \textbf{med.} & \textbf{Q3} & \textbf{max} \\
\midrule
(1) \#Words/User Story       & 231 & 26.08 & 6.56 & 11.00 & 22.00 & 25.00 & 29.00 & 51.00 \\
(3) \#Components/User Story  & 231 & 2.93  & 4.01 & 1.00  & 1.00  & 2.00  & 3.00  & 38.00 \\
(2) \#User Stories/\gls{gui} & 59  & 3.92  & 1.38 & 1.00  & 3.00  & 4.00  & 5.00  & 7.00  \\
\bottomrule
\end{tabular}
\caption[Summary statistics for \gls{gui}-\gls{us} dataset]{Summary statistics (\textit{mean} $\mu$, \textit{standard deviation} $\sigma$, \textit{minimum}, \textit{first quartile} \textbf{Q1}, \textit{median}, \textit{third quartile} \textbf{Q3}, \textit{maximum}) for the \gls{us} dataset: \textit{(1)} number of words per \gls{us}, \textit{(2)} number of corresponding \gls{gui} components per \gls{us}, and \textit{(3)} number of \glspl{us} per \gls{gui}.}
\vspace{-0.5cm}
\label{tab:us_dataset_stats_interlinking}
\end{table}

\vspace{-0.3cm}
\subsection{RQ$_{1}$: User Story Implementation Detection}

\begin{figure}[!t]
      \centering
      \includegraphics[width=\textwidth]{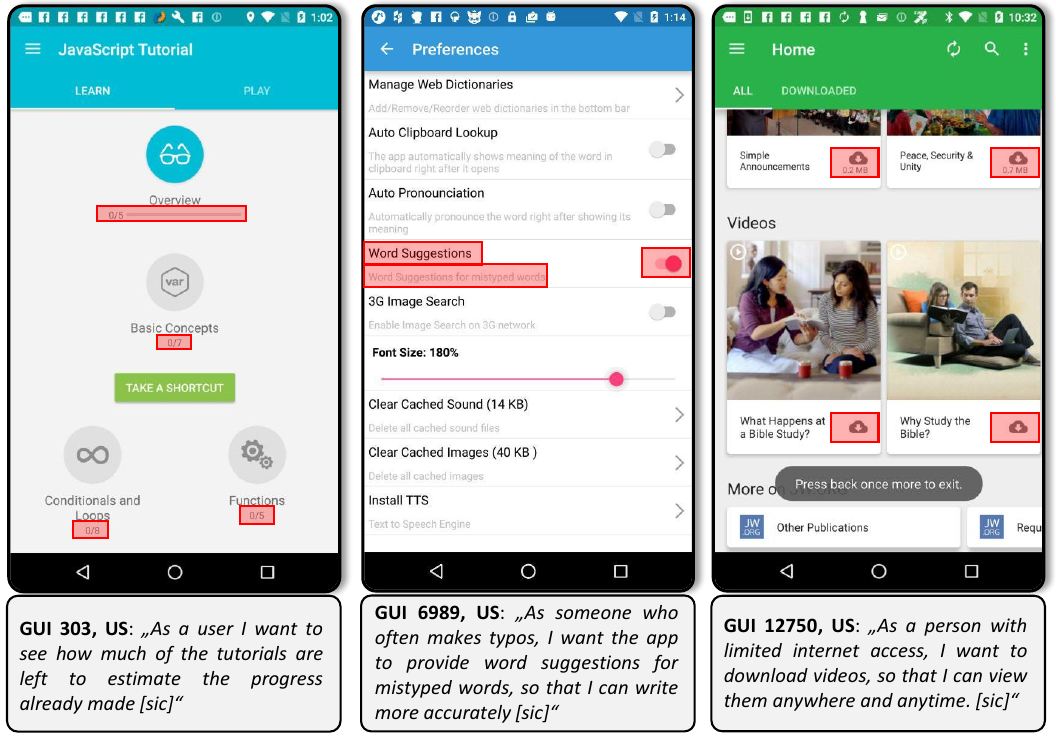}
      \caption[Examples from the dataset with \glspl{us}/\gls{gui} component annotations]{Three \textit{Rico} \citep{deka2017rico} \gls{gui} examples from our gold standard with their respective \glspl{us} and visually marked \gls{gui} components associated with the requirements.}
    \label{fig:matching_examples-interlinker}
    \vspace{-0.5cm}
\end{figure}

To answer RQ$_{1}$, we evaluated the ability of various \gls{llm}-based prompting techniques to predict whether a \gls{us} is contained in a \gls{gui}. To this end, we created a gold standard based on the collected \glspl{us} and \gls{gui} annotation pairs. First, we randomly selected five \glspl{gui} comprising 21 \glspl{us} to be employed as examples for the \gls{fs} prompting approach. The remaining 210 \gls{us}-\gls{gui}-pairs form the basis for the gold standard. This procedure ensures the avoidance of overlapping \gls{gui} abstraction data between gold standard and \gls{fs} examples, thereby avoiding bias. Next, we randomly assigned half of the examples (105) to the class \textit{Implemented (1)} and the remainder the class \textit{Not-Implemented (0)}. While the \gls{gui} data for the first class remains unchanged, in the \gls{gui} abstraction of the second class (i.e. \textit{Not-Implemented (0)}), the paired \gls{gui} component annotations were removed. For example, consider the second \gls{gui} of Figure \ref{fig:matching_examples-interlinker}. For the shown \gls{us}, the respective \gls{gui} components associated with the \gls{us} according to the gold standard are marked in the \gls{gui}. To include such an \gls{us} as a negative example in the gold standard, we would remove the respective \gls{gui} components from the \gls{gui} abstraction (two \textit{labels} and a \textit{checkbox}). Moreover, we compute \textit{precision ($P$)}, \textit{recall ($R$)}, \textit{F1-measure ($F1$)}, and \textit{accuracy ($A$)} as

\vspace{0cm}
\newlength{\midgap}
\setlength{\midgap}{-3cm} 

\newlength{\leftpull}
\setlength{\leftpull}{0.2cm} 

\noindent
\hspace*{-\leftpull}%
\begin{minipage}{\dimexpr\textwidth-2em\relax}
\centering
\setlength{\tabcolsep}{0pt}

\begin{tabular}{@{}
  p{\dimexpr0.5\linewidth-0.5\midgap\relax}
  @{\hspace{\midgap}}
  p{\dimexpr0.5\linewidth-0.5\midgap\relax}
@{}}
\coleq{eq:precision}{P=\frac{TP}{TP+FP}} &
\coleq{eq:recall}{R=\frac{TP}{TP+FN}} \\[+0.1cm]
\coleq{eq:f1}{F_{1}=2\cdot\frac{P\cdot R}{P+R}} &
\coleq{eq:accuracy}{A=\frac{TP+TN}{N}}
\end{tabular}

\end{minipage}
\vspace{0.2cm}

\glsreset{fp}

\noindent based on \gls{tp}, \gls{fp}, \gls{tn}, and \gls{fn} instances, denoting $N = TP + TN + FP + FN$. To conduct the experiments, we employed the most recent \textit{GPT-4} model\footnote{Note that at the time of implementing the approach, more powerful \glspl{mllm}, which have highly effective image understanding capabilities, were not readily available yet (such as \textit{\gls{gpt}-4o} \citep{openai_gpt4o_docs}).} \citep{openai2023gpt4} (8,192 token context length, \textit{temperature=0}, \textit{accessed in February 2024}) as our base \gls{llm}. For the \gls{fs} prompting, we evaluated one model with five (\gls{fs}$_{5}$) and another with ten examples (\gls{fs}$_{10}$), respectively. We evaluated four \gls{cot} models with varying \textit{temperature}, based on the idea that with varying \textit{temperature}, we restrict the model to provide more or less diverse explanations.
\vspace{-0.1cm}
\subsection{RQ$_{2}$: User Story \gls{gui} Component Matching}
\vspace{-0.1cm}

{\renewcommand{\arraystretch}{1.1}
\begin{table}[!t]
\footnotesize
\caption[Evaluation results of \gls{llm}-based approaches for \gls{us} implementation detection in \glspl{gui}]{Evaluation results of \gls{llm}-based approaches (\gls{zs}, \gls{fs} (\textit{\#examples}), and \gls{cot} (\textit{temperature})) for detecting \gls{us} implementation in \textit{Rico} \glspl{gui} (binary classification) showing \textit{Precision (P)}, \textit{Recall (R)}, and \textit{F1-measure (F1)} for both classes and \textit{Accuracy (A)}.}
\centering

\setlength{\tabcolsep}{\ResultsTabColSep}

\begin{tabularx}{\textwidth}{l|*{3}{>{\centering\arraybackslash}X}|*{3}{>{\centering\arraybackslash}X}|>{\centering\arraybackslash}X}
\toprule
\multicolumn{1}{c|}{\textbf{}} &
\multicolumn{3}{c|}{\textbf{US Implemented (1)}} &
\multicolumn{3}{c|}{\textbf{Not Implemented (0)}} &
\multicolumn{1}{c}{\textbf{}} \\
\cmidrule(lr){2-4}\cmidrule(lr){5-7}\cmidrule(l){8-8}
& $\mathbf{P_1}$ & $\mathbf{R_1}$ & $\mathbf{F1_1}$
& $\mathbf{P_0}$ & $\mathbf{R_0}$ & $\mathbf{F1_0}$
& $\mathbf{A}$ \\
\midrule

\textbf{Zero-Shot}      & 83.0 & 88.6 & 85.7 & 87.8 & 81.9 & 84.7 & 85.2 \\
\midrule

\rowcolor{lightgray}\textbf{Few-Shot$_{5}$}  & 82.9 & 87.6 & 85.2 & 86.9 & 81.9 & 84.3 & 84.8 \\
\textbf{Few-Shot$_{10}$}                     & 81.8 & 85.7 & 83.7 & 85.0 & 81.0 & 82.9 & 83.3 \\
\midrule

\rowcolor{lightgray}\textbf{CoT$_{t=0.0}$}     & 90.0 & 68.6 & 77.8 & 74.6 & 92.4 & 82.6 & 80.5 \\
\textbf{CoT$_{t=0.5}$}                        & 88.8 & 67.6 & 76.8 & 73.8 & 91.4 & 81.7 & 79.5 \\
\rowcolor{lightgray}\textbf{CoT$_{t=1.0}$}     & 88.8 & 75.2 & 81.4 & 78.5 & 90.5 & 84.1 & 82.9 \\
\textbf{CoT$_{t=1.3}$}                       & 81.2 & 74.3 & 77.6 & 76.3 & 82.9 & 79.5 & 78.6 \\
\bottomrule
\end{tabularx}

\label{tab:results_rq1_interlinking}
\vspace{-0.3cm}
\end{table}
}

To answer RQ$_{2}$, we evaluated the ability of several \gls{llm}-based prompting techniques to extract all \gls{gui} components relevant for fulfilling a given \gls{us}. Therefore, we employed the same gold standard as previously described. However, we used the original unchanged abstraction for each of the 210 \glspl{gui} with the labels being the annotated \gls{gui} component identifiers from the gold standard. Precisely, as the input, the model received the \gls{gui} abstraction (each \gls{gui} component marked by a numerical identifier) and a \gls{us} to predict a set of \gls{gui} component IDs relevant for the \gls{us}. Next, we compared the set of extracted \gls{gui} component identifiers with the gold standard set of identifiers and computed $P$, $R$, and $F1$ measure. We computed these metrics for each example in the dataset and averaged them over the gold standard to obtain \textit{macro} values. Therefore, \textit{macro} values represent the average of each metric over the gold standard. Although metric values for two \gls{us} examples might be equal, the absolute number of correct or erroneous classifications might differ significantly between \gls{us} depending on the \gls{gui} component set length. For example, the second \gls{us} example from Figure \ref{fig:matching_examples-interlinker} is only associated with three respective \gls{gui} components, whereas another \gls{us} from the gold standard about visualizing an overview of the daily nutrients consumed daily (see gold standard example 78 in our supplementary materials) has 20 associated \gls{gui} components. To take into account these set length differences across the gold standard and thus counter potential evaluation result inaccuracies introduced by set length, we additionally constructed binary prediction arrays to compute respective \textit{micro} values possessing correct weights, i.e. double \gls{gui} component set length leads to double the influence on the final metric result. To conduct the experiments, we utilized the identical setup of \glspl{llm} as described previously for RQ$_{1}$.

\section[Results \&\ Discussion]{Results \& Discussion}

\subsection{RQ$_{1}$: User Story Implementation Detection}

\begin{figure}[!t]
      \centering
      \includegraphics[width=\textwidth]{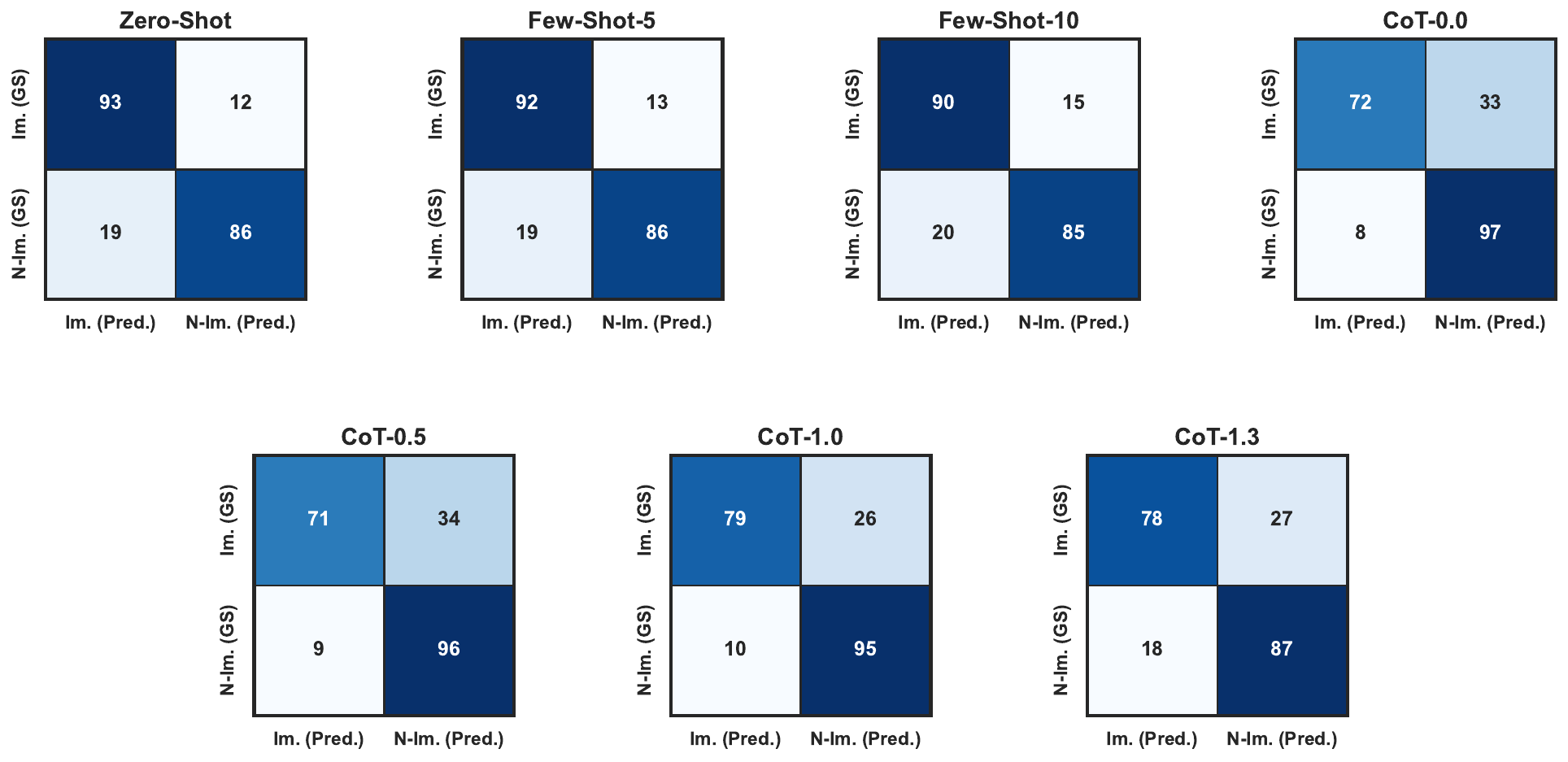}
      \caption[Confusion matrices for \gls{llm}-based approaches for \gls{us} implementation detection]{Confusion matrices for \gls{llm}-based approaches (\gls{zs}, \gls{fs} (\textit{\#examples}), and \gls{cot} (\textit{temperature})) for detecting \gls{us} implementation in \textit{Rico} \glspl{gui} (as binary classification).}
    \label{fig:interlinking_rq1_confusion}
\end{figure}

Table \ref{tab:results_rq1_interlinking} illustrates the evaluation results for RQ$_{1}$, showing the $P$, $R$, and $F1$ metrics for both classes and the accuracy $A$ over the created gold standard. Moreover, we provide the confusion matrices for each prompting approach in Figure \ref{fig:interlinking_rq1_confusion}. First, we can observe a substantially high absolute performance across all of the investigated prompting methods indicated by, for example, accuracy scores of 85.2 (\gls{zs}), 84.8 (\gls{fs}$_{5}$), and 82.9 (\gls{cot}$_{t=1.0}$). This indicates that \glspl{llm} are capable of effectively processing the semantics of the created \gls{gui} abstraction and matching it to the semantics of the functionality encompassed in the \glspl{us}. These high metric values indicate that \glspl{llm} can produce promising results for the approach. Although the \gls{zs} method seems to perform best overall according to mean values, the respective \textit{Holm}-corrected pairwise \textit{McNemar tests} \citep{adedokun2012analysis, lachenbruch2014mcnemar} between each of the prompting methods indicate no statistically significant differences (see Appendix \ref{app:interlinking} Table \ref{tab:detection-mcnemar-holm}). Moreover, the \gls{cot} methods apparently tend to be more restrictive about predicting that a \gls{us} is implemented, as indicated by the highest $P_{1}$ and $R_{0}$ values. In contrast, the \gls{zs} and \gls{fs} methods appear to be more balanced among the different metrics and both classes.

\begin{figure}[!t]
  \centering
  \includegraphics[width=\textwidth]{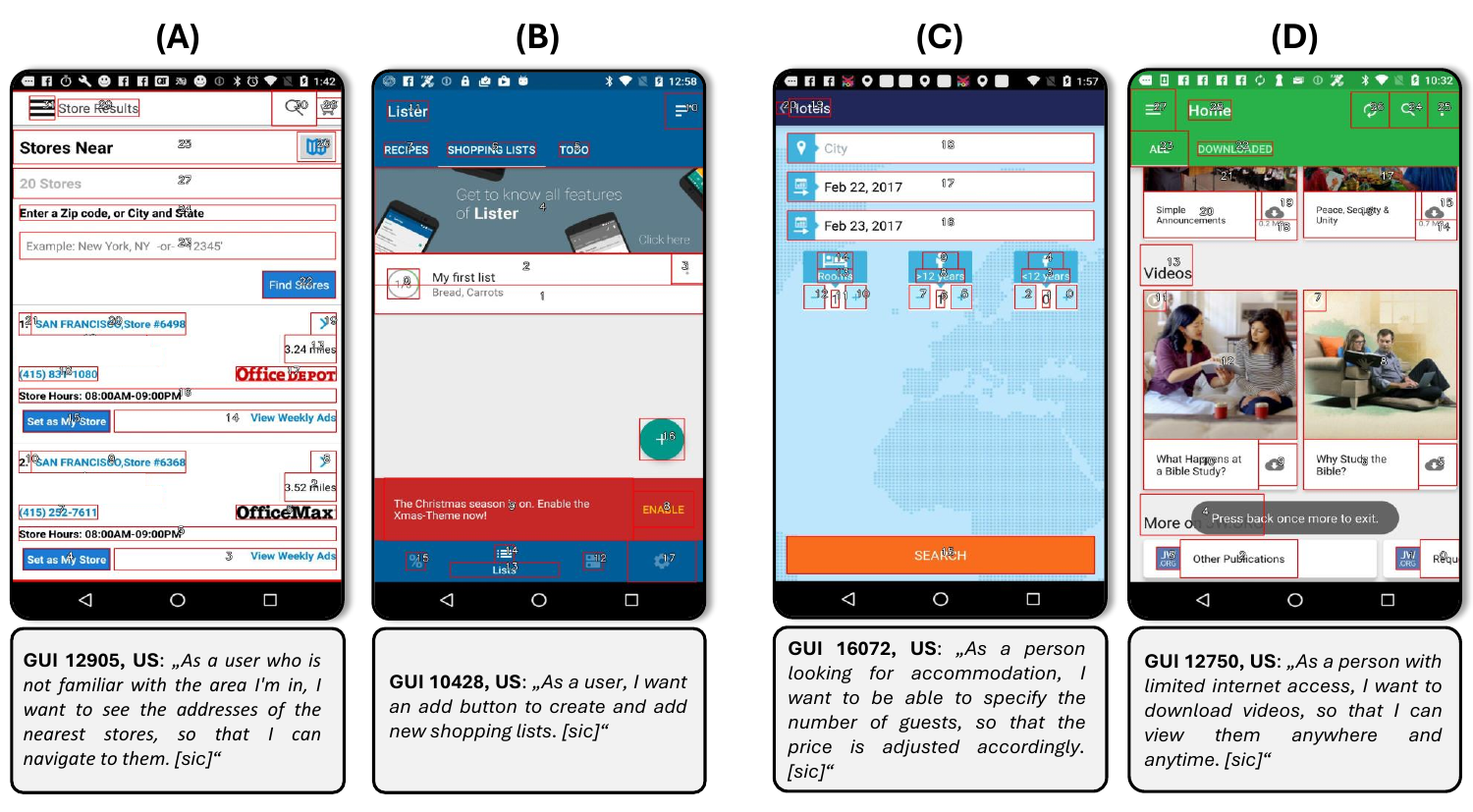}
  \caption[\gls{gui} examples with respective \glspl{us} for error analysis]{Four example \textit{Rico} \glspl{gui} taken from the gold standard with their respective \gls{us}. \textit{(A)} represents a \gls{fp} and \textit{(B)} represents a \gls{fn} instance for \gls{us} implementation detection, while \textit{(C)} represents a \gls{fn} and \textit{(D)} represents a \gls{fn}/\gls{fp} instance for the \gls{us} matching task.}
  \label{fig:interlinking_error_1}
\end{figure}

To enhance the understanding of the misclassifications made by the \gls{llm}, we conducted an error analysis and investigated the \gls{fp} and \gls{fn} instances. For the \gls{fp} instances, the main root cause for misclassifications appears to be a semantic misinterpretation of \gls{gui} components with reference to the \gls{us} by the \gls{llm}. For example, for a \gls{us} that describes providing addresses of the nearest stores the \gls{llm} identifies the component \textit{"SAN FRANCISCO, Store \#6498"} \textbf{(Button)} \texttt{(storelocator address line1)} as fulfilling the \gls{us}, although this component merely provides the city name and the detailed address fields were absent, as shown in Figure \ref{fig:interlinking_error_1}\textit{(A)}. Similarly, for the \gls{us} to enable/disable the ability to mark days as complete within a settings \gls{gui} of a fitness app, the \gls{llm} identified the \gls{gui} component \textit{"check"} \textbf{(Icon)} \texttt{(done)} as fulfilling the \gls{us}. However, this component is located in the \gls{gui} toolbar and refers to saving the overall settings.

For the \gls{fn} instances, we identified several similar main root causes. Often, detailed semantic information about the functionality of the \gls{gui} components might be absent in the created \gls{gui} abstraction, resulting in the \gls{llm} being restrictive about positively identifying the \gls{us} as fulfilled. For example, a \gls{us} requiring a donation button to easily donate money could not be detected, since the \gls{gui} component was implemented as an image without any further textual description. Hence, the image information is not accessible to the model. Similarly, a \gls{us} to add new shopping lists to a collection of lists could not be identified since the description of the \gls{gui} component \textit{"add"} \textbf{(Icon)} \texttt{(overview viewpager fab)} (Figure \ref{fig:interlinking_error_1}\textit{(B)}) is general and ambiguous for detection. These errors are mainly driven by the simplified \gls{gui} representation of used \textit{Rico} \glspl{gui}.

\begin{myrqbox}
\textbf{Answer to RQ$_1$:} \gls{llm}-based prompting approaches are effective to detect \gls{us} implementation in static \glspl{gui}, achieving high \textit{precision} (83.0 for \gls{zs}), \textit{recall} (88.6 for \gls{zs}), and \textit{F1-measure} (85.7 for \gls{zs}), while no significant pairwise differences were observed.
\end{myrqbox}

\subsection{RQ$_{2}$: User Story \gls{gui} Component Matching}

{\renewcommand{\arraystretch}{1.0}
\begin{table}[!t]
\footnotesize
\caption[Evaluation results of \gls{llm}-based approaches for \gls{gui} component matching from \glspl{us}]{Evaluation results of \gls{llm}-based approaches (\gls{zs}, \gls{fs} (\textit{\#examples}), and \gls{cot} (\textit{temperature})) for matching \gls{gui} components to \glspl{us} in \textit{Rico} \glspl{gui} with \textit{Precision (P)}, \textit{Recall (R)} and \textit{F1-measure (F1)} averaged over all instances (\textit{Macro}) or size-adjusted (\textit{Micro}).}
\centering

\setlength{\tabcolsep}{6pt}

\begin{tabularx}{\textwidth}{l|*{3}{>{\centering\arraybackslash}X}|*{3}{>{\centering\arraybackslash}X}}
\toprule
\multicolumn{1}{c|}{\textbf{}} &
\multicolumn{3}{c|}{\textbf{Macro}} &
\multicolumn{3}{c}{\textbf{Micro}} \\
\cmidrule(lr){2-4}\cmidrule(l){5-7}
& $\mathbf{P}$ & $\mathbf{R}$ & $\mathbf{F1}$ &
  $\mathbf{P}$ & $\mathbf{R}$ & $\mathbf{F1}$ \\
\midrule

\textbf{Zero-Shot$_{A}$}             & 78.4 & 81.9 & 75.5 & 62.0 & 75.5 & 68.1 \\
\rowcolor{lightgray}\textbf{Zero-Shot$_{B}$} & 71.8 & 90.3 & 74.3 & 50.2 & 85.8 & 63.3 \\
\midrule

\textbf{Few-Shot$_{5}$}              & 85.0 & 75.5 & 76.5 & 67.7 & 64.3 & 65.9 \\
\midrule

\rowcolor{lightgray}\textbf{CoT$_{t=0.0}$}     & 75.8 & 80.6 & 72.7 & 55.7 & 73.1 & 63.2 \\
\textbf{CoT$_{t=0.5}$}                & 72.1 & 78.8 & 68.8 & 49.2 & 71.0 & 58.1 \\
\rowcolor{lightgray}\textbf{CoT$_{t=1.0}$}     & 69.8 & 80.9 & 69.0 & 53.4 & 74.6 & 62.2 \\
\textbf{CoT$_{t=1.3}$}               & 67.7 & 74.0 & 65.4 & 56.2 & 64.6 & 60.1 \\
\bottomrule
\end{tabularx}

\label{tab:results_rq2_interlinking}
\end{table}
}

\begin{figure}[t]
      \centering
      \includegraphics[width=\textwidth]{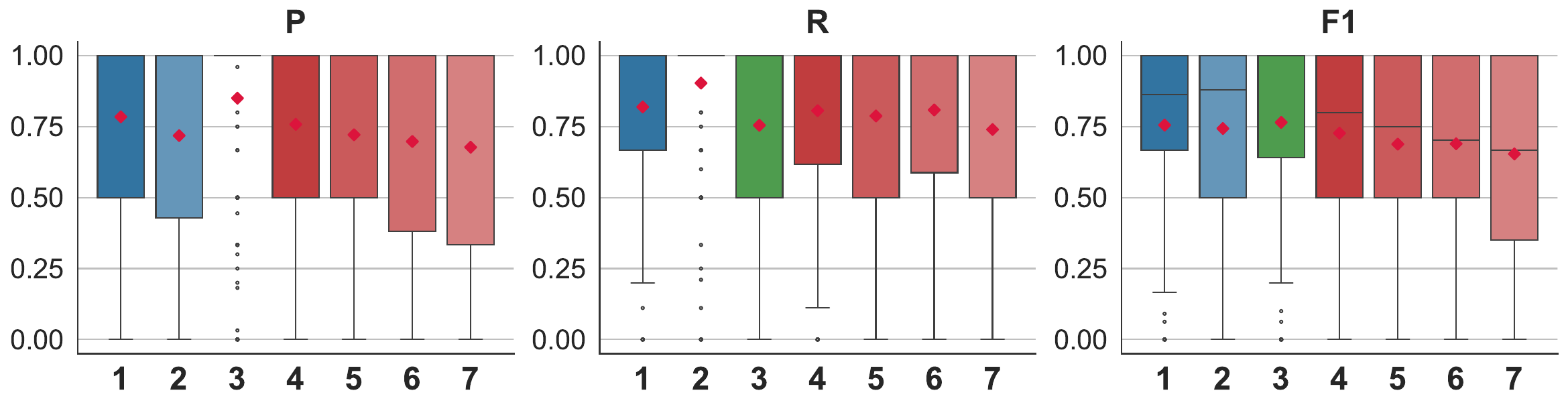}
      \caption[Boxplots for \gls{llm}-based \gls{gui} component matching effectiveness]{Boxplots for \textit{Macro} \textit{Precision (P)}, \textit{Recall (R)}, and \textit{F1-measure (F1)} across methods, with \gls{zs} (\textit{blue}), \gls{fs} (\textit{green}), and \gls{cot} (\textit{red}), in particular: \textit{(1)} \gls{zs}$_A$, \textit{(2)} \gls{zs}$_B$, \textit{(3)} \gls{fs}$_5$, \textit{(4)} \gls{cot}$_{t=0.0}$, \textit{(5)} \gls{cot}$_{t=0.5}$, \textit{(6)} \gls{cot}$_{t=1.0}$ and \textit{(7)} \gls{cot}$_{t=1.3}$, diamonds repr. means.}
    \label{fig:interlinker_boxplots}
\end{figure}

\begin{figure}[!t]
      \centering
      \includegraphics[width=\textwidth]{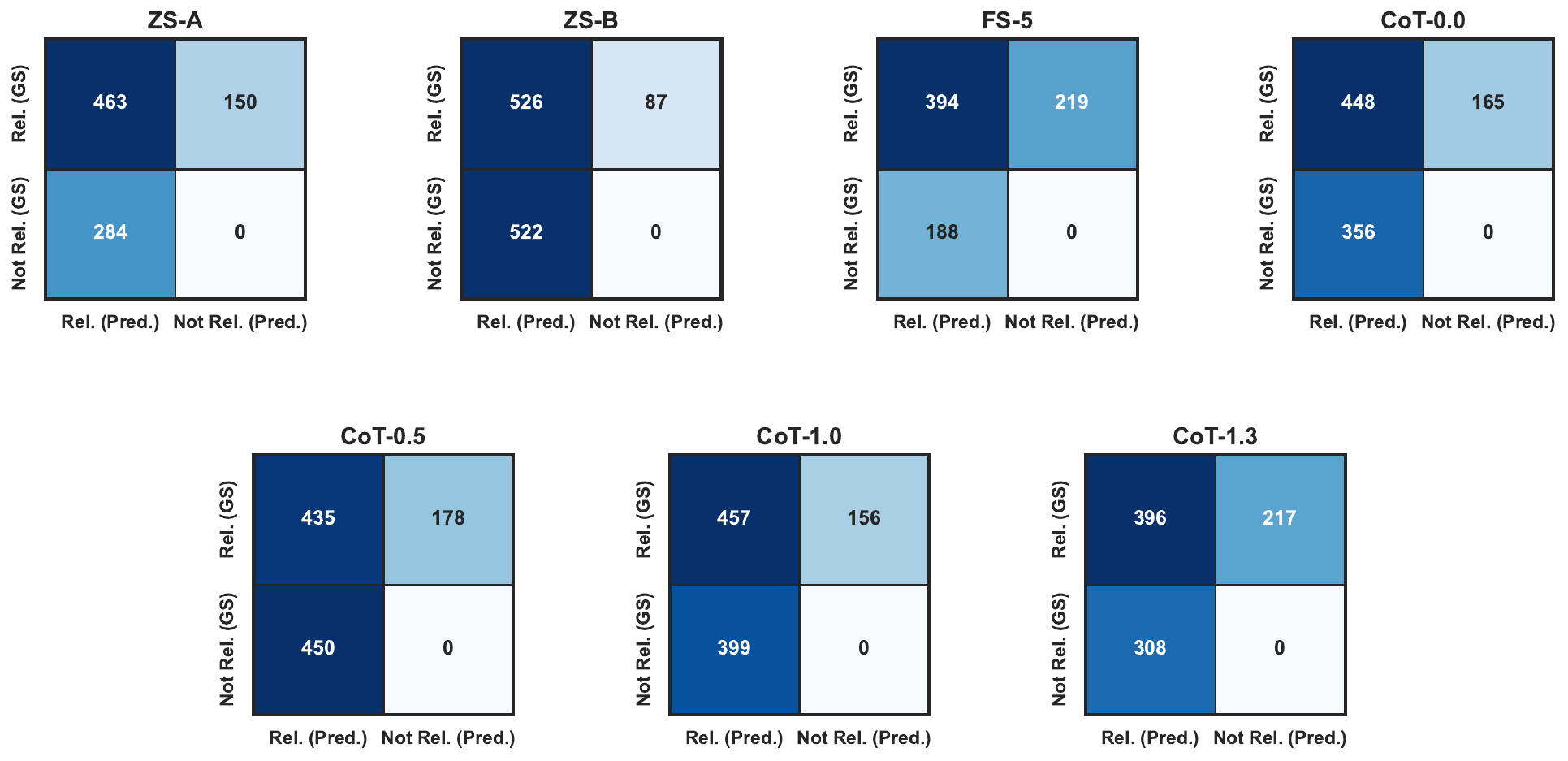}
      \caption[Confusion matrices for \gls{llm}-based approaches for \gls{gui} component matching]{Confusion matrices for \gls{llm}-based approaches (\gls{zs}, \gls{fs} (\textit{\#examples}), and \gls{cot} (\textit{temperature})) for matching \gls{gui} components to \glspl{us} within their respective \textit{Rico} \glspl{gui}. Lower right cell (\textit{Not. Rel. (GS)} and \textit{Not. Rel. (Pred.)}) always zero since \glspl{tn} were not computed (reference for comparison was the GS set instead of all \textit{Rico} \gls{gui} components).}
    \label{fig:interlinking_conf_matrix_2}
    \vspace{-0.5cm}
\end{figure}

Table \ref{tab:results_rq2_interlinking} illustrates the evaluation results for RQ$_{2}$, showing the \textit{Macro} and \textit{Micro} $P$, $R$, and $F1$ metrics. Moreover, Figure \ref{fig:interlinker_boxplots} shows boxplots for all prompting methods and across the \textit{Macro} metrics, while Figure \ref{fig:interlinking_conf_matrix_2} depicts the confusion matrices. Overall, the results indicate that the models achieve a moderate to good performance in matching \gls{gui} components to \glspl{us} shown by, for example, \textit{Micro}-$F1$ scores of 68.1 (\gls{zs}$_{A}$), 65.9 (\gls{fs}$_{5}$), and 63.2 (\gls{cot}$_{t=0.0}$). Although a direct comparison with the results of the task discussed in RQ$_{1}$ is difficult due to the difference in datasets, still the matching task can be seen as an extension of the classification task, since classification models probably perform similar computations (e.g., as indicated by explanations from \gls{cot} models) as part of their reasoning sequence. As can be observed, the performance difference of the tasks indicates that the matching task is significantly more difficult for the \glspl{llm}. However, we argue that the obtained results are promising due to the model being capable of extracting the majority of \gls{gui} components correctly, shown by the metric values. As indicated by the \textit{Holm}-corrected \textit{Wilcoxon signed-rank test} \citep{woolson2007wilcoxon} between the \textit{Macro}-$F1$ scores of the methods, all \gls{cot} prompting methods (except $t=0.0$) are significantly outperformed by both \gls{zs} and \gls{fs} approaches, whereas the differences of the \gls{zs} and \gls{fs} methods are insignificant (see Appendix \ref{app:interlinking} Table \ref{tab:matching_wilcoxon}). In addition, the \gls{zs}$_{B}$ prompting method has substantially better mean $R$ values compared to \gls{zs}$_{A}$ (and vice versa for $P$ values), indicating that the prompt extension in \gls{zs}$_{B}$ optimizes the model for not missing component matches. Figure \ref{fig:matching_examples-interlinker} shows two example \glspl{gui} from the gold standard (\textit{left} and \textit{center} \glspl{gui}) and highlighted \gls{gui} component matches as extracted by the \gls{llm}.

Moreover, we investigated \textit{low} $P$ and/or \textit{low} $R$ instances to improve the understanding of errors made by the \gls{llm}. For the cases of \textit{low} $P$, the \gls{llm} often extracted wrong \gls{gui} components that were semantically related. For example, for the \gls{us} to specify the number of guests in a hotel search \gls{gui}, the models also erroneously extract components for the number of rooms (see Figure \ref{fig:interlinking_error_1}\textit{(C)}). For the instances possessing \textit{low} $R$, similar to the misclassifications discussed earlier, we identified missing or ambiguous descriptions of \gls{gui} components as a main cause. For example, for a \gls{us} to be able to download videos to watch them offline, the \gls{gui} components offering download functionality were represented as \textbf{(Image)} \texttt{(document)}. Although the succeeding \gls{gui} component \textit{"0.7 MB"} \textbf{(Label)} \texttt{(document size)} could act as a hint, the naming as a \textit{document} and the component being marked as an image introduce ambiguity (see Figure \ref{fig:interlinking_error_1}\textit{(D)}). Finally, some cases possess ambiguity that is difficult to resolve without the stakeholder. For example, for the \gls{us} requesting to see an overview of the course (see Figure \ref{fig:matching_examples-interlinker}), the \gls{llm} extracts all listed lessons, whereas the annotation marks only the overview course item.

\begin{myrqbox}
\textbf{Answer to RQ$_2$:} \gls{llm}-based prompting approaches are effective for matching \gls{us} to their respective \gls{gui} components within static \glspl{gui}, achieving a moderate to good performance with \textit{Macro} \textit{precision} (78.4 for \gls{zs}$_{A}$), \textit{recall} (81.9 for \gls{zs}$_{A}$), and \textit{F1-measure} (75.5 for \gls{zs}$_{A}$). Significant differences were observed between \gls{zs}-\gls{cot} and \gls{fs}-\gls{cot}.
\end{myrqbox}

\section{Threats to Validity} 

\vspace{-0.2cm}

\paragraph{Internal Validity} The \textit{Rico} \gls{gui} dataset is widely adopted in research and thus indicates suitability for \gls{gui}-based evaluation tasks. Nevertheless, the dataset is influenced by the collection methods. Particularly, \textit{Rico} \glspl{gui} present a sample of free \textit{Android} applications from \textit{Google Play}. However, to avoid bias, we selected \glspl{gui} from a wide range of different domains and the initial sample was randomly drawn. Additionally, there exists no dataset combining \glspl{gui} with \glspl{us}. We therefore decided to collect \glspl{us} for existing \glspl{gui} in a lab-based annotation study, which might introduce bias and subjectivity from participants. However, we provided clear instructions and training for writing high-quality \glspl{us} and collected a large number of \glspl{us}. The exhaustiveness of \glspl{us} per \gls{gui} in our data set may be limited, as a fixed number of \glspl{us} were written by eight participants for the respective \glspl{gui}. Additional bias might have been introduced by the authors' evaluation of collected \glspl{us}. However, we conducted independent labeling and calculated \gls{iaa}. Afterwards, we discussed and resolved conflicts to reduce potential bias.

\paragraph{External Validity} Since we restricted the evaluation to \textit{Rico} \glspl{gui} with a particular representation, the evaluation results could vary for different \gls{gui} datasets and software platforms. However, \textit{Rico} represents the largest and most diverse mobile \glspl{gui} dataset available for research. In addition, we evaluated only a single \gls{llm}, which potentially restricts generalizability to other \glspl{llm}. Nevertheless, we employed a state-of-the-art \gls{llm} (at the time of implementing the approach) and tested different prompting methods.
\section{Limitations}

This work was a feasibility study, investigating the effectiveness of \gls{llm}-based approaches for verifying the implementation of \gls{nlr} (particularly \glspl{us}) in simplified, static \glspl{gui}. While the representation of \textit{Rico} \glspl{gui} in this work clearly represents a limitation, it confirmed that the profound \gls{nlu} capabilities of \glspl{llm} can be effectively employed to identify the implementation of \glspl{us} in \glspl{gui} and match the relevant \gls{gui} components.
\section{Related Work}
\label{sec:interlinking_rel_work}

Previous research investigated automated consistency or conformance checking of requirements on \glspl{gui}. For example, \cite{silva_10.1007/978-3-030-24289-3_46} present a Behavior-Driven
Development (BDD)-based approach that enables the automated testing of \glspl{us} on interactive web \glspl{gui} with browser automation tools. However, their approach relies on manually crafted ontologies, which are time-consuming and effort-demanding to create. Furthermore, they rely on manually mapping abstract \gls{gui} components to concrete \gls{gui} identifiers, which introduces brittleness and is effortful to maintain. In contrast, we investigated leveraging the powerful \gls{nlu} capabilities of modern \glspl{llm} via adopting multiple prompting-based approaches, which require no manual finetuning and can effectively process complex \gls{gui} context. 

Furthermore, \cite{mukasa2008integration} proposed the \gls{rsl}, which integrates requirements (e.g., in the form of \gls{nlr}) and \gls{gui} specifications into a conceptually coherent representation. Their specification approach enables requirements and \gls{gui} specifications to evolve simultaneously, closely bound together. Therefore, the developed system facilitates the implementation of correct \gls{fr} while also supporting high usability. While the proposed \gls{rsl} provides a novel approach for facilitating the simultaneous evolvement of requirements and their respective \gls{gui} specifications, the approach is unable to automatically provide consistency checking and requires effort-demanding modeling activities. 

\cite{moslehi2020feature} propose a feature localization approach that enables automatic traceability between individual \gls{gui} features of a software system and the corresponding source code. In particular, they rely on crowd-based screencasts for software systems as input, apply Latent Dirichlet Allocation (LDA)-based mining approaches, and locate relevant code artifacts. While their approach also provides a mapping between \gls{gui} elements and related artifacts, their focus lies on source code, while we focus on \gls{nlr}. 
\section{Conclusion}
The work presented in this chapter was driven by challenge \challtwoone{}: \textit{How can \gls{nlr}-based verification be automated for simplified, static \gls{gui} representations as a foundation for more complex \gls{gui} verification tasks?} To tackle this question, we conducted a feasibility study by employing \gls{llm}-based approaches for automatically detecting whether \gls{nlr} (particularly in the form of \glspl{us}) are implemented in a \gls{gui} and subsequently matching the relevant \gls{gui} components. While the approach relied on simplified, static \textit{Rico} \gls{gui} representations, our evaluation indicates high effectiveness for both tasks, showcasing the large potential of \gls{llm}-based approaches for automatic verification of \gls{nlr} in \glspl{gui}.

\clearpage
\newpage
\thispagestyle{empty}
\null  

\chapter{GUISpector: An MLLM Agent Framework for Automated Verification of Natural Language Requirements in GUI Applications}
\chaptermark{An MLLM Agent Framework for Verification of NLR in GUIs}
\label{cha:agent}
The previous chapter introduced an initial solution approach for challenge \challtwo{}, namely, the automated verification of \gls{nlr} in \glspl{gui}. While the previous approach was conducted as a feasibility study to investigate the potential of \gls{llm}-based approaches for verifying \gls{nlr} in \glspl{gui} with static, simplified \gls{gui} representations, in this chapter, we propose a novel approach for the automated verification of \gls{gui} applications using an \gls{mllm}-based \gls{cua} against \gls{nlr}. The work presented in this chapter has been accepted for publication and is currently in press \citep{kolthoff2025guispector}\footnote{This section is adapted from: \textbf{Kolthoff, Kristian\textsuperscript{*}}, Kretzer, Felix\textsuperscript{*}, Bartelt, Christian, Maedche, Alexander, and Ponzetto, Simone Paolo. GUISpector: An MLLM Agent Framework for Automated Verification of Natural Language Requirements in GUI Prototypes. In \emph{Proceedings of the 48th International Conference on Software Engineering: Companion Proceedings (ICSE-Companion)}, April 2026, 1-4, Rio de Janeiro, Brazil. ACM. (in press) (arXiv preprint arXiv:2510.04791). *Authors contributed equally.}. Our source code, interactive prototype, evaluation datasets, and tool demonstration video are all publicly available\footnote{Materials for this chapter, including source code and evaluation datasets, are available at \url{https://github.com/kristiankolthoff/GUISpector} and demo video at \url{https://youtu.be/JByYF6BNQeE}}.

\paragraph{Personal Contribution.} I developed the initial idea and concept for the proposed approach, while \textit{Felix Kretzer} and I conducted discussions to refine the concept. Moreover, I implemented the entire framework and tool prototype. \textit{Felix Kretzer} and I contributed equally to the evaluation design, while \textit{Felix Kretzer} created the dataset code and conducted the data collection. \textit{Felix Kretzer} and I both contributed equally to the manual annotation of the dataset. Furthermore, I conducted the data analysis and computed the results. While \textit{Felix Kretzer} wrote the experimental setup and results sections, I wrote the rest of the original manuscript and contributed all respective tables, plots, and figures.

\section{Motivation}

\newcommand{\guispector}{\textit{\gls{gui}Spector}\xspace}

The first two parts of the thesis focused on enabling rapid \gls{gui} prototyping from \gls{nlr} by proposing \gls{gui} retrieval and generation techniques. Consequently, \gls{rel} and \gls{rval} activities are facilitated, ensuring that the correct requirements are captured according to the actual needs of stakeholders. Subsequently, verification ensures that the implemented software system adheres to the created requirements specification, which is often conducted via time-consuming and effort-demanding testing of the system \citep{myers2004art}. While the previous chapter introduced \gls{llm}-based approaches for verifying \gls{nlr} in \glspl{gui} and demonstrated high effectiveness, limitations arise from employing static, simplified \textit{Rico} \citep{deka2017rico} \gls{gui} representations, which differ from real-world, complex, and highly dynamic \gls{gui} applications.

\glsreset{mcp}
\begin{figure*}[!t]
  \centering
 \includegraphics[width=1\textwidth]{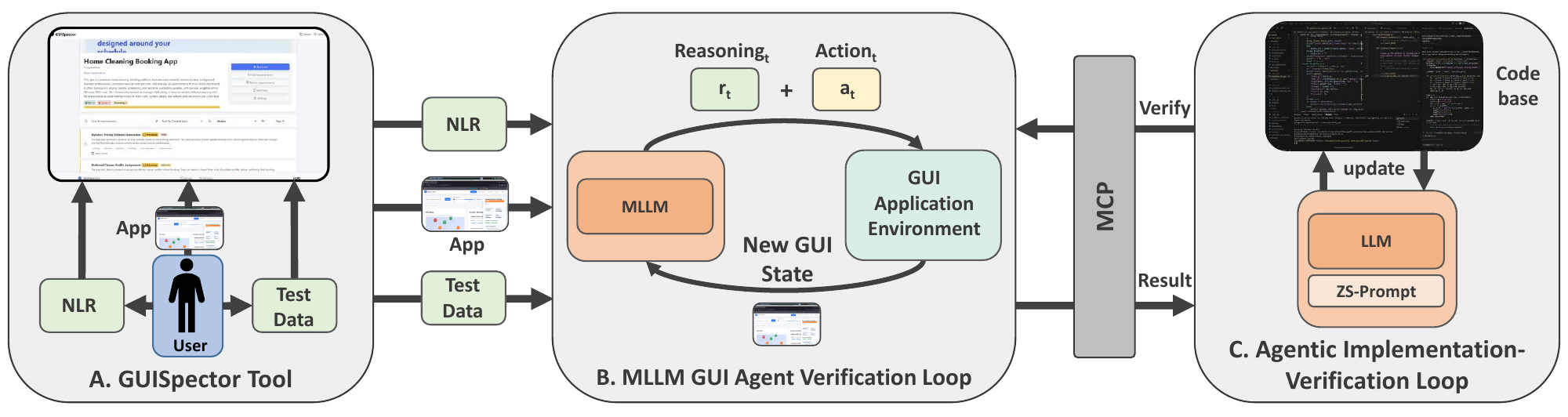}
  \caption[Overview of \guispector for \gls{mllm}-based \gls{cua} \gls{gui} verification]{Overview of \guispector consisting of \textit{(A)} our user-friendly tool, which enables to provide unstructured \gls{nlr}, the \gls{gui} application for verification and test data, \textit{(B)} the core \gls{mllm} \gls{cua} verification loop, which iteratively employs an \gls{mllm} to predict next actions, executes them in \gls{gui} environments and obtains new \gls{gui} states, and \textit{(C)} an agentic implementation-verification loop, which enables \gls{llm} programming agents to run \gls{gui} verifications in \textit{\gls{gui}Spector} via \gls{mcp} interface.}
  \label{fig:overview_summary_guispector}
\end{figure*}

As discussed in the previous chapter, many approaches have been proposed in research before to improve testing activities. For example, these include automated test scripts \citep{xie2007designing}, random exploration \citep{mao2016sapienz}, model-based techniques \citep{zeng2016automated, gu2019practical}, and scripted bug replay methods \citep{gomez2013reran, feng2022gifdroid}. While these approaches enhance the efficiency and coverage of \gls{gui} testing, they are restricted to recognizing runtime failures (system crashes) and are therefore incapable of directly verifying the correctness of \gls{nlr} in \gls{gui} applications. In addition, these methods require manual tuning or cannot fully capture the complexity and dynamics of \gls{gui} applications. More recently, \gls{llm}-based approaches gained traction for more sophisticated automated bug detection in \glspl{gui}. For example, \gls{llm}-based approaches enable the generation of tests from \gls{nl} descriptions \citep{feng2024enabling, feng2025agent}. However, these methods still rely on textual abstractions of \gls{gui} hierarchies (similar to our proposed approaches in Chapter \ref{cha:interlinking}), which renders them unable to fully assess dynamic, visual aspects of \gls{gui} applications and verify \gls{nfr} and fine-grained \gls{fr}. Therefore, the work in this chapter is driven by the leading research question of challenge \challtwotwo{}: \textit{How can \gls{nlr}-based verification be automated for highly dynamic, fully-fledged \gls{gui} applications?}

To tackle this challenge and close the gap, in this chapter, we introduce \guispector, a novel \gls{mllm}-based \gls{cua} for automated verification of \gls{gui} applications from \gls{nlr}, as shown in Figure \ref{fig:overview_summary_guispector}. In particular, the core of \guispector is the \gls{mllm} \gls{cua} verification loop (see Figure \ref{fig:overview_summary_guispector}\textit{(B)}), which iteratively enables the \gls{mllm} to provide reasoning and predict next actions, executes them in \gls{gui} environments, and obtains new \gls{gui} states. Given \gls{nlr}, the agent derives meaningful action sequences to support the verification of the requirements, thereby creating meaningful test trajectories and enabling detailed summarization of executed tests. These summaries and test results can be accessed by users from our user-friendly \guispector tool (see Figure \ref{fig:overview_summary_guispector}\textit{(A)}) and employed in a closed loop to feed back to \gls{llm}-based programming agents. Moreover, users can provide \gls{nlr}, the \gls{gui} application to test, and test data as input to our tool. Since \gls{llm}-based programming agents become more popular, \guispector provides an \gls{mcp} implementation to enable closed agentic implementation-verification loops (see Figure \ref{fig:overview_summary_guispector}\textit{(C)}), enabling agents to run automated \gls{gui} verifications after implementing changes in the codebase. Our evaluation across five \gls{gui} applications and 150 requirements with 450 \gls{ac} shows the effectiveness of the approach, achieving highly accurate automated verification scores.

\paragraph{Contributions.} With this approach, we make the following research contributions:
\begin{itemize}[leftmargin=6mm]
\setlength{\itemsep}{1pt}
    \item \textit{Novel \gls{mllm}-based \gls{cua} approach for automated verification of highly dynamic \gls{gui} applications from \gls{nlr}:} we present \guispector, a novel \gls{mllm}-based \gls{cua} approach for enabling automatic verification from \gls{nlr} in \gls{gui} applications by iteratively predicting next actions, executing them and obtaining updated \gls{gui} states.
    \item \textit{Tool prototype and \gls{mcp} interface:} we integrate the \gls{mllm}-based \gls{cua} verification approach into a novel user-friendly tool, which enables rapid adoption by practitioners and provides detailed test summaries. In addition, we provide an \gls{mcp} implementation, enabling \gls{llm}-based programming agents to access \guispector.
    \item \textit{Comprehensive evaluation and insights:} we create a novel dataset, which integrates web-based \gls{gui} applications with 150 detailed \gls{nlr} and 450 \gls{ac}. Moreover, we conduct a comprehensive experimental evaluation of the capabilities of the \gls{mllm} \gls{cua} approach, showing high effectiveness, and investigate generated test trajectories.
\end{itemize}

\section{Approach: GUISpector}

\begin{figure*}[!t]
  \centering
 \includegraphics[width=1\textwidth]{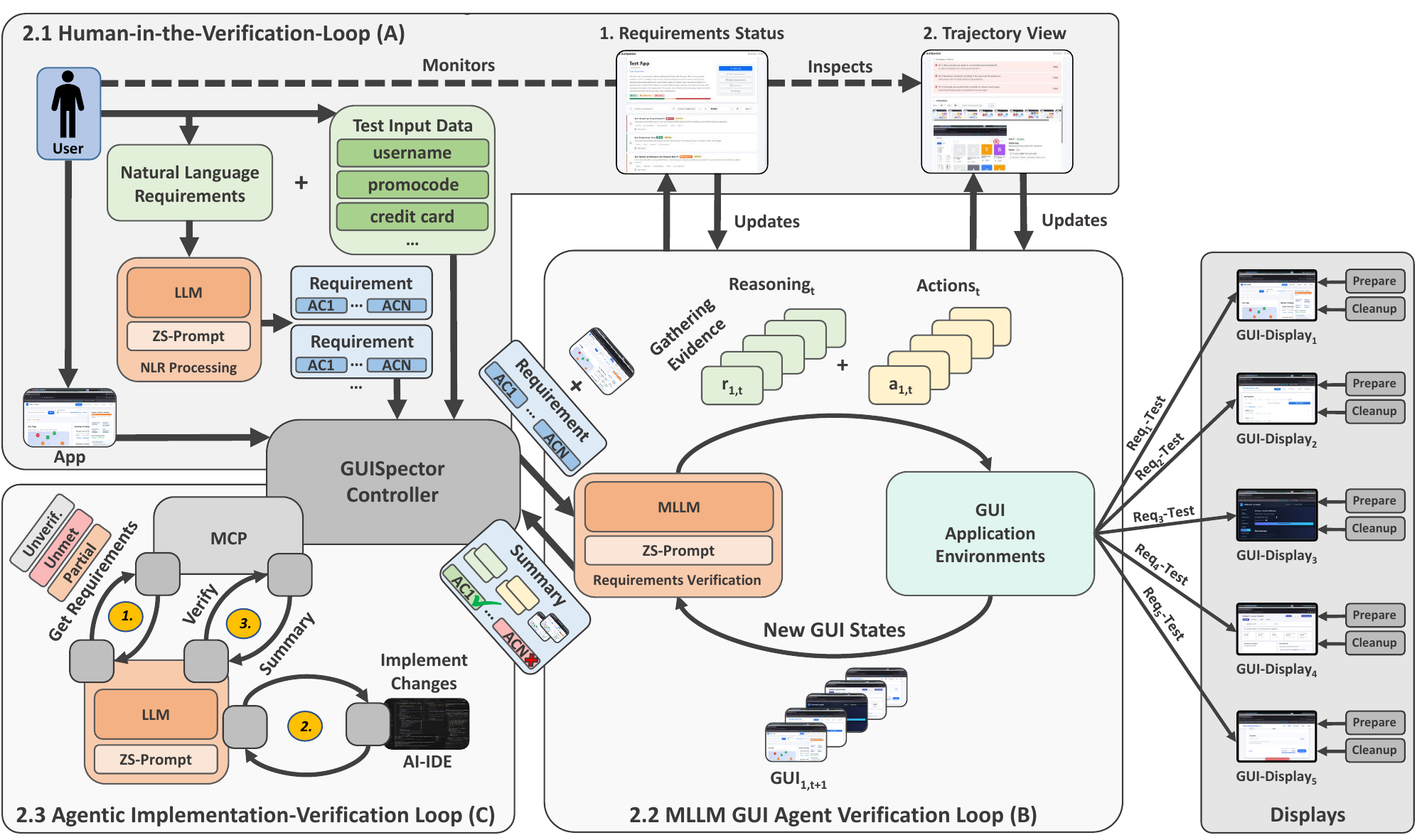}
  \caption[Overview of the \guispector architecture]{Overview of the \guispector architecture: \textit{(A)} human-in-the-verification-loop, \textit{(B)} \gls{mllm} \gls{cua} \gls{gui} verification loop with reasoning, actions and \gls{gui} state trajectories, and \textit{(C)} agentic implementation-verification loop for \gls{llm}-based programming agents.}
  \label{fig:overview_guispector}
\end{figure*}

\guispector is composed of three main components: \textit{(A)} a human-in-the-verification-loop, which enables users to set up and monitor verification runs, provide unstructured \gls{nlr} for verification, the corresponding web-based \gls{gui} application, and optional test input data. Moreover, \textit{(B)} the \gls{mllm}-based \gls{cua} verification loop, which employs the \gls{mllm} to conduct reasoning and predict actions given the current \gls{gui} state and trajectory, run actions on dedicated \gls{gui} environments, and obtain updated \gls{gui} states to close the loop. Finally, \textit{(C)} \guispector provides an \gls{mcp} implementation, enabling fully autonomous agentic implementation-verification loops for \gls{llm}-based programming agents. An overview of the architecture details of \guispector is illustrated in Figure \ref{fig:overview_guispector}.

\subsection{Human-in-the-Verification-Loop}

\begin{figure*}[!t]
  \centering
 \includegraphics[width=1\textwidth]{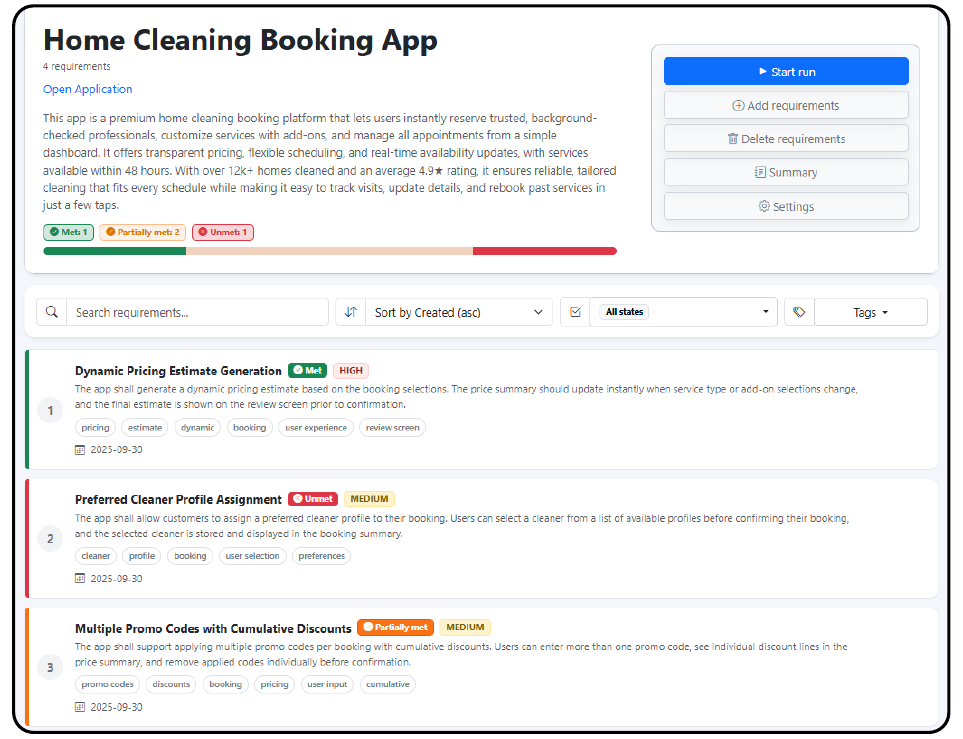}
  \caption[Tool prototype of \guispector showing a requirements collection]{Tool prototype of \guispector showing the user-friendly \gls{gui} with a collection of \textit{met}, \textit{unmet}, and \textit{partially met} \gls{nlr}. Moreover, \guispector provides options to start verification runs across all listed requirements, add unstructured requirements via an \gls{llm}-based interface, delete requirements, obtain run summaries, and update settings.}
  \label{fig:gui_spector_tool_screenshot}
\end{figure*}

To enable users to effectively run verifications with \guispector, we provide a user-friendly tool prototype and \gls{gui}. Specifically, users begin by specifying the entry point (\gls{url}) of the web application to run verification on, followed by providing structured or unstructured \gls{nlr}. Next, the \gls{nlr} are processed by employing a \gls{zs}-based \gls{llm} that creates a structured representation of the \gls{nlr}. This acts as a generic \gls{nl} interface to \guispector, and \gls{nlr} can be provided in any form. In addition, the \gls{llm} can automatically infer missing details (such as low-level \gls{ac} for high-level \gls{nlr} in the context of the application) if necessary. After successfully processing the requirements, \guispector shows a structured collection of them as an overview, including their verification status after a run has been completed, as depicted in Figure \ref{fig:gui_spector_tool_screenshot}. Furthermore, the verification trajectories created by the \gls{mllm}-based \gls{cua} of each run within each requirement can be inspected in our \gls{gui}, combined with a detailed breakdown of \gls{ac} fulfillment status. Moreover, the agent gathers evidence for each \gls{ac} decision and generates an actionable summary. Overall, users are enabled to create, organize, and manage individual verification setups.

\subsection{MLLM GUI Agent Verification Loop}

The central component of the \guispector approach is the \gls{mllm} \gls{cua} \gls{gui} verification loop, which adapts a pretrained \gls{mllm} that has additionally been finetuned for effective computer use (understanding \gls{gui} screens, recognizing \gls{gui} components, and predicting parametrized \gls{gui} actions for solving an interaction task) to autonomously run verification of \gls{gui} applications given \gls{nlr}. Specifically, we adapt the model by employing a \gls{zs} prompt, clearly instructing the model to autonomously verify the implementation of requirements in \gls{gui} applications, providing the application entry point (i.e. start \gls{url}) and expecting the model to run meaningful interactions necessary for verification. We particularly instruct the model to employ a minimal interaction sequence to optimize testing efficiency and integrate user-provided test data if necessary. Moreover, we instruct the \gls{mllm} to verify each \gls{ac} for a requirement individually, gather specific evidence during the runs, and thus justify its decisions. In particular, for each verification run, the \gls{mllm} receives the requirement to verify, accompanied by \gls{ac}, necessary test input data (such as \textit{login data}, \textit{coupon codes}, or \textit{user profile data}), the current \gls{gui} state represented as a screenshot, and the previous trajectory for improved contextual awareness. Based on this task and context representation, the \gls{mllm} subsequently provides reasoning and predicts the most likely next action to verify an \gls{ac}. Specifically, the employed \gls{cua} supports all standard actions including \textit{click}, \textit{scroll}, \textit{type}, \textit{keypress}, \textit{drag}, \textit{double click}, \textit{move}, and \textit{wait}. In addition to the action type, the agent also predicts the different action parameters (such as \textit{x}, \textit{y}, and \textit{button type} for the \textit{click} action, and \textit{text} for the \textit{type} action). Subsequently, the selected parametrized action is executed in a dedicated environment running the actual \gls{gui} application, and the updated \gls{gui} state is reported back to the \gls{mllm}, closing the loop. This iterative procedure creates a trajectory encompassing \gls{gui} states (i.e. screenshots), reasoning, and parametrized actions (\textit{GUI}$_1$, r$_1$, a$_1$, \ldots, \textit{GUI}$_n$, r$_n$, a$_n$). Finally, the \gls{mllm} is instructed to generate a \gls{json} output encompassing a detailed summary of the verification run, explanations, the predicted status of each \gls{ac} (\textit{met} or \textit{unmet}), and gathered evidence for the decision. The requirement status (\textit{met}, \textit{unmet}, or \textit{partially met}) is then computed from the individual \gls{ac} states. In particular, the requirement status is set to \textit{met} (\textit{unmet}) if all of the corresponding \gls{ac} states are \textit{met} (\textit{unmet}). Otherwise, the status is set to \textit{partially met} (at least one \textit{met} and one \textit{unmet} \gls{ac}). The generated summaries are especially helpful to manually employ as feedback to developers or \gls{llm}-based programming agents. To avoid contaminating \gls{gui} states and introducing errors, the environments are cleaned up before and after each verification.

\subsection{Agentic Implementation-Verification Loop}

While the previously introduced \gls{mllm}-based \gls{cua} verification loop in combination with the \guispector tool interface enables users to rapidly run autonomous verification of their \gls{gui} applications, \guispector additionally supports a fully automatic mode integrated into \gls{llm}-based programming agents. More recently, \gls{llm}-powered Integrated Development Environments (IDEs) such as \textit{Cursor} \citep{cursor} and \textit{Copilot} \citep{githubcopilot} gained popularity, focusing on agentic programming support by enabling developers to easily provide relevant parts of their codebases to \glspl{llm} to implement changes or entirely new functionality for increasing developer productivity. 

To facilitate seamless integration of the verification capabilities of \guispector into \gls{llm}-based programming agents, we implemented an \gls{mcp} \citep{hou2025model} server to allow external agents to interact with the verification \gls{api}. Specifically, for different verification setups created by users through the \guispector tool, these programming agents can retrieve requirements collections with their accompanying \gls{ac}, their current verification states, and start implementing patches for violated requirements. Moreover, the \gls{llm} agent can actively start verification runs for selected requirements, for example, when they are not verified yet or after a patch was implemented for a violated requirement. Subsequently, the programming agent receives the previously discussed detailed summaries with evidence gathered during the verification run, which can be employed to inform the agent to update the codebase again and potentially increase its implementation effectiveness (particularly in cases of \textit{unmet} or \textit{partially met} requirements). 

This enables a fully autonomous, iterative agent mode: the \gls{llm}-based programming agent retrieves a requirement, implements patches if necessary, runs verification, and either implements further changes if the requirement status is not identified as \textit{met} during the verification, or continues the same process for the next requirement. We created a \gls{zs} prompt that can be employed inside \gls{llm}-powered IDEs or in general for \gls{llm}-based programming agents, which instructs the \gls{llm} to follow the described autonomous implementation-verification loop, providing a novel approach for integrating automatic verification with requirements-driven, \gls{llm}-based development of \gls{gui} applications.

\subsection{Prototype Implementation}

\guispector is implemented as a \textit{Django} web application with an \textit{\gls{html}/\gls{css}/\gls{js}} frontend, combined with \textit{MySQL} \citep{mysql_server_github} for data storage. To enable efficient and parallel processing of background tasks (such as the \gls{mllm}-based \gls{cua} verification runs), we utilize the \textit{Celery} framework \citep{celery_docs_introduction} distributed among worker nodes in combination with the message broker \textit{Redis} \citep{redis_github}. Since \guispector requires dedicated environments to simultaneously run verification of \gls{gui} applications, we additionally implemented a \textit{display resource manager} based on \textit{Redis}, which enables the dynamic allocation of \textit{Xfce} desktop environments deployed on \textit{Xvfb} virtual displays. To enable execution of the parametrized \gls{gui} actions predicted by the \gls{mllm}, we utilize \textit{xdotool} for user interaction simulation. For running the \gls{mllm}-based \gls{cua}, we experimented with the \textit{OpenAI \gls{cua}} approach \citep{cua2025}. However, other agents can be easily integrated into the proposed framework. For other \gls{llm}-based tasks in \guispector (processing of \gls{nlr}), our prototype currently supports \textit{OpenAI \gls{gpt}} \citep{openai2023gpt4}, \textit{Google Gemini} \citep{team2023gemini}, and also \textit{Anthropic Claude Sonnet} \citep{claude4}.

\vspace{-0.2cm}
\section{Experimental Evaluation}

In the following, we present the comprehensive evaluation of our approach, which is based on a novel dataset combining web-based \gls{gui} applications and fine-grained \gls{nlr}. Particularly, we provide the experimental setup for the following two research questions:

\vspace{0.1cm}
\begin{itemize}
       \item \textbf{RQ$_{1}$}: \textit{How effective is the \gls{mllm}-based \gls{cua} approach for automated verification of highly dynamic \gls{gui} applications based on \gls{nlr}?} To answer this question, we create a novel dataset of web-based \gls{gui} applications combined with \textit{met}, \textit{unmet}, and \textit{partially met} \gls{nlr}. To evaluate the effectiveness of the approach, we compute standard classification metrics including \textit{precision}, \textit{recall}, and \textit{F1-measure} per class.
       \item \textbf{RQ$_{2}$}: \textit{What are the costs associated with an \gls{mllm}-based \gls{cua} approach for automated verification of highly dynamic \gls{gui} applications based on \gls{nlr}?} To answer this question, we track the costs (\textit{token consum.}, \textit{steps}, \textit{time}, \textit{monetary cost}).
\end{itemize}

\begin{figure*}[!t]
  \centering
  \frame{
    \subfloat[\centering \#Words/(\gls{nlr}/\gls{ac}) distributions]{
      \includegraphics[width=0.44\linewidth]{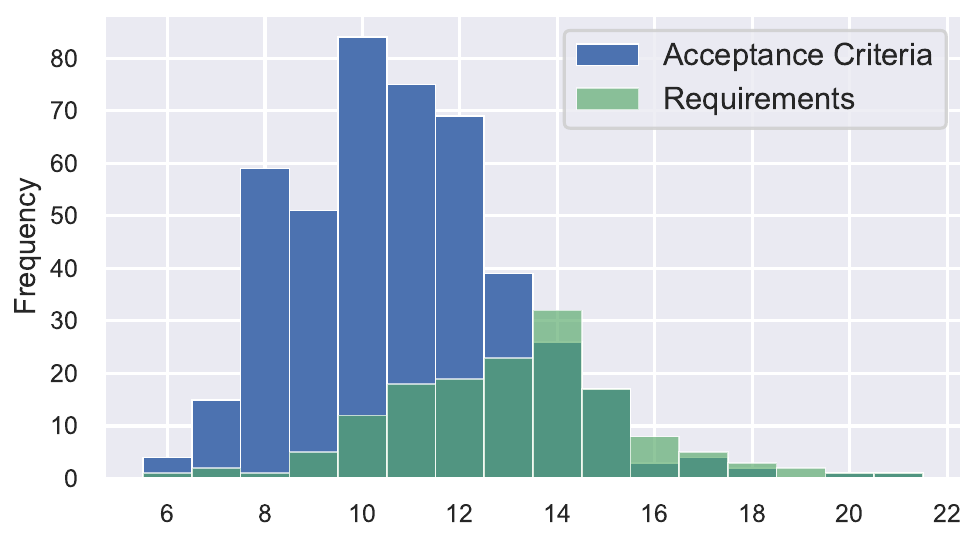}
      \label{fig:gui_spector_dataset_distributions_1}
    }
  }
  \qquad
  \frame{
    \subfloat[\centering \gls{nlr} label distributions per app]{
      \includegraphics[width=0.44\linewidth]{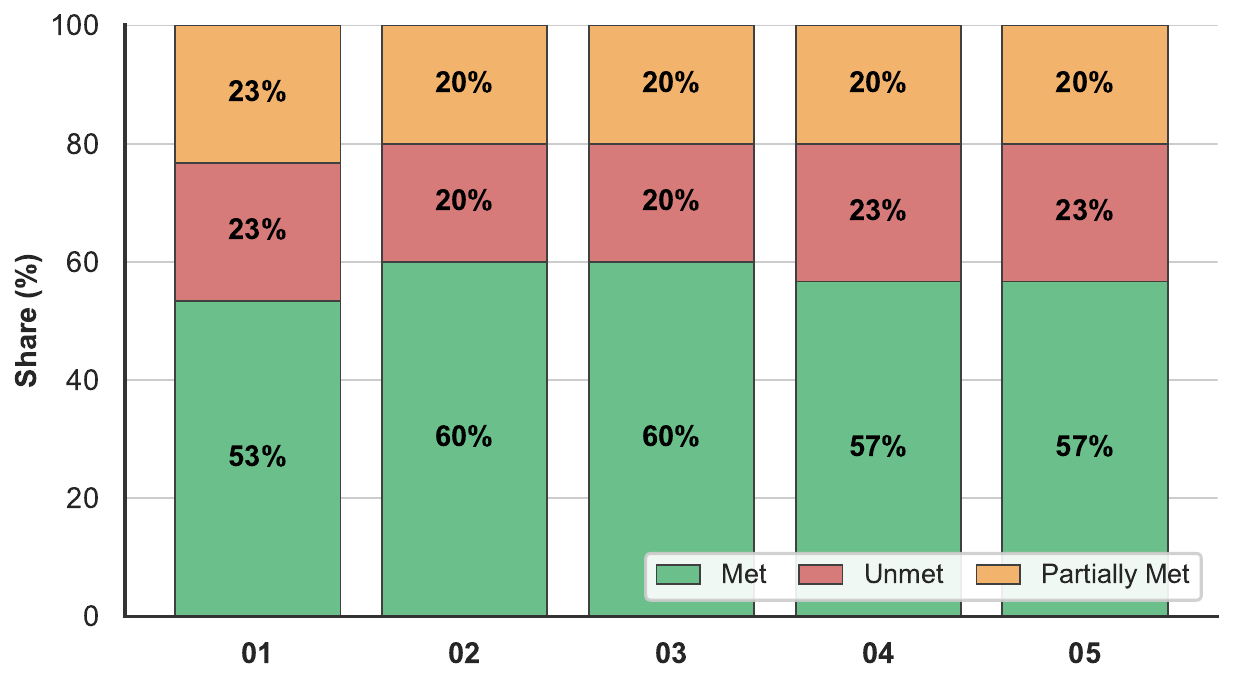}
      \label{fig:gui_spector_dataset_distributions_2}
    }
  }
  \caption[\#Words/(\gls{nlr}/\gls{ac}) and \gls{nlr} label distributions per app]{Distributions for the created dataset combining \glspl{us} with \glspl{gui}. In particular, each \gls{us} is represented by a one-to-many mapping ($\geq1$) to respective \gls{gui} components.}
  \label{fig:gui_spector_dataset_distributions}
\end{figure*}

\begin{table}[t]
\centering
\small
\setlength{\tabcolsep}{5pt}
\begin{tabular}{lcccccccc}
\toprule
\textbf{Variable} & \textbf{n} & \textbf{$\mathbf{\mu}$} & \textbf{$\mathbf{\sigma}$} & \textbf{min} & \textbf{Q1} & \textbf{med.} & \textbf{Q3} & \textbf{max} \\
\midrule
(1) \#Words/Requirement            & 150 & 13.11 & 2.51 & 6.00 & 11.00 & 13.00 & 14.00 & 21.00 \\
(2) \#Words/Acceptance Criterion   & 450 & 10.83 & 2.30 & 6.00 & 9.00  & 11.00 & 12.00 & 21.00 \\
\bottomrule
\end{tabular}
\caption[Summary statistics for \gls{nlr}-based \gls{gui} verification gold standard]{Summary statistics (\textit{mean} $\mu$, \textit{standard deviation} $\sigma$, \textit{minimum}, \textit{first quartile} \textbf{Q1}, \textit{median}, \textit{third quartile} \textbf{Q3}, \textit{maximum}) for the novel gold standard considering the following aspects: \textit{(1)} number of words per requirement, \textit{(2)} number of words per \gls{ac}.}
\label{tab:gui_spector_summary_stats}
\end{table}

\vspace{-0.4cm}
\subsection{RQ$_1$: Effectiveness of \gls{mllm} \gls{cua} for \gls{gui} Verification}

\paragraph{Gold Standard.} To evaluate the ability of the proposed approach to recognize requirements fulfillment, we first created a new dataset, due to the lack of available datasets of \glspl{gui} paired with requirements. Using \textit{Codex} \citep{codex2025}, we generated 30 functional \gls{nlr} with three \gls{ac} each for five domain-diverse applications (\textit{Park-and-Pay, Budget Tracker, Recipe Generator, Fitness Quests, Cleaning Booking}). Each requirement was also randomly assigned a status (\textit{met}, \textit{unmet}, and \textit{partially met}), with the target distribution being set to 18 \textit{met}, six \textit{unmet}, and six \textit{partially met} requirements (30 per app). In a second step, \textit{Codex} was instructed to create the corresponding apps (single \textit{\gls{html}} files with local state, no external Content Delivery Networks (CDNs)/\glspl{api}, including all \textit{\gls{css}} and \textit{\gls{js}}) and ensure the respective requirements statuses. All employed prompts, \gls{nlr}, \gls{ac}, and apps are available in our accompanying repository.

We then evaluated whether the generated apps conformed to the specifications, especially considering the different requirements statuses. Two authors independently labeled 450 \gls{ac} as \textit{met} or \textit{unmet} by running the applications and conducting manual testing. Annotations were independent but not equally blinded: one annotator had not seen the detailed specification; the other had prior exposure but did not consult it during annotation. As described before, requirement statuses were derived from \gls{ac} labels (all \gls{ac} \textit{met} = Req. \textit{met}, all \gls{ac} \textit{unmet} = Req. \textit{unmet}, otherwise = Req. \textit{partially met}).

Before resolving disagreements, we measured \gls{iaa} across 150 requirements (5 apps × 30 requirements) for three raters (\textit{\gls{llm}-Specification}, \textit{Annotator 1}, \textit{Annotator 2}), indicating very high agreement (\textit{Krippendorff’s} $\alpha:$ 0.929, ordinal, 95\% \gls{ci} [0.880, 0.966]). We adopt the agreed human evaluations as the gold standard, and eight disagreements (out of 150) occurred between the \gls{llm} and human annotators.
At the \gls{ac} level (5 apps $\times$ 30 requirements $\times$ 3 \gls{ac}), agreement between the two human annotators was similarly high (\textit{Cohen’s} $\kappa:$ 0.921, pooled, nominal, with accuracy: 96.7\%, 435 out of 450 \gls{ac}). Stratified by app, $\kappa$ ranged from 0.841 (\textit{App 1}) to 0.972 (\textit{App 2}), indicating high overall agreement with some heterogeneity. Table \ref{tab:gui_spector_summary_stats} shows summary statistics for the novel gold standard with word counts per \gls{nlr} and \gls{ac}, while Figure \ref{fig:gui_spector_dataset_distributions}\textit{(a)} shows the respective distributions and Figure \ref{fig:gui_spector_dataset_distributions}\textit{(b)} depicts the \gls{nlr} label distributions across all five apps. 

\paragraph{Evaluation Metrics.} Based on the ground truth of each \gls{nlr} and related \gls{ac}, we evaluated the ability of \guispector to identify the correct requirements statuses. Each app was evaluated in a single run. Requirements were classified into three categories (\textit{met}, \textit{unmet}, \textit{partially met}), while \gls{ac} were assigned two categories (\textit{met}, \textit{unmet}). We calculated \textit{precision}, \textit{recall}, and \textit{F1-measure} for the three-class and the binary \gls{ac} classification task.

\subsection{RQ$_2$: Cost for \gls{mllm}-based \gls{cua} \gls{gui} Verification}

\paragraph{Evaluation Metrics.} To evaluate the costs associated with \gls{mllm}-based \gls{cua} \gls{gui} verification, we track several cost metrics alongside running the experiments, including the \textit{token consumption}, \textit{steps}, \textit{time}, and \textit{monetary cost}, and report the \gls{nlr}-wise averages.

\section[Results \&\ Discussion]{Results \& Discussion}

\definecolor{lightgray}{gray}{0.92}
\newcolumntype{G}{c}

\newcommand{\EffSepRule}{%
  \cmidrule(lr){1-1}%
  \cmidrule(lr){2-4}\cmidrule(lr){5-7}\cmidrule(lr){8-10}\cmidrule(lr){11-13}\cmidrule(lr){14-16}%
}

\begin{table*}[!t]
\footnotesize
\glsreset{ac}
\caption[Evaluation results of \gls{mllm}-based \gls{cua} \gls{gui} verification]{Effectiveness results for \gls{mllm}-based \gls{cua} \gls{gui} verification for detecting \textit{met}, \textit{unmet}, and \textit{partially met} classes in \gls{gui} applications across \textit{Requirements (Req)} and \textit{\gls{ac}} in five apps with \textit{precision (P)}, \textit{recall (R)}, and \textit{F1-measure (F1)}.}
\centering

\setlength\tabcolsep{2.7pt}      
\renewcommand{\arraystretch}{1.25}

\begin{tabular}{c|ccc|ccc|ccc|ccc|ccc}
\toprule
\multicolumn{1}{c|}{} &
\multicolumn{3}{c|}{\textbf{Met (Req)}} &
\multicolumn{3}{c|}{\textbf{Unmet (Req)}} &
\multicolumn{3}{c|}{\textbf{Partial (Req)}} &
\multicolumn{3}{c|}{\textbf{Met (AC)}} &
\multicolumn{3}{c}{\textbf{Unmet (AC)}} \\
\cmidrule(lr){2-4}\cmidrule(lr){5-7}\cmidrule(lr){8-10}\cmidrule(lr){11-13}\cmidrule(lr){14-16}
\multicolumn{1}{c|}{} &
\textbf{P} & \textbf{R} & \textbf{F1} &
\textbf{P} & \textbf{R} & \textbf{F1} &
\textbf{P} & \textbf{R} & \textbf{F1} &
\textbf{P} & \textbf{R} & \textbf{F1} &
\textbf{P} & \textbf{R} & \textbf{F1} \\
\midrule

\textbf{App-1}  & 78.9 & 93.8 & 85.7 & 100.0 & 71.4 & 83.3 & 50.0 & 42.9 & 46.2 & 90.8 & 96.7 & 93.7 & 92.0 & 79.3 & 85.2 \\
\rowcolor{lightgray}
\textbf{App-2}  & 100.0 & 94.4 & 97.1 & 100.0 & 100.0 & 100.0 & 83.3 & 100.0 & 90.9 & 96.8 & 98.4 & 97.6 & 95.8 & 92.0 & 93.9 \\
\textbf{App-3}  & 93.3 & 82.4 & 87.5 & 83.3 & 83.3 & 83.3 & 62.5 & 83.3 & 71.4 & 93.1 & 90.0 & 91.5 & 79.3 & 85.2 & 82.1 \\
\rowcolor{lightgray}
\textbf{App-4}  & 100.0 & 82.4 & 90.3 & 75.0 & 85.7 & 80.0 & 50.0 & 66.7 & 57.1 & 94.6 & 89.8 & 92.2 & 82.4 & 90.3 & 86.2 \\
\textbf{App-5}  & 85.0 & 100.0 & 91.9 & 100.0 & 85.7 & 92.3 & 66.7 & 40.0 & 50.0 & 90.9 & 100.0 & 95.2 & 100.0 & 77.8 & 87.5 \\

\EffSepRule

\rowcolor{lightgray}
\textbf{Avg.}   & 91.5 & 90.6 & 90.5 & 91.7 & 85.2 & 87.8 & 62.5 & 66.6 & 63.1 & 93.3 & 95.0 & 94.0 & 89.9 & 84.9 & 87.0 \\
\textbf{SD}     &  9.3 &  7.9 &  4.4 & 11.8 & 10.2 &  8.2 & 13.8 & 25.8 & 18.3 &  2.6 &  4.8 &  2.5 &  8.8 &  6.4 &  4.3 \\
\bottomrule
\end{tabular}
\label{tab:gui_spector_rq1}
\end{table*}

\subsection{RQ$_1$: Effectiveness of \gls{mllm} \gls{cua} for \gls{gui} Verification}

\begin{figure*}[!t]
  \centering
  \includegraphics[width=\textwidth]{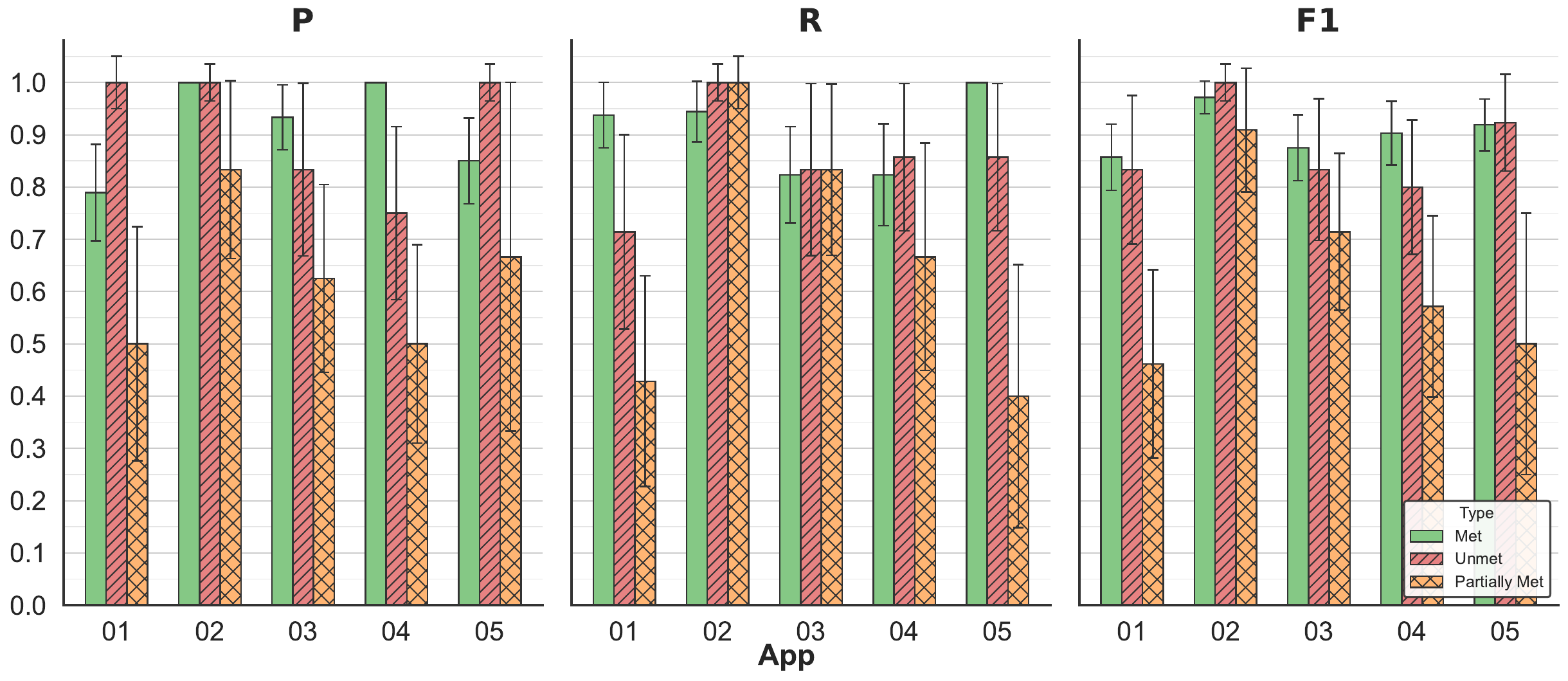}

  \caption[Bar charts for \gls{mllm}-based \gls{cua} \gls{gui} verification effectiveness on requirements level]{Bar charts (error bars represent one standard deviation) for automated \gls{mllm}-based \gls{cua} \gls{gui} verification on the requirements level across \textit{met} (\textit{green}), \textit{unmet} (\textit{red}), and \textit{partially met} (\textit{orange}) class for metrics \textit{precision (P)}, \textit{recall (R)}, \textit{F1-measure (F1)}.}
  \label{fig:gui_spector_boxplot_rq_1_1}

  \vspace{0.4cm}

  \includegraphics[width=\textwidth]{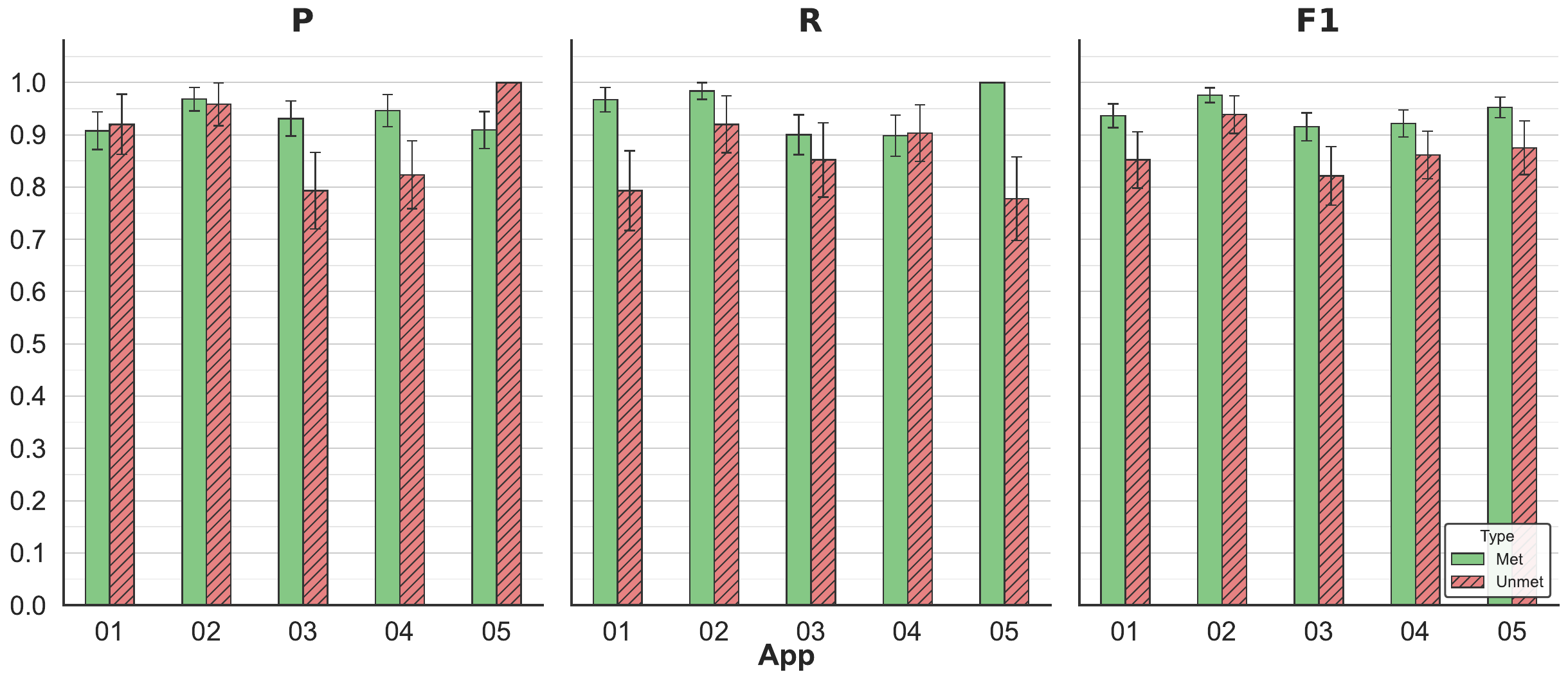}
  \glsreset{ac}
  \caption[Bar charts for \gls{mllm}-based \gls{cua} \gls{gui} verification effectiveness on \gls{ac} level]{Bar charts (error bars represent one standard deviation) for automated \gls{mllm}-based \gls{cua} \gls{gui} verification on the \gls{ac} level across \textit{met} (\textit{green}) and \textit{unmet} (\textit{red}) class (\textit{binary}) for metrics \textit{precision (P)}, \textit{recall (R)}, \textit{F1-measure (F1)}.}
  \label{fig:gui_spector_boxplot_rq_1_2}
  \vspace{0.2cm}
\end{figure*}

Table \ref{tab:gui_spector_rq1} shows a summary of per-app and average results regarding the effectiveness of \gls{mllm}-based \gls{cua} \gls{gui} verification across \gls{nlr} and \gls{ac}. In addition, Figure \ref{fig:gui_spector_boxplot_rq_1_1} illustrates respective bar charts for verification results on the requirements level, and Figure \ref{fig:gui_spector_boxplot_rq_1_2} depicts bar charts for verification results on the \gls{ac} level. For the \gls{ac} binary classification task, both classes \textit{met} (\textit{F1}=94.0) and \textit{unmet} (\textit{F1}=87.0) achieve high scores on average across the five considered applications, indicating that our approach reliably identifies fulfillment and violations at the \gls{ac} level. Considering the effectiveness on individual applications, it can be observed that \textit{App-3} (\textit{Recipe Generator}) performs worst (with the lowest \textit{F1} scores for \textit{met} (\textit{F1}=91.5) and \textit{unmet} (\textit{F1}=82.1)), while \textit{App-2} (\textit{Budget Tracker}) performs best (with the highest \textit{F1} scores for \textit{met} (\textit{F1}=97.6) and \textit{unmet} (\textit{F1}=93.9)).

\begin{figure*}[!t]
  \centering
 \includegraphics[width=1\textwidth]{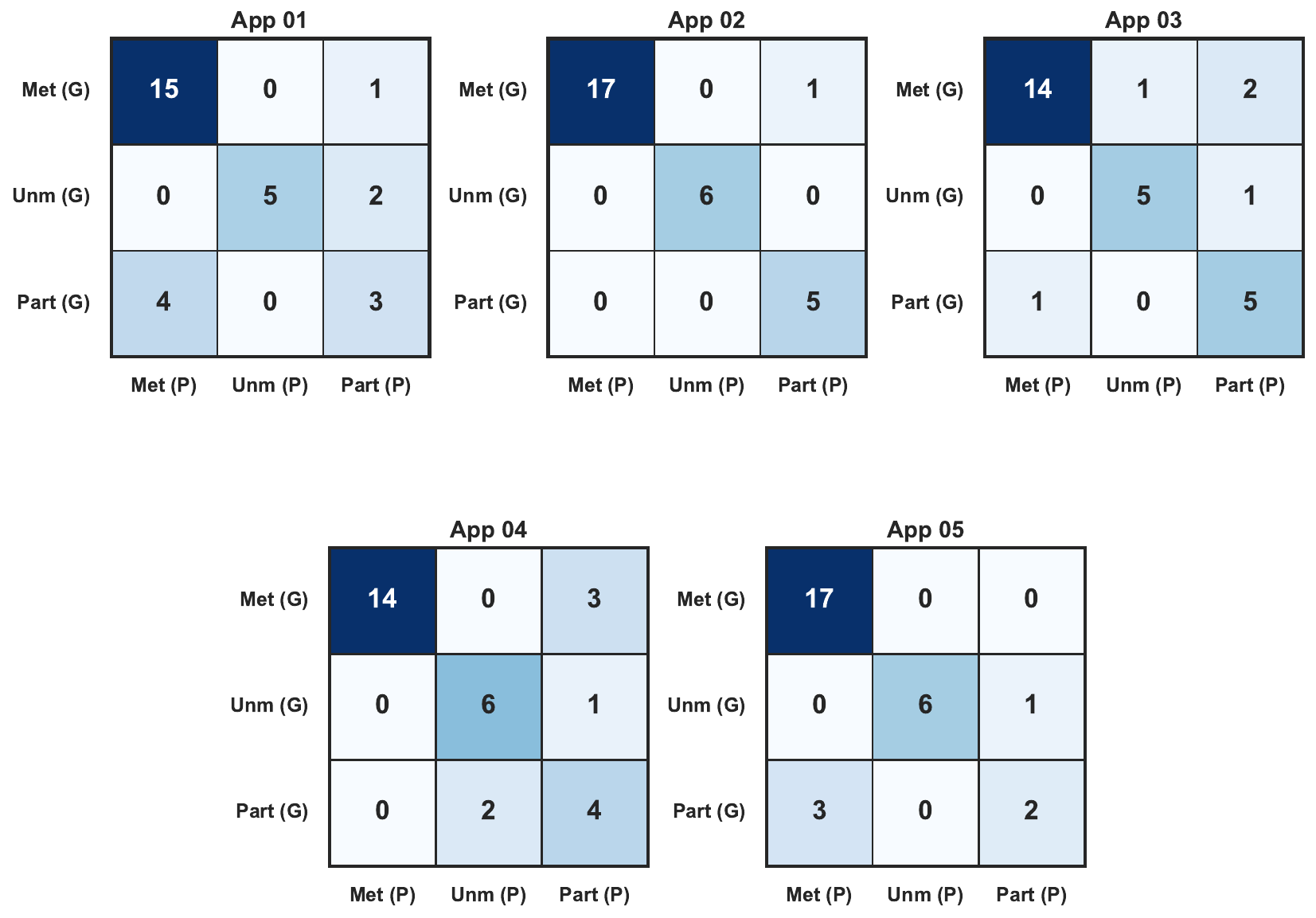}
  \caption[Confusion matrices for \gls{mllm}-based \gls{cua} \gls{gui} verification on the requirements level]{Confusion matrices for \gls{mllm}-based \gls{cua} \gls{gui} verification on the requirements level across five apps and three classes \textit{met}, \textit{unmet (Unm)}, and \textit{partially met (Part)}, showing highest error rate on the \textit{partially met} class. G = Gold standard, P = Prediction.}
  \label{fig:gui_spector_confusion_matrix_rq_1}
  \vspace{-0.2cm}
\end{figure*}

For the three-class requirements task, \textit{F1} scores remain high for \textit{met} (\textit{F1}=90.5) and \textit{unmet} (\textit{F1}=87.8) on average across the five considered applications, whereas \textit{partially met} is slightly lower (\textit{F1}=63.1), which may result from boundary confusions due to greater semantic ambiguity. For example, for the \textit{met} and \textit{unmet} classes on the requirements level, a single \gls{fp} prediction by the model on the \gls{ac} level leads to an erroneous prediction on the requirements level (also \gls{fp}), rendering this the most fragile class of the three. Figure \ref{fig:gui_spector_confusion_matrix_rq_1} illustrates the confusion matrices on the requirements level and across all five apps. In particular, it highlights the highest error rate on the \textit{partially met} class, which is often erroneously predicted by the model (reducing \textit{precision}, rightmost column in the confusion matrices) or erroneously not recognized (reducing \textit{recall}, lowest row in the confusion matrices). Overall, the obtained results in our experimental evaluation across both \gls{ac} and requirements levels indicate that our \gls{mllm}-based \gls{cua} \gls{gui} verification approach possesses high effectiveness.

\begin{figure*}[!t]
  \centering
 \includegraphics[width=1\textwidth]{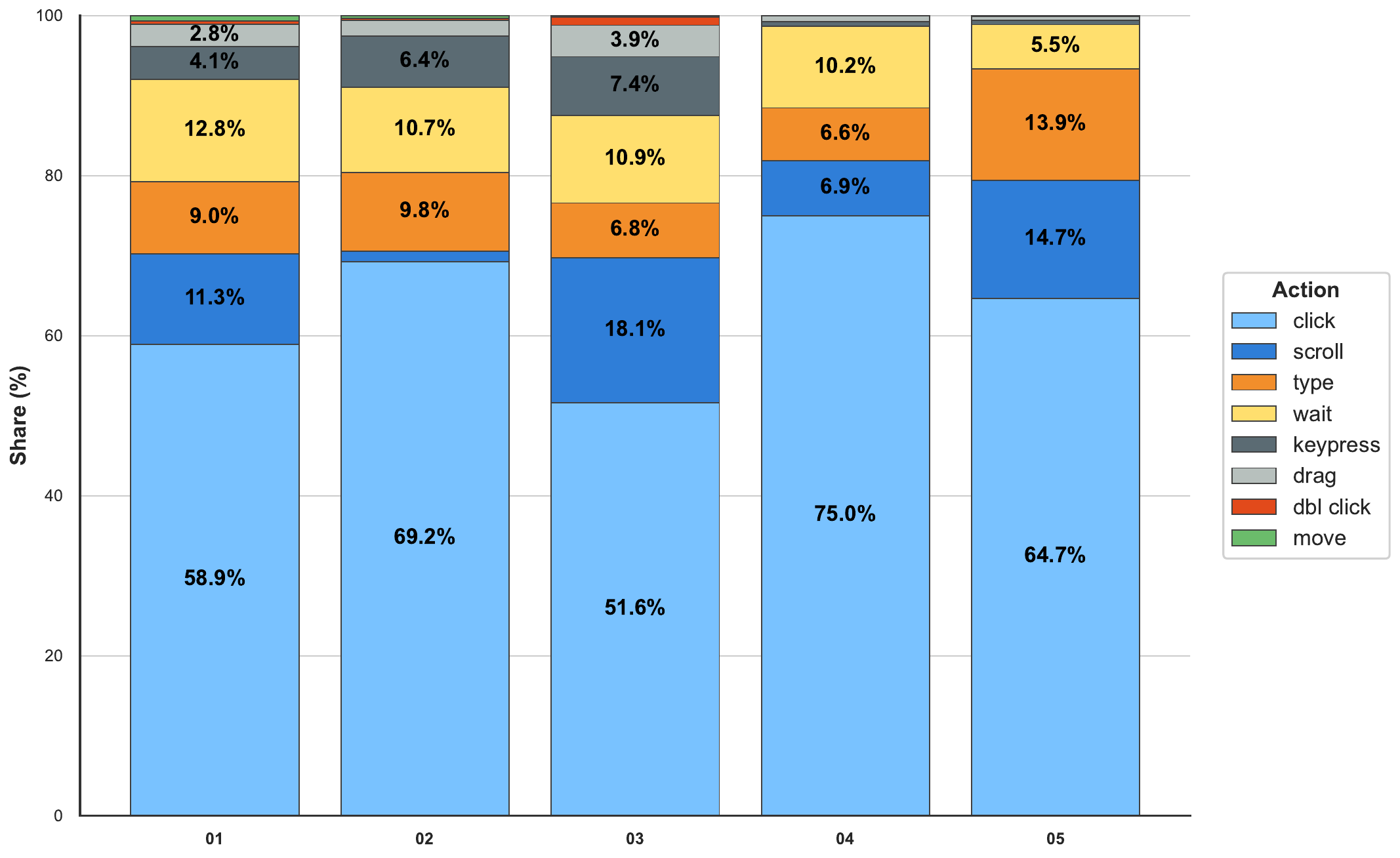}
  \caption[Stacked bar charts with predicted \gls{gui} action distributions across apps]{Stacked bar charts showing the distributions of the predicted and executed \gls{mllm}-based \gls{cua} \gls{gui} actions (including \textit{click}, \textit{scroll}, \textit{type}, \textit{keypress}, \textit{drag}, \textit{double click}, \textit{move}, and \textit{wait}) across the five \gls{gui} apps considered in our experimental evaluation runs.}
  \label{fig:gui_spector_action_dist}
  \vspace{-0.4cm}
\end{figure*}

Next, we conducted an additional analysis of the trajectories obtained during the verification runs over the requirements in our gold standard to better understand the behavior of the \gls{mllm}-based \gls{cua} \gls{gui} verification approach. Figure \ref{fig:gui_spector_action_dist} shows a series of stacked bar charts with distributions of the predicted and executed \gls{mllm}-based \gls{cua} \gls{gui} actions across all five applications. Notably, as expected, the most frequently predicted \gls{gui} action is \textit{click}, which represents one of the most common interaction mechanisms on \glspl{gui} overall and is also supported by many different \gls{gui} components. The second most frequent action is represented by \textit{scroll}. While the action \textit{click} is also commonly utilized to conduct high-level navigation in applications (via the navigation drawer or menu), \textit{scroll} is required to access the entire content on a fixed, local page. However, \textit{App-2} represents an exception since \textit{scroll} is among the least frequently utilized actions, representing that vertical or horizontal scrolling was not necessary and the content was entirely visible within the window. The third most frequently predicted action on average is \textit{wait}, which is utilized, for example, when content was not yet rendered in the previous \gls{gui} state and therefore no meaningful reasoning could be generated. Subsequently, \textit{type} and \textit{keypress} actions follow, which are often required for providing custom input data, but less common compared to clickable \gls{gui} components. Actions \textit{drag}, \textit{double click}, and \textit{move} were the least frequently predicted and often only required for very special cases (such as drag-and-drop of list items or calendar entries).

\begin{figure*}[!t]
  \centering
 \includegraphics[width=1\textwidth]{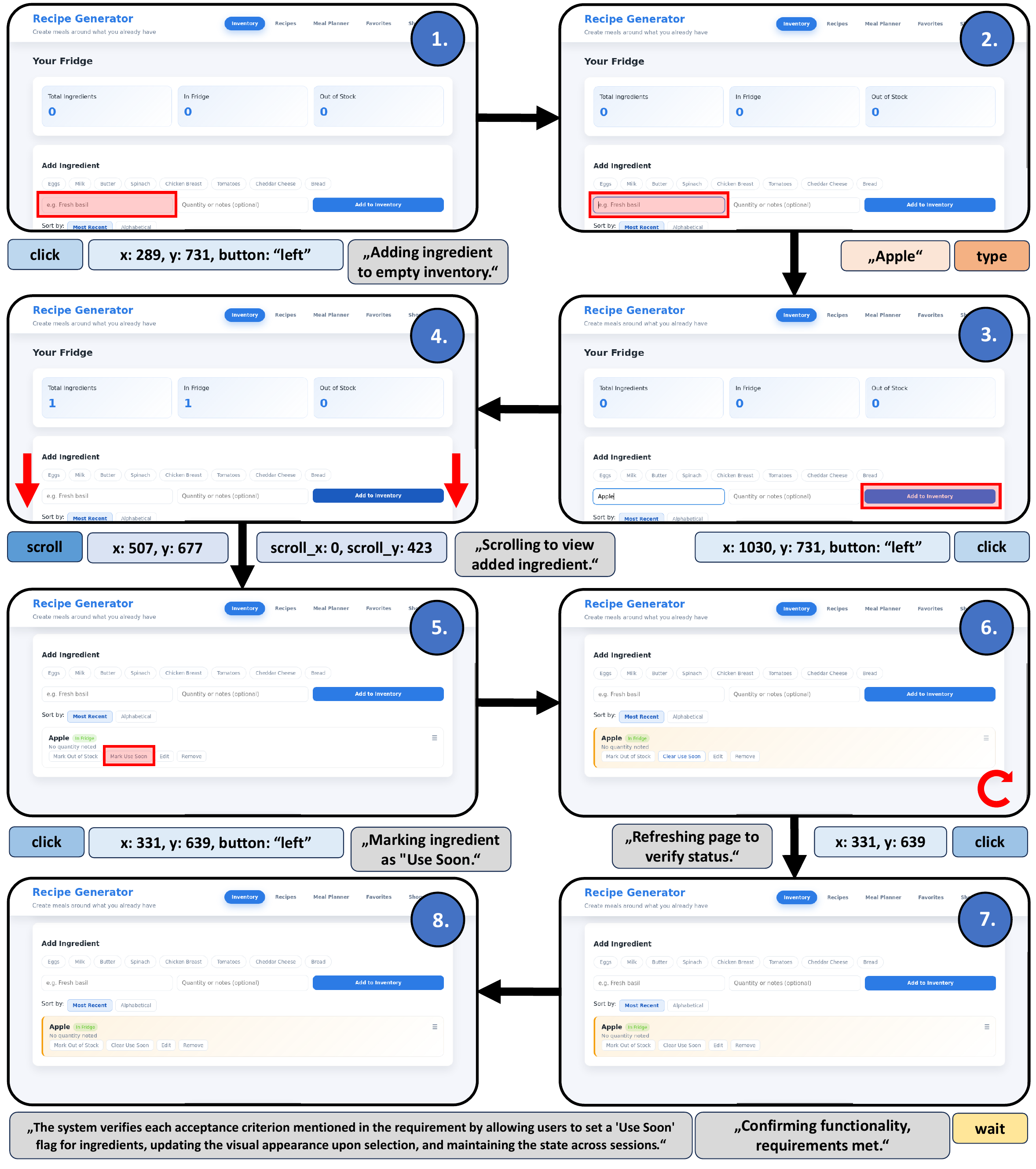}
  \caption[Example trajectory extracted from the gold standard run]{Example trajectory extracted from the gold standard run encompassing \gls{gui} states (i.e. screenshots), optional reasoning, and parametrized \gls{gui} actions from \textit{App-3} (\textit{Recipe Generator}) and the \gls{nlr}: \textit{``The system shall enable users to mark ingredients with a 'Use Soon' flag for prioritization.[sic]''} with \gls{ac}$_1$: \textit{``Ingredient cards provide a 'Use Soon' control.[sic]''}, \gls{ac}$_2$: \textit{``Activating the control highlights the ingredient with a distinct visual style.[sic]''}, and \gls{ac}$_3$: \textit{``The 'Use Soon' state is saved and restored via browser storage.[sic]''}.}
  \label{fig:gui_spector_example_trajectory}
\end{figure*}

Moreover, in Figure \ref{fig:gui_spector_example_trajectory}, we provide an example trajectory extracted from an experiment run over the gold standard, particularly for the high-level \gls{nlr} of being able to flag an ingredient with \textit{Use Soon} with three \gls{ac}. In addition, we provide annotations for the predicted parametrized actions at each state and optional reasoning. First, the agent clicks on the \textit{text input field} to add an ingredient to the inventory list (as indicated by the reasoning). After the \textit{text input field} is focused, the agent types \textit{``Apple''} and clicks on the \textit{Add to Inventory} button. Next, since the current view does not fully capture the entire page content, the agent predicts a downward \textit{scroll} action, identifies and clicks the \textit{Mark Use Soon} button on the created ingredient card (\gls{ac}$_1$), identifies the distinct visual style (\gls{ac}$_2$), and confirms the state after refreshing the page (\gls{ac}$_3$), therefore fulfilling all three \gls{ac}. At the end, the agent provides a detailed summary and gathered evidence per \gls{ac}.

\begin{myrqbox}
\textbf{Answer to RQ$_1$:} The proposed automated \gls{mllm}-based \gls{cua} \gls{gui} verification approach shows high effectiveness on the \gls{ac} level (\textit{met} (\textit{F1}=94.0) and \textit{unmet} (\textit{F1}=87.0)) and requirements level (\textit{met} (\textit{F1}=90.5), \textit{unmet} (\textit{F1}=87.8), and \textit{partially met} (\textit{F1}=63.1)).
\end{myrqbox}

\subsection{RQ$_2$: Cost for \gls{mllm}-based \gls{cua} \gls{gui} Verification}

\newcommand{\PerfSepRule}{%
  \cmidrule(lr){1-1}%
  \cmidrule(lr){2-3}\cmidrule(lr){4-5}\cmidrule(lr){6-7}\cmidrule(lr){8-9}\cmidrule(lr){10-11}%
}

\begin{table}[!t]
\footnotesize
\caption[Cost-related performance metrics for \gls{mllm}-based \gls{cua} \gls{gui} verification from \gls{nlr}]{Performance metrics across applications (averaged over requirements): \textit{\#steps}, \textit{time (s)}, \textit{token consumption} \textit{(k)}, and \textit{cost (\$)} (OpenAI \gls{cua}: \$3/M input, \$12/M output).}
\centering
\setlength\tabcolsep{7pt}
\renewcommand{\arraystretch}{1.15}

\begin{tabular}{c|cc|cc|cc|cc|cc}
\toprule
\multicolumn{1}{c|}{} &
\multicolumn{2}{c|}{\textbf{\#Steps}} &
\multicolumn{2}{c|}{\textbf{Time (s)}} &
\multicolumn{2}{c|}{\textbf{\#In-Tok. (k)}} &
\multicolumn{2}{c|}{\textbf{\#Out-Tok. (k)}} &
\multicolumn{2}{c}{\textbf{Cost (\$)}} \\
\cmidrule(lr){2-3}\cmidrule(lr){4-5}\cmidrule(lr){6-7}\cmidrule(lr){8-9}\cmidrule(lr){10-11}
\multicolumn{1}{c|}{} &
\textbf{Avg.} & \textbf{SD} &
\textbf{Avg.} & \textbf{SD} &
\textbf{Avg.} & \textbf{SD} &
\textbf{Avg.} & \textbf{SD} &
\textbf{Avg.} & \textbf{SD} \\
\midrule

\textbf{App-1}  & 25.1 & 17.4 & 336.0 & 291.2 & 229.8 & 165.5 & 2.394 & 1.323 & 0.689 & 0.029 \\
\rowcolor{lightgray}
\textbf{App-2}  & 21.1 & 14.9 & 322.3 & 316.0 & 189.7 & 127.8 & 1.984 & 0.817 & 0.569 & 0.024 \\
\textbf{App-3}  & 19.5 & 16.6 & 223.1 & 240.5 & 174.6 & 149.2 & 2.002 & 1.020 & 0.524 & 0.024 \\
\rowcolor{lightgray}
\textbf{App-4}  & 22.3 & 12.4 & 293.4 & 182.3 & 202.7 & 107.7 & 2.115 & 0.822 & 0.608 & 0.025 \\
\textbf{App-5}  & 34.2 & 19.8 & 412.9 & 266.7 & 311.9 & 188.0 & 2.850 & 1.399 & 0.936 & 0.034 \\

\PerfSepRule

\rowcolor{lightgray}
\textbf{Avg.}   & 24.4 & 16.2 & 317.6 & 259.3 & 221.7 & 147.6 & 2.269 & 1.076 & 0.665 & 0.027 \\
\textbf{SD}     &  5.8 &  2.8 &  68.8 &  51.4 &  54.3 &  31.4 & 0.364 & 0.274 & 0.163 & 0.004 \\
\bottomrule
\end{tabular}
\label{tab:gui_spector_performance_rq_2}
\end{table}

\begin{figure*}[!tbp]
  \centering
 \includegraphics[width=1\textwidth]{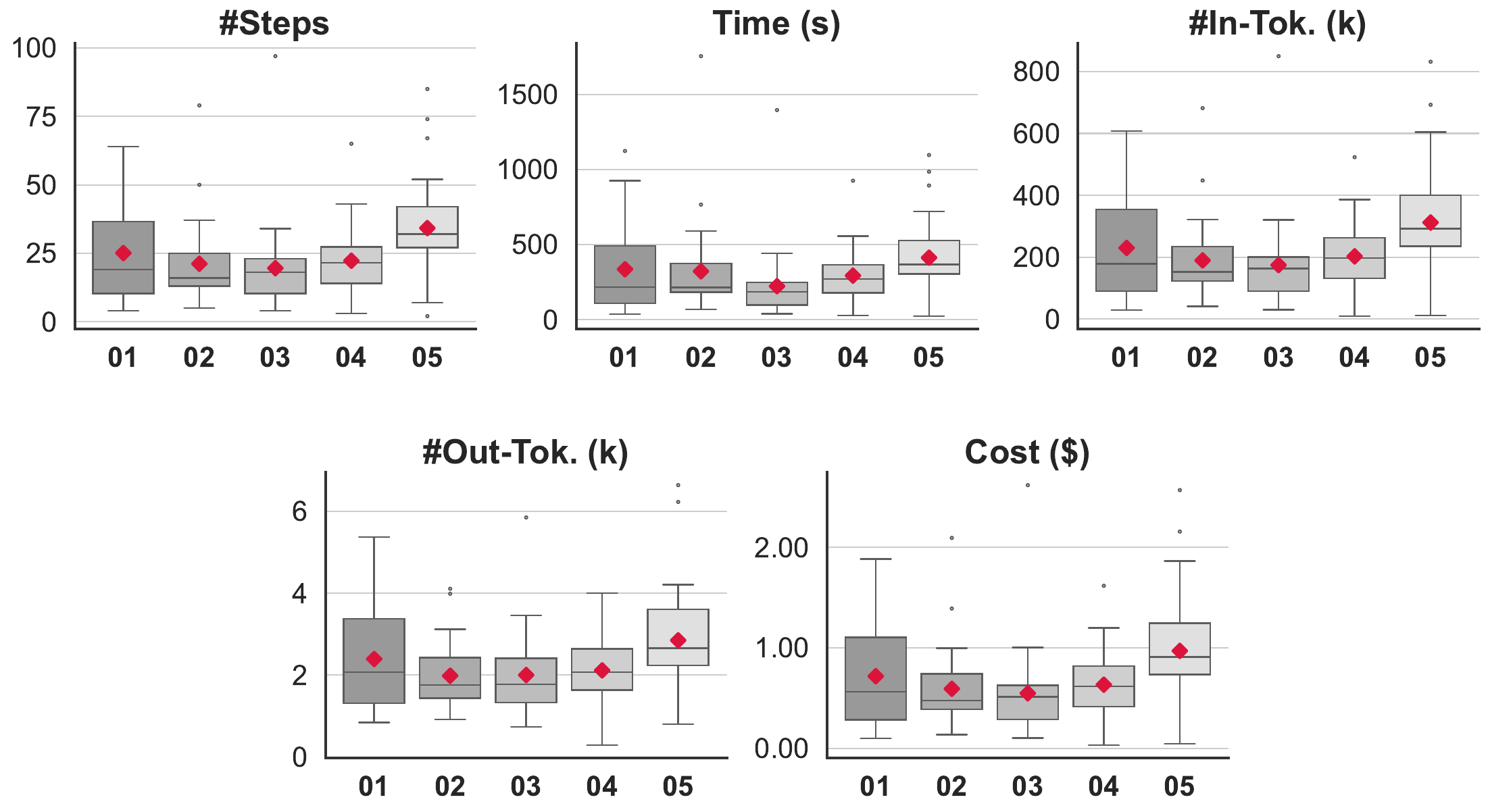}
  \caption[Boxplots for cost-related performance metrics]{Boxplots for the performance metrics across all five considered \gls{gui} applications (averaged over all requirements per application): \textit{\#steps}, \textit{time (s)}, \textit{token consumption} \textit{(k)}, \textit{cost (\$)} (OpenAI \gls{cua}: \$3/M input, \$12/M output), red triangles represent means.}
  \label{fig:gui_spector_performance_rq_2_boxplots}
  \vspace{-0.4cm}
\end{figure*}

To answer this research question, we gathered several cost-related metrics during the experiment runs shown in Table \ref{tab:gui_spector_performance_rq_2} and Figure \ref{fig:gui_spector_performance_rq_2_boxplots}. Execution effort varies by app (\textit{\#steps} from 19.5 for \textit{App-3} to 34.2 for \textit{App-5} on average). The observed spread is expected because each run starts from a clean state and certain apps require longer prerequisite flows before a requirement can be verified (e.g., always configuring a cleaning appointment prior to editing it). The elapsed \textit{time} per requirement likewise differs across apps for the same reasons. Although per-requirement runtime may, in some instances, exceed that of human testers, the framework can make up for it by supporting straightforward parallelization.
Costs are dominated by \gls{mllm} input tokens, with an average of \$0.67 per requirement verification. We expect reductions by merging related verification runs.

\begin{myrqbox}
\textbf{Answer to RQ$_2$:} The costs range from \$0.524 (min. average) to \$0.936 (max. average) and from 22.3 (min. average) to 34.2 (max. average) \#steps per requirement verification.
\end{myrqbox}

\section{Threats to Validity}

\paragraph{Internal Validity.} Due to the absence of any dataset combining \gls{gui} implementations with requirements, the dataset utilized in our evaluation relies on requirements, \gls{ac}, and \gls{gui} implementations generated by the \textit{Codex} \citep{codex2025} model, which might introduce bias and errors. Furthermore, the employed \gls{mllm} \gls{cua} is based on the \textit{\gls{gpt}-4o} \citep{gpt-4o} model. While the model used to create the dataset differs from the agent model used in our verification approach, they are from a similar model family (\textit{OpenAI} \gls{gpt}). However, the length of requirements and \gls{ac} indicate a good level of complexity. Moreover, two paper authors independently investigated the requirements for correctness, verified the requirements status in the implemented system, and resolved annotation conflicts, showing a very high \gls{iaa}, which indicates a high-quality annotation.

\paragraph{External Validity.} While our evaluation dataset encompasses five different applications from diverse domains, a larger dataset with even more applications from more domains could improve the generalizability of the results. Moreover, the requirements are generated by the \textit{Codex} \citep{codex2025} model. Therefore, the considered requirements in our evaluation might differ from real-world requirements, which potentially restricts generalizability. However, two paper authors independently investigated the requirements manually, which revealed a high level of quality and annotation correctness.

\section{Limitations}

While our approach enables automated verification of \gls{nlr} in highly dynamic \gls{gui} applications, it also carries substantial costs through the usage of an \gls{mllm}-based \gls{cua}. However, the long runtime of tests (up to several minutes on average across the evaluation dataset) can be mitigated by heavy parallelization through increasing the number of worker nodes in \textit{Celery}. Furthermore, the monetary (\gls{api}) costs entailed by employing a proprietary, state-of-the-art \gls{mllm}-based \gls{cua} might be considered high. However, running these tests manually with developers or writing automation scripts with developers also entails high costs. In addition, in this work, we focused on investigating the general potential and effectiveness of \gls{mllm}-based \gls{cua} for \gls{gui} verification, while we neglected optimization. However, we observed high potential for efficiency improvements: many requirements encompass similar sub-trajectories, which could be exploited for efficient merges, which could save a large number of verification steps. Moreover, an external \gls{mllm} memory could optimize the predicted action sequences for efficiency.
\section{Related Work}

Since we already discussed related work regarding testing and verification of \gls{gui} applications previously (see Chapter \ref{cha:interlinking} Section \ref{sec:interlinking_rel_work}), we provide a brief summarization. Specifically, many \gls{gui} testing approaches have been proposed before in research including automated test scripts \citep{xie2007designing}, random exploration \citep{mao2016sapienz}, model-based techniques \citep{zeng2016automated, gu2019practical}, and scripted bug replay methods \citep{gomez2013reran, feng2022gifdroid}. However, with our proposed \guispector approach, we enable \gls{nlr}-based verification of \gls{gui} applications, while previous methods focused on identifying runtime failures (system crashes). Therefore, our approach requires profound \gls{nlu} capabilities to process provided \gls{nlr} and effectively run verifications in \gls{gui} applications. While recent \gls{llm}-based methods have been proposed that run tests based on \gls{nl} descriptions \citep{feng2024enabling, feng2025agent}, they rely on textual \gls{gui} hierarchy abstractions, which oversimplify the \gls{gui} state and limit the capabilities for detecting requirement violations related to visual and dynamic aspects. In addition, with the proposed \guispector approach, to the best of our knowledge, we are the first to enable fully autonomous \gls{gui} implementation-verification loops for \gls{llm}-based programming agents with an \gls{mcp} server and provide a user-friendly tool.
\section{Conclusion}

The work presented in this chapter was driven by challenge \challtwotwo{}: \textit{How can \gls{nlr}-based verification be automated for highly dynamic, fully-fledged \gls{gui} applications?} To tackle this question, in this work, we proposed \guispector, a novel \gls{mllm}-based \gls{cua} \gls{gui} verification approach capable of processing complex \gls{nlr}. Moreover, we provide a tool prototype implementation of the approach and an \gls{mcp} server, which enables fully autonomous implementation-verification loops for \gls{llm}-based programming agents. Finally, the experimental evaluation indicates the high effectiveness of our approach, achieving high verification accuracy across \gls{ac} and requirements level verification tasks.

\part{\sc{Discussion and Conclusion}}
\label{part:discussion}
\clearpage
\newpage
\thispagestyle{empty}
\null  

\chapter{Discussion}
\label{cha:discussion}

While we presented our solution approaches for the different challenges for automated \gls{nlr}-based \gls{gui} prototyping and verification in previous chapters, in this chapter, we provide a brief discussion about interesting aspects spanning the entire work of the thesis.

\section{\gls{gui} Retrieval and \gls{gui} Generation Trade-offs}

To address the challenge of automated \gls{gui} prototyping from \gls{nlr}, we initially proposed \gls{gui} retrieval approaches (see Chapter \ref{cha:nl_gui_retrieval}, \ref{cha:gui_rerank}, and \ref{cha:self_elicitation}), followed by \gls{llm}-based \gls{gui} generation approaches (see Chapter \ref{cha:zs_gui_generation}, \ref{cha:closing}, and \ref{cha:guide}). This raises the question whether one approach is generally better and should be preferred with respect to \gls{nlr}-based \gls{gui} prototyping. While retrieval approaches benefit from the low computational cost and speed (low latency), enabling the computation of millions of relevance scores within seconds (although with less representational power for both \gls{nlr} and \gls{gui} prototypes), they are restricted to the content of associated \gls{gui} repositories, which are static \gls{gui} representations untailored to arbitrary, custom \gls{nlr}. In contrast, this represents a crucial advantage of \gls{llm}-based \gls{gui} generation approaches, which --- given the impressive \gls{nlu} capabilities of modern \glspl{llm} --- can adjust and directly tailor generated \gls{gui} prototypes to particular details formulated in the requirements. However, both retrieval and generation approaches are inherently limited regarding their potential domain coverage, given their data-driven nature (\gls{gui} repository for retrieval and pretraining data for \glspl{llm}), while \gls{llm}-based generation possesses higher adaptation capabilities.

However, we argue that retrieval-based approaches are particularly beneficial as \gls{gui} prototyping support in domains covered well by the underlying \gls{gui} repository, since these approaches can retrieve and show relevant \gls{gui} prototypes rapidly and enable immediate comparison of many similar \glspl{gui}, which potentially improves elicitation and helps to identify additional relevant requirements. While \gls{llm}-based approaches could be parallelized to generate many variations, this entails much higher cost and still carries latency. However, for requirements that are less commonly known or represent many deviations from more standardized \glspl{gui}, \gls{llm}-based approaches should be preferred. An \gls{llm}-based approach could be created to decide which approach to utilize per requirement. In addition, \gls{gui} retrieval and \gls{llm}-based generation approaches can also be effectively integrated. In particular, we presented the \gls{ragg} approach (see Chapter \ref{cha:zs_gui_generation}), which enables the incorporation of external \gls{gui} prototyping knowledge into the \gls{llm}, which is especially helpful in domains not well covered in the pretraining data of the \gls{llm}. Moreover, retrieval-based methods could be utilized initially to discover broadly relevant \gls{gui} prototypes, which are subsequently adapted with an \gls{llm}. To summarize, \gls{gui} retrieval and \gls{gui} generation approaches both have advantages and disadvantages, which should be carefully considered and weighed when choosing an appropriate approach.

\vspace{-0.2cm}
\section{\gls{llm}-based Approaches Augment Human Capabilities}
\label{discussion:human}
\vspace{-0.1cm}
While we presented several approaches for \gls{gui} prototyping automation, we focused on integrating \glspl{llm} into existing \gls{gui} prototyping workflows to augment the capabilities of requirements analysts and prototype developers (such as approaches in Chapter \ref{cha:closing} and \ref{cha:guide}). Particularly, we focused on reducing labor-intensive, manual tasks, enabling analysts and prototype developers to focus more on the actual creative aspects. Moreover, while \gls{llm}-based approaches encompass tremendous \gls{nlu} capabilities, they still produce erroneous outputs, emphasizing the importance of close collaboration between humans and \glspl{llm} for \gls{gui} prototyping. With approaches such as presented in Chapter \ref{cha:closing}, we provide a framework that integrates \glspl{llm} closely into \gls{gui} prototyping environments, while ensuring human control in the generation process and for subsequent \gls{gui} editing.

\vspace{-0.2cm}
\section{Cost-Effectiveness of \gls{llm}-based Approaches}
\vspace{-0.1cm}
While \gls{llm}-based approaches for automated \gls{gui} prototyping (such as \gls{mllm}-based \gls{gui} reranking, feature prediction, and \gls{zs}-based \gls{gui} generation) and verification are highly effective, they also entail high costs, as shown across several evaluations, including long \textit{runtime} (especially for verification) and high \textit{monetary cost}. In particular, \gls{api} costs play a crucial role, since state-of-the-art \glspl{llm} are typically proprietary models with dedicated, paid \glspl{api}. However, the gains in effectiveness across different tasks are also high, therefore requiring careful consideration when choosing an \gls{llm}-based approach. This highlights the importance of research for effective \glspl{slm}.

\vspace{-0.2cm}
\section{Evaluation Challenges for \gls{gui} Prototyping Approaches}
\vspace{-0.1cm}
One crucial methodological challenge for evaluating automated \gls{gui} prototyping approaches is the subjectivity encompassed in deciding whether \gls{gui} prototypes are \textit{relevant}, \textit{accurate}, or \textit{good} with respect to a provided \gls{nlr}. While certain aspects are more objective (such as the number of \gls{gui} components contained in a \gls{gui}), others are more subjective (such as requirements that are expressed through that \gls{gui}). To embrace and mitigate this subjectivity when rating \glspl{gui}, we conducted different types of evaluations (such as relevance annotations with \gls{ir} metrics, user studies, and Likert scale perception) and included large numbers of distinct annotators, typically via high-quality crowdsourcing.

\vspace{-0.2cm}
\section{Hallucinations in LLM-based Approaches}
\vspace{-0.1cm}
Hallucinations represent a popular issue and central reliability risk occurring in \glspl{llm}. In particular, hallucinations refer to generated \gls{llm} output that appears fluent and plausible while being factually incorrect or not grounded in the relevant sources \citep{ji2023survey, huang2025survey, kalai2025language}. Prior research differentiates between \textit{faithfulness}, referring to model responses that are not supported by or contradict the provided input material, and \textit{factuality}, referring to generated outputs that are factually incorrect with respect to external world knowledge \citep{ji2023survey}. 

In this thesis, we proposed several approaches that are built upon and rely heavily on \glspl{llm}, which renders hallucinations as a relevant issue to be considered. In general, the investigated tasks and proposed \gls{llm}-based approaches differ from many typical settings that are particularly prone to hallucinations. First, the models are not primarily employed for factual question answering with an objectively correct output, where the validity of the generated model response depends on fetching facts from the internal model parameters. In contrast, the answers are often grounded through concrete artifacts such as \glspl{gui}, requirements, or design specifications. In addition, many of the considered tasks, such as \gls{gui} relevance judgment given an \gls{nlr}, are inherently subjective and therefore lack an objectively correct answer (also multiple human raters would judge differently). Moreover, our approaches often include validation steps (e.g., automatically validating if generated \textit{Material Design} component configurations adhere to the available components and represent valid configurations) or emphasize human-in-the-loop (see Section \ref{discussion:human}), ensuring that generated outputs are verified by humans. In the following, we provide more detailed descriptions for each of the \gls{llm}-based approaches.

For \textit{\gls{gui}-ReRank} (see Chapter \ref{cha:gui_rerank}), hallucinations could mainly occur in the \gls{gui} description generation phase or while decomposing the requirements. For example, the generated descriptions might include functionality which is not present in the \gls{gui}. However, in this task, the \gls{llm} output is directly grounded to the \gls{gui} screenshots, rather than retrieving facts from the internal model parameters. In addition, we manually inspected a random sample of the descriptions and requirements decompositions, which did not reveal hallucinated fragments. In contrast, \gls{gui} relevance judging or reranking represents a subjective task, not possessing an objectively correct answer. Therefore, we do not consider deviations from the ground truth as hallucinations.

In \textit{\gls{ser}\gls{gui}} (see Chapter \ref{cha:self_elicitation}), hallucination risks are primarily related to the \gls{llm}-based \gls{gui} feature recommendations, given the \gls{nlr}, an initially selected \gls{gui}, and already specified \gls{gui} features. Since these recommendations are intended to stimulate \gls{rel} with the user rather than automatically creating final \gls{gui} specifications, unsupported or less relevant recommendations are mitigated through human-in-the-loop. In addition, the recommended \gls{gui} features are grounded through the provided context. Moreover, the recommendation task considered is also subjective in nature. During the user study, we did not observe any recommendations that were entirely out of scope.

For the proposed \gls{zs} approaches for \gls{gui} generation (see Chapter \ref{cha:zs_gui_generation}), the hallucination risks lie mainly in whether the created \gls{gui} prototypes are sufficiently grounded in the \gls{nlr}. For example, the employed \gls{llm} might introduce \gls{gui} elements or design choices that have not explicitly been stated in the \gls{nlr}. However, since our approach focuses particularly on the elicitation phase, exploring potential requirements options and design choices derived by the \gls{llm} are valuable for stimulation. In addition, we considered abstracted, high-level \gls{nlr}, which apparently forces the \gls{llm} to rely on assumptions for generating concrete low-level \gls{gui} prototypes. However, since the prototypes are intended to be employed in discussions with stakeholders, we would incorporate human feedback and mitigate \gls{llm} assumptions. Moreover, the proposed techniques provide grounding for the generation process (e.g., \gls{ragg}, which provides concrete \gls{gui} screenshots relevant to the \gls{nlr} in the context), and we evaluated the prototypes with human judgments, trying to assess the subjective nature of \gls{gui} prototype quality.

For our proposed \textit{Figma} \gls{gui} prototyping support approaches (see Chapter \ref{cha:closing} and \ref{cha:guide}), the hallucination risks mainly concern the correct generation of \textit{Material Design} components and automatic derivation of additional requirements from high-level descriptions. For example, an \gls{llm} might create component configurations with attributes, icons, and layouts that are not supported by the employed component library. In addition, the implemented components might not adequately represent the intended functionality given the \gls{nlr}. For the requirements derivation, the \gls{llm} could propose low-level requirements irrelevant to the high-level description. However, to mitigate these risks, we constrain the generation process with multiple strategies. For example, we provide the \textit{Material Design} component library specification via a two-stage \gls{rag} approach and existing \gls{gui} context to the \gls{llm}, enabling grounding of the responses. Moreover, we automatically validate the generated components against the specification, restricting the \gls{llm} output to solely valid component configurations. Finally, we propose a human-in-the-loop approach, which enables developers or designers to rapidly verify and adapt the created components or derived requirements.

Considering \gls{llm}-based static \gls{gui} verification (i.e. predicting the implementation status of a \gls{us} in a \gls{gui} and extracting \gls{gui} components, see Chapter \ref{cha:interlinking}), the main hallucination risks lie in false predictions (e.g., classifying a \gls{us} as implemented while it is not fulfilled in the \gls{gui} or extracting false \gls{gui} components). While both tasks involve a higher degree of factuality compared to tasks discussed earlier, the tasks still encompass a certain level of ambiguity or subjectivity. In addition, the \gls{llm} responses are grounded in the provided context of the \gls{gui} and the \gls{us}. While for higher temperature settings some \gls{cot} matching approaches created random output sequences (resulting in \textit{F1}=0), we did not observe hallucinations for the rest of the evaluation dataset. Errors stem mainly from underspecified textual \gls{gui} representations. In addition, this approach is constructed again as human-in-the-loop, ensuring that humans conduct output verification.

Finally, in \textit{\gls{gui}Spector} (see Chapter \ref{cha:agent}), the primary hallucination risks concern unsupported \gls{nlr} verification predictions. Hallucinations could occur when the \gls{mllm} agent runs trajectories in the apps irrelevant to the \gls{nlr}. However, the \gls{llm} verification result is grounded in specific \gls{gui} screenshots, executed actions, and overall evidence collected during the verification run. To mitigate hallucination risks, \textit{\gls{gui}Spector} follows a human-in-the-verification-loop approach, allowing users to inspect verification runs.

\chapter{Conclusion}
\label{cha:conclusion}

\gls{gui} prototyping represents an important activity for stimulating \gls{rel} and ensuring that captured requirements reflect the actual needs of stakeholders (\gls{rval}) via tangible, visual prototypes. However, while effective, traditional \gls{gui} prototyping approaches are time-consuming and costly. Furthermore, verifying that the implemented \gls{gui} adheres to its specified requirements is another related challenge. While critically important, traditional \gls{gui} testing methods lack the ability to process complex \gls{nlr}. The leading challenge behind automating these activities is the representation of the complex, unrestricted \gls{nlr} on the one hand --- which requires strong \gls{nlu} capabilities --- and the similarly complex \gls{gui} prototypes, encompassing multi-dimensional aspects and being highly dynamic. With the advent of powerful \glspl{llm}, which are effective across various \gls{nlp} tasks, many automation approaches became feasible. Therefore, in this work, we explored the challenges of automated \gls{nlr}-based \gls{gui} prototyping and verification mainly with \gls{llm}-based approaches. Specifically, we addressed the following seven sub-challenges:

\vspace{0.3cm}
\noindent
(\challoneone{}) \textit{\gls{nlr}-based \gls{gui} Retrieval Gap:} when considering automation of \gls{gui} prototyping from \gls{nlr}, an initial solution approach is represented by employing \gls{ir} methods. While existing research has proposed several \gls{gui} retrieval approaches (such as \textit{Swire} \citep{huang2019swire} and \textit{GUIFetch} \citep{behrang2018guifetch}), they largely focused on different input representations (such as \gls{gui} images and hand-drawn sketches), while neglecting \gls{nlr}. Therefore, employing traditional \gls{ir} methods (such as \gls{tfidf} and \gls{bm25}) provides potential. However, \gls{gui} prototypes require an effective textual representation for enabling \gls{ir} methods. Moreover, there is a gap between \gls{nlr} and textual representations of \gls{gui} prototypes, which represents a challenge for traditional text-based \gls{ir} scoring models. To tackle this challenge, we proposed a \gls{bert}-\gls{ltr} model, which we trained and finetuned on \gls{gui} relevance data with respect to \gls{nlr} obtained through crowdsourcing platforms. By leveraging the \gls{plm} \gls{bert} --- with its state-of-the-art \gls{nlu} capabilities before the advent of \glspl{llm} --- we effectively reduced the gap between text representations of \gls{nlr} and \gls{gui} prototypes, as indicated by the effectiveness improvements on the novel \gls{gui} retrieval gold standard. Furthermore, the created dataset enabled the training of a model that learned improved relevance scoring for the particular \gls{gui} problem at hand.

\vspace{0.3cm}
\noindent
(\challonetwo{}) \textit{Semantic Representation Gap in \gls{gui} Retrieval and Reranking:} while the previously discussed \gls{bert}-\gls{ltr} models reduced the gap between the \gls{nlr} and textual \gls{gui} prototype representations to improve \gls{gui} ranking effectiveness, both representations rely on simplifications and therefore cannot fully capture their semantics. In particular, the \gls{gui} text representations utilized in the previous approach are based on an extraction method exploiting the \gls{gui} hierarchy data. While this method already enhances the text representation compared to relying solely on displayed text, it simultaneously neglects the multi-dimensional characteristics of \gls{gui} prototypes, such as functional and non-functional aspects, including \gls{gui} components and their inter-relationship, displayed texts, icons, layouts, and the overall design. In addition, the expressiveness of \gls{nlr} is restricted, particularly with respect to negation or exclusion criteria, which are neglected by earlier approaches. Hence, the main challenge is represented by the semantic gap between \gls{nlr} and multi-dimensional \gls{gui} prototypes. To tackle this challenge, we proposed an \gls{mllm}-based \gls{gui} reranking approach, which can utilize multi-dimensional, detailed descriptions for \gls{gui} prototypes derived by an \gls{mllm}, or \gls{gui} screenshots directly, to substantially improve the \gls{gui} ranking effectiveness, clearly outperforming \gls{bert}-\gls{ltr} models. To integrate the multi-dimensional \gls{gui} representation and complex \gls{nlr} already for the retrieval phase, we additionally proposed embedding-based constrained \gls{gui} retrieval. The proposed retrieval and reranking methods enable the processing of a semantically richer \gls{nlr} and \gls{gui} prototype representation, with effectiveness gains shown for reranking. Finally, we integrated both approaches into a novel unified tool.

\vspace{0.6cm}
\noindent
(\challonethree{}) \textit{\gls{gui} Prototyping Automation for Self-Elicitation:} the previously discussed \gls{nlr}-based \gls{gui} retrieval and reranking approaches enable the rapid mapping of \gls{nlr} to corresponding \gls{gui} prototypes. While the novel data-driven \gls{gui} prototyping tool \textit{\gls{rawi}} facilitates prototyping via \gls{gui} retrieval, it lacks guidance for the elicitation process integrated with \gls{gui} prototyping. Furthermore, while previous research introduced the notion of \gls{ser} \citep{rietz2019ladderbot}, their approach focuses on the simple \textit{laddering} interview technique \citep{corbridge1994laddering} to enable stakeholders to uncover their own requirements. We previously established the importance of visual representations of the requirements in the form of \gls{gui} prototypes to mitigate ambiguity and misunderstandings. Hence, since their approach relies completely on \gls{nlr}, it cannot profit from the benefits of prototyping. To enable automated \gls{ser} integrated with the benefits of \gls{gui} prototyping, we proposed \textit{\gls{ser}\gls{gui}}, a novel approach that integrates \gls{gui} retrieval methods with a guided dialogue-based chat interface and \gls{llm}-based feature recommendations to stimulate elicitation. Moreover, users are enabled to conduct feature-level searches, integrate aspects from different \glspl{gui} into a coherent \gls{gui} prototype specification, and benefit from a \gls{gui} retrieval approach extended by feature-based reranking. We demonstrated the effectiveness of the approach through a user study, indicating high effectiveness for feature recommendation, feature visualization, and reranking, representing integral components for supporting \gls{ser} with automated \gls{gui} prototyping. Our evaluation also indicates the high perceived usability of \textit{\gls{ser}\gls{gui}} based on the \gls{sus}.

\newpage
\noindent
(\challonefour{}) \textit{Efficient and Effective Adaptation of \glspl{llm} for \gls{nlr}-based \gls{gui} Generation:} while \gls{gui} retrieval approaches provide a cost-efficient solution for challenge \challone{}, they carry the disadvantage of providing static \gls{gui} prototypes with limited support for special \gls{nlr}, requiring \gls{gui} adaptation. With the advent of \glspl{llm} -- possessing impressive \gls{nlu} and text generation capabilities -- formulating the rapid mapping of \gls{nlr} to \gls{gui} prototypes as a translation and text generation problem became feasible. Previous research proposed approaches to train \glspl{llm} from scratch \citep{brie2023evaluating} and finetune pretrained \glspl{llm} \citep{feng2023designing} to produce low-fidelity \glspl{gui}. However, these approaches solely support the creation of low-fidelity \glspl{gui} and require resource-intensive training or finetuning. Therefore, we investigated how to more efficiently optimize \glspl{llm} for effectively generating high-fidelity \glspl{gui} through several \gls{zs}-prompting-based methods. In particular, we proposed \gls{ragg}, which is a novel method that integrates previously developed \gls{gui} retrieval methods with the generative capabilities of modern \glspl{llm}, enabling the \gls{llm} to leverage relevant external \gls{gui} prototyping knowledge. Furthermore, we adapted \gls{sc} for \gls{gui} prototyping, which employs the \gls{llm} itself in an iterative prototyping and feedback approach. Our large-scale evaluation, which is based on over 20,400 \gls{gui} annotations from 101 \gls{uiux} experts, demonstrated the effectiveness of the proposed \gls{zs} prompting methods over several \gls{zs} baselines to efficiently improve \gls{llm}-based \gls{gui} generation.

\vspace{0.3cm}
\noindent
(\challonefive{}) \textit{Efficient Adaptation of \gls{nlr}-based \gls{gui} Generation via \gls{llm}s for Editable Prototypes:} while pretrained \glspl{llm} offer high effectiveness for generating \gls{gui} prototypes in common representation languages (such as \gls{html}), which \glspl{llm} have been heavily exposed to during pretraining, these representations are not well suited for direct customization or editing, typically conducted in visual graphical prototyping editors. Moreover, these editors rely on proprietary \gls{gui} prototype representations, on which \glspl{llm} are not trained. Therefore, this raises the challenge of how to efficiently adapt \glspl{llm} to generate proprietary \gls{gui} representations that can directly be integrated into common \gls{gui} prototyping editors and workflows. To tackle this challenge, we proposed a novel two-stage \gls{rag} approach, which enables the efficient integration of external \gls{gui} component libraries by initially solely providing an abstracted library view to the \gls{llm} for selection, followed by retrieving the full \gls{gui} component specifications for selected components, which are employed for generating proprietary \gls{gui} representations that can directly be integrated into the popular prototyping environment \textit{Figma}. Furthermore, we integrated this approach into an \gls{llm}-based prototyping assistant plugin, which is also capable of identifying the implementation status of requirements and matching components to respective \gls{gui} components. The evaluation of our approach demonstrated substantial prototyping effectiveness improvements, clearly outperforming prototyping workflows without \gls{llm} integration. Our evaluation demonstrated that the proposed two-stage \gls{rag}-based approach substantially increases total \gls{llm} token usage efficiency.

\vspace{0.3cm}
\noindent
(\challtwoone{}) \textit{\gls{nlr} Fulfillment Verification in Static \glspl{gui}:} the discussed \gls{gui} retrieval and generation approaches facilitate rapid \gls{gui} prototyping, which benefits \gls{rel} and \gls{rval}. However, another time-consuming and often manual activity is \gls{sver}, ensuring that the implemented system adheres to its specification. Among verification activities, \gls{gui} testing represents a particularly effort-demanding task \citep{memon2001comprehensive, memon2007event}. Previous research mainly focused on improving automated \gls{gui} testing efficiency and coverage \citep{xie2007designing, mao2016sapienz, zeng2016automated}, enabling the detection of system crashes. However, verifying the correctness of an implemented \gls{gui} application against a collection of complex \gls{nlr} was neglected. Such testing activities are often still manual tasks or require automation scripts, which themselves are effort-demanding to create and require constant maintenance. Moreover, the advent of \glspl{llm} and continuous improvements in \gls{nlu} capabilities provide large potential to address this challenge and increase automation. Therefore, to tackle this challenge, we conducted an initial feasibility study by employing \gls{llm}-based methods to automatically verify requirements implementation in static \textit{Rico} \glspl{gui}. Our evaluation demonstrated high performance, with the approach achieving high effectiveness in identifying whether a specific requirement is implemented in a \gls{gui} and which \gls{gui} components correspond to that particular requirement.

\vspace{0.3cm}
\noindent
(\challtwotwo{}) \textit{\gls{nlr} Fulfillment Verification in Fully-Fledged, Dynamic \gls{gui} Applications:} while the previously discussed feasibility study showed promising results, it represents a simplified scenario with static \textit{Rico} \glspl{gui}, which differ from real-world, highly dynamic \gls{gui} applications. Therefore, we extended the approach to tackle the verification of \gls{nlr} for web-based \gls{gui} applications. In particular, we proposed an \gls{mllm}-based \gls{cua} approach for \gls{nlr}-based \gls{gui} verification, which leverages the image understanding and reasoning capabilities of state-of-the-art \glspl{mllm} to autonomously predict meaningful interactions with the \gls{gui}, given the current trajectory consisting of \gls{gui} states (screenshots), reasoning, and predicted actions. In addition, we implemented an \gls{mcp} server for enabling fully autonomous implementation-verification loops for \gls{llm}-based programming agents. Our evaluation demonstrates the effectiveness of \gls{mllm}-based \gls{cua} for \gls{gui} verification, achieving high scores at both the requirements and \gls{ac} levels.

\vspace{0.3cm}\noindent
To summarize, in this thesis, we identified key challenges in automated \gls{nlr}-based \gls{gui} prototyping and \gls{gui} verification, necessitating substantial \gls{nlu} capabilities. We proposed novel \gls{gui} retrieval approaches based on \glspl{plm} (\gls{bert}) and improved the effectiveness with \gls{mllm}-based \gls{gui} reranking. Moreover, we investigated \gls{llm}-based approaches for \gls{gui} generation. Finally, we proposed an automated \gls{gui} verification approach via an \gls{mllm}-based \gls{cua}. An integral component of the proposed methods is the use of pretrained \glspl{llm}, enabling effective processing of \gls{nlr} and \gls{gui} prototypes.

\vspace{0.3cm}\noindent
Given the work proposed in this thesis, important ideas for future work and potential improvements are identified. As presented in our discussion, while the pretrained \glspl{llm} are highly effective across the investigated problems, they simultaneously entail high costs. Reducing the costs associated with running these models and increasing their efficiency represents an important direction for future research. Adapting \glspl{slm} to improve task-specific performance (such as \gls{nlr}-based \gls{gui} prototyping and \gls{gui} verification) -- while acknowledging the reduction of more general capabilities of the model -- offers substantial potential to enhance the efficiency while potentially maintaining high task effectiveness.

\cleardoublepage
\phantomsection
\addcontentsline{toc}{part}{Bibliography}
\bibliographystyle{acl_natbib} 
\bibliography{bibtex/bib/ref}


\appendix
\makeatletter
\setlength{\@fptop}{0pt}
\makeatother

\clearpage
\newpage
\thispagestyle{empty}
\null  

\chapter{Published Resources}
\label{appendix:published}
We provide an overview of the resources and artifacts published in the context of this thesis, including code, datasets, experimental results, and easily reusable tool prototypes.

\renewcommand{\arraystretch}{0.92} 
\setlength{\tabcolsep}{5.8pt}      

\begin{table}[h!]
\centering
\begin{tabular}{cllp{4.2cm}c}
\toprule
\textbf{Chap.} & \textbf{Resource Name} & \textbf{Type} & \textbf{Location} & \textbf{Challenges} \\
\midrule
3 & \gls{rawi} &
\begin{tabular}[t]{@{}l@{}}Code+\\Dataset\end{tabular} &
\small{\url{https://github.com/kristiankolthoff/RaWi}} &
\challoneone{} \\

3 & GUI2WiRe & Code &
\small{\url{https://github.com/kristiankolthoff/GUI2WiReTool}} &
\challoneone{} \\

4 & \gls{gui}-ReRank & \begin{tabular}[t]{@{}l@{}}Code+\\Dataset\end{tabular} &
\small{\url{https://github.com/kristiankolthoff/GUI-ReRank}} &
\challonetwo{} \\

5 & \gls{ser}\gls{gui} & Code &
\small{\url{https://github.com/kristiankolthoff/SERGUI-Prototyping}} &
\challonethree{} \\

\midrule
6 & \gls{zs}-Prompting & \begin{tabular}[t]{@{}l@{}}Code+\\Dataset\end{tabular} &
\small{\url{https://github.com/kristiankolthoff/ZS-Prompting}} &
\challonefour{}\\

7 & \gls{gui}-\gls{us}-Loop & Code &
\small{\url{https://github.com/kristiankolthoff/Closing-the-Loop-US-GUI}} &
\challonefive{} \\

8 & GUIDE & Code &
\small{\url{https://github.com/kristiankolthoff/GUIDE-Prototyping}} &
\challonefive{} \\

\midrule
9 & Interlinking & \begin{tabular}[t]{@{}l@{}}Code+\\Dataset\end{tabular} &
\small{\url{https://github.com/kristiankolthoff/IUS-GUI-Prototyping}} &
\challtwoone{} \\

10 & GUISpector & \begin{tabular}[t]{@{}l@{}}Code+\\Dataset\end{tabular} &
\small{\url{https://github.com/kristiankolthoff/GUISpector}} &
\challtwotwo{} \\

\end{tabular}
\caption[Overview of published resources and artifacts]{Overview of published resources and artifacts in the context of this thesis.}
\label{tab:resources}
\end{table}


\newpage

\noindent In addition to the above resources, we also created and published several videos accompanying our research paper publications, introducing not only the respective novel approaches and techniques, but also the functionality of the developed research tool prototypes.
\newlength{\ResourceTabColSep}
\setlength{\ResourceTabColSep}{11.5pt} 

\begin{table}[h!]
\centering
\setlength{\tabcolsep}{\ResourceTabColSep}
\begin{tabular}{clp{4.2cm}c}
\toprule
\textbf{Chap.} & \textbf{Resource Name} & \textbf{Location} & \textbf{Challenges} \\
\midrule

3 & GUI2WiRe &
\small{\url{https://youtu.be/2nN-Xr2Hk7I}} &
\challoneone{} \\

4 & \gls{gui}-ReRank &
\small{\url{https://youtu.be/_7x9UCh82ug}} &
\challonetwo{} \\

5 & \gls{ser}\gls{gui} &
\small{\url{https://youtu.be/pzAAB9Uht80}} &
\challonethree{} \\

\midrule

7 & \gls{gui}-\gls{us}-Loop &
\small{\url{https://youtu.be/QQd007gJLHQ}} &
\challonefive{} \\

8 & GUIDE &
\small{\url{https://youtu.be/C9RbhMxqpTU}} &
\challonefive{} \\

\midrule

10 & GUISpector &
\small{\url{https://youtu.be/JByYF6BNQeE}} &
\challtwotwo{} \\

\end{tabular}
\caption[Overview of published approach demonstration videos]{Overview of published approach demo videos in the context of this thesis.}
\label{tab:resources}
\end{table}

\chapter{Appendix for Chapter \ref{cha:nl_gui_retrieval}}
\label{chaptper:app-rawi}
\section{Experimental Details}

\setlength{\ResourceTabColSep}{4.8pt} 

\begin{table}[!htbp]
\centering
\setlength{\tabcolsep}{\ResourceTabColSep}
\small
\begin{tabular}{llccccccc}
\toprule
\textbf{Model 1} & \textbf{Model 2} & \textbf{AP} & \textbf{MRR} & \textbf{H@1} & \textbf{H@3} & \textbf{H@5} & \textbf{H@10} & \textbf{H@15} \\
\midrule
\textbf{BM25} & \textbf{nBoW} & \textbf{<0.001*} & 1.000 & 1.000 & 1.000 & 0.914 & 1.000 & 1.000 \\
\textbf{BM25} & \textbf{TF-IDF} & \textbf{0.034*} & 0.494 & 1.000 & 0.118 & 0.300 & 0.149 & 1.000 \\
\textbf{PRF} & \textbf{BM25} & 1.000 & 1.000 & 1.000 & 1.000 & 1.000 & 1.000 & 1.000 \\
\textbf{PRF (s)} & \textbf{BM25} & 0.608 & 1.000 & 1.000 & 1.000 & 1.000 & 1.000 & 0.809 \\
\textbf{PRF (w)} & \textbf{BM25} & 1.000 & 1.000 & 1.000 & 1.000 & 1.000 & 1.000 & 1.000 \\
\textbf{PRF (sw)} & \textbf{BM25} & 1.000 & 1.000 & 1.000 & 1.000 & 1.000 & 1.000 & 1.000 \\
\textbf{SBERT} & \textbf{BM25} & 0.382 & 0.636 & 1.000 & 0.927 & 0.669 & \textbf{0.092*} & 0.169 \\
\textbf{BERT-LTR (1)} & \textbf{BM25} & 0.050 & \textbf{0.044*} & 1.000 & 0.315 & \textbf{0.012*} & \textbf{0.007*} & 0.053 \\
\textbf{BERT-LTR (2)} & \textbf{BM25} & \textbf{0.016*} & 0.131 & 1.000 & 0.106 & \textbf{<0.001*} & \textbf{0.003*} & 0.053 \\
\textbf{BERT-LTR (3)} & \textbf{BM25} & \textbf{0.039*} & 0.168 & 1.000 & 0.206 & \textbf{0.012*} & \textbf{0.001*} & 0.053 \\
\textbf{BERT-LTR (1)} & \textbf{SBERT} & 0.388 & 0.572 & 1.000 & 1.000 & 0.272 & 1.000 & 1.000 \\
\textbf{BERT-LTR (2)} & \textbf{SBERT} & 0.305 & 0.530 & 1.000 & 0.927 & \textbf{0.011*} & 1.000 & 1.000 \\
\textbf{BERT-LTR (3)} & \textbf{SBERT} & 0.388 & 0.572 & 1.000 & 1.000 & 0.265 & 0.182 & 1.000 \\
\bottomrule
\end{tabular}
\caption[\textit{Holm}-corrected $p$-values for comparing \gls{nlr}-based \gls{gui} retrieval approaches (1)]{\textit{Holm}-corrected $p$-values from \textit{Wilcoxon signed-rank test} over\gls{nlr}-based \gls{gui} retrieval gold standard with \textit{\gls{ap}}, \textit{\gls{mrr}}, and \textit{HITS@k}. Particularly, we test whether \textit{model 1} is greater than \textit{model 2}. \textbf{Bold*} indicates significance at $\alpha=0.05$ after \textit{Holm} correction.}
\label{tab:appendix-rawi-holm-1}
\end{table}

\setlength{\ResourceTabColSep}{13.5pt} 
\begin{table}[!htbp]
\centering
\setlength{\tabcolsep}{\ResourceTabColSep}
\small
\begin{tabular}{llcccc}
\toprule
\textbf{Model 1} & \textbf{Model 2} & \textbf{P@3} & \textbf{P@5} & \textbf{P@7} & \textbf{P@10} \\
\midrule
\textbf{BM25} & \textbf{nBoW} & 1.000 & 0.309 & 0.486 & 0.306 \\
\textbf{BM25} & \textbf{TF-IDF} & 0.149 & \textbf{0.028*} & \textbf{0.012*} & \textbf{0.011*} \\
\textbf{PRF} & \textbf{BM25} & 1.000 & 1.000 & 0.946 & 0.448 \\
\textbf{PRF (s)} & \textbf{BM25} & 1.000 & 1.000 & 0.946 & 0.251 \\
\textbf{PRF (w)} & \textbf{BM25} & 1.000 & 1.000 & 0.946 & 0.459 \\
\textbf{PRF (sw)} & \textbf{BM25} & 1.000 & 1.000 & 0.946 & 0.306 \\
\textbf{SBERT} & \textbf{BM25} & 0.948 & 0.797 & 0.195 & \textbf{0.013*} \\
\textbf{BERT-LTR (1)} & \textbf{BM25} & 0.407 & 0.052 & \textbf{0.031*} & \textbf{0.008*} \\
\textbf{BERT-LTR (2)} & \textbf{BM25} & 0.065 & 0.089 & \textbf{0.037*} & \textbf{0.002*} \\
\textbf{BERT-LTR (3)} & \textbf{BM25} & 0.706 & 0.061 & \textbf{0.025*} & \textbf{<0.001*} \\
\textbf{BERT-LTR (1)} & \textbf{SBERT} & 1.000 & 0.608 & 0.946 & 0.577 \\
\textbf{BERT-LTR (2)} & \textbf{SBERT} & 0.368 & 1.000 & 0.946 & 0.459 \\
\textbf{BERT-LTR (3)} & \textbf{SBERT} & 1.000 & 0.282 & 0.764 & 0.441 \\
\bottomrule
\end{tabular}
\caption[\textit{Holm}-corrected $p$-values for comparing \gls{nlr}-based \gls{gui} retrieval approaches (2)]{\textit{Holm}-corrected $p$-values from \textit{Wilcoxon signed-rank test} over \gls{nlr}-based \gls{gui} retrieval gold standard with \textit{precision at rank k (P@k)}. Particularly, we test whether \textit{Model 1} is greater than \textit{Model 2}. \textbf{Bold*} indicates significance at $\alpha=0.05$ after \textit{Holm} correction.}
\label{tab:appendix-rawi-holm-2}
\end{table}

\setlength{\ResourceTabColSep}{7.5pt} 
\begin{table}[!htbp]
\centering
\setlength{\tabcolsep}{\ResourceTabColSep}
\small
\begin{tabular}{llcccc}
\toprule
\textbf{Model 1} & \textbf{Model 2} & \textbf{NDCG@3} & \textbf{NDCG@5} & \textbf{NDCG@10} & \textbf{NDCG@15} \\
\midrule
\textbf{BM25} & \textbf{nBoW} & 0.877 & 0.213 & \textbf{<0.001*} & \textbf{<0.001*} \\
\textbf{BM25} & \textbf{TF-IDF} & 0.054 & \textbf{0.012*} & \textbf{0.002*} & \textbf{0.001*} \\
\textbf{PRF} & \textbf{BM25} & 0.969 & 0.816 & 1.000 & 0.942 \\
\textbf{PRF (s)} & \textbf{BM25} & 0.969 & 0.525 & 0.455 & 0.204 \\
\textbf{PRF (w)} & \textbf{BM25} & 0.877 & 1.000 & 1.000 & 1.000 \\
\textbf{PRF (sw)} & \textbf{BM25} & 0.969 & 1.000 & 1.000 & 1.000 \\
\textbf{SBERT} & \textbf{BM25} & 0.384 & 0.213 & \textbf{0.004*} & \textbf{0.004*} \\
\textbf{BERT-LTR (1)} & \textbf{BM25} & \textbf{0.042*} & \textbf{0.007*} & \textbf{<0.001*} & \textbf{<0.001*} \\
\textbf{BERT-LTR (2)} & \textbf{BM25} & \textbf{0.011*} & \textbf{0.007*} & \textbf{<0.001*} & \textbf{<0.001*} \\
\textbf{BERT-LTR (3)} & \textbf{BM25} & 0.090 & \textbf{0.012*} & \textbf{<0.001*} & \textbf{<0.001*} \\
\textbf{BERT-LTR (1)} & \textbf{SBERT} & 0.247 & 0.213 & 0.706 & 0.424 \\
\textbf{BERT-LTR (2)} & \textbf{SBERT} & 0.091 & 0.213 & 0.455 & 0.165 \\
\textbf{BERT-LTR (3)} & \textbf{SBERT} & 0.495 & 0.228 & 0.409 & 0.399 \\
\bottomrule
\end{tabular}
\caption[\textit{Holm}-corrected $p$-values for comparing \gls{nlr}-based \gls{gui} retrieval approaches (3)]{\textit{Holm}-corrected $p$-values from \textit{Wilcoxon signed-rank test} over \gls{nlr}-based \gls{gui} retrieval gold standard with metric \textit{\gls{ndcg}@k}. Particularly, we test whether \textit{Model 1} is greater than \textit{Model 2}. \textbf{Bold*} indicates significance at $\alpha=0.05$ after \textit{Holm} correction.}
\label{tab:appendix-rawi-holm-3}
\end{table}


\vspace*{-\baselineskip}
\begin{table}[!t]
\centering
\small

\newlength{\colw}
\setlength{\colw}{1.2cm} 

\setlength\tabcolsep{5pt}
\begin{tabular}{>{\centering\arraybackslash}p{\colw}cccccccccc}
\toprule
\textbf{Minute} &
$\boldsymbol{\overline{C}_A}$ &
$\boldsymbol{\tilde{C}_A}$ &
$\boldsymbol{\overline{C}_B}$ &
$\boldsymbol{\tilde{C}_B}$ &
$\boldsymbol{\Delta\overline{C}}$ &
$\boldsymbol{\Delta\tilde{C}}$ &
$\boldsymbol{\delta}$ &
$\mathbf{W}$ &
\textit{p} &
$\textit{p}_{\text{Holm}}$ \\
\midrule

\textbf{3} &
15.87 & 17.00 &
5.05  & 4.00  &
10.82 & 12.00 &
0.807 &
692.0 &
<0.001 &
\textbf{<0.001*} \\

\textbf{4} &
17.68 & 19.00 &
7.18  & 6.50  &
10.50 & 12.00 &
0.785 &
685.5 &
<0.001 &
\textbf{<0.001*} \\

\textbf{5} &
19.74 & 22.00 &
9.82  & 9.00  &
9.92  & 10.50 &
0.756 &
656.5 &
<0.001 &
\textbf{<0.001*} \\

\textbf{6} &
21.16 & 23.00 &
12.55 & 11.50 &
8.61  & 8.00  &
0.672 &
659.0 &
<0.001 &
\textbf{<0.001*} \\

\textbf{7} &
22.97 & 24.00 &
15.11 & 14.00 &
7.87  & 7.00  &
0.659 &
688.5 &
<0.001 &
\textbf{<0.001*} \\

\bottomrule
\end{tabular}
\caption[\textit{Holm}-corrected $p$-values for comparing \textit{\gls{rawi}} against \textit{Mockplus} on \textit{\#\gls{gui}-comps} metric]{\textbf{Holm}-corrected \textit{p}-values from pairwise one-sided \textit{Wilcoxon signed-rank tests} comparing tool \textit{A} (\textit{\gls{rawi}}) with tool \textit{B} (\textit{Mockplus}) on paired task data per minute on the \textit{\#\gls{gui}-comps} metric
($\alpha=0.05$). $\overline{C}$ and $\tilde{C}$ represent the mean and median values, respectively. $\Delta\overline{C}$ and $\Delta\tilde{C}$ denote mean and median differences. $\delta$ denotes Cliff’s Delta. \textbf{Bold} values with \textbf{*} indicate significance at $\alpha=0.05$ after applying the \textit{Holm} correction.
}
\label{tab:productivity_1}
\end{table}

\vspace*{-\baselineskip}
\begin{table}[!t]
\centering
\small

\setlength{\colw}{1.2cm} 

\setlength\tabcolsep{5pt}
\begin{tabular}{>{\centering\arraybackslash}p{\colw}cccccccccc}
\toprule
\textbf{Minute} &
$\boldsymbol{\overline{C}_A}$ &
$\boldsymbol{\tilde{C}_A}$ &
$\boldsymbol{\overline{C}_B}$ &
$\boldsymbol{\tilde{C}_B}$ &
$\boldsymbol{\Delta\overline{C}}$ &
$\boldsymbol{\Delta\tilde{C}}$ &
$\boldsymbol{\delta}$ &
$\mathbf{W}$ &
\textit{p} &
$\textit{p}_{\text{Holm}}$ \\
\midrule

\textbf{3} &
15.50 & 16.50 &
4.76  & 4.00  &
10.74 & 12.50 &
0.810 &
692.0 &
<0.001 &
\textbf{<0.001*} \\

\textbf{4} &
16.89 & 18.00 &
6.45  & 6.00  &
10.45 & 11.50 &
0.810 &
693.0 &
<0.001 &
\textbf{<0.001*} \\

\textbf{5} &
18.97 & 19.00 &
8.71  & 7.50  &
10.26 & 10.50 &
0.805 &
661.5 &
<0.001 &
\textbf{<0.001*} \\

\textbf{6} &
20.45 & 21.50 &
10.45 & 9.50  &
10.00 & 9.00  &
0.834 &
732.0 &
<0.001 &
\textbf{<0.001*} \\

\textbf{7} &
22.11 & 23.50 &
12.89 & 12.00 &
9.21  & 9.00  &
0.808 &
695.0 &
<0.001 &
\textbf{<0.001*} \\

\bottomrule
\end{tabular}
\caption[\textit{Holm}-corrected $p$-values for comparing \textit{\gls{rawi}} against \textit{Mockplus} on \textit{\#\gls{gui}-comps-div} metric]{\textbf{Holm}-corrected \textit{p}-values from pairwise one-sided \textit{Wilcoxon signed-rank tests} comparing tool \textit{A} (\textit{\gls{rawi}}) with tool \textit{B} (\textit{Mockplus}) on paired task data per minute on the \textit{\#\gls{gui}-comps-div} metric
($\alpha=0.05$). $\overline{C}$ and $\tilde{C}$ represent the mean and median values, respectively. $\Delta\overline{C}$ and $\Delta\tilde{C}$ denote mean and median differences, respectively. $\delta$ denotes Cliff’s Delta. \textbf{Bold} values with \textbf{*} indicate significance at $\alpha=0.05$ after applying the \textit{Holm} correction.
}
\label{tab:productivity_2}
\end{table}

\vspace*{-\baselineskip}
\begin{table}[!t]
\centering
\small

\setlength{\colw}{1.2cm} 

\setlength\tabcolsep{5pt}
\begin{tabular}{>{\centering\arraybackslash}p{\colw}cccccccccc}
\toprule
\textbf{Minute} &
$\boldsymbol{\overline{C}_A}$ &
$\boldsymbol{\tilde{C}_A}$ &
$\boldsymbol{\overline{C}_B}$ &
$\boldsymbol{\tilde{C}_B}$ &
$\boldsymbol{\Delta\overline{C}}$ &
$\boldsymbol{\Delta\tilde{C}}$ &
$\boldsymbol{\delta}$ &
$\mathbf{W}$ &
\textit{p} &
$\textit{p}_{\text{Holm}}$ \\
\midrule

\textbf{3} &
10.45 & 11.00 &
0.29  & 2.00  &
10.16 & 10.50 &
0.680 &
656.0 &
<0.001 &
\textbf{<0.001*} \\

\textbf{4} &
12.97 & 13.00 &
3.32  & 3.00  &
9.66  & 10.00 &
0.660 &
591.5 &
<0.001 &
\textbf{<0.001*} \\

\textbf{5} &
15.68 & 14.50 &
6.76  & 6.50  &
8.92  & 8.50  &
0.605 &
598.5 &
<0.001 &
\textbf{<0.001*} \\

\textbf{6} &
17.74 & 18.00 &
10.50 & 10.50 &
7.24  & 6.50  &
0.512 &
578.5 &
<0.001 &
\textbf{<0.001*} \\

\textbf{7} &
19.76 & 20.00 &
13.18 & 13.00 &
6.58  & 5.00  &
0.482 &
545.0 &
<0.001 &
\textbf{<0.001*} \\

\bottomrule
\end{tabular}
\caption[\textit{Holm}-corrected $p$-values for comparing \textit{\gls{rawi}} against \textit{Mockplus} on \textit{\#\gls{gui}-comps-neg-div} metric]{\textbf{Holm}-corrected \textit{p}-values (and raw \textit{p}-values) from pairwise one-sided \textit{Wilcoxon signed-rank tests} comparing tool \textit{A} (\textit{\gls{rawi}}) with tool \textit{B} (\textit{Mockplus}) on paired task data per minute on the metric \textit{\#\gls{gui}-comps} adjusted by the \textit{\#\gls{gui}-comps-neg} metric
($\alpha=0.05$). $\overline{C}$ and $\tilde{C}$ represent the mean and median values, respectively. $\Delta\overline{C}$ and $\Delta\tilde{C}$ denote mean and median differences, respectively. $\delta$ denotes Cliff’s Delta. \textbf{Bold} values with \textbf{*} indicate significance at $\alpha=0.05$ after applying the \textit{Holm} correction.
}
\label{tab:productivity_3}
\end{table}

\begin{table}[!t]
\centering
\small
\setlength\tabcolsep{9.1pt}

\begin{tabular}{lrrrrrr}
\toprule
\textbf{Term} & $\boldsymbol{\beta}$ & \textbf{SE} & $\boldsymbol{z}$ & \textit{p} & \textbf{CI$_L$} & \textbf{CI$_U$} \\
\midrule
\textbf{Intercept}                & 12.435  & 1.050 & 11.838  & <0.001 & 10.376  & 14.494 \\
\textbf{Approach B}               & -11.090 & 0.874 & -12.691 & <0.001 & -12.803 & -9.377 \\
\textbf{Task 2}                   & -0.188  & 0.505 & -0.372  & 0.710  & -1.178  & 0.802 \\
\textbf{GUI 2}                    & 2.216   & 0.504 & 4.394   & <0.001 & 1.227   & 3.204 \\
\textbf{Condition 2}              & 4.640   & 1.153 & 4.024   & <0.001 & 2.380   & 6.900 \\
\textbf{Condition 3}              & 4.460   & 1.153 & 3.868   & <0.001 & 2.200   & 6.720 \\
\textbf{Condition 4}              & 0.468   & 1.223 & 0.382   & 0.702  & -1.930  & 2.865 \\
\textbf{Time}                     & 1.768   & 0.252 & 7.013   & <0.001 & 1.274   & 2.263 \\
\textbf{Approach B $\times$ Time} & 0.779   & 0.357 & 2.184   & 0.029  & 0.080   & 1.478 \\
\bottomrule
\end{tabular}

\caption[Mixed-effects model for \textit{\#\gls{gui}-comps}]{
Results of the linear mixed-effects model for the \textit{\#\gls{gui}-comps} productivity metric. The model includes \textit{approach}, \textit{time}, \textit{task}, \textit{GUI}, and \textit{condition} as fixed effects, as well as an interaction term between \textit{approach} and \textit{time} ($\textit{\#\gls{gui}-comps} \sim \text{Approach} \times \text{Time}_c
+ \text{Task} + \text{GUI} + \text{Condition}
+ (1 \mid \text{Participant})$). Participant was modeled as a random intercept to account for repeated measurements. Coefficients ($\beta$), standard errors (SE), Wald $z$ statistics, \textit{p}-values, and 95\% confidence intervals are reported. Approach A (\textit{\gls{rawi}}), task 1, GUI 1, and condition 1 serve as reference levels. Time was centered at minute 3 (representing the starting point of collection). Approach \textit{B} represents \textit{Mockplus}.
}

\label{tab:mixed_effects_1}
\end{table}

\begin{table}[!t]
\centering
\small
\setlength\tabcolsep{9.1pt}

\begin{tabular}{lrrrrrr}
\toprule
\textbf{Term} & $\boldsymbol{\beta}$ & \textbf{SE} & $\boldsymbol{z}$ & \textit{p} & \textbf{CI$_L$} & \textbf{CI$_U$} \\
\midrule
\textbf{Intercept}                & 12.090  & 0.912 & 13.263  & <0.001 & 10.303  & 13.877 \\
\textbf{Approach B}               & -10.851 & 0.791 & -13.718 & <0.001 & -12.401 & -9.301 \\
\textbf{Task 2}                   & 0.371   & 0.457 & 0.812   & 0.417  & -0.525  & 1.267 \\
\textbf{GUI 2}                    & 1.795   & 0.456 & 3.932   & <0.001 & 0.900   & 2.689 \\
\textbf{Condition 2}              & 4.600   & 0.971 & 4.737   & <0.001 & 2.697   & 6.503 \\
\textbf{Condition 3}              & 3.690   & 0.971 & 3.800   & <0.001 & 1.787   & 5.593 \\
\textbf{Condition 4}              & 0.412   & 1.030 & 0.401   & 0.689  & -1.606  & 2.431 \\
\textbf{Time}                     & 1.676   & 0.228 & 7.345   & <0.001 & 1.229   & 2.124 \\
\textbf{Approach B $\times$ Time} & 0.350   & 0.323 & 1.084   & 0.278  & -0.283  & 0.983 \\
\bottomrule
\end{tabular}

\caption[Mixed-effects model for \textit{\#\gls{gui}-comps-div}]{
Results of the linear mixed-effects model for the \textit{\#\gls{gui}-comps-div} productivity metric. The model includes \textit{approach}, \textit{time}, \textit{task}, \textit{GUI}, and \textit{condition} as fixed effects, as well as an interaction term between \textit{approach} and \textit{time} ($\textit{\#\gls{gui}-comps-div} \sim \text{Approach} \times \text{Time}_c
+ \text{Task} + \text{GUI} + \text{Condition}
+ (1 \mid \text{Participant})$). Participant was modeled as a random intercept to account for repeated measurements. Coefficients ($\beta$), standard errors (SE), Wald $z$ statistics, \textit{p}-values, and 95\% confidence intervals are reported. Approach A (\textit{\gls{rawi}}), task 1, GUI 1, and condition 1 serve as reference levels. Time was centered at minute 3 (representing the starting point of collection). Approach \textit{B} represents \textit{Mockplus}.
}

\label{tab:mixed_effects_2}
\end{table}

\begin{table}[!t]
\centering
\small
\setlength\tabcolsep{9.1pt}

\begin{tabular}{lrrrrrr}
\toprule
\textbf{Term} & $\boldsymbol{\beta}$ & \textbf{SE} & $\boldsymbol{z}$ & \textit{p} & \textbf{CI$_L$} & \textbf{CI$_U$} \\
\midrule
\textbf{Intercept}                & 5.621  & 0.934 & 6.018  & <0.001 & 3.790  & 7.452 \\
\textbf{Approach B}               & -0.737 & 0.725 & -1.016 & 0.310  & -2.158 & 0.685 \\
\textbf{Task 2}                   & 1.194  & 0.419 & 2.850  & 0.004  & 0.373  & 2.016 \\
\textbf{GUI 2}                    & 1.042  & 0.418 & 2.490  & 0.013  & 0.222  & 1.862 \\
\textbf{Condition 2}              & -1.600 & 1.068 & -1.498 & 0.134  & -3.694 & 0.494 \\
\textbf{Condition 3}              & -0.880 & 1.068 & -0.824 & 0.410  & -2.974 & 1.214 \\
\textbf{Condition 4}              & -3.562 & 1.133 & -3.144 & 0.002  & -5.783 & -1.342 \\
\textbf{Time}                     & -0.571 & 0.209 & -2.729 & 0.006  & -0.981 & -0.161 \\
\textbf{Approach B $\times$ Time} & -0.179 & 0.296 & -0.605 & 0.545  & -0.759 & 0.401 \\
\bottomrule
\end{tabular}

\caption[Mixed-effects model for \textit{\#\gls{gui}-comps-neg-div}]{
Results of the linear mixed-effects model for the \textit{\#\gls{gui}-comps-neg-div} productivity metric. The model includes \textit{approach}, \textit{time}, \textit{task}, \textit{GUI}, and \textit{condition} as fixed effects, as well as an interaction term between \textit{approach} and \textit{time} ($\textit{\#\gls{gui}-comps-neg-div} \sim \text{Approach} \times \text{Time}_c
+ \text{Task} + \text{GUI} + \text{Condition}
+ (1 \mid \text{Participant})$). Participant was modeled as a random intercept to account for repeated measurements. Coefficients ($\beta$), standard errors (SE), Wald $z$ statistics, \textit{p}-values, and 95\% confidence intervals are reported. Approach A (\textit{\gls{rawi}}), task 1, GUI 1, and condition 1 serve as reference levels. Time was centered at minute 3 (representing the starting point). Approach \textit{B} represents \textit{Mockplus}.
}

\label{tab:mixed_effects_3}
\end{table}

\chapter{Appendix for Chapter \ref{cha:gui_rerank}}
\label{chapter:app-gui-rerank}
\section{Experimental Details}

\begin{table}[!htbp]
\centering
\small
\setlength\tabcolsep{3.7pt}
\begin{tabular}{llccccccc}
\toprule
\textbf{Model 1} & \textbf{Model 2} & \textbf{AP} & \textbf{MRR} & \textbf{H@1} & \textbf{H@3} & \textbf{H@5} & \textbf{H@10} & \textbf{H@15} \\
\midrule
\textbf{G4.1 (I)} & \textbf{BERT-LTR-1} & \textbf{<0.001*} & \textbf{<0.001*} & \textbf{<0.001*} & \textbf{<0.001*} & \textbf{0.001*} & 1.000 & 1.000 \\
\textbf{G4.1 (I)} & \textbf{BERT-LTR-2} & \textbf{<0.001*} & \textbf{<0.001*} & \textbf{<0.001*} & \textbf{<0.001*} & \textbf{0.012*} & 1.000 & 1.000 \\
\textbf{G4.1 (I)} & \textbf{BERT-LTR-3} & \textbf{<0.001*} & \textbf{<0.001*} & \textbf{<0.001*} & \textbf{<0.001*} & \textbf{0.001*} & 1.000 & 1.000 \\
\textbf{G4.1 (T)} & \textbf{BERT-LTR-1} & \textbf{<0.001*} & \textbf{<0.001*} & \textbf{<0.001*} & \textbf{<0.001*} & \textbf{0.001*} & 1.000 & 1.000 \\
\textbf{G4.1 (T)} & \textbf{BERT-LTR-2} & \textbf{<0.001*} & \textbf{<0.001*} & \textbf{<0.001*} & \textbf{<0.001*} & \textbf{0.012*} & 1.000 & 1.000 \\
\textbf{G4.1 (T)} & \textbf{BERT-LTR-3} & \textbf{<0.001*} & \textbf{<0.001*} & \textbf{<0.001*} & \textbf{<0.001*} & \textbf{0.001*} & 1.000 & 1.000 \\
\textbf{G4.1 (I)} & \textbf{G4.1 (T)} & 0.050 & 0.785 & 1.000 & 0.841 & 1.000 & 1.000 & 1.000 \\
\textbf{G4.1 (I)} & \textbf{G4.1M (I)} & \textbf{0.043*} & 0.785 & 1.000 & 0.317 & 1.000 & 1.000 & 1.000 \\
\textbf{G4.1 (I)} & \textbf{G4.1N (I)} & \textbf{<0.001*} & \textbf{<0.001*} & \textbf{<0.001*} & \textbf{0.034*} & 0.793 & 1.000 & 1.000 \\
\textbf{G4.1M (I)} & \textbf{G4.1N (I)} & \textbf{<0.001*} & \textbf{<0.001*} & \textbf{<0.001*} & 0.087 & 0.793 & 1.000 & 1.000 \\
\textbf{G4.1 (T)} & \textbf{G4.1M (T)} & 0.055 & 0.709 & 1.000 & 0.236 & 1.000 & 1.000 & 1.000 \\
\textbf{G4.1 (T)} & \textbf{G4.1N (T)} & \textbf{<0.001*} & \textbf{0.009*} & \textbf{0.026*} & \textbf{0.009*} & 0.159 & 1.000 & 1.000 \\
\textbf{G4.1M (T)} & \textbf{G4.1N (T)} & \textbf{<0.001*} & \textbf{0.020*} & \textbf{0.037*} & 0.087 & 0.159 & 1.000 & 1.000 \\
\bottomrule
\end{tabular}
\caption[\textit{Holm}-corrected $p$-values for comparing \gls{mllm}-based \gls{gui} reranking approaches (1)]{\textit{Holm}-corrected $p$-values from \textit{Wilcoxon signed-rank test} with \textit{\gls{ap}}, \textit{\gls{mrr}}, and \textit{HITS@k}. Particularly, we test whether \textit{model 1} is greater than \textit{model 2}. \textbf{Bold*} indicates significance at $\alpha=0.05$ after \textit{Holm} correction. Abbreviations: I=Image, T=Text, M=Mini, N=Nano, G4.1=GPT-4.1.}
\label{tab:gui-rerank-appendix-llm-holm-1}
\end{table}

\begin{table}[!htbp]
\centering
\setlength\tabcolsep{11.5pt}
\small
\begin{tabular}{llcccc}
\toprule
\textbf{Model 1} & \textbf{Model 2} & \textbf{P@3} & \textbf{P@5} & \textbf{P@7} & \textbf{P@10} \\
\midrule
\textbf{G4.1 (I)} & \textbf{BERT-LTR-1} & \textbf{<0.001*} & \textbf{<0.001*} & \textbf{<0.001*} & \textbf{<0.001*} \\
\textbf{G4.1 (I)} & \textbf{BERT-LTR-2} & \textbf{<0.001*} & \textbf{<0.001*} & \textbf{<0.001*} & \textbf{<0.001*} \\
\textbf{G4.1 (I)} & \textbf{BERT-LTR-3} & \textbf{<0.001*} & \textbf{<0.001*} & \textbf{<0.001*} & \textbf{<0.001*} \\
\textbf{G4.1 (T)} & \textbf{BERT-LTR-1} & \textbf{<0.001*} & \textbf{<0.001*} & \textbf{<0.001*} & \textbf{<0.001*} \\
\textbf{G4.1 (T)} & \textbf{BERT-LTR-2} & \textbf{<0.001*} & \textbf{<0.001*} & \textbf{<0.001*} & \textbf{<0.001*} \\
\textbf{G4.1 (T)} & \textbf{BERT-LTR-3} & \textbf{<0.001*} & \textbf{<0.001*} & \textbf{<0.001*} & \textbf{<0.001*} \\
\textbf{G4.1 (I)} & \textbf{G4.1 (T)} & 0.057 & 0.159 & 0.129 & 0.198 \\
\textbf{G4.1 (I)} & \textbf{G4.1M (I)} & \textbf{0.041*} & 0.082 & \textbf{0.035*} & 0.223 \\
\textbf{G4.1 (I)} & \textbf{G4.1N (I)} & \textbf{<0.001*} & \textbf{<0.001*} & \textbf{<0.001*} & \textbf{<0.001*} \\
\textbf{G4.1M (I)} & \textbf{G4.1N (I)} & \textbf{<0.001*} & \textbf{<0.001*} & \textbf{<0.001*} & \textbf{<0.001*} \\
\textbf{G4.1 (T)} & \textbf{G4.1M (T)} & \textbf{0.008*} & 0.296 & 0.312 & 0.223 \\
\textbf{G4.1 (T)} & \textbf{G4.1N (T)} & \textbf{<0.001*} & \textbf{<0.001*} & \textbf{<0.001*} & \textbf{<0.001*} \\
\textbf{G4.1M (T)} & \textbf{G4.1N (T)} & \textbf{<0.001*} & \textbf{<0.001*} & \textbf{<0.001*} & \textbf{<0.001*} \\
\bottomrule
\end{tabular}
\caption[\textit{Holm}-corrected $p$-values for comparing \gls{mllm}-based \gls{gui} reranking approaches (2)]{\textit{Holm}-corrected $p$-values from \textit{Wilcoxon signed-rank test} with \textit{precision at rank k (P@k)}. Particularly, we test whether \textit{model 1} is greater than \textit{model 2}. \textbf{Bold*} indicates significance at $\alpha=0.05$ after \textit{Holm} correction. Abbreviations: I=Image, T=Text, M=Mini, N=Nano, G4.1=GPT-4.1.}
\label{tab:gui-rerank-appendix-llm-holm-2}
\end{table}

\begin{table}[!htbp]
\centering
\setlength\tabcolsep{7.1pt}
\small
\begin{tabular}{llcccc}
\toprule
\textbf{Model 1} & \textbf{Model 2} & \textbf{NDCG@3} & \textbf{NDCG@5} & \textbf{NDCG@10} & \textbf{NDCG@15} \\
\midrule
\textbf{G4.1 (I)} & \textbf{BERT-LTR-1} & \textbf{<0.001*} & \textbf{<0.001*} & \textbf{<0.001*} & \textbf{<0.001*} \\
\textbf{G4.1 (I)} & \textbf{BERT-LTR-2} & \textbf{<0.001*} & \textbf{<0.001*} & \textbf{<0.001*} & \textbf{<0.001*} \\
\textbf{G4.1 (I)} & \textbf{BERT-LTR-3} & \textbf{<0.001*} & \textbf{<0.001*} & \textbf{<0.001*} & \textbf{<0.001*} \\
\textbf{G4.1 (T)} & \textbf{BERT-LTR-1} & \textbf{<0.001*} & \textbf{<0.001*} & \textbf{<0.001*} & \textbf{<0.001*} \\
\textbf{G4.1 (T)} & \textbf{BERT-LTR-2} & \textbf{<0.001*} & \textbf{<0.001*} & \textbf{<0.001*} & \textbf{<0.001*} \\
\textbf{G4.1 (T)} & \textbf{BERT-LTR-3} & \textbf{<0.001*} & \textbf{<0.001*} & \textbf{<0.001*} & \textbf{<0.001*} \\
\textbf{G4.1 (I)} & \textbf{G4.1 (T)} & \textbf{0.036*} & \textbf{0.011*} & \textbf{0.028*} & \textbf{0.036*} \\
\textbf{G4.1 (I)} & \textbf{G4.1M (I)} & \textbf{0.036*} & \textbf{0.011*} & \textbf{0.047*} & 0.059 \\
\textbf{G4.1 (I)} & \textbf{G4.1N (I)} & \textbf{<0.001*} & \textbf{<0.001*} & \textbf{<0.001*} & \textbf{<0.001*} \\
\textbf{G4.1M (I)} & \textbf{G4.1N (I)} & \textbf{<0.001*} & \textbf{<0.001*} & \textbf{<0.001*} & \textbf{<0.001*} \\
\textbf{G4.1 (T)} & \textbf{G4.1M (T)} & \textbf{0.036*} & 0.054 & \textbf{0.006*} & \textbf{0.041*} \\
\textbf{G4.1 (T)} & \textbf{G4.1N (T)} & \textbf{<0.001*} & \textbf{<0.001*} & \textbf{<0.001*} & \textbf{<0.001*} \\
\textbf{G4.1M (T)} & \textbf{G4.1N (T)} & \textbf{<0.001*} & \textbf{<0.001*} & \textbf{<0.001*} & \textbf{<0.001*} \\
\bottomrule
\end{tabular}
\caption[\textit{Holm}-corrected $p$-values for comparing \gls{mllm}-based \gls{gui} reranking approaches (3)]{\textit{Holm}-corrected $p$-values from \textit{Wilcoxon signed-rank test} with metric \textit{\gls{ndcg}@k}. Particularly, we test whether \textit{model 1} is greater than \textit{model 2}. \textbf{Bold*} indicates significance at $\alpha=0.05$ after \textit{Holm} correction. Abbreviations: I=Image, T=Text, M=Mini, N=Nano, G4.1=GPT-4.1.}
\label{tab:gui-rerank-appendix-llm-holm-3}
\end{table}

\chapter{Appendix for Chapter \ref{cha:zs_gui_generation}}
\label{app:zs-generation}
\section{Experimental Details}

\begin{table}[!htbp]
\centering
\setlength\tabcolsep{16.5pt}
\small
\begin{tabular}{clccc}
\toprule
\textbf{Metric} & \textbf{Description} & \textbf{$n$} & \textbf{$\chi^2$} & \textbf{$p$} \\
\midrule
\textbf{A} & \textbf{Feature Completion}        & 50 & 23.81 & \textbf{<0.001} \\
\textbf{B} & \textbf{Feature Extensiveness}     & 50 & 92.49 & \textbf{<0.001} \\
\textbf{C} & \textbf{Feature Implementation}   & 50 & 21.00 & \textbf{<0.001} \\
\textbf{D} & \textbf{Information Organization} & 50 & 16.15 & \textbf{0.003}  \\
\textbf{E} & \textbf{Visual Appeal}             & 50 & 37.45 & \textbf{<0.001} \\
\textbf{F} & \textbf{Minimal Errors}            & 50 & 15.98 & \textbf{0.003}  \\
\textbf{G} & \textbf{Overall Satisfaction}      & 50 & 25.96 & \textbf{<0.001} \\
\textbf{H} & \textbf{Complete App}              & 50 & 25.51 & \textbf{<0.001} \\
\bottomrule
\end{tabular}
\caption[\textit{Friedman test} results for \gls{gui} generation (1)]
{Results of the \textit{Friedman test} on full dataset (50 \glspl{gui}), assessing overall differences between \gls{zs} prompting (\gls{zs} baselines, \gls{pd} and \gls{sc}) across metrics.
$p$-values indicate whether at least one model differs significantly from the others for a given metric.}
\label{tab:analysisA_friedman}
\end{table}

\begin{table}[!htbp]
\centering
\setlength\tabcolsep{16.5pt}
\small
\begin{tabular}{clccc}
\toprule
\textbf{Metric} & \textbf{Description} & \textbf{$n$} & \textbf{$\chi^2$} & \textbf{$p$} \\
\midrule
\textbf{A} & \textbf{Feature Completion}        & 15 & 16.30 & \textbf{0.006} \\
\textbf{B} & \textbf{Feature Extensiveness}     & 15 & 28.28 & \textbf{<0.001} \\
\textbf{C} & \textbf{Feature Implementation}   & 15 & 13.55 & \textbf{0.019} \\
\textbf{D} & \textbf{Information Organization} & 15 & 4.19  & 0.523 \\
\textbf{E} & \textbf{Visual Appeal}             & 15 & 21.96 & \textbf{<0.001} \\
\textbf{F} & \textbf{Minimal Errors}            & 15 & 8.48  & 0.132 \\
\textbf{G} & \textbf{Overall Satisfaction}      & 15 & 13.19 & \textbf{0.022} \\
\textbf{H} & \textbf{Complete App}              & 15 & 24.91 & \textbf{<0.001} \\
\bottomrule
\end{tabular}
\caption[\textit{Friedman test} results for \gls{gui} generation (2)]
{Results of the \textit{Friedman test} on \textit{cleaned dataset} (15 \glspl{gui}), assessing overall differences between \gls{zs} prompting (including \gls{ragg}) across evaluation metrics.
$p$-values indicate whether at least one model differs significantly from the others for a given metric.}
\label{tab:analysisB_friedman}
\end{table}

\begin{table}[!htbp]
\centering
\setlength\tabcolsep{3.5pt}
\small
\begin{tabular}{llcccccccc}
\toprule
\textbf{Model 1} & \textbf{Model 2} &
\textbf{A} & \textbf{B} & \textbf{C} & \textbf{D} &
\textbf{E} & \textbf{F} & \textbf{G} & \textbf{H} \\
\midrule

\textbf{ZS} & \textbf{PD1} &
1.000 & \textbf{<0.001*} & 1.000 & 0.142 & 1.000 & 0.096 & 1.000 & 0.867 \\

\textbf{ZS} & \textbf{PD2} &
1.000 & 0.317 & 1.000 & 0.142 & 0.268 & 0.490 & 1.000 & 0.316 \\

\textbf{ZS} & \textbf{SC} &
\textbf{0.003*} & \textbf{<0.001*} & \textbf{0.028*} & 0.872 &
\textbf{0.003*} & 0.639 & \textbf{0.003*} & \textbf{0.036*} \\

\textbf{ZS-CoT} & \textbf{PD1} &
1.000 & \textbf{<0.001*} & 1.000 & 0.142 & 1.000 & 0.459 & 0.703 & 0.200 \\

\textbf{ZS-CoT} & \textbf{PD2} &
1.000 & 0.100 & 1.000 & 0.142 & 1.000 & 0.639 & 1.000 & 0.927 \\

\textbf{ZS-CoT} & \textbf{SC} &
\textbf{0.002*} & \textbf{<0.001*} & \textbf{0.033*} & 0.697 &
\textbf{0.001*} & 0.490 & \textbf{<0.001*} & \textbf{0.001*} \\

\textbf{PD1} & \textbf{PD2} &
1.000 & \textbf{0.002*} & 1.000 & 0.872 & 0.274 & 0.639 & 1.000 & 0.269 \\

\textbf{PD1} & \textbf{SC} &
\textbf{0.002*} & \textbf{<0.001*} & \textbf{0.009*} & \textbf{0.016*} &
\textbf{0.004*} & \textbf{0.012*} & \textbf{0.002*} & 0.172 \\

\textbf{PD2} & \textbf{SC} &
\textbf{0.002*} & \textbf{<0.001*} & \textbf{0.002*} & \textbf{0.010*} &
\textbf{<0.001*} & \textbf{0.017*} & \textbf{<0.001*} & \textbf{<0.001*} \\

\bottomrule
\end{tabular}
\caption[\textit{Holm}-corrected post-hoc comparisons for \gls{gui} generation (1)]
{\textit{Holm}-corrected $p$-values from paired post-hoc \textit{Wilcoxon signed-rank tests} (two-sided) on full dataset (50 \glspl{gui}).
Abbreviations: \textit{A=Feature Completion, B=Feature Extensiveness, C=Feature Implementation, D=Information Organization,
E=Visual Appeal, F=Minimal Errors, G=Overall Satisfaction, H=Complete App}.
PD1=Prompt Decomposition (\gls{zs}), PD2=Prompt Decomposition (\gls{zs}-\gls{cot}) and \gls{sc} ($k=4$).
\textbf{Bold*} indicates significance at $\alpha=0.05$ after \textit{Holm} correction.}
\label{tab:analysisA_posthoc_holm}
\end{table}

\begin{table}[!htbp]
\centering
\setlength\tabcolsep{8.5pt}
\small
\begin{tabular}{llcccccc}
\toprule
\textbf{Model 1} & \textbf{Model 2} &
\textbf{A} & \textbf{B} & \textbf{C} &
\textbf{E} & \textbf{G} & \textbf{H} \\
\midrule

\textbf{ZS} & \textbf{RAGG} &
\textbf{0.020*} & \textbf{0.029*} & \textbf{0.019*} &
\textbf{0.006*} & \textbf{0.022*} & \textbf{0.006*} \\

\textbf{ZS-CoT} & \textbf{RAGG} &
\textbf{0.021*} & \textbf{0.023*} & 0.062 &
\textbf{0.018*} & 0.050 & \textbf{0.012*} \\

\textbf{PD1} & \textbf{RAGG} &
0.181 & 0.485 & \textbf{0.050*} &
0.091 & 0.335 & 0.064 \\

\textbf{PD2} & \textbf{RAGG} &
0.181 & 0.113 & 0.062 &
\textbf{0.025*} & 0.147 & \textbf{0.006*} \\

\textbf{SC} & \textbf{RAGG} &
0.793 & 0.485 & 0.414 &
0.301 & 0.572 & \textbf{0.015*} \\

\bottomrule
\end{tabular}
\caption[\textit{Holm}-corrected post-hoc comparisons for \gls{gui} generation (2)]
{\textit{Holm}-corrected $p$-values from paired post-hoc \textit{Wilcoxon signed-rank tests} (two-sided) on \textit{cleaned dataset} (15 \glspl{gui}).
Abbreviations: \textit{A=Feature Completion, B=Feature Extensiveness, C=Feature Implementation, D=Information Organization,
E=Visual Appeal, F=Minimal Errors, G=Overall Satisfaction, H=Complete App}.
PD1=Prompt Decomposition (\gls{zs}), PD2=Prompt Decomposition (\gls{zs}-\gls{cot}), \gls{sc} ($k=4$) and \gls{ragg} ($k=7$).
\textbf{Bold*} indicates significance at $\alpha=0.05$ after \textit{Holm} correction.}
\label{tab:analysisB_posthoc_holm}
\end{table}

\begin{table}[!htbp]
\centering
\setlength\tabcolsep{2.6pt}
\small
\begin{tabular}{llcccccccc}
\toprule
\textbf{Model 1} & \textbf{Model 2} &
\textbf{A} & \textbf{B} & \textbf{C} & \textbf{D} &
\textbf{E} & \textbf{F} & \textbf{G} & \textbf{H} \\
\midrule

\textbf{RAGG ($k=1$)} & \textbf{RAGG ($k=3$)} &
0.104 & 0.420 & 0.416 & 1.000 &
0.759 & 1.000 & 0.545 & 0.750 \\

\textbf{RAGG ($k=1$)} & \textbf{RAGG ($k=5$)} &
\textbf{0.026*} & 0.174 & 0.097 & 1.000 &
0.704 & 0.275 & 0.545 & 0.601 \\

\textbf{RAGG ($k=1$)} & \textbf{RAGG ($k=7$)} &
\textbf{0.011*} & 0.174 & \textbf{0.021*} & 0.481 &
\textbf{0.032*} & 0.447 & 0.159 & \textbf{0.041*} \\

\textbf{RAGG ($k=3$)} & \textbf{RAGG ($k=5$)} &
0.670 & 1.000 & 0.422 & 1.000 &
0.759 & 0.529 & 0.714 & 0.694 \\

\textbf{RAGG ($k=3$)} & \textbf{RAGG ($k=7$)} &
0.206 & 1.000 & \textbf{0.021*} & 0.217 &
0.056 & 0.529 & 0.396 & 0.098 \\

\textbf{RAGG ($k=5$)} & \textbf{RAGG ($k=7$)} &
0.206 & 1.000 & 0.086 & 0.934 &
0.385 & 1.000 & 0.714 & 0.545 \\

\bottomrule
\end{tabular}
\caption[\textit{Holm}-corrected \textit{Wilcoxon signed-rank tests} for \gls{ragg} with varying $k$]
{\textit{Holm}-corrected $p$-values from paired post-hoc \textit{Wilcoxon signed-rank tests} (two-sided) comparing \gls{ragg} with different numbers of retrieved examples ($k = 1,3,5,7$).
Abbreviations: \textit{A=Feature Completion, B=Feature Extensiveness, C=Feature Implementation, D=Information Organization,
E=Visual Appeal, F=Minimal Errors, G=Overall Satisfaction, H=Complete App}.
\textbf{Bold*} indicates significance at $\alpha=0.05$ \textit{Holm} correction.}
\label{tab:ragg_k_posthoc_holm}
\end{table}

\begin{table}[!t]
\centering
\setlength\tabcolsep{2.8pt}
\small
\begin{tabular}{llcccccccc}
\toprule
\textbf{Model 1} & \textbf{Model 2} &
\textbf{A} & \textbf{B} & \textbf{C} & \textbf{D} &
\textbf{E} & \textbf{F} & \textbf{G} & \textbf{H} \\
\midrule

\textbf{SC ($k=0$)} & \textbf{SC ($k=1$)} &
\textbf{0.044*} & \textbf{<0.001*} & \textbf{0.030*} & 1.000 &
\textbf{0.031*} & 0.593 & \textbf{0.005*} & \textbf{0.046*} \\

\textbf{SC ($k=0$)} & \textbf{SC ($k=2$)} &
\textbf{0.004*} & \textbf{<0.001*} & \textbf{0.007*} & 0.154 &
\textbf{<0.001*} & 0.695 & \textbf{<0.001*} & \textbf{<0.001*} \\

\textbf{SC ($k=0$)} & \textbf{SC ($k=3$)} &
\textbf{0.002*} & \textbf{<0.001*} & \textbf{0.030*} & 0.981 &
\textbf{<0.001*} & 0.689 & \textbf{<0.001*} & \textbf{<0.001*} \\

\textbf{SC ($k=0$)} & \textbf{SC ($k=4$)} &
\textbf{0.004*} & \textbf{<0.001*} & \textbf{0.030*} & 1.000 &
\textbf{0.004*} & 1.000 & \textbf{0.004*} & \textbf{0.031*} \\

\textbf{SC ($k=1$)} & \textbf{SC ($k=2$)} &
1.000 & \textbf{0.002*} & 1.000 & 0.211 &
\textbf{0.045*} & 1.000 & 0.126 & \textbf{0.004*} \\

\textbf{SC ($k=1$)} & \textbf{SC ($k=3$)} &
0.693 & \textbf{<0.001*} & 0.853 & 0.981 &
\textbf{0.045*} & 1.000 & 0.126 & \textbf{0.028*} \\

\textbf{SC ($k=1$)} & \textbf{SC ($k=4$)} &
1.000 & \textbf{0.002*} & 1.000 & 1.000 &
0.530 & 1.000 & 1.000 & 0.492 \\

\textbf{SC ($k=2$)} & \textbf{SC ($k=3$)} &
1.000 & 0.088 & 1.000 & 1.000 &
1.000 & 1.000 & 1.000 & 0.492 \\

\textbf{SC ($k=2$)} & \textbf{SC ($k=4$)} &
1.000 & 0.134 & 1.000 & 0.981 &
1.000 & 1.000 & 0.289 & \textbf{0.036*} \\

\textbf{SC ($k=3$)} & \textbf{SC ($k=4$)} &
1.000 & 0.768 & 1.000 & 1.000 &
1.000 & 1.000 & 0.640 & 0.248 \\

\bottomrule
\end{tabular}
\caption[\textit{Holm}-corrected \textit{Wilcoxon signed-rank tests} for \gls{sc} with varying $k$]
{\textit{Holm}-corrected $p$-values from paired post-hoc \textit{Wilcoxon signed-rank tests} (two-sided) comparing \gls{sc} with different loop counts ($k = 0,1,2,3,4$), where $k=0$ corresponds to the underlying \gls{zs} baseline (as base prototype input to the \gls{sc} approach).
Abbreviations: \textit{A=Feature Completion, B=Feature Extensiveness, C=Feature Implementation, D=Information Organization,
E=Visual Appeal, F=Minimal Errors, G=Overall Satisfaction, H=Complete App}.
\textbf{Bold*} indicates significance at $\alpha=0.05$ after \textit{Holm} correction (within each metric across pairwise $k$ comparisons).}
\label{tab:sc_k_posthoc_holm}
\end{table}

\begin{table}[H]
\centering
\setlength\tabcolsep{4pt}
\small
\begin{tabular}{lcccccccc}
\toprule
\textbf{Method} &
\textbf{A} & \textbf{B} & \textbf{C} & \textbf{D} &
\textbf{E} & \textbf{F} & \textbf{G} & \textbf{H} \\
\midrule

\textbf{ZS-CoT (+ Content)} &
\textbf{0.050*} & \textbf{0.010*} & 0.302 & 0.059 &
\textbf{0.002*} & \textbf{0.045*} & \textbf{0.004*} & \textbf{<0.001*}\\

\textbf{PD-ZS (+ Content)} &
\textbf{0.013*} & 0.065 & 0.132 & 0.468 &
\textbf{0.003*} & \textbf{0.040*} & \textbf{0.008*} & \textbf{0.005*} \\

\textbf{RAGG (+ Content)} &
0.194 & 0.282 & 0.266 & 0.779 &
0.235 & 0.228 & 0.127 & 0.325 \\

\textbf{SC (+ Content)} &
0.467 & 0.357 & 0.131 & 0.240 &
0.051 & 0.068 & 0.101 & \textbf{0.032*} \\

\bottomrule
\end{tabular}
\caption[Results for \textit{Wilcoxon signed-rank tests} on content generation]
{Paired \textit{Wilcoxon signed-rank tests} (two-sided) comparing prompting techniques (\gls{zs}-\gls{cot}, \gls{pd}-\gls{zs}, \gls{ragg} and \gls{sc}) \textit{with} and \textit{without} content generation.
Abbreviations: \textit{A=Feature Completion, B=Feature Extensiveness, C=Feature Implementation, D=Information Organization,
E=Visual Appeal, F=Minimal Errors, G=Overall Satisfaction, H=Complete App}. \gls{ragg} and \gls{sc} both use $k=3$.
\textbf{Bold*} indicates significance at $\alpha = 0.05$.}
\label{tab:content_diversification_wilcoxon}
\end{table}





\chapter{Appendix for Chapter \ref{cha:interlinking}}
\label{app:interlinking}

\section{Experimental Details}

\begin{table}[!htbp]
\centering
\small
\setlength\tabcolsep{10.7pt}
\begin{tabular}{llcccccc}
\toprule
\textbf{Model 1} & \textbf{Model 2} &
$\mathbf{n_{10}}$ & $\mathbf{n_{01}}$ & \textbf{Disc.} &
$\boldsymbol{\chi^2}$ & \textit{p} & \textit{p}$_{\text{Holm}}$ \\
\midrule
\textbf{ZS} & \textbf{FS-5}  &  6 &  5 & 11 & 0.000 & 1.000 & 1.000 \\
\textbf{ZS} & \textbf{FS-10} &  9 &  5 & 14 & 0.643 & 0.423 & 1.000 \\
\textbf{ZS} & \textbf{CoT-0.0} & 21 & 11 & 32 & 2.531 & 0.112 & 1.000 \\
\textbf{ZS} & \textbf{CoT-0.5} & 23 & 11 & 34 & 3.559 & 0.059 & 1.000 \\
\textbf{ZS} & \textbf{CoT-1.0} & 16 & 11 & 27 & 0.593 & 0.441 & 1.000 \\
\textbf{ZS} & \textbf{CoT-1.3} & 22 &  8 & 30 & 5.633 & 0.018 & 0.370 \\
\textbf{FS-5} & \textbf{FS-10} & 5 & 2 & 7 & 0.571 & 0.450 & 1.000 \\
\textbf{FS-5} & \textbf{CoT-0.0} & 21 & 12 & 33 & 1.939 & 0.164 & 1.000 \\
\textbf{FS-5} & \textbf{CoT-0.5} & 21 & 10 & 31 & 3.226 & 0.072 & 1.000 \\
\textbf{FS-5} & \textbf{CoT-1.0} & 16 & 12 & 28 & 0.321 & 0.571 & 1.000 \\
\textbf{FS-5} & \textbf{CoT-1.3} & 25 & 12 & 37 & 3.892 & 0.049 & 0.970 \\
\textbf{FS-10} & \textbf{CoT-0.0} & 21 & 15 & 36 & 0.694 & 0.405 & 1.000 \\
\textbf{FS-10} & \textbf{CoT-0.5} & 21 & 13 & 34 & 1.441 & 0.230 & 1.000 \\
\textbf{FS-10} & \textbf{CoT-1.0} & 16 & 15 & 31 & 0.000 & 1.000 & 1.000 \\
\textbf{FS-10} & \textbf{CoT-1.3} & 24 & 14 & 38 & 2.132 & 0.144 & 1.000 \\
\textbf{CoT-0.0} & \textbf{CoT-0.5} &  8 &  6 & 14 & 0.071 & 0.789 & 1.000 \\
\textbf{CoT-0.0} & \textbf{CoT-1.0} &  5 & 10 & 15 & 1.067 & 0.302 & 1.000 \\
\textbf{CoT-0.0} & \textbf{CoT-1.3} & 18 & 14 & 32 & 0.281 & 0.596 & 1.000 \\
\textbf{CoT-0.5} & \textbf{CoT-1.0} &  7 & 14 & 21 & 1.714 & 0.190 & 1.000 \\
\textbf{CoT-0.5} & \textbf{CoT-1.3} & 20 & 18 & 38 & 0.026 & 0.871 & 1.000 \\
\textbf{CoT-1.0} & \textbf{CoT-1.3} & 22 & 13 & 35 & 1.829 & 0.176 & 1.000 \\
\bottomrule
\end{tabular}

\caption[\textit{Holm}-corrected $p$-values from the \textit{McNemar tests} for comparing \gls{llm} prompting methods for \gls{us} implementation detection in \glspl{gui}]{\textit{Holm}-corrected $p$-values from the \textit{McNemar} tests (with continuity correction) on paired correctness for US detection. Particularly, we test whether \textit{Model 1} differs from \textit{Model 2}. $n_{10}$: \textit{Model 1} correct/\textit{Model 2} incorrect; $n_{01}$: \textit{Model 1} incorrect/\textit{Model 2} correct, Disc.\ $=n_{10}+n_{01}$. Abbrev.: ZS=Zero-Shot, FS=Few-Shot, CoT=Chain-of-Thought.}
\label{tab:detection-mcnemar-holm}
\end{table}

\vspace*{-\baselineskip}
\begin{table}[!t]
\centering
\small
\setlength\tabcolsep{7.5pt}
\begin{tabular}{llcccccc}
\toprule
\textbf{Model 1} & \textbf{Model 2} &
$\mathbf{\overline{F1}_1}$ & $\mathbf{\overline{F1}_2}$ &
$\boldsymbol{\Delta\overline{F1}}$ & $\mathbf{W}$ & \textit{p} & $\textit{p}_{\text{Holm}}$ \\
\midrule
\textbf{ZS$_A$}  & \textbf{ZS$_B$}   & 0.755 & 0.743 &  0.012 & 2053.5 & 0.278  & 1.000 \\
\textbf{ZS$_A$}  & \textbf{FS-5}     & 0.755 & 0.765 & -0.009 & 2483.5 & 0.558  & 1.000 \\
\textbf{ZS$_A$}  & \textbf{CoT-0.0}  & 0.755 & 0.727 &  0.028 & 1893.0 & 0.0632 & 0.505 \\
\textbf{ZS$_A$}  & \textbf{CoT-0.5}  & 0.755 & 0.688 &  0.067 & 2950.0 & <0.001 & \textbf{0.000948*} \\
\textbf{ZS$_A$}  & \textbf{CoT-1.0}  & 0.755 & 0.690 &  0.066 & 3313.0 & <0.001 & \textbf{0.00602*} \\
\textbf{ZS$_A$}  & \textbf{CoT-1.3}  & 0.755 & 0.654 &  0.102 & 4309.5 & <0.001 & \textbf{<0.001*} \\

\textbf{ZS$_B$}  & \textbf{FS-5}     & 0.743 & 0.765 & -0.022 & 3683.0 & 0.735  & 1.000 \\
\textbf{ZS$_B$}  & \textbf{CoT-0.0}  & 0.743 & 0.727 &  0.016 & 3344.5 & 0.147  & 0.736 \\
\textbf{ZS$_B$}  & \textbf{CoT-0.5}  & 0.743 & 0.688 &  0.055 & 4184.0 & 0.00152 & \textbf{0.0198*} \\
\textbf{ZS$_B$}  & \textbf{CoT-1.0}  & 0.743 & 0.690 &  0.054 & 4861.0 & 0.00408 & \textbf{0.0449*} \\
\textbf{ZS$_B$}  & \textbf{CoT-1.3}  & 0.743 & 0.654 &  0.089 & 5447.0 & <0.001 & \textbf{0.000833*} \\

\textbf{FS-5}    & \textbf{CoT-0.0}  & 0.765 & 0.727 &  0.038 & 3795.5 & 0.0214 & 0.214 \\
\textbf{FS-5}    & \textbf{CoT-0.5}  & 0.765 & 0.688 &  0.077 & 4303.0 & <0.001 & \textbf{0.00699*} \\
\textbf{FS-5}    & \textbf{CoT-1.0}  & 0.765 & 0.690 &  0.075 & 5166.0 & 0.00122 & \textbf{0.0171*} \\
\textbf{FS-5}    & \textbf{CoT-1.3}  & 0.765 & 0.654 &  0.111 & 5316.0 & <0.001 & \textbf{<0.001*} \\

\textbf{CoT-0.0} & \textbf{CoT-0.5}  & 0.727 & 0.688 &  0.039 & 2552.0 & 0.00348 & \textbf{0.0418*} \\
\textbf{CoT-0.0} & \textbf{CoT-1.0}  & 0.727 & 0.690 &  0.037 & 2625.0 & 0.0462  & 0.416 \\
\textbf{CoT-0.0} & \textbf{CoT-1.3}  & 0.727 & 0.654 &  0.073 & 3804.5 & <0.001 & \textbf{0.00416*} \\
\textbf{CoT-0.5} & \textbf{CoT-1.0}  & 0.688 & 0.690 & -0.001 & 2510.5 & 0.651  & 1.000 \\
\textbf{CoT-0.5} & \textbf{CoT-1.3}  & 0.688 & 0.654 &  0.034 & 3824.5 & 0.0859 & 0.515 \\
\textbf{CoT-1.0} & \textbf{CoT-1.3}  & 0.690 & 0.654 &  0.036 & 4412.5 & 0.0651 & 0.505 \\
\bottomrule
\end{tabular}

\caption[\textit{Holm}-corrected $p$-values from pairwise one-sided \textit{Wilcoxon signed-rank tests} on paired per-US \textit{F1} scores for \gls{gui} component matching]{\textit{Holm}-corrected $p$-values from pairwise one-sided \textit{Wilcoxon signed-rank tests} on paired per-US \textit{F1} scores, testing whether \textit{Model 1} has greater $F1$ than \textit{Model 2}. $\Delta\overline{F1}=\overline{F1}_1-\overline{F1}_2$. \textbf{Bold} values with \textbf{*} indicate significance at $\alpha=0.05$ after applying \textit{Holm} correction. Abbreviations: ZS=Zero-Shot, FS=Few-Shot and CoT=Chain-of-Thought.} 
\label{tab:matching_wilcoxon}
\end{table}







\end{document}